\documentclass{mn2e}
\usepackage{times}
\usepackage{amssymb}
\usepackage{epsfig}
\usepackage{lscape}
\usepackage{enumerate}

\begin{document}
\hsize=6truein

\title[Properties of SMGs as revealed by CANDELS]{The properties of (sub)millimetre-selected galaxies as revealed by CANDELS HST WFC3/IR imaging in GOODS-South}

\author[T.A.~Targett, et al.]
{T.A. Targett$^{1}$\thanks{Email: tat@roe.ac.uk}, J.S. Dunlop$^{1}$, M. Cirasuolo$^{1}$, R.J. McLure$^{1}$, V.A. Bruce$^{1}$, A. Fontana$^{2}$, \and A. Galametz$^{2}$, D. Paris$^{2}$, R. Dav\'{e}$^{3}$, A. Dekel$^{4}$, S.M. Faber$^{5}$, H.C. Ferguson$^{6}$, N.A. Grogin$^{6}$, \and J.S. Kartaltepe$^{7}$, D.D. Kocevski$^{5}$, A.M. Koekemoer$^{6}$, P. Kurczynski$^{8}$, K. Lai$^{5}$, J. Lotz$^{6}$\\
\footnotesize\\
$^{1}$ SUPA\thanks{Scottish Universities Physics Alliance}, 
Institute for Astronomy, University of Edinburgh, 
Royal Observatory, Edinburgh, EH9 3HJ\\
$^{2}$ INAF - Osservatorio Astronomico di Roma, Via Frascati 33, 00040 Monteporzio, Italy\\
$^{3}$ Steward Observatory, University of Arizona, 933 North Cherry Avenue, Tucson, AZ 85721, USA\\
$^{4}$ Racah Institute of Physics, The Hebrew University, Jerusalem 91904, Israel\\
$^{5}$ University of California Observatories/Lick Observatory, University of California, Santa Cruz, CA 95064, USA\\
$^{6}$ Space Telescope Science Institute, 3700 San Martin Drive, Baltimore, MD 21218, USA\\
$^{7}$ National Optical Astronomy Observatory, 950 N. Cherry Ave., Tucson, AZ, 85719, USA\\
$^{8}$ Department of Physics and Astronomy, Rutgers University, Piscataway, NJ 08854, USA}
\maketitle

\vspace*{-1.0cm}

\begin{abstract}
We have exploited the {\it Hubble Space Telescope} ({\it HST}) CANDELS $J$ and $H$-band WFC3/IR imaging to study the properties of (sub)millimetre galaxies within the GOODS-South field. After using the deep radio (VLA 1.4\,GHz) and {\it Spitzer} (IRAC 8\,$\mu$m) imaging to identify galaxy counterparts for the (sub)millimetre sources, we have then utilised the new CANDELS WFC3/IR imaging in two ways. First, the addition of new deep near-infrared photometry from both {\it HST} and (at $K$-band) the VLT to the existing GOODS-South database has enabled us to derive improved photometric redshifts and stellar masses, confirming that the (sub)millimetre sources are massive ($\langle {M}_{\star} \rangle = 2.2\times10^{11} \pm 0.2\,{\rm M_{\odot}}$) galaxies at $z \simeq 1-3$. Second, we have exploited the depth and resolution of the WFC3/IR imaging to determine the sizes and morphologies of the galaxies at rest-frame optical wavelengths $\lambda_{\rm{rest}} > 4000$\AA. Specifically, we have fitted two-dimensional axi-symmetric galaxy models to the WFC3/IR images, varying luminosity, axial ratio, half-light radius $r_{1/2}$, and S\'{e}rsic index $n$. Crucially, the wavelength and depth of the WFC3/IR imaging enables modelling of the mass-dominant galaxy, rather than the blue high surface-brightness features which often dominate optical (rest-frame ultraviolet) images of (sub)millimetre galaxies, and can confuse visual morphological classification. As a result of this analysis we find that $>95$\% of the rest-frame optical light in almost all of the (sub)millimetre galaxies is well-described by either a single exponential disk ($n \simeq 1$), or a multiple-component system in which the dominant constituent is disk-like. We demonstrate that this conclusion is completely consistent with the results of recent high-quality ground-based $K$-band imaging sampling even longer rest-frame wavelengths, and explain why. These massive disk galaxies are reasonably extended ($\langle r_{1/2} \rangle = 4.5 \pm 0.5$\,kpc; median $r_{1/2}=4.0$\,kpc), consistent with the sizes of other massive star-forming disks at $z \simeq 2$. In many cases we find evidence of blue clumps within the sources, with the mass-dominant disk component becoming more significant at longer wavelengths. Finally, only a minority of the sources show evidence for a major galaxy-galaxy interaction. Taken together, these results support the view that most (sub)millimetre galaxies at $z \simeq 2$ are simply the most extreme examples of normal star-forming galaxies at that era. Interestingly, the only two bulge-dominated galaxies are also the two lowest-redshift sources in the sample ($z \simeq 1$), a result which may reflect the structural evolution of high-mass galaxies in general.
\end{abstract}


\newpage

\section{INTRODUCTION}

Since (sub)millimetre galaxies were first discovered with the SCUBA camera on the James Clerk Maxwell Telescope in the late 1990s, it has been known that they are violently star-forming galaxies at high redshifts. Indeed, while effective sub-mm/mm imaging took a long time to arrive, the strongly negative k-correction provided by the spectral energy distribution (SED) of dust-enshrouded star formation (i.e. a modified black body -- e.g. Hughes et al. 1993; Blain et al. 2002) resulted in the first sub-mm galaxies being discovered at rather high redshifts (i.e. $z \simeq 2 - 4$; Hughes et al. 1998). Interestingly, this meant that the supporting multi-frequency data available at the time was of insufficient quality to adequately reveal the true nature of these sources, and it has taken over a decade of effort with ever-improving radio surveys, optical-infrared imaging/spectroscopy with ground-based 8-10\,m telescopes, and deep {\it Spitzer} and {\it Hubble Space Telescope} ({\it HST}) imaging to begin to clarify the physical nature of (sub)millimetre galaxies, and their role in the process of galaxy formation.

This is still a work in progress. For example, it is only very recently that a spectroscopic redshift was finally measured for the brightest sub-mm galaxy discovered towards the HDF in the first unbiased sub-mm survey at 850\,$\mu$m (HDF850.1; Hughes et al. 1998; Downes et al. 1999; Dunlop et al. 2004; Cowie et al. 2009) revealing it to lie at $z = 5.183$ (Walter et al. 2012). This measurement, based on CO lines, has been made possible by recent advances in sub-mm/mm spectroscopy (e.g. Riechers et al. 2010, 2012, and Hodge et al. 2012 for morphology), a technique which should soon enable complete spectroscopic redshift distributions to be established for the (now substantial) samples of (sub)millimetre sources which have been uncovered by continuum surveys with SCUBA, MAMBO, LABOCA and AzTEC (e.g. Coppin et al. 2006; Greve et al. 2008; Weiss et al. 2009; Austermann et al. 2010). With the advent of the Atacama Large Millimetre Array (ALMA), high-resolution sub-mm/mm imaging is set to become straightforward, and detailed sub-mm/mm spectroscopy will soon be routine.

However, as we now enter a new era in sub-mm/mm observational astronomy, it is important to note that major advances in our knowledge of (sub)millimetre galaxies have already been gained from studies at optical-mid-infrared wavelengths. For example, aided by the improved positional accuracy provided by deep VLA 1.4-GHz imaging, it has already proved possible to identify optical/near-infrared counterparts for $\simeq 80$\% of the sub-mm/mm galaxies in bright samples, and hence to determine the basic form of their redshift distribution. These studies indicate that (sub)millimetre galaxies peak in number density around $z \simeq 2$ (e.g. Chapman et al. 2005; Wardlow et al. 2011; Micha{\l}owski et al. 2012b; Schael et al. 2013), although there is some evidence that the brightest (sub)millimetre galaxies lie predominantly at $z > 3$ (e.g. Smolcic et al. 2012; Koprowski et al. 2013), and the full effects of large-scale structure have yet to be properly clarified (e.g. Micha{\l}owski et al. 2012b; Scott et al. 2012; Shimizu, Yoshida \& Okamoto 2012).

Moreover, while one might argue that redshift measurements for (sub)millimetre sources may increasingly be made at sub-mm/mm wavelengths, there are some crucial properties of (sub)millimetre-selected galaxies which can {\it only} be determined from deep optical--mid-infrared data. These include their stellar masses, formation history, and the spatial distribution of their existing stellar populations. Measurement of these physical quantities is of crucial importance if we are to understand the nature of (sub)millimetre galaxies as a function of luminosity and redshift, and place them in the context of the general galaxy population at comparable redshift/stellar-mass. Importantly, this is now an achievable goal, as the near/mid-infrared study of more moderate star-forming galaxies has now reached out to comparable redshifts (e.g. Daddi et al. 2007; Elbaz et al. 2011) and the construction of significant and complete mass-selected samples down to $M_{\star} \simeq 5 \times 10^{10}\,{\rm M_{\odot}}$ is now possible out to $z \simeq 3$ (e.g. Bruce et al. 2012).

The most important recent advance in this context is the provision of deep, high-quality, near-infrared imaging provided by WFC3/IR on {\it HST}. In particular, the CANDELS survey (Grogin et al. 2011; Koekemoer et al. 2011) is providing homogeneous WFC3/IR $J_{125}$ and $H_{160}$ imaging of five fields which overlap significantly with existing sub-mm/mm surveys. Such imaging is of general importance for the study of galaxies at $z \simeq 1-3$, as it offers high quality {\it rest-frame optical} imaging at these redshifts where, until recently, our knowledge of galaxy morphologies and colours was largely based on the rest-frame ultraviolet. However, it is arguably of special importance for (sub)millimetre galaxies, where much of the rest-frame UV light may be obscured by dust, and hence morphological k-corrections can be potentially dramatic.

In this paper we exploit the CANDELS data in the GOODS-South field to attempt to improve our knowledge of the physical properties of (sub)millimetre galaxies. Of the five CANDELS fields, GOODS-South currently offers the best overlap between WFC3/IR imaging and existing samples of sub-mm/mm-selected sources (although this situation will soon change as a result of the SCUBA2 Cosmology Redshift Survey\footnote{www.jach.hawaii.edu/JCMT/surveys/Cosmology.html}), with the added advantages of very deep {\it Spitzer}, {\it HST} ACS, and VLT imaging, and high-density spectroscopy (e.g. Vanzella et al. 2008). Our aim was to utilise the WFC3/IR imaging (Grogin et al. 2011) to first improve the measurement of the photometric redshifts and stellar masses of the known LABOCA/AzTEC sources in this field (Weiss et al. 2009; Scott et al. 2010; Wardlow et al. 2011; Yun et al. 2012), and then to determine their rest-frame optical morphologies. Crucially, because of the complete, homogeneous, and unbiased nature of the CANDELS imaging, we have also been able to compare the rest-frame optical morphologies of the (sub)millimetre galaxies to those displayed by the general population of galaxies at comparable redshift and stellar mass (Bruce et al. 2012).

This latter point is one of the key advantages offered by CANDELS, alongside, of course, the major improvement in depth and angular-resolution offered by WFC3/IR over previous near-infrared imagers. To date, deep, near-infrared imaging has only been obtained for a small number of (sub)millimetre galaxies (e.g. Smail et al. 2003), and generally with inadequate resolution for the reliable extraction of basic morphological parameters. In an attempt to exploit the resolution advantage offered by {\it HST}, Swinbank et al. (2010) reported the results of {\it HST} NICMOS+ACS imaging of a sub-sample of (sub)millimetre galaxies. While S\'{e}rsic indices for individual galaxies were not reported, median values of $n_{i}=1.8 \pm 1.0$ and $n_{H}=1.4 \pm 0.8$ were given, and additional measurements of S\'{e}rsic index based on a stacked NICMOS imaging found $n_{H}=2.0\pm0.5$. Combining these results with measurements of half-light radii and asymmetry, Swinbank et al. (2010) suggested that the stellar structure of (sub)millimetre galaxies was best described by a spheroid/elliptical galaxy light distribution, despite the low median S\'{e}rsic index. More recently, Targett et al. (2011) analysed deep, high-quality ground-based $K$-band images of complete subsamples of (sub)millimetre galaxies at redshifts $z \simeq 2-3$. While the use of $K$-band ensures imaging longward of the 4000\AA/Balmer break out to $z \simeq 4$, the ground-based images have poorer resolution than {\it HST} data. Nevertheless, Targett et al. (2011) found the galaxies they studied to be reasonably extended ($\langle r_{1/2} \rangle = 3.4 \pm 0.3$\,kpc; median $r_{1/2}=3.1$\,kpc) exponential disks ($\langle n \rangle = 1.44 \pm 0.16$; median $n = 1.08$). Thus, while there is improving agreement over the basic sizes of (sub)millimetre galaxies, there remains controversy over their morphologies.

There is also an ongoing debate over the stellar masses of (sub)millimetre galaxies, and whether or not their extreme star-formation rates are triggered predominately by major galaxy-galaxy mergers. As discussed in Dunlop (2011) and Micha{\l}owski et al. (2012a), this strikes to the heart of the nature of (sub)millimetre galaxies, with important implications for refining current models of galaxy formation. For a long time it was believed that (sub)millimetre galaxies were the consequence of major galaxy mergers, in part because the brightest ULIRGs uncovered by IRAS in the local Universe appeared to be merger driven (e.g. Sanders \& Mirabel 1996). However, over recent years it has become clear that the specific star-formation rate ($sSFR$=star-formation rate/stellar mass) of the general star-forming galaxy population is a factor $\simeq 20$ higher at $z \simeq 2$ than at the present day (Daddi et al. 2007; Elbaz et al. 2011; Karim et al. 2011), with $sSFR \simeq 2-4\,{\rm Gyr^{-1}}$ (Gonzalez et al. 2010). Consequently, provided they have stellar masses $M_{\star} > 10^{11}\,{\rm M_{\odot}}$, it can be argued that the extreme star-formation rates inferred for (sub)millimetre galaxies are not ``unexpected'' at $z \simeq 2$, and that the situation has been confused by their relatively early discovery. This realization has focussed increased attention on the accurate determination of the stellar masses of (sub)millimetre galaxies (e.g. Dye et al. 2008; Micha{\l}owski et al. 2010a,b; Hainline et al. 2011; Micha{\l}owski et al. 2012a) and the evidence for galaxy interactions (e.g. Tacconi et al. 2008; Engel et al. 2010; Alaghband-Zadeh et al. 2012). 

Importantly, some theoretical models appear to require galaxy interactions, and even the additional boost provided by a top-heavy stellar initial mass function (IMF) to produce sufficient numbers of high-redshift (sub)millimetre sources (Baugh et al. 2005), but this may simply reflect an underlying inability of some semi-analytic models to produce adequate numbers of high-mass galaxies at $z \simeq 2-3$ (Swinbank et al. 2008). Models involving AGN feedback have improved in this respect (e.g. Bower et al. 2006; Croton et al. 2006), and indeed some authors have argued that the (sub)millimetre galaxy population can be reconciled with the high-mass end of the ``main sequence'' of galaxy formation at $z \simeq 2 - 3$, and are fed by smooth infall rather than major mergers (e.g. Dav\'{e} et al. 2010, and see also Finlator et al. 2006; Fardal et al. 2007; Dekel et al. 2009). Moreover, it has been argued that major mergers are not sufficiently common in the redshift range of interest to explain the (sub)millimetre population, and that in any case the ability of galaxy mergers to significantly enhance the already high specific star-formation rates of galaxies at $z \simeq 2 - 3$ is relatively modest, $\simeq 10-25$\% (e.g. Cen 2012, Kaviraj et al. 2012). Of course the true situation may be complex, and recently arguments have been advanced that the (sub)millimetre galaxy population may contain a mix of both massive ``normal'' star-forming galaxies and a subset of extreme objects boosted by interactions (e.g. Hayward et al. 2012).

Clearly, therefore, further observational clarification of the masses and morphologies of (sub)millimetre galaxies is of value, especially if the results can be robustly compared with the properties of the general galaxy population. This is facilitated here by the work of Bruce et al. (2012), who utilised comparable CANDELS imaging to study the rest-frame optical morphologies of $\simeq 200$ massive ($M_{\star}\ge10^{11}\,{\rm M_{\odot}}$) galaxies in the the redshift range $1<z<3$. This work has shown that the CANDELS ``Wide'' data are of sufficient quality and depth to enable the robust measurement of the morphologies of such massive galaxies out to $z \simeq 3$, and has revealed the emergence of a predominantly disk-dominated population by $z \ge 2$. This pattern is confirmed at somewhat lower stellar masses by Law et al. (2012a) who, from an {\it HST} study of 306 star-forming galaxies at $1.5 < z < 3.6$ with $M_{\star} = 10^{9}-10^{11}\,{\rm M_{\odot}}$, found that star-forming galaxies at these redshifts are best described by an exponential (disky) surface brightness profile. However, while there appears to be a growing consensus that massive galaxies at $z > 2$ are disk-like, there is an interesting debate over how these high-redshift disks compare to their local counterparts, since they appear to display a peaked axial-ratio distribution (e.g. Law et al. 2012a; Bruce et al. 2012), and have substantially higher velocity dispersions than low-redshift disk galaxies (e.g. F\"{o}rster-Schreiber et al. 2009; Law et al. 2012b). There is also considerable debate over the prevalence and potential importance (or otherwise) of clumps in these high-redshift star-forming disks (e.g. Wuyts et al. 2012), and the extent to which they are predicted by current models of galaxy formation (e.g. Ceverino et al. 2012; Genel et al. 2012). Whatever their origin and importance, it is clear that such high-surface brightness, and generally blue (e.g. Guo et al. 2012) clumps mitigate against the effective morphological classification of high-redshift star-forming galaxies from optical imaging (which samples the rest-frame near-ultraviolet at $z > 1.5$).

It is clearly important to establish where (sub)millimetre galaxies fit into this picture. For example, as star-forming galaxies they might be expected to be extended disks, and yet, based on their high stellar masses it is hard to avoid the conclusion that they are the progenitors of present-day giant elliptical galaxies. Here we attempt to move this subject forward using the new WFC3/IR CANDELS imaging to study the properties of the AzTEC and LABOCA-selected galaxies in the GOODS-South field. For the first time, we can study the rest-frame optical morphologies of (sub)millimetre galaxies in detail, due to the combination of high angular-resolution (0.1\,arcsec) and depth ($H_{160} = 26.5$; AB mag, 5-$\sigma$) provided by the CANDELS WFC3/IR imaging. In addition, the existing rich multi-frequency data in this well-established field provides a significant number of spectroscopic redshifts, and enables the most accurate photometric redshifts achieved for (sub)millimetre galaxies to date (with consequently reduced uncertainties in derived stellar masses). We note that it is important to distinguish between the evidence for or against merger-driven starbursts and the nature of the dominant underlying galaxy; in this paper our focus is on determining the latter and comparing the stellar masses and morphologies of (sub)millimetre galaxies to the properties of the general galaxy population at $z \simeq 2 -3$. We do also consider the imaging evidence for major galaxy-galaxy interactions, but such work is arguably better pursued with the velocity information gleaned from optical and/or mm spectroscopy. It does, however, need to be performed in an unbiased manner including both (sub)millimetre bright and non (sub)millimetre-detected massive galaxies at $z \simeq 2$, before definitive conclusions can be drawn regarding the importance of major mergers in producing the (sub)millimetre galaxy population.

This paper is structured as follows. In Section 2 we summarise the available (sub)millimetre survey data within GOODS-South, the new CANDELS {\it HST} imaging, and the additional key supporting data available within this field. Next, in Section 3, we explain how we determined the radio and IRAC IDs, and hence WFC3/IR $H$-band galaxy counterparts for the (sub)millimetre sources. Then, in Section 4 we describe the two-dimensional modelling used to extract the basic morphological properties of the galaxies, and the SED fitting to the multi-frequency data used to determine redshifts and stellar masses. In Section 5 we present the results of our analysis, and also place our findings in the context of studies of other galaxy populations at both high and low redshift. Our main conclusions are summarised in Section 6. Throughout we quote magnitudes in the AB system (Oke 1974), and assume a cosmological model with $H_0 = 70\,{\rm km s^{-1} Mpc^{-1}}$, $\Omega_{\Lambda} = 0.7$, and $\Omega_m = 0.3$.

\begin{table*}
 \begin{center}
  \caption{Positional information on the 24-source combined AzTEC+LABOCA GOODS-South (sub)millimetre galaxy sample utilised in this paper, and the adopted galaxy counterparts in the {\it HST} CANDELS $H_{160}$ imaging. The first column gives the name of the AzTEC source from Scott et al. (2010), with the next two columns giving the AzTEC position (AzTEC.GS19 is tabulated twice here because it has two alternative IDs, as explained in Section 3). The names of sources with statistically secure ($P < 0.1$) counterparts are highlighted in bold. Column 4 then gives the source name from the LESS survey (Weiss et al. 2009), where here the name itself contains the original LABOCA survey position. Column 5 gives the de-boosted flux densities of the (sub)milimetre sources at 1.1\,mm (AzTEC) and 870\,${\rm \mu m}$ (LABOCA). Columns 6 and 7 then give the positions of the $H_{160}$ galaxy counterpart we have selected for morphological study. These galaxy identifications were selected as described in Section 3. Column 8 then gives the probability that the adopted galaxy counterpart is merely a chance association, while column 9 indicates whether the galaxy identification was selected on the basis of the radio or {\it Spitzer} imaging, and the source of the quoted $P$ value; (a) our identification process, or (b) from Yun et al. (2012).}
  \begin{tabular}{lllllllll}
\hline\hline
AzTEC             & RA          & Dec.          & LABOCA                    & Flux        & CANDELS      & CANDELS      & $P$   & Type\\
ID                & (J2000)     & (J2000)       & ID                        & (mJy)       & RA  (J2000)  & Dec (J2000)  &       &\\
\hline
\bf{AzTEC.GS03}   & 03:32:47.86 & $-$27:54:19.3 & \bf{LESSJ033248.1-275414} & 4.8$\pm$0.6 & 03:32:47.992 & -27:54:16.42 & 0.017 & a - 1.4\,GHz\\
\bf{AzTEC.GS06}   & 03:32:25.73 & $-$27:52:19.4 & \bf{LESSJ033225.7-275228} & 3.6$\pm$0.5 & 03:32:25.258 & -27:52:30.51 & 0.062 & a - 1.4\,GHz\\
\bf{AzTEC.GS08}   & 03:32:05.12 & $-$27:46:45.8 & \bf{LESSJ033205.1-274652} & 3.4$\pm$0.6 & 03:32:04.873 & -27:46:47.37 & 0.015 & a - 1.4\,GHz\\
\bf{AzTEC.GS11}   & 03:32:15.79 & $-$27:50:36.8 &                           & 3.3$\pm$0.6 & 03:32:15.319 & -27:50:37.12 & 0.067 & a - 1.4\,GHz\\
\bf{AzTEC.GS12}   & 03:32:29.13 & $-$27:56:13.8 & \bf{LESSJ033229.3-275619} & 5.1$\pm$1.4 & 03:32:29.290 & -27:56:19.47 & 0.032 & b - 1.4\,GHz\\
\bf{AzTEC.GS13}   & 03:32:11.91 & $-$27:46:16.9 &                           & 3.1$\pm$0.6 & 03:32:11.908 & -27:46:15.51 & 0.009 & b - 1.4\,GHz\\
\bf{AzTEC.GS16}   & 03:32:37.67 & $-$27:44:01.8 &                           & 2.7$\pm$0.5 & 03:32:38.006 & -27:44:00.57 & 0.066 & a - 1.4\,GHz\\
\bf{AzTEC.GS17}   & 03:32:22.31 & $-$27:48:16.4 &                           & 2.9$\pm$0.6 & 03:32:22.566 & -27:48:14.83 & 0.026 & a/b - 8/24\,${\rm \mu m}$\\
\bf{AzTEC.GS18}   & 03:32:43.58 & $-$27:46:36.9 & \bf{LESSJ033243.6-274644} & 3.1$\pm$0.6 & 03:32:43.536 & -27:46:38.90 & 0.035 & b - 1.4\,GHz\\
\bf{AzTEC.GS19-1} & 03:32:23.21 & $-$27:41:28.8 &                           & 2.6$\pm$0.5 & 03:32:22.879 & -27:41:24.96 & 0.074 & a - 1.4\,GHz\\
\bf{AzTEC.GS19-2} & 03:32:23.21 & $-$27:41:28.8 &                           & 2.6$\pm$0.5 & 03:32:22.708 & -27:41:26.39 & 0.088 & a - 1.4\,GHz\\
\bf{AzTEC.GS21}   & 03:32:47.60 & $-$27:44:49.3 &                           & 2.7$\pm$0.6 & 03:32:47.592 & -27:44:52.26 & 0.026 & a - 1.4\,GHz\\
AzTEC.GS22        & 03:32:12.60 & $-$27:42:57.9 &                           & 2.1$\pm$0.6 & 03:32:12.545 & -27:43:05.99 & 0.116 & b - 1.4\,GHz\\
\bf{AzTEC.GS23}   & 03:32:21.37 & $-$27:56:28.1 & \bf{LESSJ033221.3-275623} & 4.7$\pm$1.4 & 03:32:21.574 & -27:56:23.94 & 0.054 & a - 1.4\,GHz\\
AzTEC.GS24        & 03:32:34.76 & $-$27:49:43.1 &                           & 2.3$\pm$0.6 & 03:32:34.163 & -27:49:39.52 & 0.141 & b - 1.4\,GHz\\
AzTEC.GS26        & 03:32:15.79 & $-$27:43:36.6 &                           & 2.2$\pm$0.5 & 03:32:16.281 & -27:43:43.41 & 0.624 & a - 8\,${\rm \mu m}$\\
AzTEC.GS27        & 03:32:42.42 & $-$27:41:51.9 &                           & 2.2$\pm$0.6 & 03:32:41.632 & -27:41:51.45 & 0.623 & a - 8\,${\rm \mu m}$\\
AzTEC.GS28        & 03:32:42.71 & $-$27:52:06.8 &                           & 2.1$\pm$0.5 & 03:32:41.853 & -27:52:02.47 & 0.734 & a - 8\,${\rm \mu m}$\\
\bf{AzTEC.GS30}   & 03:32:20.94 & $-$27:42:40.8 &                           & 1.8$\pm$0.5 & 03:32:20.658 & -27:42:34.40 & 0.082 & b - 1.4\,GHz\\
AzTEC.GS34        & 03:32:29.77 & $-$27:43:13.1 &                           & 1.7$\pm$0.5 & 03:32:29.479 & -27:43:22.10 & 0.171 & a - 1.4\,GHz\\
\bf{AzTEC.GS35}   & 03:32:26.90 & $-$27:40:52.1 &                           & 2.1$\pm$0.6 & 03:32:27.181 & -27:40:51.42 & 0.008 & b - 1.4\,GHz\\
\bf{AzTEC.GS38}   & 03:32:09.26 & $-$27:42:45.5 &                           & 1.7$\pm$0.6 & 03:32:09.705 & -27:42:48.10 & 0.015 & a - 1.4\,GHz\\
                  &             &               & \bf{LESSJ033217.6-275230} & 6.3$\pm$1.3 & 03:32:17.617 & -27:52:28.52 & 0.016 & a - 8\,${\rm \mu m}$\\
                  &             &               & \bf{LESSJ033219.0-275219} & 9.1$\pm$1.2 & 03:32:19.032 & -27:52:13.97 & 0.001 & b - 1.4\,GHz\\
                  &             &               & \bf{LESSJ033243.3-275517} & 5.2$\pm$1.4 & 03:32:43.179 & -27:55:14.49 & 0.019 & a - 1.4\,GHz\\
\hline
  \end{tabular}
 \end{center}
\end{table*}

\section{DATA}

\subsection{(Sub)millimetre data}
The full 30$\times$30\,arcmin ECDFS has been mapped at a wavelength of 870\,${\rm \mu m}$ by the LABOCA ECDFS (sub)millimetre Survey (LESS) (Weiss et al. 2009). The 12-m diameter of the APEX telescope (Gusten et al. 2006) delivers a 19.2\,arcsec FWHM beam at this wavelength. The LESS image has a uniform depth of $\sigma_{870} \simeq 1.2 {\rm \, mJy\,beam^{-1}}$ (as measured in the map, and hence including confusion noise). 

A 512\,arcmin$^{2}$ 1.1\,mm map of GOODS-South has now also been obtained with the AzTEC camera on the ASTE, which delivers a 30\,arcsec FWHM beam at this wavelength (Scott et al. 2010). Within the deepest central 270\,arcmin$^{2}$ region of the AzTEC map, the rms noise ranges from 0.48 to 0.73${\rm \, mJy\,beam^{-1}}$, making this GOODS-South millimetre map the deepest contiguous region ever mapped at 1.1\,mm.
 
The (sub)millimetre galaxy sample considered here constitutes the subset of published $870\, {\rm \mu m}$ (LABOCA) and 1.1\,mm (AzTEC) sources which lie within the areas covered by the {\it HST} WFC3/IR imaging of the CANDELS deep (10-orbit depth), wide (2-orbit depth), and Early Release Science (ERS, 2-orbit depth) surveys (total area $\simeq$164\,arcmin$^{2}$) centred on the southern field of the Great Observatories Origins Deep Survey (GOODS; Dickinson et al. 2004) at ${\rm RA\, 03^h \, 32^m \, 30^s}$, ${\rm Dec\, -28^{\circ}\, 48^{\prime}}$ (J2000).

Our final GOODS-South (sub)millimetre sample comprises the 24 sources listed in Table 1 (AzTEC.GS19 is tabulated twice here because it has two alternative IDs, as explained in Section 3). Because the AzTEC imaging is slightly deeper, only 3 of the sources are uniquely supplied by LESS (as indicated in Table 1). There are, however, 6 sources in common, and here the LESS imaging is still of value as it supplies a somewhat better position due to the smaller beam size of the shorter-wavelength survey.
 
\subsection{CANDELS data}
The Cosmic Assembly Near-IR Deep Extragalactic Legacy Survey (CANDELS; Grogin et al. 2011; Koekemoer et al. 2011) is an {\it HST} Multi-Cycle Treasury Program designed to image portions of five legacy fields (GOODS-N, GOODS-S, COSMOS, UDS, and EGS) with the Wide Field Camera 3 (WFC3/IR) in the near-infrared $J_{125}$ and $H_{160}$ bands. The CANDELS data have been reduced and drizzled to a 0.06\,arcsec pixel scale (Koekemoer et al. 2011) to create the high-resolution (FWHM $\simeq 0.2$\, arcsec) mosaics. Despite the variation in exposure times across the field, even the shallowest region reaches a 5-$\sigma$ AB limiting magnitude of 26.5 in the $H_{160}$-band, two magnitudes deeper than the existing ground-based ISAAC coverage at the same wavelength (Retzlaff et al. 2010).

\subsection{Supporting multi-frequency data}

\subsubsection{Radio: VLA 1.4-GHz imaging}
As demonstrated by the follow-up of the sub-mm/mm surveys undertaken with SCUBA, IRAM, LABOCA and AzTEC, very deep VLA imaging is necessary (and frequently sufficient) for the successful identification of the galaxy counterparts of sources detected with the large beams delivered by single-dish far-infrared/(sub)millimetre facilities (e.g. Ivison et al. 2002, 2007; Micha{\l}owski et al. 2012b; Yun et al. 2012). This works for three reasons. First, star-forming galaxies produce copious quantities of synchrotron emission. Second, even in the deepest available radio maps, 1.4\,GHz sources still have sufficiently low surface-density that positional coincidences within a `reasonable' search radius are generally statistically rare. Third, if a robust radio counterpart is found, the $\simeq$1\,arcsec positional accuracy provided by the VLA at $1.4\,$GHz essentially always yields an unambiguous optical/infrared galaxy counterpart for further study.

Very deep ($\sigma_{1.4} \simeq 7.5\,{\rm \mu Jy}$), high-resolution 1.4\,GHz imaging of GOODS-South is available at the centre of the ECDFS as described by Kellermann et al. (2008) and Miller et al. (2008). We adopt the radio catalogue from Dunlop et al. (2010) who, using the techniques described by Ibar et al. (2009), created a radio catalogue down to a 4-$\sigma$ limit of 30\,${\rm \mu Jy}$ (at which depth the cumulative source density on the sky is $\rm \simeq 0.8\, arcmin^{-2}$). This catalogue was then searched for potential radio counterparts to the sub-mm/mm sources using the method described in Section 3. We note that even deeper VLA 1.4\,GHz imaging in GOODS-South has recently been presented by Yun et al. (2012), who have also searched for the counterparts to the AzTEC sources discussed here (using the same statistical technique). Where appropriate, we compare and cross-check our galaxy identifications in Section 3.

\subsubsection{Optical: HST ACS imaging}
Deep optical imaging over the GOODS-South field was obtained (prior to CANDELS) with the Advanced Camera for Surveys (ACS) on board {\it HST} in 4 different filters: $F435W\,(B_{435})$, $F606W\,(V_{606})$, $F775W\,(i_{775})$ and $F850LP\,(z_{850})$. These data were taken as part of GOODS (Giavalisco et al. 2004) and reach 5-$\sigma$ limiting (AB) magnitudes (within a 1\,arcsec diameter aperture) of 26.9, 26.9, 26.1, and 25.8, respectively.

\subsubsection{Ground-based $K_s$-band VLT imaging}
Ground-based near-infrared imaging of almost all of the CANDELS GOODS-South field was completed several years ago with the ISAAC camera on the VLT (Retzlaff et al. 2010). These data cover $\simeq$143\,arcmin$^{2}$ within the $\simeq$164\,arcmin$^{2}$ CANDELS GOODS-South field. Although ground-based, these data remain of value because the {\it HST} cannot image longward of the $H$-band. In addition, the image quality of the ISAAC data is excellent, with a median FWHM PSF value of $\simeq 0.5$\,arcsec (variable from $\simeq 0.37$ to $\simeq 0.7$\,arcsec across the field; Bouwens et al. 2008). Partly as a result of this good image quality, the point-source sensitivity of this imaging is deep, reaching a 5-$\sigma$ detection level of $K_s \simeq 24.7$ (AB mag). 

As part of the larger CANDELS project, even deeper ground-based $K_s$-band imaging of GOODS-South (covering $\simeq 88$\,arcmin$^{2}$, within the CANDELS imaging region) has recently been obtained with the HAWK-I imager on the VLT. These data have been reduced and processed as described in Fontana et al. (2012). Again taken in excellent seeing ($\simeq 0.45$\,arcsec), this imaging reaches a 5-$\sigma$ limiting point-source depth of $K_s \simeq 26.0$\,(AB mag), comparable to the CANDELS {\it HST} $H_{160}$-band data.

\subsubsection{Mid-infrared: Spitzer imaging}
Again as part of GOODS, ultra-deep {\it Spitzer} imaging with the Infrared Array Camera (IRAC: Fazio et al. 2004) has been obtained over the whole of the CANDELS GOODS-South field, in all 4 IRAC channels (3.6, 4.5, 5.6, and 8.0\,${\rm \mu m}$). The IRAC 5-$\sigma$ detection limits are $S_{3.6} \simeq 25.9$, $S_{4.5} \simeq 25.5$, $S_{5.6} \simeq 23.3$, $S_{8.0} \simeq 22.9$ (AB mag). 

The 24\,${\rm \mu m}$ {\it Spitzer} MIPS data originally obtained as part of GOODS has been augmented and incorporated within the {\it Spitzer} Far-Infrared Deep Extragalactic Legacy (FIDEL)\footnote{PI M. Dickinson, see {\tt http://www.noao.edu/noao/fidel/}} survey (Magnelli et al. 2009), and reach 5-$\sigma$ detection limits of $S_{24} \simeq 30\,{\rm \mu Jy}$.

\begin{center}
\begin{figure*}
\begin{tabular}{lccc}
\epsfig{file=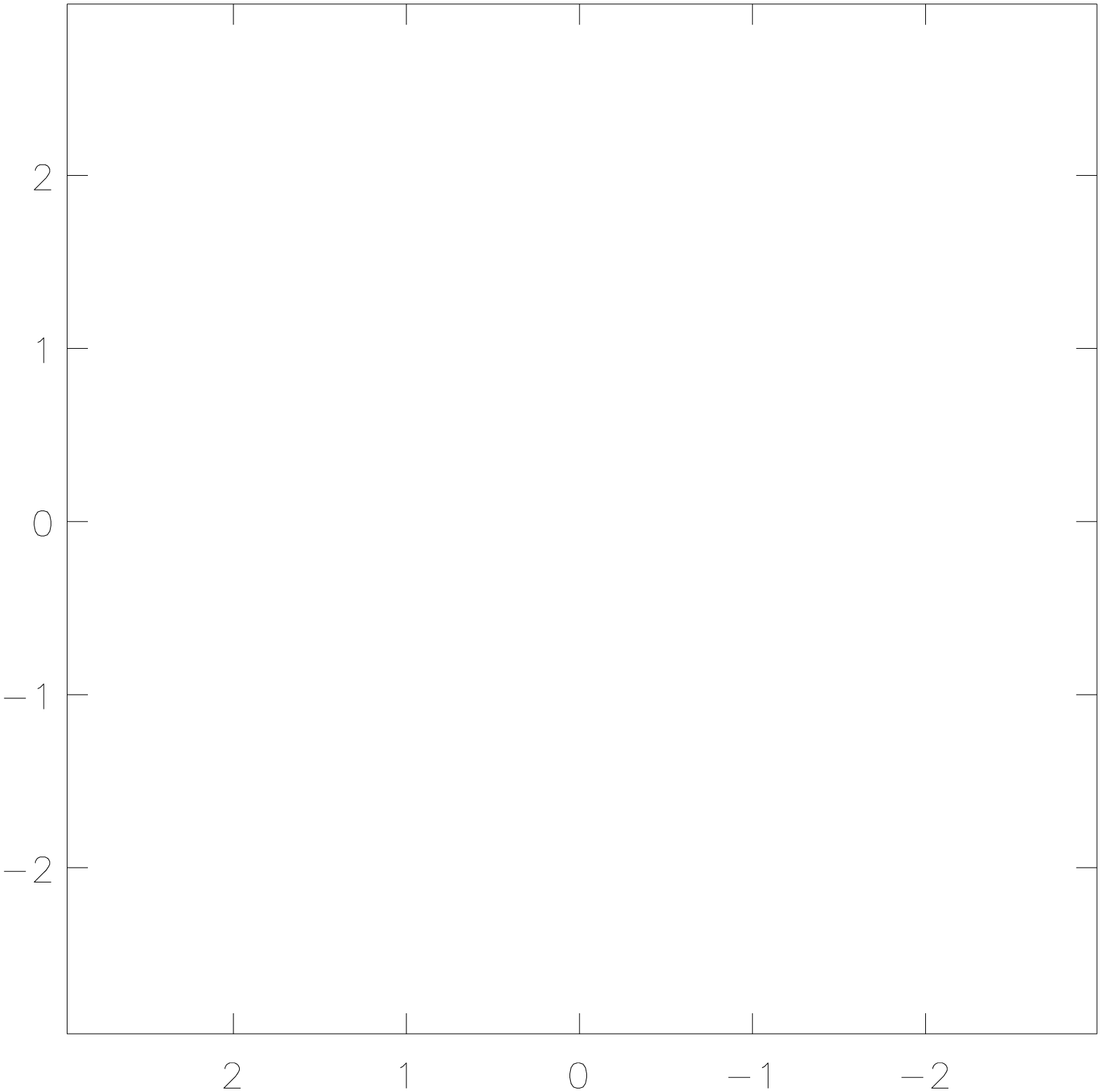,width=0.228\textwidth}
\hspace{-4.175cm}
\epsfig{file=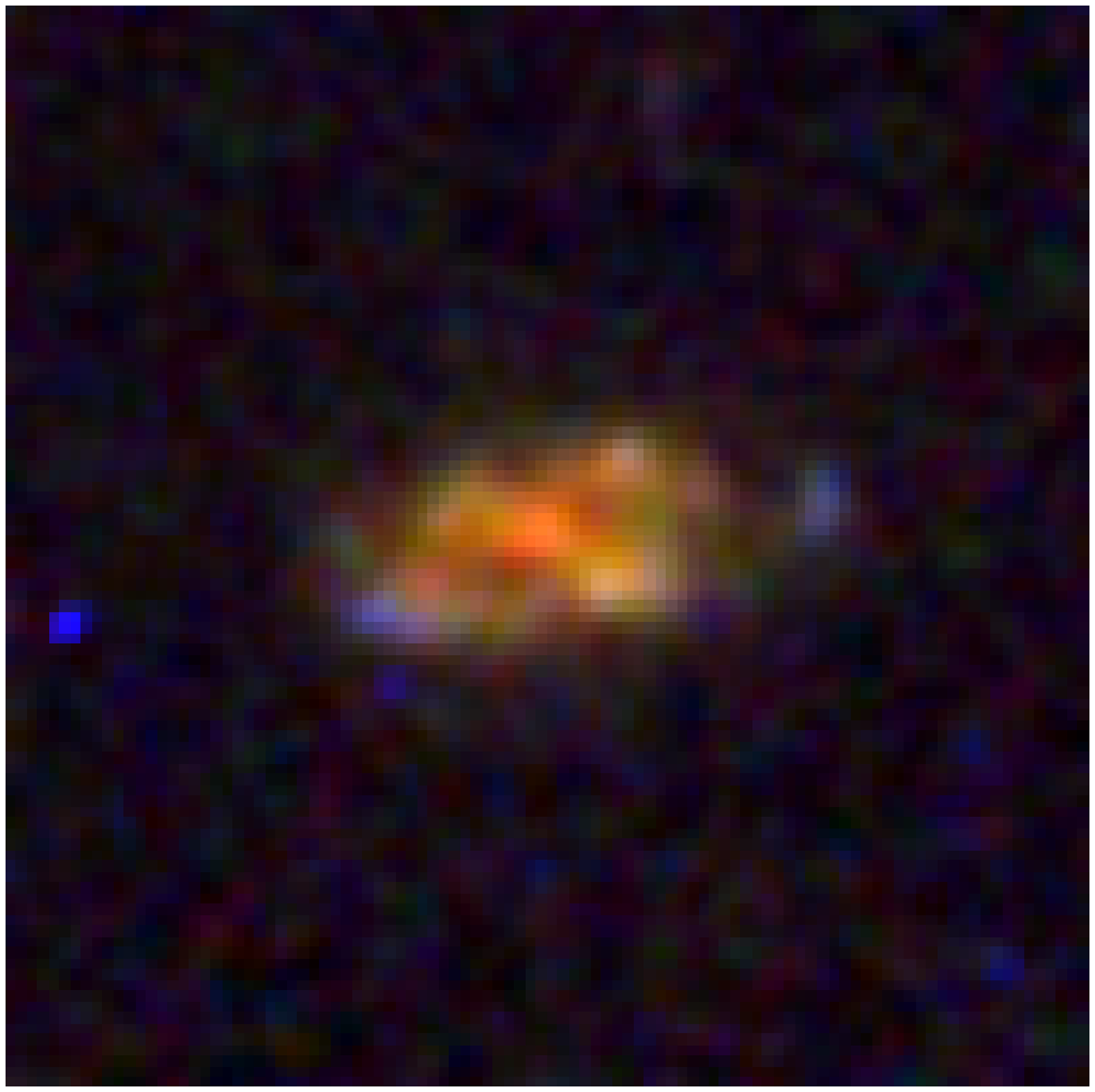,width=0.229\textwidth}&
\epsfig{file=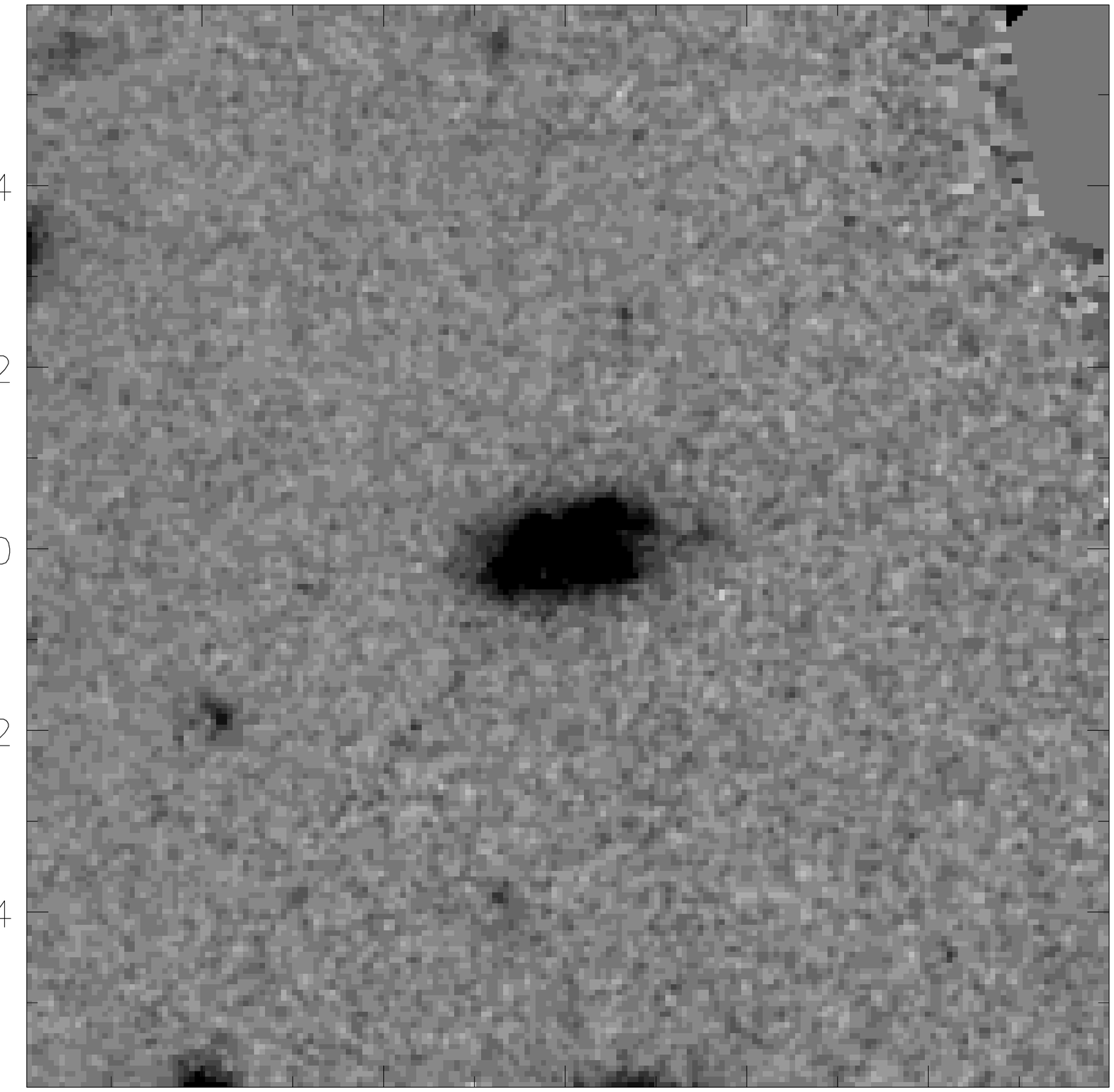,width=0.228\textwidth}&
\epsfig{file=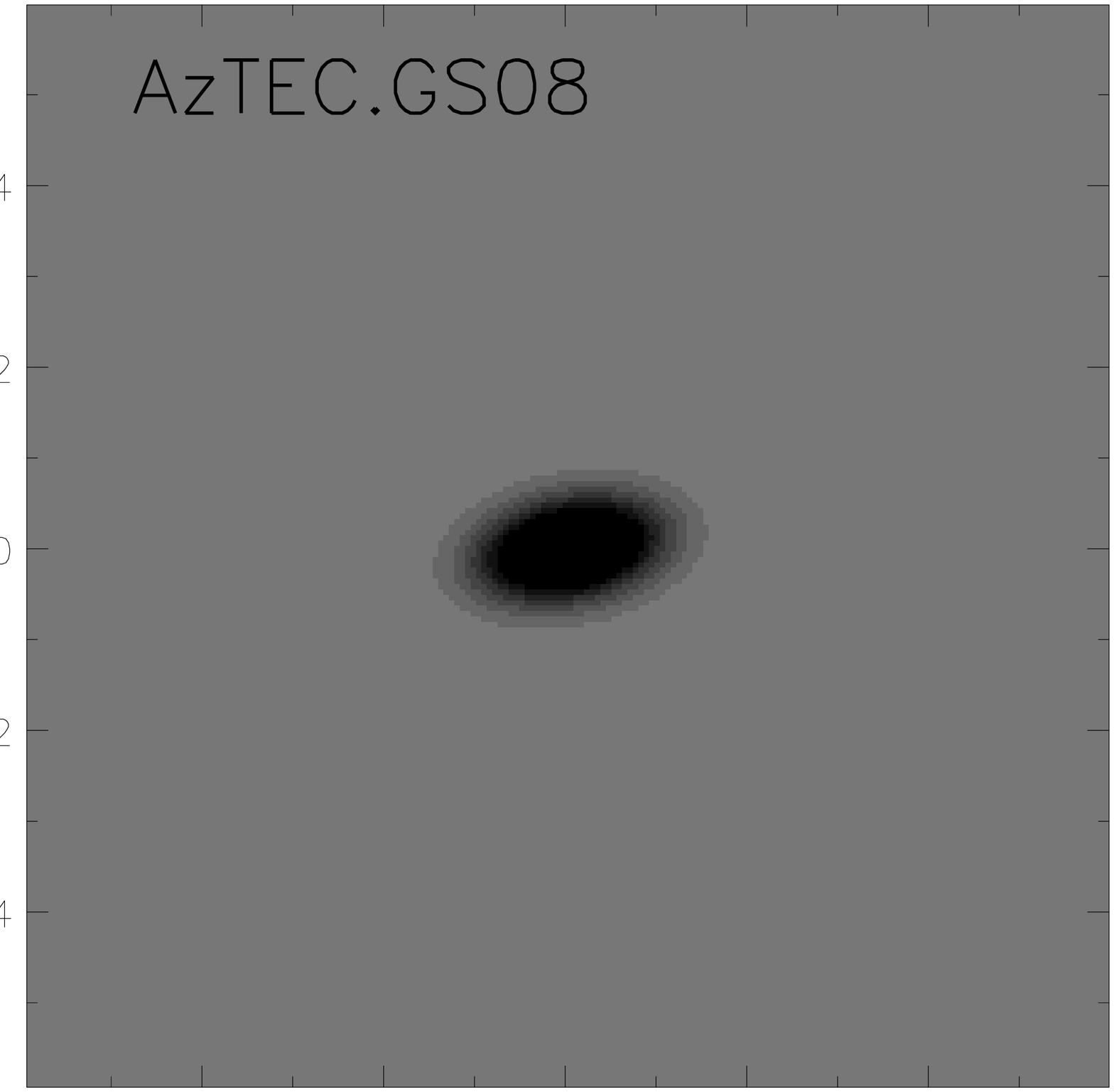,width=0.228\textwidth}&
\epsfig{file=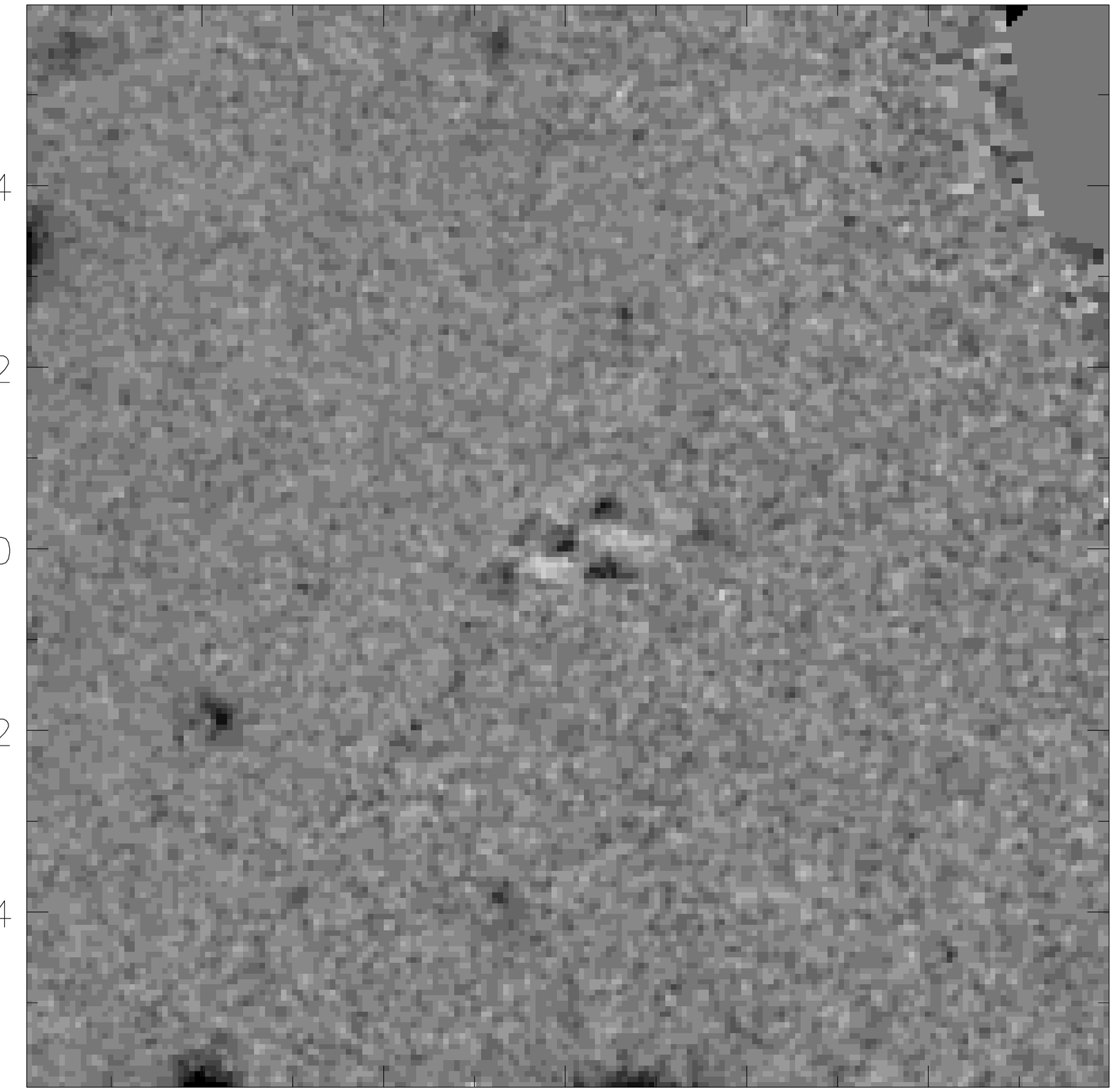,width=0.228\textwidth}\\
\\
\epsfig{file=TAT2012A_FIG01_00.ps,width=0.228\textwidth}
\hspace{-4.175cm}
\epsfig{file=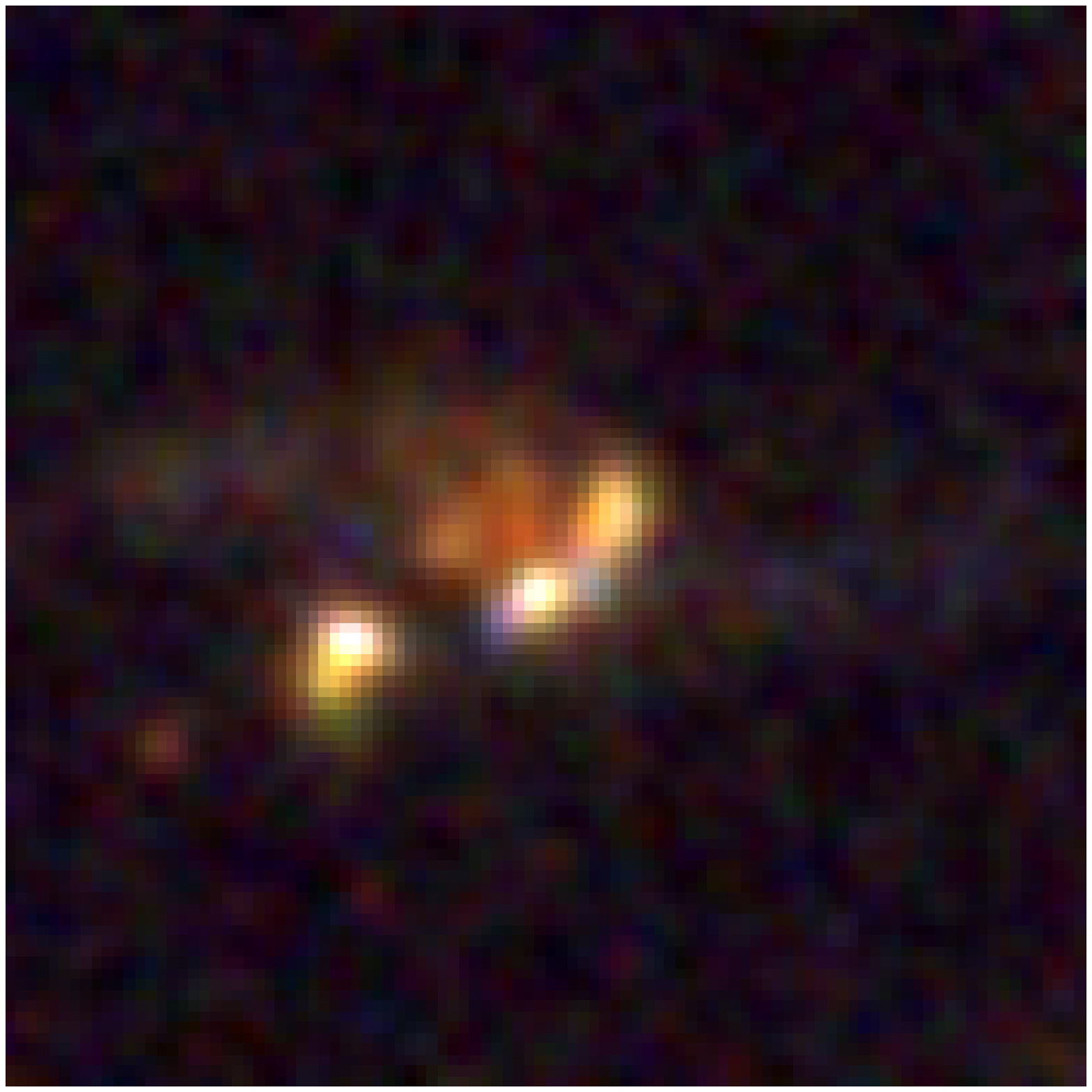,width=0.228\textwidth}&
\epsfig{file=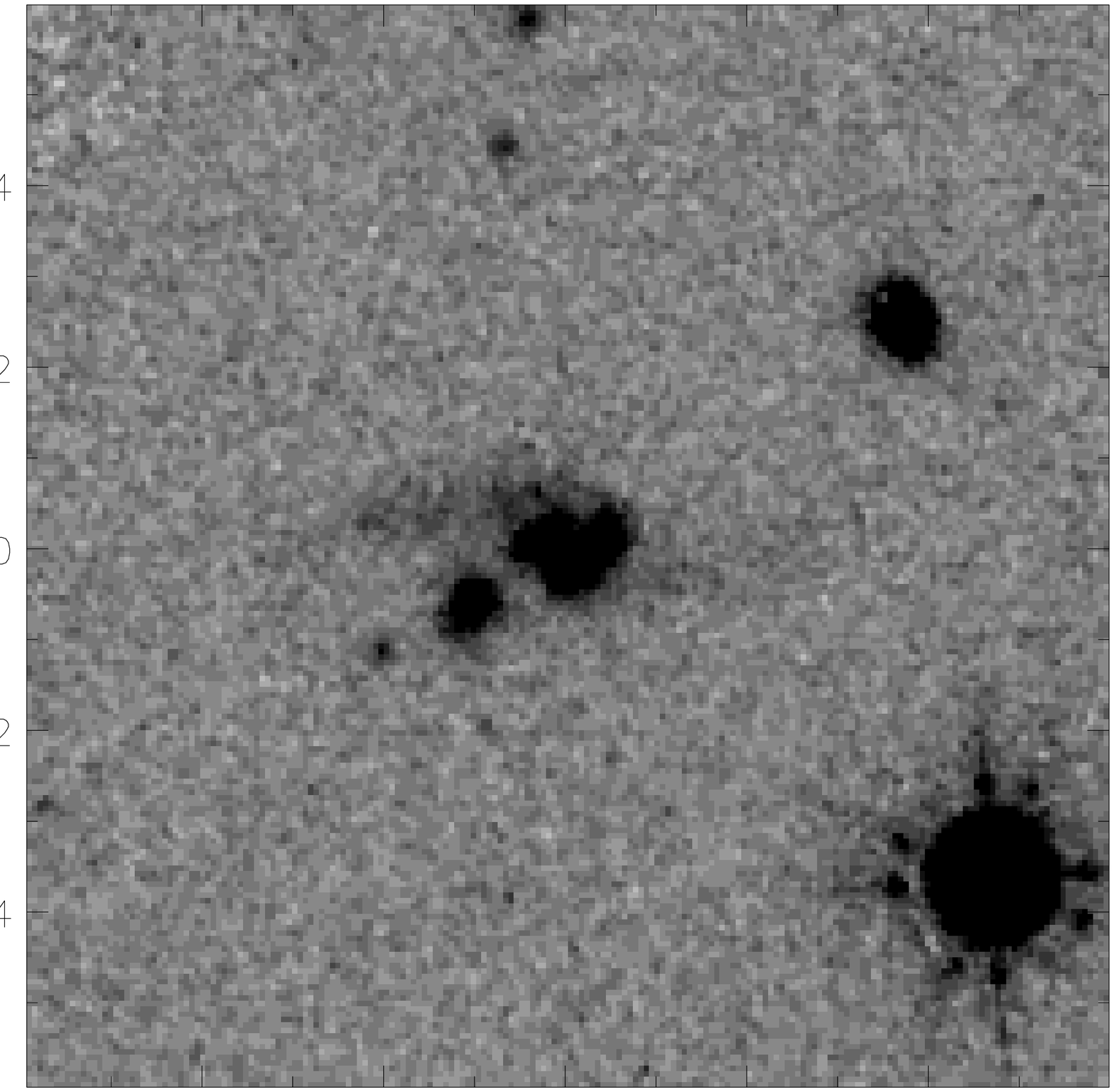,width=0.228\textwidth}&
\epsfig{file=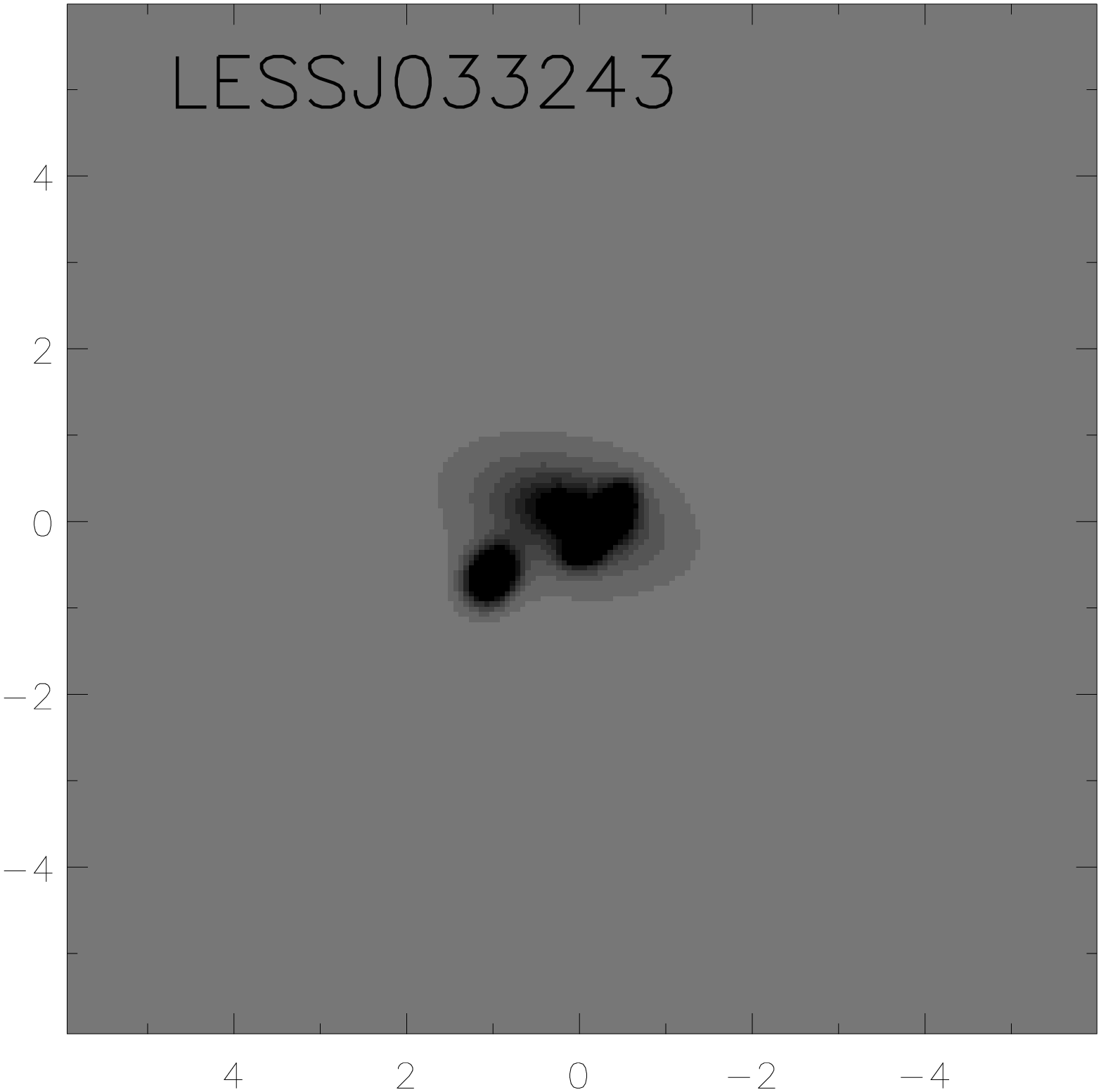,width=0.228\textwidth}&
\epsfig{file=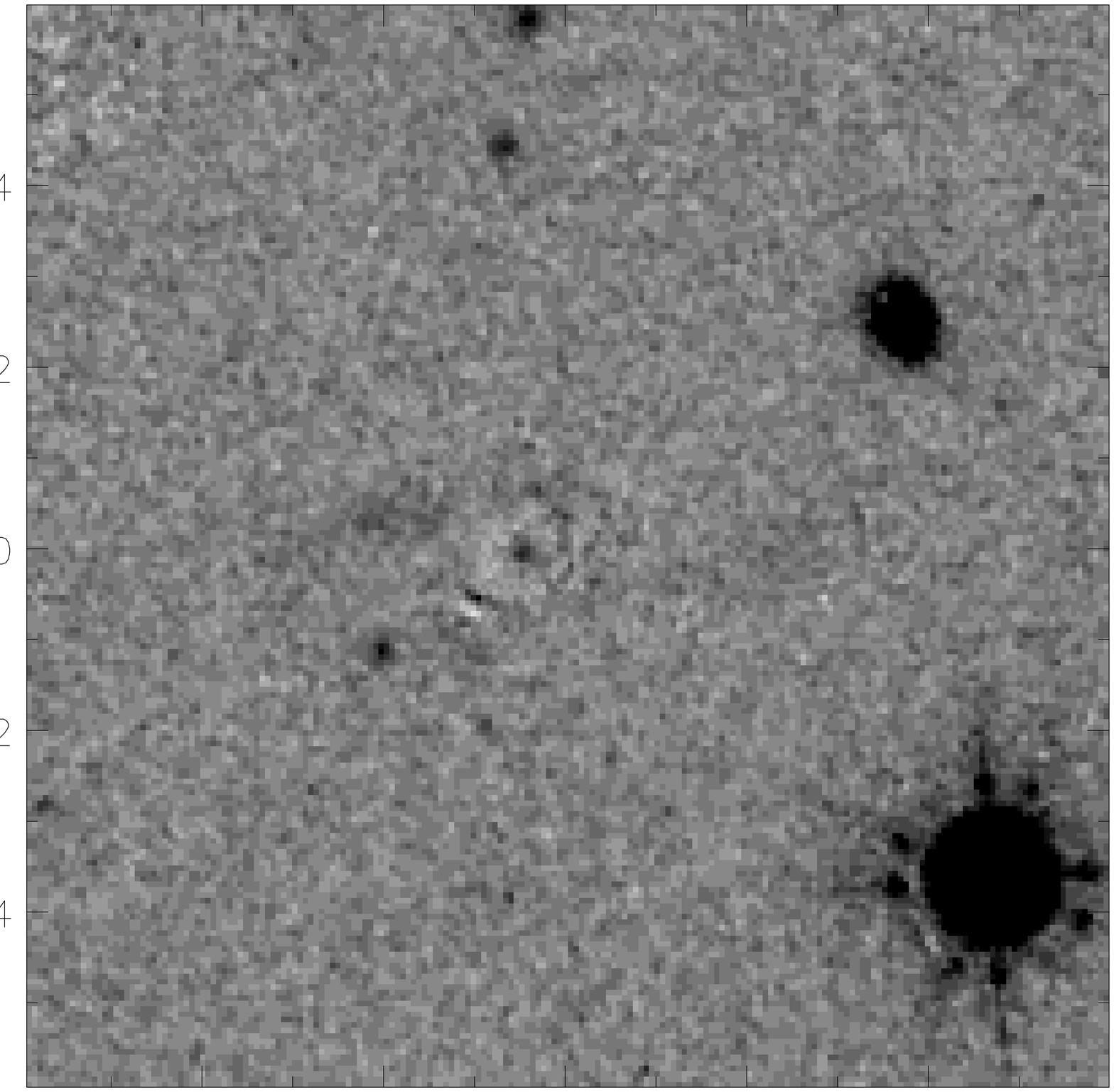,width=0.228\textwidth}\\
\end{tabular}
\caption{Examples of the results of two-dimensional modelling of what appears to be a single-component clumpy disk galaxy (AzTEC.GS08; upper row) and a multi-component system (LESSJ033243; lower row). The left-hand panels show colour images (6$^{\prime \prime}$ $\times$6$^{\prime \prime}$) based on the $V_{606}+i_{775}$, $J_{125}$ and $H_{160}$ images. The next column shows grey-scale $H_{160}$-band CANDELS postage-stamp images (12$^{\prime \prime}$ $\times$12$^{\prime \prime}$) centred on the (sub)millimetre galaxy counterparts. The third column shows the best-fitting two-dimensional models. The right-hand panels show the residual images after subtraction of the best-fitting models from the data. The panels in columns 2, 3 and 4 are shown on the same greyscale.}
\end{figure*}
\end{center}

\section{SELECTION OF GALAXY COUNTERPARTS}

\subsection{Radio identifications}

For the reasons given in subsection 2.3.1, the primary method adopted to achieve the improved positional information required to connect the (sub)millimetre sources to unique galaxy counterparts in the optical/near-infrared imaging is via the identification of statistically-significant counterparts in the deep radio 1.4\,GHz imaging. Here this is even more challenging than in the surveys undertaken with SCUBA on the JCMT (e.g. Ivison et al. 2007) because of the larger size of the AzTEC and LABOCA beams in the GOODS-South surveys (FWHM=30\,arcsec at 1.1\,mm and FWHM=19.2\,arcsec at 870\,${\rm \mu m}$ respectively).

However, the radio data described above in subsection 2.3.1 are certainly deep enough to find counterparts to the vast majority of the (sub)millimetre sources, especially when supplemented by the even deeper radio imaging of Yun et al. (2012), and the only real issue is establishing an acceptable level of statistical security.

The anticipated uncertainty in the (sub)millimetre positions is given by

$$\sigma_{pos} (= \Delta \alpha = \Delta \delta) = \frac{0.6 \theta}{S/N}$$

\noindent
where $\theta$ is the {\it FWHM} of the (sub)millimetre beam, and $S/N$ is the statistically-deboosted signal:noise ratio of the (sub)millimetre source. To be conservative, we adopted a $S/N$ value of 3, and hence searched for counterparts with a 2.5-$\sigma$ search radius of 15\,arcsec around the AzTEC sources and 10\,arcsec around the LABOCA sources (a radius expected to contain 95\% of all genuine radio counterparts).

Having located all potential radio counterparts within the search radii, we then calculated the probability ($P$) that each potential counterpart could have been found simply by chance using the method described by Downes et al. (1986) (see also Dunlop et al. 1989). This involves first calculating the raw Poisson probability that a radio source of the observed 1.4\,GHz flux density would be discovered by chance at the measured distance from the nominal (sub)millimetre source position, and then correcting this probability for the number of ways such a statistical coincidence could have been discovered given the available search parameter space defined by the maximum search radii, the radio-source number density at the limiting search flux-density ($N(S_{1.4} > 30\, {\rm \mu Jy}) \simeq0.8$\,arcmin$^{-2}$), and the form of the radio source counts over the flux-density range of interest (here we adopt a power-law index of $1.4$). For the deep data under study here, this correction is often substantial, typically increasing $P$ by a factor of a few. This technique has been applied previously to estimate the robustness of radio identifications for SCUBA sources (e.g. Ivison et al. 2002, 2007), and BLAST sources (Dunlop et al. 2010), and is also the technique adopted by Yun et al. (2012) in their own work on the radio identification of the GOODS-South AzTEC sources.

The calculated value of $P$ is the probability that the observed association is the result of chance. Even a very low value of $P$ does not {\it prove} that the radio source {\it is} the (sub)millimetre source. Nevertheless, a low value of $P$ clearly does imply that the radio source is likely related to the(sub)millimetre source in some way. This could be true for several reasons. First, the radio source could be {\it the single} true counterpart of the (sub)millimetre source. Alternatively it could be one of a group of two or more galaxies which contribute to the (sub)millimetre flux-density peak (of particular relevance here, given the large beam-sizes). Third, the statistical result could be a consequence of some secondary association (e.g. clustering with the true (sub)millimetre sources, or the result of gravitational lensing). Despite these potential concerns over statistical interpretation, we re-emphasise that our search for radio counterparts is not motivated purely by their statistical rarity and good positional accuracy, but also by the physical evidence that dust-enshrouded star-forming galaxies also produce copious quantities of synchrotron emission (resulting in the well-known far-infrared:radio luminosity correlation -- e.g. Helou et al. 1985; de Jong et al. 1985). A sensible hypothesis, therefore, is that an apparently statistically secure radio counterpart to a (sub)millimetre source is at least a significant contributor to the observed (sub)millimetre emission.

As summarised in the final two columns of Table 1, the radio imaging yields counterparts for 19 out of the 24 (sub)millimetre sources, all but 3 of which have $P < 0.1$. The sum of all the $P$ values for the 19 radio identifications suggests, statistically, that only one of these sources may have been mis-identified.

\subsection{Spitzer ${\bf 8{\rm \mu m}}$ identifications}
This leaves five sources which lack a galaxy identification. To identify the most probable counterparts for these we follow recent work by Micha{\l}owski et al. (2012b) and Koprowski et al. (2013) which confirms that the brightest {\it Spitzer} $8{\rm \mu m}$ source within the search radii (as defined above) can be reliably adopted as the correct galaxy counterpart, even when a radio source (or indeed a 24\,$\mu m$ counterpart) is absent (possibly because the source lies at a high redshift where the k-correction from the steep-spectrum synchrotron spectrum has rendered the radio counterpart undetectable). This appears to work because sub-mm/mm galaxies are among the most massive at any epoch $z > 1$ and even at, say, $z \simeq 4$, the observed $8\,{\rm \mu m}$ band continues to sample the rest-frame SED of a galaxy close to the peak emission of its mass-dominant stellar population (i.e. at $\lambda_{rest} \simeq 1.6$\,$\mu$m). 

Indeed for 18 out of the 19 sources in Table 1 which have radio identifications as described above, selecting the brightest $8\,{\rm \mu m}$ counterpart would have delivered the same optical/near-infrared galaxy identification, and the only reason we have listed these objects as primarily radio identifications is that the $P$ values are lower (because the surface density of radio sources at the limit of the radio imaging is significantly lower than that of the $8\,{\rm \mu m}$ sources at the limit of the (deep) GOODS-South {\it Spitzer} imaging).

We thus searched for, and successfully located, 8\,$\mu$m counterparts for the remaining five sources. Two of these are statistically secure, with $P < 0.05$, but the remaining three have rather high $P$ values, and must be regarded as only tentative identifications.

\subsection{Final galaxy sample}

Table 1 thus summarises the 25 proposed identifications to the 24 (sub)millimetre sources which are the subject of the further morphological and SED analyses presented in this paper. The reason these numbers do not match is simply that, as indicated in Table 1, we have chosen to retain the two alternative galaxy counterparts for AzTEC.GS19 because they are both equally probable, and at least one such source blend is not unexpected in this sample given the large size of the AzTEC beam. Of our 24 (sub)millimetre sources, 18/24 have secure galaxy identifications ($P < 0.1$) while 6/24 are statistically insecure ($P > 0.1$). While it would be desirable to insist on a significance threshold of $P < 0.05$, we here adopt $P < 0.1$ as appropriate given the large search radius required by the AzTEC and LABOCA beams (12 sources still have $P< 0.05$). Despite the fact that a sum of the $P$ values indicates that at most only 3 sources may have been misidentified, the less secure identifications ($P > 0.1$) are explicitly flagged as a subset in all the subsequent analysis presented in this paper (it transpires that our results are unaffected by their inclusion or exclusion).

Of the galaxy identifications, two objects, AzTEC.GS12 and AzTEC.GS28, were found to be unresolved in the $H$-band imaging, rendering them unsuitable for two-dimensional modelling, although we retained them in the sample for SED fitting (note that AzTEC.GS28 has the formally least secure galaxy identification in our whole sample, so may well have been misidentified; AzTEC.GS12 is presumably an AGN, as it has a secure radio counterpart). Finally we note that AzTEC.GS03, although a secure radio identification, appears too faint in the $H$-band (and all shorter wavelengths) for morphological modelling, although its solid IRAC detections still allowed meaningful SED fitting.

We are thus left with a sample of 21 (sub)millimetre sources, with 22 galaxy counterparts suitable for two-dimensional modelling as described in the next section.

\section{MODEL FITTING}

\subsection{Two-dimensional modelling fitting}

Morphological parameters for the (sub)millimetre galaxy sample were obtained from the GOODS-South deep CANDELS $H$-band data using {\sc galfit} (Peng et al. 2002, 2010). The {\sc galfit} code employs $\chi^2$ minimisation to create a best-fit numerical model, where free parameters are tuned to match the light distribution of a source. For comparison with the data the model obviously needs to be convolved with the appropriate imaging point spread function (PSF); here the PSF for the CANDELS imaging was derived using the {\sc iraf} package {\sc psf}, fitting to several stars from the science image. The resulting models were then centroided, averaged, and processed with look-up tables to reproduce any asymmetries. 

The results of the two-dimensional modelling for all 22 viable (sub)millimetre galaxy counterparts are presented in Table 2, with examples of models and model-subtracted residuals for single and multi-component fits shown in Fig.~1 (see Fig.~A1 in the Appendix for the model fits to all objects). We find that $>95$\% of the rest-frame optical light in almost all of the (sub)millimetre galaxies is best-described by either a single-component exponential disk (10/22 objects), or multiple component systems where the dominant constituent is disk-like (10/22).

For comparison with existing studies of larger samples of massive galaxies, and as a test of S\'{e}rsic index measurements obtained from lower-resolution data (i.e. Targett et al. 2011), all multiple-component sources were also fitted with a single-component model (SCP) forced to attempt to reproduce the entire source (see SCP entries in Table 2). We find that both the average size and S\'{e}rsic index of these single-component fits match those of the dominant component in the multi-object fits to within a typical accuracy of 5\%. This similarity highlights the dominant nature of the disk-like component in the multiple component systems. It also suggests that single S\'{e}rsic fits to complex systems in moderate-resolution data (e.g. the $\simeq 0.4$\,arcsec seeing ground-based imaging analysed by Targett et al. 2011), while missing the fine-structure seen at {\it HST} resolution, can successfully recover the basic morphology of the underlying mass-dominant galaxy (see the further analysis of this issue presented in Section 5.1).

Among the 10 multiple-component systems we find an even division between objects whose secondary components display a S\'{e}rsic index consistent with a disky morphology (possibly indicating a companion, possibly interacting or merging galaxy), and objects with secondary components which appear to be best described by extremely low S\'{e}rsic indices outside the normal range displayed by galaxies. The anomalous S\'{e}rsic index of these secondary components could be indicative of regions of intense star-formation or giant clumps believed to arise in star-forming galaxies at high-redshift (Ceverino et al. 2012, Hopkins et al. 2012). This separation into galactic and non-galactic secondary components is not possible via visual classification, and demonstrates another advantage of performing detailed multiple-component variable-S\'{e}rsic index model fitting to complex objects.

\begin{table*}
 \begin{center}
  \caption{Results from two-dimensional modelling of the CANDELS $H_{160}$ galaxy counterparts (including sub-components where significant, indicated by multiple enteries under source name with suffix a, b, c, d etc.) 
to the AzTEC and LABOCA selected (sub)millimetre galaxies in CANDELS GOODS-South. All multiple-component sources were also fitted with a single-component model forced to attempt to reproduce the entire source (entries labelled SCP, shown just below the results from the separate components of the source name listed above). Columns 1 \& 2 give the source name and sub-component identifier. Sources with statistically secure ($P < 0.1$) counterparts 
have their names written in bold. Column 3 lists the total model host-integrated $H_{160}$-band magnitude (AB). Column 4 lists the semi-major axis scalelength (half-light radius) of the host galaxy fit in kpc. Column 5 lists the value of the S\'{e}rsic index ($n$). Column 6 gives the axial ratio of the host galaxy. Column 7 gives the value of reduced $\chi^{2}_{\nu}$ for each model fit, where multiple-component fits share a common $\chi^{2}_{\nu}$ for the entire combined fit.}
  \begin{tabular}{llcccccc}
\hline\hline
Name & Comp. & $H_{160}$ & $r_{1/2}$ & S\'{e}rsic & Axial & $\chi^{2}_{\nu}$\\
     &       & (AB mag) & (kpc)   & ($n$)       & ratio & \\
\hline
\bf{AzTEC.GS06}   & a   & 23.64 & 3.5  & 0.8 & 0.34 & 0.997 \\
                  & b   & 24.49 & 11.8 & 9.4 & 0.72 & 0.997 \\
                  & SCP & 23.76 & 3.3  & 0.7 & 0.76 & 1.030 \\
\bf{AzTEC.GS08}   &     & 21.92 & 6.3  & 1.0 & 0.51 & 1.196 \\
\bf{AzTEC.GS11}   &     & 23.91 & 3.8  & 0.7 & 0.48 & 1.270 \\
\bf{AzTEC.GS12}   &     & unresolved & &     &      &       \\
\bf{AzTEC.GS13}   & a   & 24.13 & 3.1  & 1.2 & 0.79 & 0.885 \\
                  & b   & 24.81 & 3.2  & 0.8 & 0.59 & 0.885 \\
                  & SCP & 23.76 & 4.6  & 0.7 & 0.53 & 0.916 \\
\bf{AzTEC.GS16}   &     & 24.39 & 4.6  & 1.0 & 0.93 & 0.954 \\
\bf{AzTEC.GS17}   &     & 23.31 & 1.6  & 1.4 & 0.96 & 1.150 \\
\bf{AzTEC.GS18}   &     & 24.38 & 5.3  & 1.1 & 0.49 & 0.962 \\
\bf{AzTEC.GS19-1} &     & 22.18 & 4.1  & 0.9 & 0.72 & 1.717 \\
\bf{AzTEC.GS19-2} &     & 23.15 & 2.8  & 1.2 & 0.65 & 1.717 \\
\bf{AzTEC.GS21}   & a   & 22.94 & 2.6  & 1.3 & 0.65 & 2.048 \\
                  & b   & 24.25 & 3.2  & 0.3 & 0.56 & 2.048 \\
	          & c   & 24.50 & 2.3  & 0.2 & 0.34 & 2.048 \\
                  & SCP & 22.42 & 3.7  & 1.5 & 0.82 & 2.635 \\
AzTEC.GS22        & a   & 23.70 & 4.5  & 1.2 & 0.40 & 1.036 \\
	          & b   & 24.22 & 3.5  & 0.3 & 0.44 & 1.036 \\
                  & SCP & 23.32 & 6.0  & 0.7 & 0.27 & 1.065 \\
\bf{AzTEC.GS23}   & a   & 22.67 & 4.6  & 1.3 & 0.67 & 1.193 \\
                  & b   & 23.18 & 3.8  & 0.7 & 0.31 & 1.593 \\
                  & c   & 23.28 & 3.0  & 1.0 & 0.62 & 1.593 \\
                  & SCP & 21.54 & 5.9  & 1.1 & 0.67 & 1.558 \\
AzTEC.GS24        &     & 24.24 & 1.7  & 1.0 & 0.59 & 1.913 \\
AzTEC.GS26        &     & 23.01 & 6.6  & 1.3 & 0.30 & 1.148 \\
AzTEC.GS27        & a   & 20.64 & 3.7  & 4.0 & 0.94 & 1.875 \\
                  & b   & 21.70 & 7.0  & 1.0 & 0.73 & 1.875 \\
AzTEC.GS28        &     & unresolved & &     &      &       \\
\bf{AzTEC.GS30}   & a   & 21.93 & 6.3  & 1.4 & 0.66 & 1.457 \\
	          & b   & 24.51 & 3.5  & 0.2 & 0.22 & 1.457 \\
                  & SCP & 21.88 & 5.4  & 1.1 & 0.68 & 1.824 \\
AzTEC.GS34        &     & 21.92 & 1.6  & 1.9 & 0.84 & 1.193 \\
\bf{AzTEC.GS35}   & a   & 23.69 & 1.8  & 1.3 & 0.48 & 1.426 \\
                  & b   & 23.76 & 0.8  & 0.9 & 0.55 & 1.426 \\
                  & SCP & 23.05 & 2.8  & 0.7 & 0.81 & 2.789 \\
\bf{AzTEC.GS38}   & a   & 19.32 & 4.4  & 4.0 & 0.95 & 2.219 \\
                  & b   & 21.46 & 7.2  & 1.0 & 0.62 & 2.219 \\
\bf{LESSJ033217}  & a   & 20.16 & 5.9  & 1.0 & 0.82 & 1.274 \\
	          & b   & 22.65 & 1.4  & 0.3 & 0.28 & 1.274 \\
                  & SCP & 19.88 & 6.4  & 2.0 & 0.78 & 1.489 \\
\bf{LESSJ033219}  & a   & 24.31 & 2.4  & 1.4 & 0.48 & 1.058 \\
	          & b   & 24.78 & 1.1  & 1.3 & 0.48 & 1.058 \\
\bf{LESSJ033243}  & a   & 22.25 & 7.6  & 1.1 & 0.53 & 1.120 \\
	          & b   & 23.47 & 1.6  & 0.8 & 0.66 & 1.120 \\
	          & c   & 24.17 & 1.6  & 0.4 & 0.38 & 1.120 \\
	          & d   & 24.37 & 1.1  & 0.4 & 0.38 & 1.120 \\
                  & SCP & 22.04 & 7.5  & 0.7 & 0.38 & 2.686 \\
\hline
  \end{tabular}
 \end{center}
\end{table*}

In order to further check the accuracy and reliability of model parameter recovery, we have also measured the Petrosian radius and $r_{1/2}$ via a curve-of-growth analysis directly from the $H_{160}$-band data. The Petrosian radius was defined as the point at which the ratio of the surface brightness in an annulus at radius $r$ to the mean surface brightness within an aperture of radius $r$ equals $0.2$ (Blanton et al. 2001, Yasuda et al. 2001). It was found that the values of $r_{1/2}$ as determined from the two-dimensional modelling, curve-of-growth, and Petrosian techniques were consistent to within 10\%.

\begin{center}
\begin{figure*}
\begin{tabular}{cc}
\epsfig{file=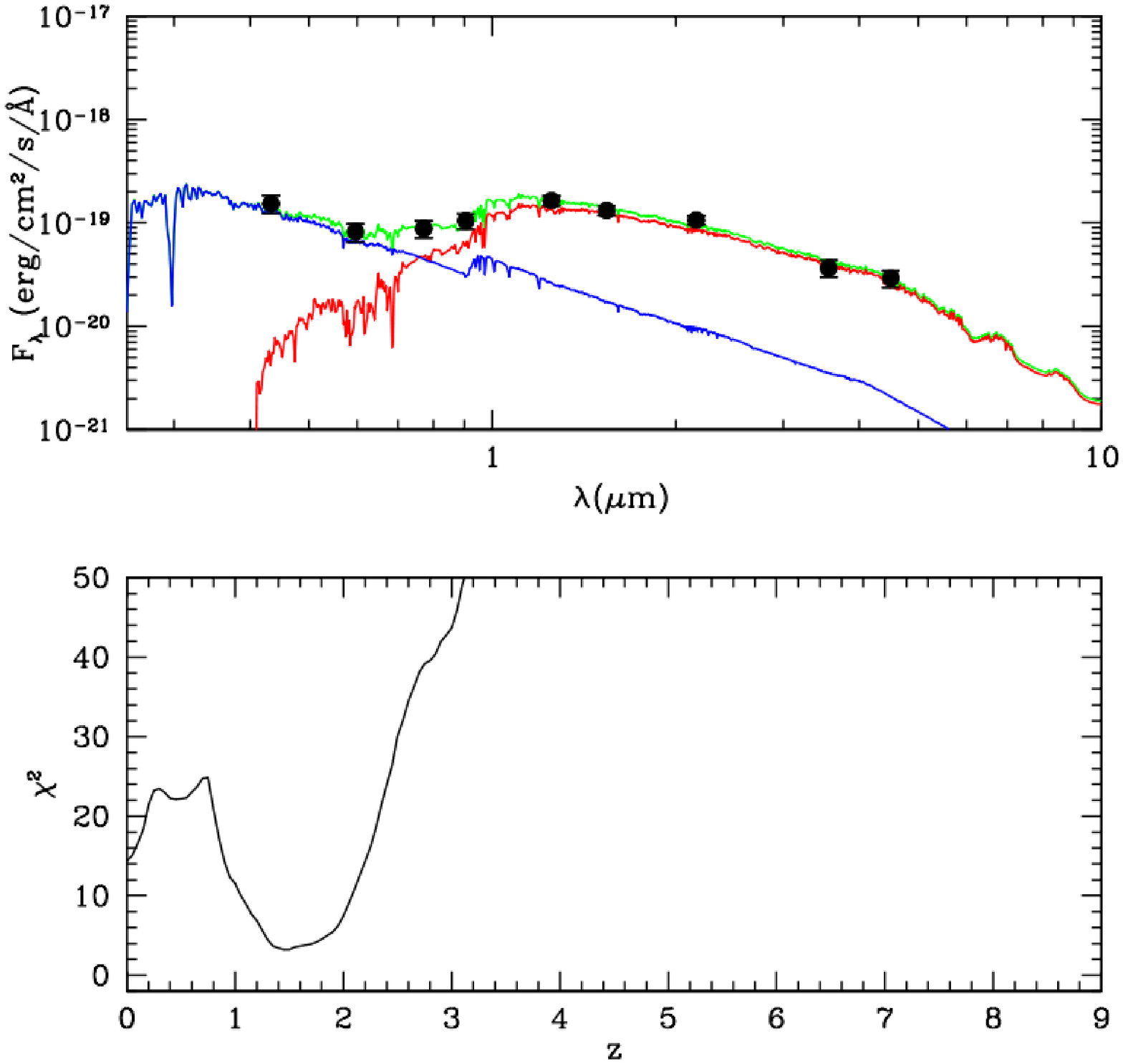,width=0.45\textwidth}&
\epsfig{file=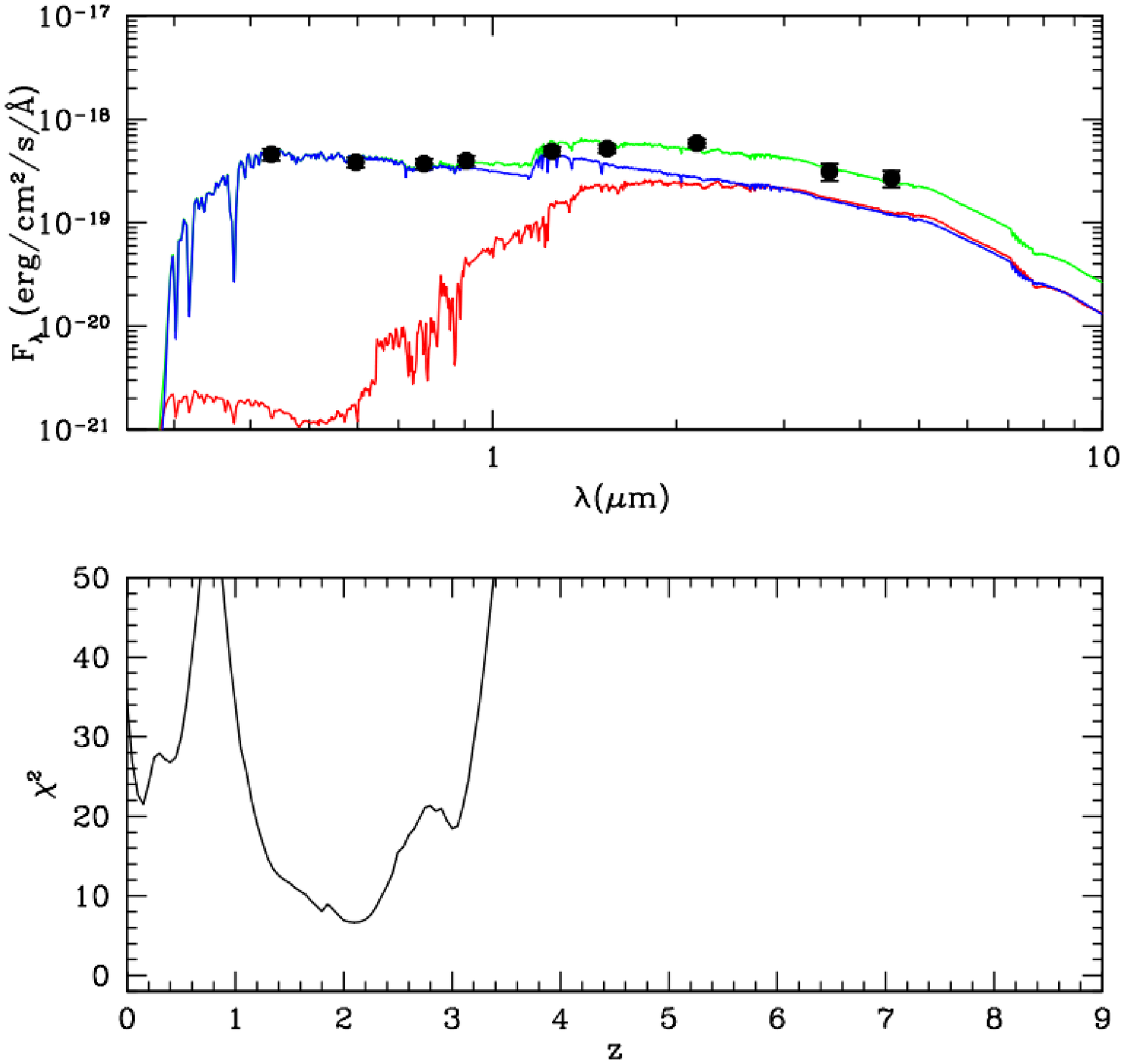,width=0.45\textwidth}\\
\\
\epsfig{file=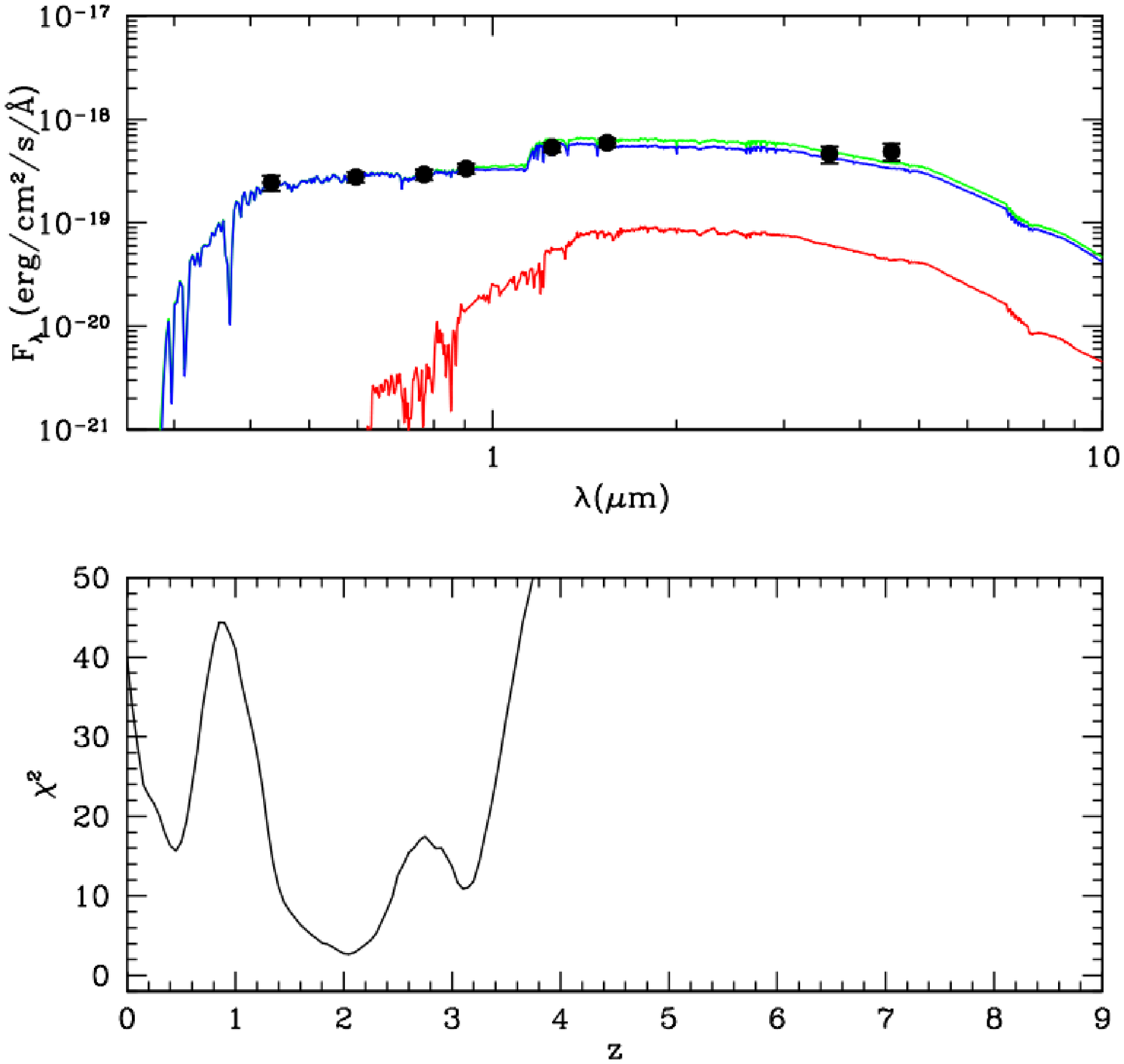,width=0.45\textwidth}&
\epsfig{file=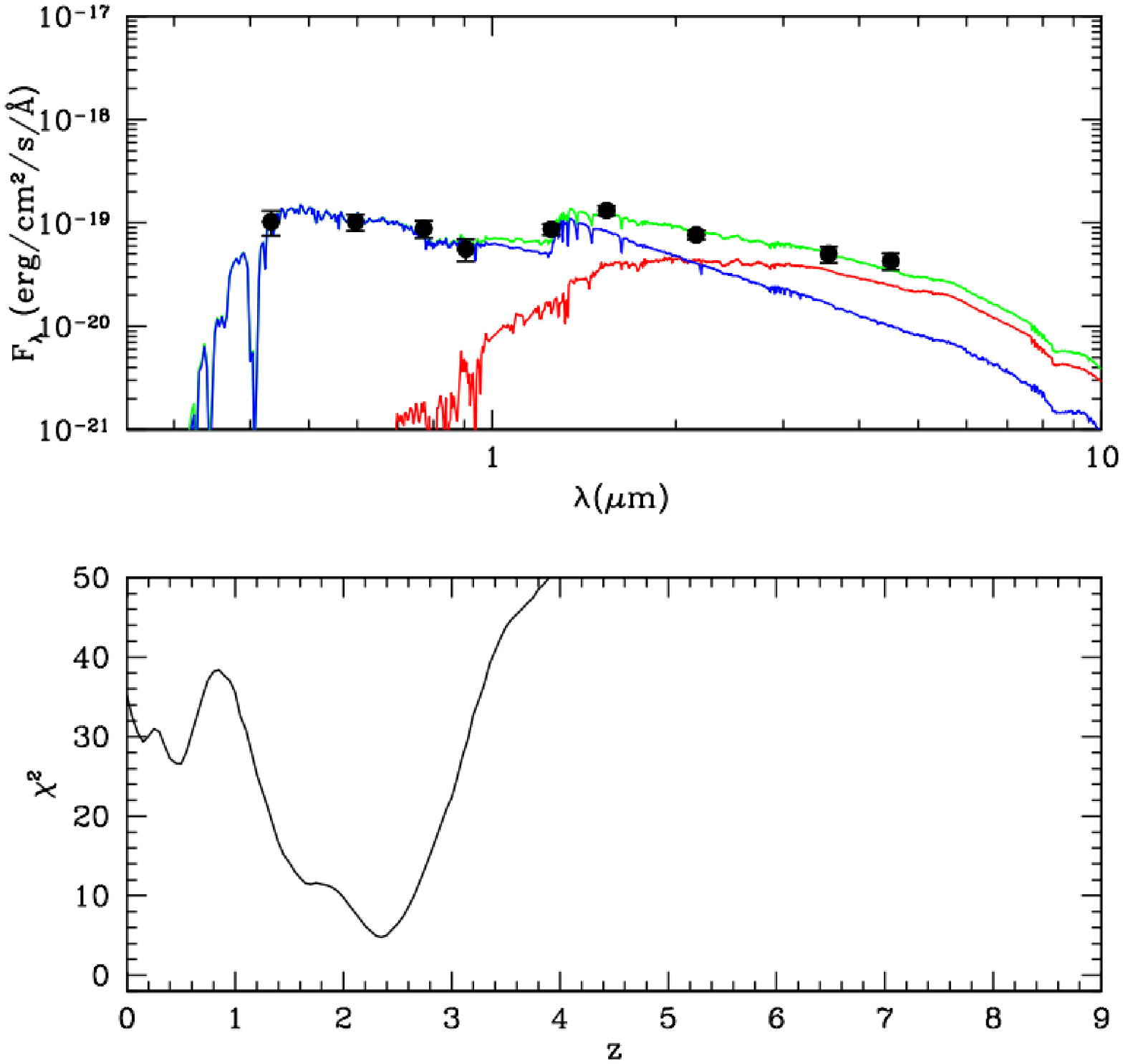,width=0.45\textwidth}\\
\end{tabular}
\caption{Example plots for two-component SED fitting to LESSJ033219 (top left), LESSJ033243 (top right), AzTEC.GS08 (bottom left), and AzTEC.GS13 (bottom right). For each object the upper panel shows $F_{\lambda}$ versus observed wavelength $\lambda$, with the red line showing the contribution from the older stellar population, the blue line showing the contribution from the most recent burst of star formation, and the green line the combination of the two. For each source the lower panel shows the value of $\chi^{2}$ as a function of redshift (marginalized over all other parameters varied during the SED fitting).}
\end{figure*}
\end{center}

\subsection{Reliability and limitations of modelling results}
The ability of {\sc galfit} to recover accurate values of scalelength and S\'{e}rsic index from the CANDELS data was examined via two-dimensional modelling of synthetic galaxies. With luminosity, axial ratio, position angle, and redshift set from the mean values of the (sub)millimetre galaxy sample, model galaxies where generated with scalelengths $r_{1/2}$ over the range 1-20\,kpc for both $n=1$ (exponential disc) and $n=4$ (de Vaucouleurs spheroid) light profiles. These synthetic sources were inserted into an image, with noise levels based upon the average rms-per-pixel values derived from the CANDELS data. To ensure a realistic level of uncertainty in the accuracy of the PSF used to model a given galaxy, the synthetic sources were convolved with a secondary PSF derived from alternative stars on the science image. Morphological parameters recovered from these artificial galaxies via {\sc galfit} modelling matched their input values to within (at worst) an accuracy of 10\%.

\subsection{Image stacking}
Despite the excellent depth of the CANDELS imaging (5-$\sigma$ AB limiting magnitude of 26.5), we explore the extent to which extended low-surface brightness emission could still have remained undetected in faint or compact (sub)millimetre galaxies in the sample via stacking. Initially, all the individual galaxy images were centroided, scaled to the average luminosity, aligned to the same position angle, and mean-combined using the IRAF package IMCOMBINE. While GALFIT modelling of this initial stack found a disk-like (albeit higher than average) S\'{e}rsic index, it was apparent that the asymmetric nature of the multi-component sources was contaminating the stacked image. In order to remove this contamination from any real extended low-surface brightness emission revealed by the stack, all multi-component sources were processed through GALFIT to remove secondary components from the brightest primary source before stacking. The surface-brightness profile for the final (sub)millimetre galaxy stack is shown in Fig.~3, with the derived best-fitting GALFIT model parameters given in Table 3. The best-fitting GALFIT model to the stacked image yields a S\'{e}rsic index ($n=1.54$) and a half-light radius ($r_{1/2}=4.6$\,kpc), consistent with the average values for the sample. Fig.~3 also includes the results of attempting to fit the image stack with a de Vaucouleurs spheroidal model (i.e. with fixed $n = 4$), simply to demonstrate how strongly such a model can be excluded.

\begin{figure}
 \begin{center}
  \epsfig{file=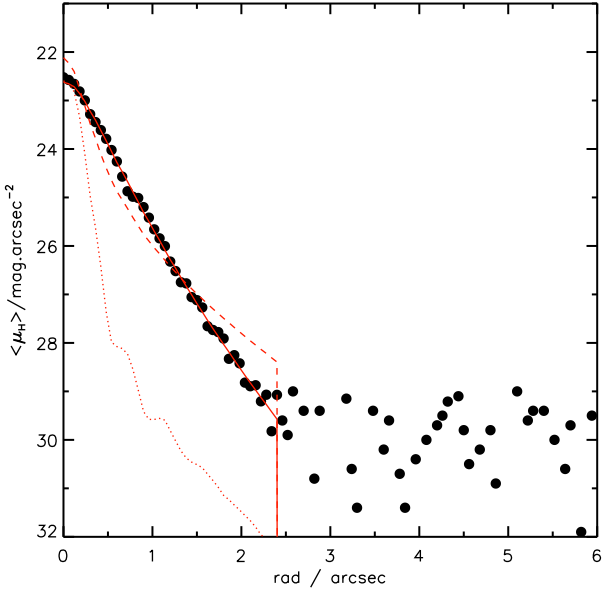,width=0.5\textwidth}
  \caption{The surface-brightness profile derived from the stack of the $H_{160}$-band images of the (sub)millimetre galaxies. The data are indicated by the black points. The surface-brightness profile of the best-fitting axi-symmetric model (S\'{e}rsic index $n=1.54$) is shown by the solid line, after convolution with the PSF (indicated by the dotted line). The dashed line shows the surface-brightness profile produced by the best-fitting spheroidal model in which S\'{e}rsic index is fixed at $n=4$; clearly this fit is completely unacceptable. The model profiles are plotted out to the radius at which the signal:noise level in the stacked images allows them to be constrained.}
 \end{center}
\end{figure}

\begin{table}
 \begin{center}
  \caption{The results of two-dimensional modelling to the stacked (sub)millimetre galaxies.}
 \begin{tabular}{lcccc}
\hline\hline
Name       & $H_{160}$ & $r_{1/2}$ & S\'{e}rsic & Axial\\
           & (AB mag) & (kpc)   & ($n$)       & ratio\\
\hline
IMAGE STACK & 22.65    & 4.6     & 1.54         & 0.9\\
\hline
  \end{tabular}
 \end{center}
\end{table}

\subsection{SED fitting}

The extensive and deep multi-frequency data in the GOODS-South field, now further enriched with the deep CANDELS near-infrared imaging, allows high-quality SED fits to the (sub)millimetre galaxy counterparts to determine redshifts, stellar masses, and other model parameters. The photometry for the SED fitting was extracted via 2-arcsec diameter aperture photometry at the position of each CANDELS $H_{160}$ counterpart in the {\it HST} ACS $F435W\,(B_{435})$, $F606W\,(V_{606})$, $F775W\,(i_{775})$ and $F850LP\,(z_{850})$ bands, the CANDELS $F125W$ $J$-band and $F160W$ $H$-band, and in all 4 IRAC channels (3.6, 4.5, 5.6, and 8.0\,${\rm \mu m}$). The 2\,arcsec aperture did not require any significant aperture correction in the optical and near-infrared {\it HST} imaging, while the IRAC data were de-confused to comparable resolution using the CANDELS $H_{160}$ image as a template (as described in McLure et al. 2011).

The SED fitting procedure we then applied to derive photometric redshifts and stellar masses is based largely on the public package {\sc hyperz} (Bolzonella, Miralles, \& Pell\'{o} 2000). The observed photometry was fitted with synthetic galaxy templates generated with the Bruzual \& Charlot (2003) stellar population models. We used a variety of star-formation histories: single and double-component instantaneous bursts, and exponentially-declining star-formation. Following the analysis of (sub)millimetre galaxy counterparts presented in Micha{\l}owski et al. (2012a), we assumed solar metallicity and a Chabrier (2003) IMF. 
At each redshift step we allowed galaxy templates with ages ranging from 1\,Myr to the age of the Universe at that redshift. For dust reddening, we adopted the prescription from Calzetti et al. (2000), within the range $0 \le A_V \le 2$, and included absorption due to HI clouds along the line-of-sight according to Madau et al. (1996). The results of the multi-frequency SED fitting for the (sub)millimetre galaxies are presented in Table 4, with examples of the SED fits shown in Fig.~2, including AzTEC.GS08 and LESSJ033243, the two example objects already presented in Fig.~1.

\begin{figure}
 \epsfig{file=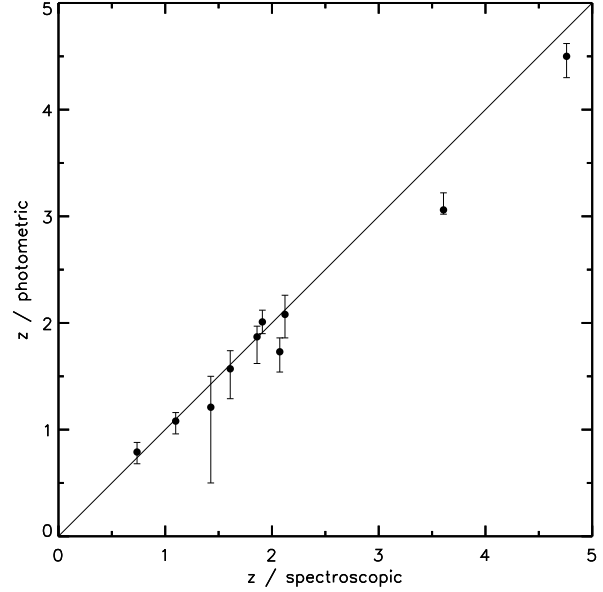,width=0.5\textwidth}
  \caption{A comparison of the photometric and spectroscopic redshifts for the 10 sources in our sample which already possess good-quality optical spectroscopy 
(from the literature sources given in Table 4). The error bars represent the 1-$\sigma$ uncertainties in $z_{phot}$ as derived from the SED fits and 
tabulated in Table 4.}
\end{figure}

\begin{table}
 \begin{center}
  \caption{The results of SED fitting to the optical--mid-infrared photometry of the AzTEC and LABOCA selected (sub)millimetre selected galaxies in CANDELS GOODS-South. Column 1 gives the source name. Columns 2, 3, 4, and 5 list the photometric redshifts, the 1-$\sigma$ uncertainty in the photometric redshifts, the stellar masses, and $\chi^2$ for the best-fitting SED template. Column 6 gives the spectroscopic redshift where available, and the reference from which this redshift was obtained: {\bf a} from Vanzella et al. (2008), {\bf b} from Huang et al. (2012), {\bf c} from Balestra et al. (2010), {\bf d} from Straughn et al. (2009), {\bf e} from Kurk et al. (2009), and {\bf f} from Kriek et al. (2007). Since spectroscopic redshifts are available for 10 sources (i.e. essentially half the sample), we show a plot of $z_{spec}$ versus $z_{phot}$ for these objects in Fig.~4, which demonstrates the accuracy of the photometric redshifts as derived from the available high-quality photometry.}
   \begin{tabular}{llccrl}
\hline\hline
Name         & z    & $\Delta$z   & $\log_{10}(\frac{M_{\star}}{\rm M_{\odot}})$& $\chi^{2}$ & $\rm z_{spec}$\\
\hline
AzTEC.GS03   & 2.39 & 2.15$-$2.70 & 11.03 & 13.87 & \\
AzTEC.GS06   & 2.13 & 2.00$-$2.27 & 11.59 & 1.88  & \\
AzTEC.GS08   & 2.04 & 1.85$-$2.20 & 11.78 & 2.72  & \\
AzTEC.GS11   & 1.69 & 1.50$-$1.95 & 11.23 & 4.37  & \\
AzTEC.GS12   & 4.50 & 4.38$-$4.70 & 11.18 & 3.08  & 4.760$\rm ^{a}$\\
AzTEC.GS13   & 2.38 & 2.25$-$2.48 & 11.16 & 4.76  & \\
AzTEC.GS16   & 1.80 & 1.55$-$2.20 & 11.04 & 1.61  & \\
AzTEC.GS17   & 2.40 & 2.20$-$3.00 & 11.38 & 2.64  & \\
AzTEC.GS18   & 2.80 & 2.45$-$2.95 & 11.09 & 7.45  & \\
AzTEC.GS19-1 & 1.62 & 1.27$-$1.97 & 11.18 & 2.73  & \\
AzTEC.GS19-2 & 1.83 & 1.55$-$2.10 & 11.28 & 0.63  & \\
AzTEC.GS21   & 2.01 & 1.90$-$2.12 & 11.28 & 3.01  & 1.910$\rm ^{a}$\\
AzTEC.GS22   & 1.87 & 1.77$-$2.12 & 11.06 & 4.16  & 1.860$\rm ^{b}$\\
AzTEC.GS23   & 1.73 & 1.60$-$1.92 & 11.82 & 3.47  & 2.073$\rm ^{c}$\\
AzTEC.GS24   & 2.18 & 2.07$-$2.27 & 11.25 & 19.4  & \\
AzTEC.GS26   & 2.00 & 1.90$-$2.15 & 11.70 & 2.57  & \\
AzTEC.GS27   & 1.21 & 0.92$-$1.92 & 11.33 & 0.33  & 1.427$\rm ^{a}$\\
AzTEC.GS28   & 3.06 & 2.90$-$3.10 & 11.51 & 6.39  & 3.607$\rm ^{d}$\\
AzTEC.GS30   & 1.57 & 1.47$-$1.70 & 11.25 & 5.36  & \\
AzTEC.GS34   & 1.57 & 1.40$-$1.85 & 11.38 & 1.89  & 1.609$\rm ^{e}$\\
AzTEC.GS35   & 2.17 & 1.75$-$2.38 & 11.31 & 0.40  & \\
AzTEC.GS38   & 0.79 & 0.70$-$0.90 & 11.56 & 1.94  & 0.736$\rm ^{a}$\\
LESSJ033217  & 1.08 & 1.00$-$1.20 & 11.36 & 2.61  & 1.098$\rm ^{a}$\\
LESSJ033219  & 1.65 & 1.40$-$2.10 & 11.09 & 3.26  & \\
LESSJ033243  & 2.08 & 1.90$-$2.30 & 11.53 & 1.80  & 2.122$\rm ^{f}$\\
\hline
  \end{tabular}
 \end{center}
\end{table}

In Table 4 we also give the spectroscopic redshifts for those sources for which robust spectroscopic results are available in the literature. Due the intensive nature of the spectroscopic campaigns in GOODS-South, almost half of our (sub)millimetre sample (i.e. 10 sources) have spectroscopic redshifts (an unusually good situation for a (sub)millimetre galaxy sample). In Fig.~4 we plot $z_{phot}$ v $z_{spec}$ for these 10 sources, which serves to demonstrate the quality of the photometric redshifts derived from the unusually high-quality multi-frequency photometry now available in the GOODS-South field. Clearly this figure provides confidence in the photometric redshifts derived for the remaining sources.

\section{DISCUSSION}

\subsection{Size and Morphology}
We find that $>95$\% of the rest-frame optical light in almost all (20/22 objects) of the (sub)millimetre galaxies is well-described by either a single-component exponential disk (10/22 objects), or multiple-component systems where the dominant constituent is disk-like (10/22 objects). This leaves two sources (AzTEC.GS27 and AzTEC.GS38) which are best fitted by a disk+bulge model with a dominant bulge component, but interestingly one of these (AzTEC.GS27) is one of the 3 least secure identifications in the sample, while the other (AzTEC.GS38) is the only galaxy in the sample with a confirmed redshift $z < 1$. In summary, all of the (sub)millimetre galaxies in our sample at $z > 1.5$ appear to be disks.

To place these results in context, the distribution of S\'{e}rsic indices for our (sub)millimetre galaxies is plotted in Fig.~5 along with the distribution of S\'{e}rsic indices derived by Bruce et al. (2012) for a sample of 83 comparably-massive ($M_{\star} > 10^{11} {\rm M_{\odot}}$) galaxies at $z \simeq 2$ from $H_{160}$ CANDELS imaging of essentially identical quality and depth. Bruce et al. (2012) have already noted that a significant fraction of massive galaxies at these redshifts appear to be disk-dominated, and that a substantial majority of these are star-forming. However, this figure shows that the (sub)millimetre selected galaxies studied here are {\it particularly disk-like} in the sense that their S\'{e}rsic indices lie very close to unity. Indeed, it appears that (sub)millimetre selection is a good way of pre-selecting a substantial fraction of the most disk-dominated massive galaxies at these redshifts.

\begin{center}
 \begin{figure}
  \epsfig{file=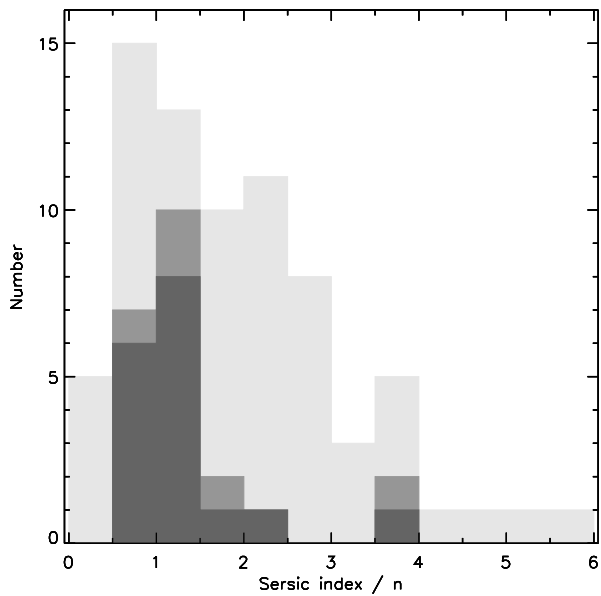,width=0.5\textwidth}
  \caption{The S\'{e}rsic index distribution for the 22 primary (sub)millimetre galaxy counterparts as derived from our two-dimensional modelling. The grey shading from darkest to lightest indicates {\bf i)} the secure (i.e. $P<0.1$) (sub)millimetre galaxy identifications, {\bf ii)} the unsecure (i.e. $P>0.1$) identifications and, for comparison {\bf iii)} the S\'{e}rsic index distribution displayed by a complete mass-selected sample of 83 comparably massive galaxies at $2 < z < 3$, as deduced by Bruce et al. (2012) from directly-comparable CANDELS $H_{160}$ imaging of the UDS field.}
 \end{figure}
\end{center}

Comparing these new results to those already in the literature requires consideration of the average values. For completeness, in objects which required multiple-component fits we determined separate average scalelengths and S\'{e}rsic indices using results from both the dominant component, and a single component fit to the whole system. We find that both of these methods produce virtually identical results (average agreement within 5\%), indicating that (sub)millimetre galaxy hosts are reasonably extended ($\langle r_{1/2} \rangle = 4.3 \pm 0.5$\,kpc; median $r_{1/2}=4.1$\,kpc) disk galaxies ($\langle n \rangle = 1.1 \pm 0.1$; median $n = 1.0$). These values were calculated excluding the relatively low-redshift bulge-dominated sources AzTEC.GS27 and AzTEC.GS38 as explained above. If the two bulge-dominated sources are included, the conclusion obviously remains virtually unchanged with no change in scalelength, and only a slight increase in average S\'{e}rsic index ($\langle r_{1/2} \rangle = 4.3 \pm 0.5$\,kpc, $\langle n \rangle = 1.4 \pm 0.1$).

These new measurements of scale-length and S\'{e}rsic index for our (sub)millimetre galaxy sample are in excellent agreement with those derived from ground-based $K$-band data by Targett et al. (2011). A further exploration of of the extent to which {\it HST} and ground-based imaging yield comparable results is presented in subsection 5.1.1 below.

The axial ratio (b/a) distribution of our (sub)millimetre galaxy sample is plotted in Fig.~6. This provides an alternative (and model-independent) test of galaxy morphology. Although a detailed statistical comparison is limited by small sample size, it can be seen that the distribution of axial ratios for the (sub)millimetre galaxies is very broad and flat, extending to values $\simeq 0.2$, as expected for a population of disk galaxies (Sandage et al. 1970). In contrast, the axial ratio distribution for a mass-selected sample of galaxies with $M_{\star} > 10^{11}\,{\rm M_{\odot}}$ in the redshift range $2 < z < 3$ from Bruce et al. (2012) also plotted in Fig.~6, is peaked around $(b/a)_{peak} \simeq 0.7$ with tails extending to $(b/a) = 0.1$ and $(b/a) = 1.0$.

 \begin{figure}
  \epsfig{file=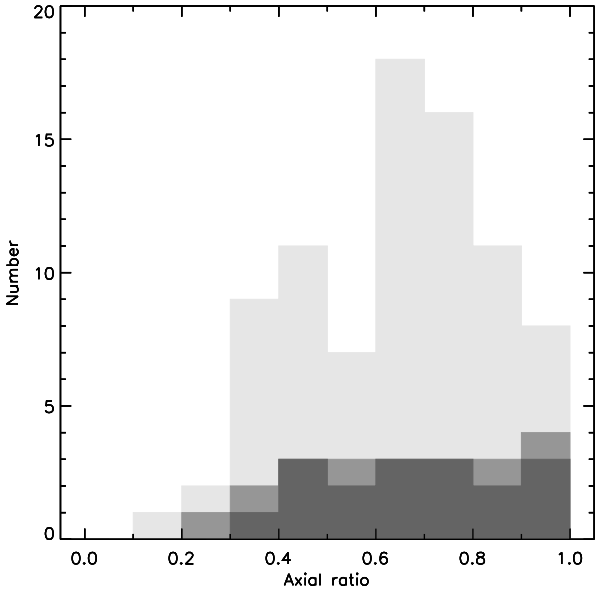,width=0.5\textwidth}
  \caption{The distribution of axial ratios (b/a) for the 22 primary (sub)millimetre galaxy counterparts as derived from our two-dimensional modelling. The grey shading from darkest to lightest indicates {\bf i)} the 17 secure (sub)millimetre galaxy identifications, {\bf ii)} the 5 less secure (sub)millimetre galaxy identifications and, {\bf iii)} for comparison, the axial-ratio distribution displayed by the 83 comparably-massive galaxies in the CANDELS UDS field at $2 < z < 3$, as deduced by Bruce et al. (2012), also using {\sc galfit} applied to the directly-comparable CANDELS UDS $H_{160}$ imaging.}
 \end{figure}

Thus, while there has been much debate in the literature concerning why the axial-ratio distribution of disk galaxies at $z \simeq 2$ appears more peaked than expected from a population of randomly-oriented thin disks (e.g., Bruce et al. 2012; Law et al. 2012a,b) the picture presented by the (sub)millimetre galaxies seems to be relatively straightforward; judged either by typical S\'{e}rsic index or axial ratio distribution, {\it (sub)millimetre galaxies are massive, star-forming disks}. We stress that this conclusion stands irrespective of what the mechanism for triggering or fuelling the star-formation activity in these galaxies might be (a point discussed further below).

\subsubsection{Comparison to ground-based observations}

In conjunction with the CANDELS {\it HST} imaging, deep ground-based $K_s$-band imaging of GOODS-South has recently been obtained with the HAWK-I imager on the VLT. As described in section 2.3.3, these data were taken in excellent seeing ({\it FWHM} $\simeq$ 0.45\,arcsec) and reach a detection limit of $\simeq$26.0 (AB mag, 5-$\sigma$, in a 1\,arcsec diameter aperture), a depth comparable to that achieved by the CANDELS {\it HST} $H$-band imaging. The HAWK-I data have been reduced and processed as described in Fontana et al. (2012). 

This new deep HAWK-I imaging allows us to make a {\it direct} test of the extent to which the basic morphological parameters derived from modelling the high-resolution WFC3/IR $H_{160}$ imaging agree or disagree from those that would be derived with {\sc galfit} from poorer-resolution ground-based imaging available (albeit at slightly longer wavelengths). We have thus fitted axi-symmetric models to the $K_s$-band images of the fourteen (sub)millimetre galaxies in our sample which lie within the HAWK-I imaging. The lower resolution of the ground-based data does not allow multiple component fits to the more complex systems, and therefore might be expected to be more comparable to the single-component model fits (see Table 2) attempted to the CANDELS data. 

The results of this $K_s$-band modelling are summarised in Table 5. The modelling was carried out in the same manner as described for the modelling of the $H_{160}$ imaging in Section 4. We find that the {\it average} half-light radii and S\'{e}rsic index of both the single and dominant-component $H_{160}$-band model fits match those derived from the $K_s$-band images to within 5\%. A comparison of the {\it HST} and ground-based results for individual objects is shown in Fig.~7. These results suggest that, despite the extra detail revealed by the {\it HST} imaging, the basic sizes and morphologies of (sub)millimetre galaxies can be successfully recovered from the very best ground-based $K_s$-band imaging, explaining the similarity between the results presented here and those reported by Targett et al. (2011), and consistent with our finding that single-component fits to the high-resolution CANDELS images correctly recover the basic morphology of the mass-dominant disk-like components.

\begin{table} 
 \begin{center}
\caption{Results from two-dimensional modelling of AzTEC and LABOCA selected (sub)millimetre galaxies, as derived from the VLT HAWK-I $K_s$-band imaging. Column 1 gives the source name. Column 2 lists the total host-integrated $K_s$-band mag (AB). Column 3 lists the semi-major axis scalelength (half-light radius) of the host galaxy fit in kiloparsec. Column 4 lists the best-fitting value of the S\'{e}rsic index ($n$). Column 5 gives the axial ratio of the host galaxy. Column 6 gives the value of reduced $\chi^{2}_{\nu}$ for each model fit.}
  \begin{tabular}{llllllll}
\hline\hline
Name & $K$-mag & $r_{1/2}$ & S\'{e}rsic $n$ & Axial ratio& $\chi^{2}_{\nu}$\\
\hline
AzTEC.GS16   & 23.11 & 5.1 & 1.5 & 0.62 & 0.914\\
AzTEC.GS17   & 21.89 & 1.4 & 1.7 & 0.85 & 1.036\\
AzTEC.GS18   & 23.28 & 3.1 & 1.0 & 0.70 & 0.999\\
AzTEC.GS19-1 & 21.85 & 2.3 & 1.5 & 0.66 & 0.787\\
AzTEC.GS19-2 & 21.23 & 3.7 & 0.5 & 0.65 & 0.787\\
AzTEC.GS21   & 21.37 & 3.4 & 1.4 & 0.73 & 1.298\\
AzTEC.GS22   & 22.65 & 3.7 & 1.2 & 0.40 & 0.967\\
AzTEC.GS24   & 24.07 & 1.5 & 0.8 & 0.50 & 1.068\\
AzTEC.GS26   & 21.61 & 8.2 & 1.7 & 0.36 & 1.021\\
AzTEC.GS27   & 20.05 & 2.1 & 3.5 & 0.87 & 1.082\\
AzTEC.GS30   & 20.77 & 4.5 & 1.3 & 0.70 & 1.077\\
AzTEC.GS34   & 21.11 & 1.4 & 1.2 & 0.81 & 0.858\\
AzTEC.GS35   & 21.94 & 3.3 & 0.7 & 0.74 & 0.734\\
AzTEC.GS38   & 18.32 & 4.1 & 4.6 & 0.99 & 6.542\\
\hline
  \end{tabular}
 \end{center}
\end{table}

\begin{center}
 \begin{figure*}
 \begin{tabular}{cc}
  \epsfig{file=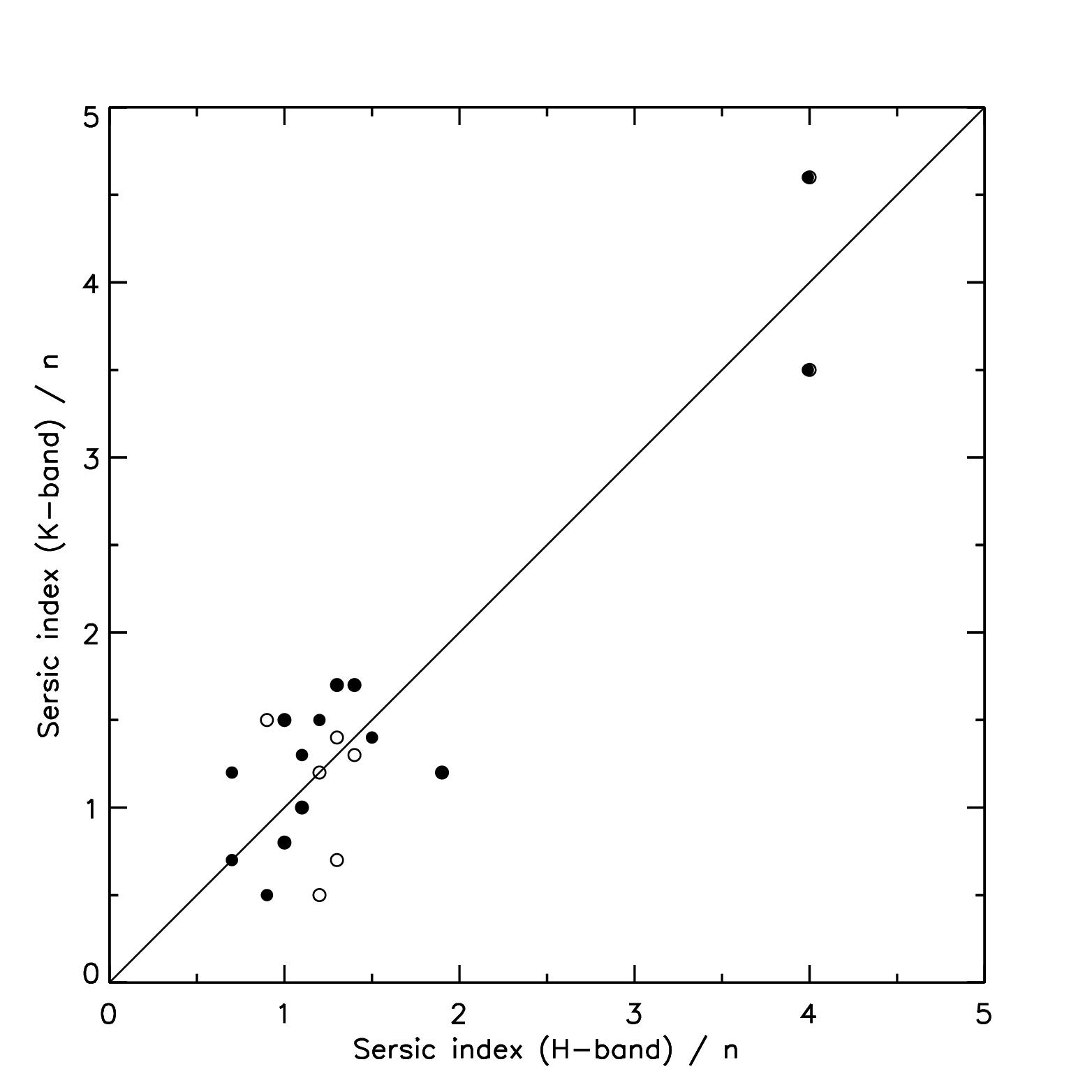,width=0.5\textwidth}&
  \epsfig{file=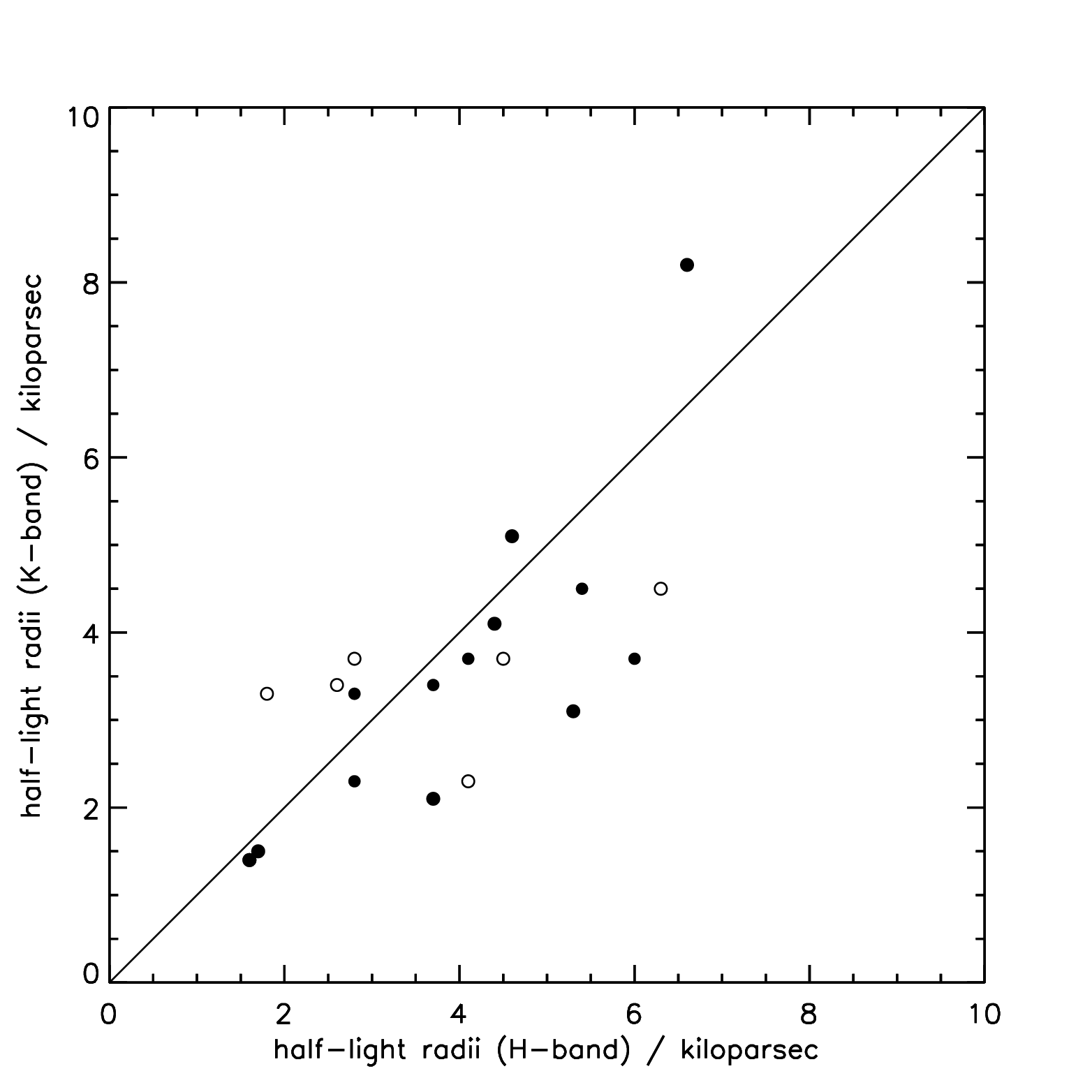,width=0.5\textwidth}\\
 \end{tabular}
  \caption{Comparison of two-dimensional model S\'{e}rsic index (left) and semi-major half-light radii (right) from (sub)millimetre galaxies in the $H_{160}$-band {\it HST} CANDELS data and the $K_s$-band HAWK-I VLT imaging. The values shown for the WFC3/IR imaging represent those of the single-component fits to the (sub)millimetre galaxies (filled circles), although the results remain unchanged (on average) if the dominant model component is adopted (open circles).}
\end{figure*}
\end{center}

\subsubsection{Comparison to ULIRGs in CANDELS}

Recently, Kartaltepe et al. (2012) have studied the morphologies of a sample of 52 Ultra Luminous Infrared Galaxies (ULIRGs) in the CANDELS GOODS-South field, selected at 100 and 160\,$\mu m$ from Herschel PACS imaging. Based primarily on visual classification, Kartaltepe et al. (2012) reported that the majority of the GOODS-South ULIRGs were mergers or interacting sources, with a relatively small fraction of isolated disks. 

However, it is not clear that these results are in conflict with those found here for the (sub)millimetre-selected galaxies, and in fact it can be argued they are perfectly consistent. First, there is relatively little overlap between the two samples; the ULIRGs were selected at much shorter wavelengths and the resulting sample is therefore inevitably dominated by hotter and/or lower-redshift objects, and such a sample may well contain a higher fraction of compact merger-driven starbursts and/or more significant bulges. There are in fact only six objects in common (AzTEC.GS06, GS08, GS17, GS21, GS24 and LESSJ033219), and Kartaltepe et al. (2012) classify three of these as isolated disks, and three as irregular, a result very comparable to our own findings (albeit based on only a small subsample of our sources). Second, it is clear that visual classification inevitably places a stronger emphasis on high surface brightness features and morphological peculiarities than the multi-component modelling approach adopted here; several of our massive disk galaxies (especially those with clumps and/or companions) might well be classified as irregular and/or disturbed via purely visual classification. Third, Kartaltepe et al. (2012) did also use {\sc galfit} to determine scalelengths and S\'{e}rsic indices for their ULIRGs, and their results are not really very different from our own; they report a median $R_e = 3.3$\,kpc, making the ULIRGs only $\simeq 20$\% more compact than our (sub)millimetre galaxies (median $R_e = 4.0$\,kpc), and a median S\'{e}rsic index $n = 1.4$ (also see Bussmann et al. 2011, who find $n < 2$ for the majority of a sample of ULIRGS at $z\sim2$). This value is not as low as our own median S\'{e}rsic $n = 1.1$, but still consistent with a disk-dominated population).

\subsection{Redshifts and stellar masses}

The redshift distribution of our (sub)millimetre galaxy sample is plotted in Fig.~8. We emphasise that $\simeq 50$\% of these redshifts are spectroscopic, (including the one galaxy at $z > 4.5$, AzTEC.GS12) and that Fig.~4 provides confidence that the photometric redshifts are among the most accurate ever derived for (sub)millimetre galaxies. This redshift distribution is thus essentially complete and unbiased, and is broadly as expected given studies of other, larger, (sub)millimetre-selected samples of galaxies (e.g. Chapman et al. 2005; Wardlow et al. 2011; Micha{\l}owksi et al. 2012b). Viewed together, Figs~3, 5, 6, and 8 provide compelling evidence that (sub)millimetre selection primarily picks out star-forming disk galaxies at the redshift corresponding to the peak epoch of star-formation density in cosmic history.

These disks are also all massive. As tabulated in Table 4, our SED fitting, assuming a Chabrier (2003) IMF, reveals all of the (sub)millimetre galaxies to have stellar masses $M_{\star} > 10^{11}\,{\rm M_{\odot}}$, with an average stellar mass $\langle \rm{M}_{\star} \rangle = 2.2\times10^{11} \pm 0.2\, {\rm M_{\odot}}$. Again these results are consistent with much of the recent literature (e.g. Borys et al. 2005; Dye et al. 2008; Targett et al. 2011; Micha{\l}owski et al. 2012a); the reader is referred to Micha{\l}owski et al. (2012a) for a detailed discussion of how the the choice of stellar population model, star-formation history and IMF can influence the estimation of stellar mass (the key points are that adoption of a Salpeter (1955) IMF increases stellar mass by a factor $\simeq 2$, while adoption of a single-component star-formation history generally reduces stellar mass by a comparable factor). As discussed in the next sub-section, the stellar masses and sizes of the (sub)millimetre galaxies appear to be entirely plausible in the context of mass-selected galaxy samples at comparable redshifts.
 
\subsection{The size-mass relation}
The average half-light radius and stellar mass of our (sub)millimetre galaxies is plotted in Fig.~9, overlaid on the distribution displayed by the general population of massive disk-dominated galaxies at $2 < z < 3$ as derived by Bruce et al. (2012). The curves on this plot indicate the size-mass relations observed for early- and late-type SDSS galaxies at low-redshifts from Shen et al. (2003) (see figure caption for details). 

The Bruce et al. (2012) results illustrate the generally observed evolution towards smaller sizes at a given mass with increasing redshift; massive disk galaxies at $ z\simeq 2$ seem to range in size from objects which are still (just) consistent with the local relation to galaxies almost an order of magnitude more compact. Viewed in this context, the (sub)millimetre galaxies appear reasonably extended for their mass ($\langle r_{1/2} \rangle = 4.5 \pm 0.5$\,kpc; median $r_{1/2}=4.0$\,kpc), consistent with the sizes of the larger objects in the Bruce et al. (2012) sample, and indeed with other massive star-forming disks at $z \simeq 2$ (Law et al. 2012a) 
and other (sub)millimetre galaxies (e.g. Targett et al. 2011; Mosleh et al. 2011). In fact, the (sub)millimetre galaxies are, on average, only a factor $< 1.5$ smaller 
than present-day galaxies of comparable stellar mass. 

Clearly these results are at least consistent with the view that that most (sub)millimetre galaxies are simply drawn from the high-mass end of the normal star-forming disk galaxy population at $z \simeq 2$ (as proposed, theoretically, by Dav\'{e} et al. 2010). We now briefly consider whether our imaging provides any significant evidence for or against the viewpoint that the high star-formation rates displayed by the (sub)millimetre galaxies are the result of (or are significantly enhanced by) major galaxy-galaxy interactions (as, for example, inferred from the theoretical work of Narayanan et al. 2010a,b, and arguably supported by the kinematical observational studies of Tacconi et al. 2008; Engel et al. 2010; Alaghband-Zadeh et al. 2012). 

 \begin{figure}
  \epsfig{file=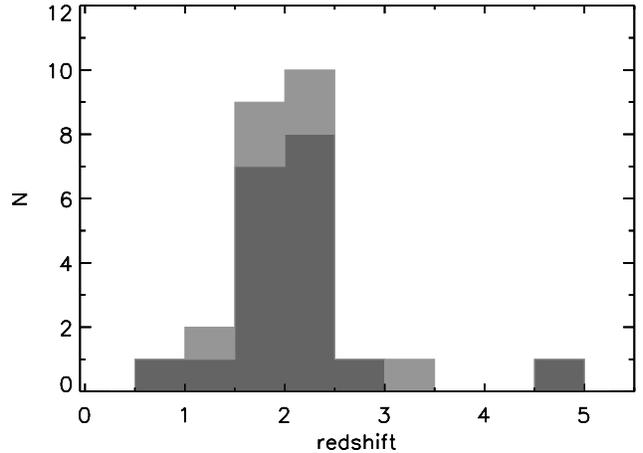,width=0.52\textwidth}
  \caption{The redshift distribution for our full sample of (sub)millimetre galaxies. The secure (i.e. $P<0.1$) identifications are shown with dark grey shading, while the less secure (i.e. $P>0.1$) identifications are indicated by light grey shading. Ten of these redshifts are the result of optical spectroscopy as summarised in Table 4, with the remainder being based on the SED modelling of the optical--near-infrared photometry as described in Section 4.2 (the reliability of which is supported by Fig.~4).}
 \end{figure}

\subsection{Clumps, companions, and interactions?}

Our results show that (sub)millimetre galaxies are large, massive star-forming disk galaxies at $z \simeq 1 - 3$. However, at least in principle, this does not necessarily prove that they all lie on the so-called ``main-sequence'' of star-forming galaxies at $z \simeq 2$ (e.g. Daddi et al. 2007; Elbaz et al. 2011; Wuyts et al. 2011). Even if essentially all (sub)millimetre galaxies are massive disks, at least some of their star-formation activity could be externally triggered. Indeed, as argued by Hayward et al. (2012) the (sub)millimetre galaxy population could contain a mix of objects including galaxies at the high-mass end of the star-forming main-sequence, along with other objects whose star-formation may have received a short-term boost from a galaxy-galaxy interaction (and potentially include blends of multiple (sub)millimetre bright sources).

The issue of galaxy interactions is arguably best addressed with the velocity information provided by spectroscopy (although even then the interpretation of the data at this epoch of generally violent star-formation activity is not straightforward). However, we can at least attempt to quantify the fraction of potential major galaxy mergers in our sample from our multi-wavelength broad-band imaging data, and our multi-component S\'{e}rsic modelling.

One simple way to do this is to investigate the relative $H$-band flux density ratio of the primary and brightest secondary components seen in the WFC3/IR imaging of each object. This is shown (plotted in terms of magnitude difference) in Fig.~10. Here it can be seen that only $\simeq 1/3$ of the (sub)millimetre disk galaxies have even a potential companion object which is more than $\simeq 0.1$ times as bright as the dominant component. Treating $H$-band brightness as the best proxy we have for the stellar mass of individual sub-components, and bearing in mind that there is no evidence that all the brighter companion objects actually lie at the same redshifts as the dominant (sub)millimetre disk-galaxy counterparts, it seems clear that only a minority of the (sub)millimetre galaxies (probably $< 25$\%) can be viewed as potentially involved in a major (i.e. $> 1:3$ mass ratio) galaxy-galaxy merger. This provides further support for the conclusion reached by Micha{\l}owski et al. (2012a), based on the independent calculation of specific star-formation rates, that most (sub)millimetre galaxies are not primarily the result of extreme merger-driven starbursts, but in fact lie on the main-sequence of star-forming galaxies. with star-formation rates broadly as expected at $z \simeq 2-3$ given their redshifts and stellar masses.

Nevertheless, it is clear that, in {\it HST} imaging of this depth and quality, secondary components can be uncovered at some level for most of the galaxies in our sample. The key question is whether these components are generally significant, or indeed unusual at this epoch of maximum star-formation activity in the gas-rich young universe.

\begin{figure}
  \epsfig{file=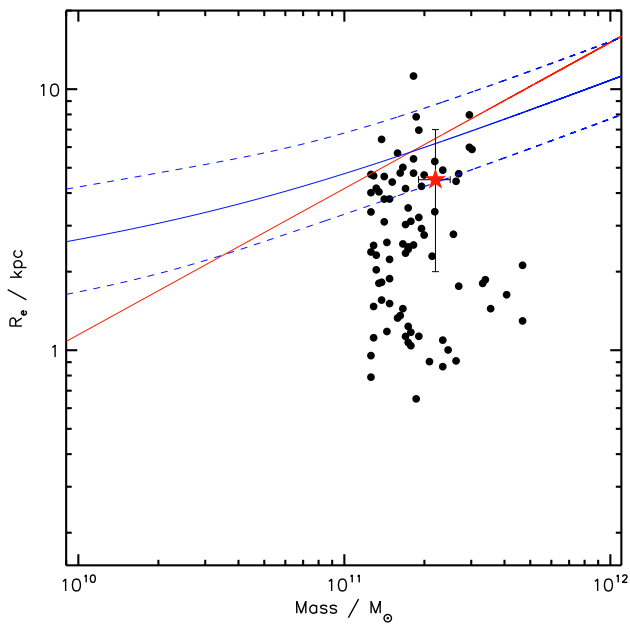,width=0.52\textwidth}
  \caption{The location of our (sub)millimetre galaxy sample on the size-mass plane is indicated by the red star (with the error bar indicating the 1-$\sigma$ range of the full size distribution). The black circles indicate the positions of the massive disk galaxies at $z>2$ studied by Bruce et al. (2012). The solid blue and red lines show the size-mass relations deduced by Shen et al. (2003) for local early- and late-type galaxies respectively, with the dashed blue lines representing the 1-$\sigma$ scatter for the late-type galaxies.}
 \end{figure}

We can attempt to investigate this issue in two ways, first by reviewing the results of our attempts to model any apparently significant multiple-component structures in the $H_{160}$ imaging (as illustrated in Fig.~1, and shown for all objects in the Appendix in Fig.~A1), and second by considering the multi-frequency imaging of each object (presented in Fig.~A2).

As discussed in Section 4.1, we attempted to model all apparently significant substructures or companions using variable S\'{e}rsic-index galaxy models. This was done both to help obtain satisfactory model fits to the primary galaxy identification, and to see if any useful information could be gleaned from the modelling of the secondary sources. As can be seen from the summary of the modelling presented in Table 2, in practice models were fitted to secondary components for approximately half of the (sub)millimetre sources. It can also be seen that the modelling was largely confined to secondary sources/components with $H_{160} < 24.5$. This is simply because the modelling of fainter subcomponents was neither necessary nor practical, and as a consequence of this it transpires that all 7 of the brighter secondary sources in the left-hand peak of Fig.~10 were modelled, while only 2 of the relatively fainter secondary sources were modelled, with the flux-densities of most of the fainter clumps/companions being derived from aperture photometry. An interesting feature of the results presented in Table 2 is that the modelling of many of the fainter secondary sources delivered very low values of the S\'{e}rsic index $n < 0.5$. A good example is the case of LESSJ033243, already highlighted in the second row of plots in Fig.~1. Here the brightest secondary source appears to be a disk galaxy with $n = 0.8$, while the two even fainter objects (which appear to be distorted clumps), have $n = 0.4$. As a further example, the two subsidiary components of AzTEC.GS21 have $n=0.3$ and $n=0.2$. By contrast, no primary galaxy identification in the sample has $n < 0.8$. This suggests that S\'{e}rsic index can possibly be used to differentiate between genuine galaxy companions, and distorted clumps of excess emission. Some of the latter may be dwarf irregular galaxies, but they could equally just be viewed as "weather". If we apply the criterion that any genuine candidate for a major galaxy-galaxy interaction (or indeed significant (sub)millimetre source blending) must have a secondary component with a "sensible" S\'{e}rsic index, then we are left with only 6 examples of disk galaxies with potential galaxy companions (AzTEC.GS13, AzTEC.GS19, AzTEC.GS23, AzTEC.GS35, LESS0J33219, LESSJ033243), again representing $\simeq 25$\% of the sample.

Second, moving away from the detailed modelling, a review of the multi-wavelength imaging presented in Fig.~A2 reinforces the view that most complex substructure is blue, becoming less and less evident at longer wavelengths as the mass-dominant primary galaxy identification becomes progressively more important. Thus, while at rest-frame UV wavelengths many of the galaxies in our sample look complex, clumpy and sometimes distorted, it is clear that much of this structure must involve relatively insignificant fractions of the stellar mass of the galaxy which hosts the (sub)millimetre emission. This sort of behaviour is well illustrated by AzTEC.GS08 (the first example shown in Fig.~1) where the multi-wavelength imaging shown in Fig.~A2 shows that the clumps in the galaxy (which are too faint in $H_{160}$ to merit individual modelling) are very blue; viewed in the optical AzTEC.GS08 would certainly not be reported as a single isolated disk galaxy, and even at $J_{125}$ it could easily be mis-interpreted as a merger or disturbed system. Indeed, if the signal:noise in our CANDELS WFC3/IR $H$-band imaging is artificially degraded to the level typically achieved with NICMOS (e.g. Swinbank et al. 2010) it becomes clear that, even at $H$-band, imaging of inadequate depth would fail to properly uncover the underlying mass-dominant disk galaxy, instead simply revealing the highest surface-brightness features. This is perhaps not surprising; at $z \simeq 2.5$ the $H$-band is sampling the rest-frame $B$-band. Ideally, 
superior results could be obtained by moving to $K$ or IRAC wavelengths to determine the morphology of the mass-dominant galaxy. However, imaging at the necessary 
angular resolution at these longer wavelengths is not available with the present generation of instruments, but will be possible in the future with $\simeq 4\,{\rm \mu m}$ imaging 
on the the {\it James Webb Space Telescope}.

We close by noting that the prevalence of clumpy/complex structures in star-forming galaxies at $z \simeq 2$ is certainly not confined to (sub)millimetre galaxies. Indeed, as shown for example by Wuyts et al. (2012), most lower-mass star-forming galaxies at these epochs display clumpy, blue substructure, the origin and importance of which remains the subject of vigorous theoretical debate (e.g. Ceverino et al. 2012; Genel et al. 2012). In this sense, while it can be argued that $\simeq 25$\% of the (sub)millimetre galaxies in our sample could be involved in a significant galaxy-galaxy interaction, in general the galaxies selected at (sub)millimetre wavelengths do not seem unusual for star-forming galaxies at $z \simeq 2-3$ (except that they are among the most massive and largest galaxies at that epoch). There is in fact now considerable doubt over whether major mergers are common enough at $z \simeq 2 - 3$ to explain the prevalence of (sub)millimetre galaxies at this epoch (e.g. Hopkins et al. 2010). Moreover, as pointed out by Cen et al. (2012), that even for those $z \simeq 2$ (sub)millimetre galaxies which arguably do show some evidence of companions or tidal interactions, it does not necessarily follow that interactions and/or mergers are the primary cause of their high star-formation rates. Against the backdrop of our relatively peaceful present-day Universe, it is clear that galaxy-galaxy mergers can have a dramatic effect (e.g. in the production of the low-redshift ULIRGs first uncovered by the IRAS satellite), but amid the general frenzy of star-forming activity in the gas-rich universe at $z \simeq 2$ (e.g. Dekel et al. 2009) the {\it relative} impact of mergers on the star-formation rate of massive galaxies may be largely inconsequential.

\begin{figure}
 \epsfig{file=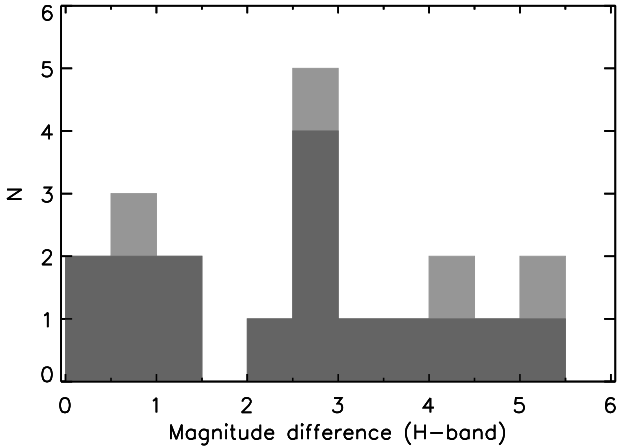,width=0.52\textwidth}
  \caption{The magnitude difference between the brightest and second-brightest $H$-band components for all 20 disk-dominated (sub)millimetre galaxies. As in previous figures, for completeness, the lighter grey shading is used to indicate the less reliable galaxy identifications (although clearly the basic result shown here is unaffected by this minor uncertainty). Only $\simeq 1/3$ of the (sub)millimetre galaxies have even a potential companion galaxy contributing $>10$\% of the $H$-band flux density of the primary mass-dominant disk galaxy.} 
  \end{figure}

\section{CONCLUSIONS}

We have exploited the {\it Hubble Space Telescope} CANDELS $J$ and $H$-band WFC3/IR imaging to study the properties of (sub)millimetre galaxies in the GOODS-South field which have been revealed by the AzTEC 1.1\,mm and LABOCA 870\,$\mu$m surveys of the Extended Chandra Deep Field South). After using the deep radio (VLA 1.4\,GHz) and {\it Spitzer} (IRAC 8\,$\mu$m) imaging to identify galaxy counterparts for the (sub)millimetre galaxies, we have then utilised the new CANDELS WFC3/IR imaging in two ways. 

First, the addition of new deep near-infrared photometry from both {\it HST} and (at $K$-band) HAWK-I on the VLT to the existing extensive GOODS-South multi-frequency database has enabled us to derive improved photometric redshifts and stellar masses for all the (sub)millimetre sources. Our results confirm that the (sub)millimetre sources are massive ($\langle {M}_{\star} \rangle = 2.2\times10^{11} \pm 0.2\,{\rm M_{\odot}}$) galaxies at $z \simeq 1-3$.
 
Second, we have exploited the depth and superior angular resolution of the WFC3/IR $H_{160}$ imaging to determine the sizes and morphologies of the galaxies at rest-frame optical wavelengths $\lambda_{\rm{rest}} > 4000$\AA. Crucially, the wavelength and depth of the WFC3/IR imaging enables modelling of the mass-dominant galaxy, rather than the blue high-surface brightness features which often dominate optical (rest-frame ultraviolet) images of sub-mm galaxies, and can confuse visual morphological classification. As a result of this analysis we find that $>95$\% of the rest-frame optical light in almost all of the (sub)millimetre galaxies is well-described by either a single exponential disk ($n \simeq 1$), or a multiple-component system in which the dominant constituent is disk-like. We demonstrate that this conclusion is completely consistent with the results of recent high-quality ground-based $K$-band imaging studies of (sub)millimetre galaxies sampling even longer rest-frame wavelengths (Targett et al. 2011), and explain why. We also briefly compare our findings with the results of recent morphological studies of high-redshift ULIRGs selected at much shorter ($\simeq 10\times$) far-infrared wavelengths (Kartaltepe et al. 2012).

The massive disk galaxies which host strong (sub)millimetre emission are reasonably extended ($\langle r_{1/2} \rangle = 4.5 \pm 0.5$\,kpc; median $r_{1/2}=4.0$\,kpc), consistent with the sizes of other massive star-forming disks at $z \simeq 2$. In many cases we find evidence of blue clumps within the sources, with the mass-dominant disk component becoming more significant at longer wavelengths. Interestingly, the only two bulge-dominated sub-mm galaxies are also the two lowest-redshift sources in the sample ($z \simeq 1$), a result which may reflect the structural evolution of high-mass galaxies in general.

Our main result that (sub)millimetre galaxies are almost universally massive, star-forming disk galaxies does not exclude the possibility that some fraction of these objects have their star-formation boosted by interactions/mergers. However, our imaging data indicate that only a minority ($< 25$\%) of the (sub)millimetre sources in the GOODS-South sample could be regarded as involved in a major galaxy-galaxy interaction. Taken together, our results support the view that most sub-mm galaxies at $z \simeq 2$ are simply the most extreme examples of normal star-forming galaxies at that era. 

\section*{ACKNOWLEDGEMENTS}

TAT and JSD acknowledge the support of the European Research Council via the award of an Advanced Grant. JSD and RJM acknowledge the support of the Royal Society via a Wolfson Research Merit award and a University Research Fellowship respectively. This work is based on observations taken by the CANDELS Multi-Cycle Treasury Program with the NASA/ESA HST, which is operated by the Association of Universities for Research in Astronomy, Inc., under NASA contract NAS5-26555. This work is based [in part] on observations made with the Spitzer Space Telescope, which is operated by the Jet Propulsion Laboratory, California Institute of Technology under a contract with NASA. The National Radio Astronomy Observatory is a facility of the National Science Foundation operated under cooperative agreement by Associated Universities, Inc. Based on observations made with ESO Telescopes at the La Silla Paranal Observatory under programme ID 181.A-0717(D) \& 186.A-0898(B).

\appendix

\section{Images and galaxy model fits}

In this Appendix we first show the CANDELS $H_{160}$ images of all the galaxies in our (sub)millimetre sample, along with the model fits and residual images (Fig.~A1), and then also provide a multi-wavelength montage for each object to reveal the wavelength-dependence of the structure seen in the imaging (Fig.~A2).

\begin{center}
\begin{figure*}
\begin{tabular}{ccc}
\epsfig{file=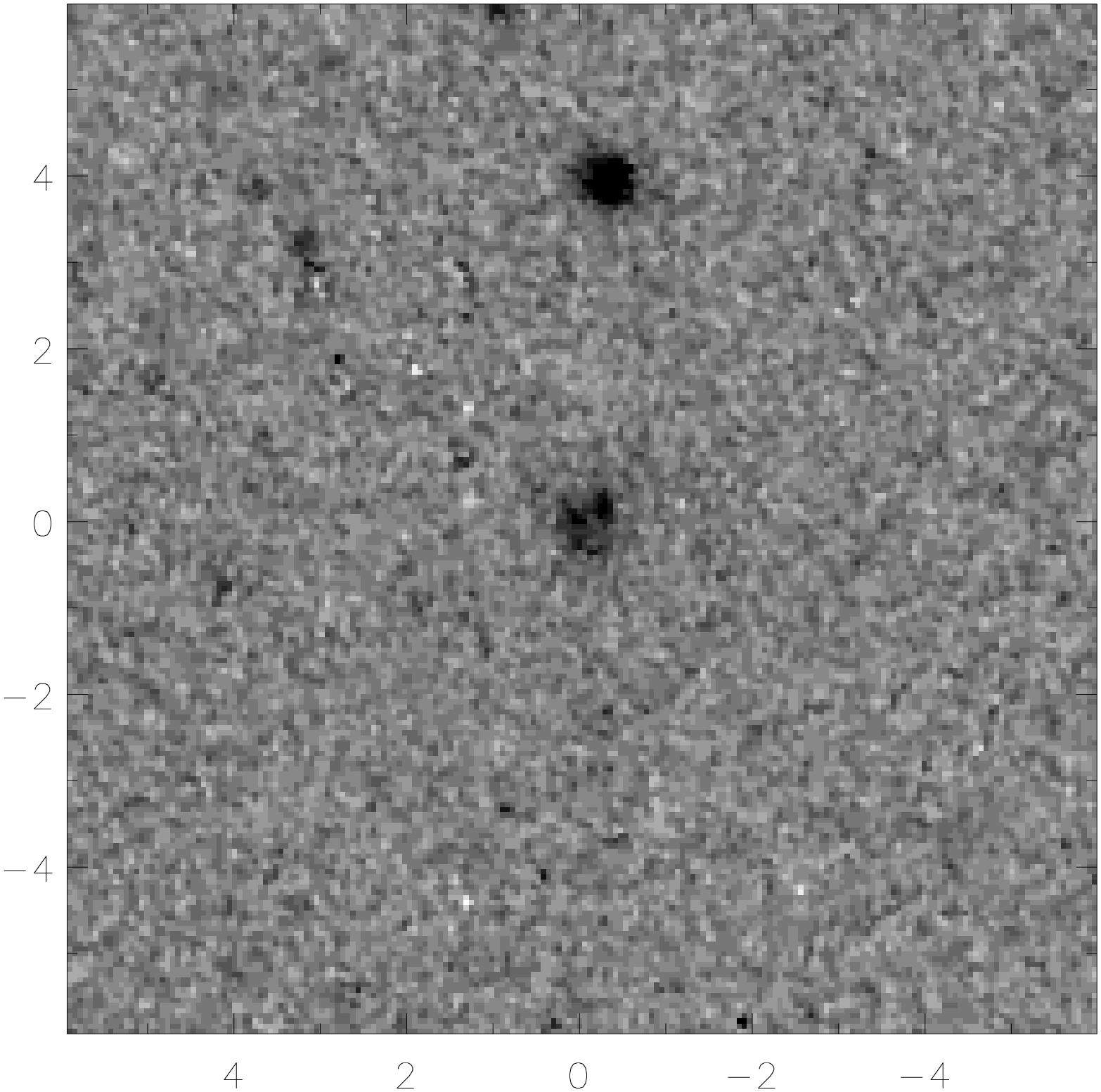,width=0.3\textwidth}&
\epsfig{file=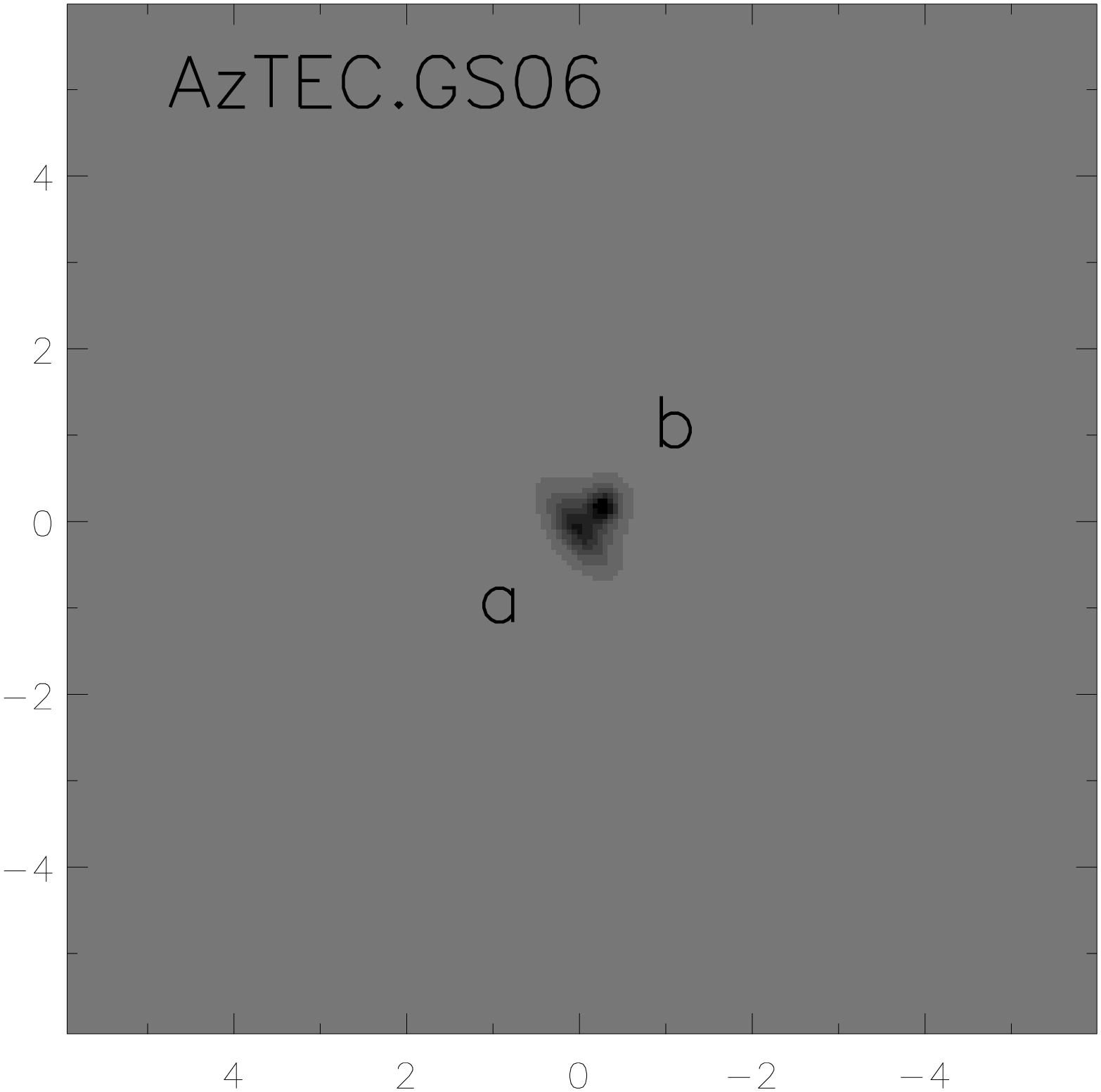,width=0.3\textwidth}&
\epsfig{file=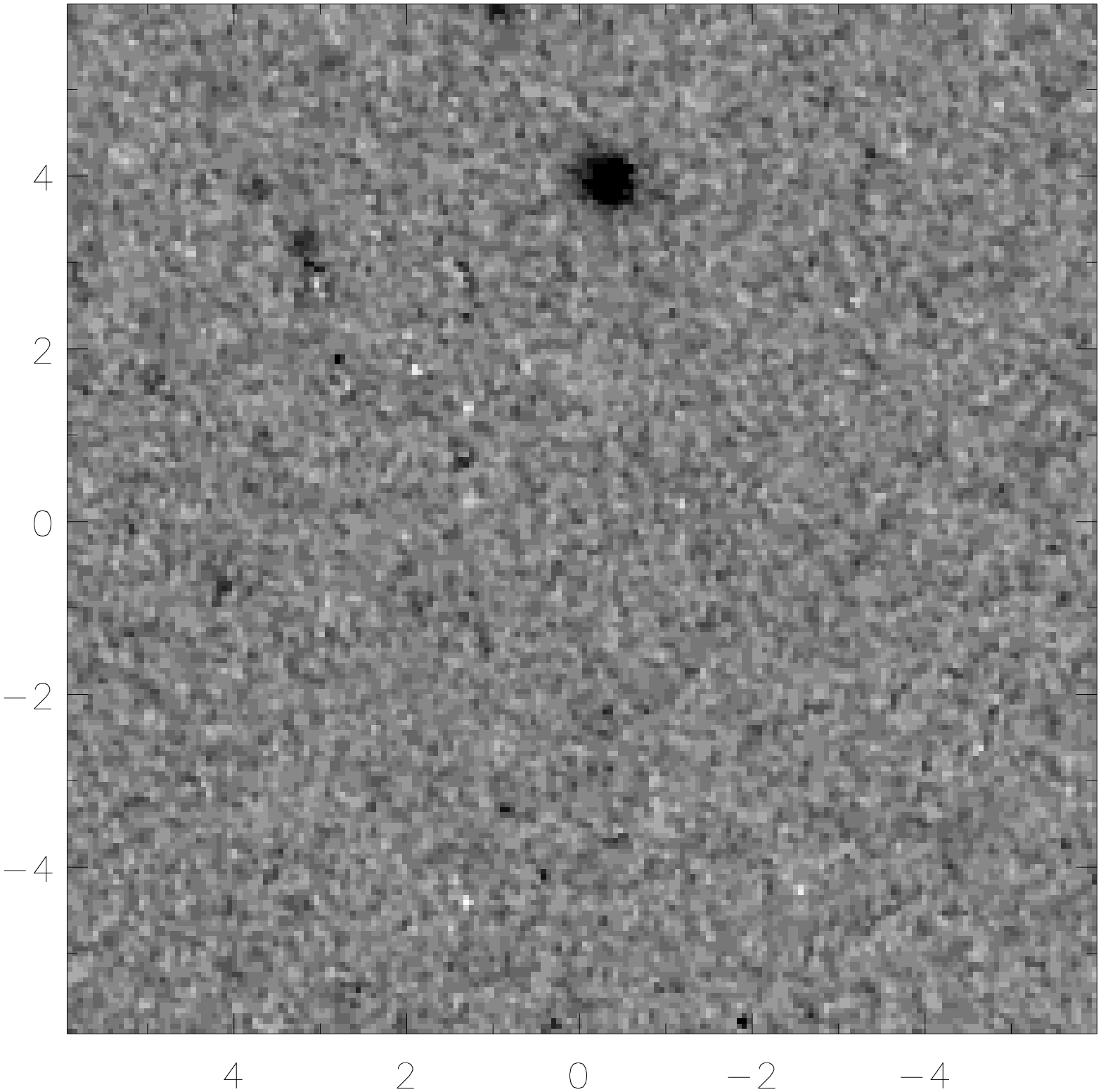,width=0.3\textwidth}\\
\\
\epsfig{file=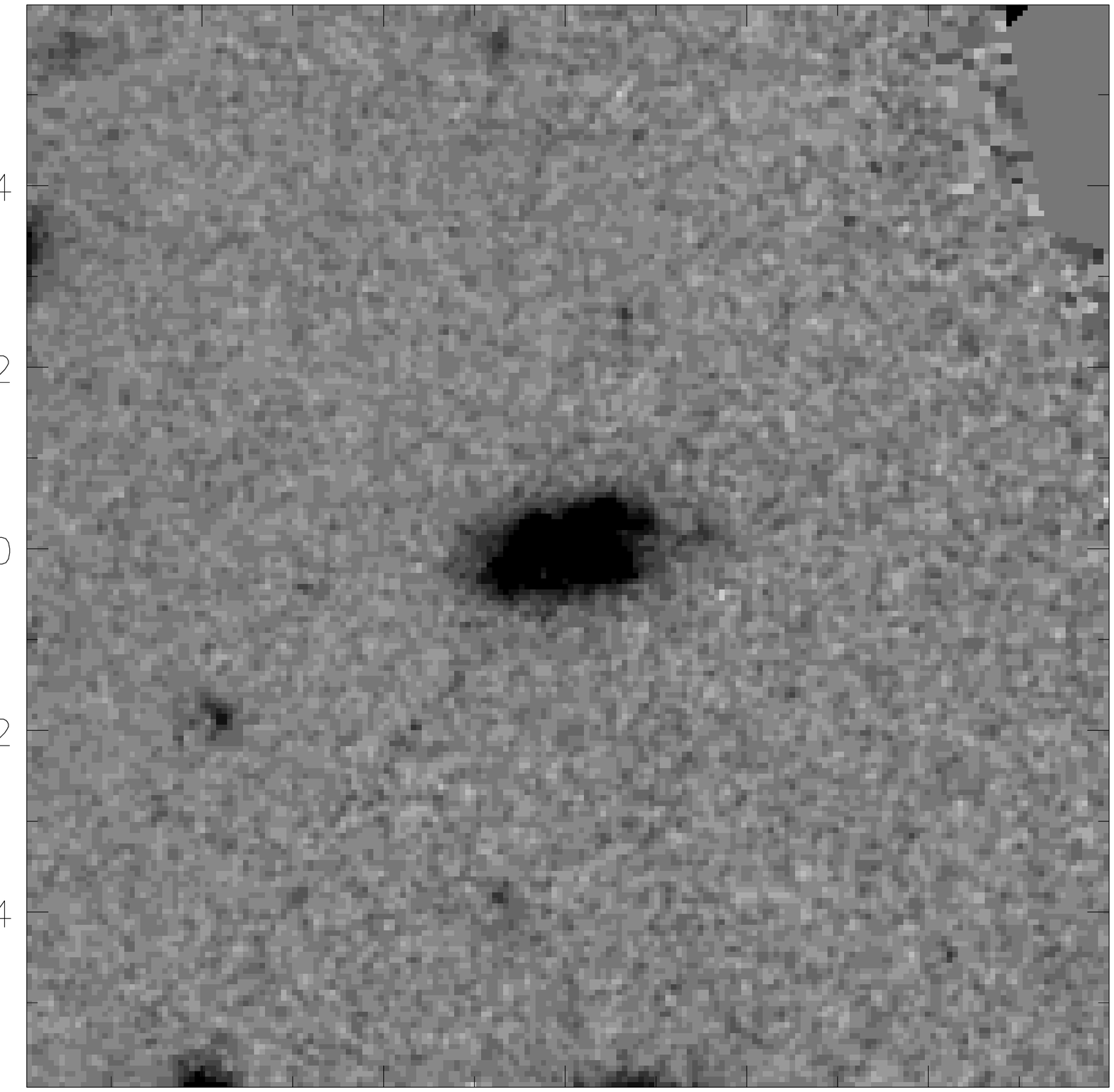,width=0.3\textwidth}&
\epsfig{file=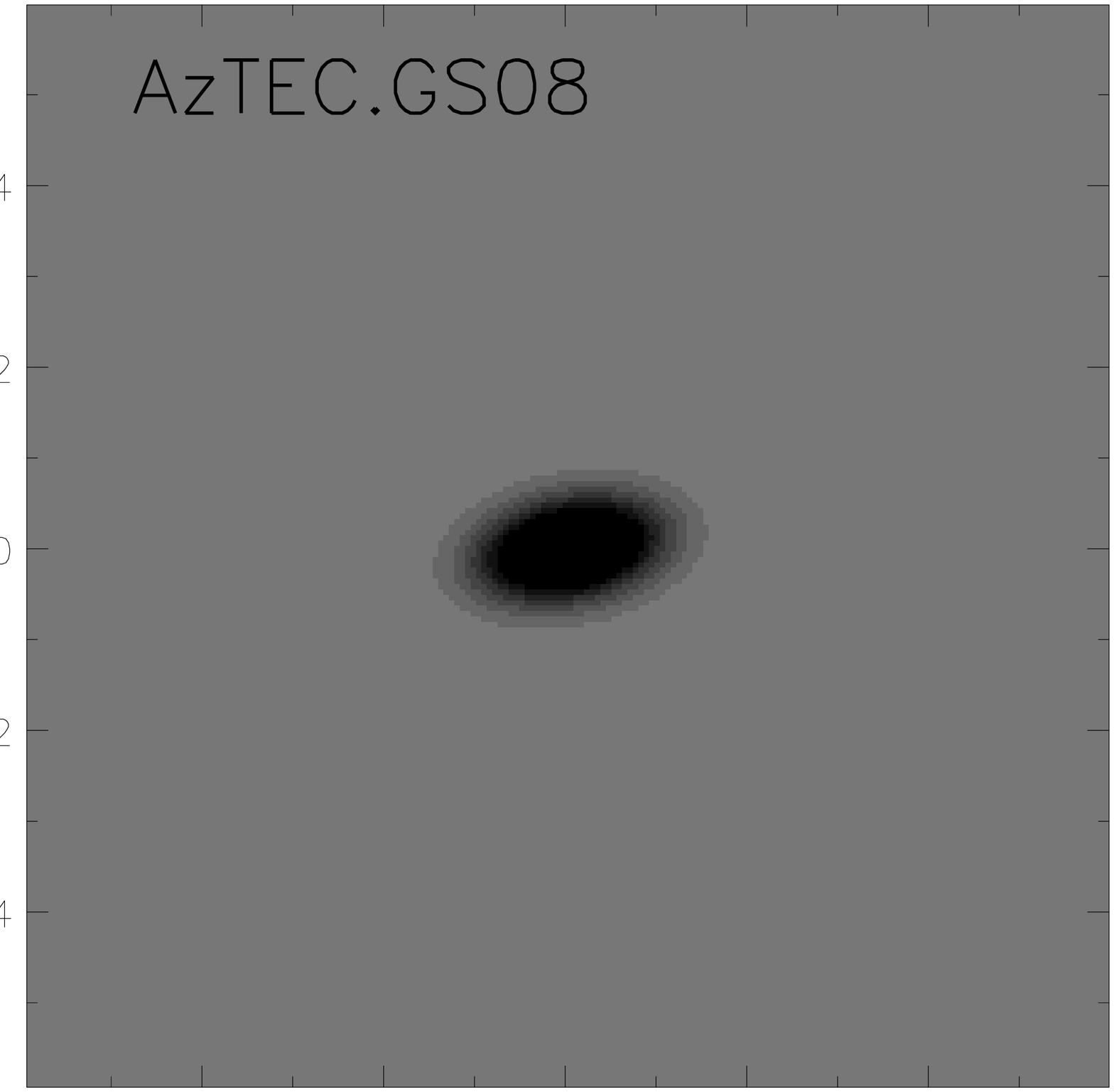,width=0.3\textwidth}&
\epsfig{file=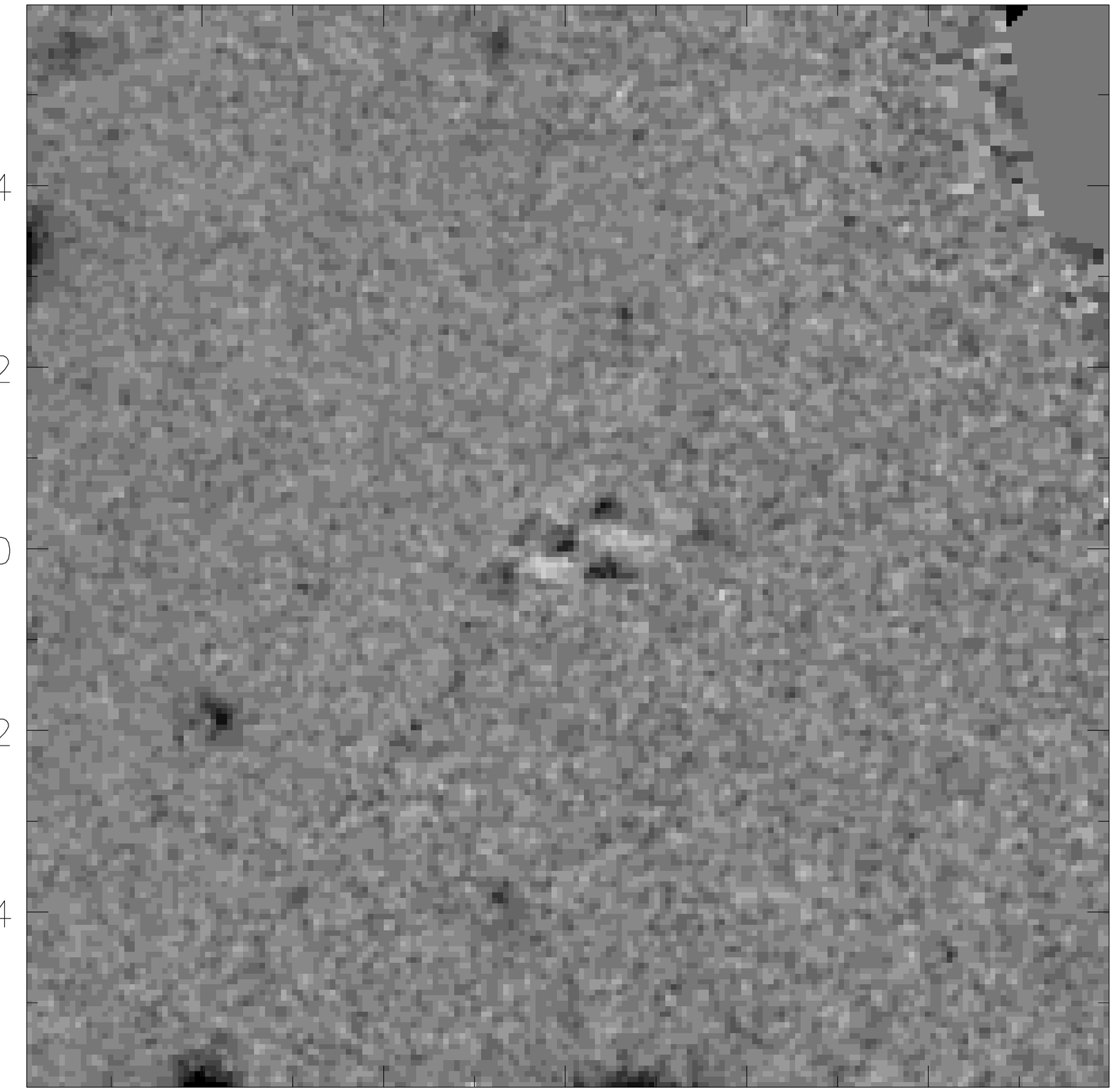,width=0.3\textwidth}\\
\\
\epsfig{file=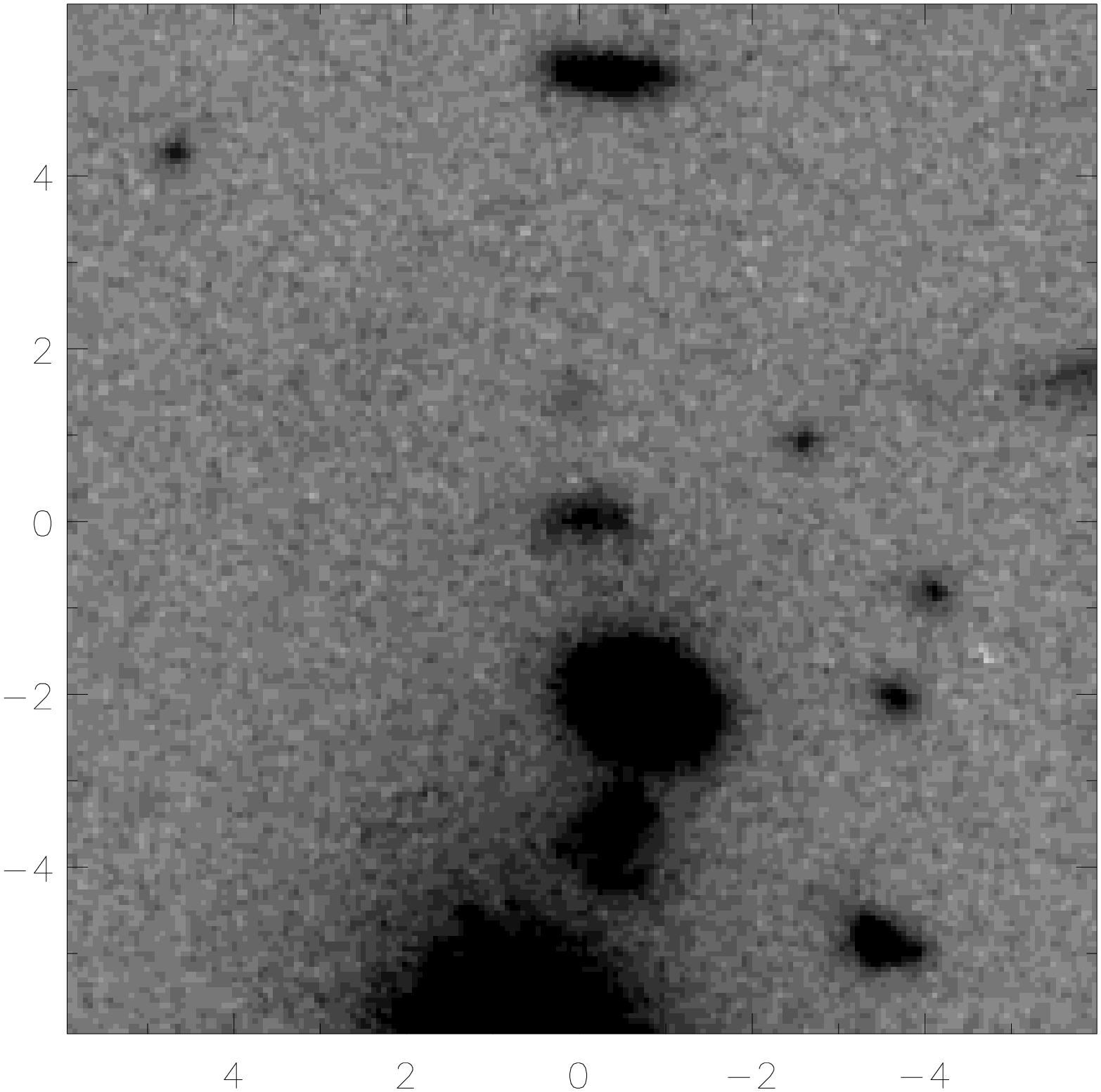,width=0.3\textwidth}&
\epsfig{file=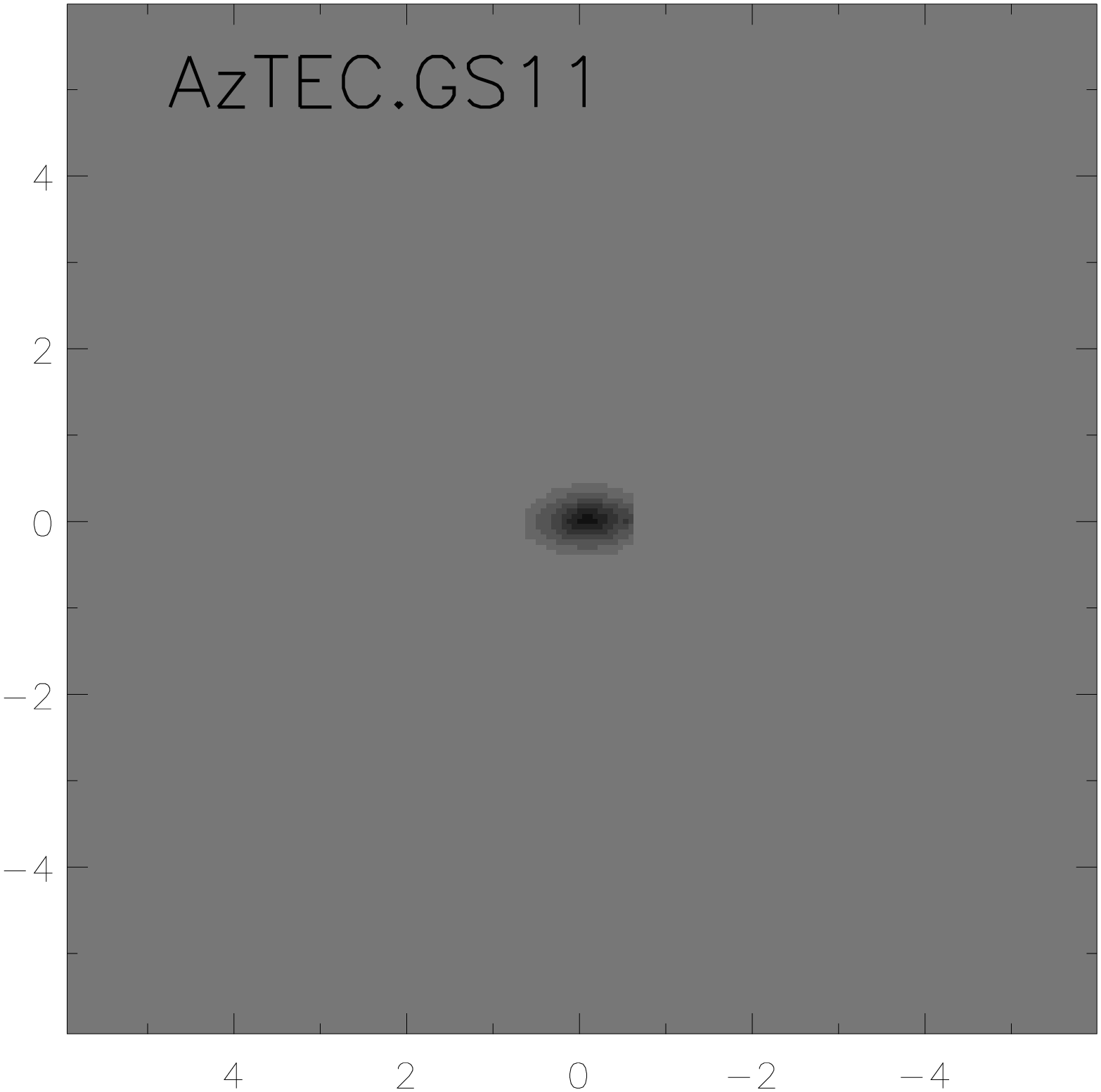,width=0.3\textwidth}&
\epsfig{file=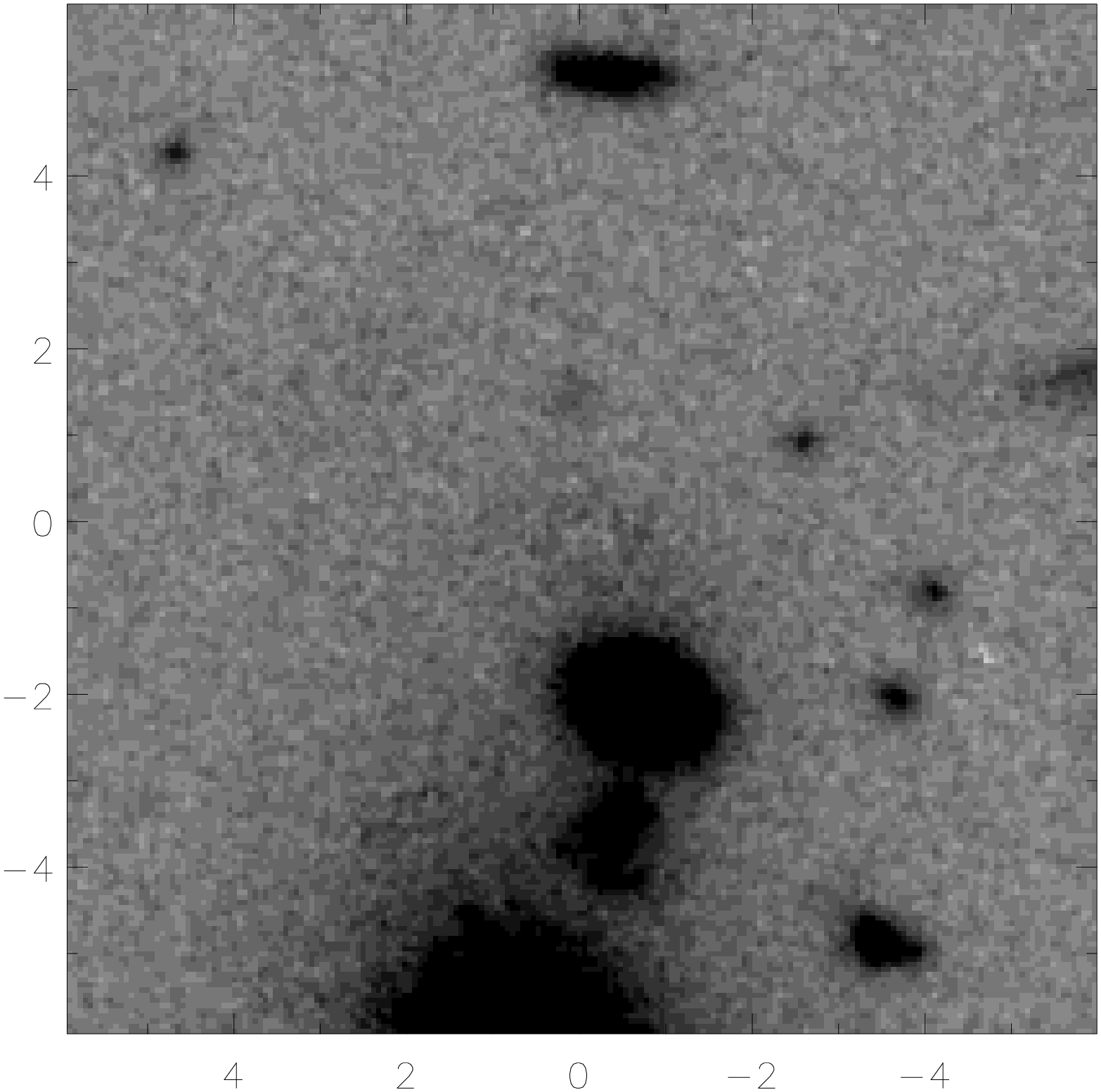,width=0.3\textwidth}\\
\\
\epsfig{file=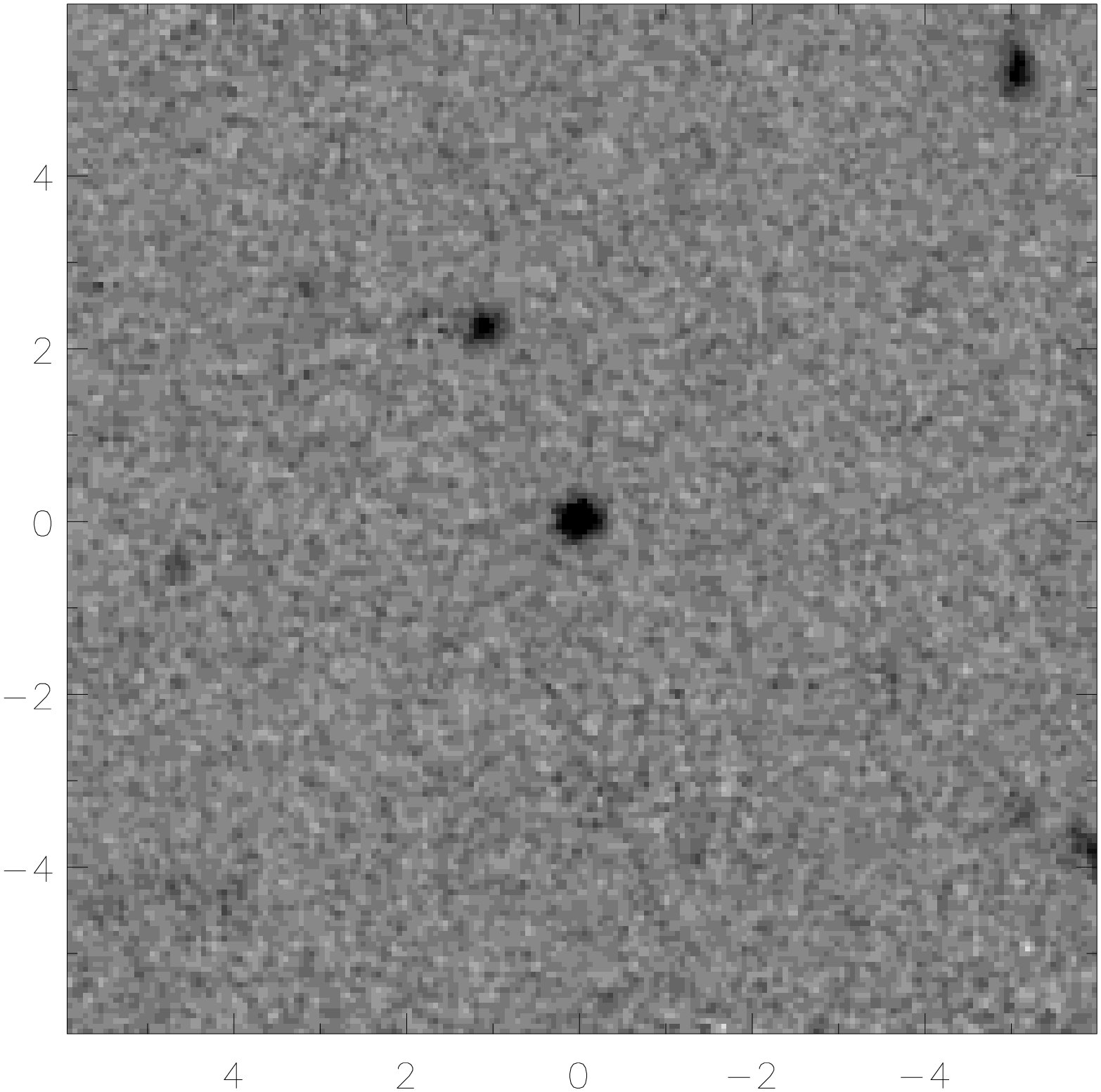,width=0.3\textwidth}&
\epsfig{file=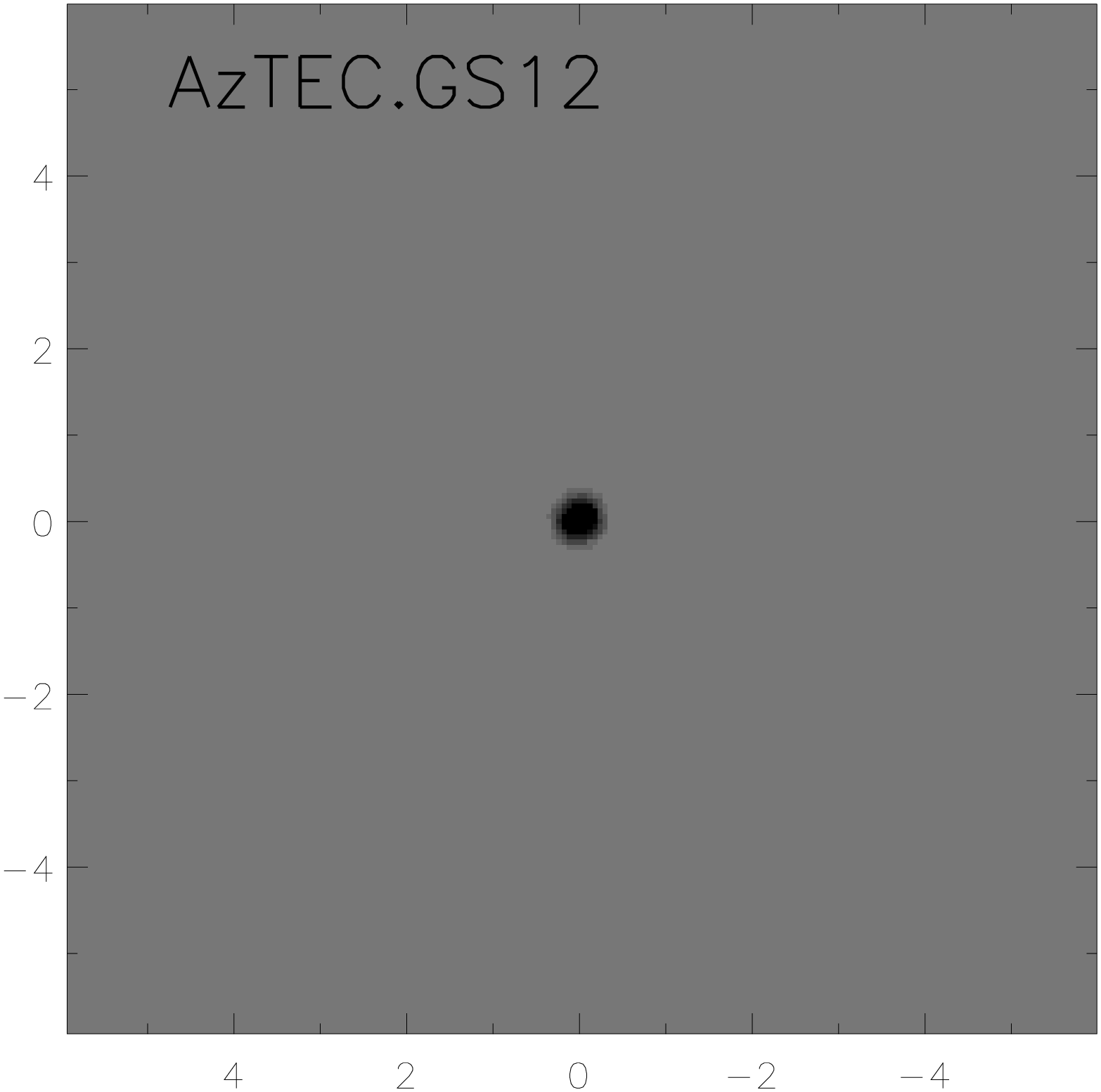,width=0.3\textwidth}&
\epsfig{file=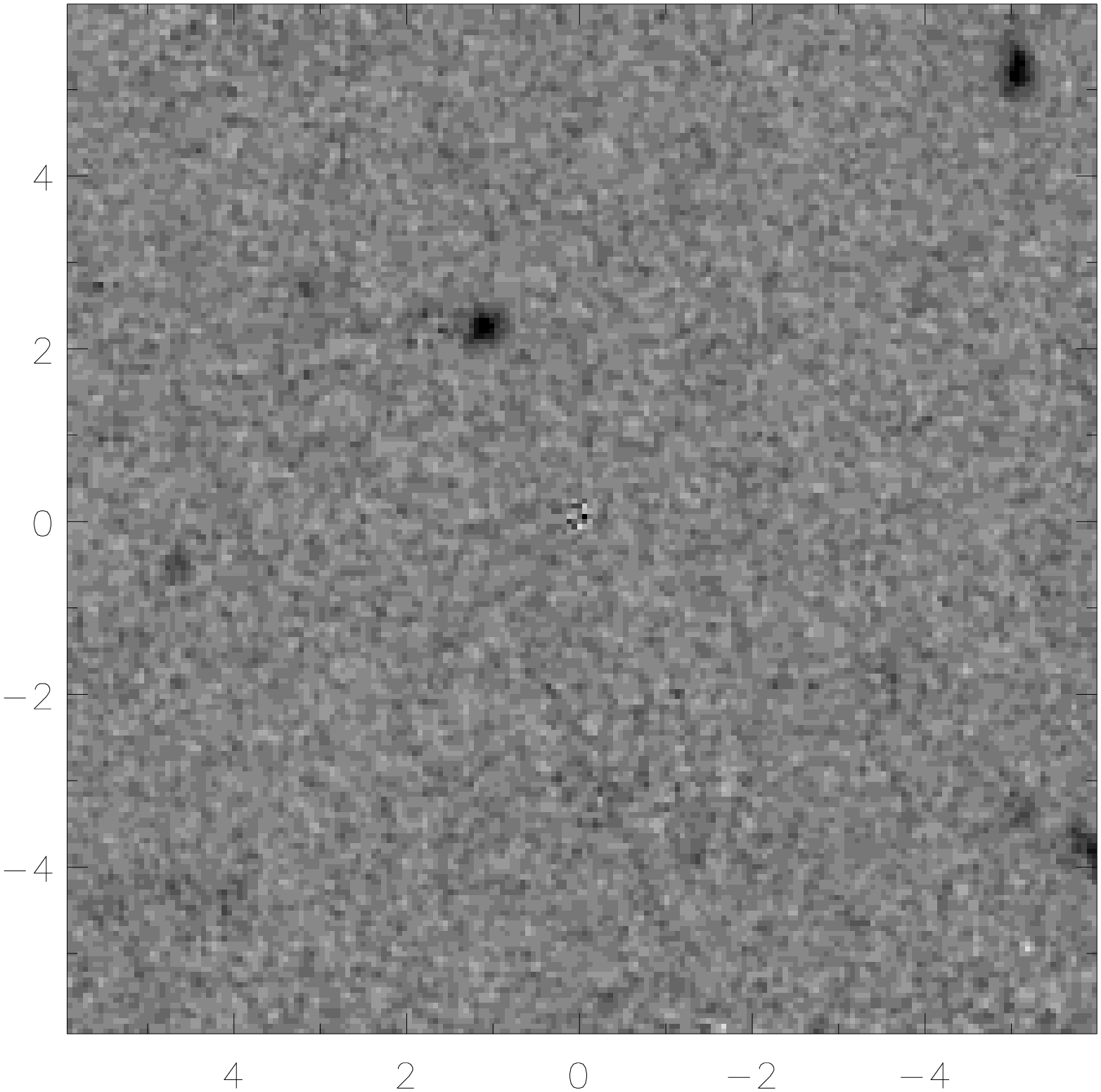,width=0.3\textwidth}\\
\end{tabular}
\caption{Two-dimensional modelling of the AzTEC and LABOCA selected (sub)millimetre galaxies in CANDELS GOODS-South. The left-hand panel shows the WFC3/IR $H_{160}$-band image centred on the (sub)millimetre galaxy counterpart. The centre panel shows the best-fitting two-dimensional galaxy model (convolved with the {\it HST} PSF). The right-hand panel shows the residual image after subtraction of the model from the data. All panels are 12.0$^{\prime \prime}$ $\times$12.0$^{\prime \prime}$, and the images are shown with a linear greyscale in which black corresponds to 10-$\sigma$ above, and white to 1-$\sigma$ below the median sky background value.}
\end{figure*}
\end{center}

\begin{center}
\begin{figure*}
\begin{tabular}{ccc}
\epsfig{file=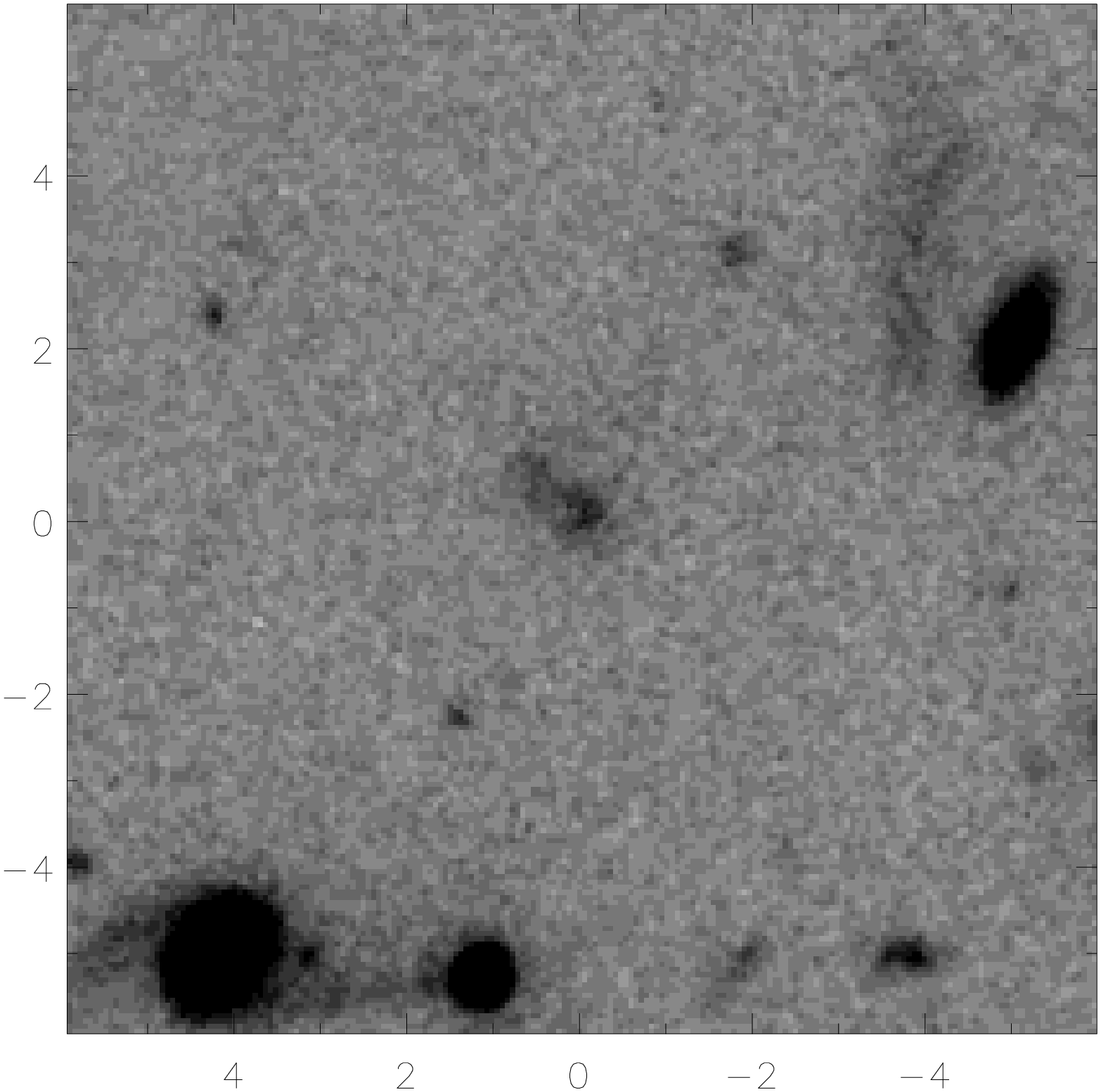,width=0.3\textwidth}&
\epsfig{file=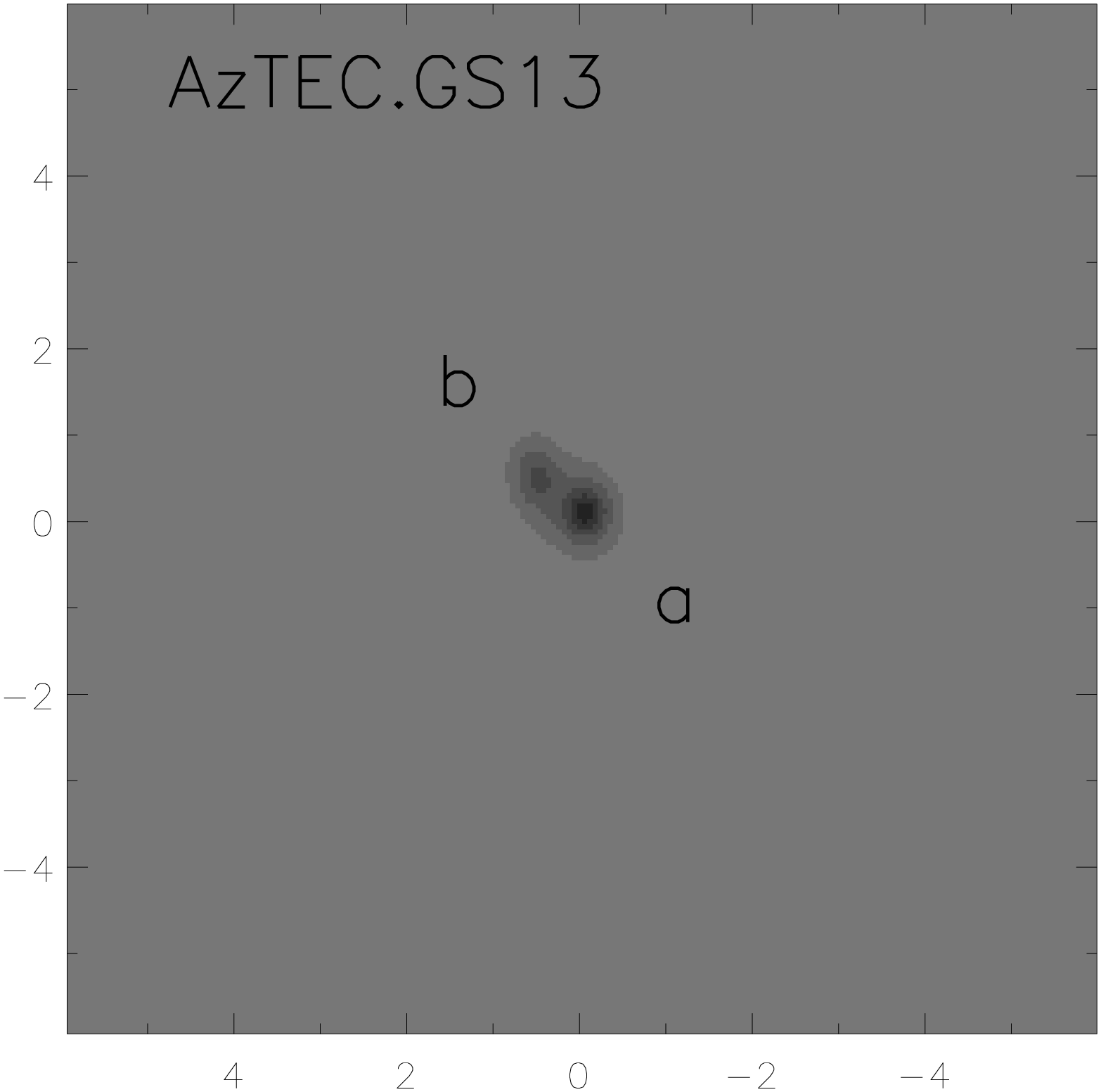,width=0.3\textwidth}&
\epsfig{file=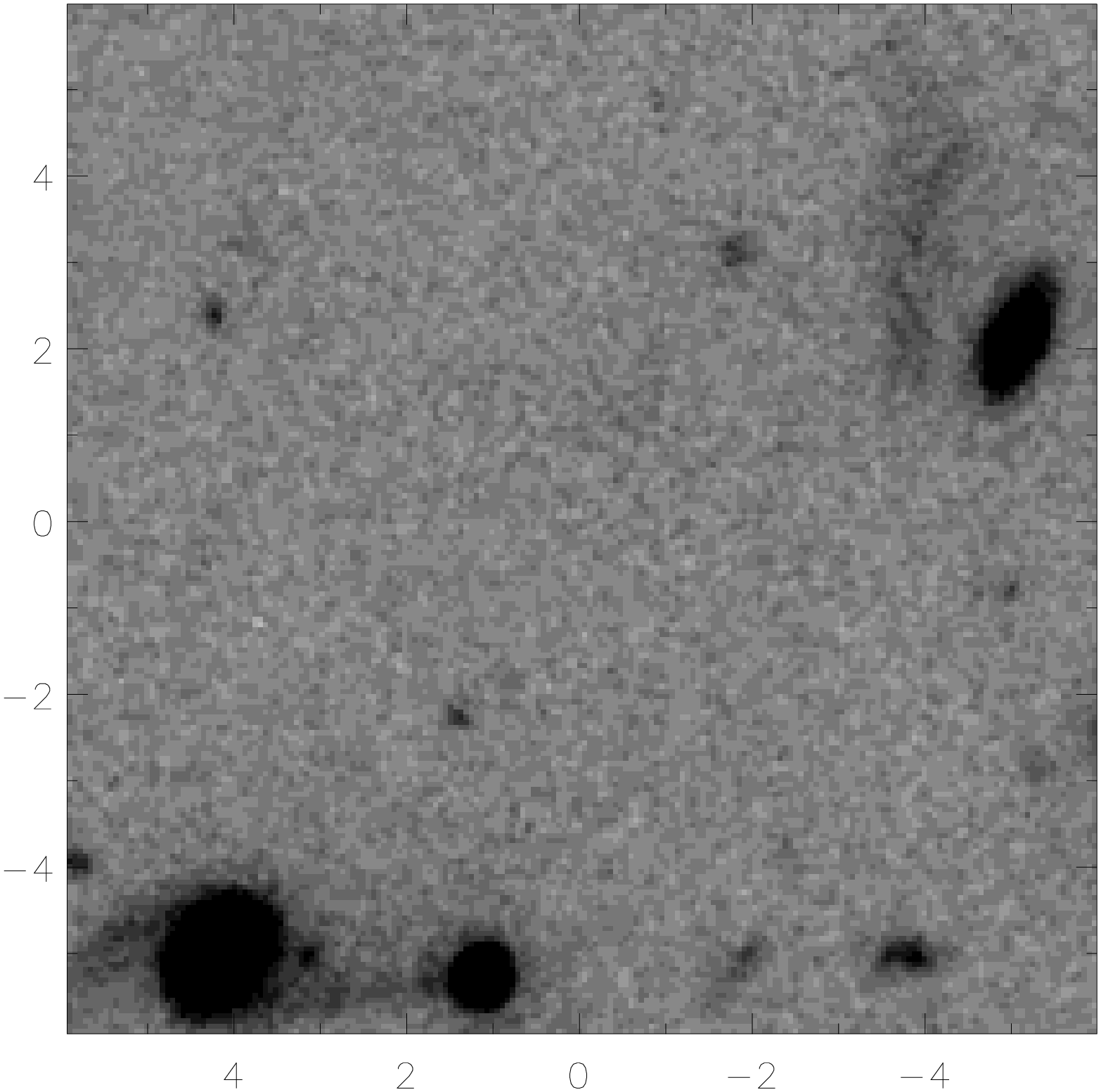,width=0.3\textwidth}\\
\\
\epsfig{file=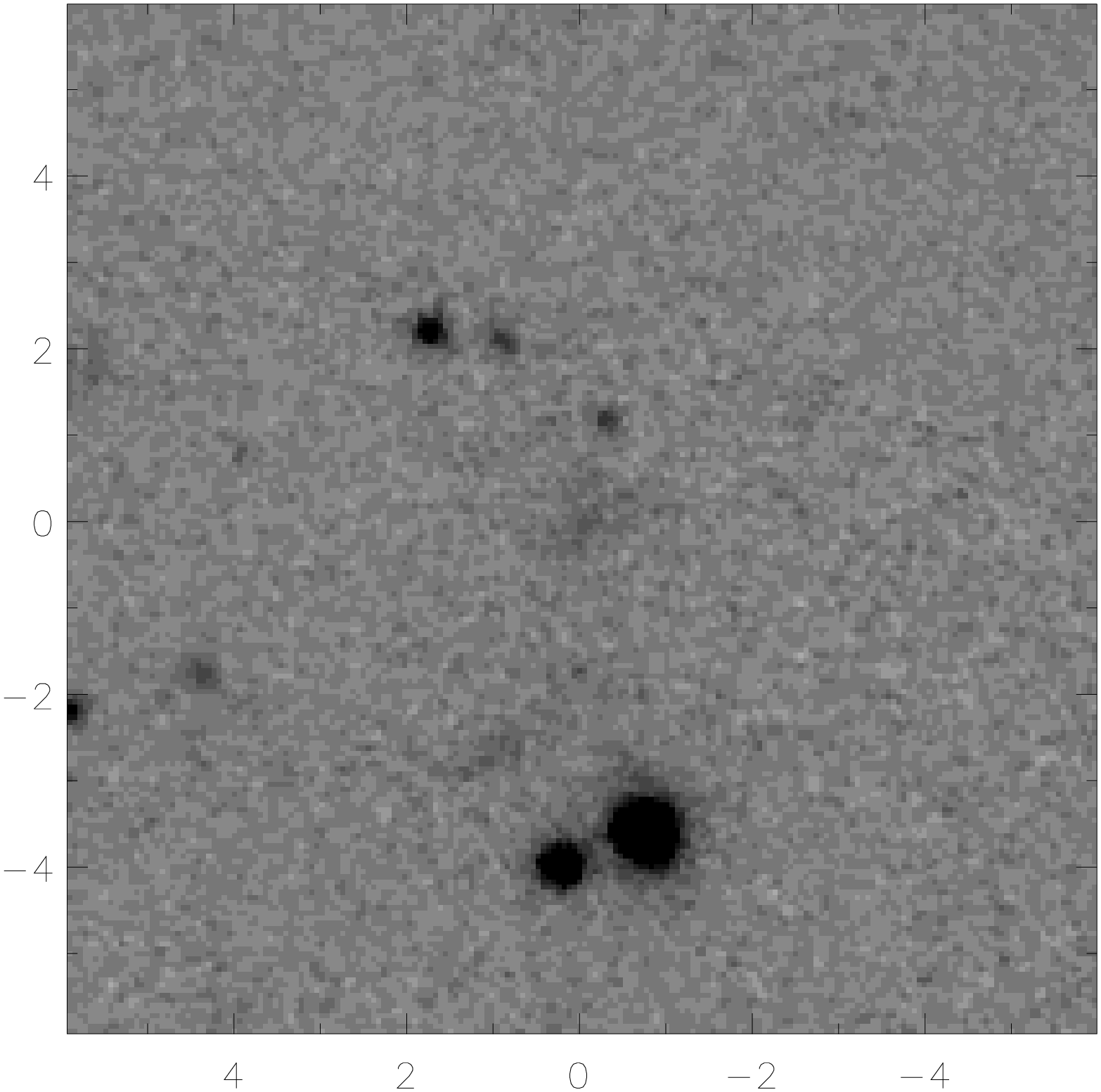,width=0.3\textwidth}&
\epsfig{file=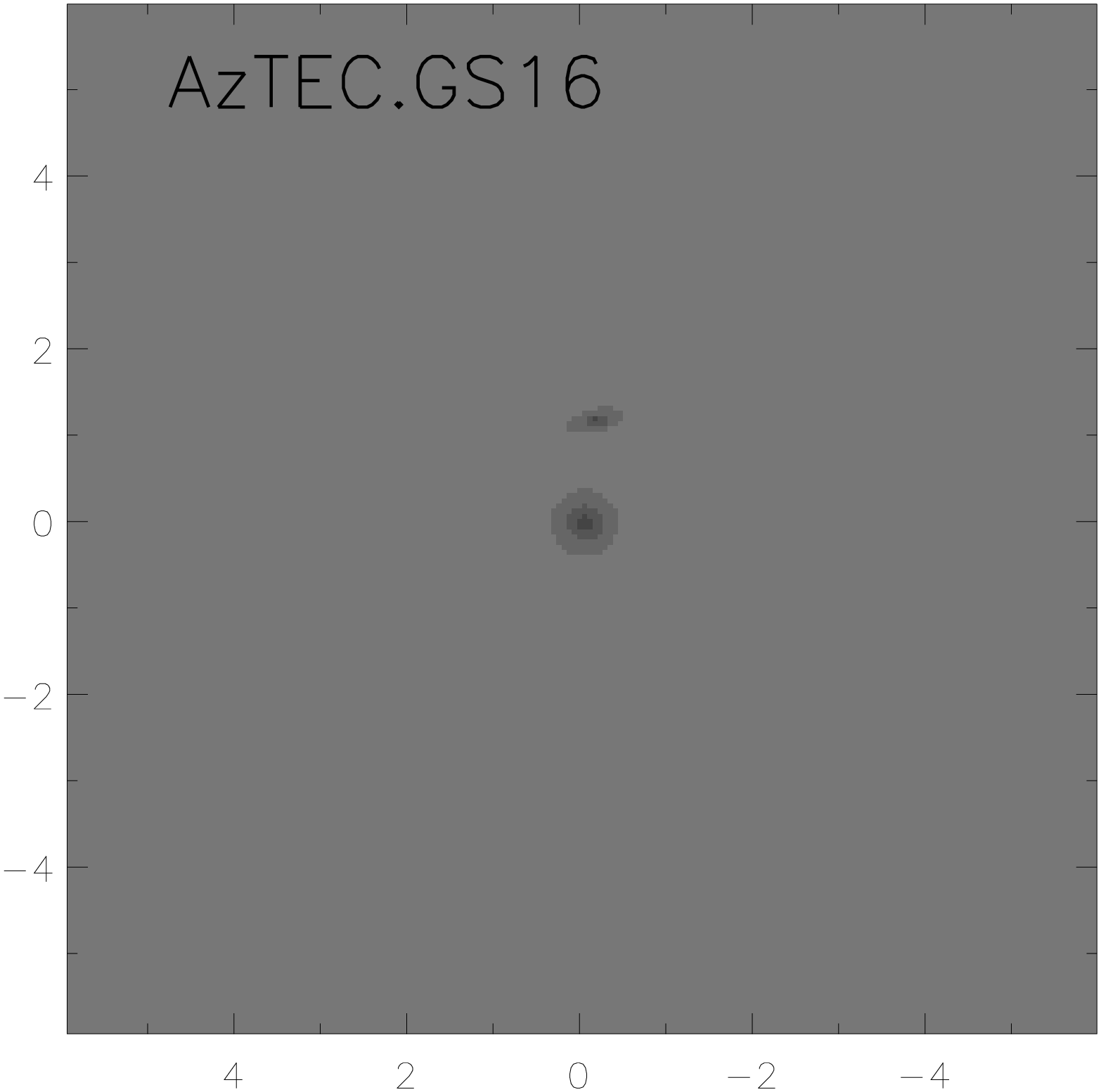,width=0.3\textwidth}&
\epsfig{file=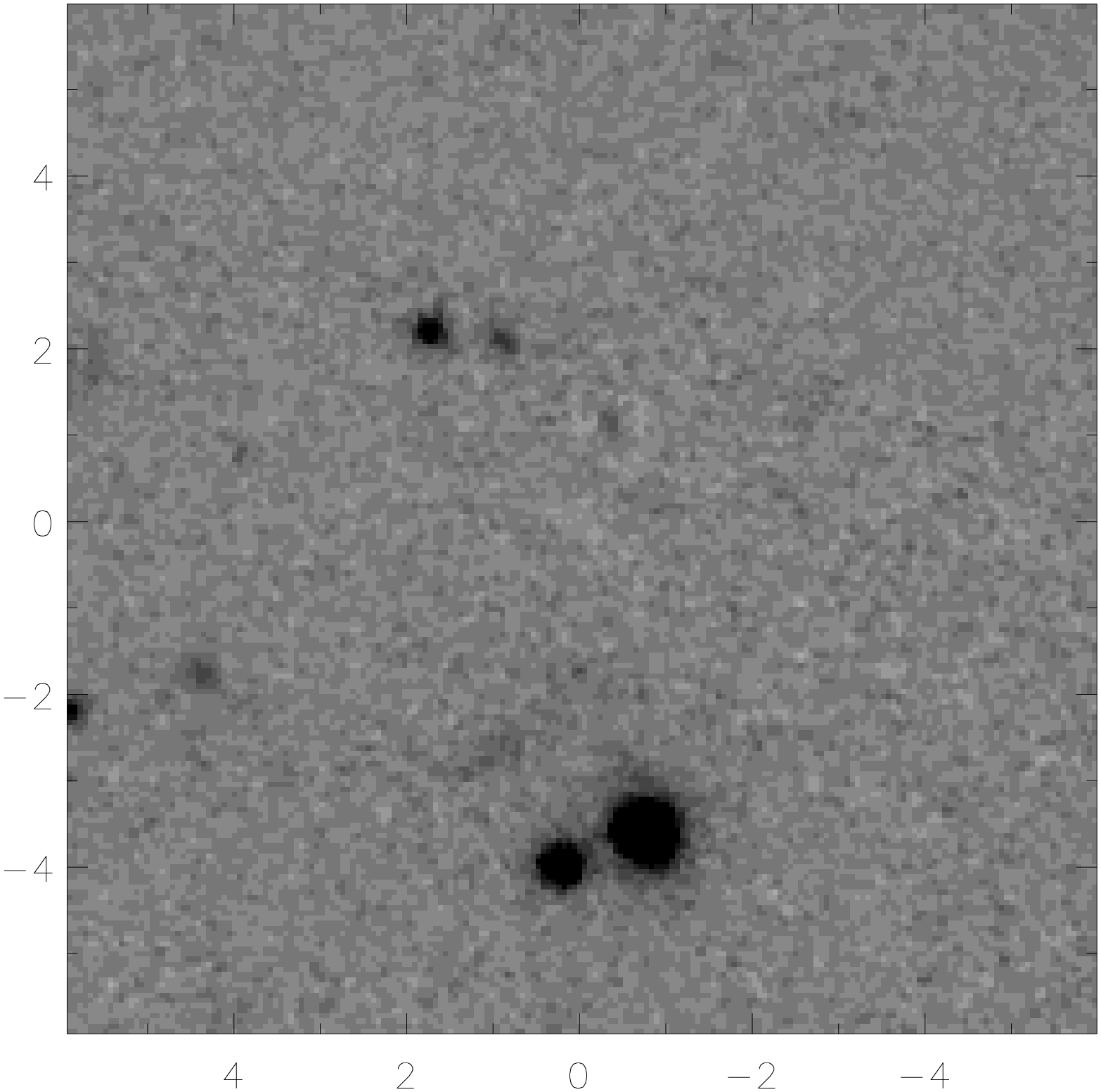,width=0.3\textwidth}\\
\\
\epsfig{file=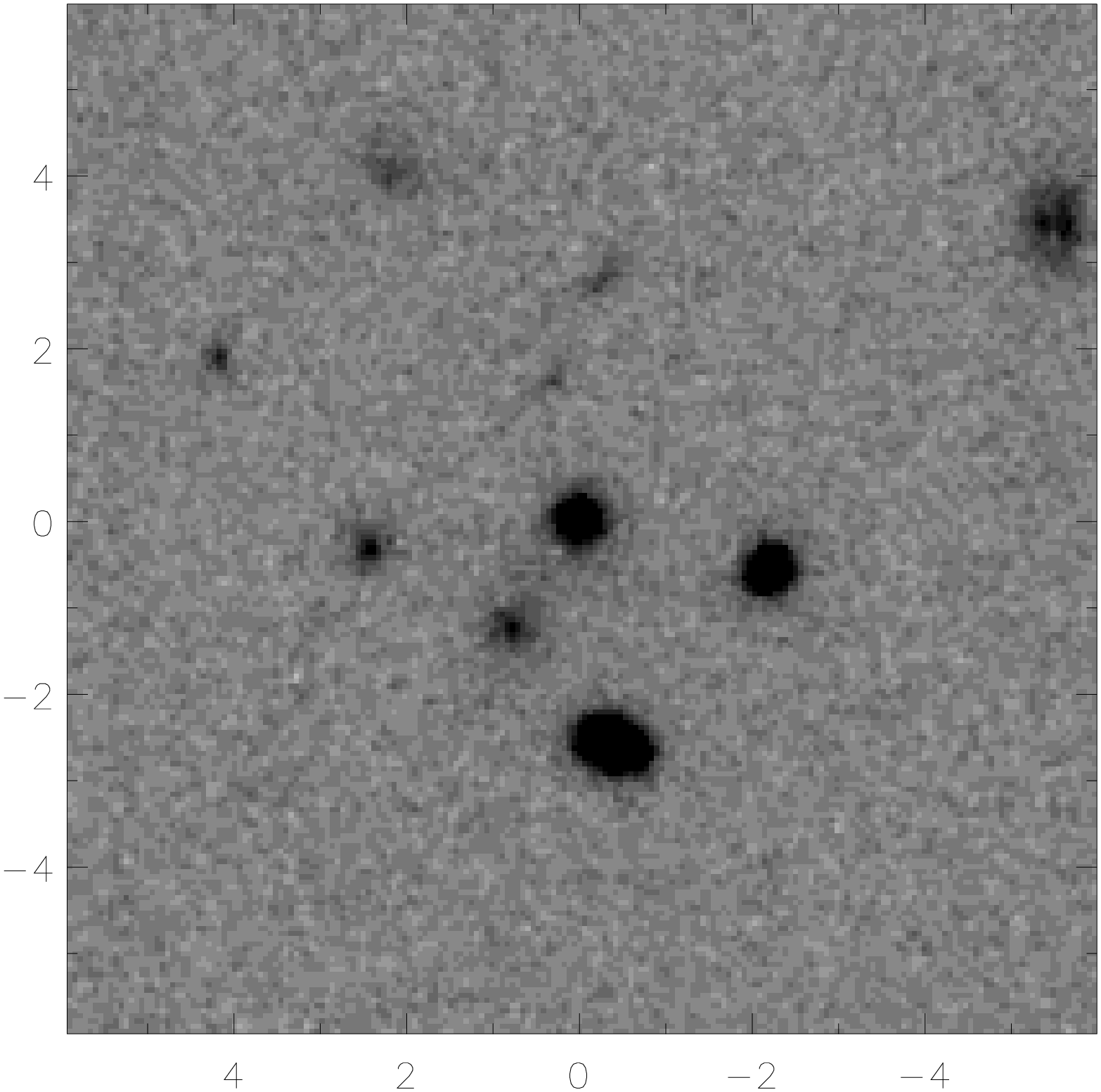,width=0.3\textwidth}&
\epsfig{file=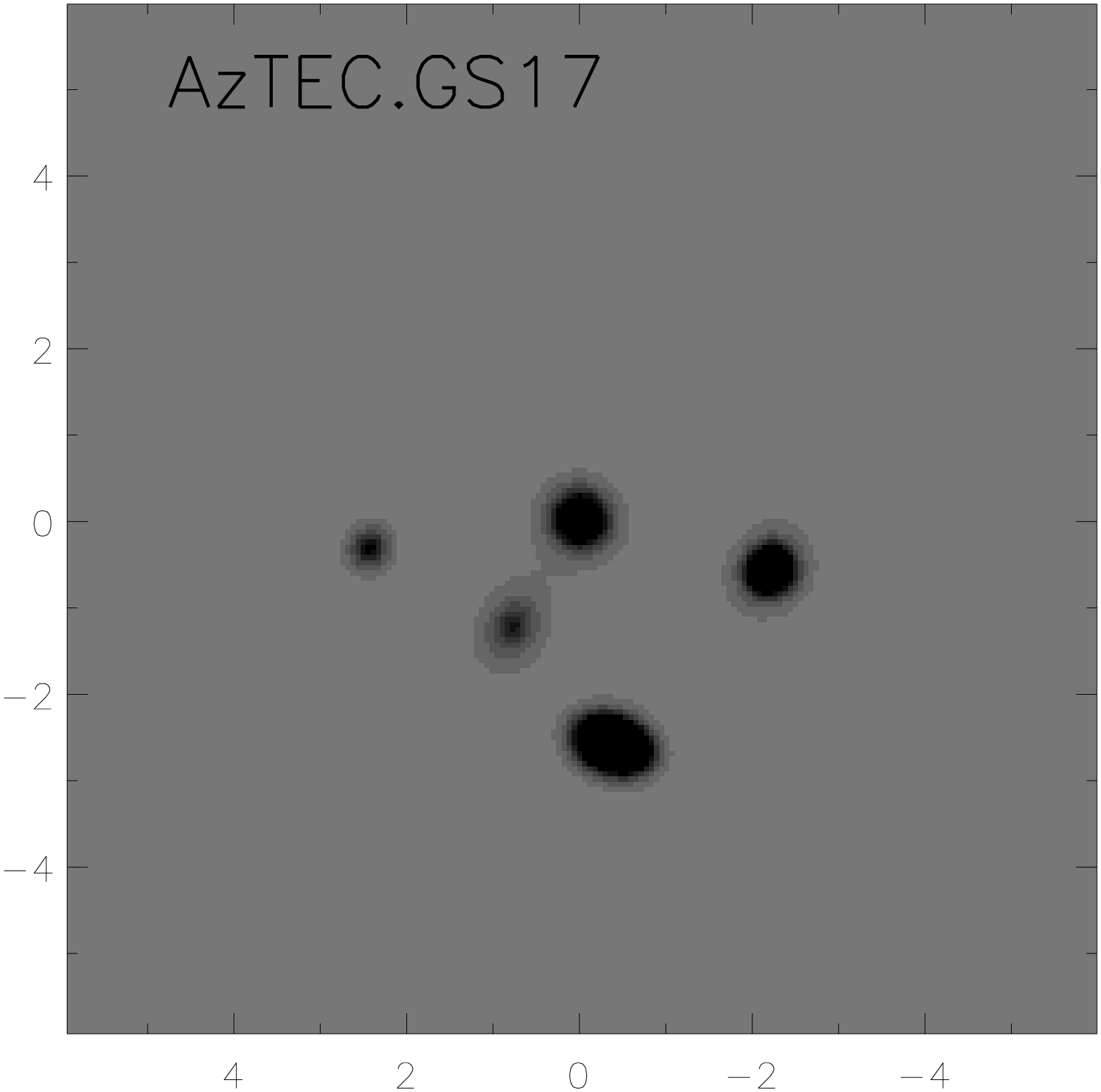,width=0.3\textwidth}&
\epsfig{file=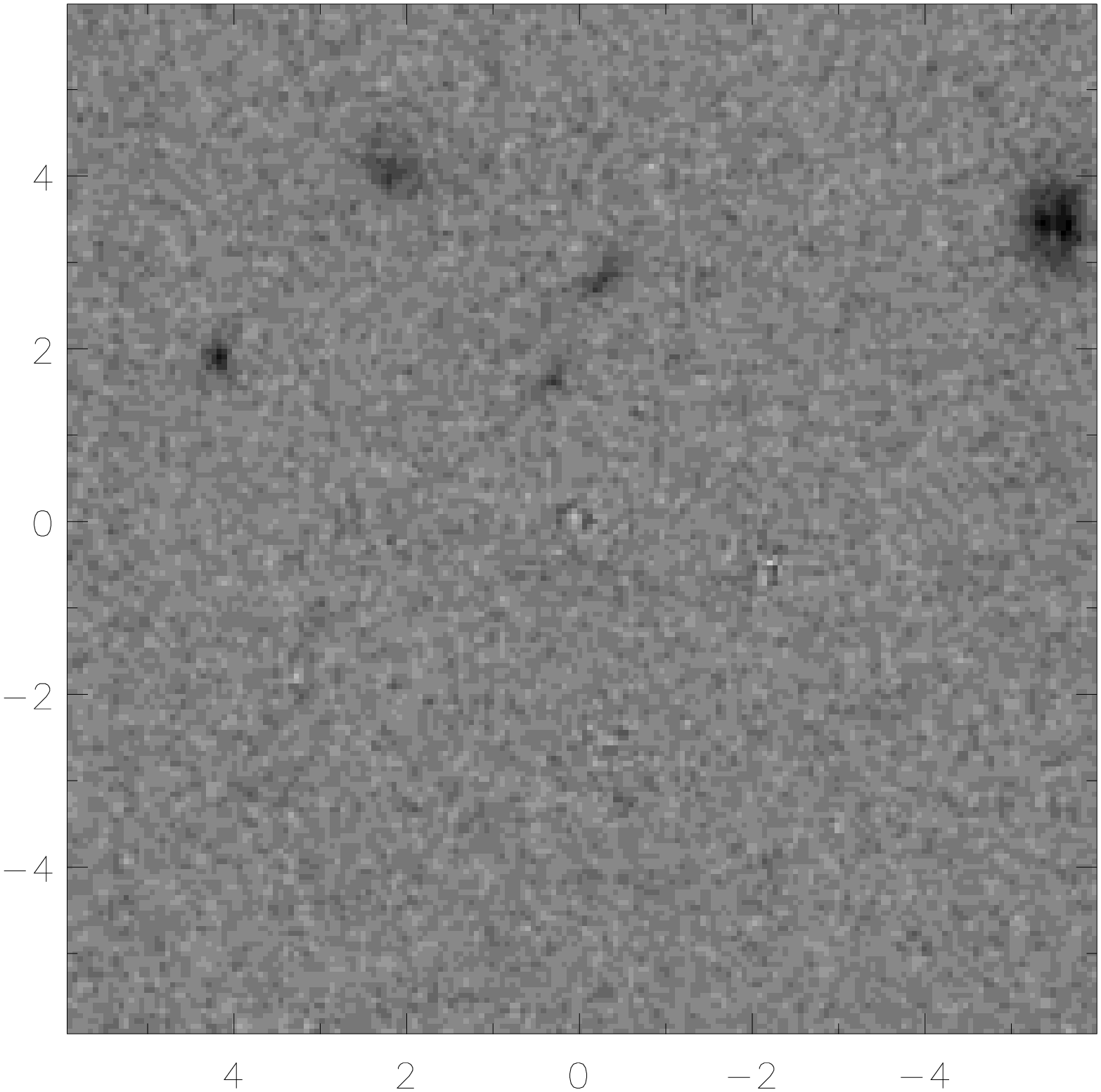,width=0.3\textwidth}\\
\\
\epsfig{file=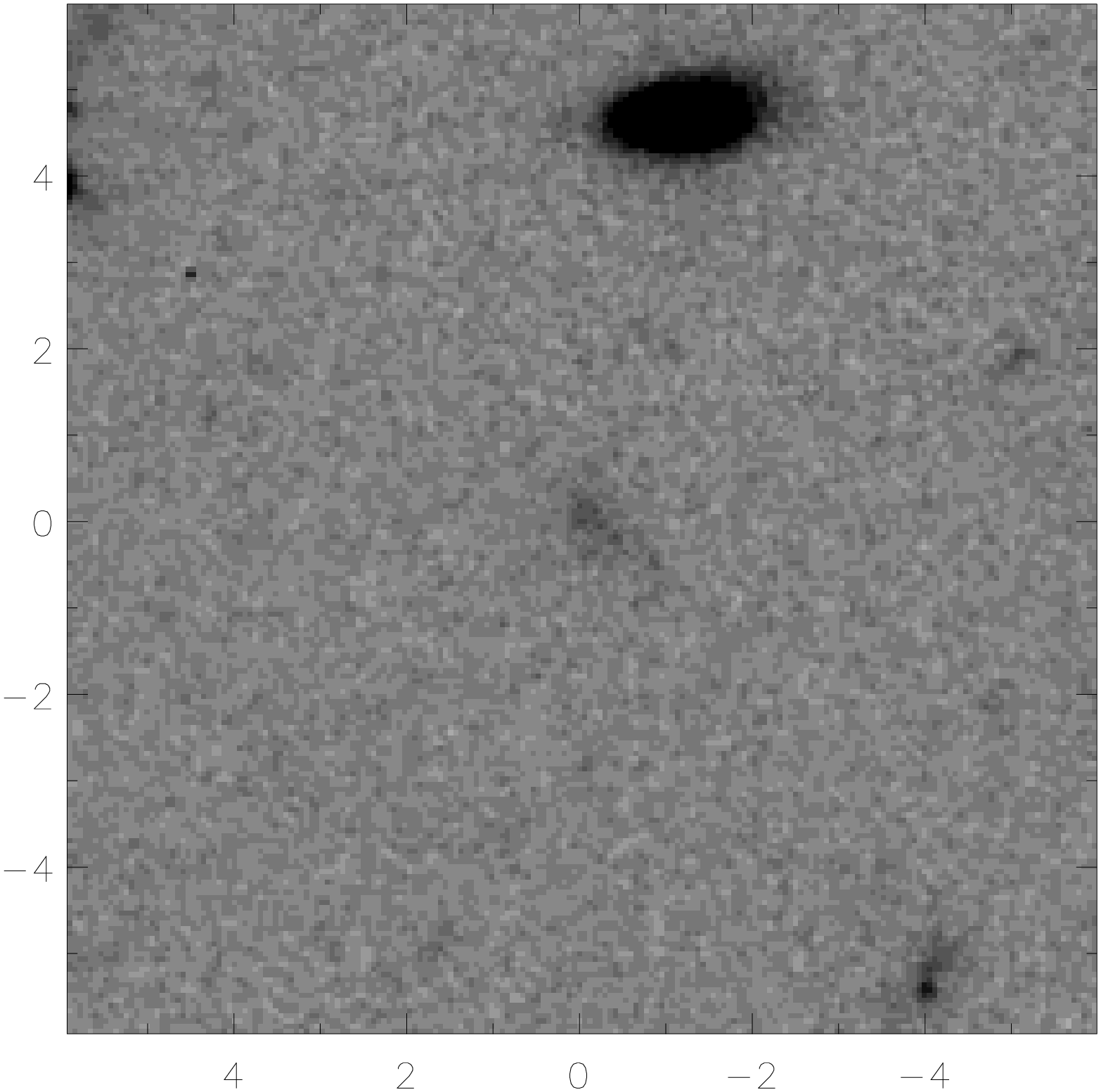,width=0.3\textwidth}&
\epsfig{file=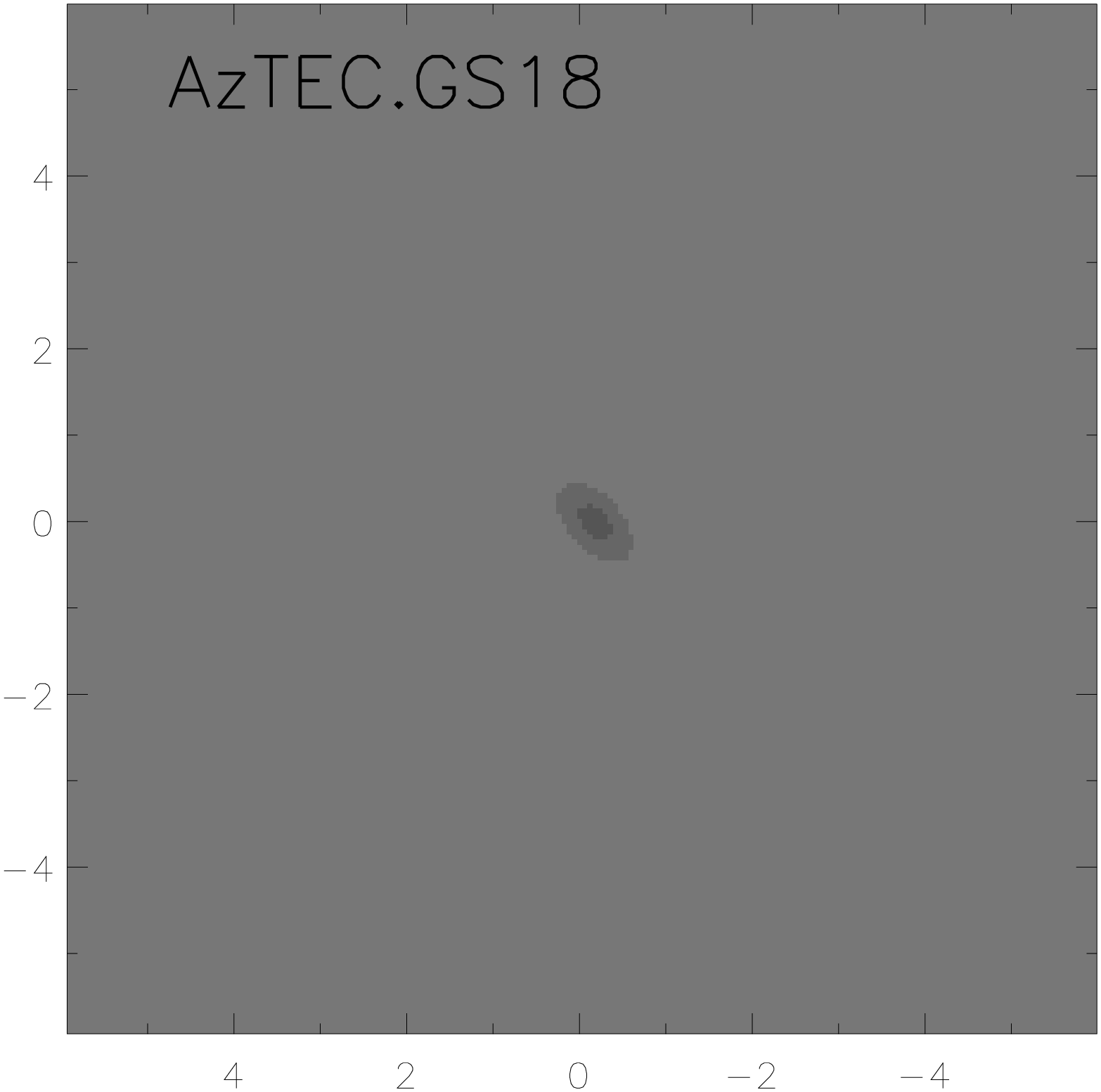,width=0.3\textwidth}&
\epsfig{file=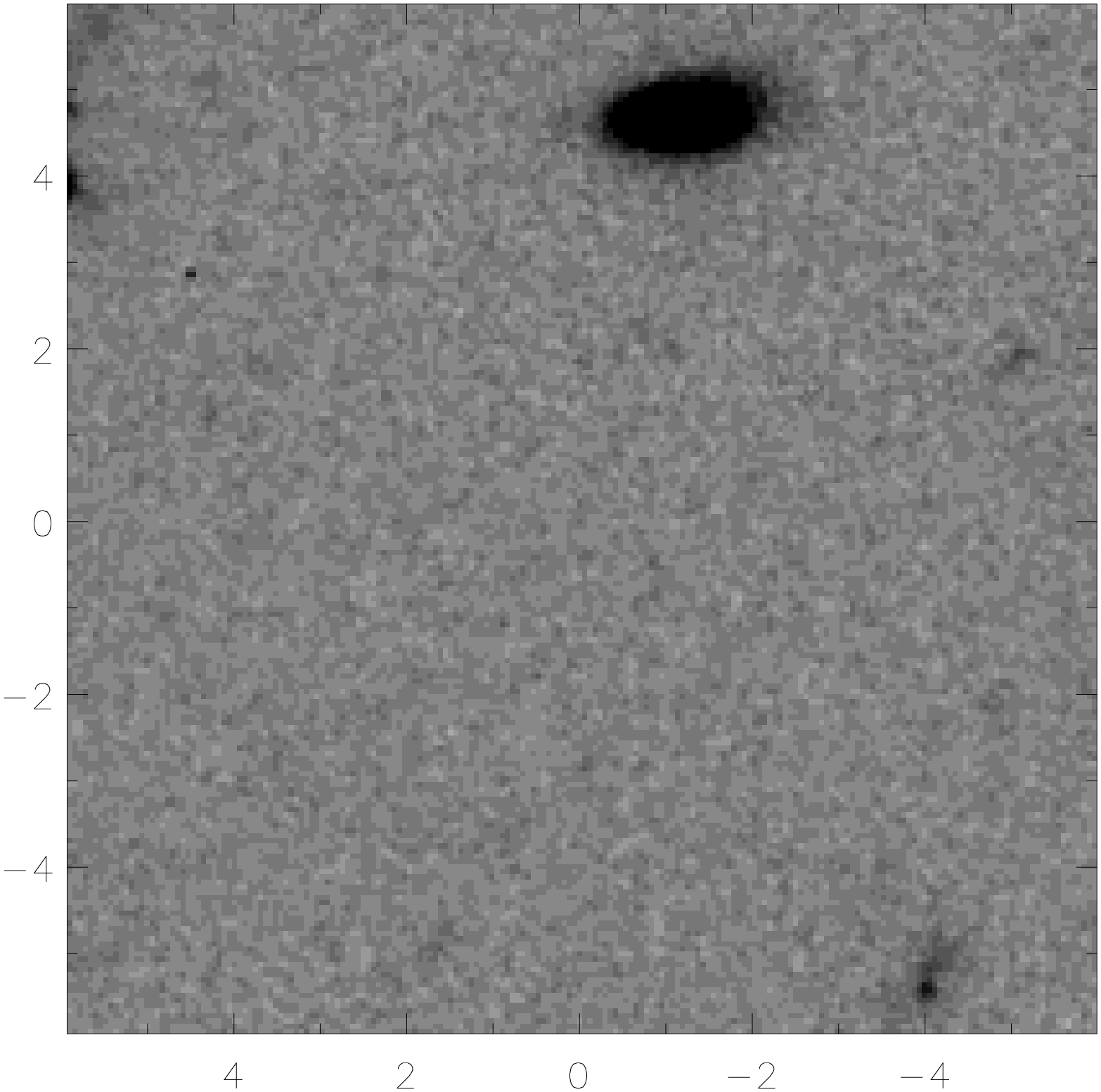,width=0.3\textwidth}\\
\end{tabular}
\addtocounter{figure}{-1}
\caption{- continued}
\end{figure*}
\end{center}


\begin{center}
\begin{figure*}
\begin{tabular}{ccc}
\epsfig{file=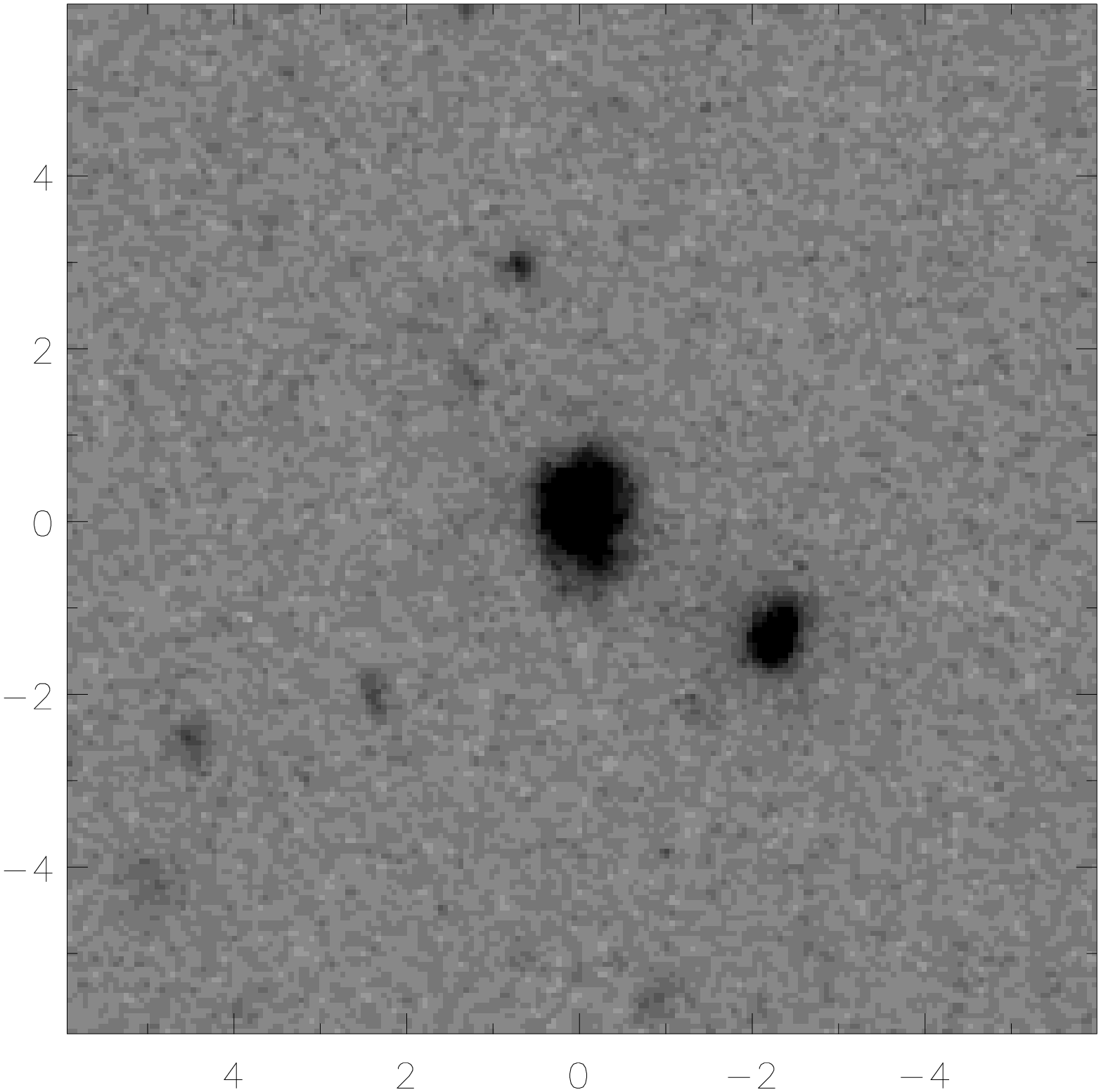,width=0.3\textwidth}&
\epsfig{file=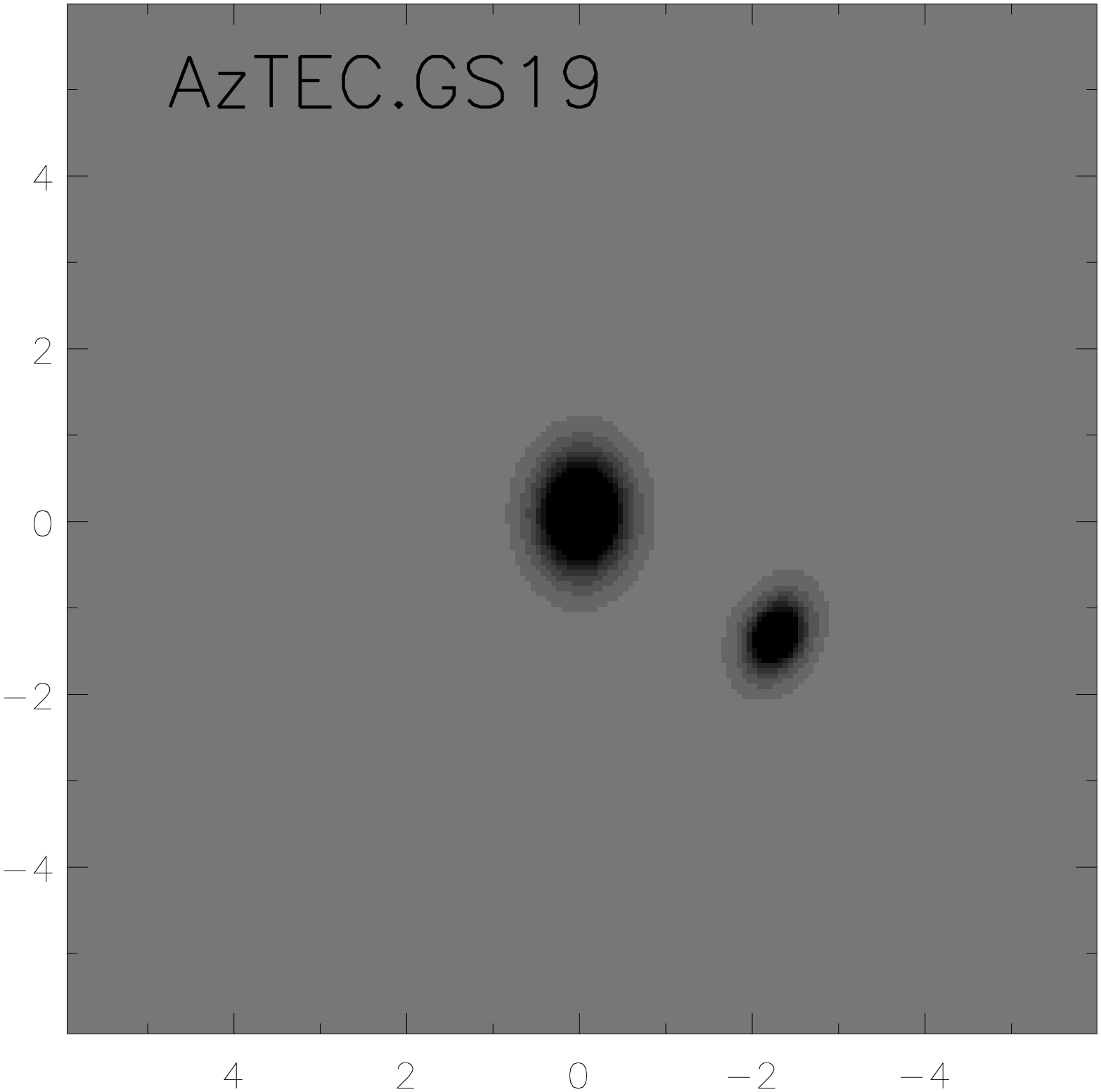,width=0.3\textwidth}&
\epsfig{file=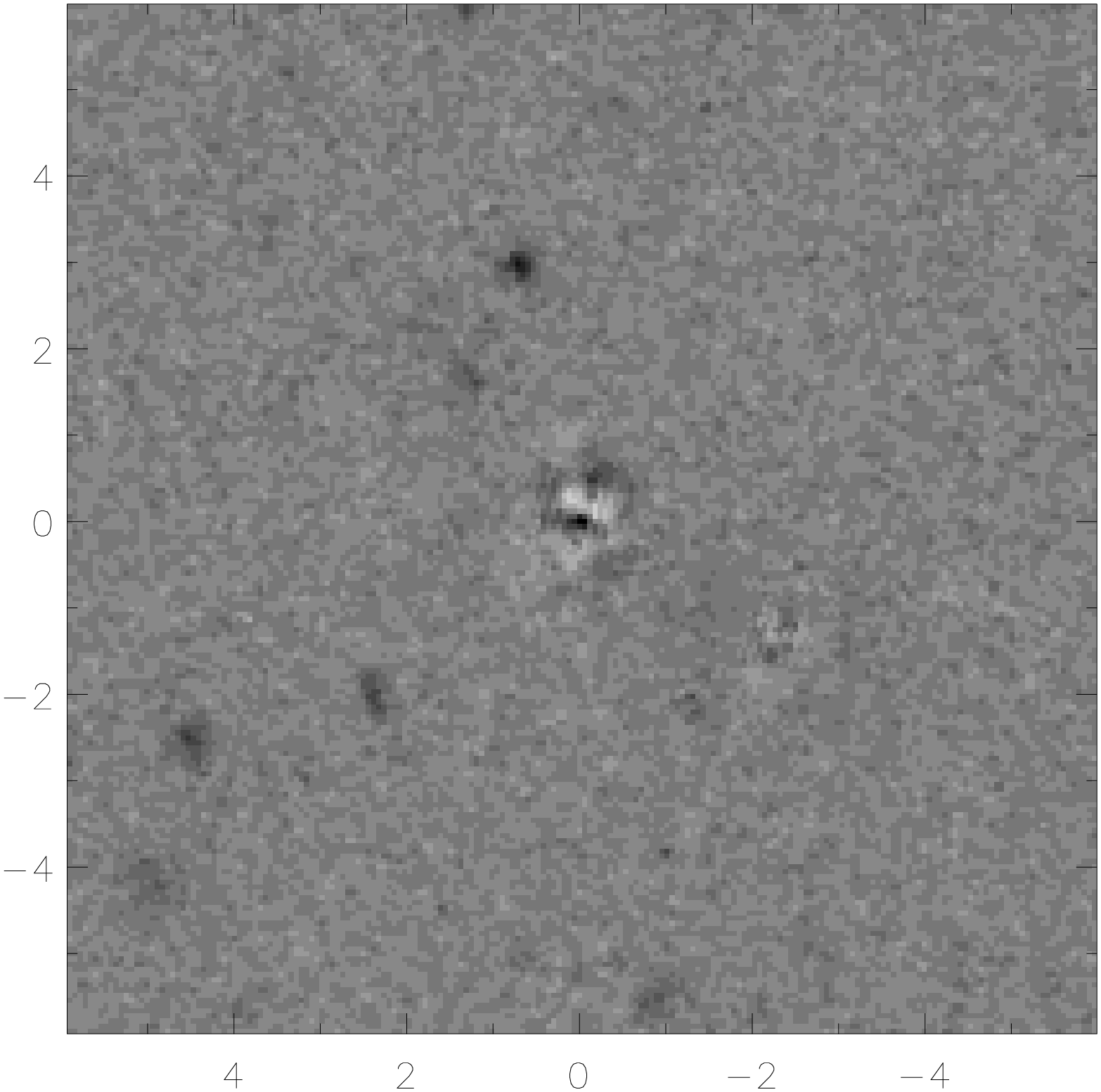,width=0.3\textwidth}\\
\\
\epsfig{file=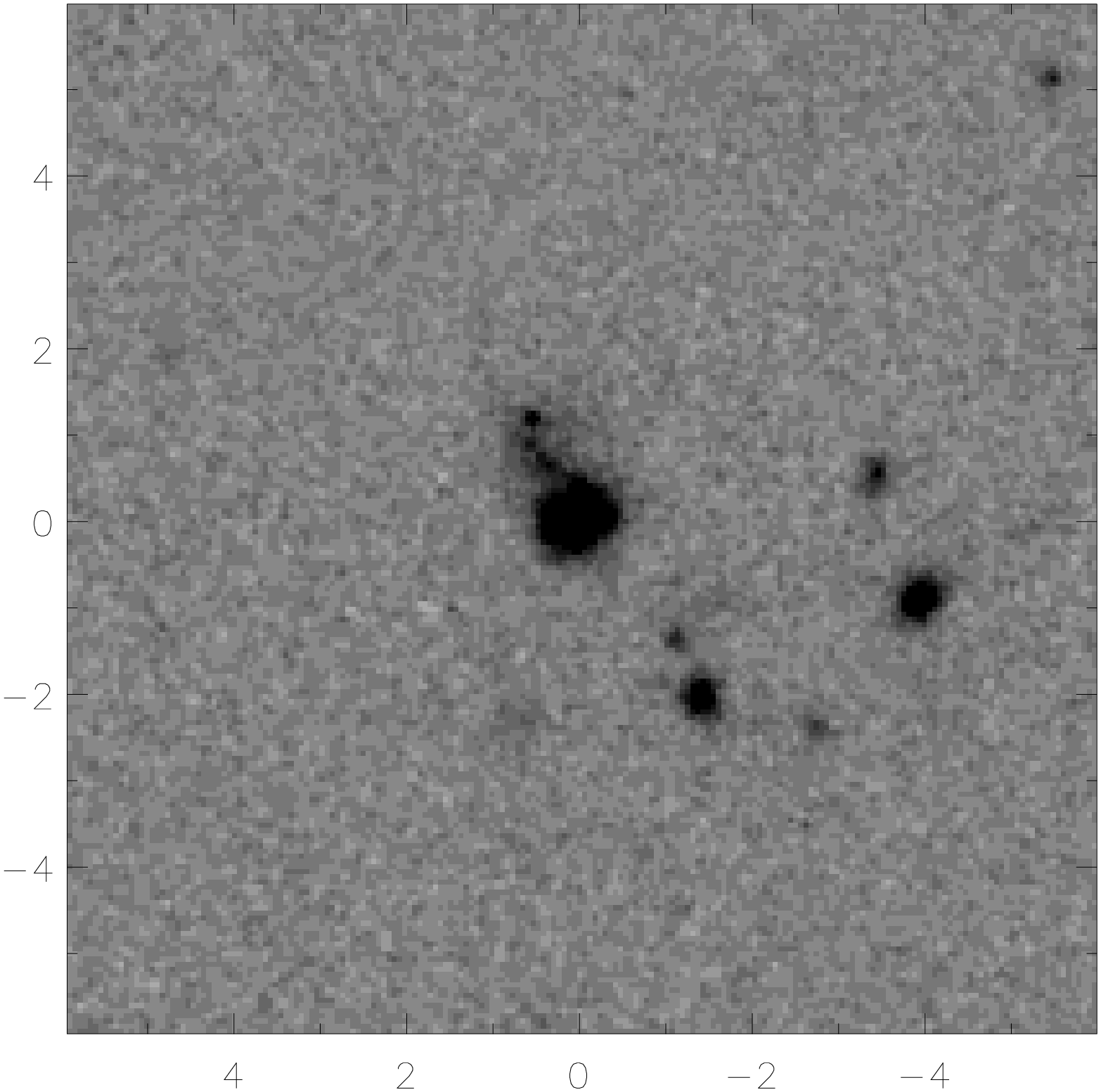,width=0.3\textwidth}&
\epsfig{file=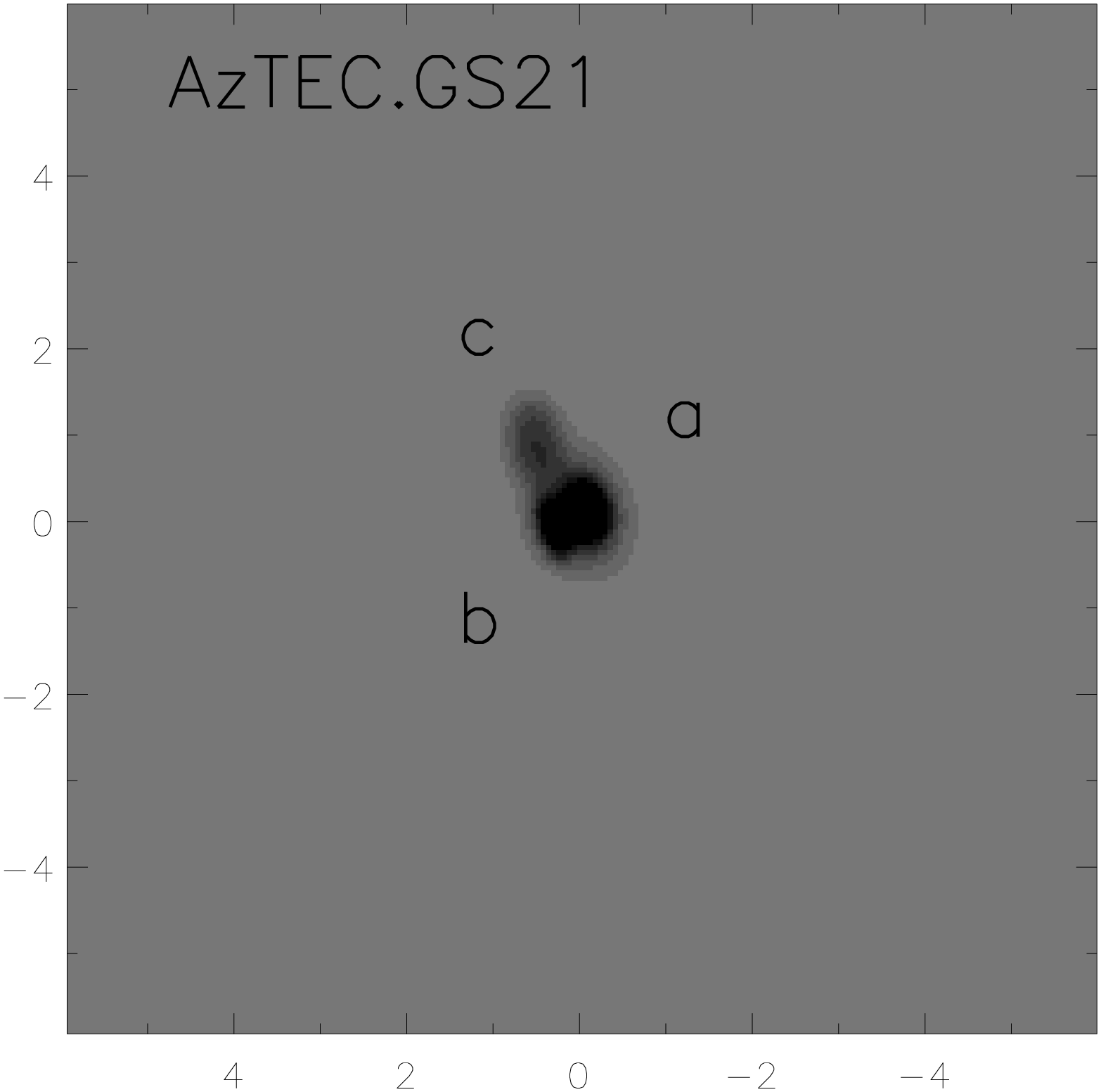,width=0.3\textwidth}&
\epsfig{file=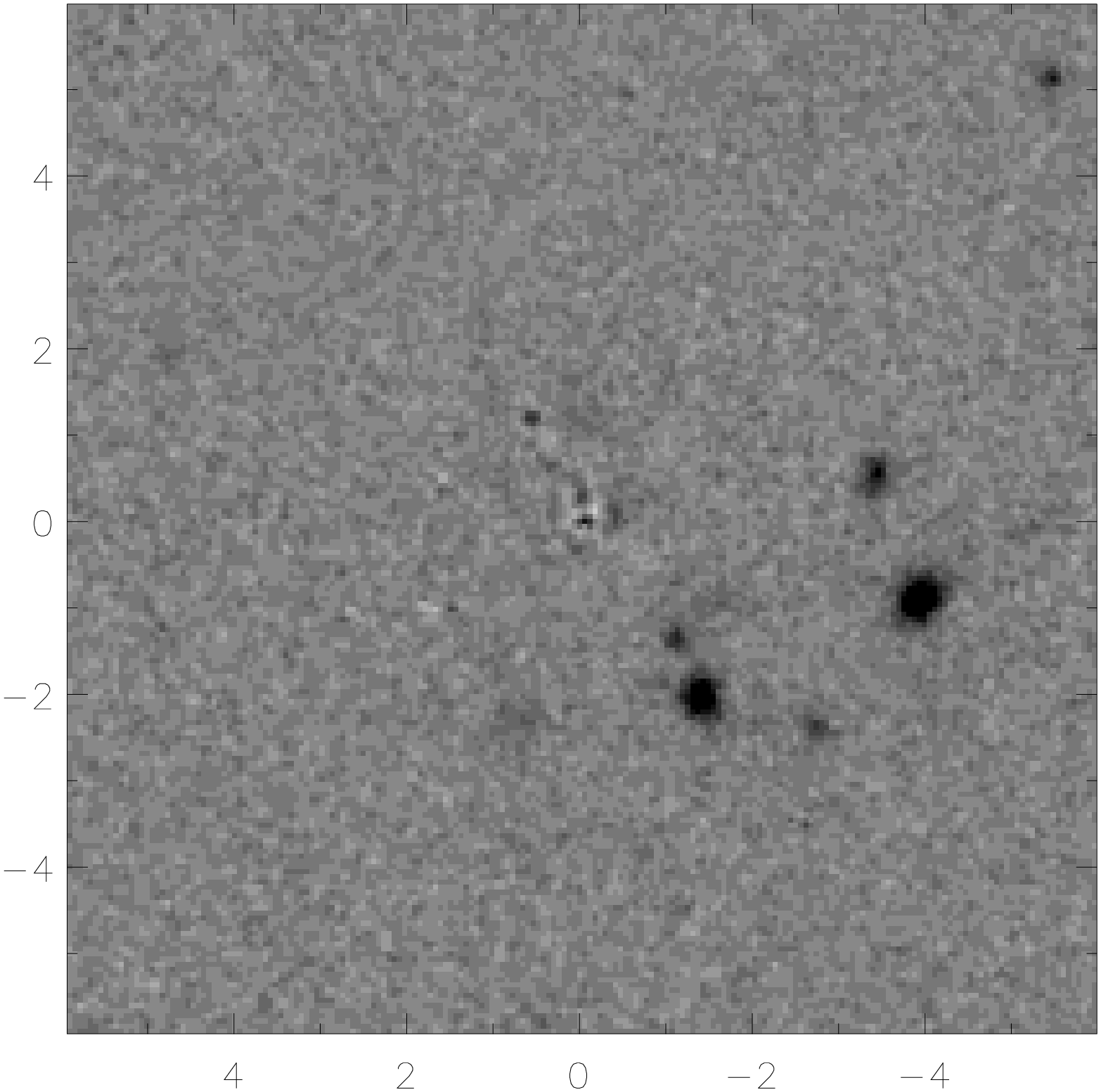,width=0.3\textwidth}\\
\\
\epsfig{file=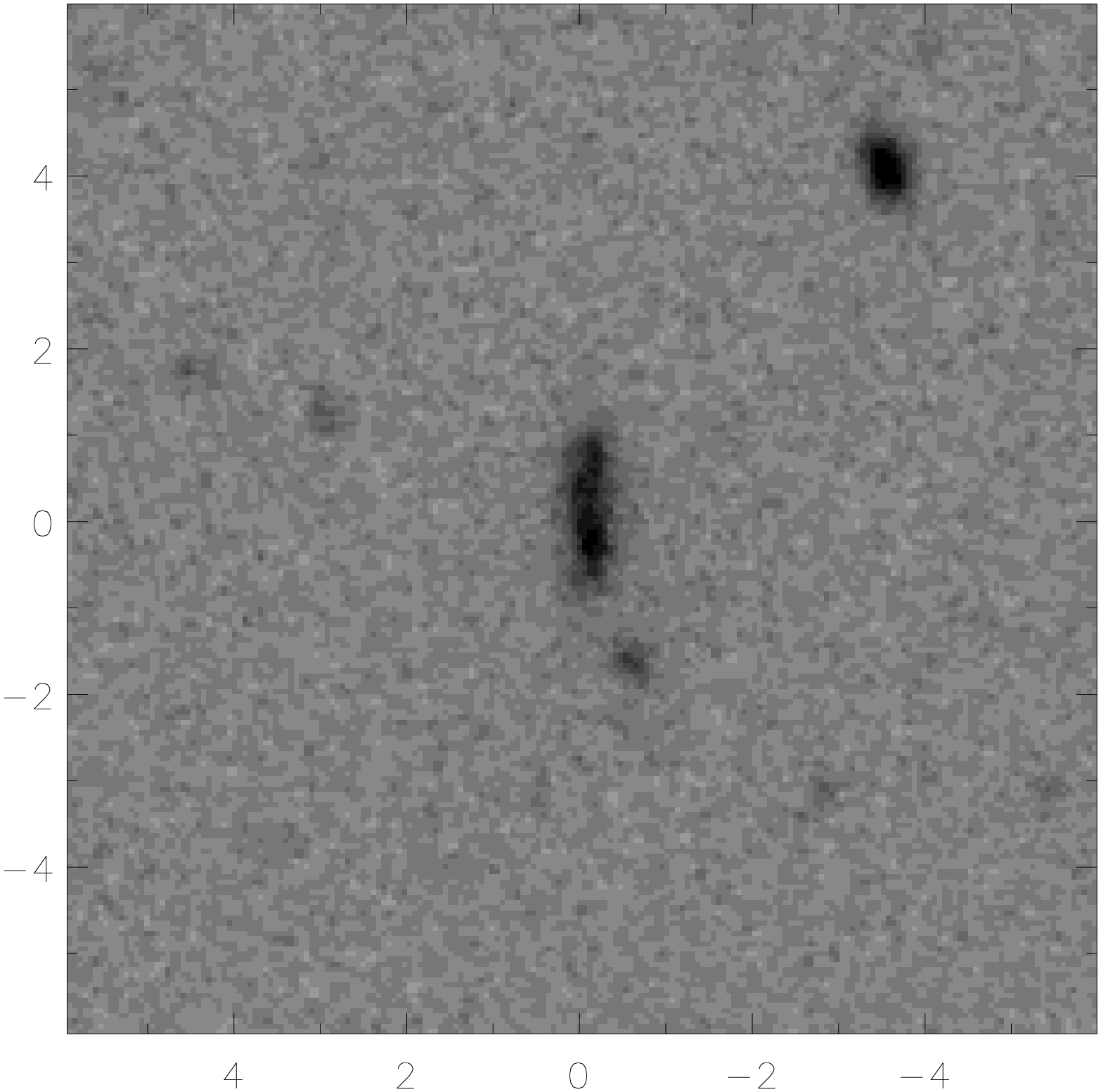,width=0.3\textwidth}&
\epsfig{file=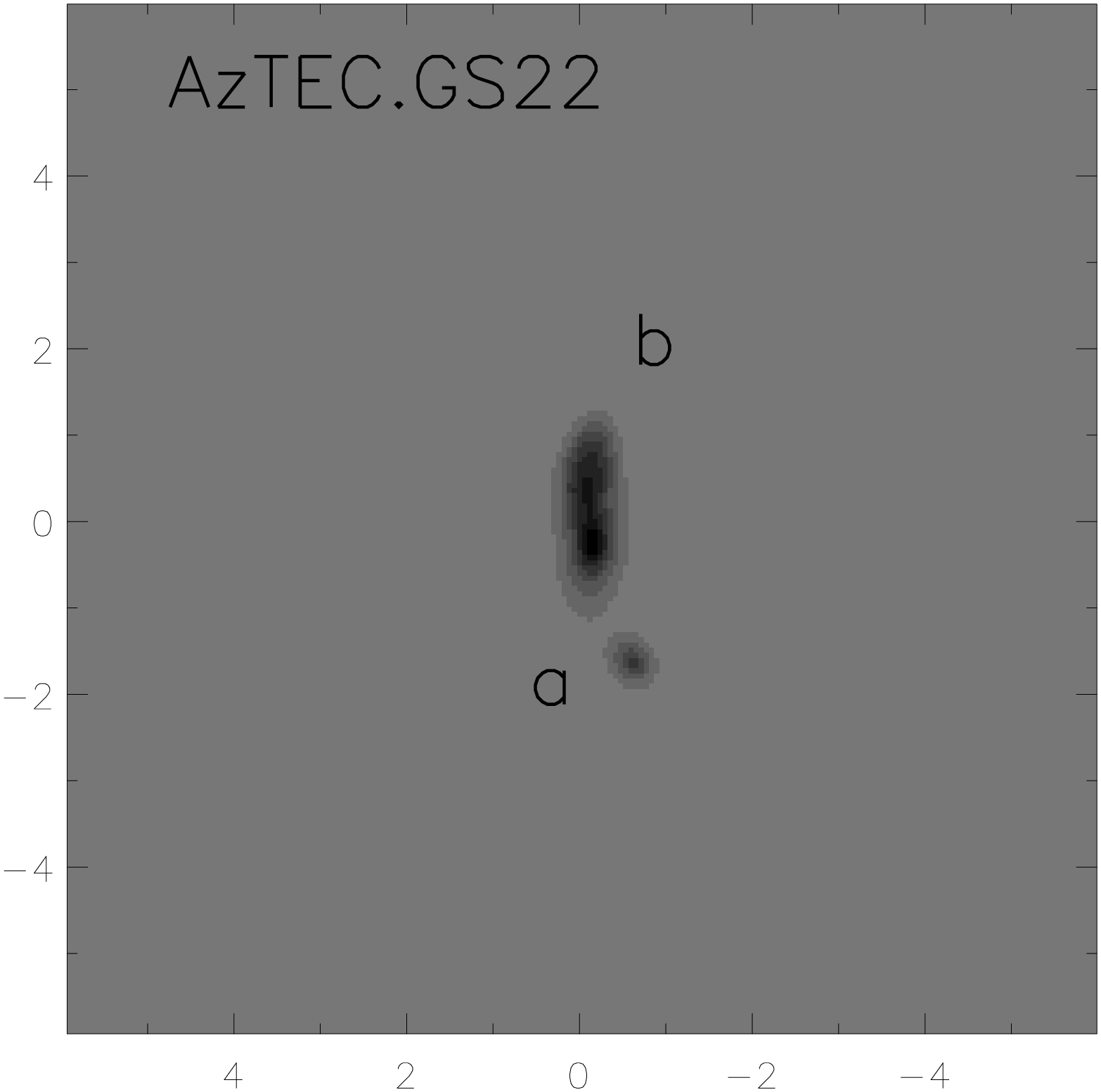,width=0.3\textwidth}&
\epsfig{file=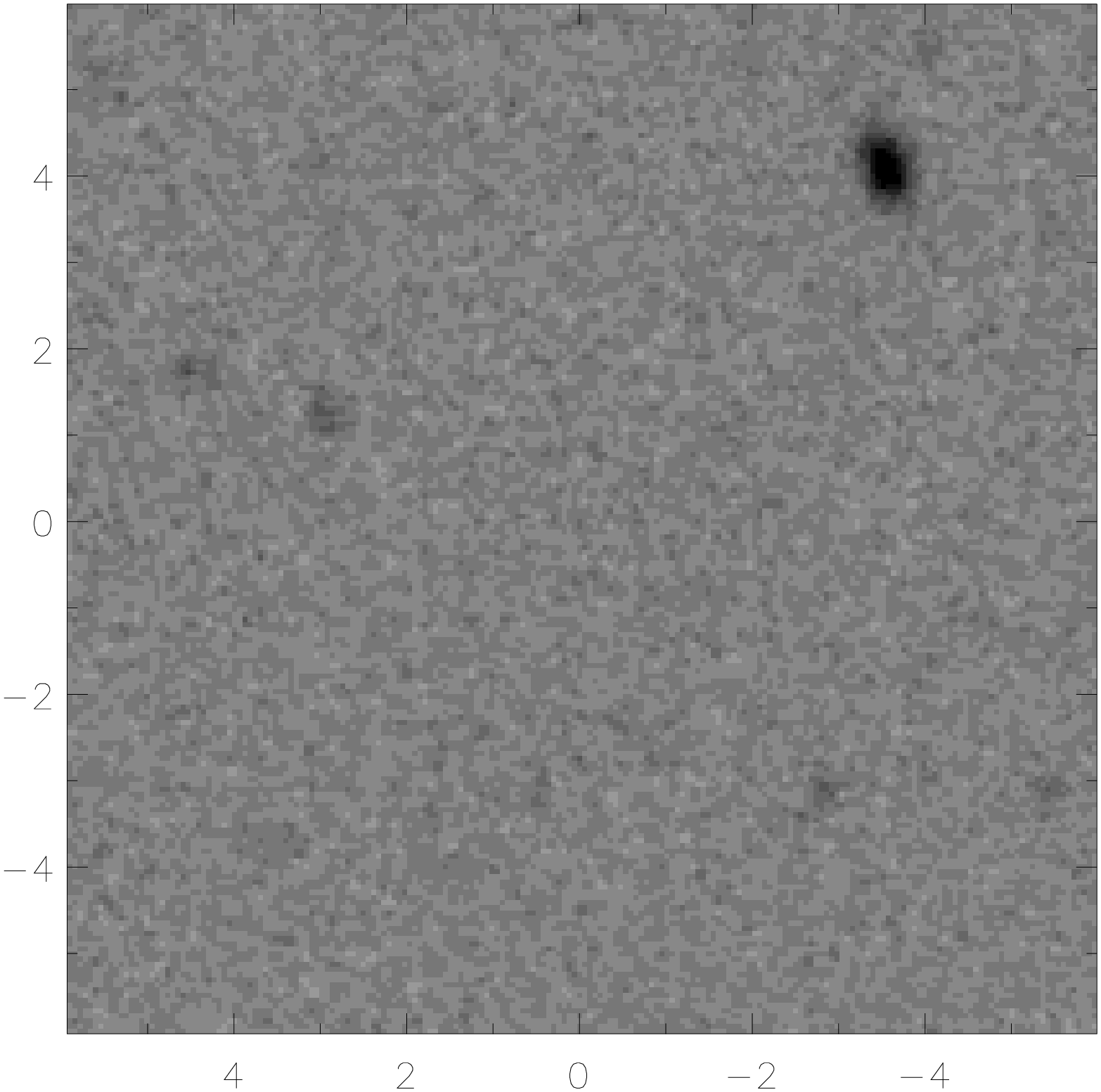,width=0.3\textwidth}\\
\\
\epsfig{file=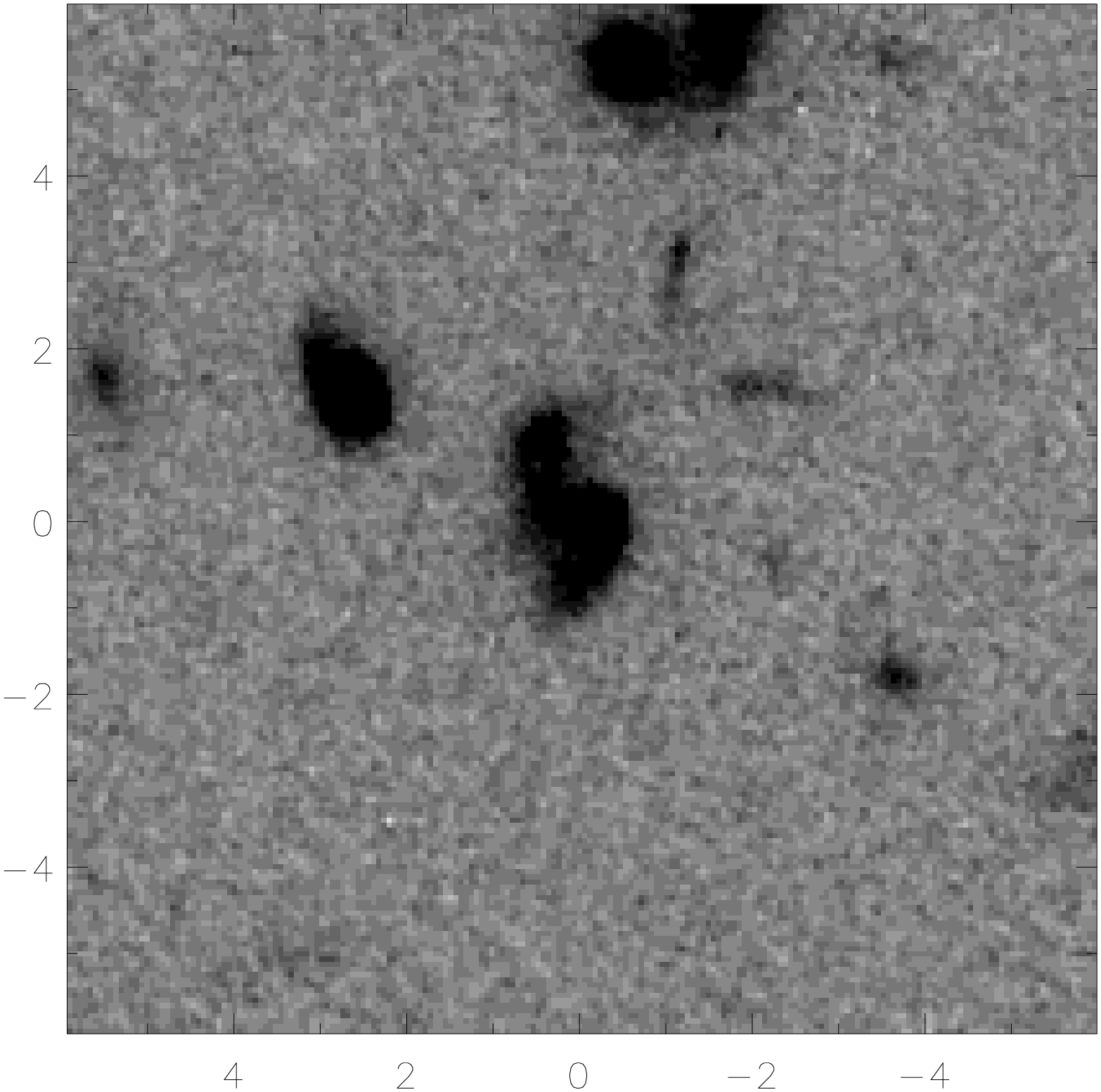,width=0.3\textwidth}&
\epsfig{file=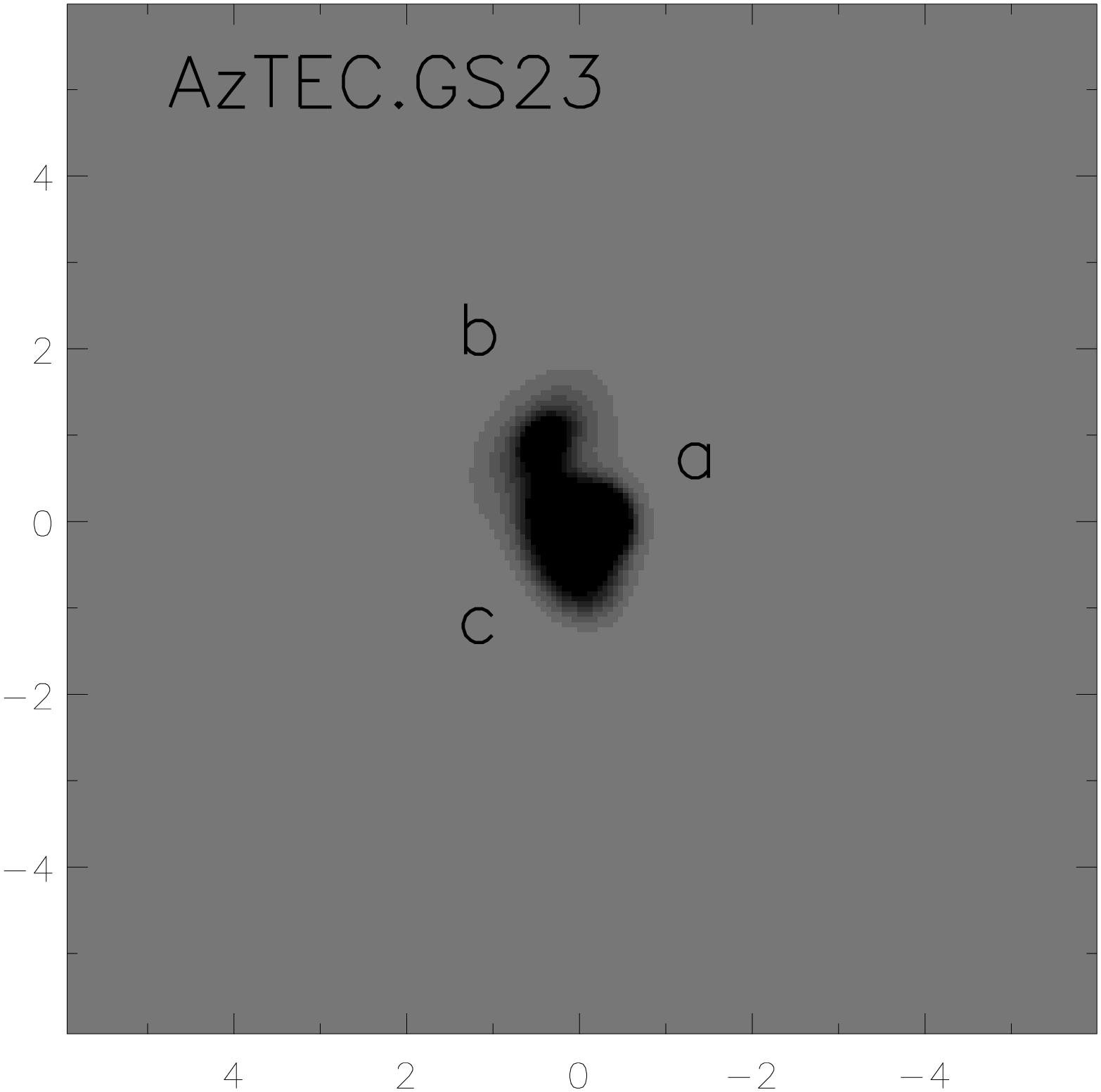,width=0.3\textwidth}&
\epsfig{file=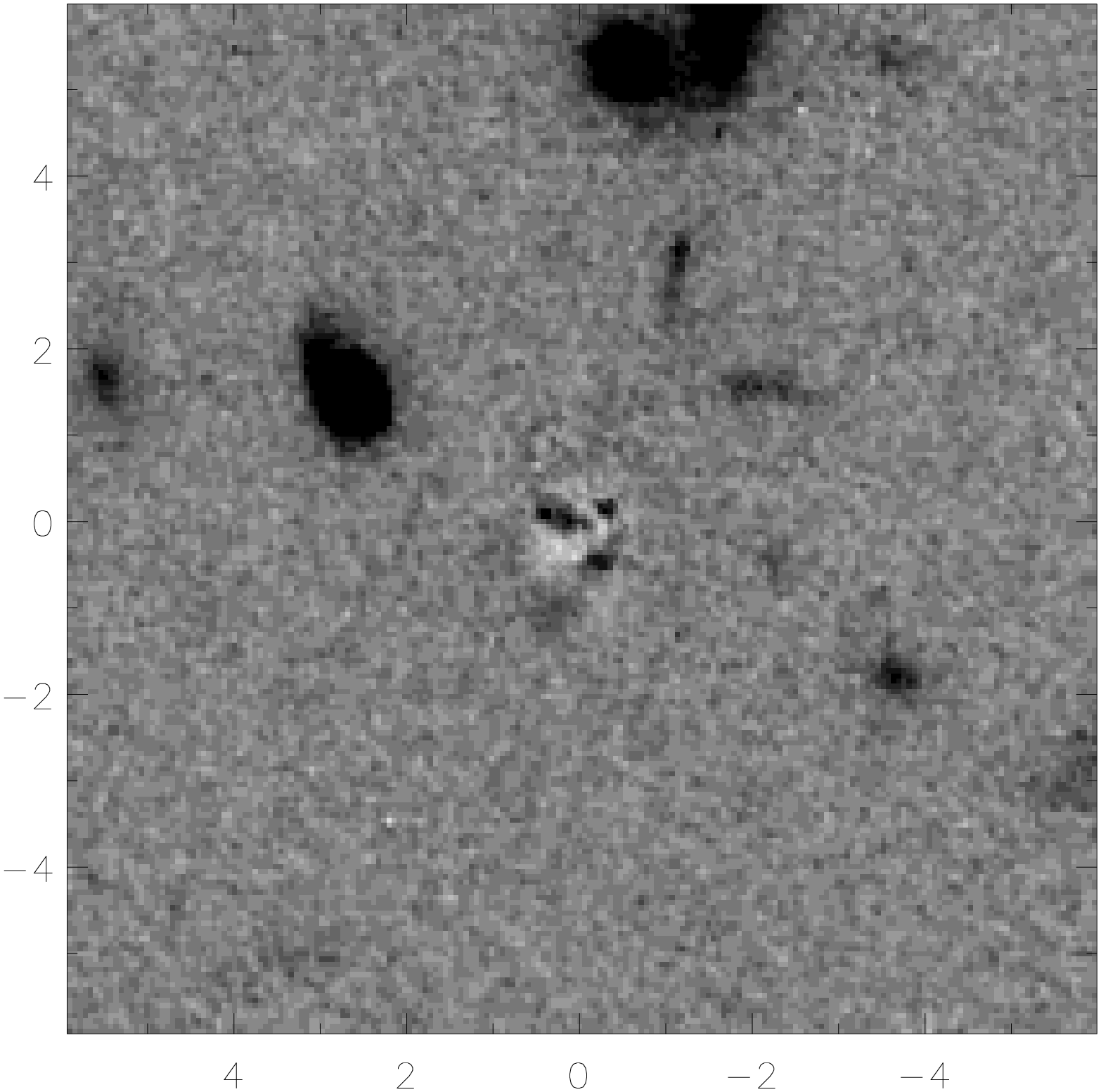,width=0.3\textwidth}\\
\end{tabular}
\addtocounter{figure}{-1}
\caption{- continued}
\end{figure*}
\end{center}


\begin{center}
\begin{figure*}
\begin{tabular}{ccc}
\epsfig{file=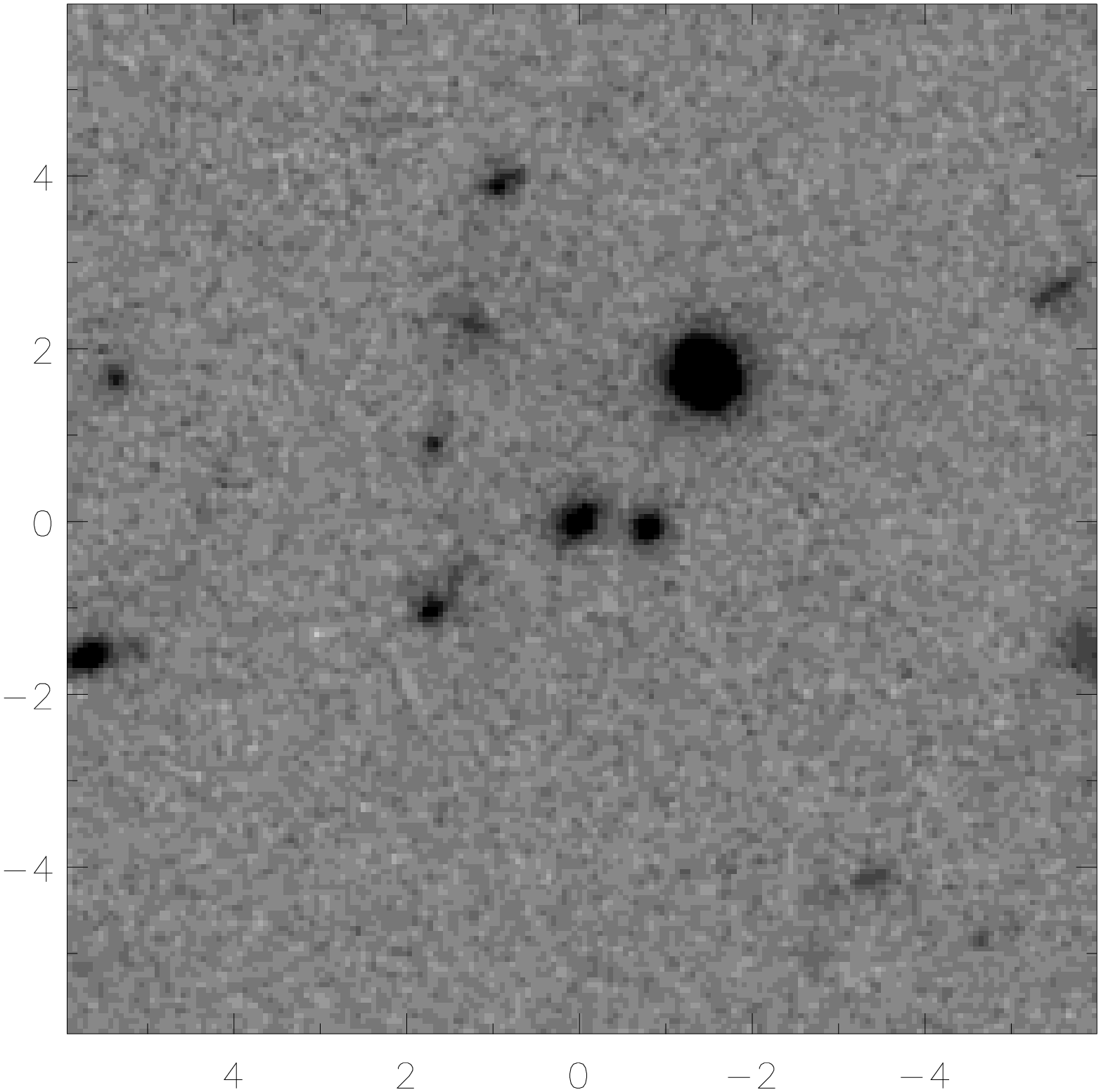,width=0.3\textwidth}&
\epsfig{file=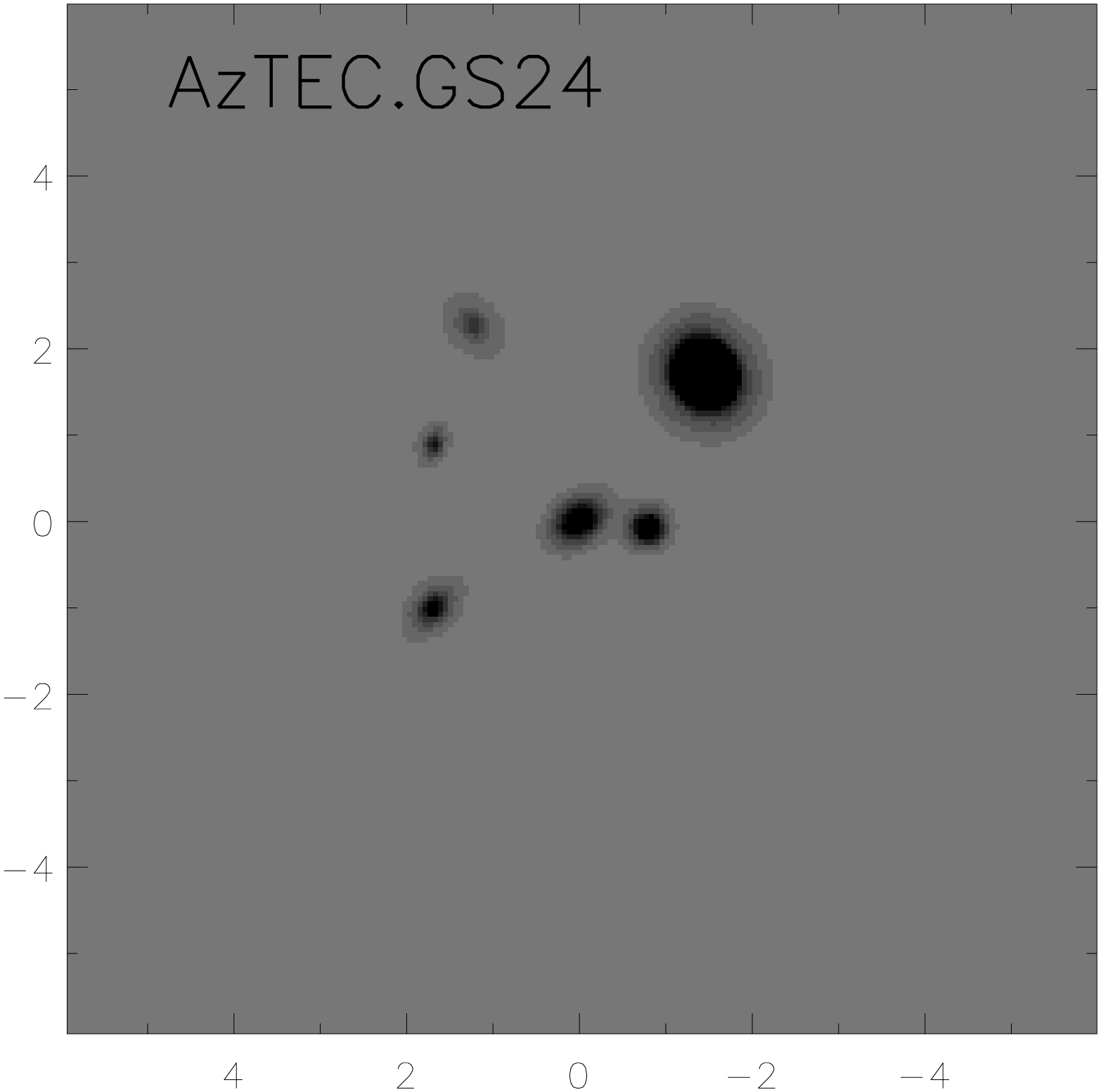,width=0.3\textwidth}&
\epsfig{file=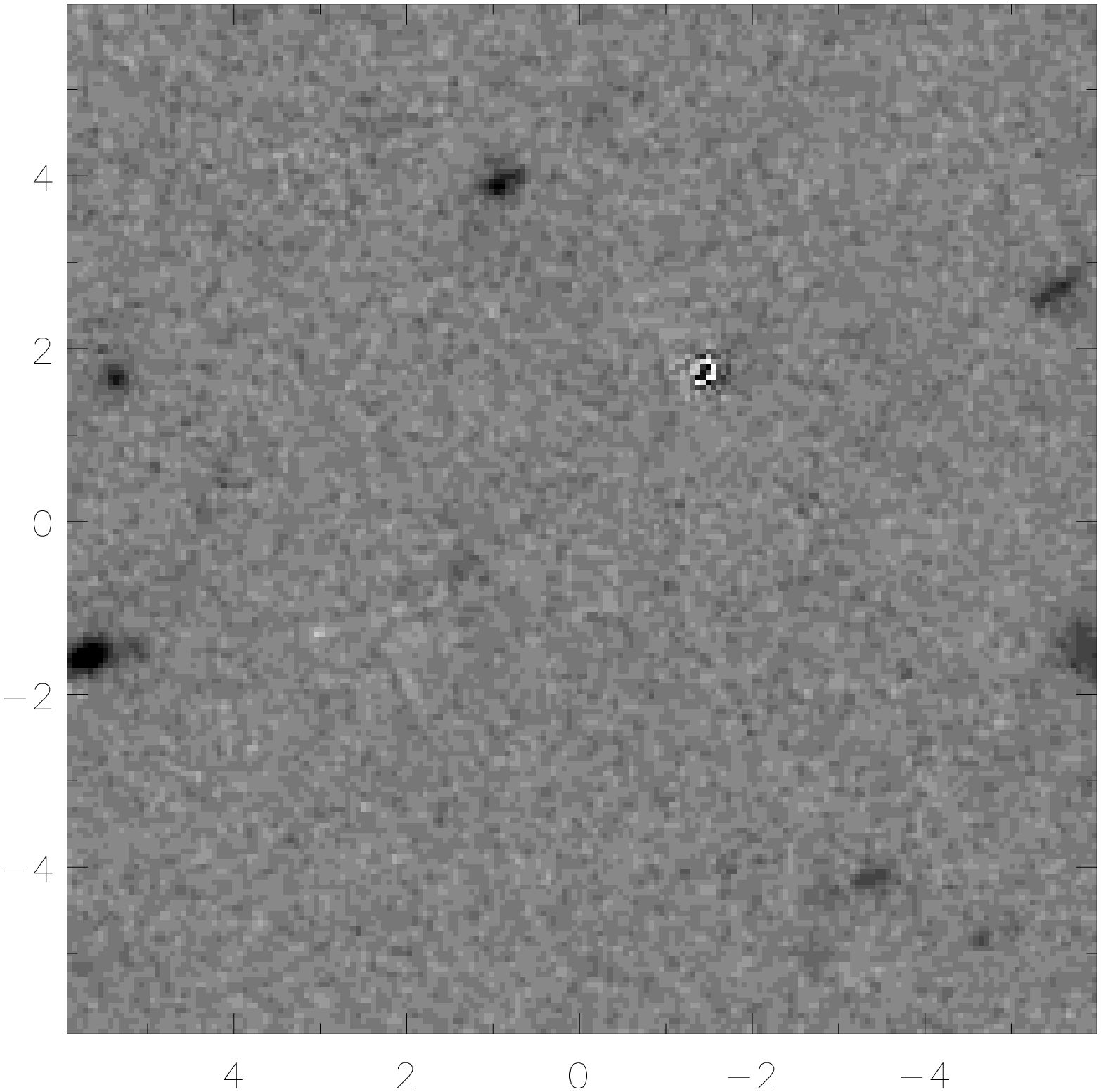,width=0.3\textwidth}\\
\\
\epsfig{file=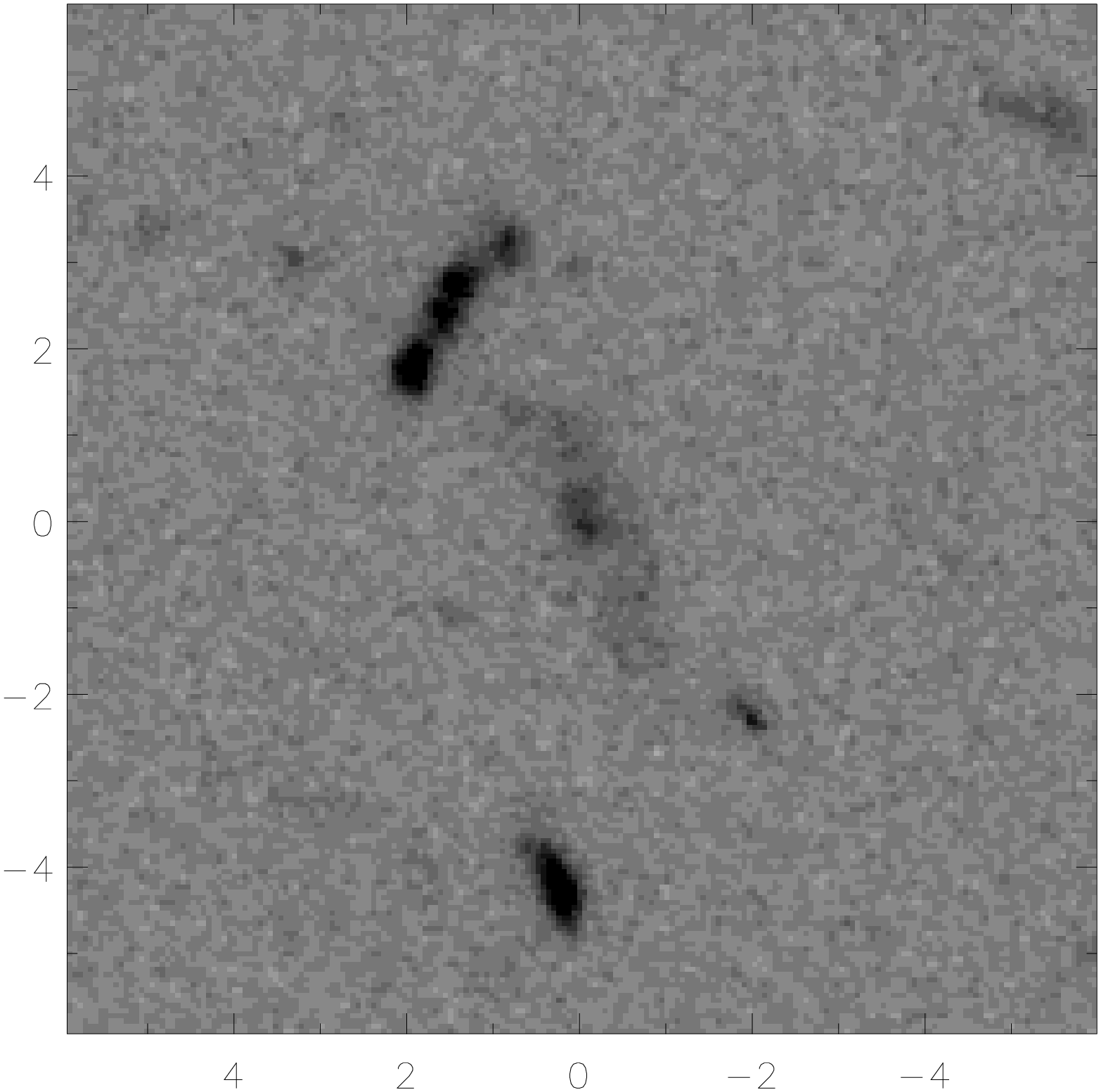,width=0.3\textwidth}&
\epsfig{file=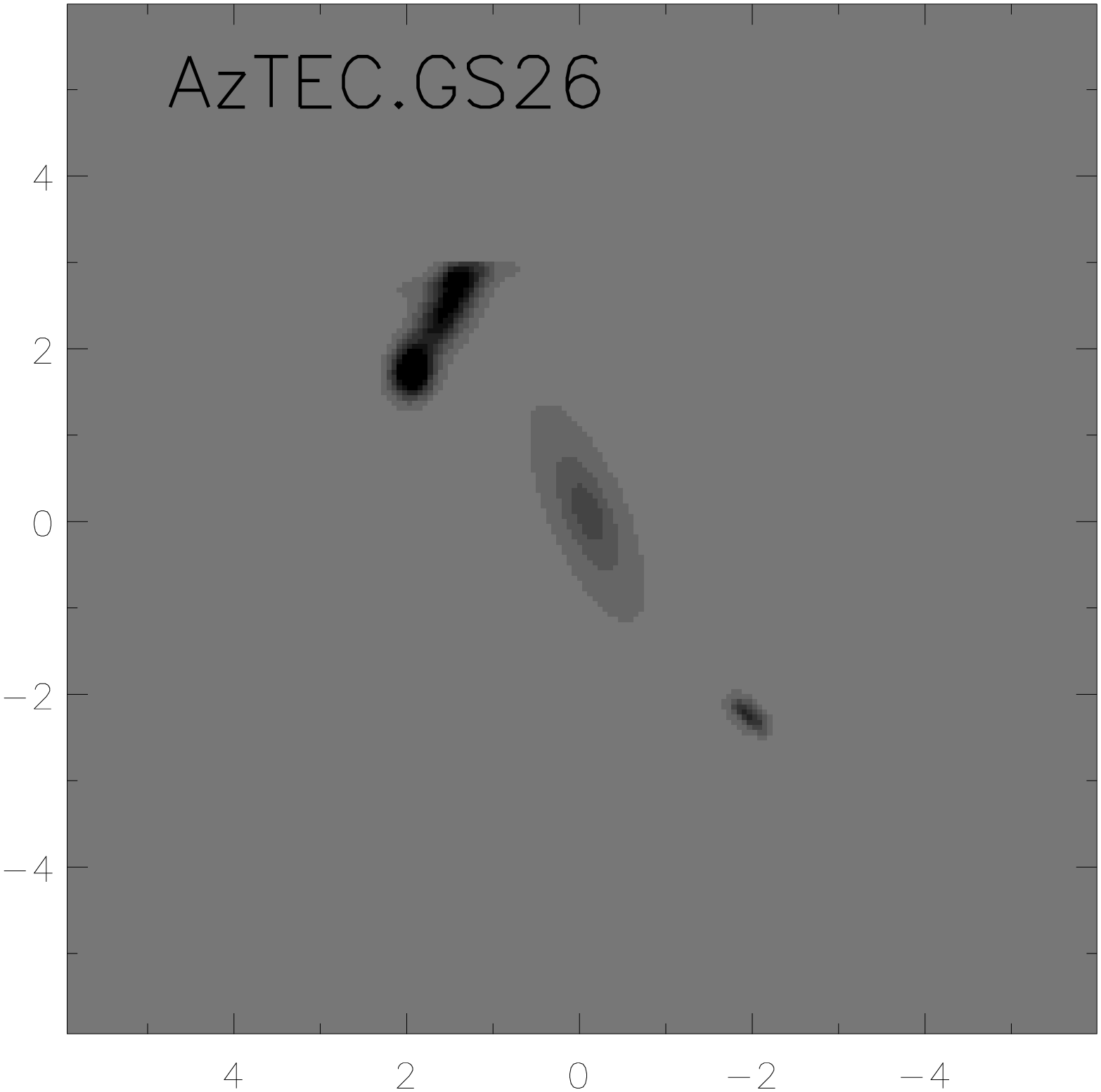,width=0.3\textwidth}&
\epsfig{file=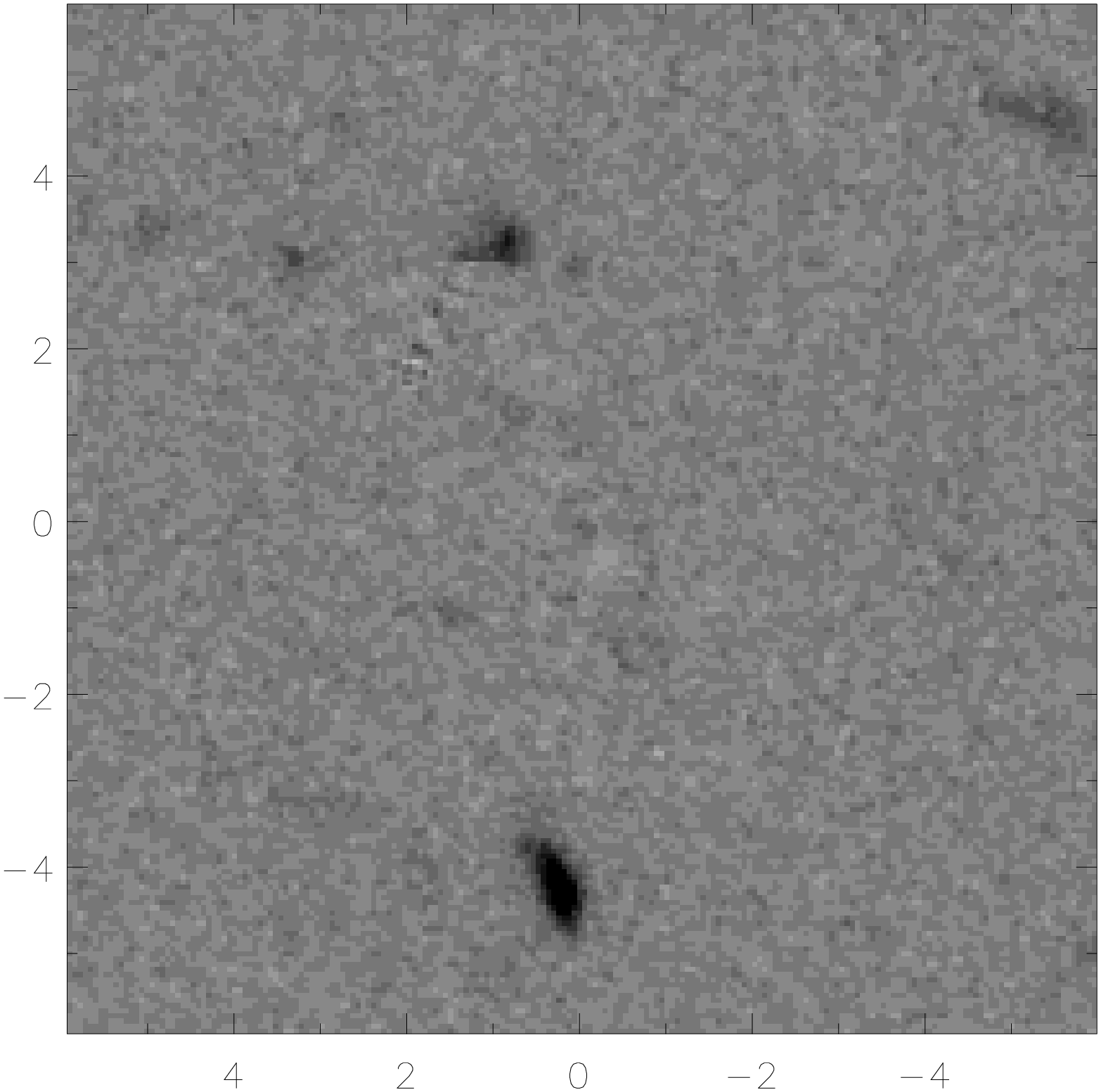,width=0.3\textwidth}\\
\\
\epsfig{file=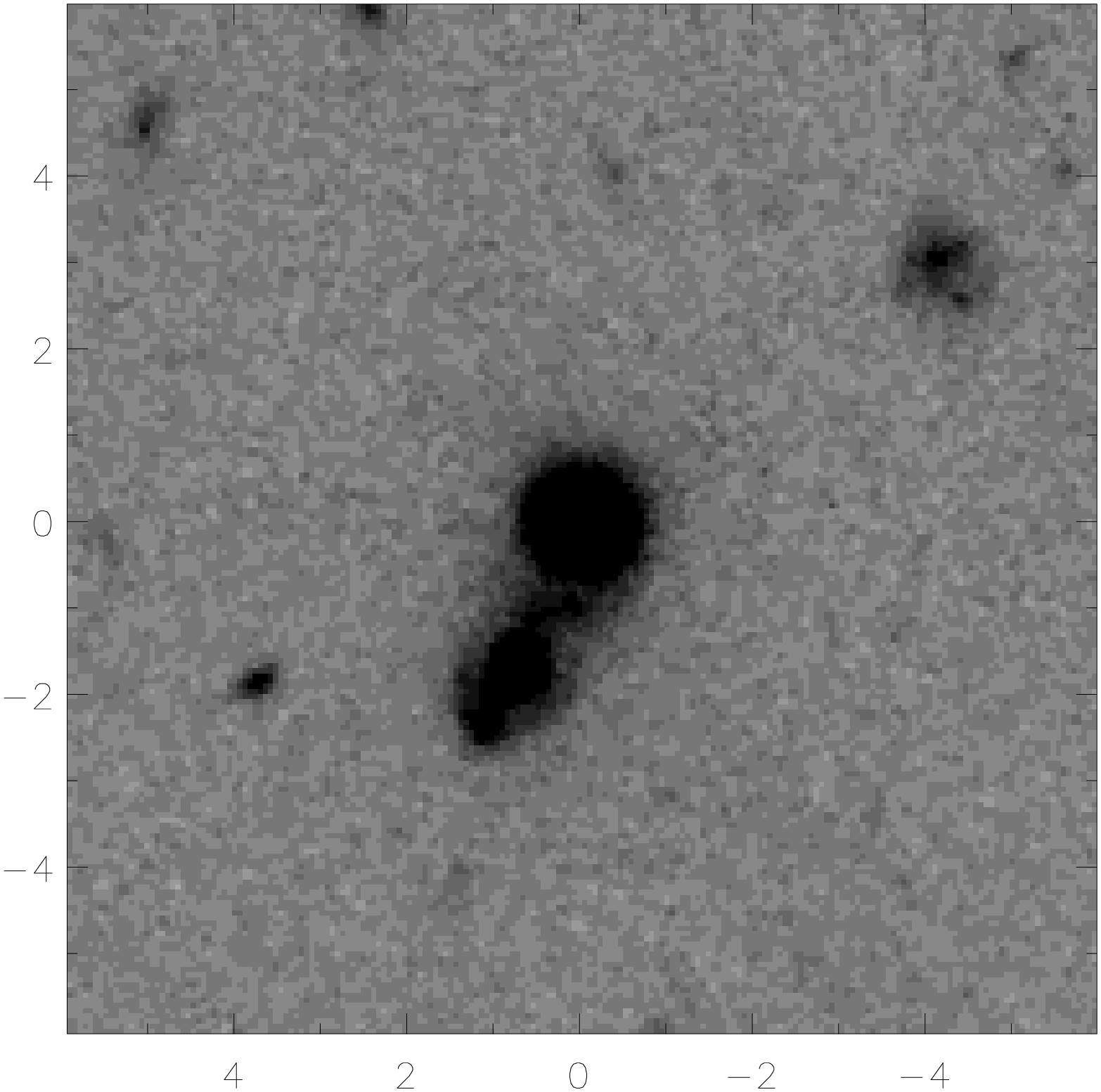,width=0.3\textwidth}&
\epsfig{file=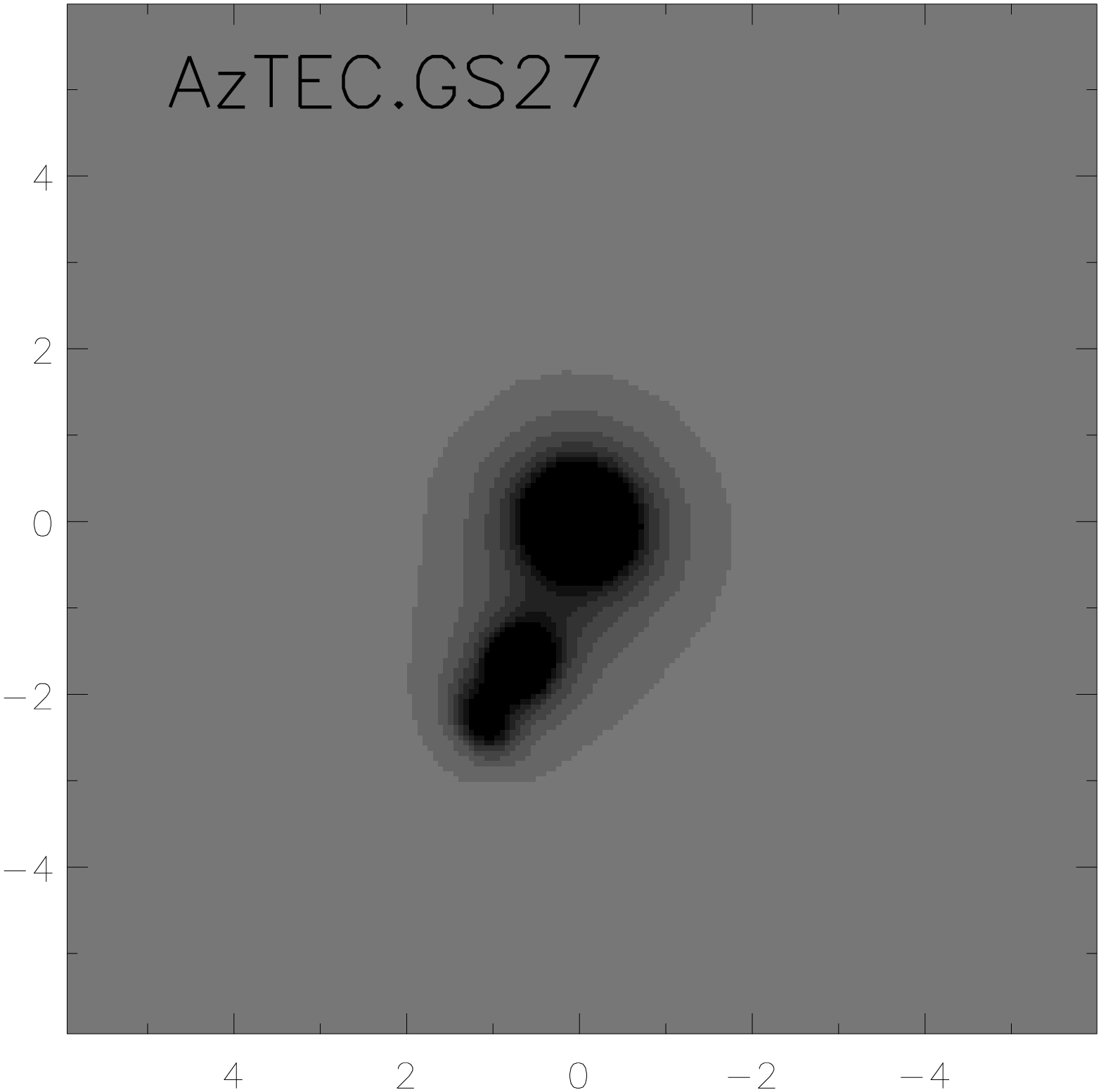,width=0.3\textwidth}&
\epsfig{file=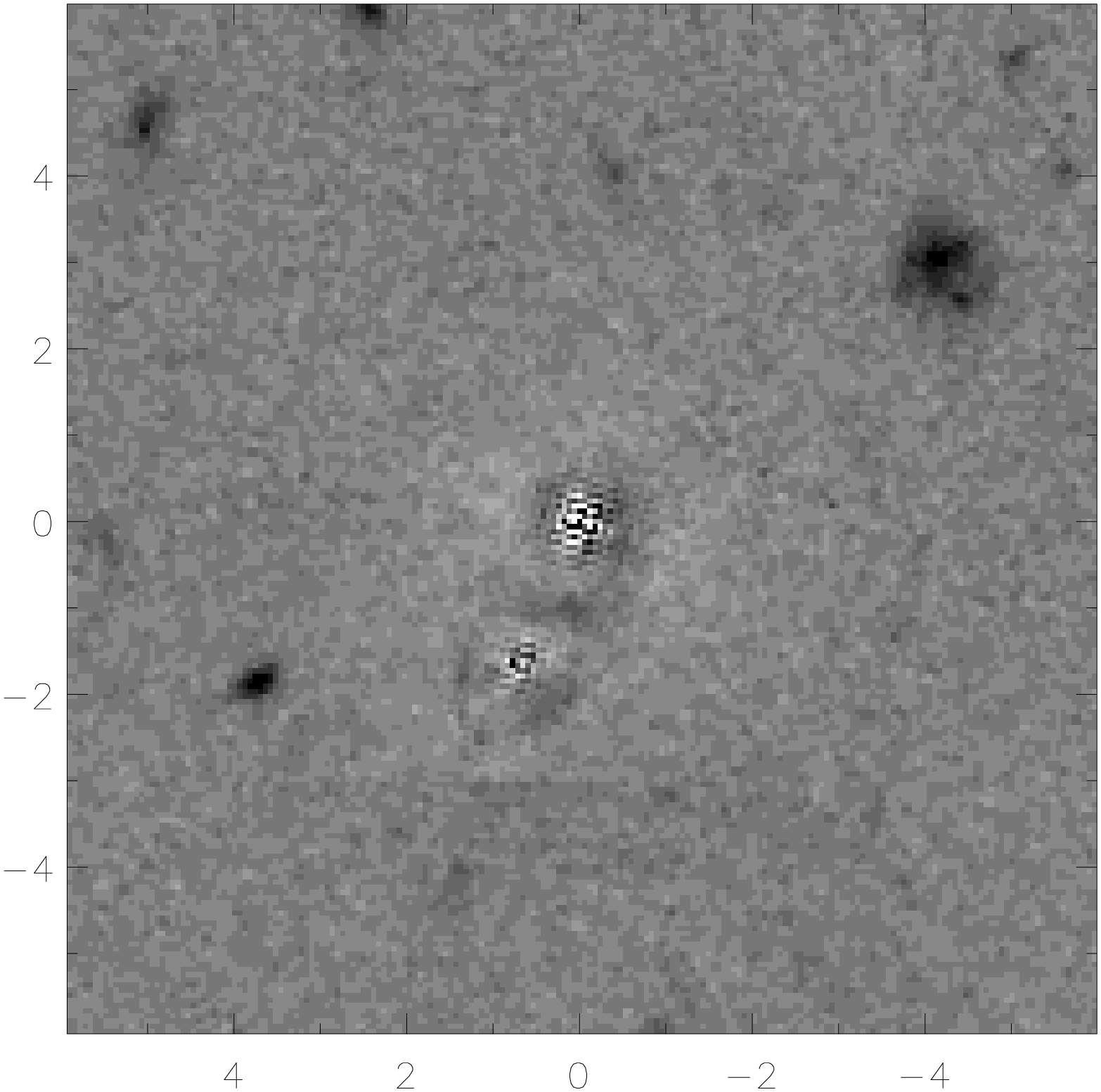,width=0.3\textwidth}\\
\\
\epsfig{file=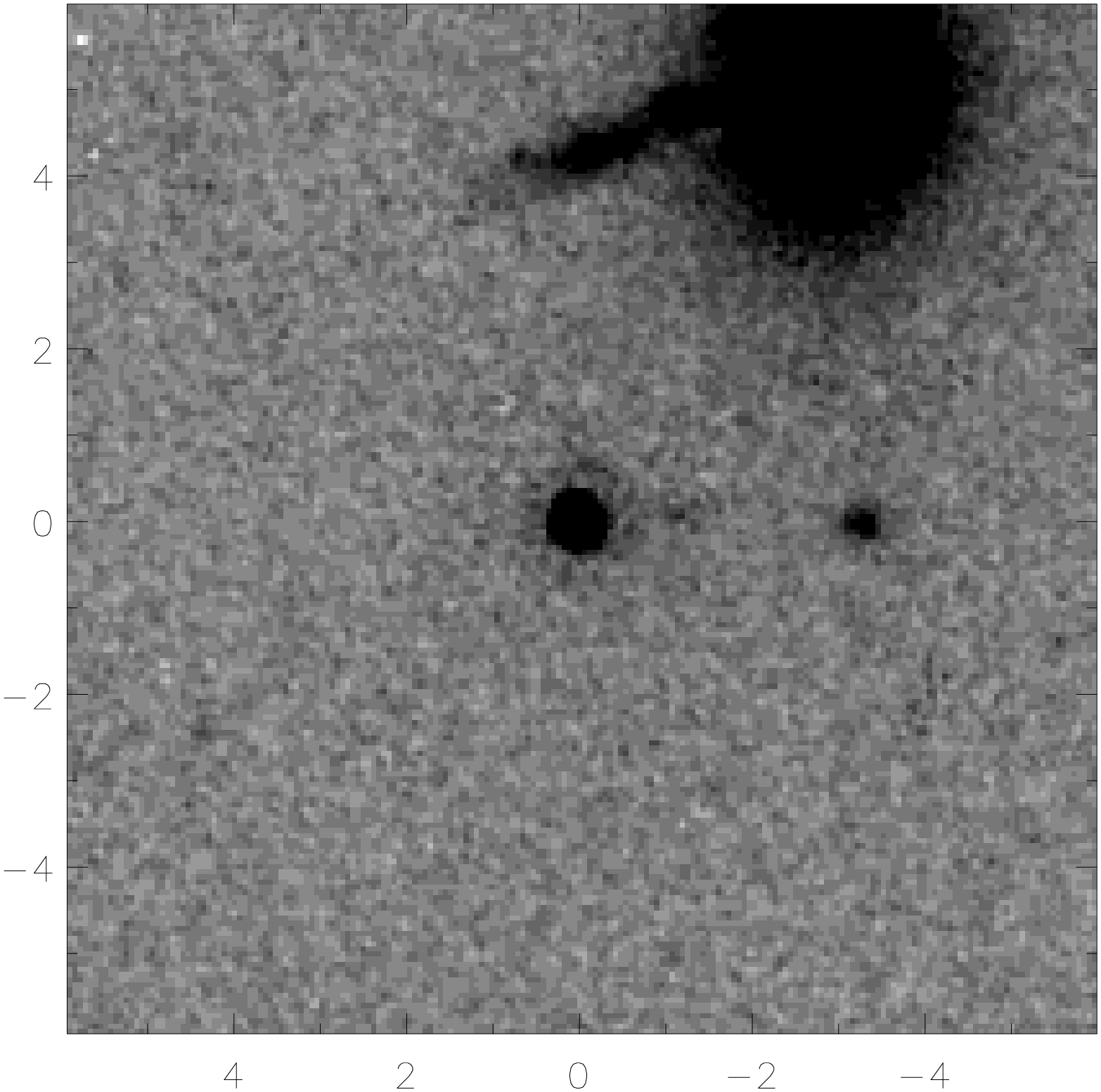,width=0.3\textwidth}&
\epsfig{file=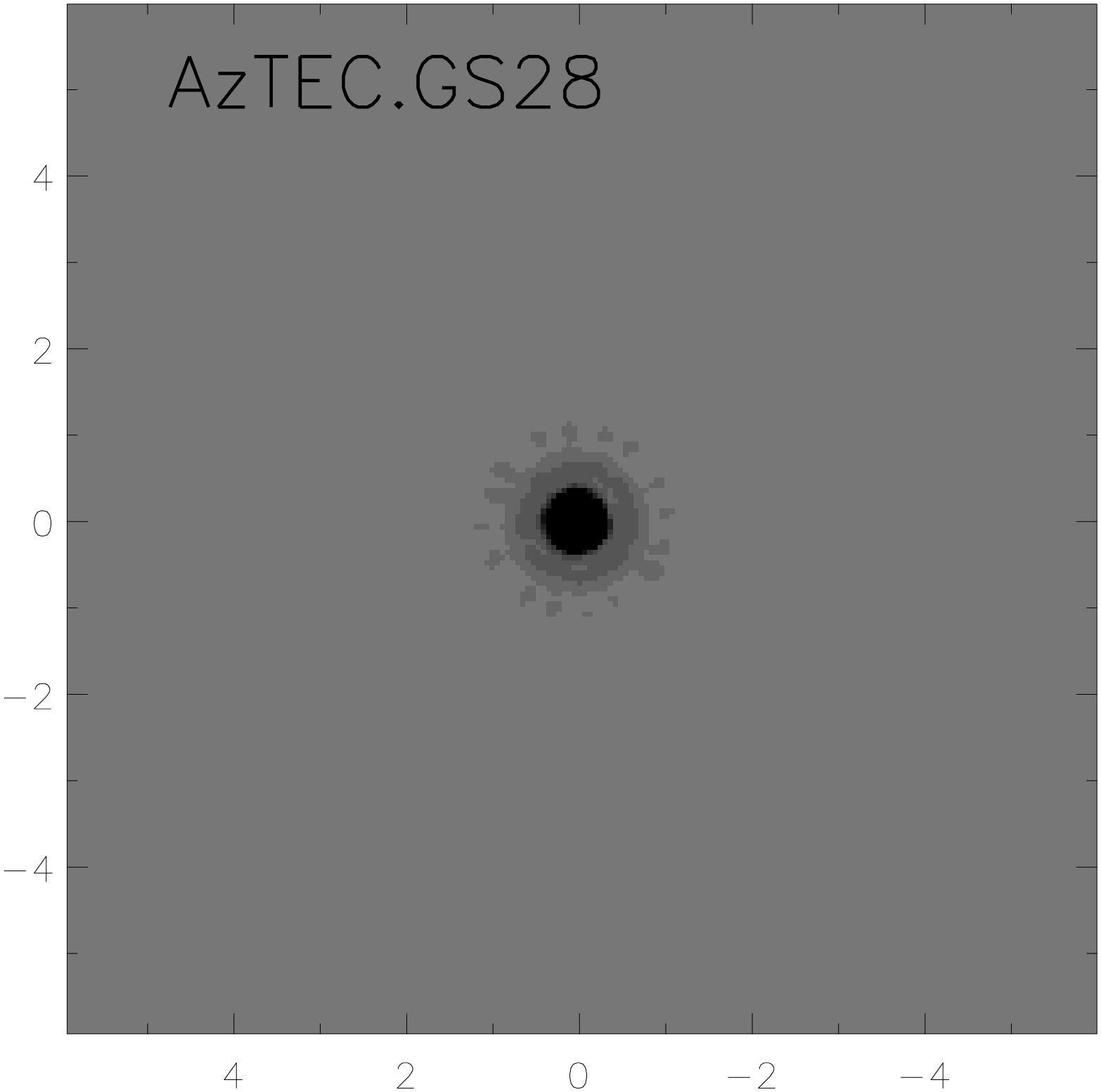,width=0.3\textwidth}&
\epsfig{file=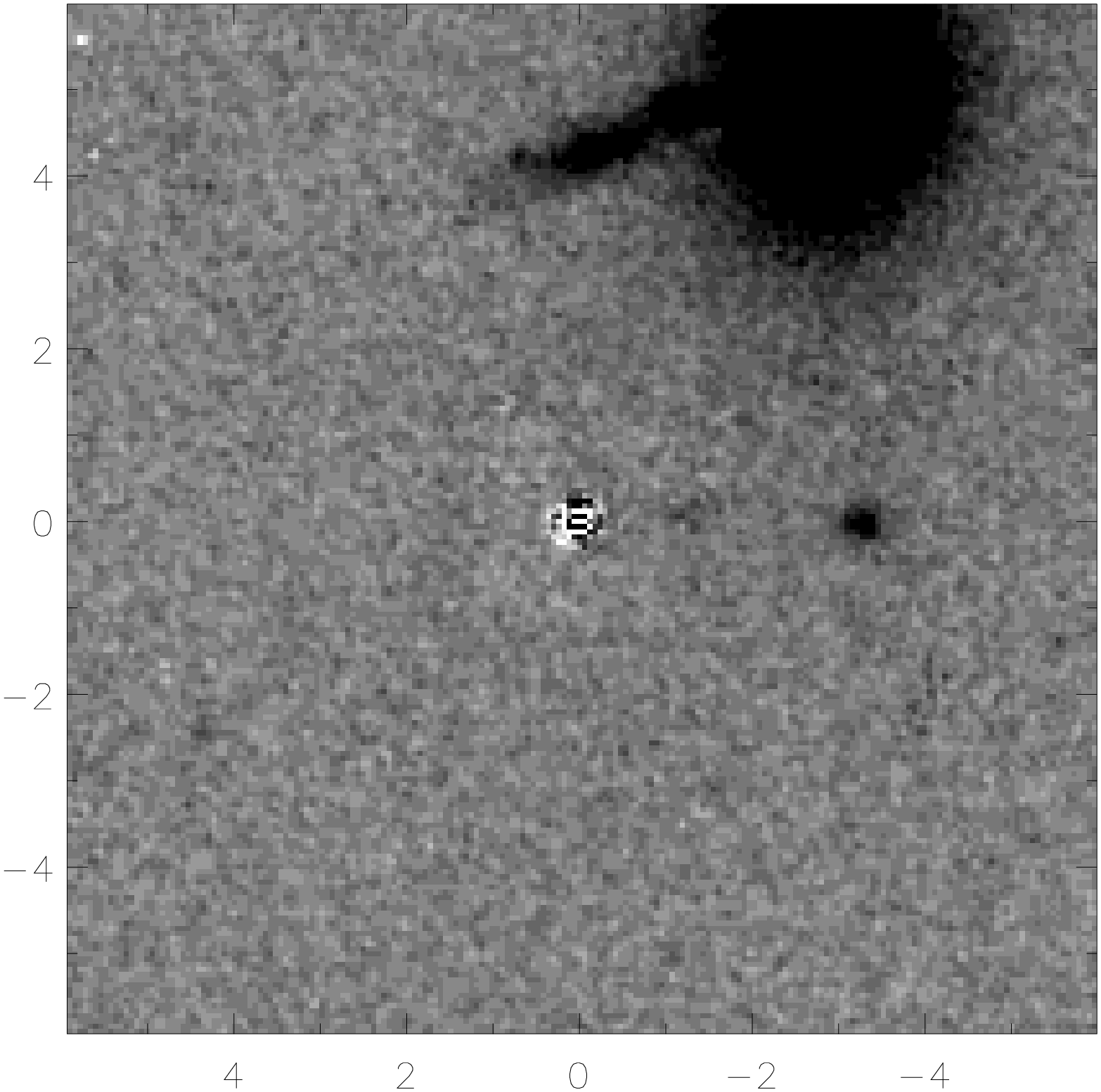,width=0.3\textwidth}\\
\end{tabular}
\addtocounter{figure}{-1}
\caption{- continued}
\end{figure*}
\end{center}


\begin{center}
\begin{figure*}
\begin{tabular}{ccc}
\epsfig{file=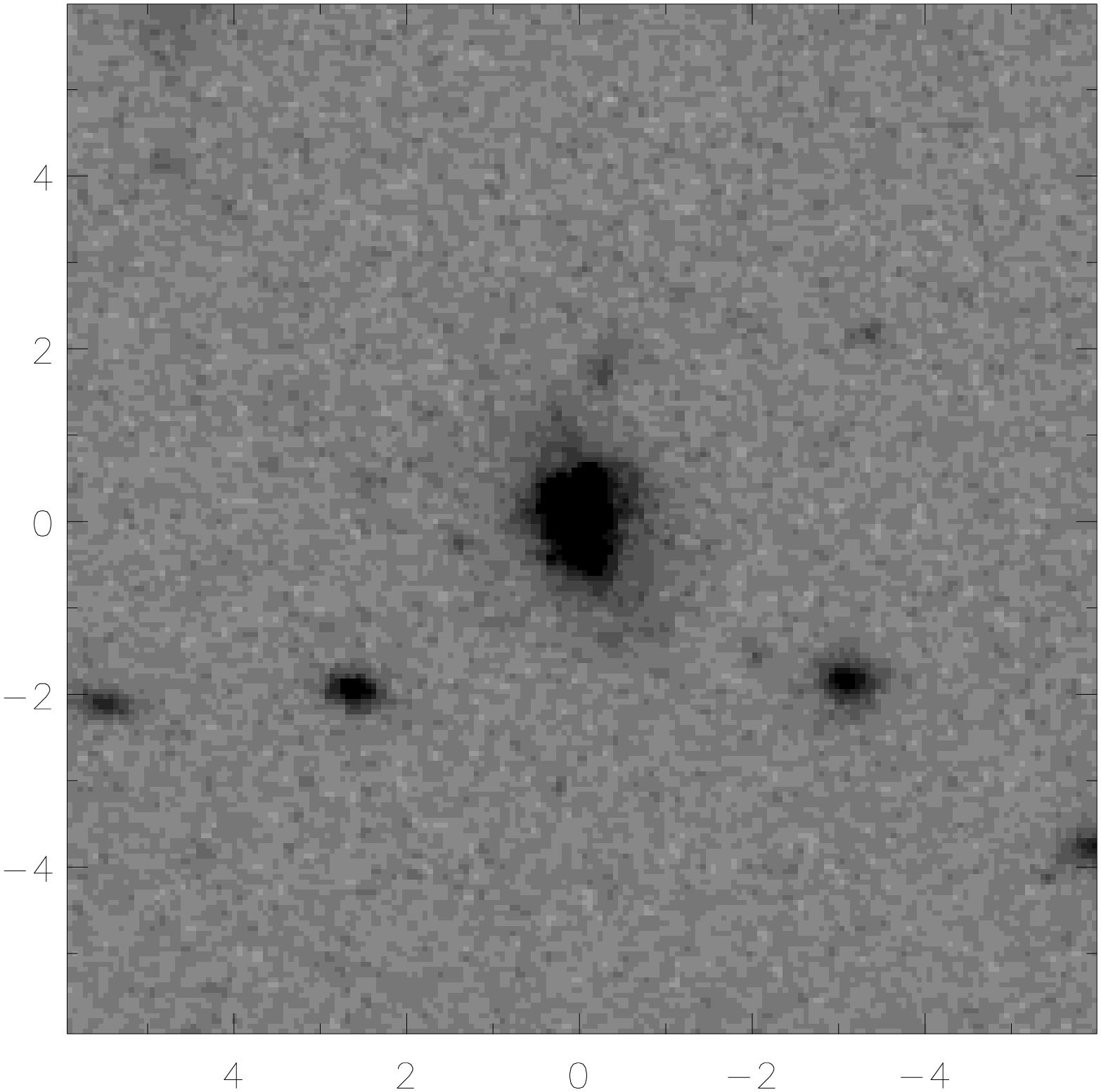,width=0.3\textwidth}&
\epsfig{file=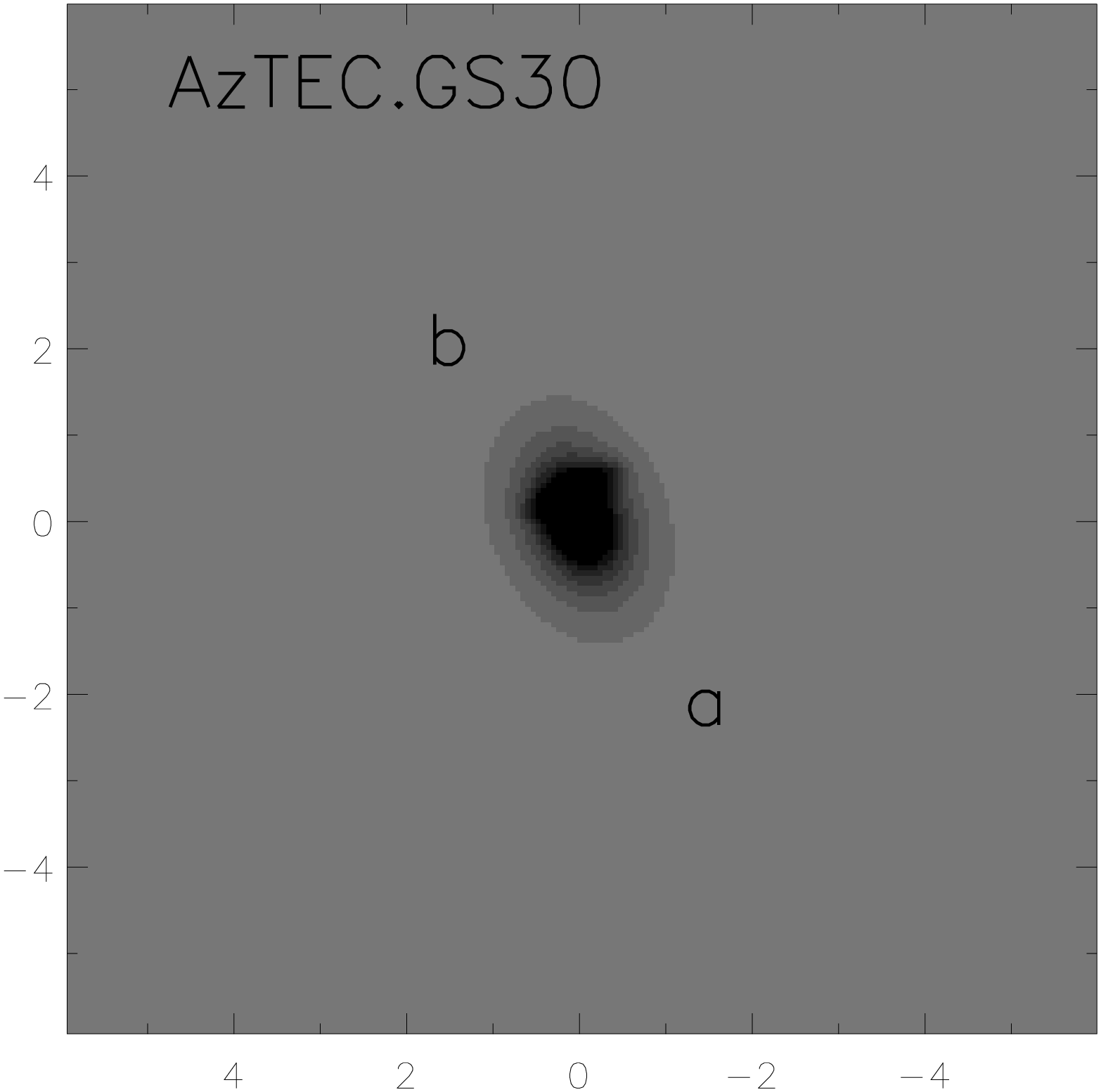,width=0.3\textwidth}&
\epsfig{file=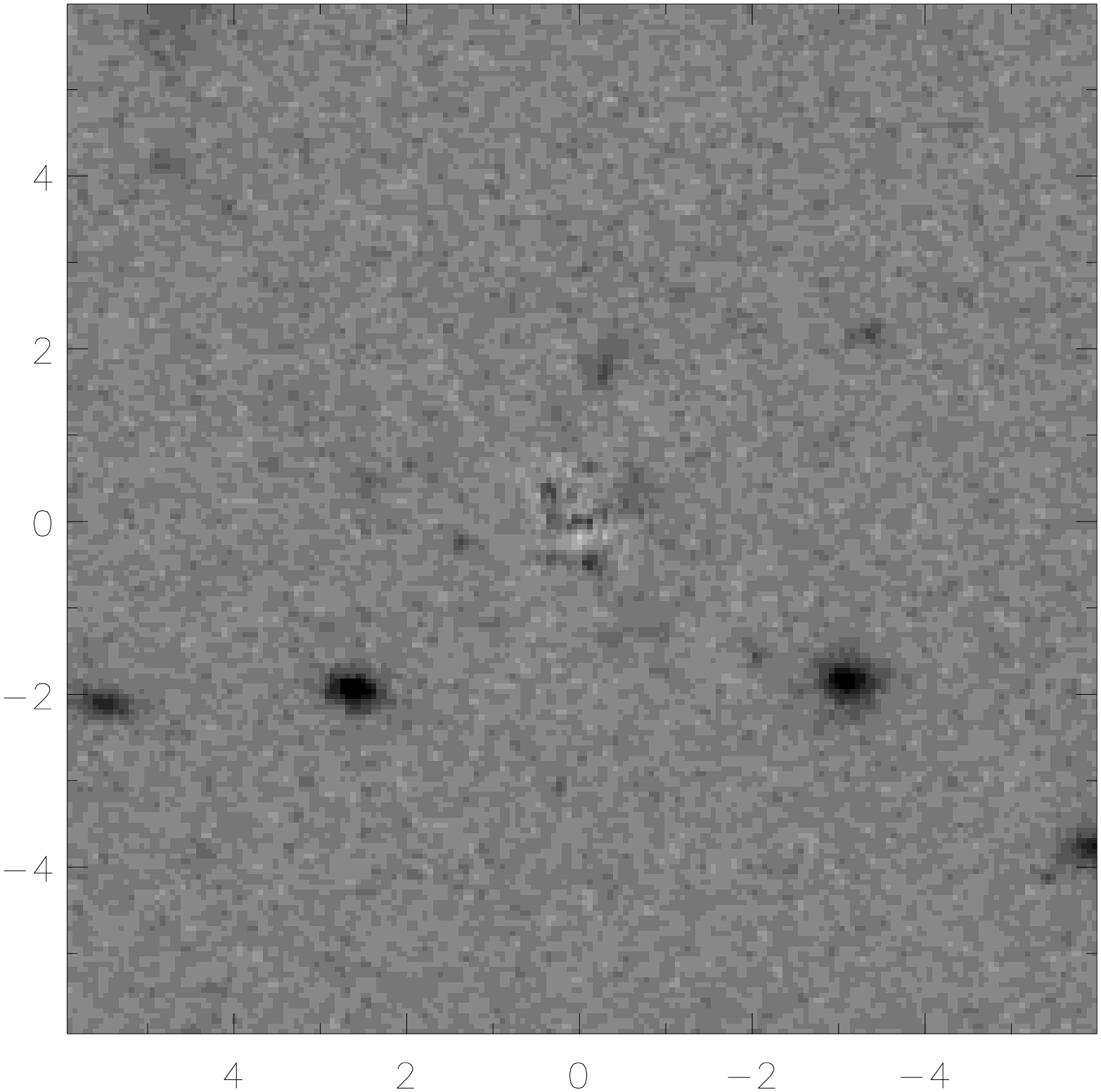,width=0.3\textwidth}\\
\\
\epsfig{file=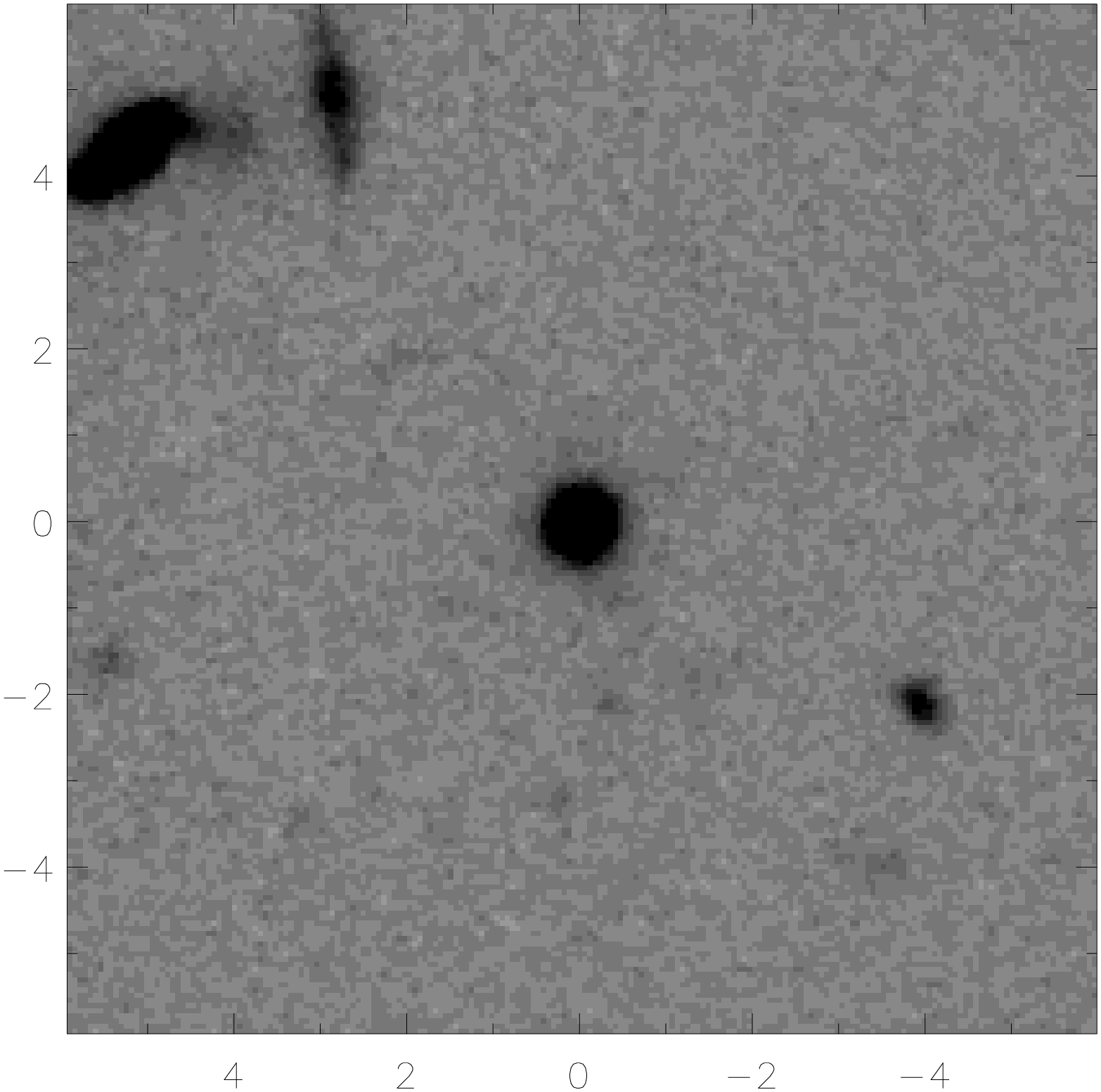,width=0.3\textwidth}&
\epsfig{file=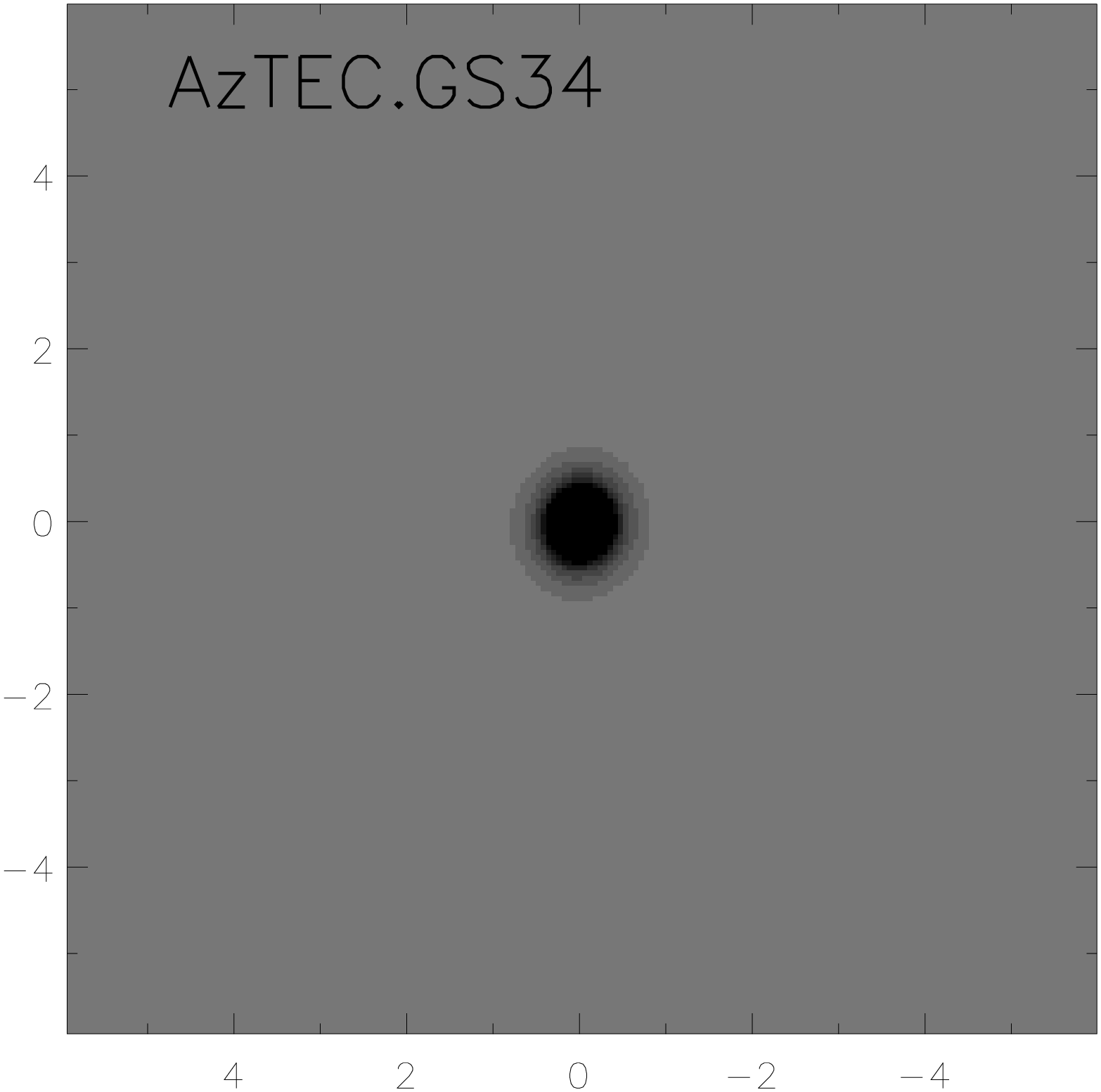,width=0.3\textwidth}&
\epsfig{file=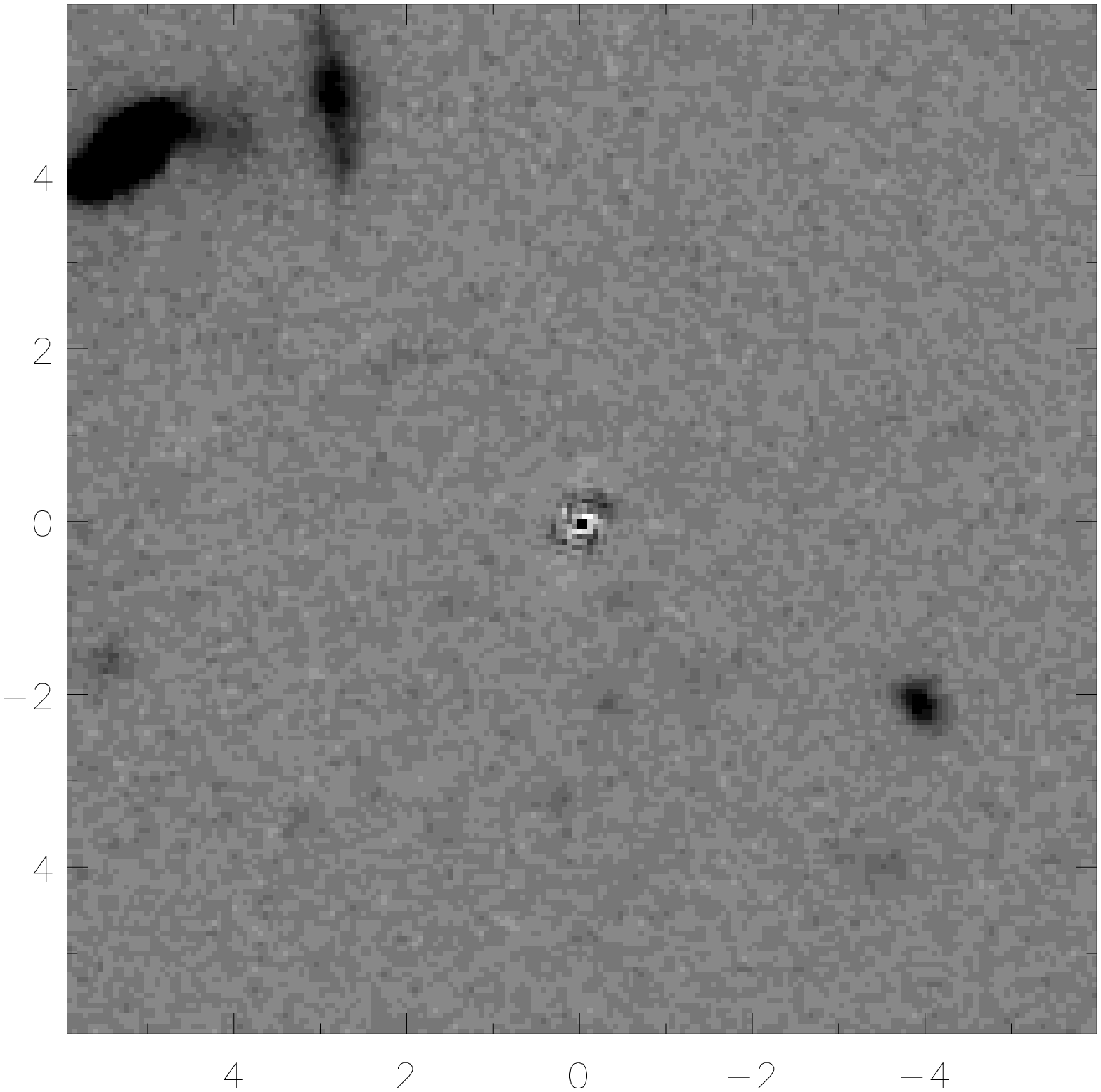,width=0.3\textwidth}\\
\\
\epsfig{file=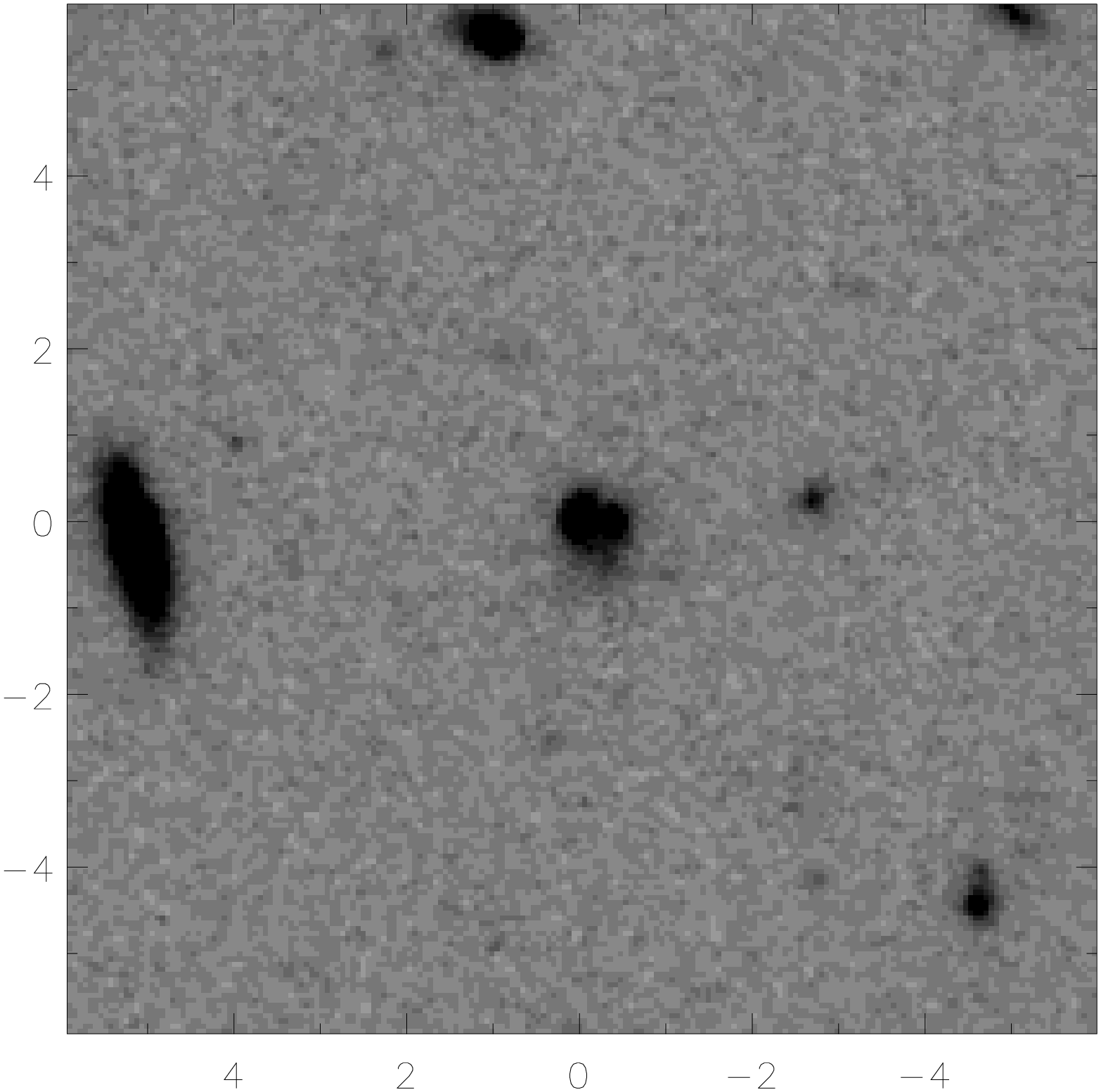,width=0.3\textwidth}&
\epsfig{file=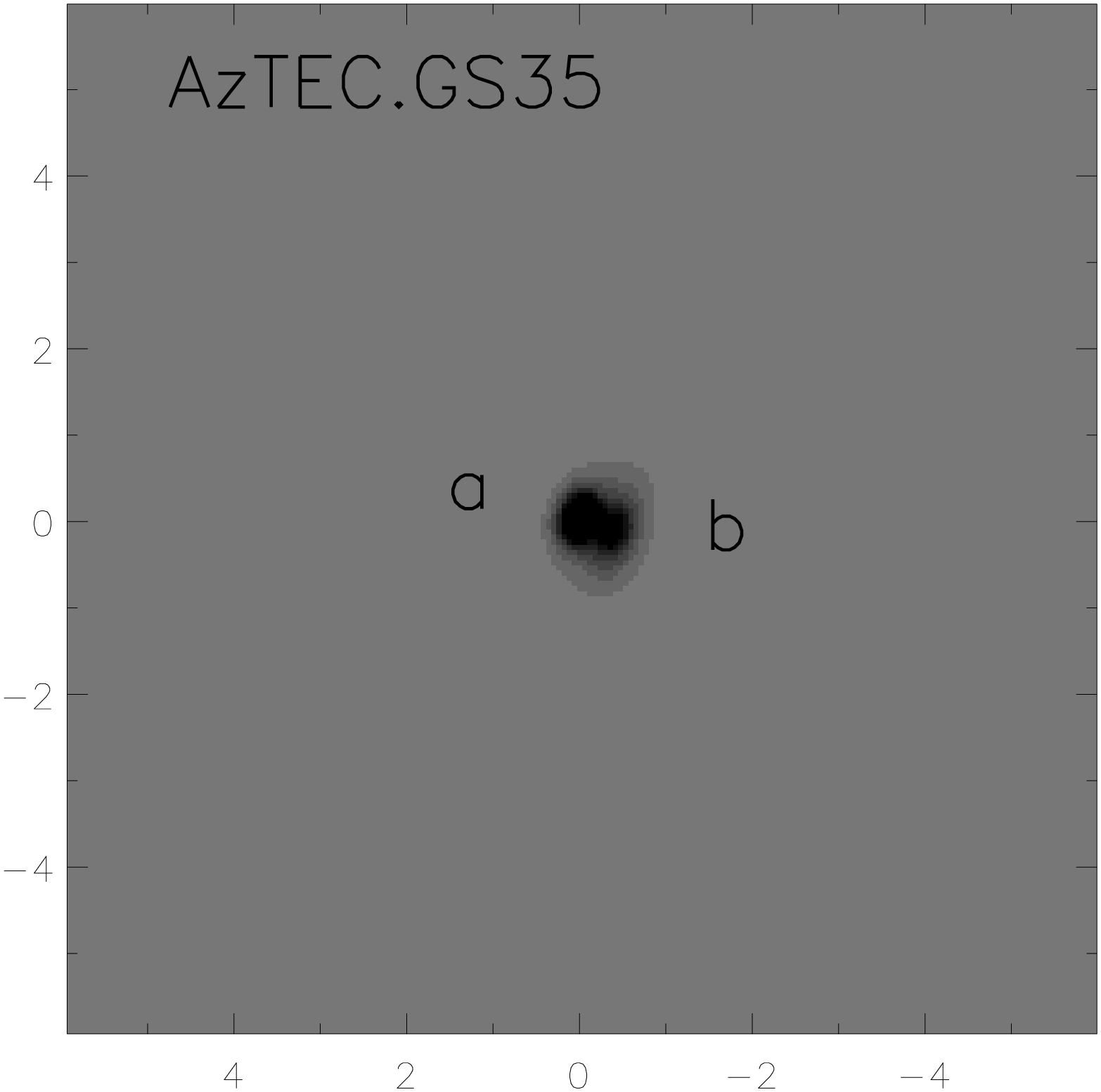,width=0.3\textwidth}&
\epsfig{file=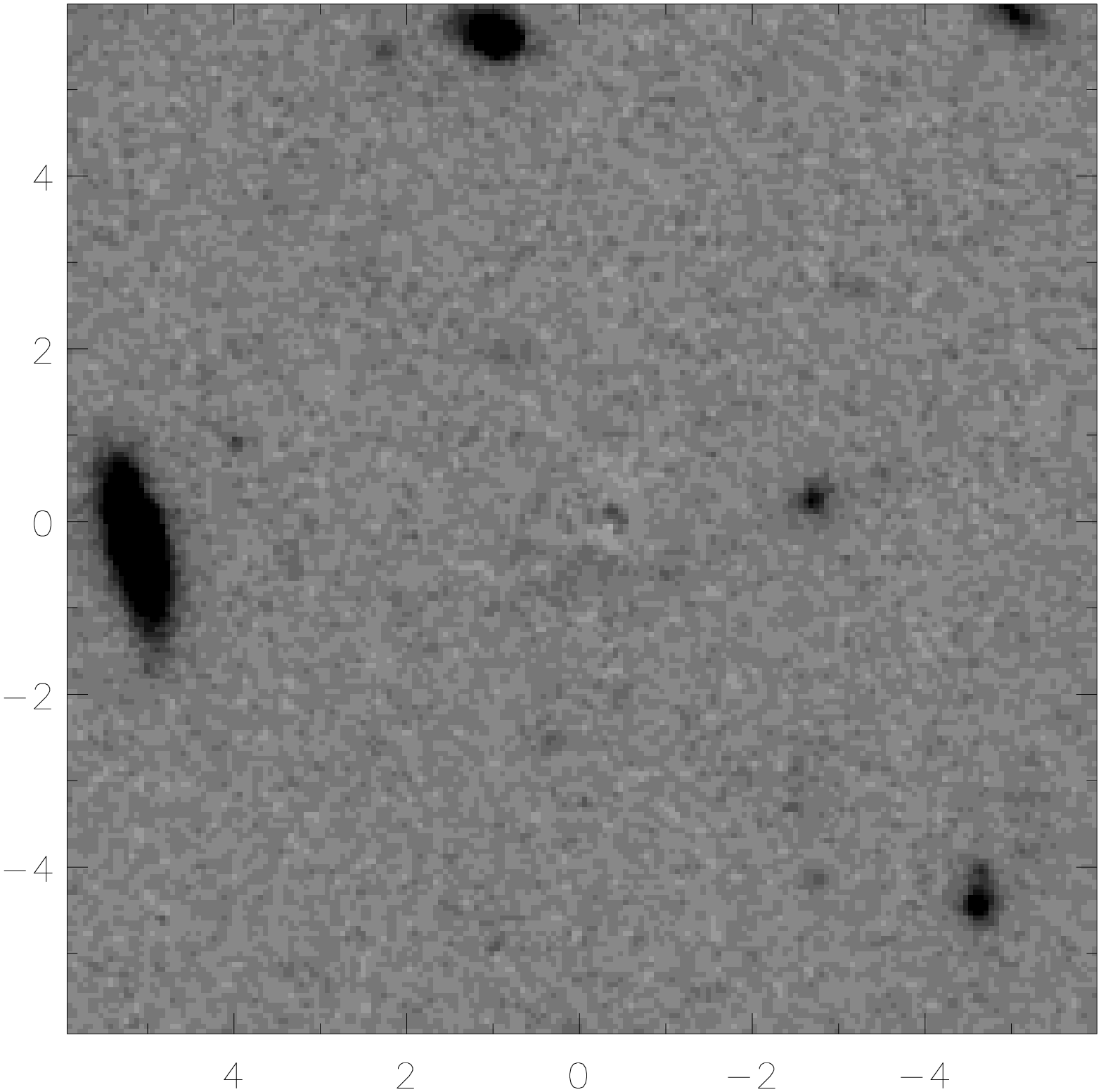,width=0.3\textwidth}\\
\\
\epsfig{file=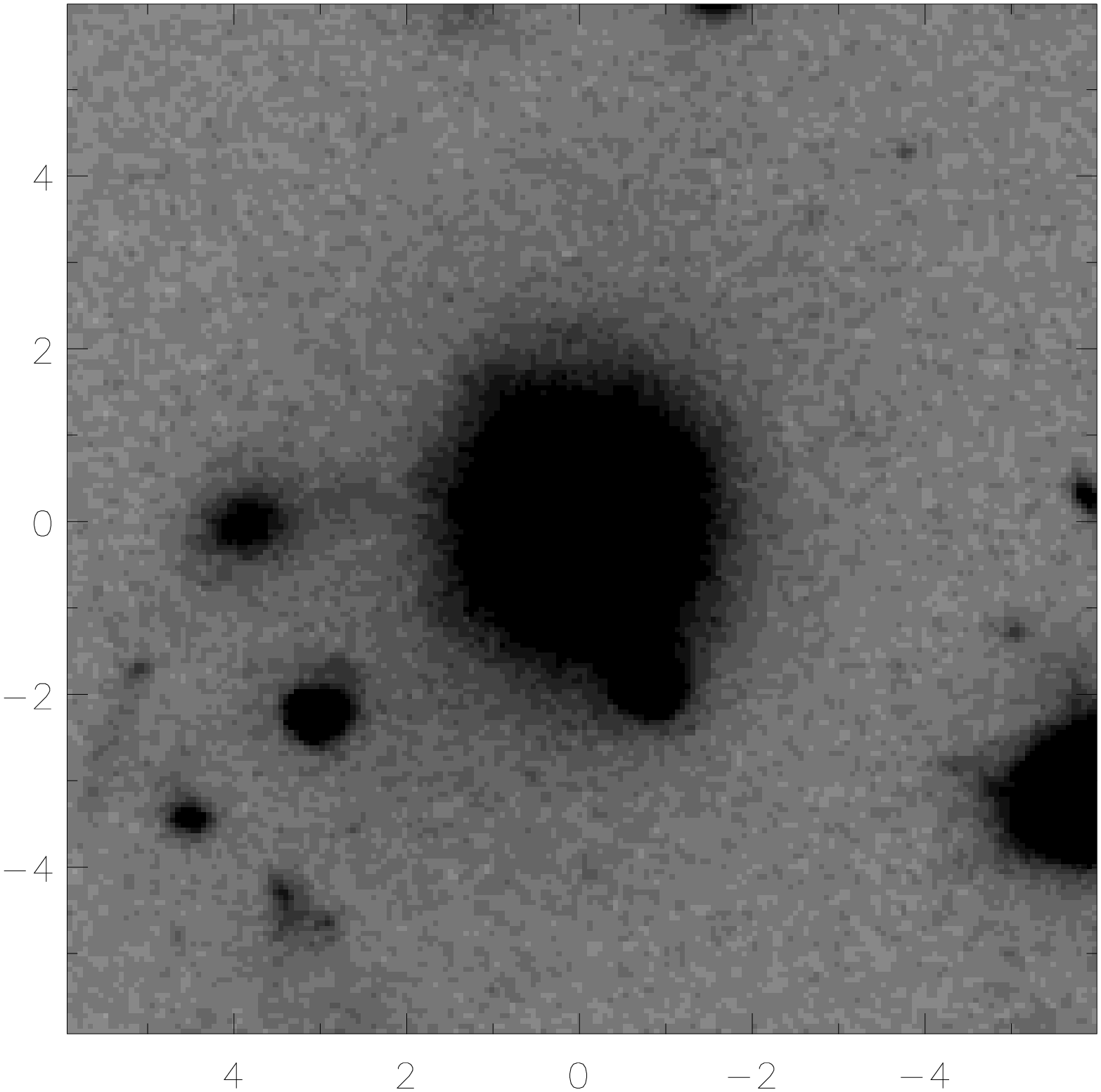,width=0.3\textwidth}&
\epsfig{file=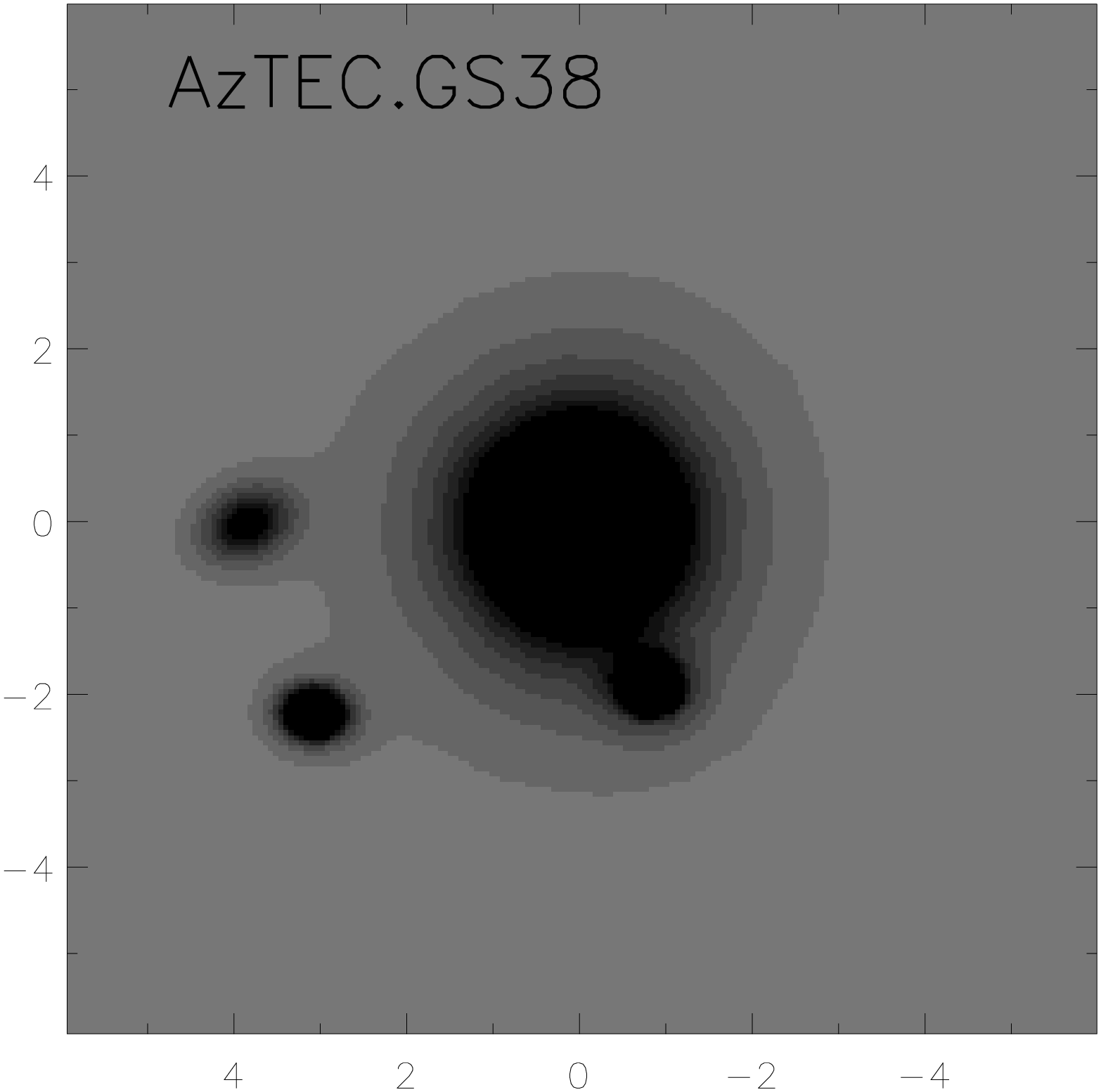,width=0.3\textwidth}&
\epsfig{file=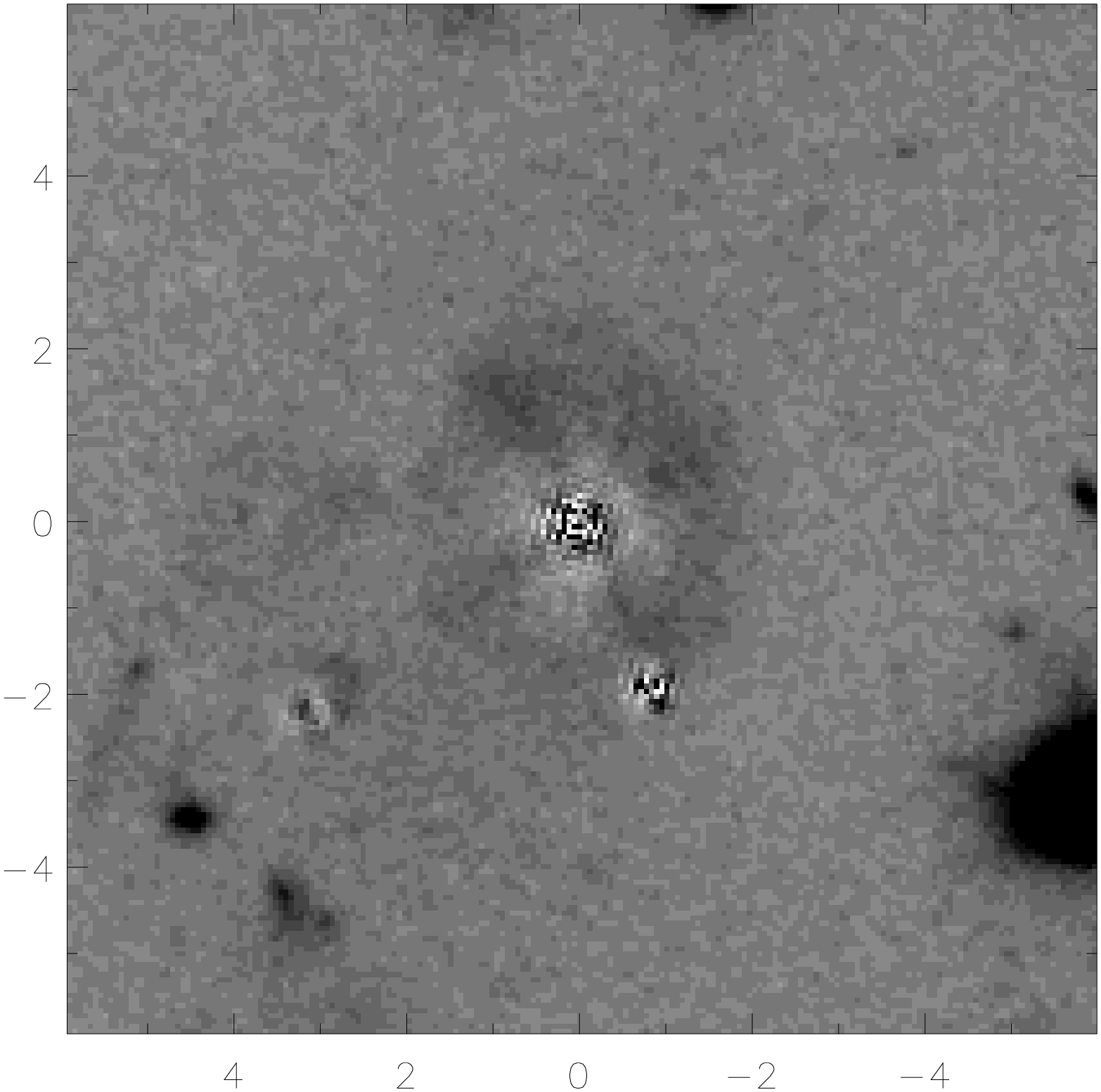,width=0.3\textwidth}\\
\end{tabular}
\addtocounter{figure}{-1}
\caption{- continued}
\end{figure*}
\end{center}


\begin{center}
\begin{figure*}
\begin{tabular}{ccc}
\epsfig{file=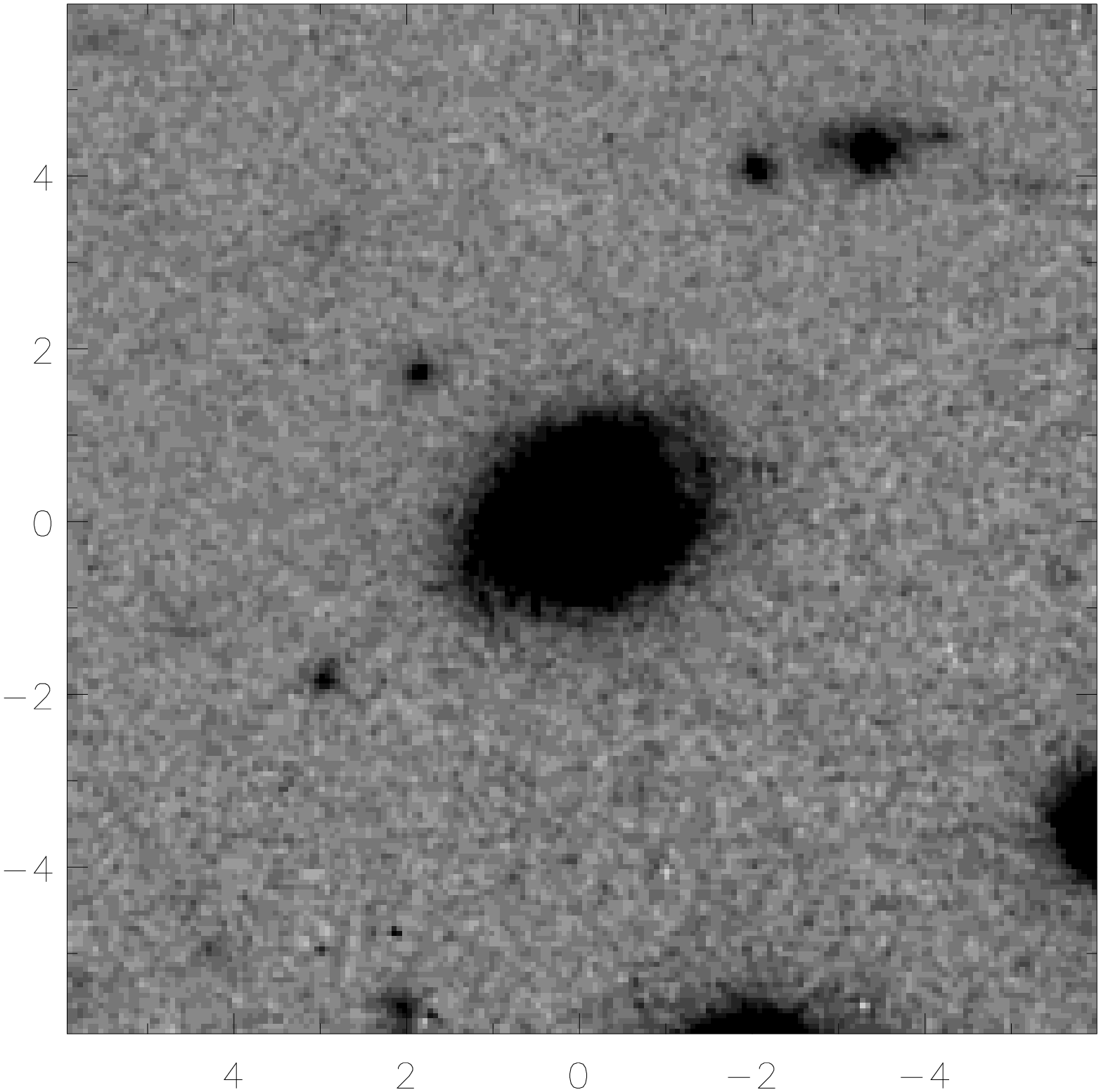,width=0.3\textwidth}&
\epsfig{file=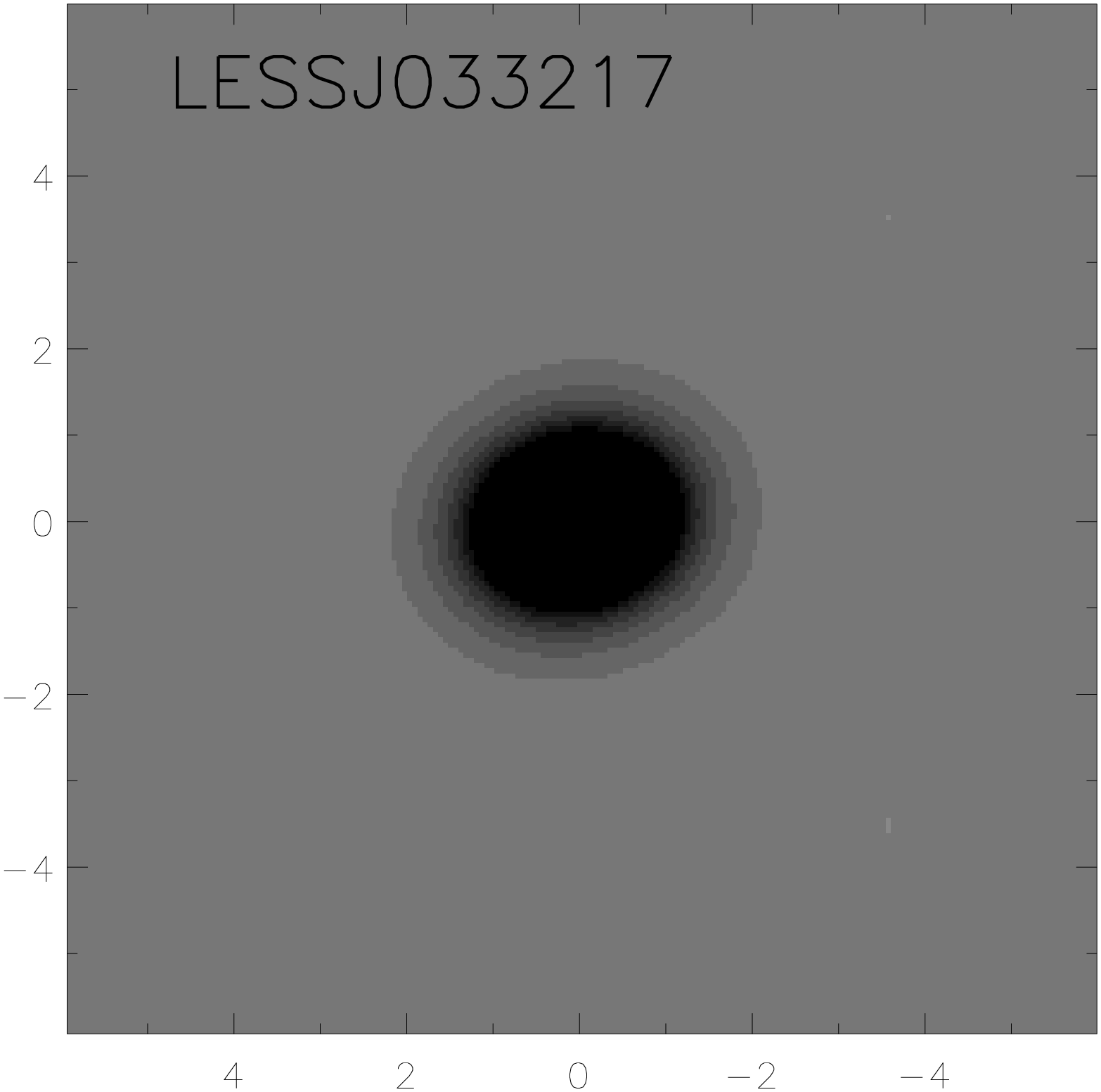,width=0.3\textwidth}&
\epsfig{file=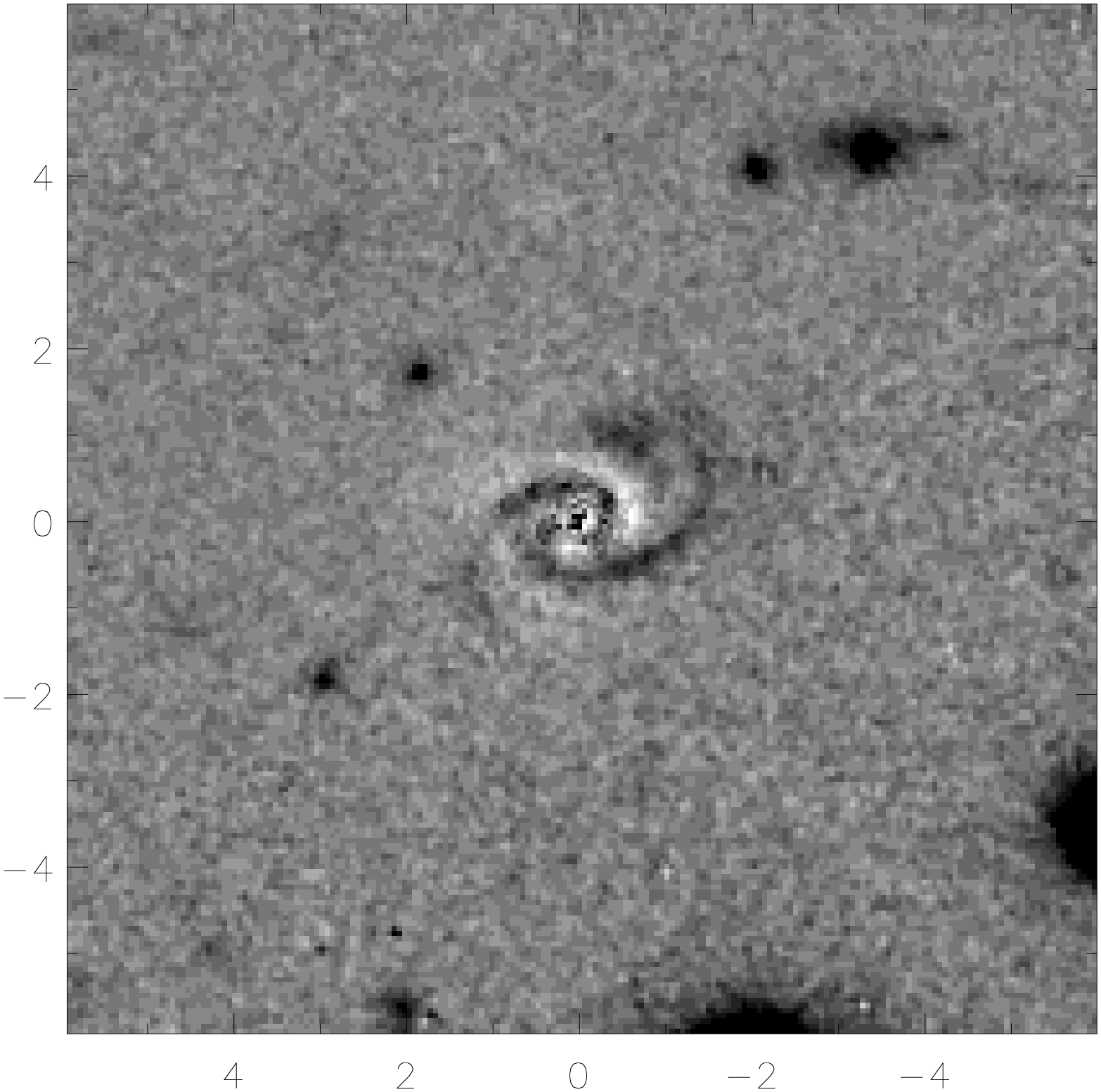,width=0.3\textwidth}\\
\\
\epsfig{file=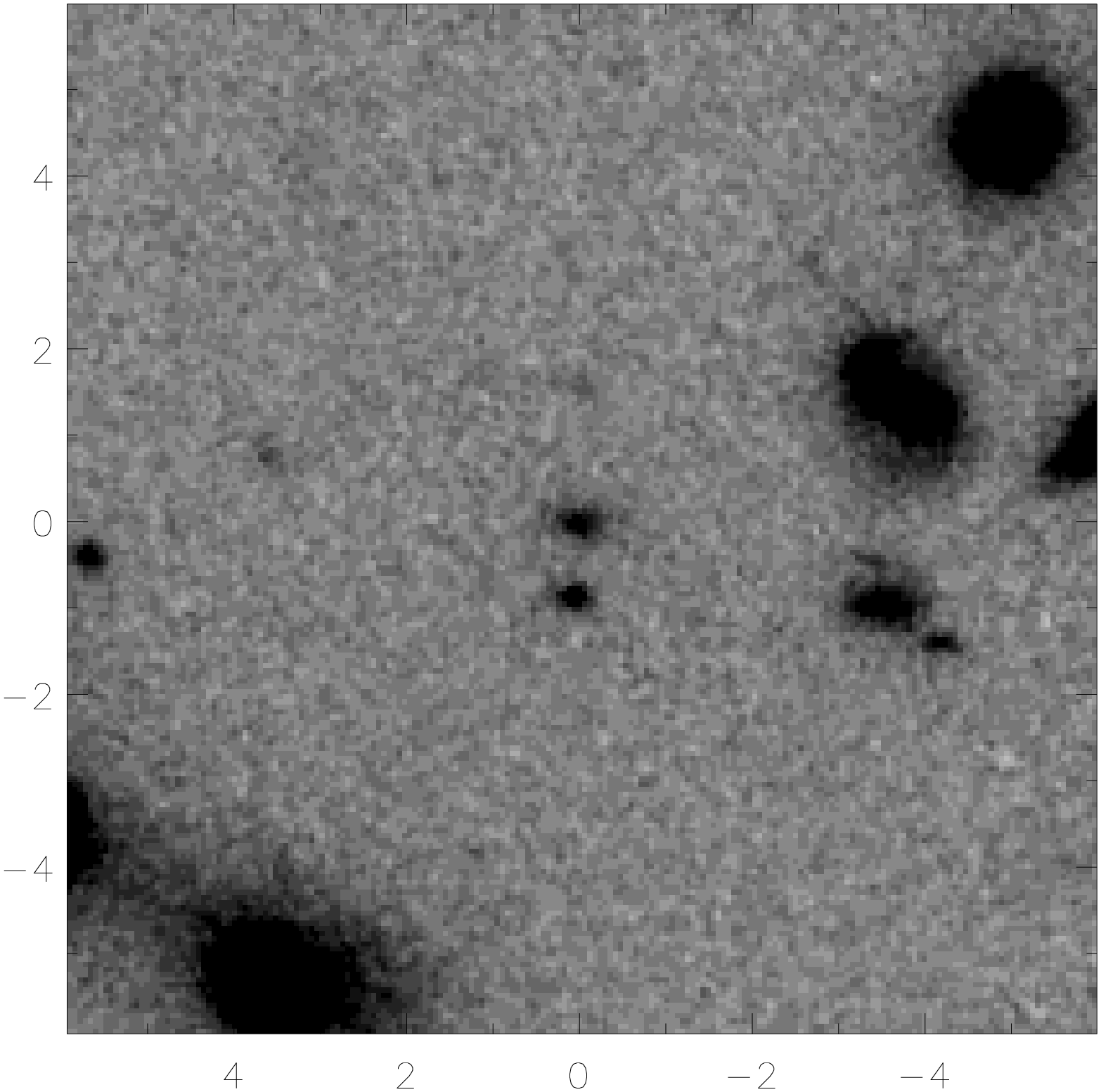,width=0.3\textwidth}&
\epsfig{file=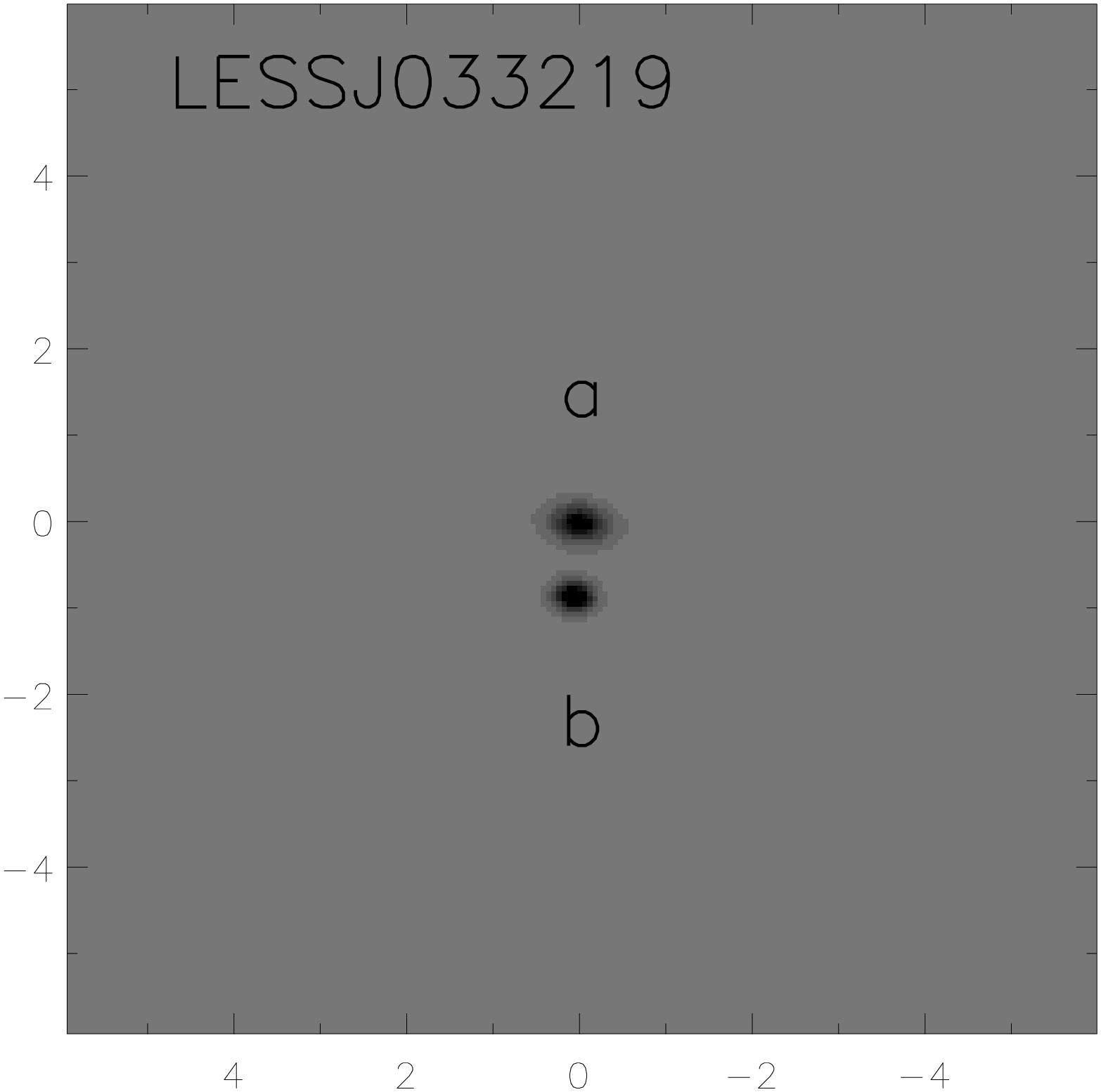,width=0.3\textwidth}&
\epsfig{file=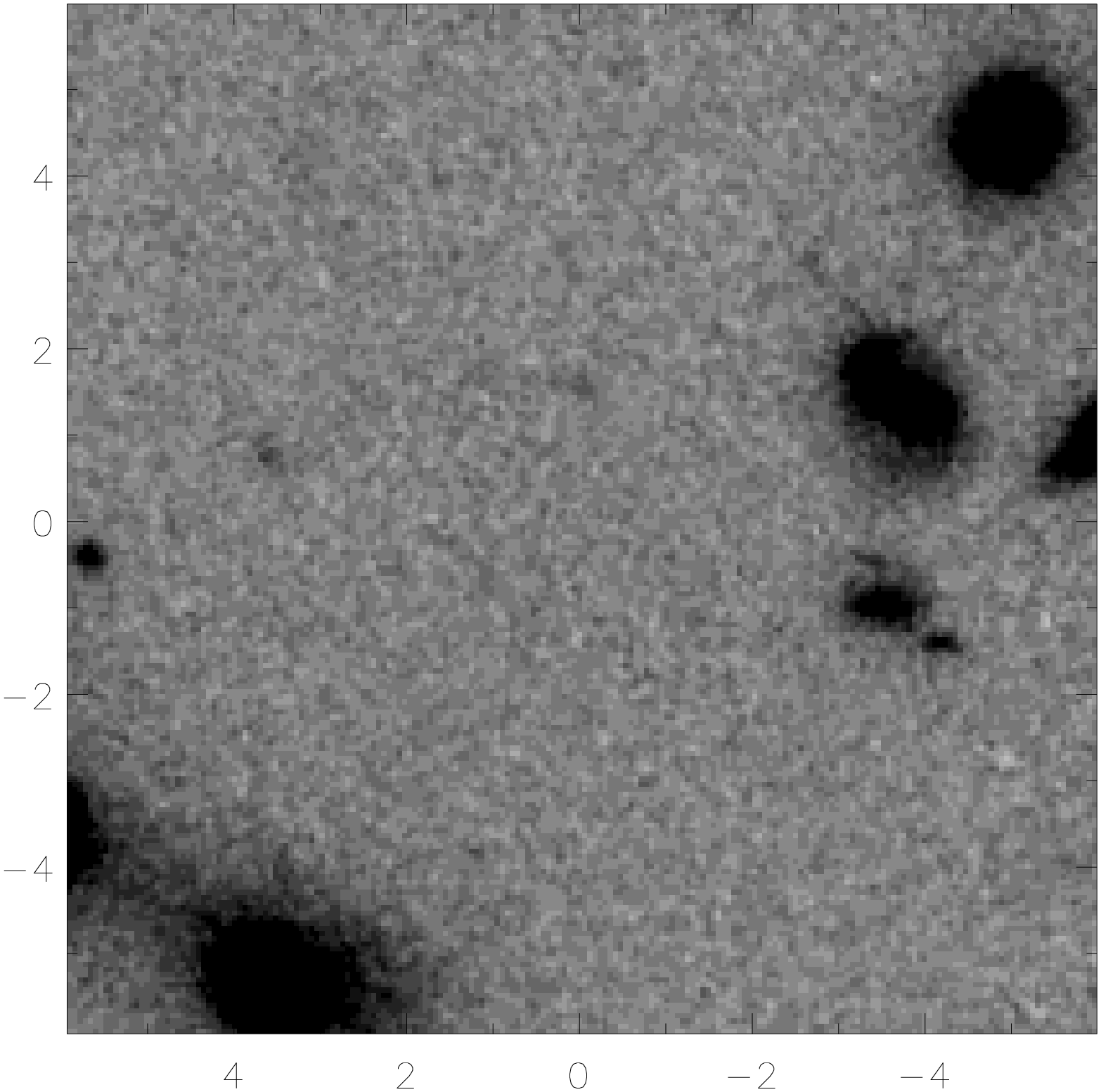,width=0.3\textwidth}\\
\\
\epsfig{file=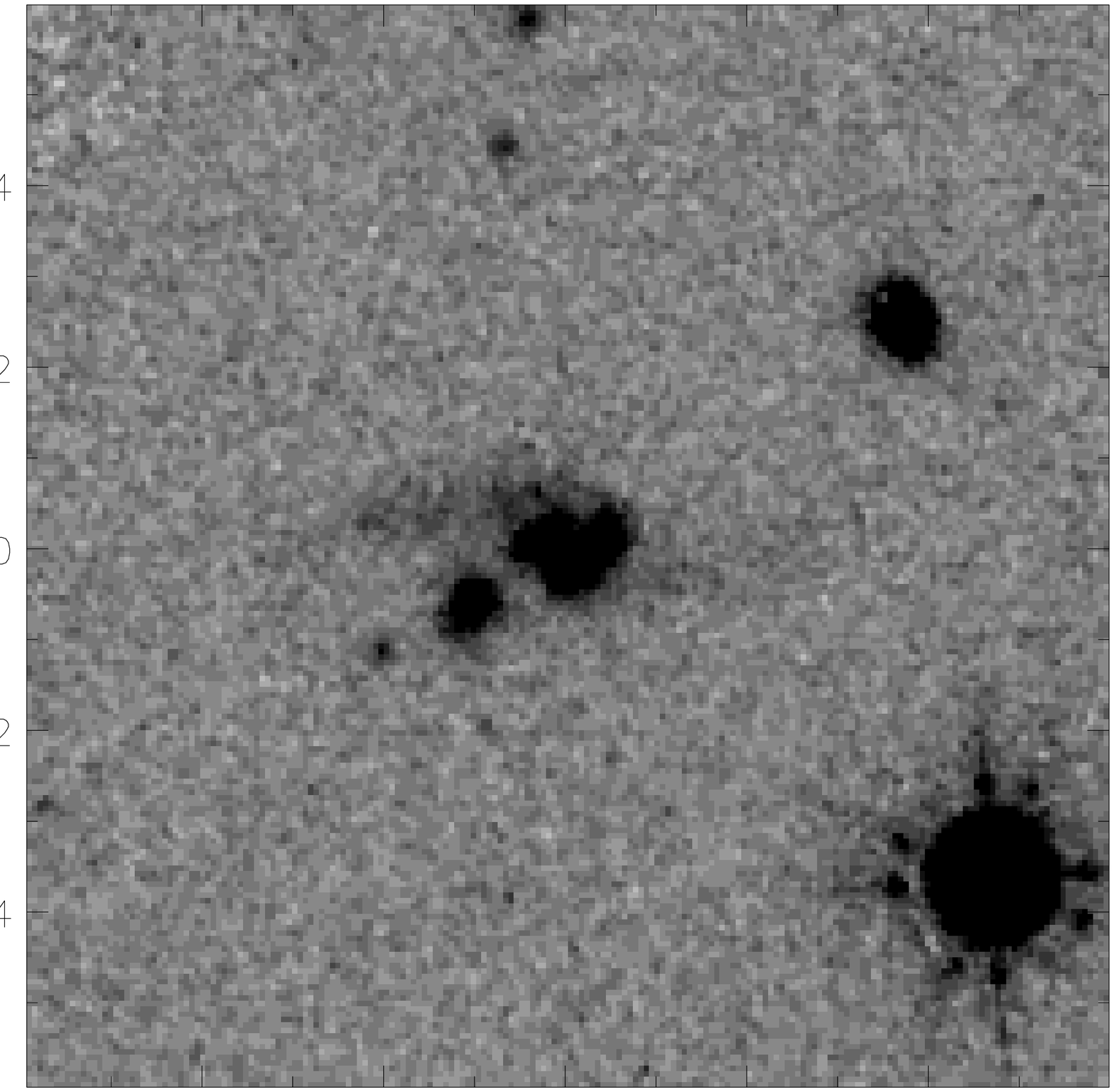,width=0.3\textwidth}&
\epsfig{file=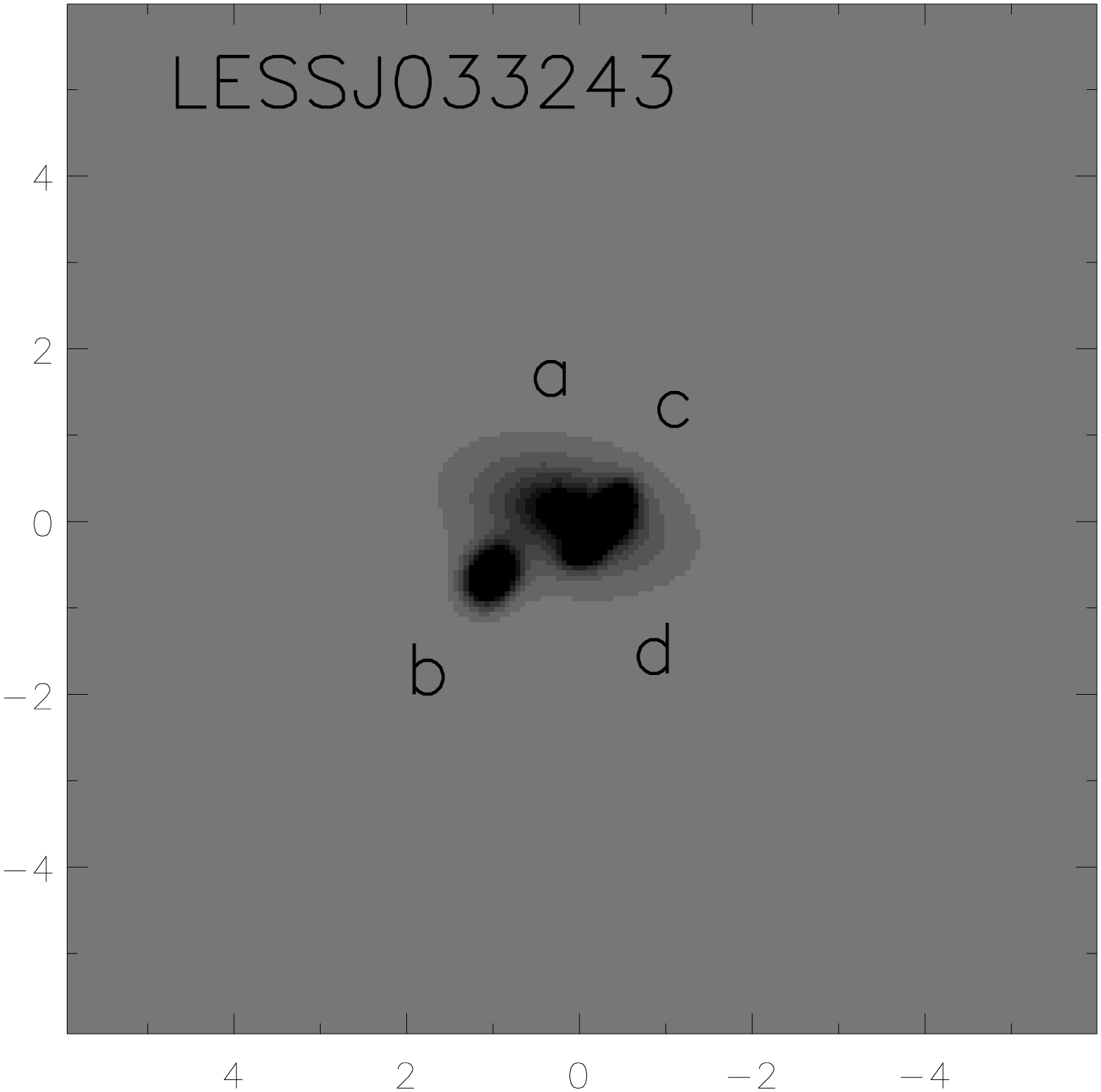,width=0.3\textwidth}&
\epsfig{file=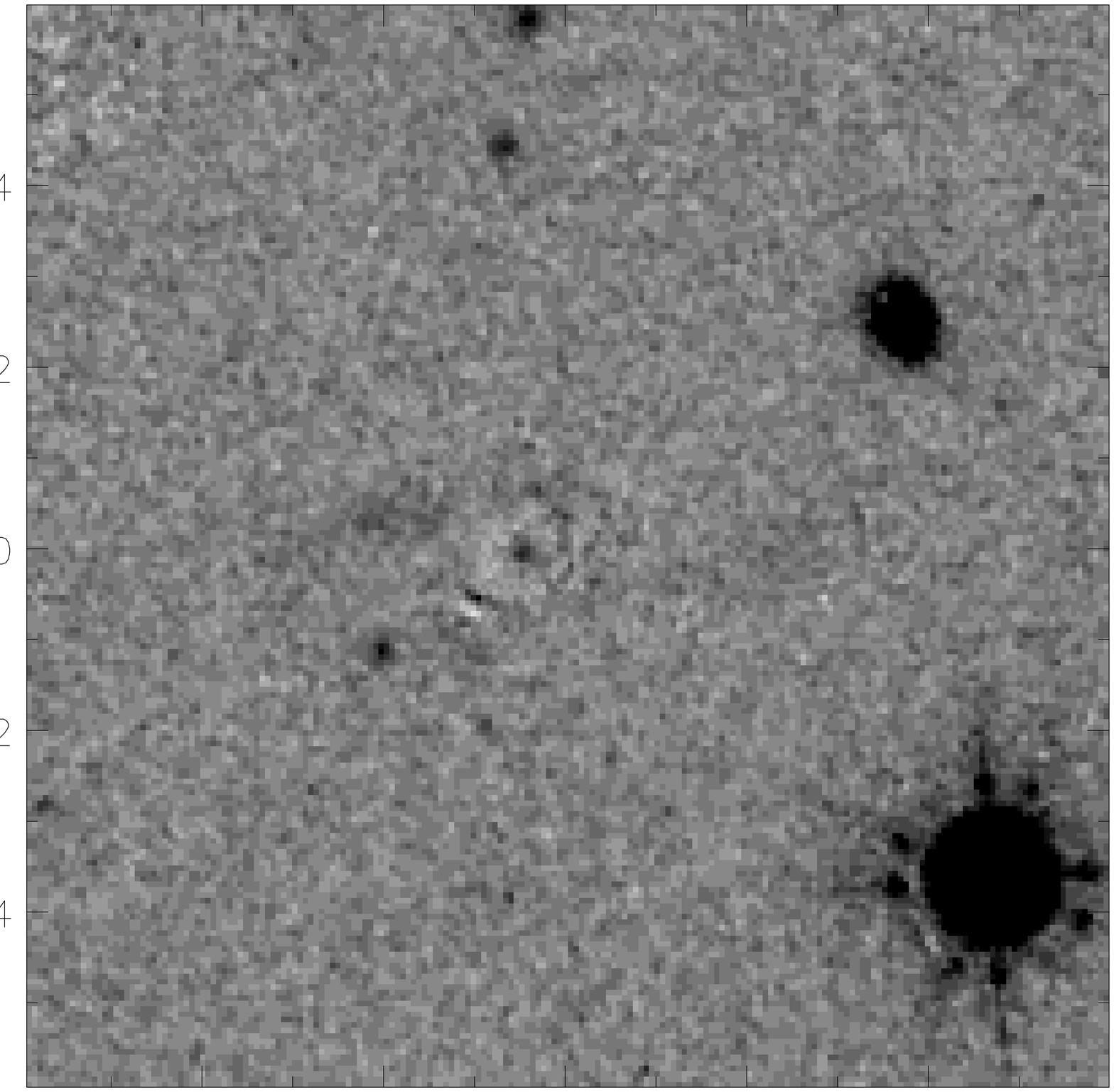,width=0.3\textwidth}\\
\end{tabular}
\addtocounter{figure}{-1}
\caption{- continued}
\end{figure*}
\end{center}

\begin{center}
\begin{figure*}
\vbox to220mm{\vfil
\begin{tabular}{cccccccc}
\epsfig{file=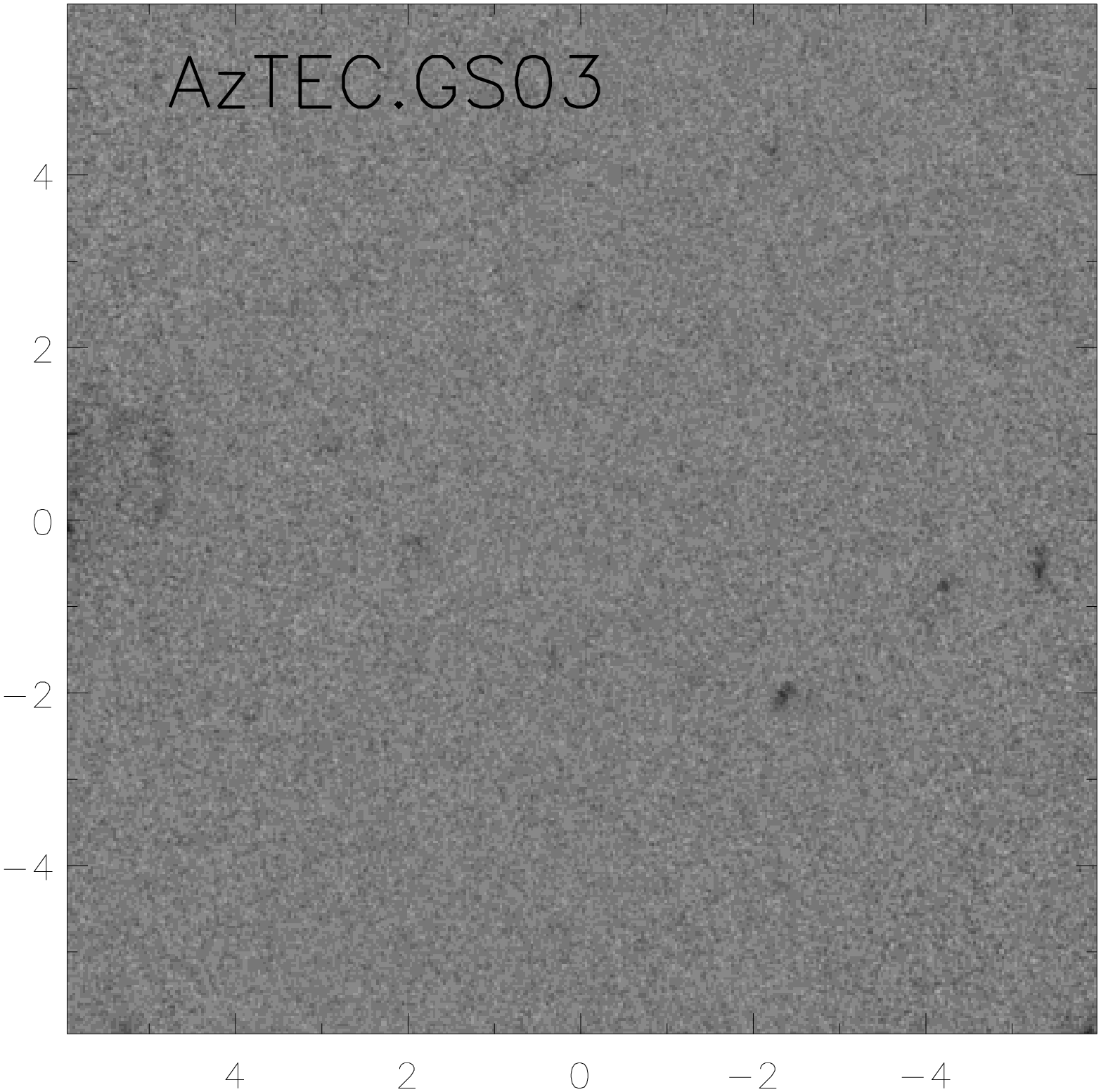,width=0.20\textwidth}&
\epsfig{file=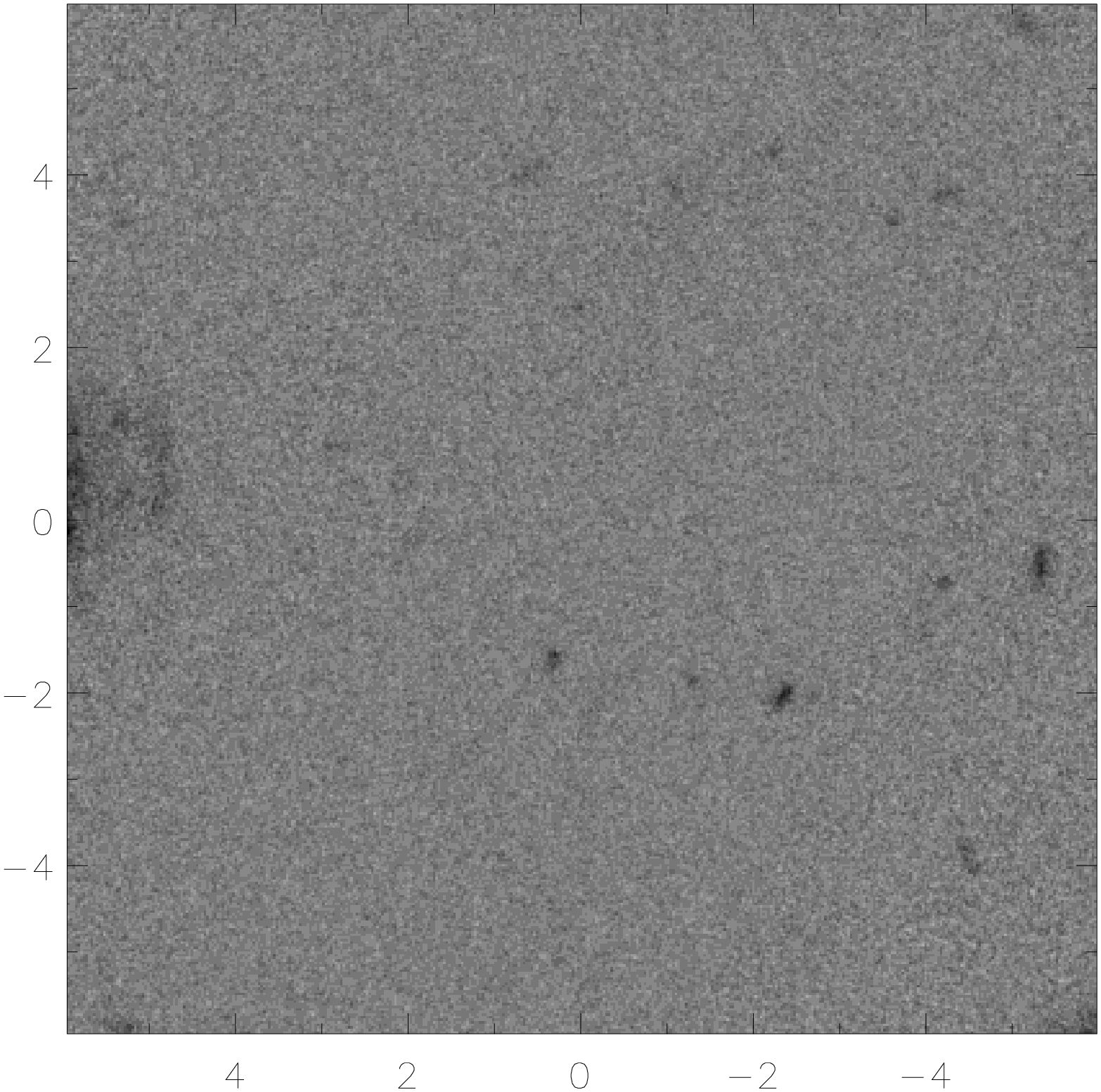,width=0.20\textwidth}&
\epsfig{file=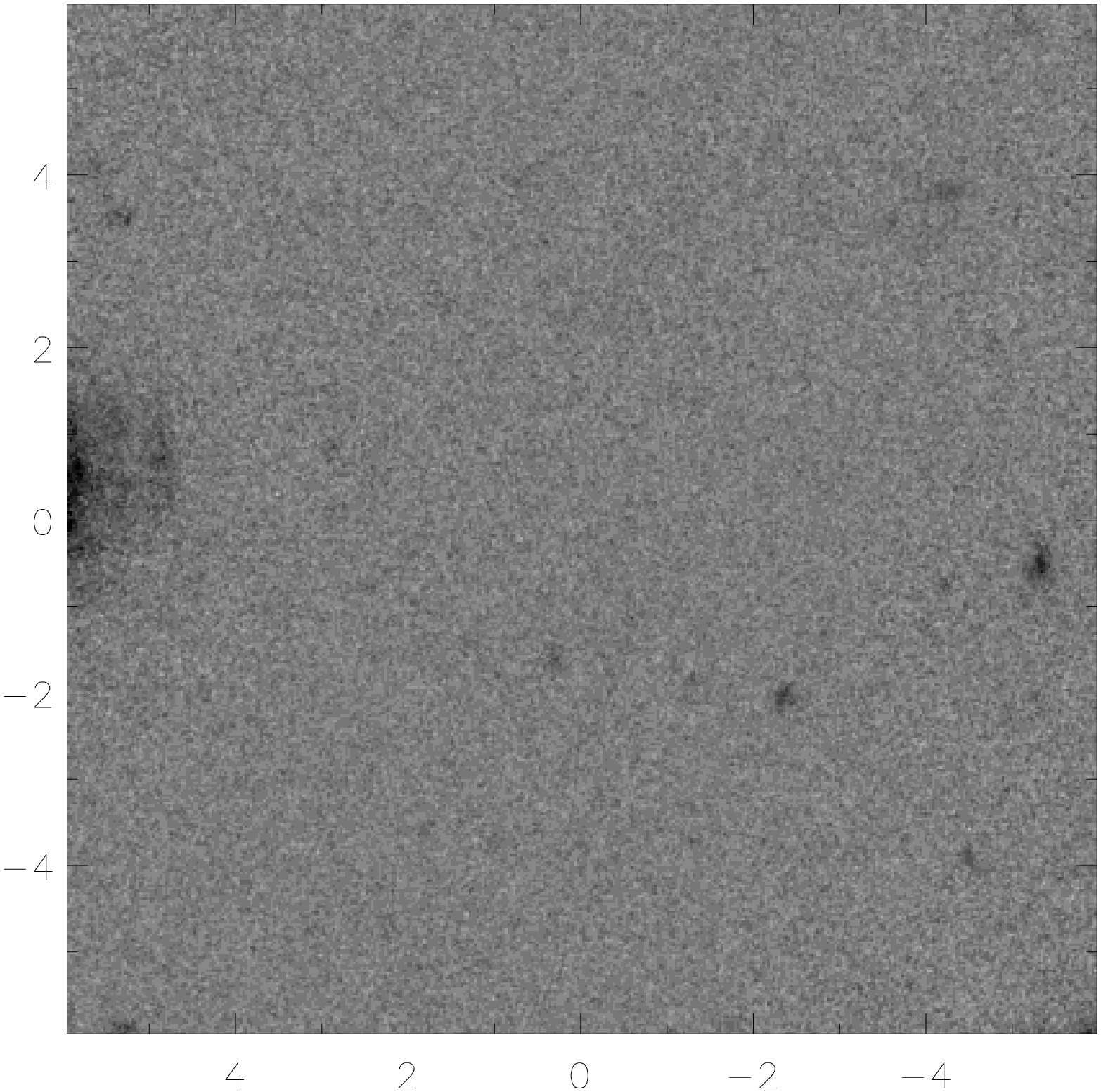,width=0.20\textwidth}&
\epsfig{file=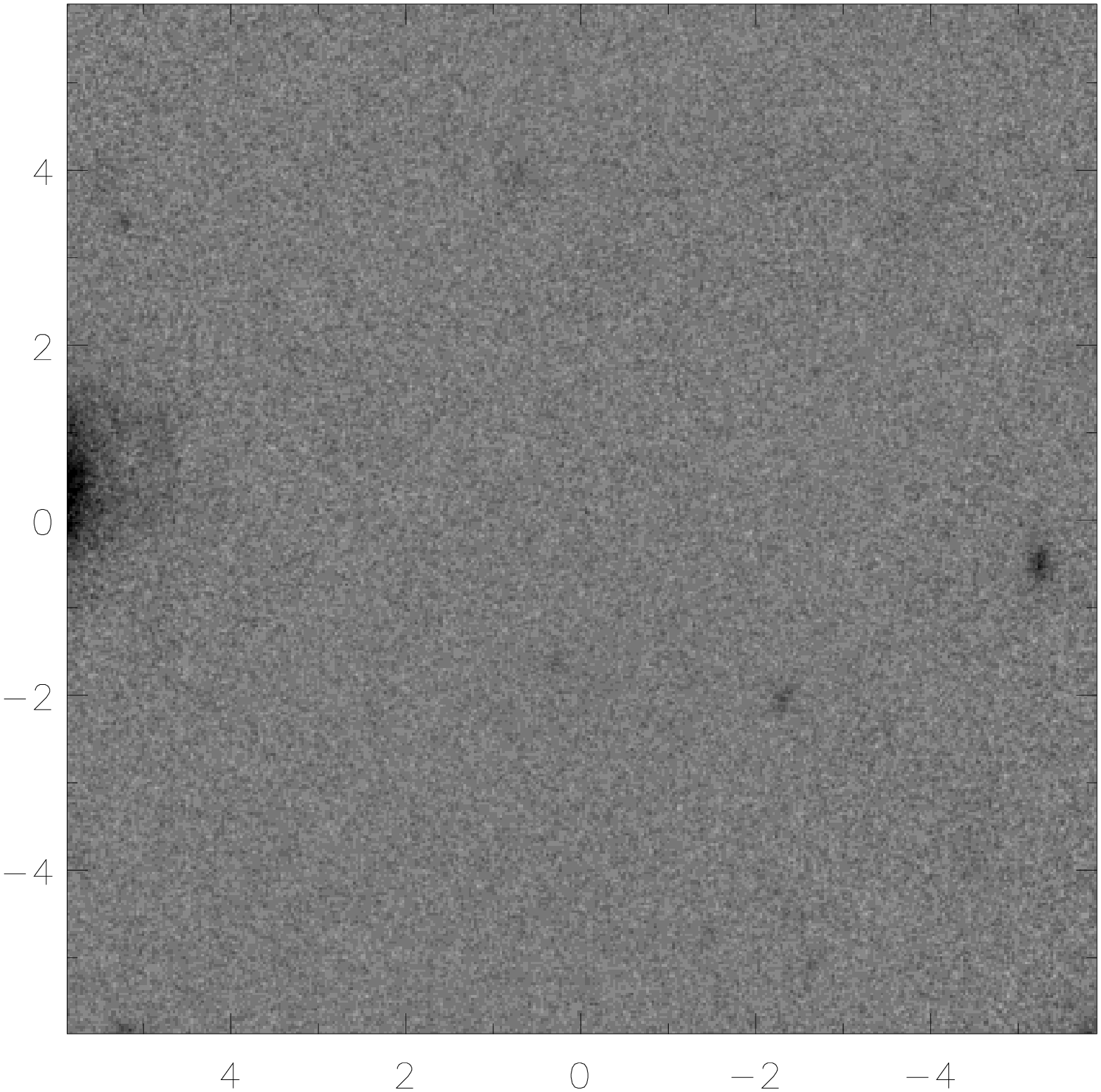,width=0.20\textwidth}\\
\epsfig{file=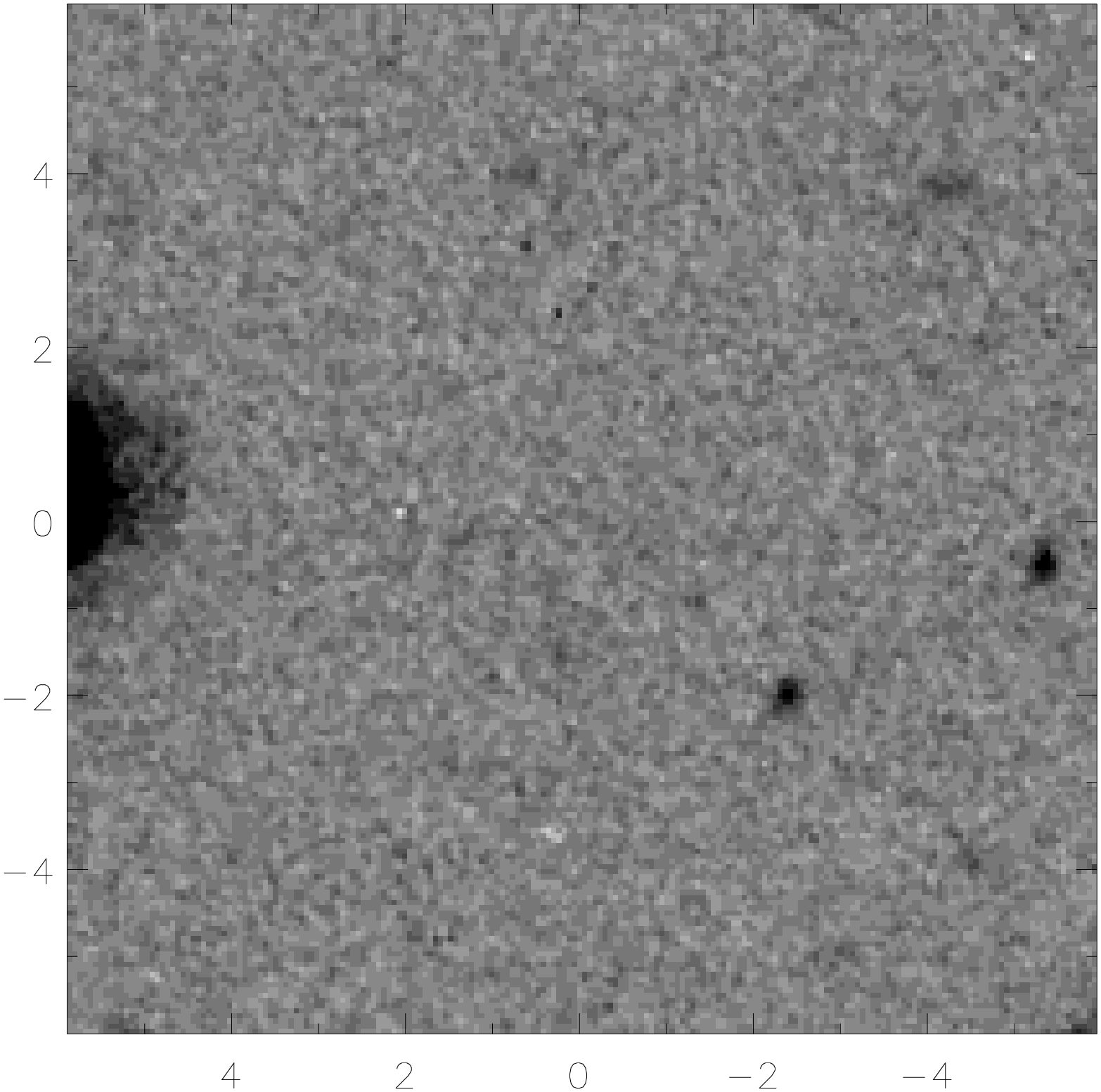,width=0.20\textwidth}&
\epsfig{file=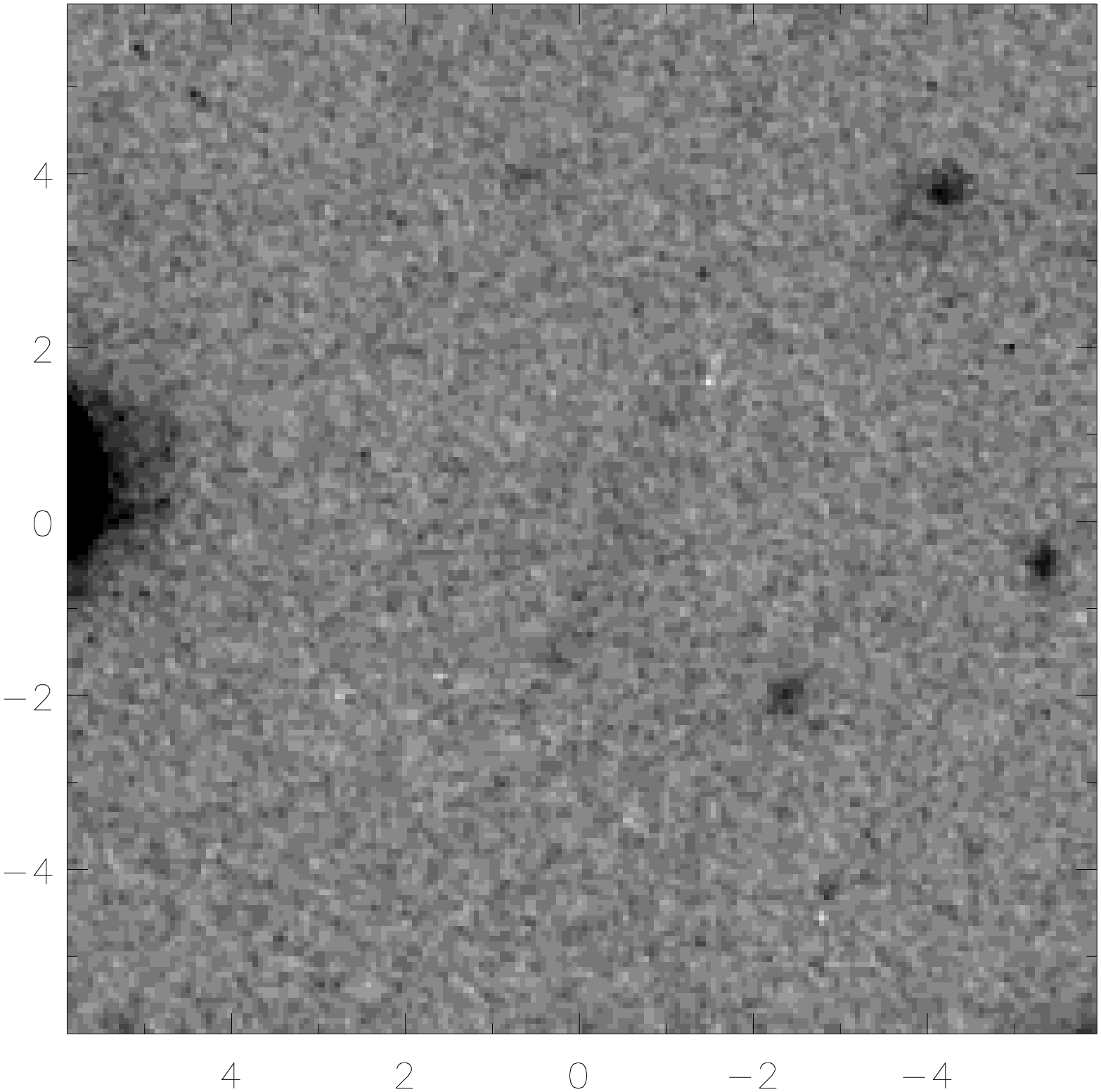,width=0.20\textwidth}&
\epsfig{file=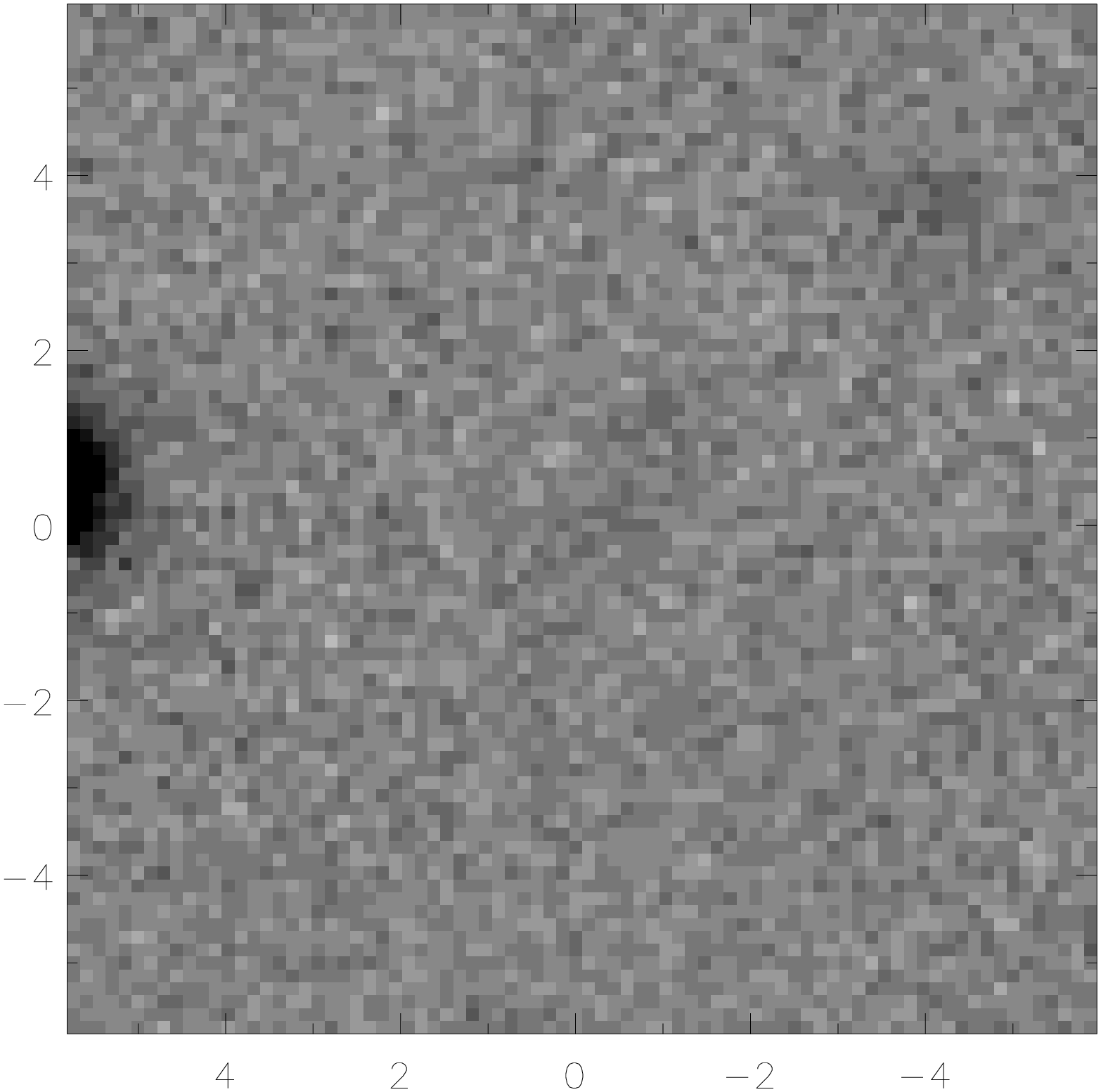,width=0.20\textwidth}&
\epsfig{file=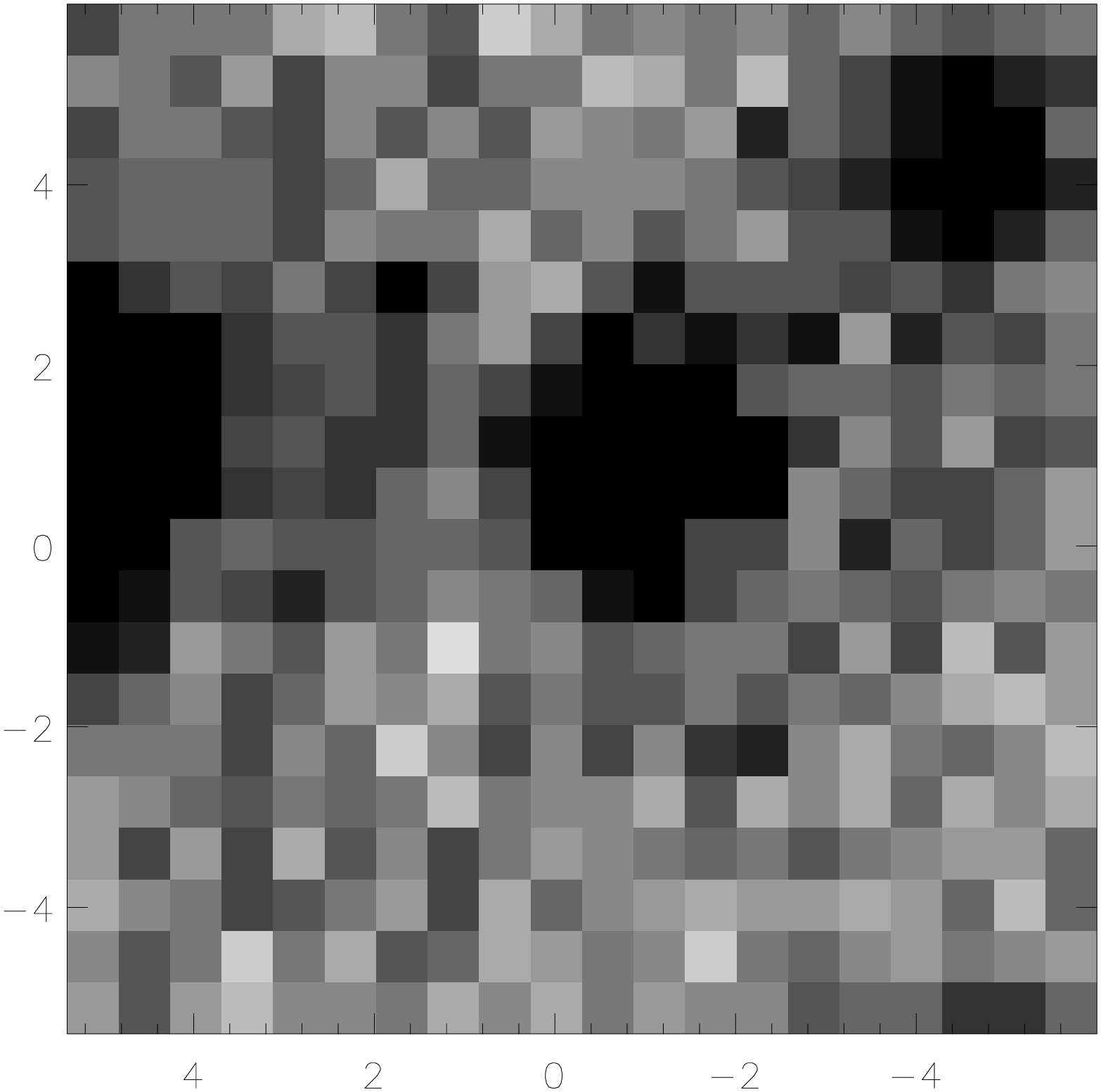,width=0.20\textwidth}\\
\\
\epsfig{file=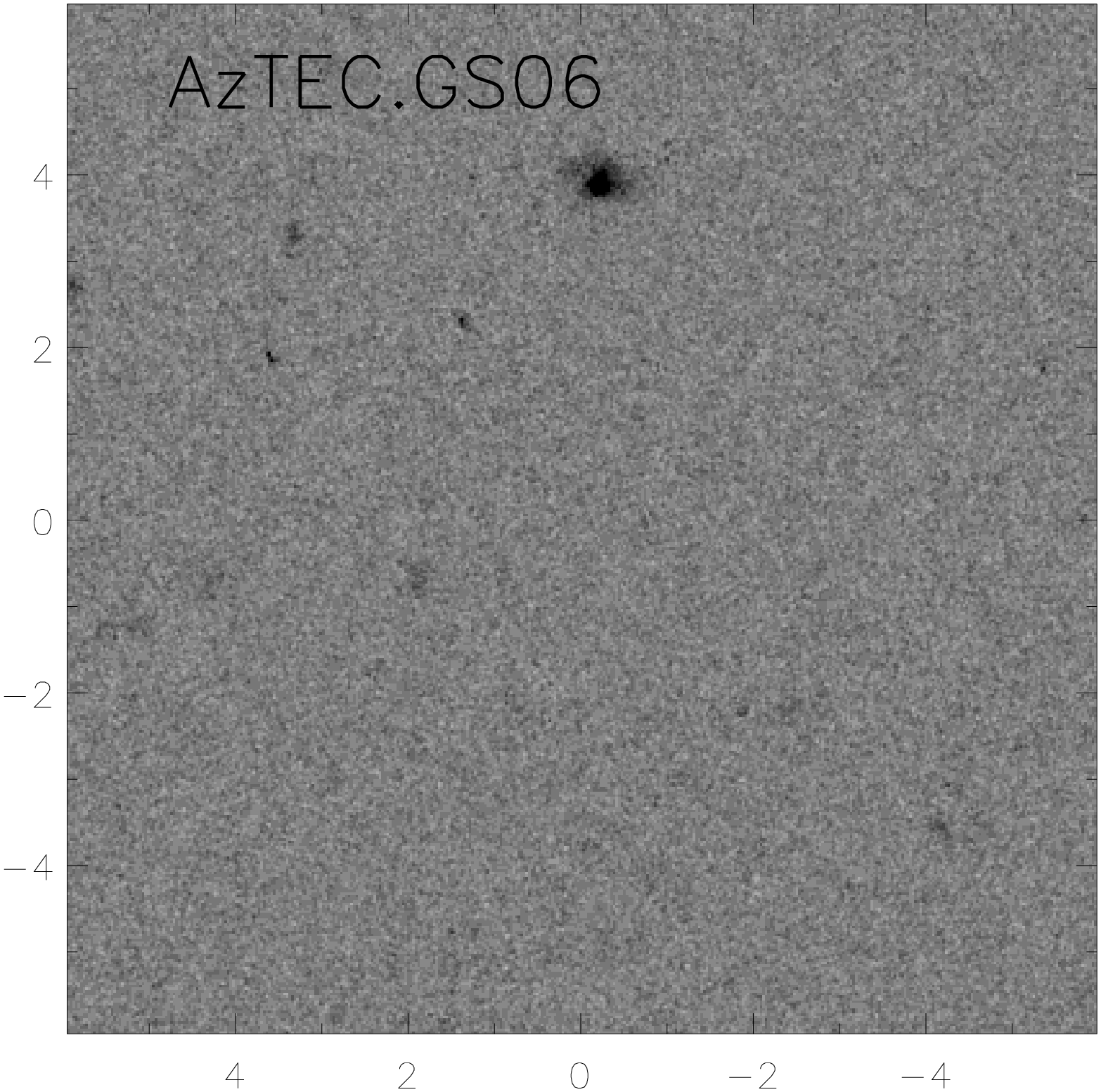,width=0.20\textwidth}&
\epsfig{file=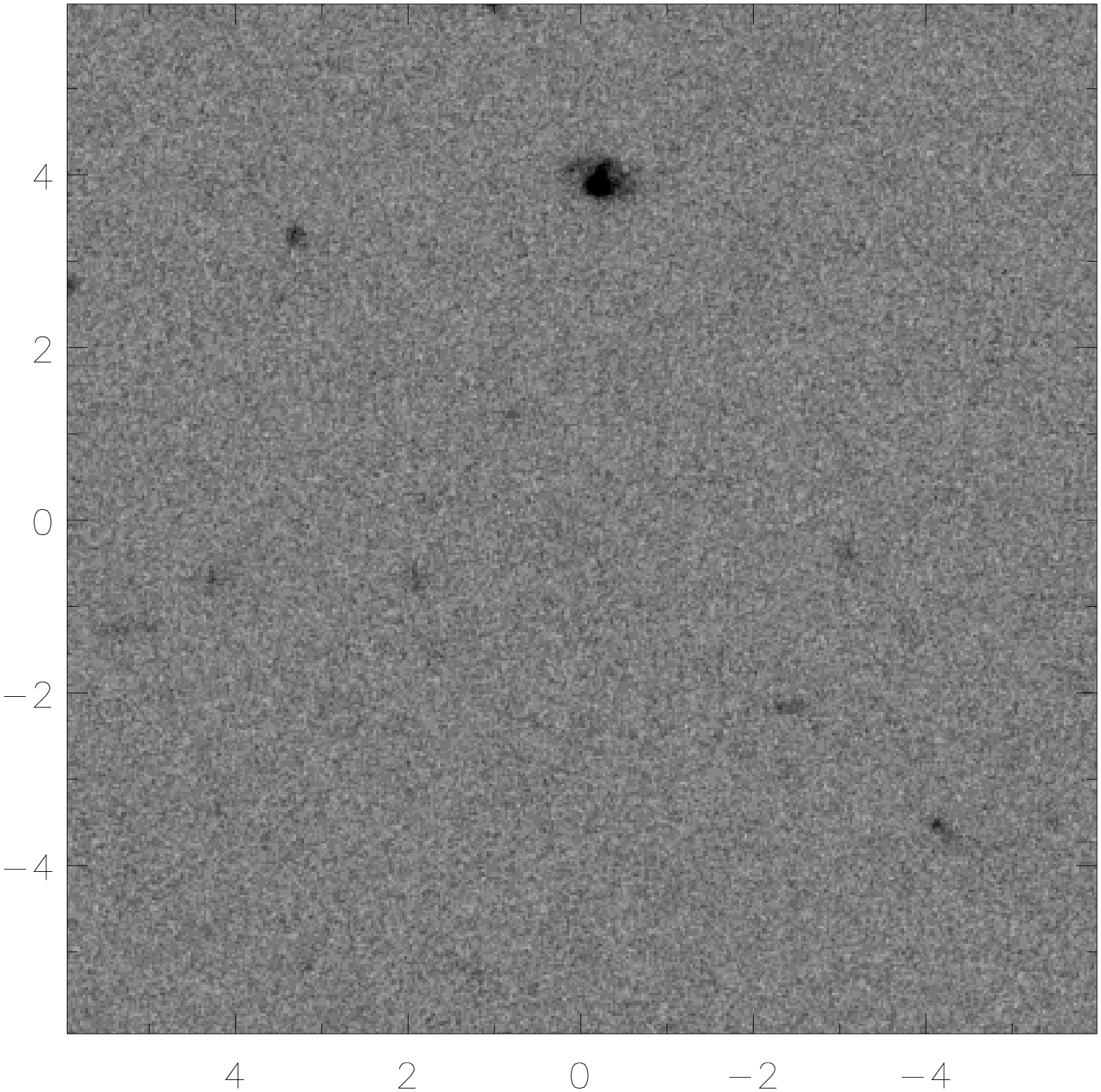,width=0.20\textwidth}&
\epsfig{file=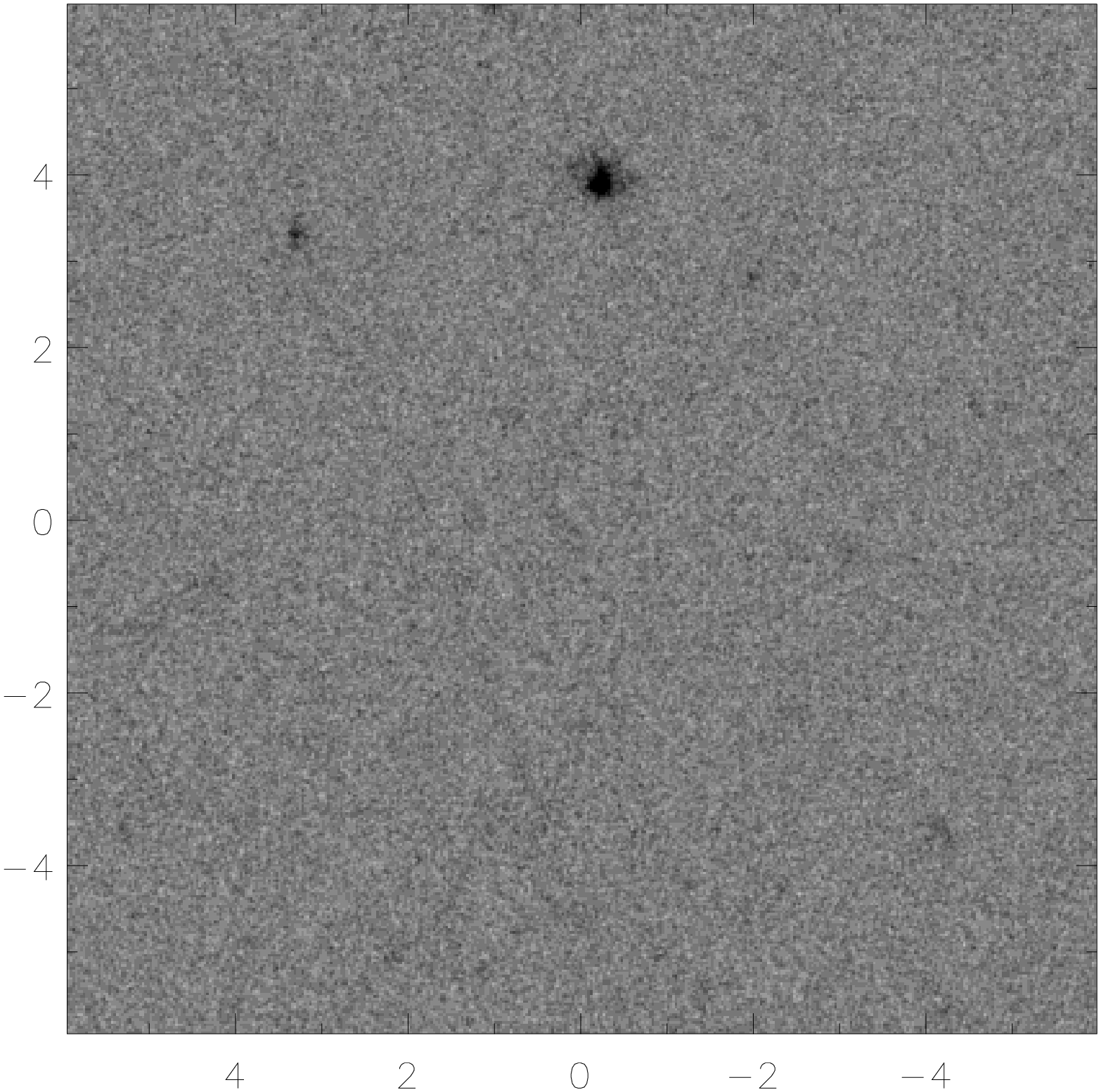,width=0.20\textwidth}&
\epsfig{file=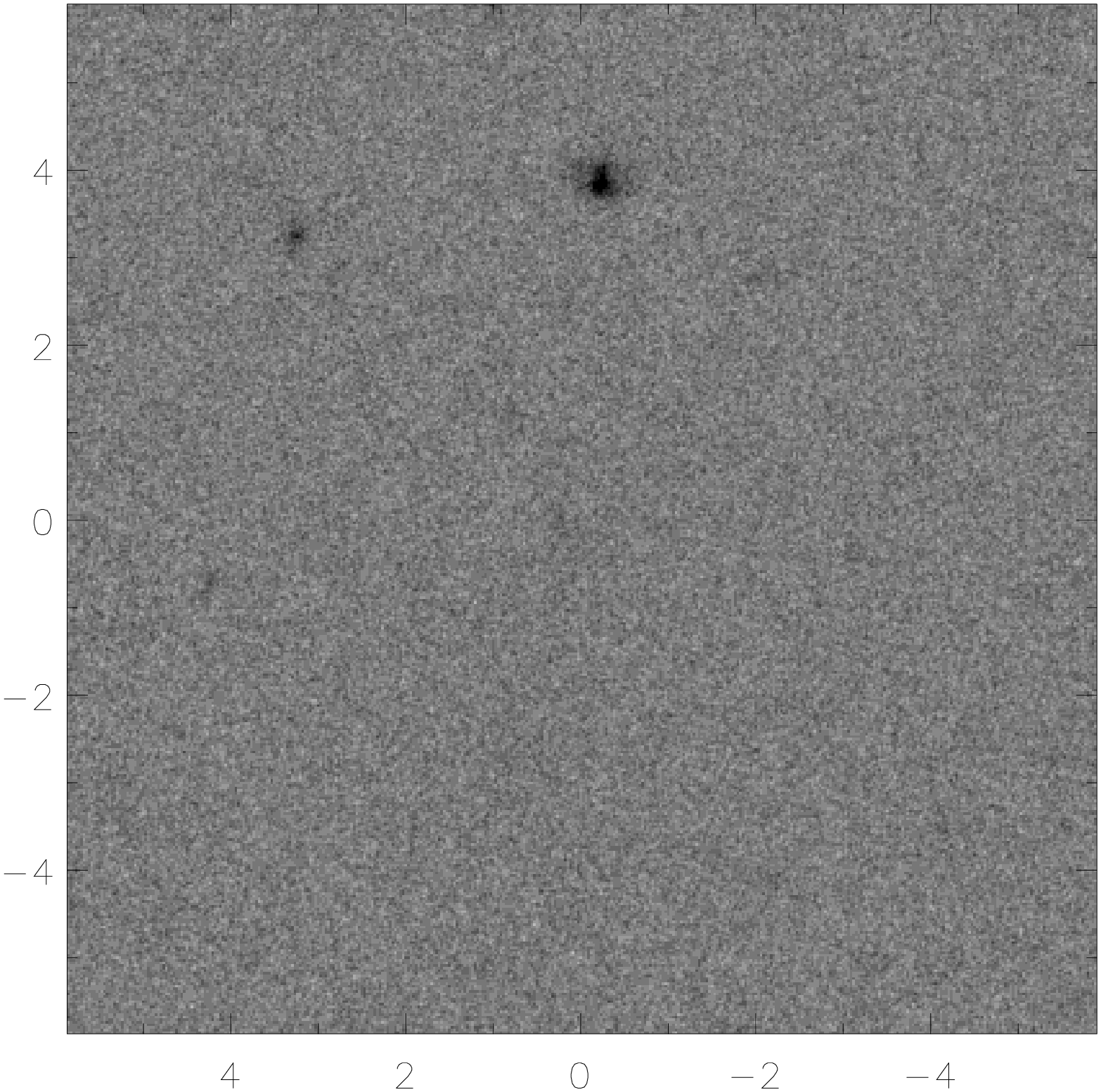,width=0.20\textwidth}\\
\epsfig{file=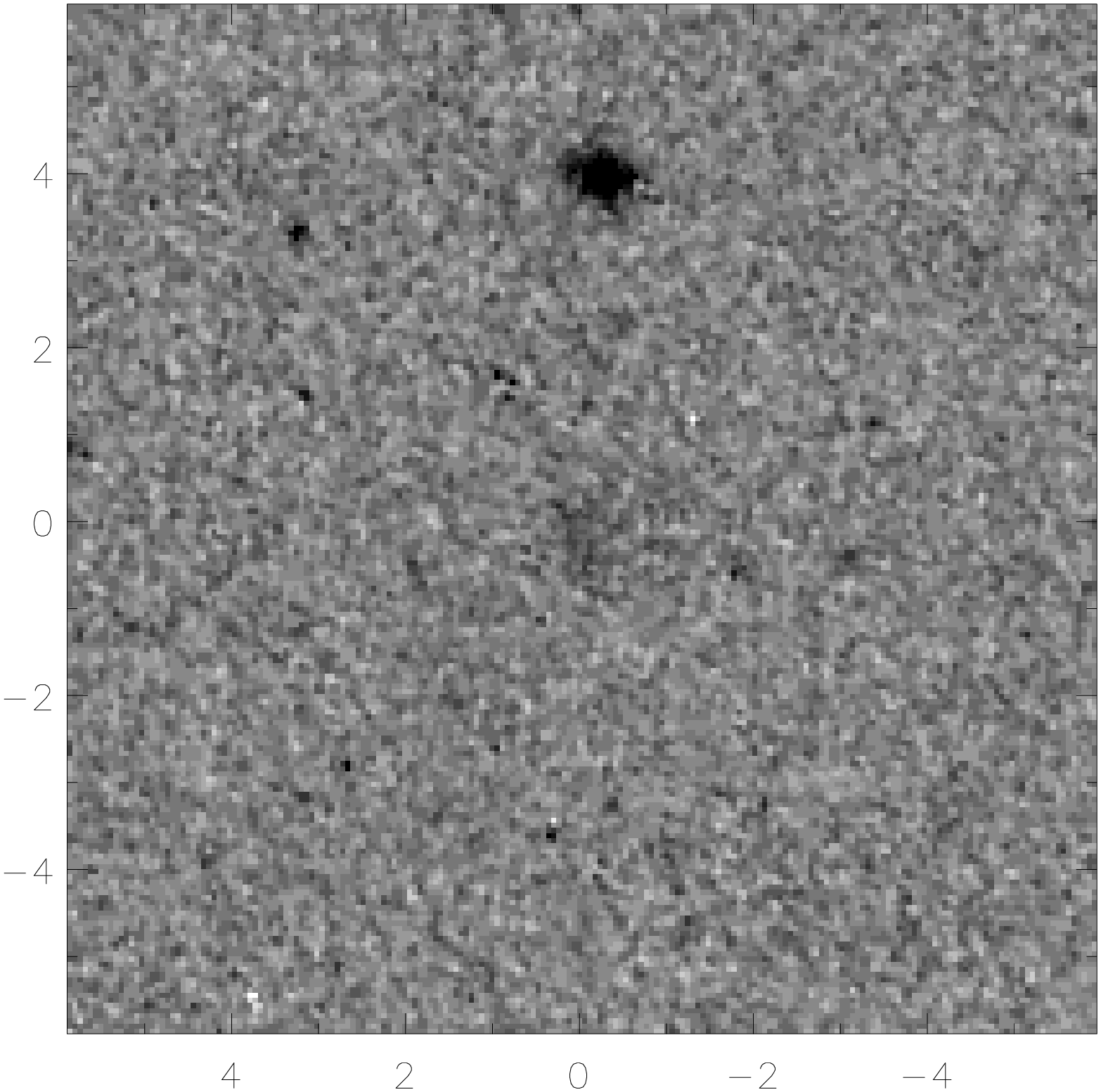,width=0.20\textwidth}&
\epsfig{file=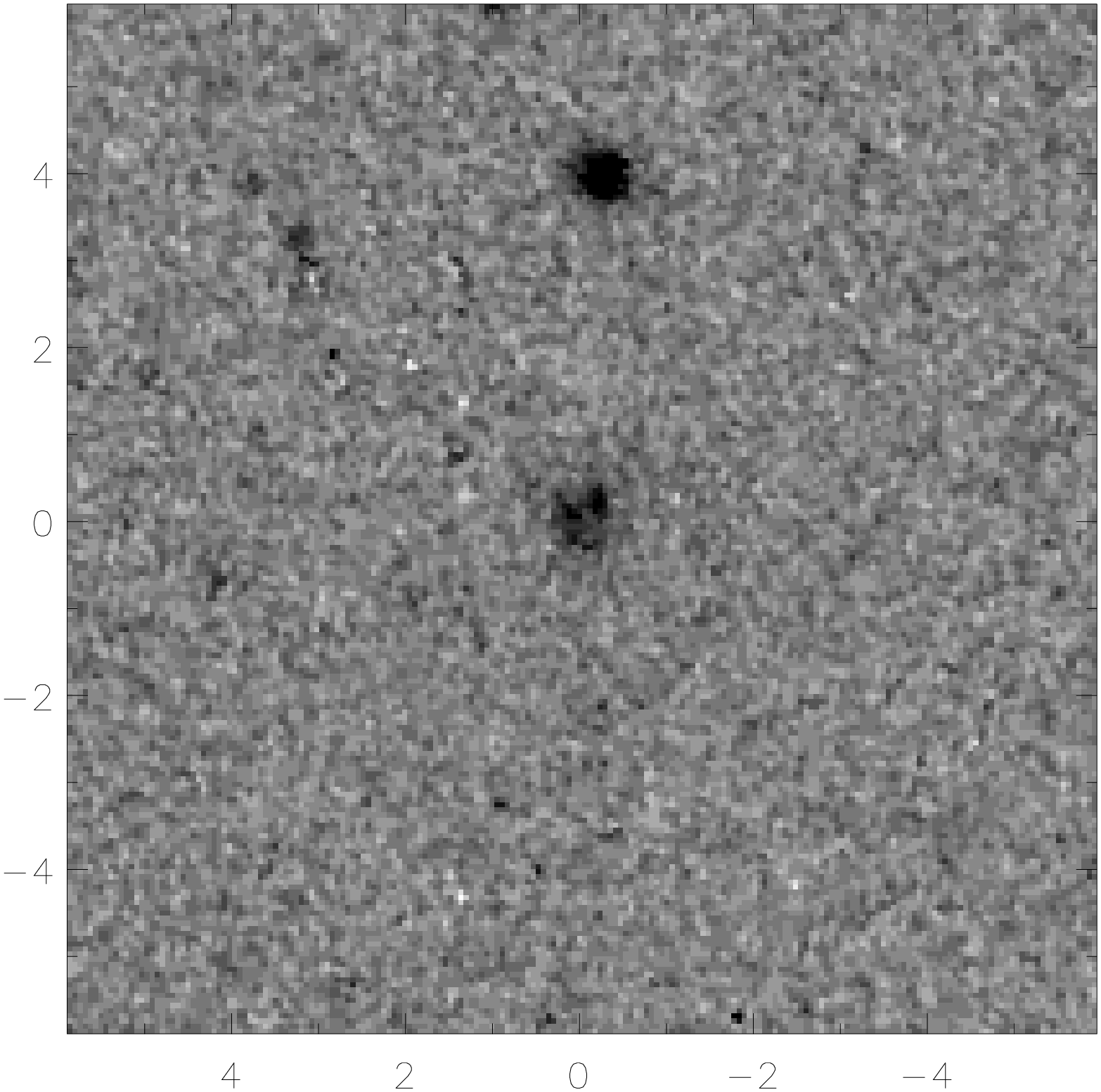,width=0.20\textwidth}&
\epsfig{file=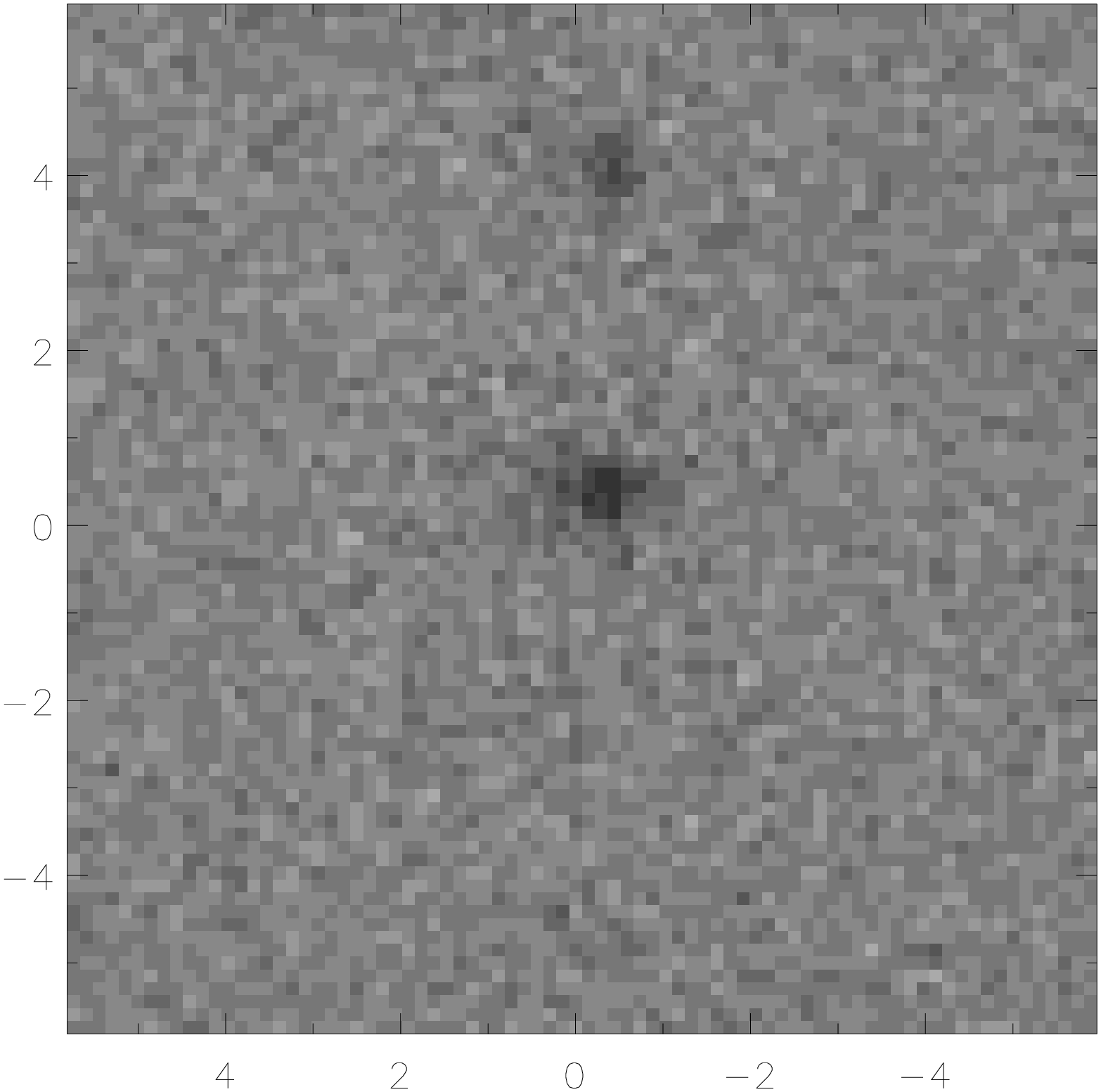,width=0.20\textwidth}&
\epsfig{file=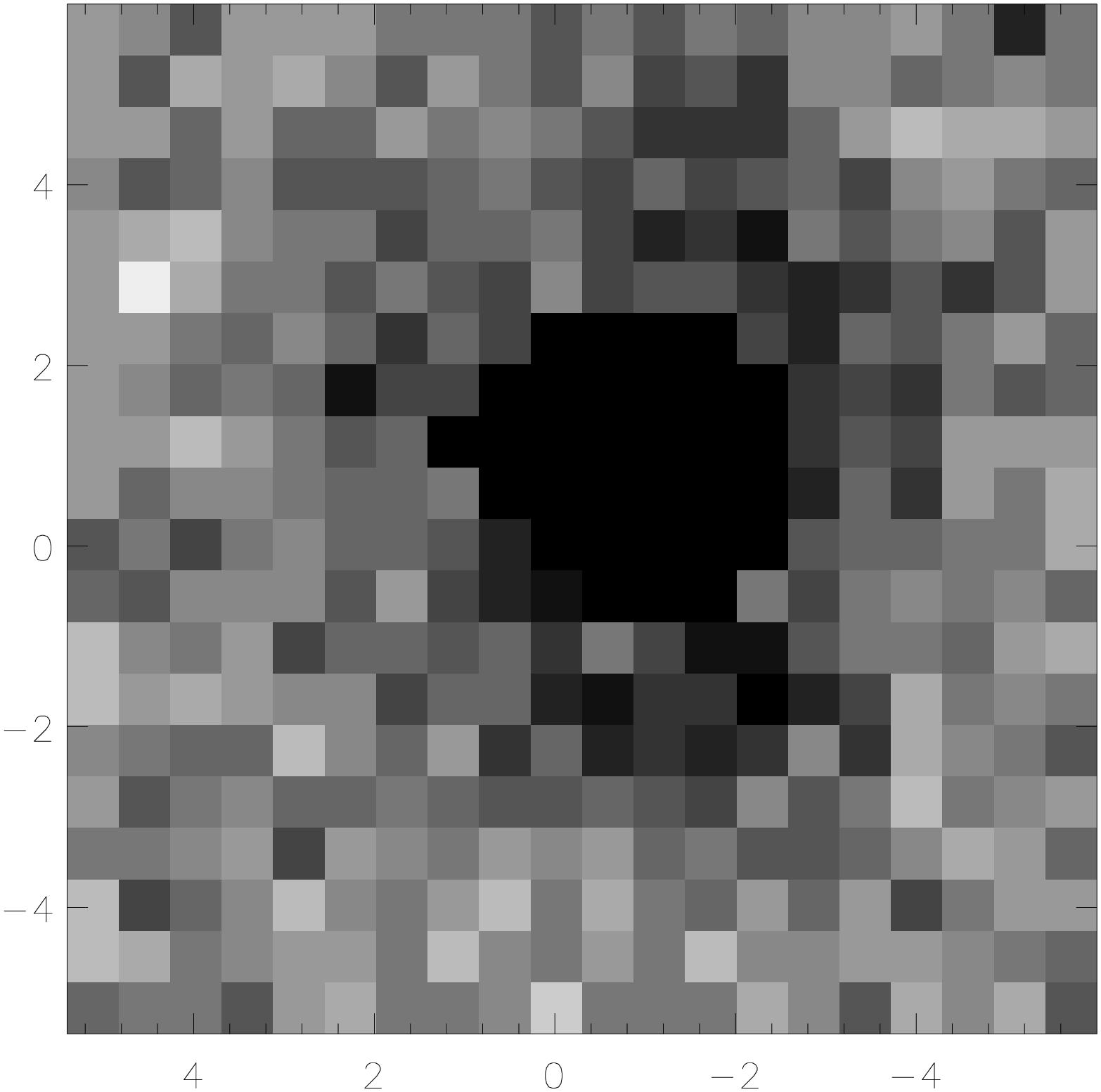,width=0.20\textwidth}\\
\\
\epsfig{file=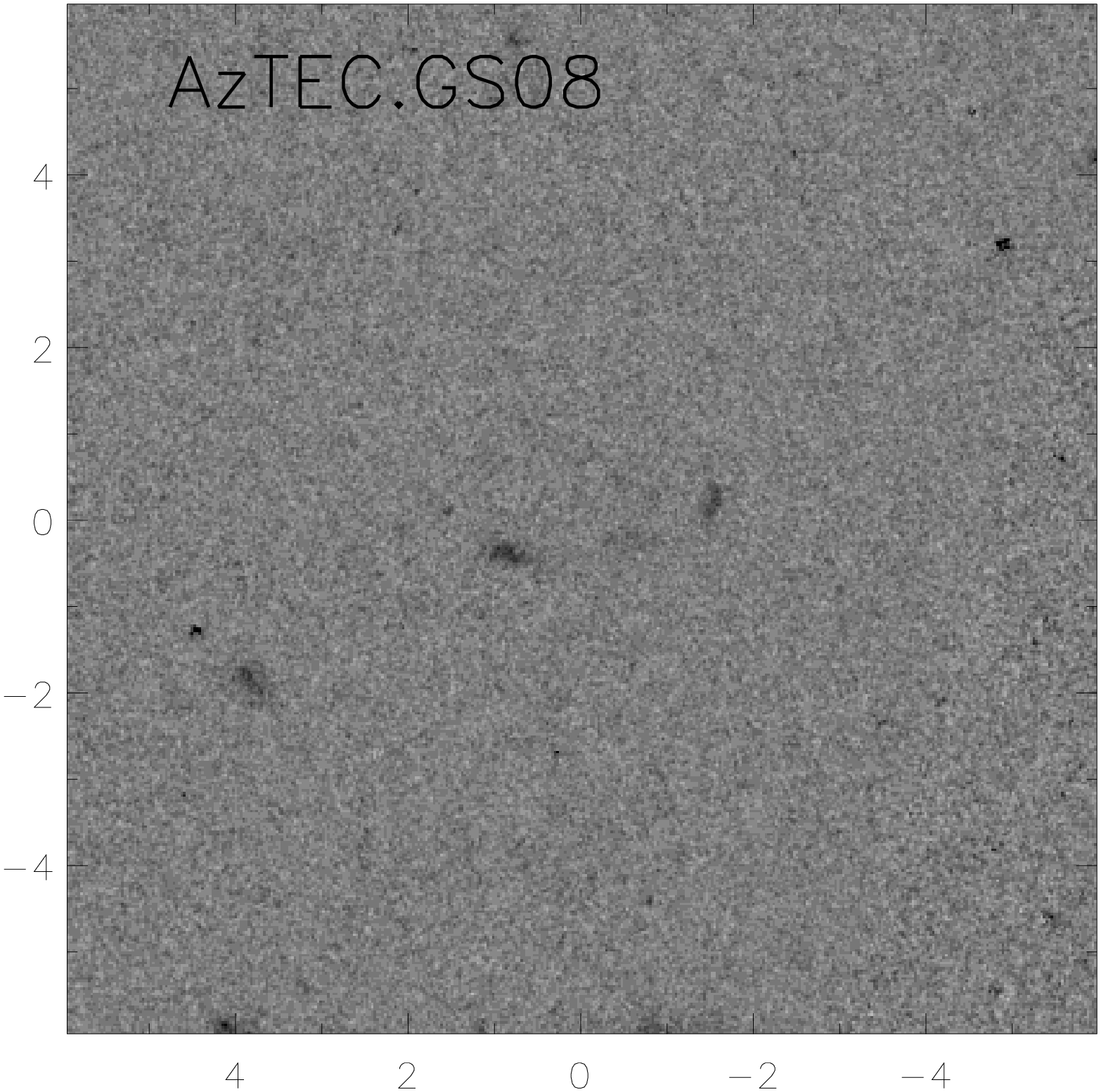,width=0.20\textwidth}&
\epsfig{file=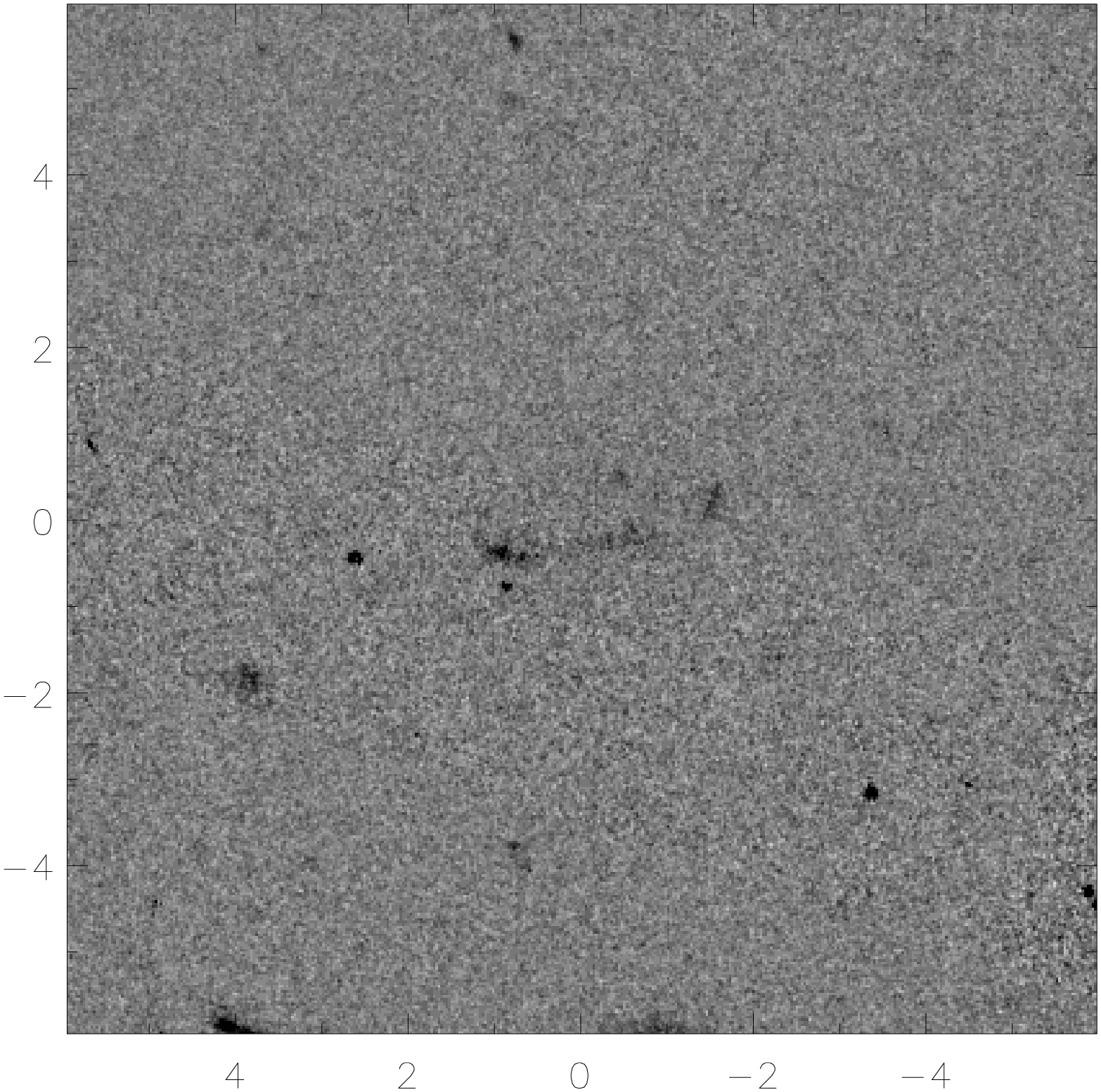,width=0.20\textwidth}&
\epsfig{file=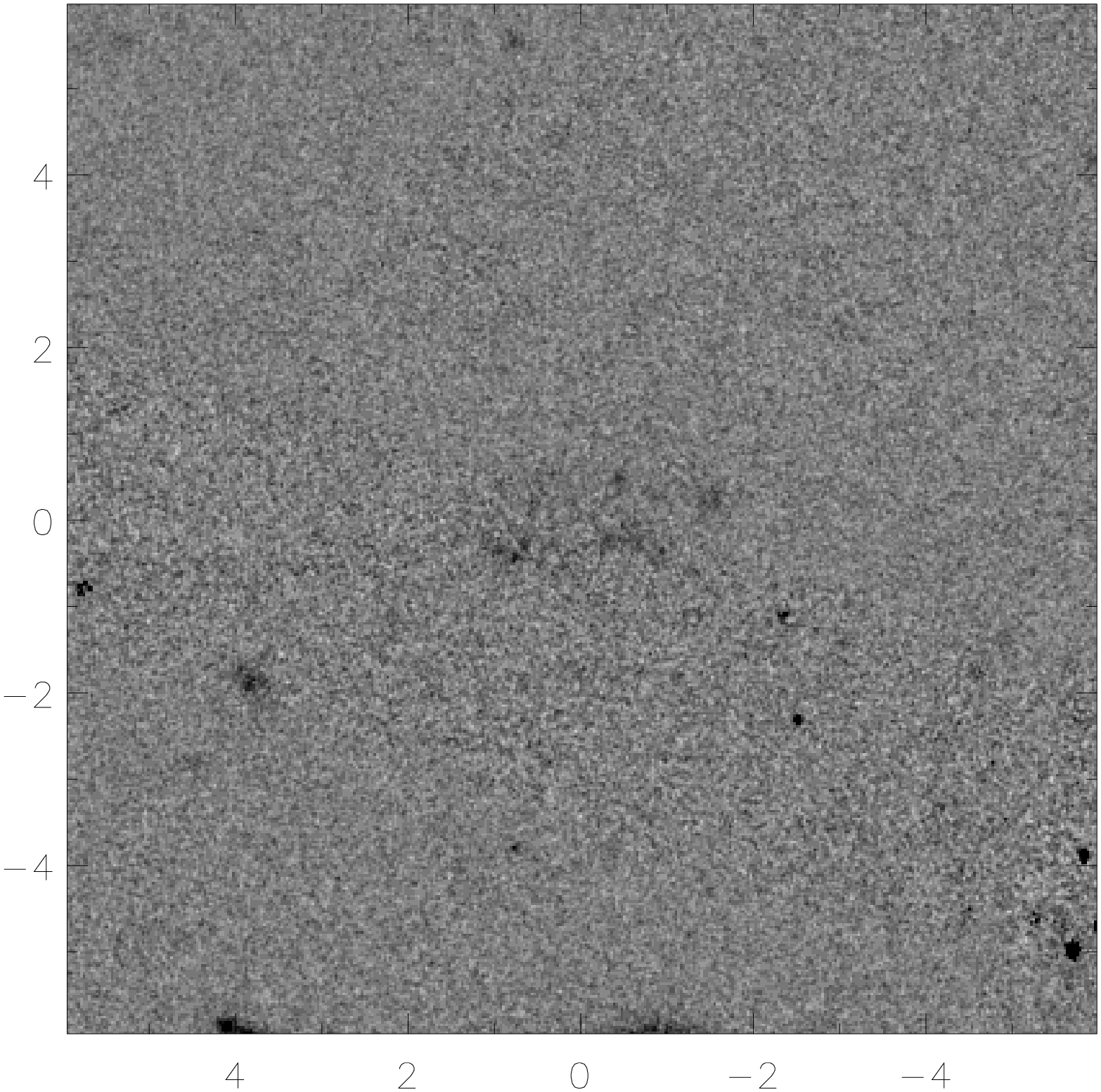,width=0.20\textwidth}&
\epsfig{file=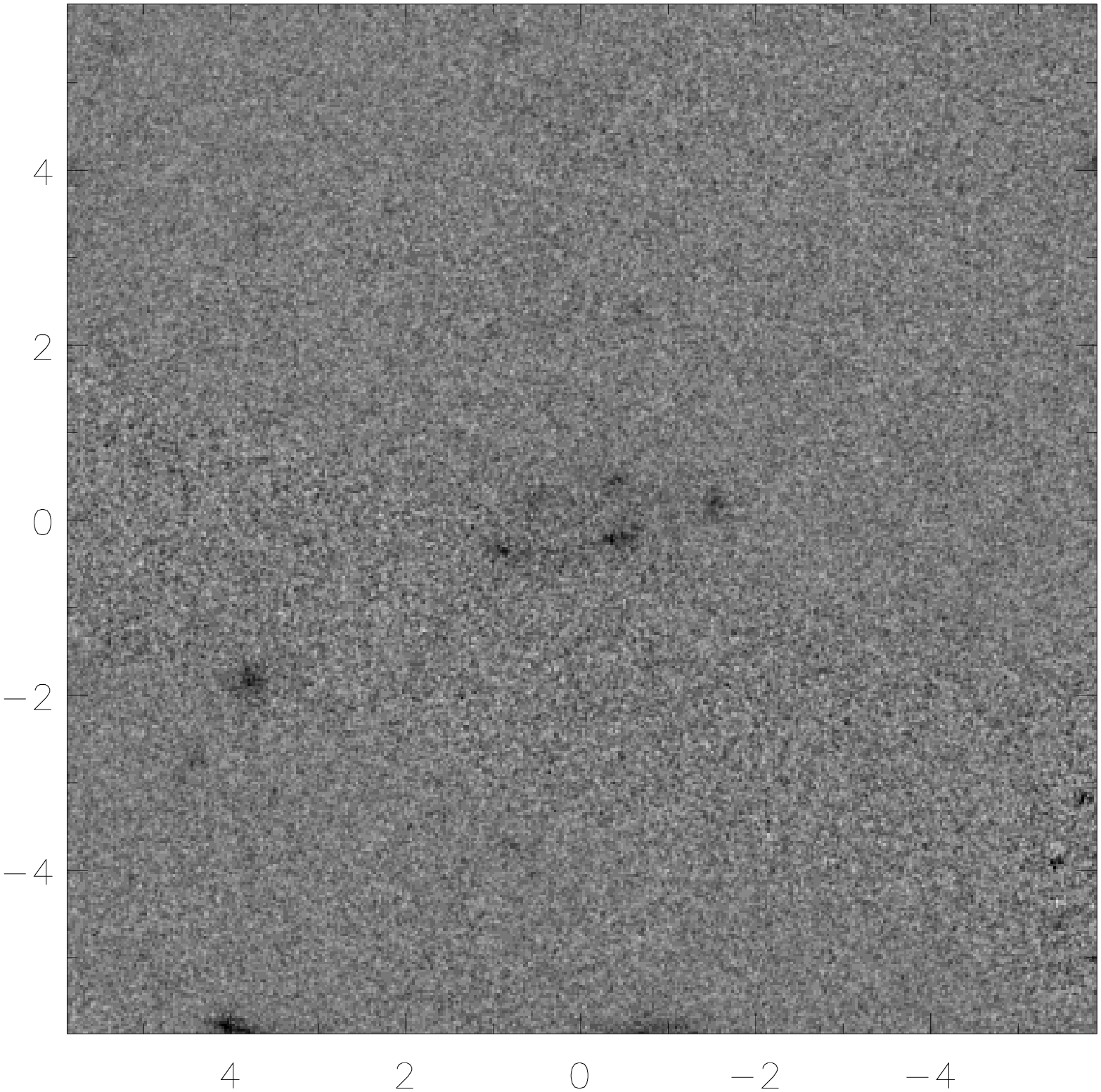,width=0.20\textwidth}\\
\epsfig{file=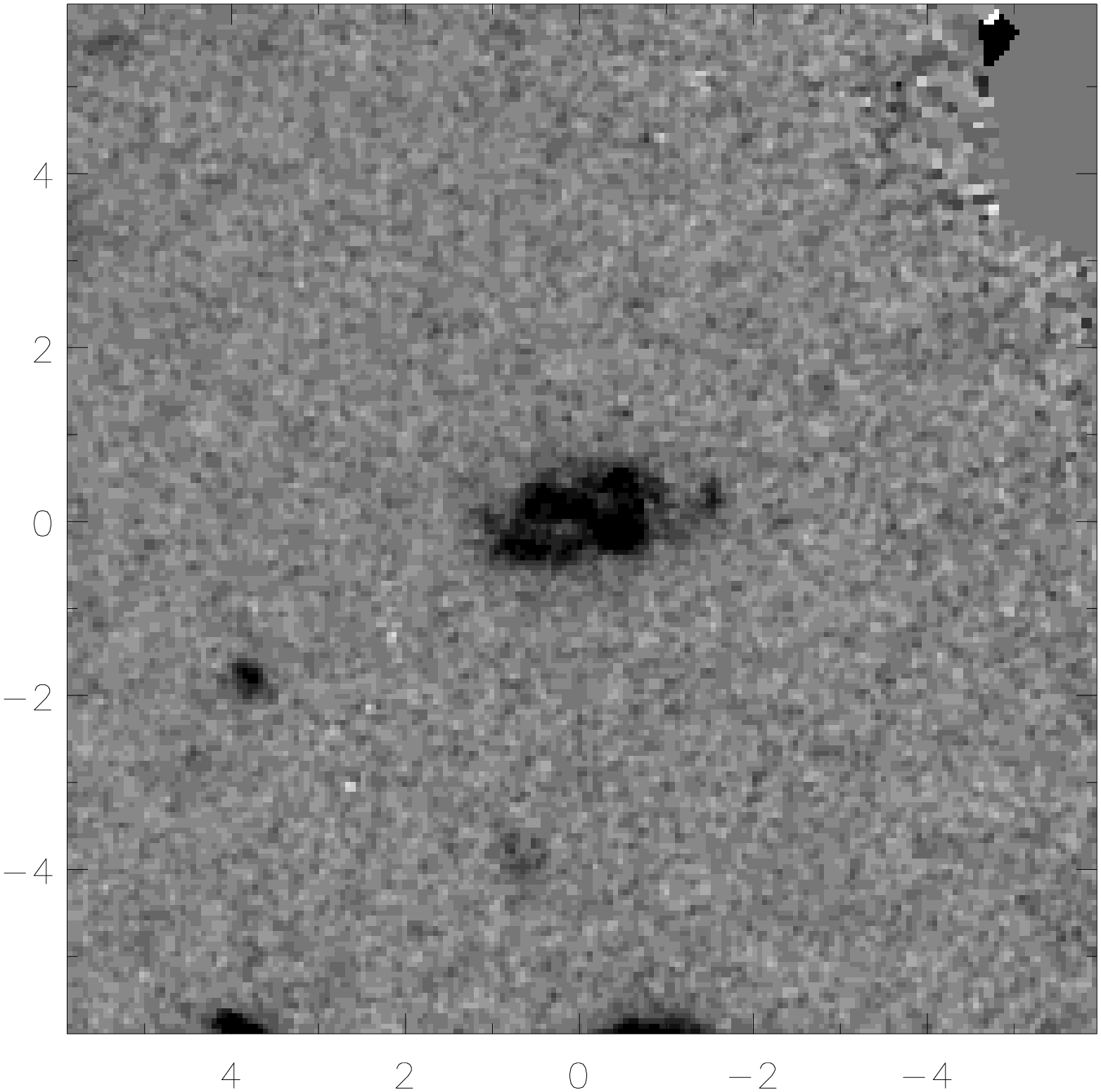,width=0.20\textwidth}&
\epsfig{file=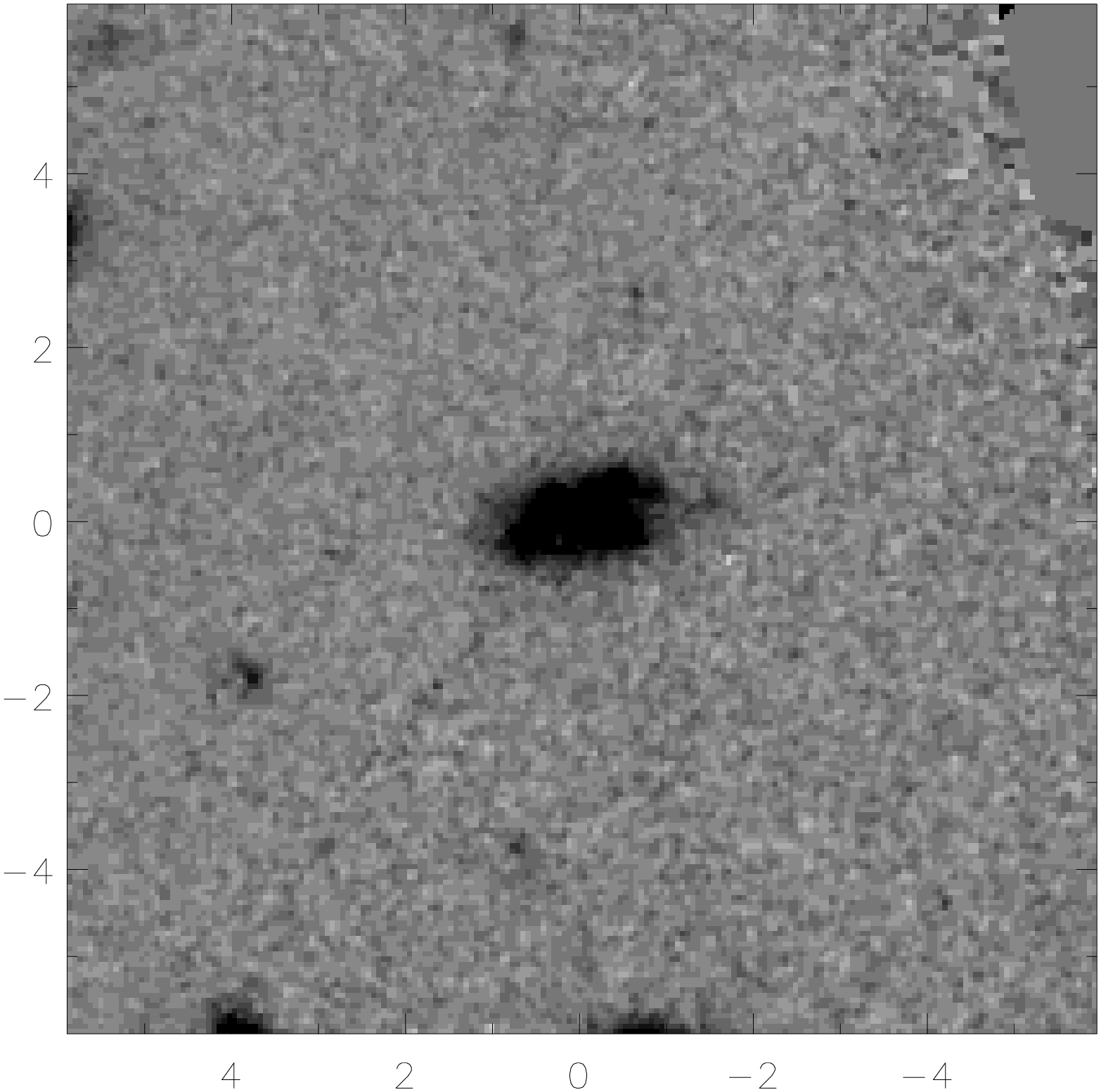,width=0.20\textwidth}&
\epsfig{file=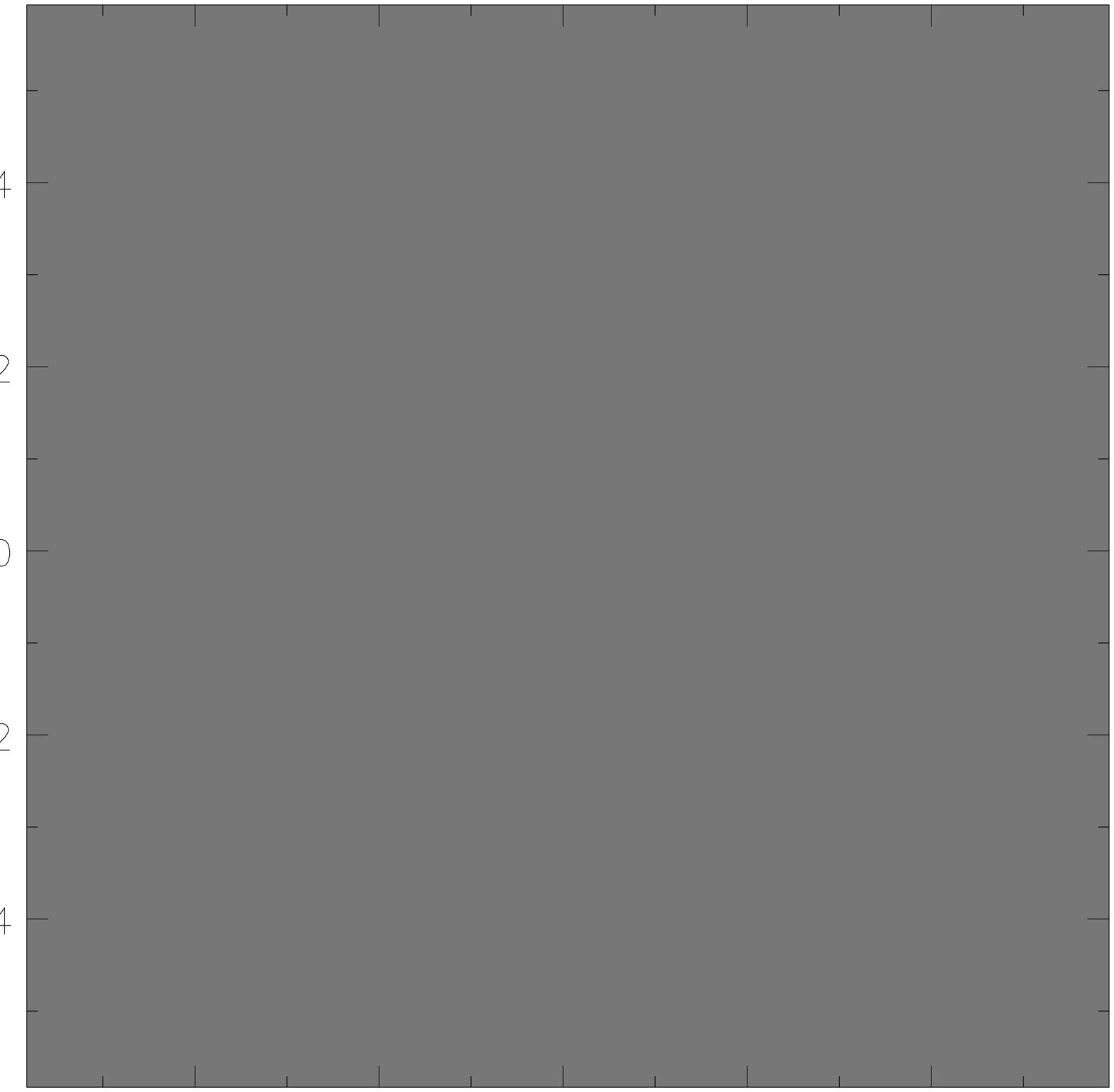,width=0.20\textwidth}&
\epsfig{file=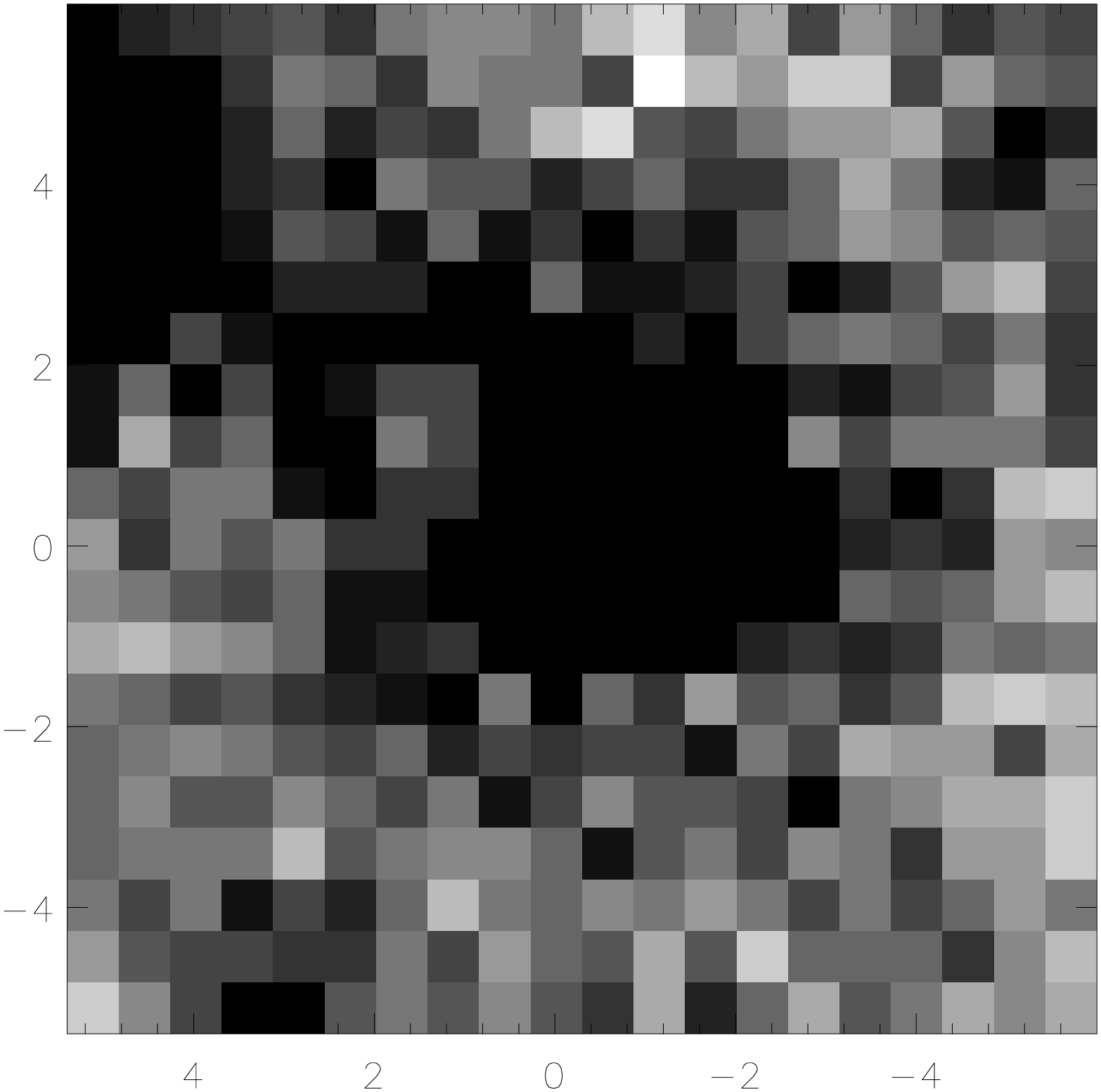,width=0.20\textwidth}\\
\end{tabular}
\caption{Optical (upper row: $B$ (HST), $V$ (HST), $i$ (HST), $z$-band (HST)) and near-infrared (lower row: $J$ (HST), $H$ (HST), $K$-band (VLT), $3.6{\rm \mu m}$ (Spitzer)) imaging of (sub)millimetre selected galaxies in CANDELS GOODS-South. All panels are 12.0$^{\prime \prime}$ $\times$12.0$^{\prime \prime}$, and the images are shown with a linear greyscale in which black corresponds to 10-$\sigma$ above, and white to 1-$\sigma$ below the median sky value.}
\vfil}
\end{figure*}
\end{center}


\begin{center}
\begin{figure*}
\vbox to220mm{\vfil
\begin{tabular}{cccccccc}
\epsfig{file=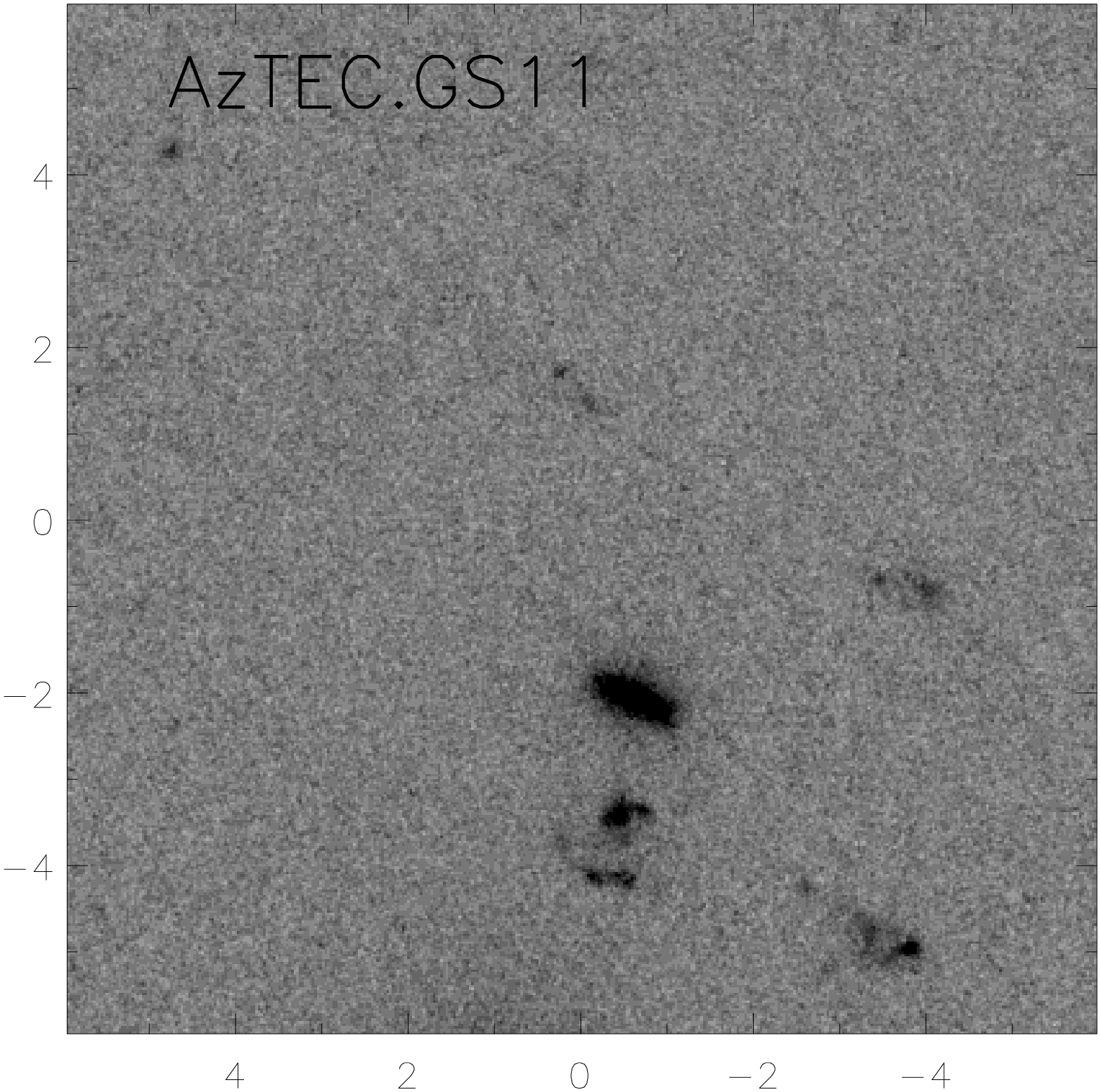,width=0.20\textwidth}&
\epsfig{file=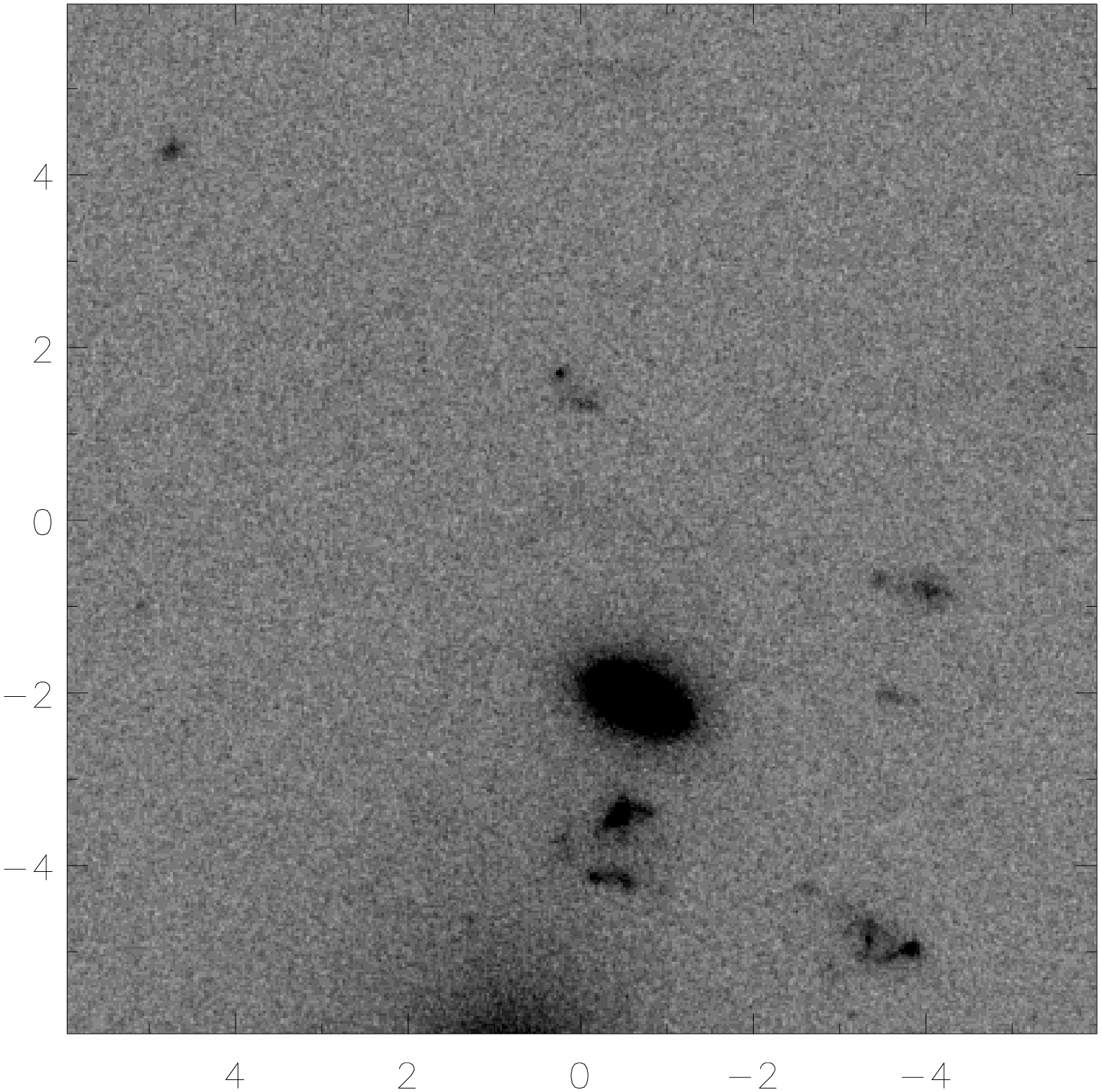,width=0.20\textwidth}&
\epsfig{file=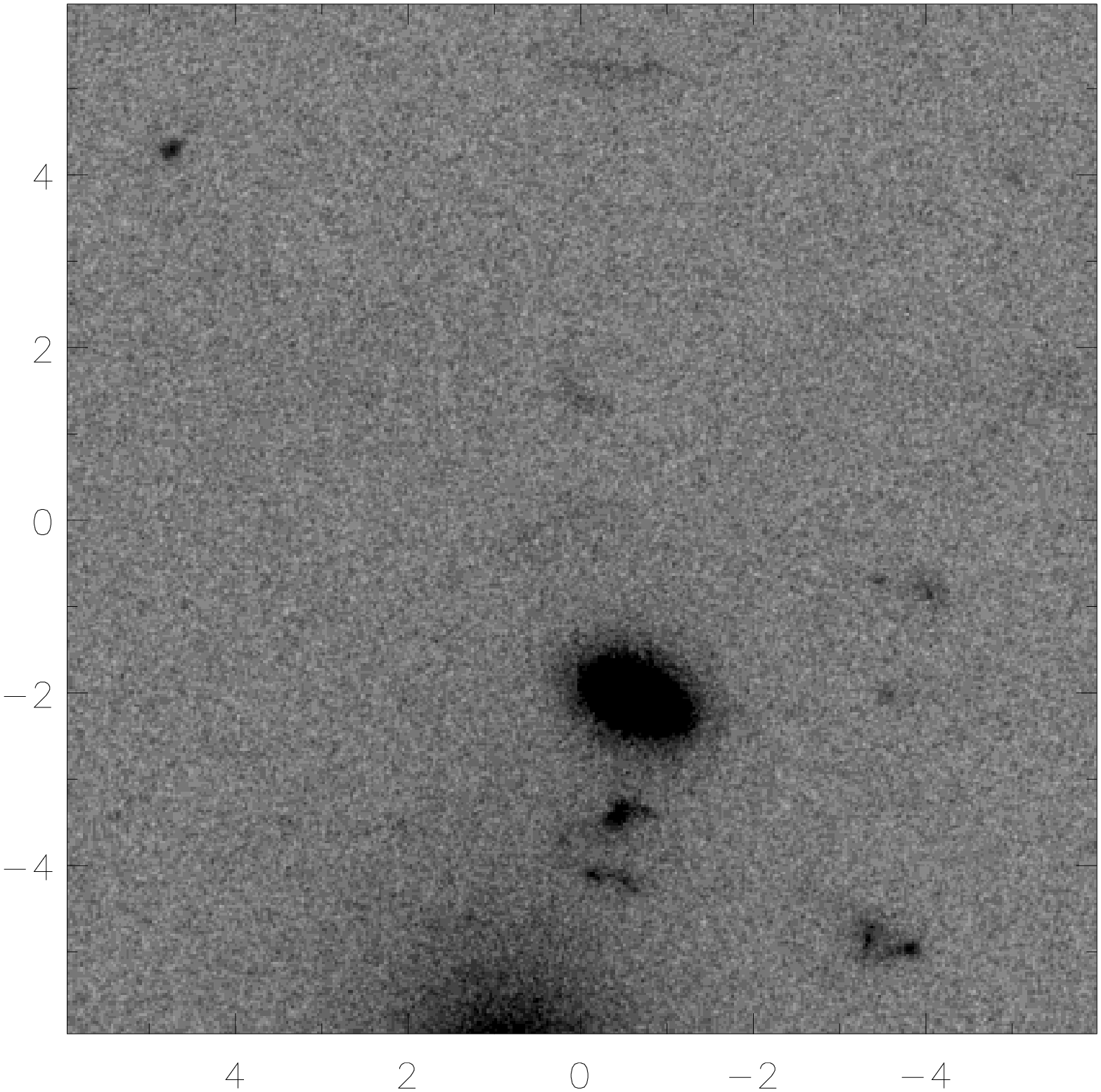,width=0.20\textwidth}&
\epsfig{file=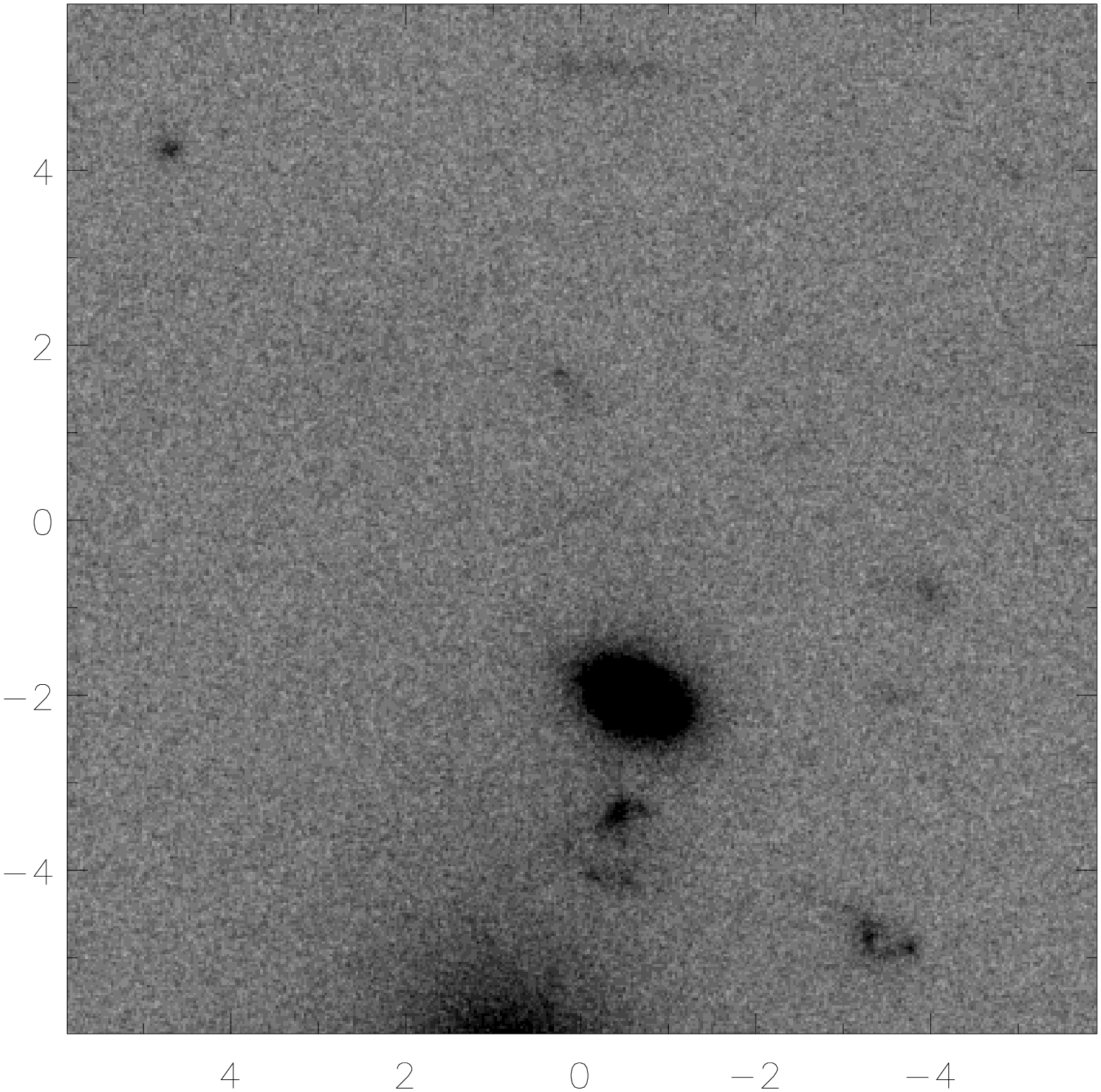,width=0.20\textwidth}\\
\epsfig{file=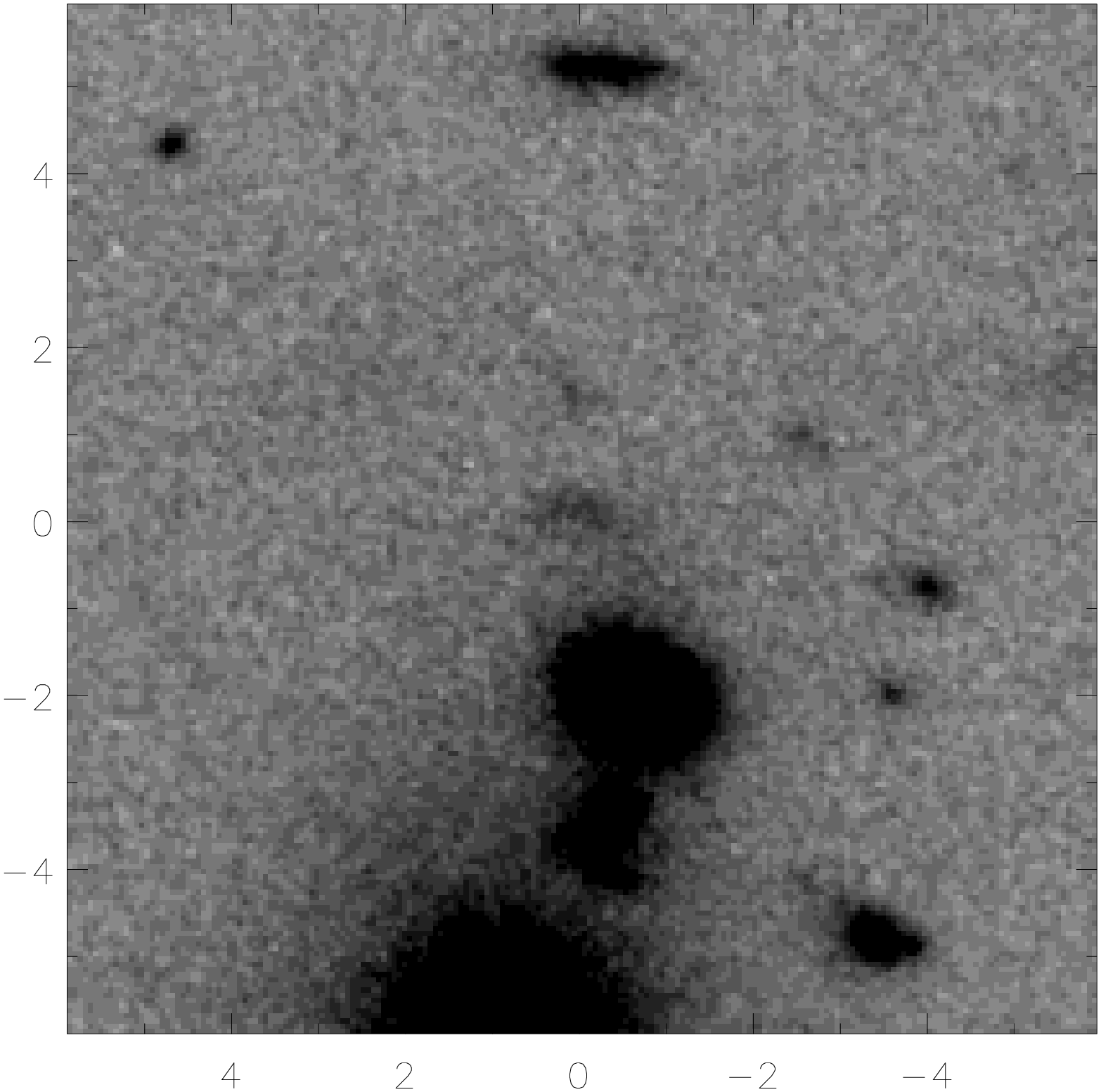,width=0.20\textwidth}&
\epsfig{file=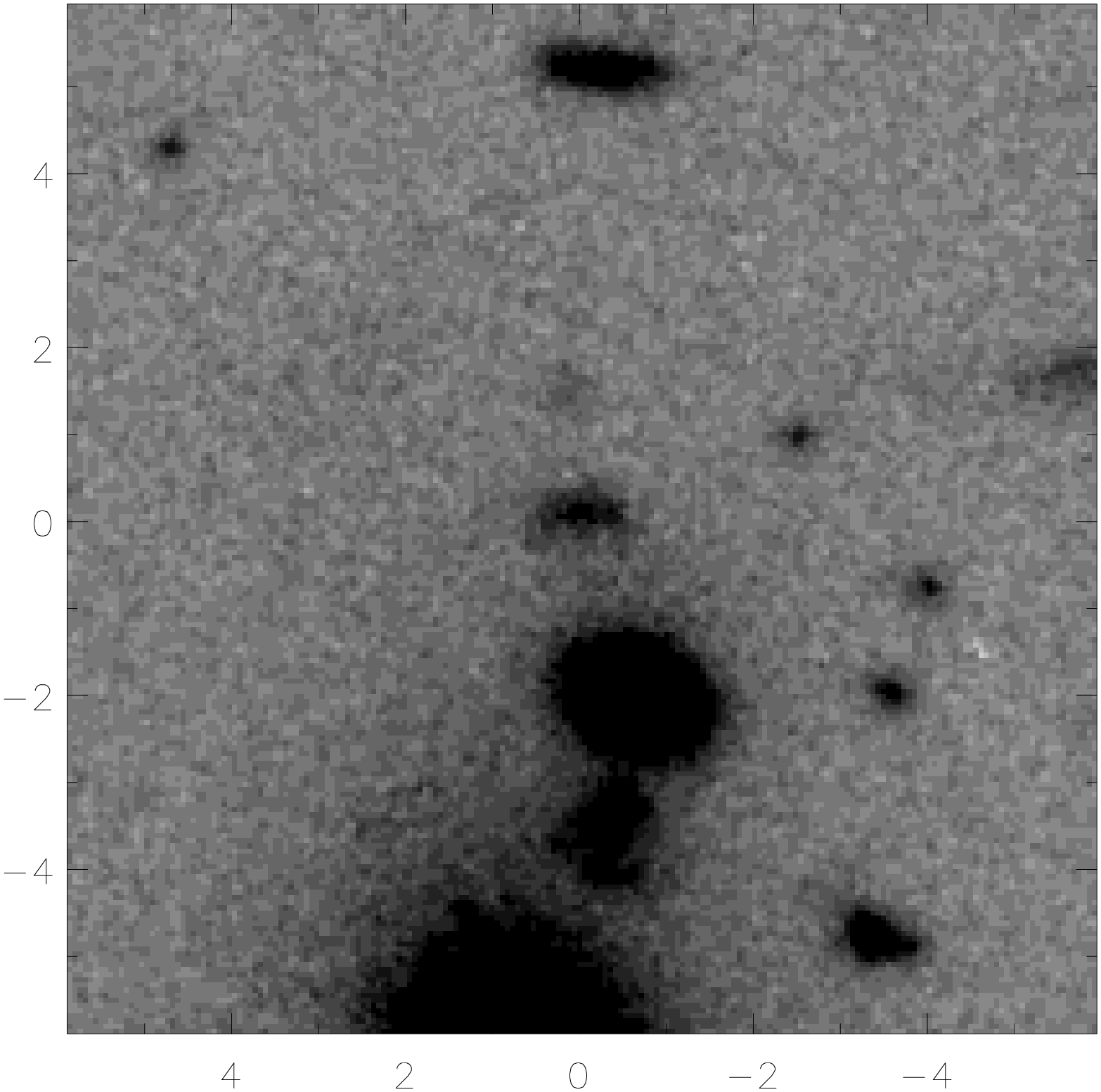,width=0.20\textwidth}&
\epsfig{file=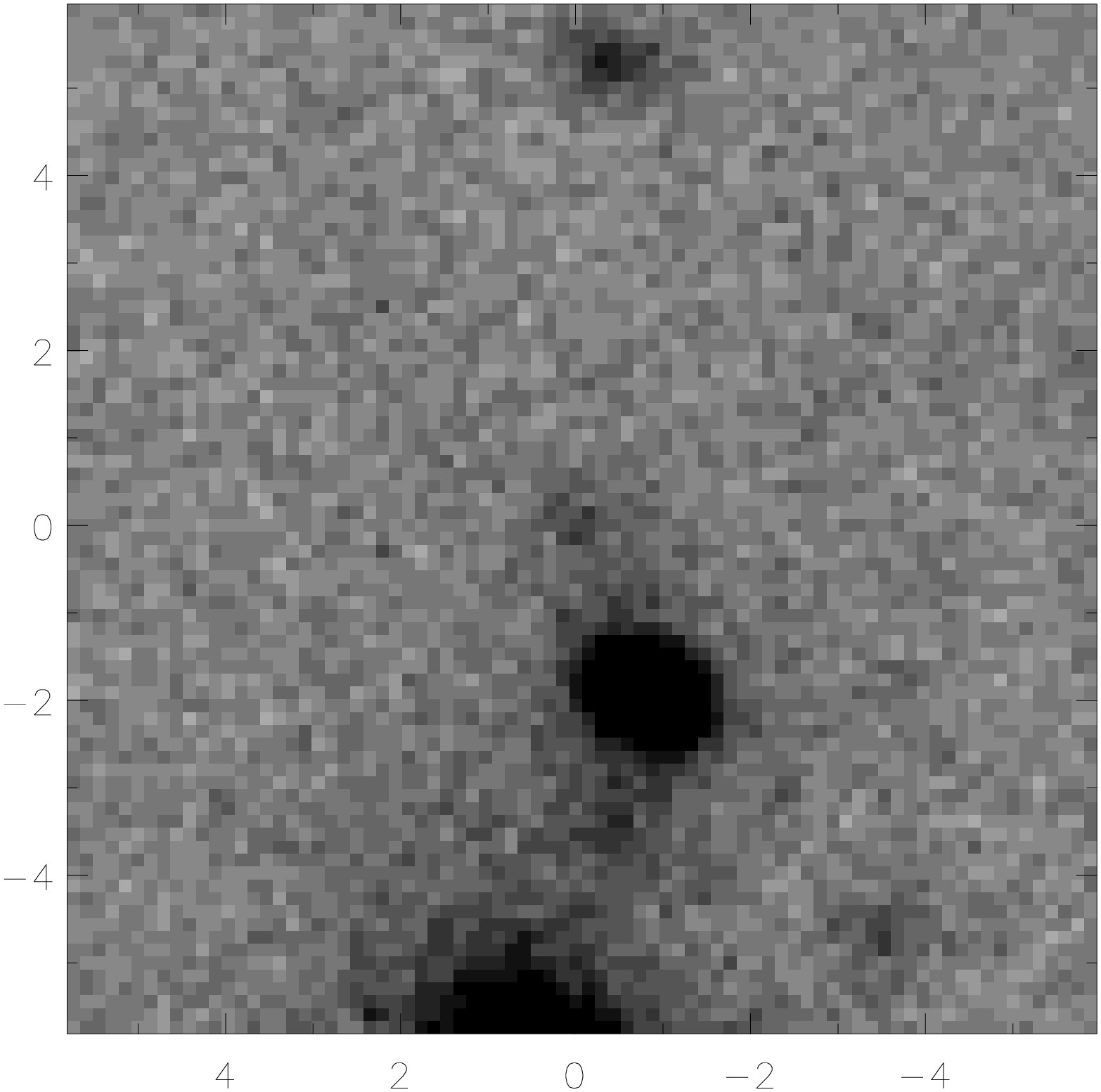,width=0.20\textwidth}&
\epsfig{file=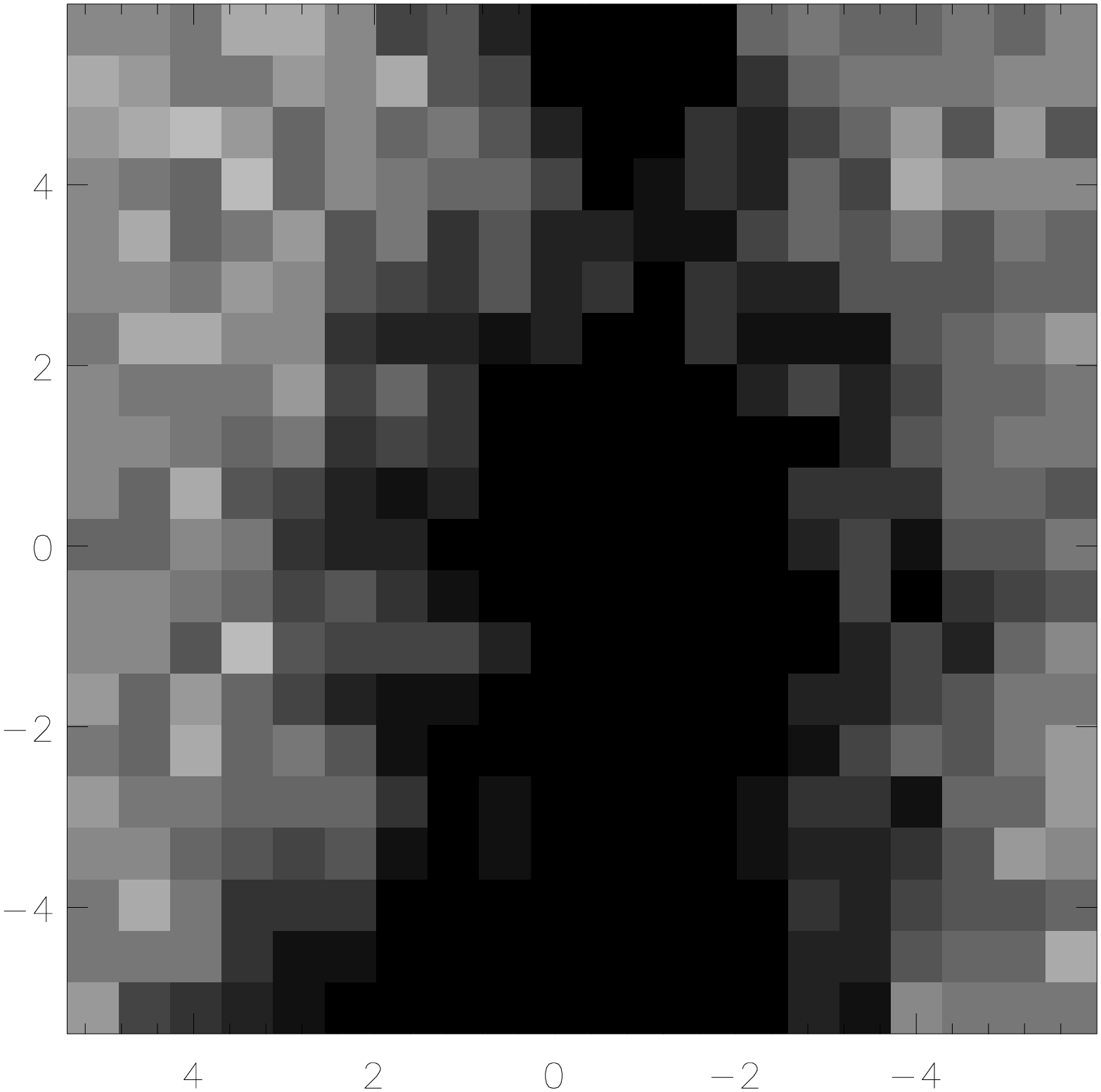,width=0.20\textwidth}\\
\\
\epsfig{file=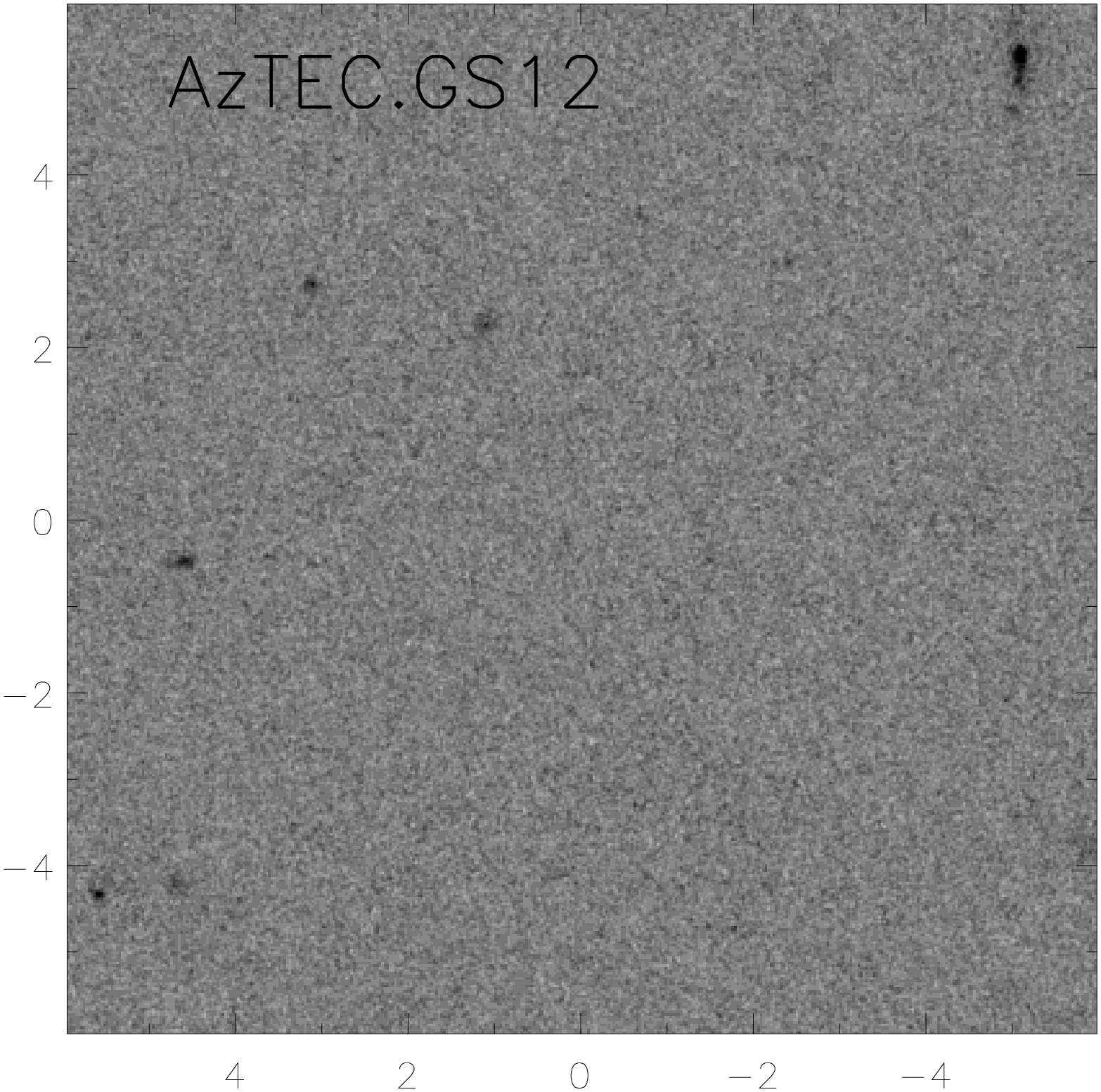,width=0.20\textwidth}&
\epsfig{file=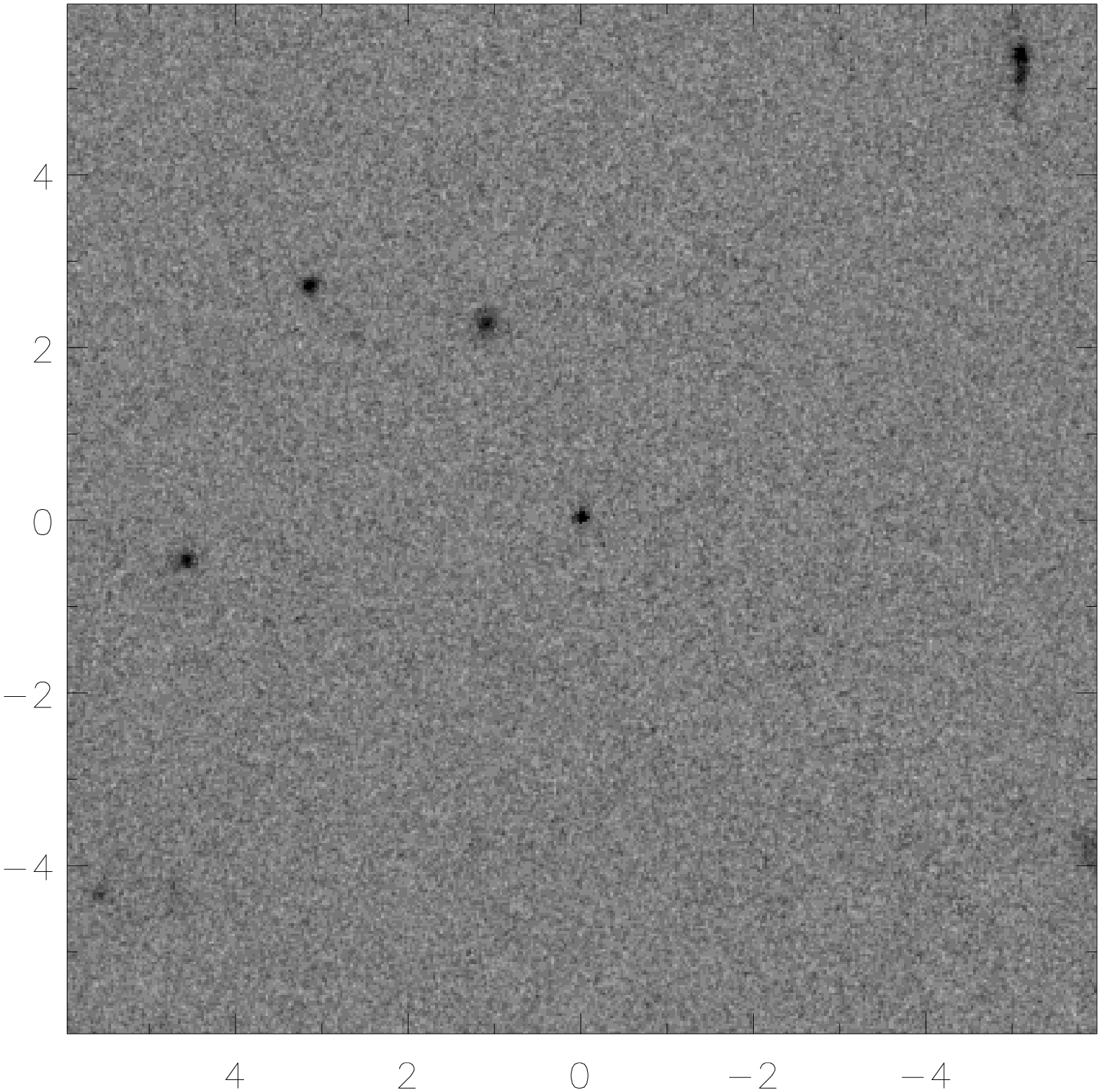,width=0.20\textwidth}&
\epsfig{file=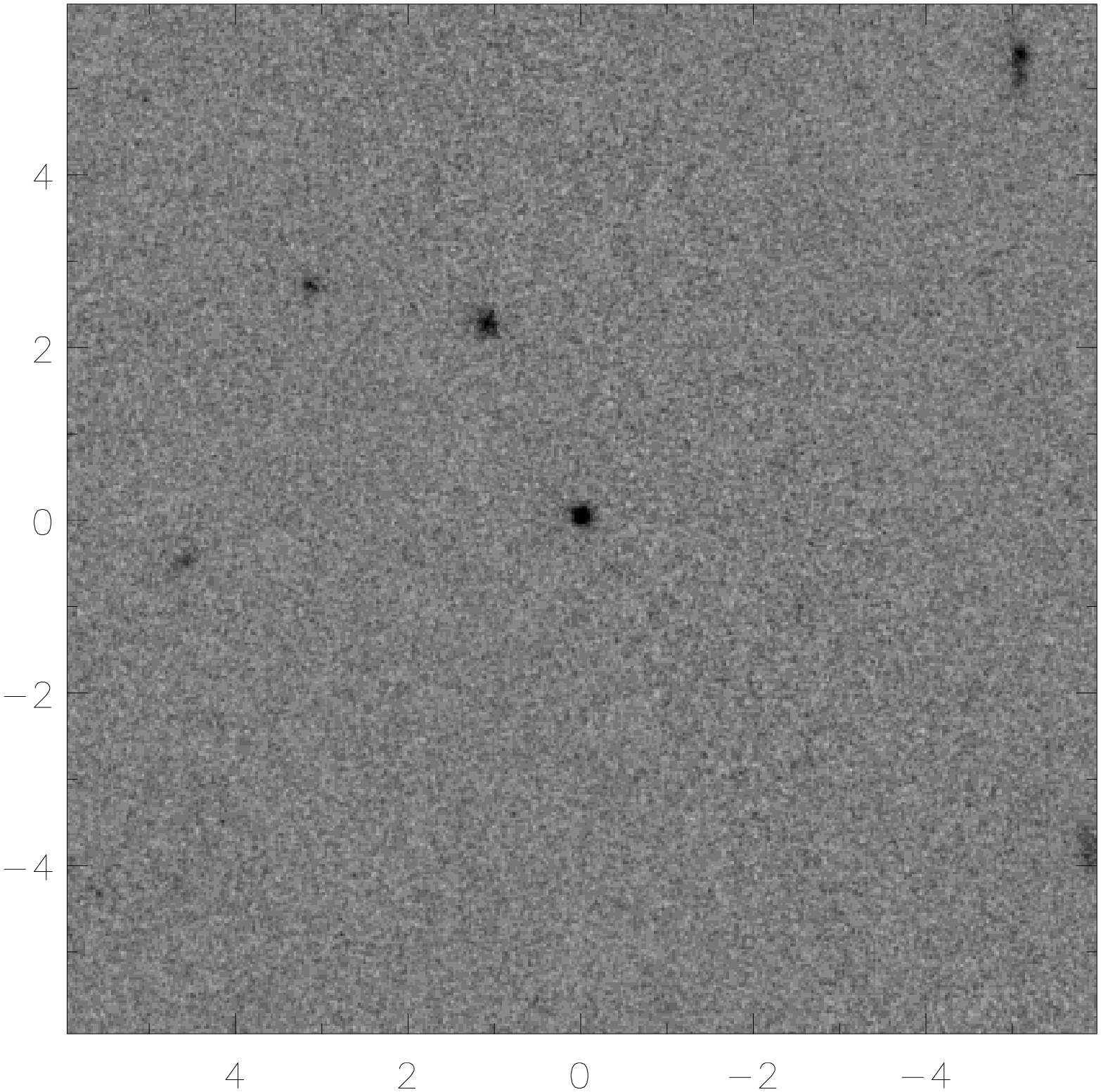,width=0.20\textwidth}&
\epsfig{file=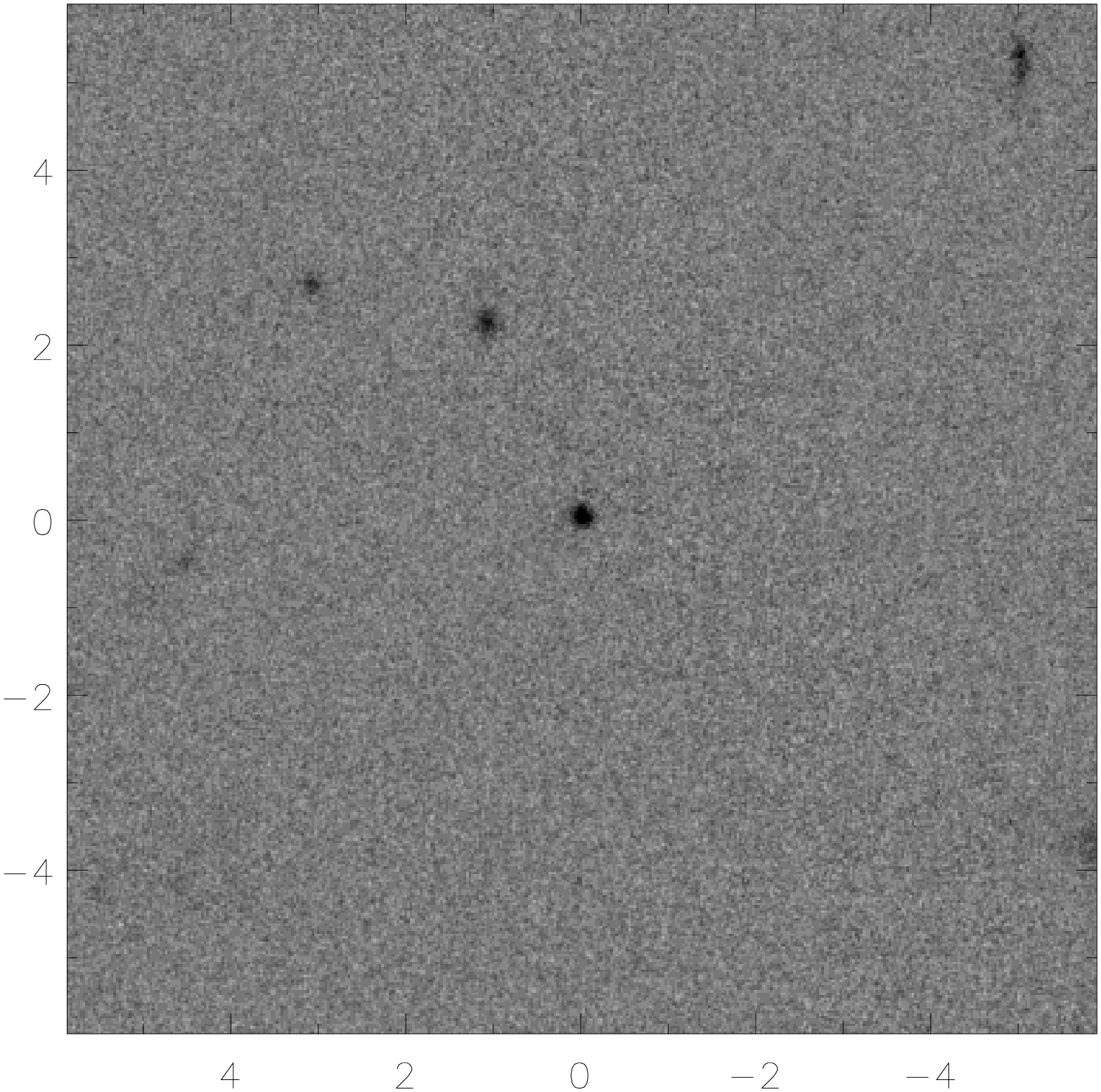,width=0.20\textwidth}\\
\epsfig{file=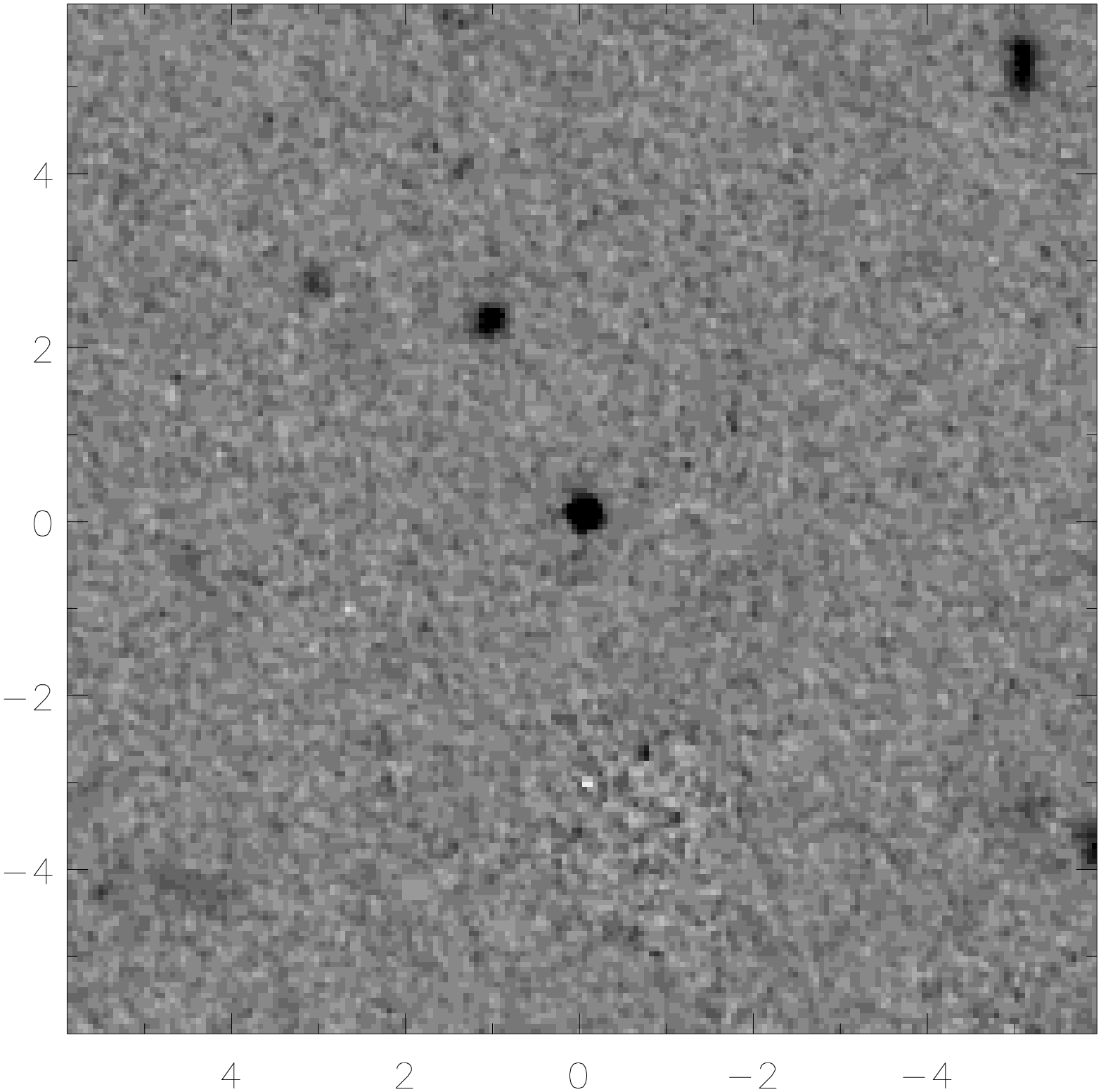,width=0.20\textwidth}&
\epsfig{file=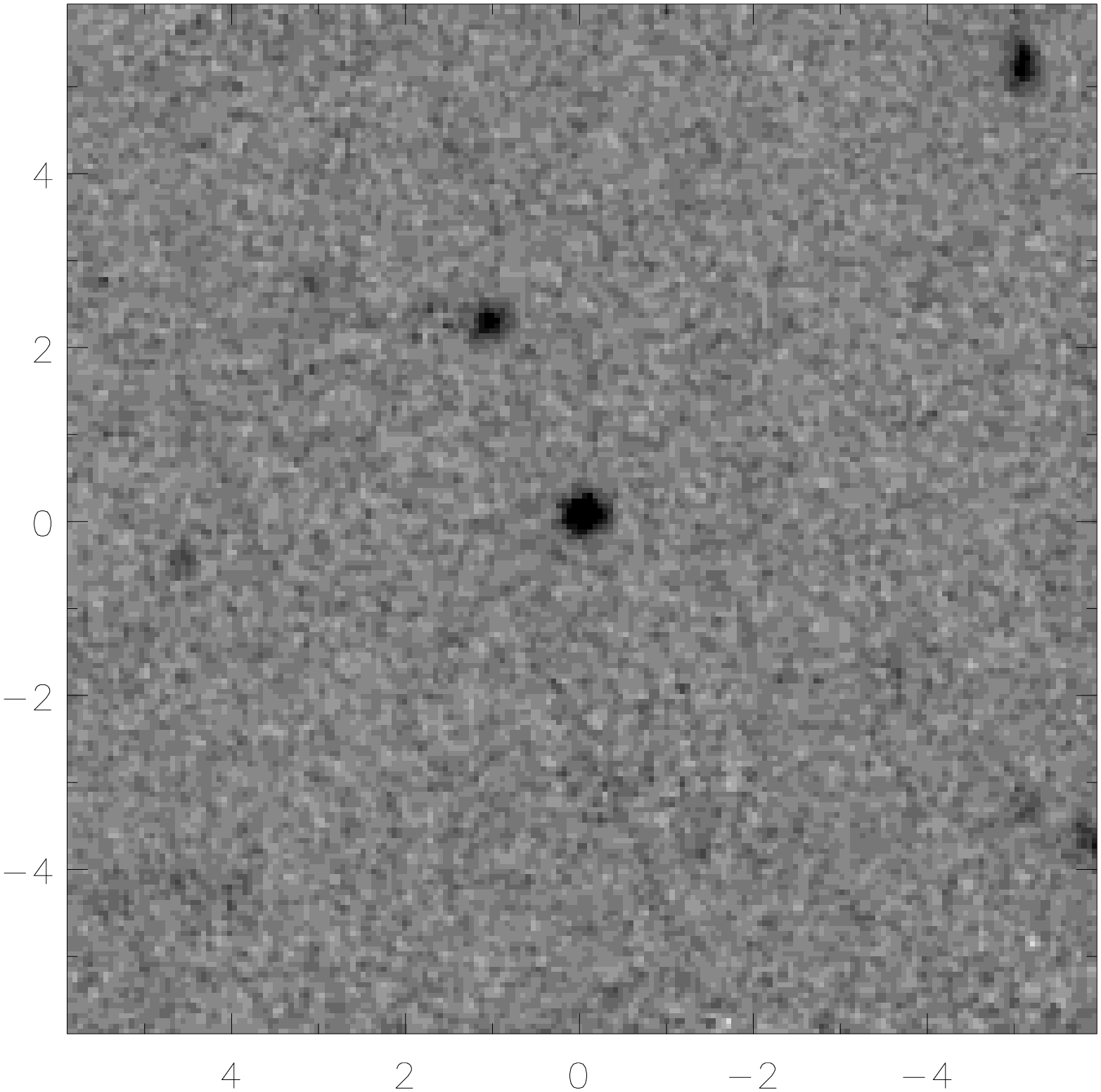,width=0.20\textwidth}&
\epsfig{file=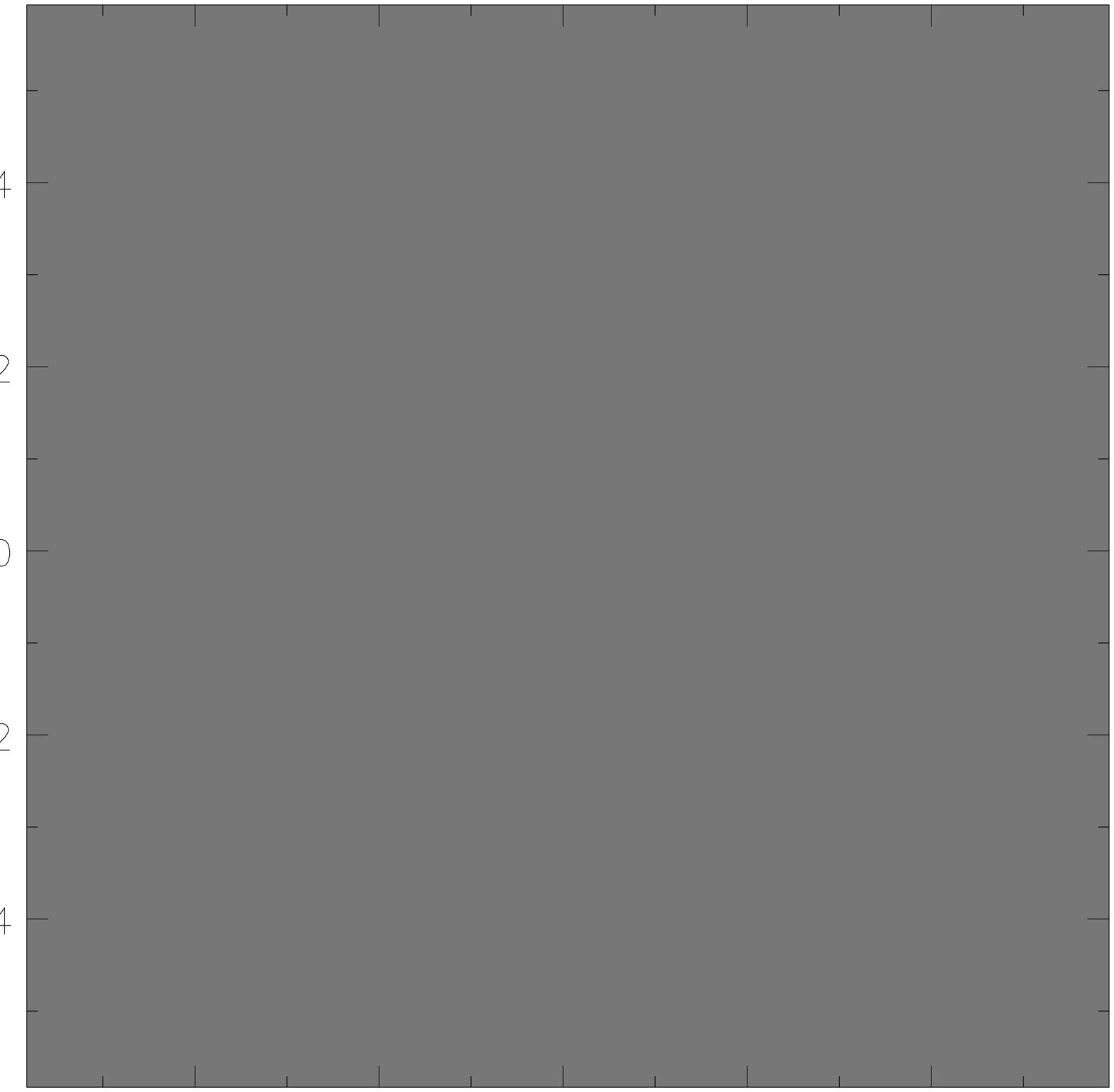,width=0.20\textwidth}&
\epsfig{file=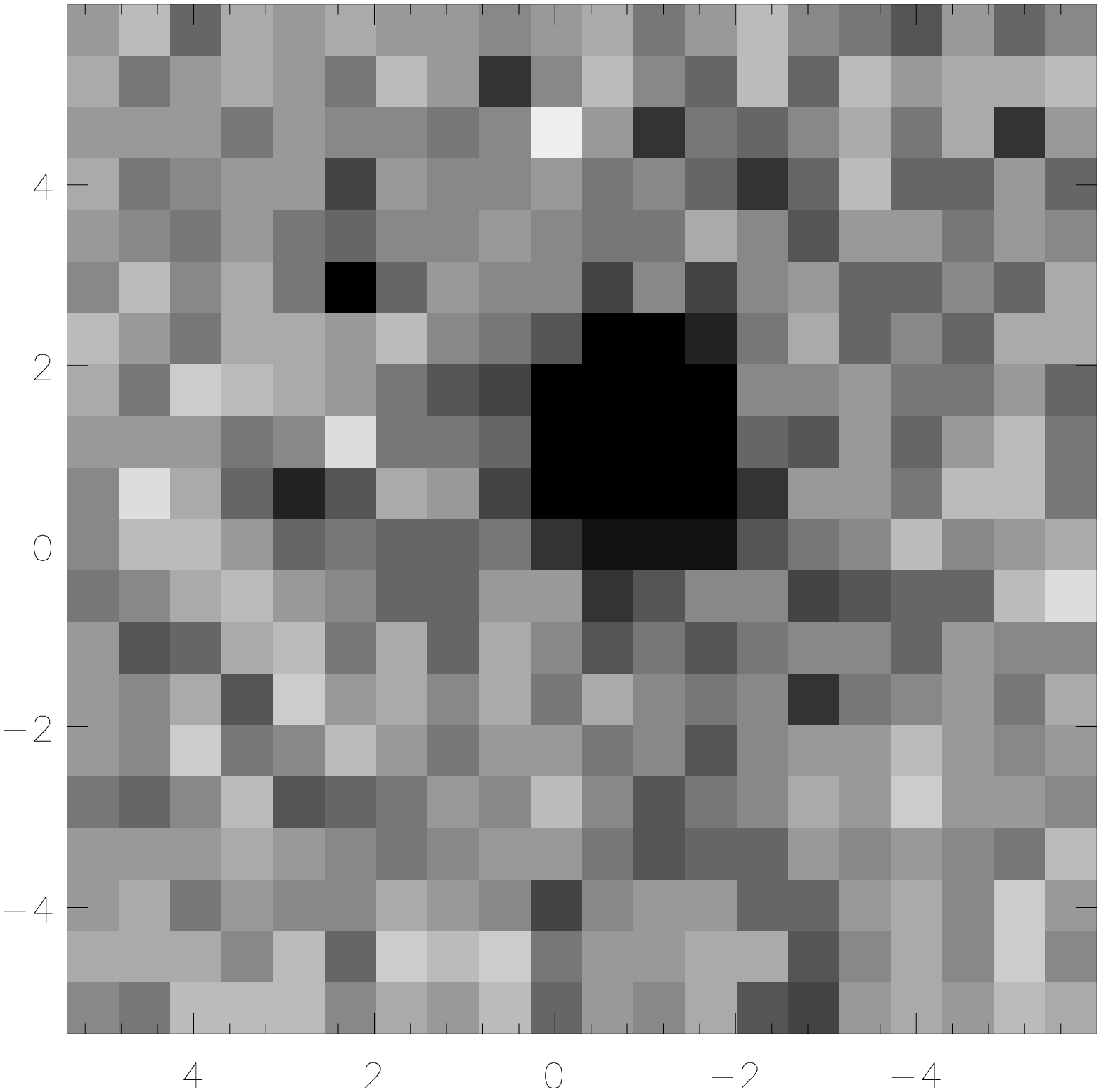,width=0.20\textwidth}\\
\\
\epsfig{file=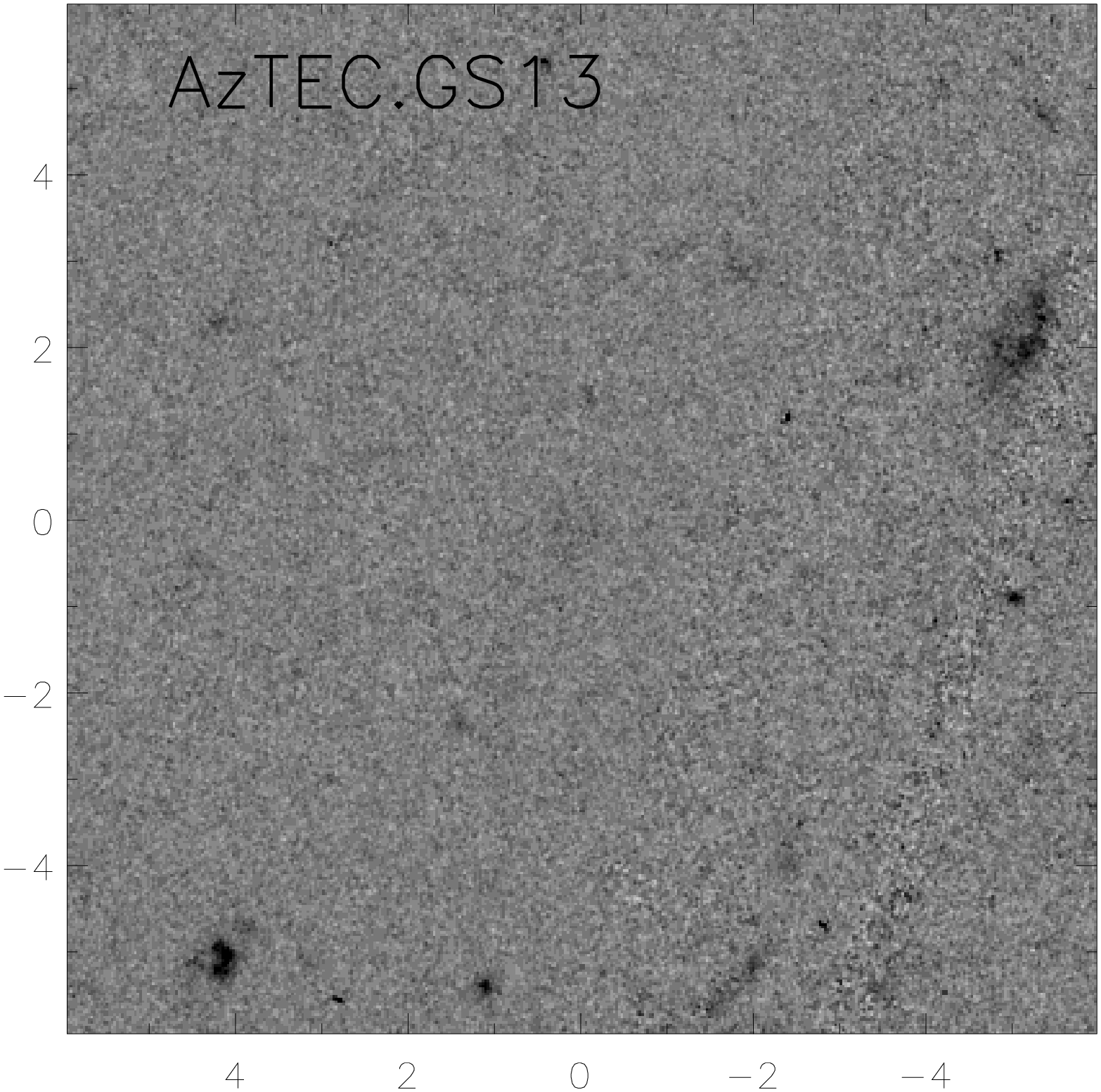,width=0.20\textwidth}&
\epsfig{file=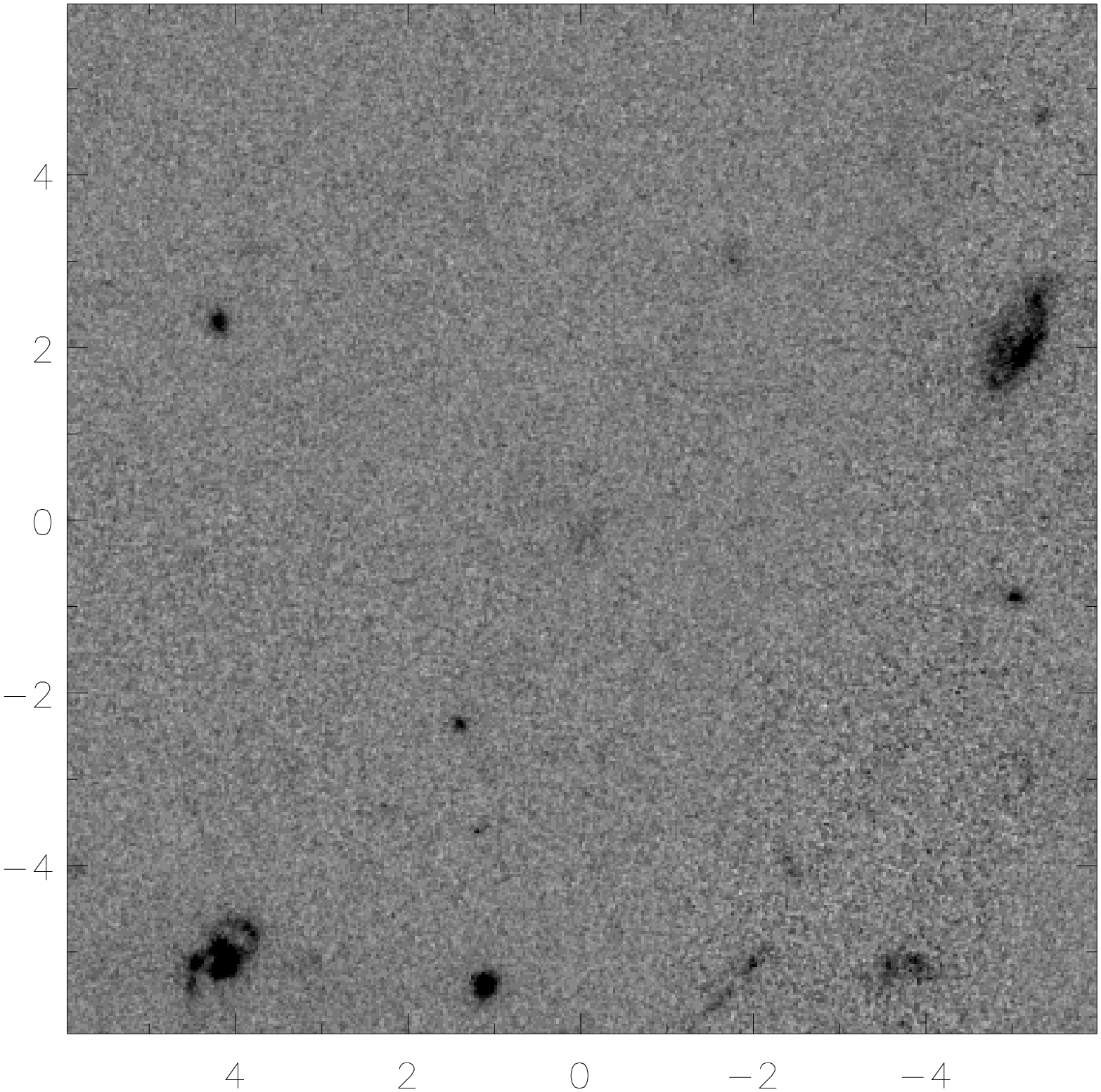,width=0.20\textwidth}&
\epsfig{file=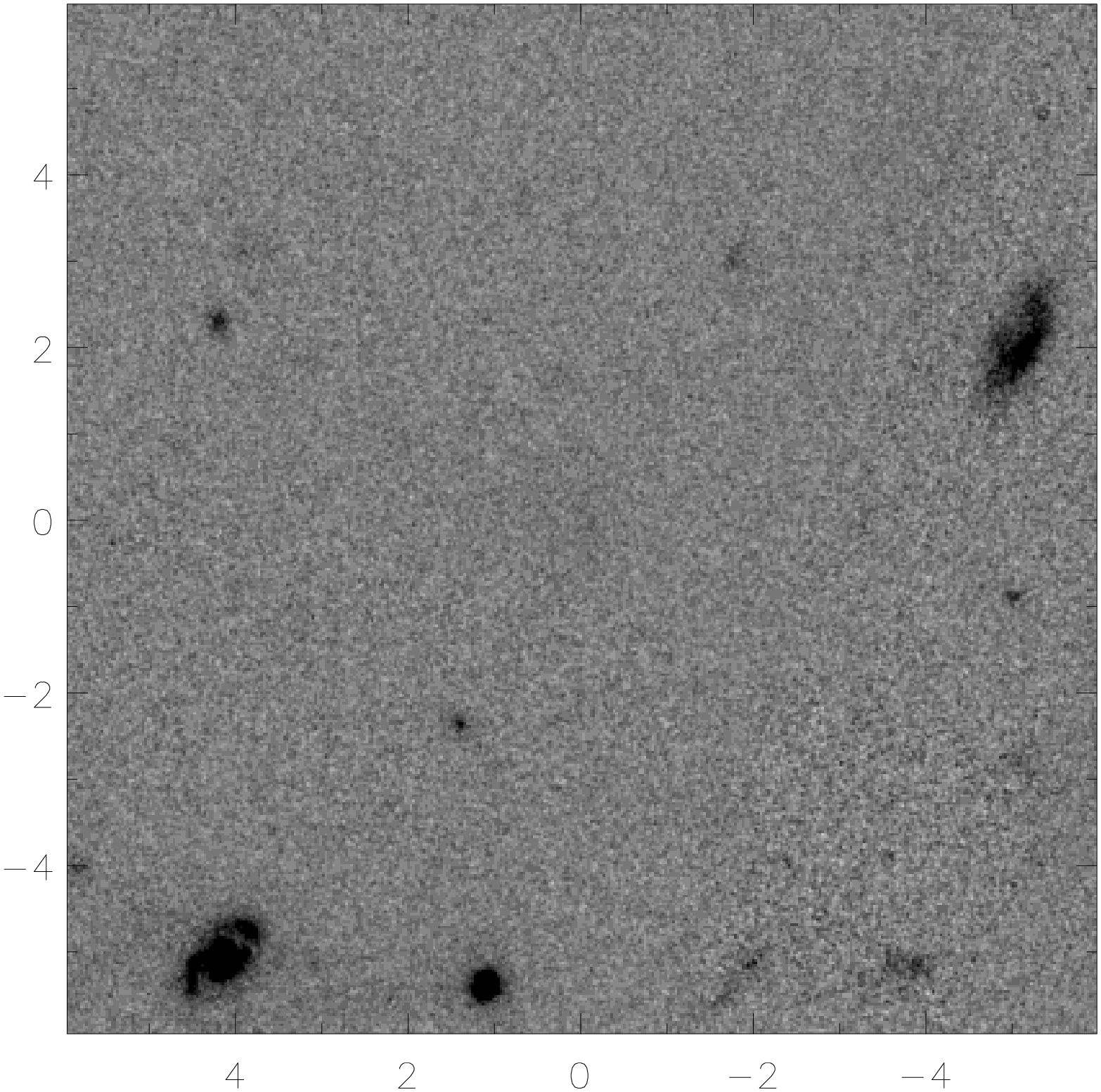,width=0.20\textwidth}&
\epsfig{file=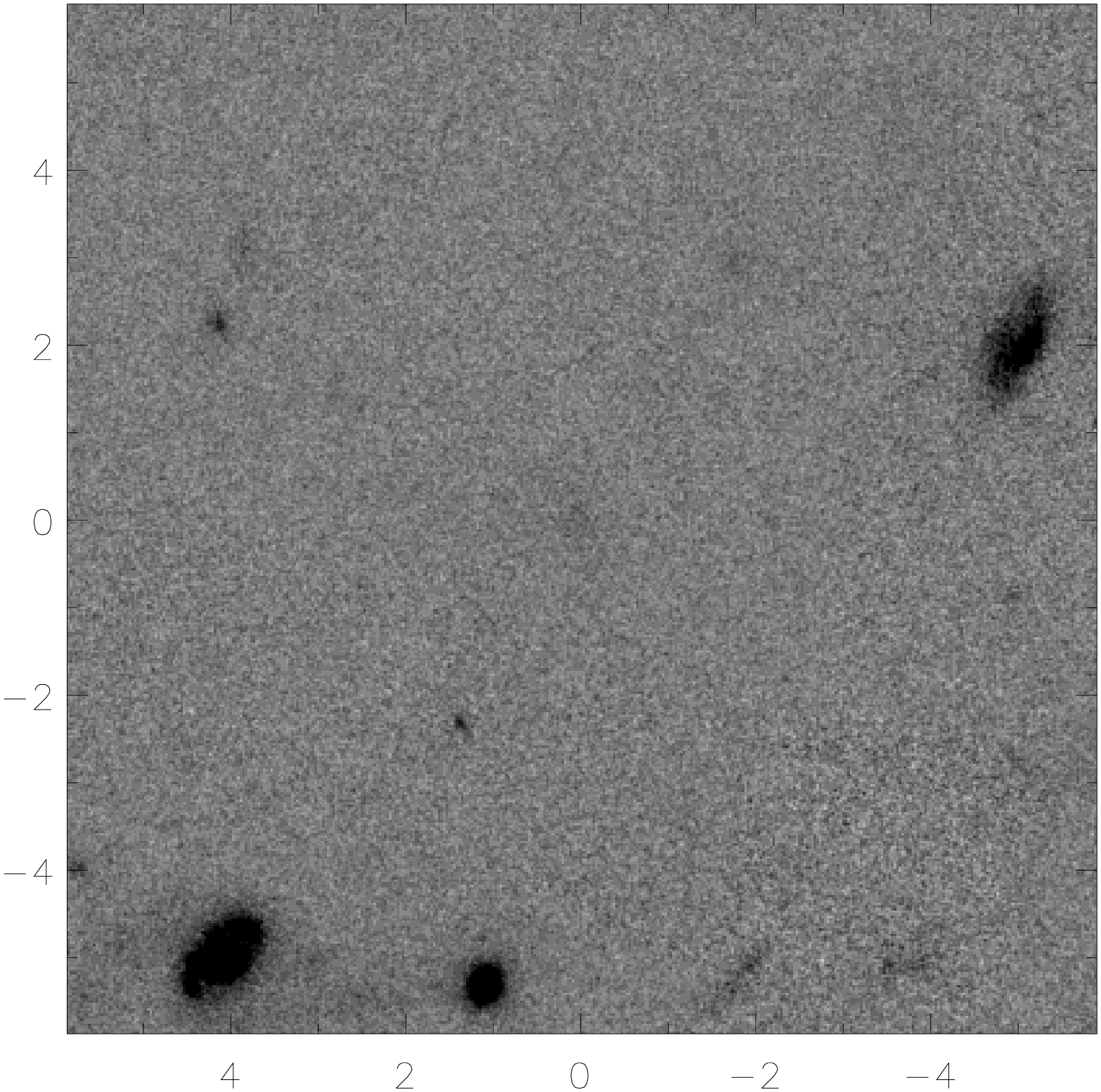,width=0.20\textwidth}\\
\epsfig{file=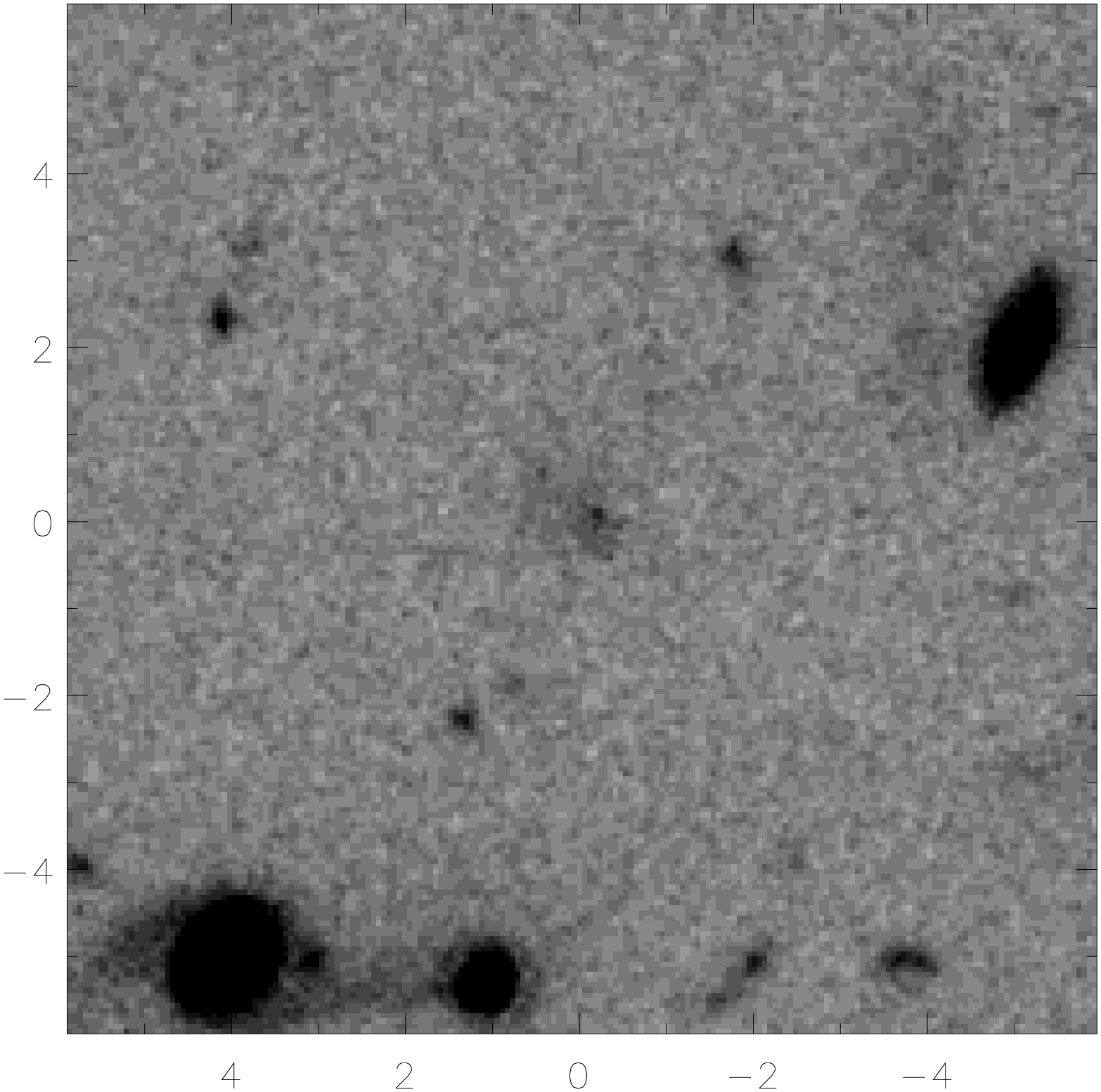,width=0.20\textwidth}&
\epsfig{file=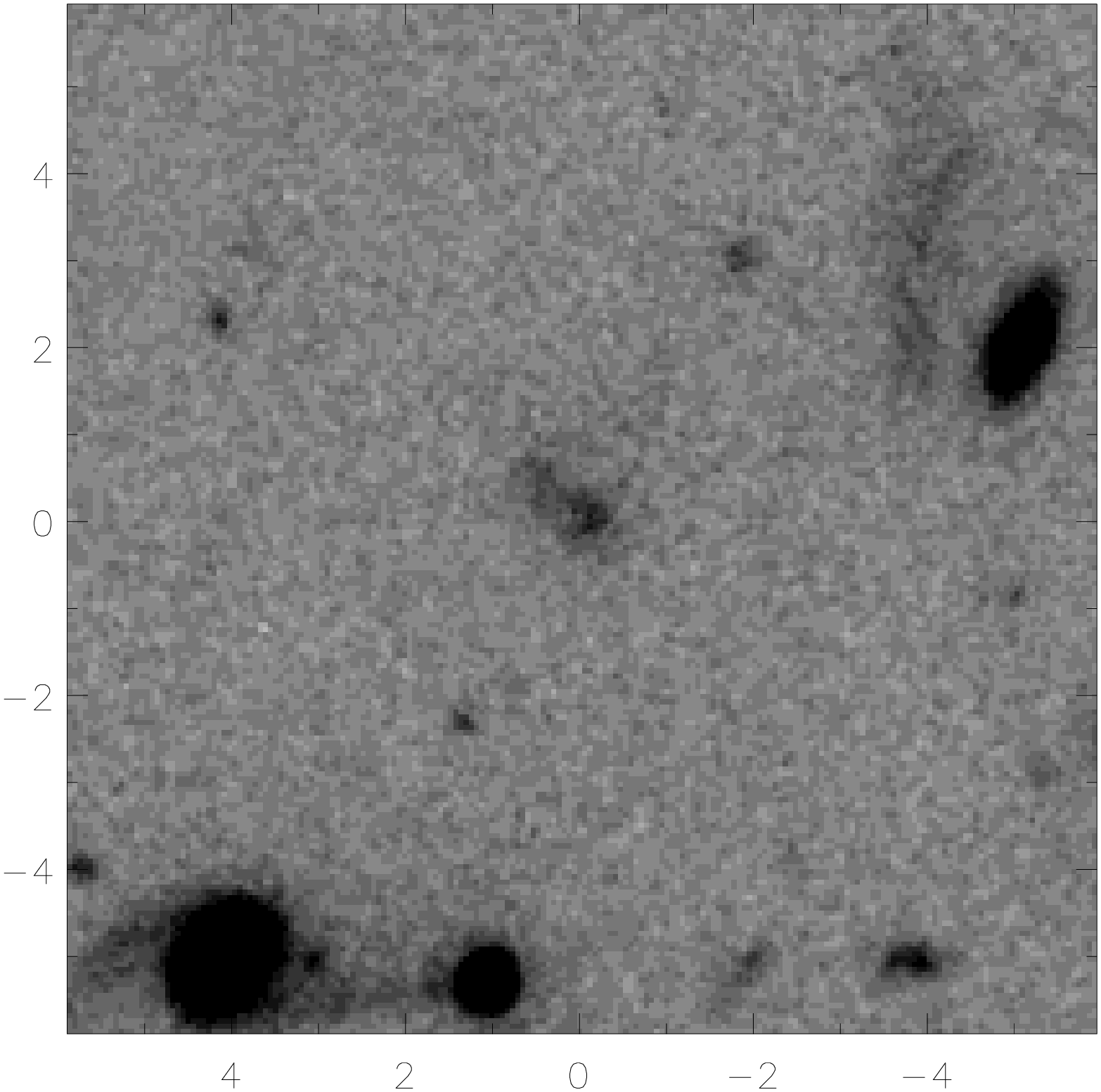,width=0.20\textwidth}&
\epsfig{file=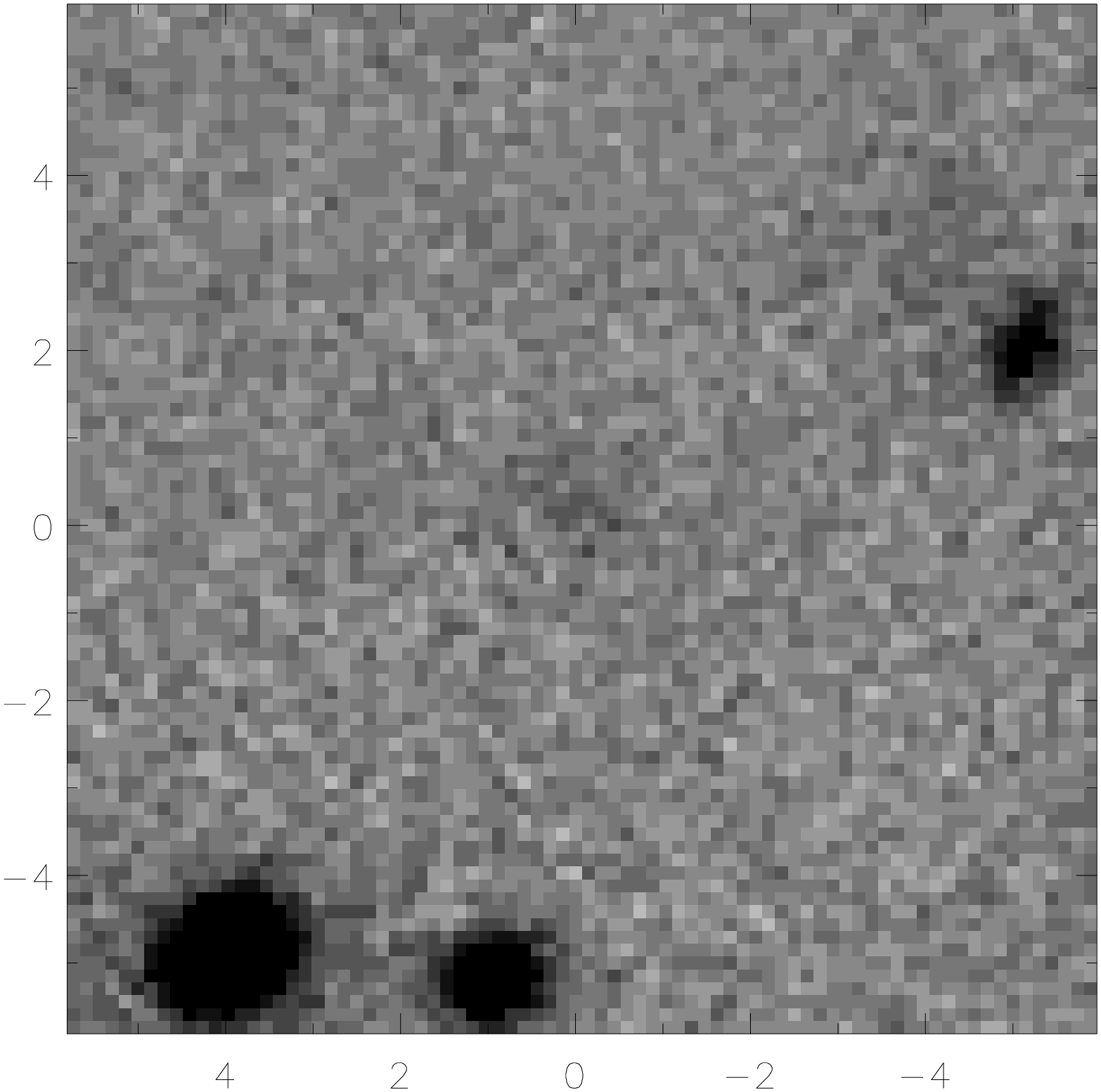,width=0.20\textwidth}&
\epsfig{file=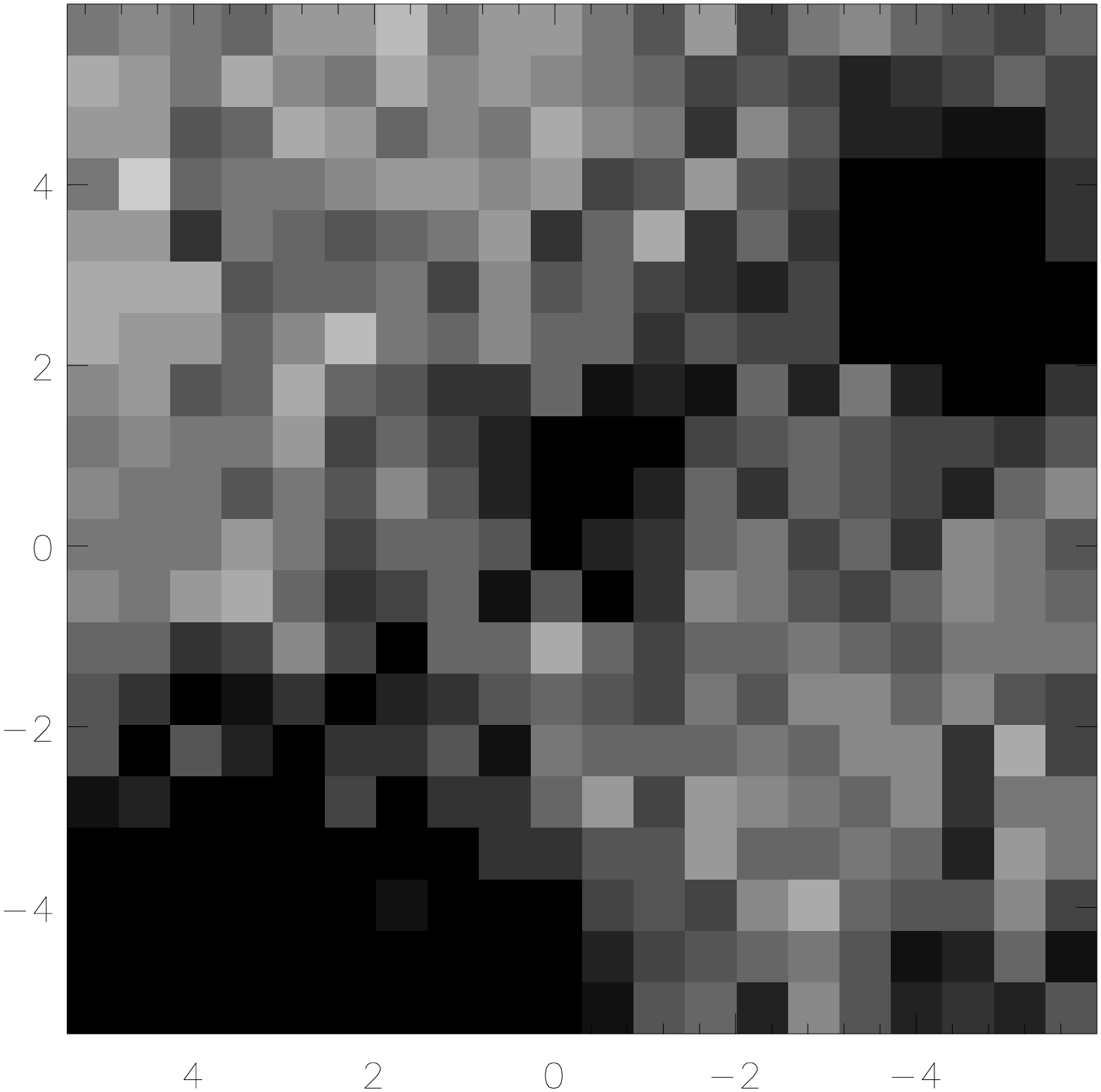,width=0.20\textwidth}\\
\end{tabular}
\addtocounter{figure}{-1}
\caption{- continued}
\vfil}
\end{figure*}
\end{center}


\begin{center}
\begin{figure*}
\vbox to220mm{\vfil
\begin{tabular}{cccccccc}
\epsfig{file=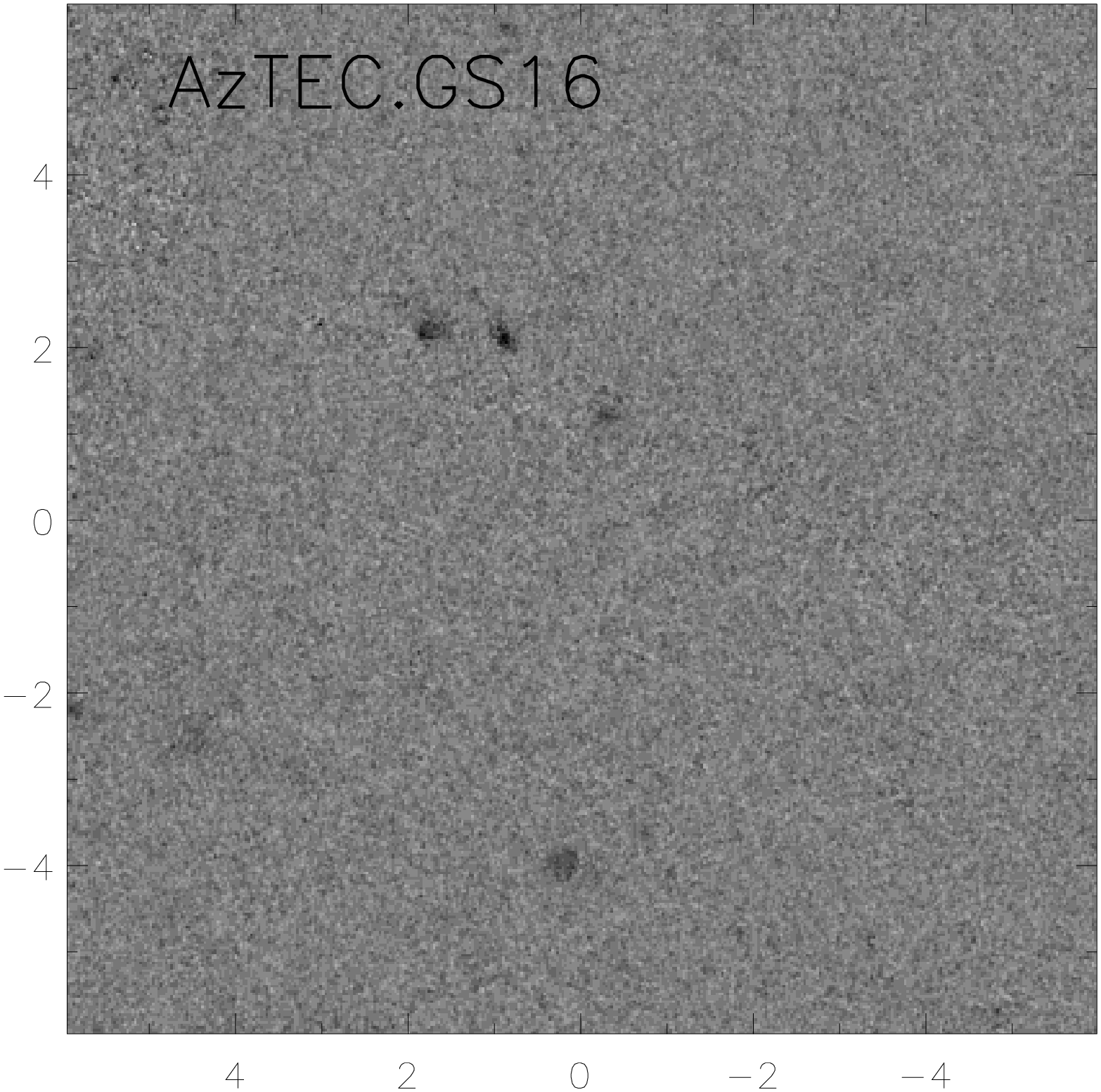,width=0.20\textwidth}&
\epsfig{file=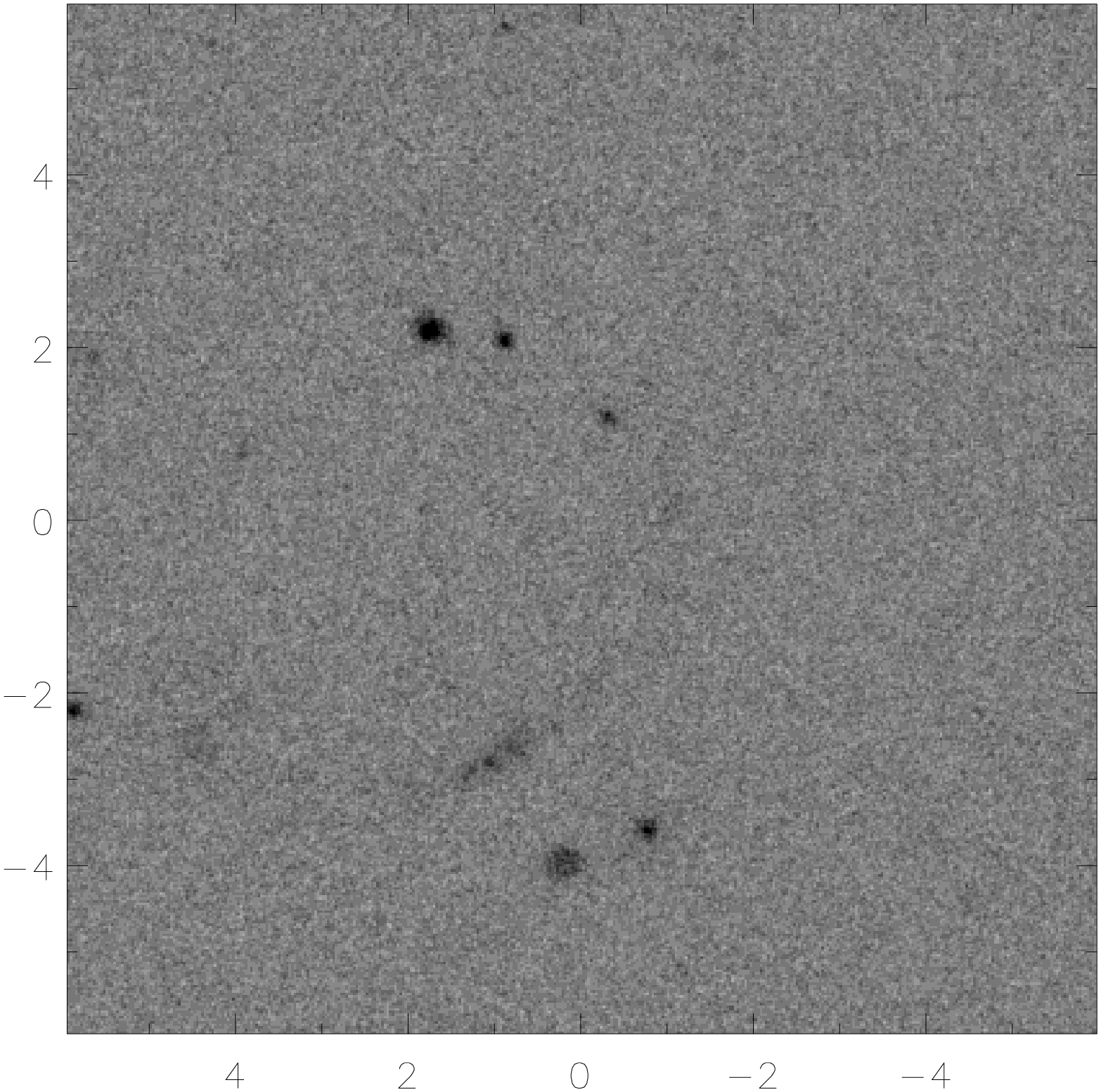,width=0.20\textwidth}&
\epsfig{file=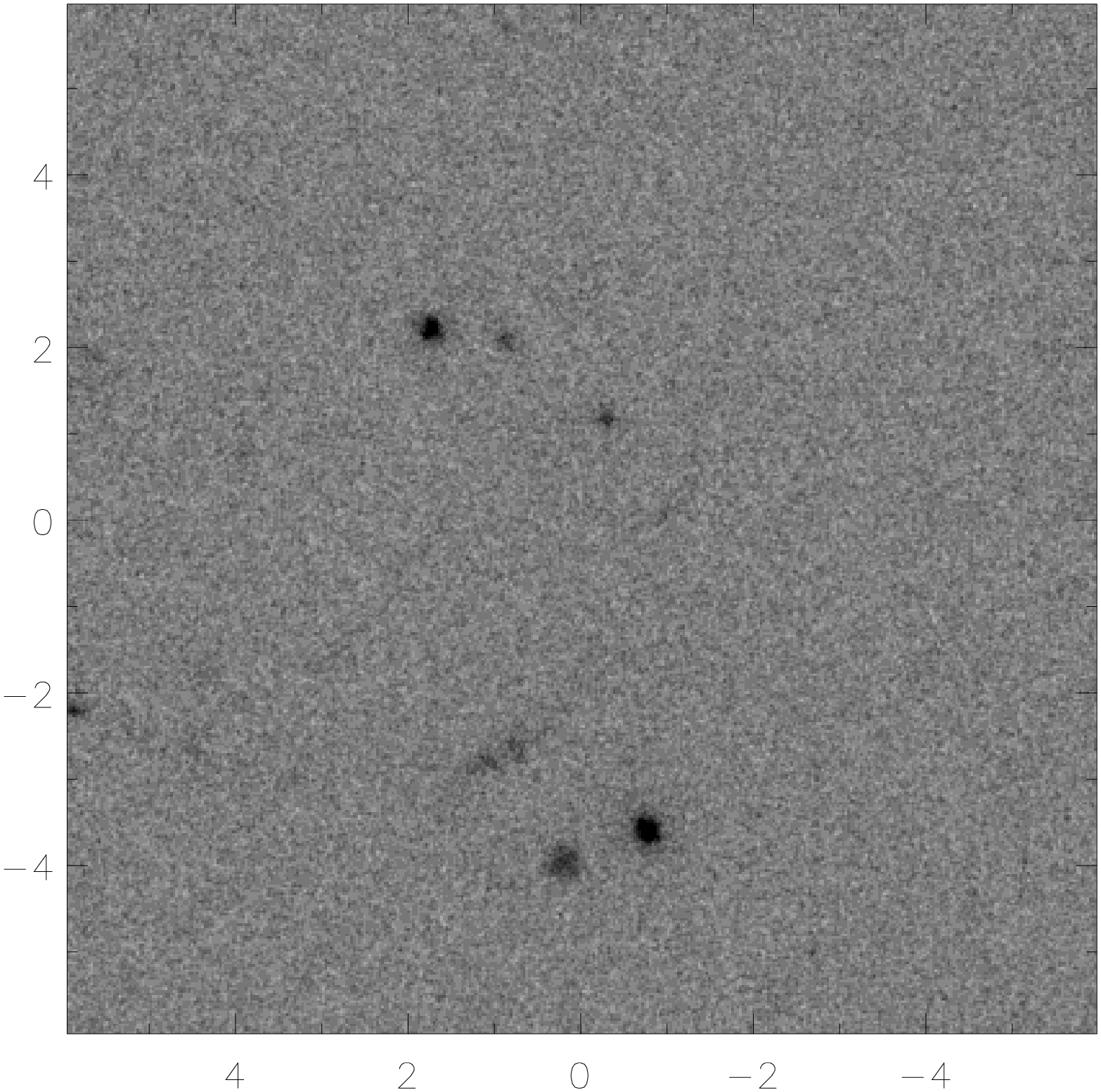,width=0.20\textwidth}&
\epsfig{file=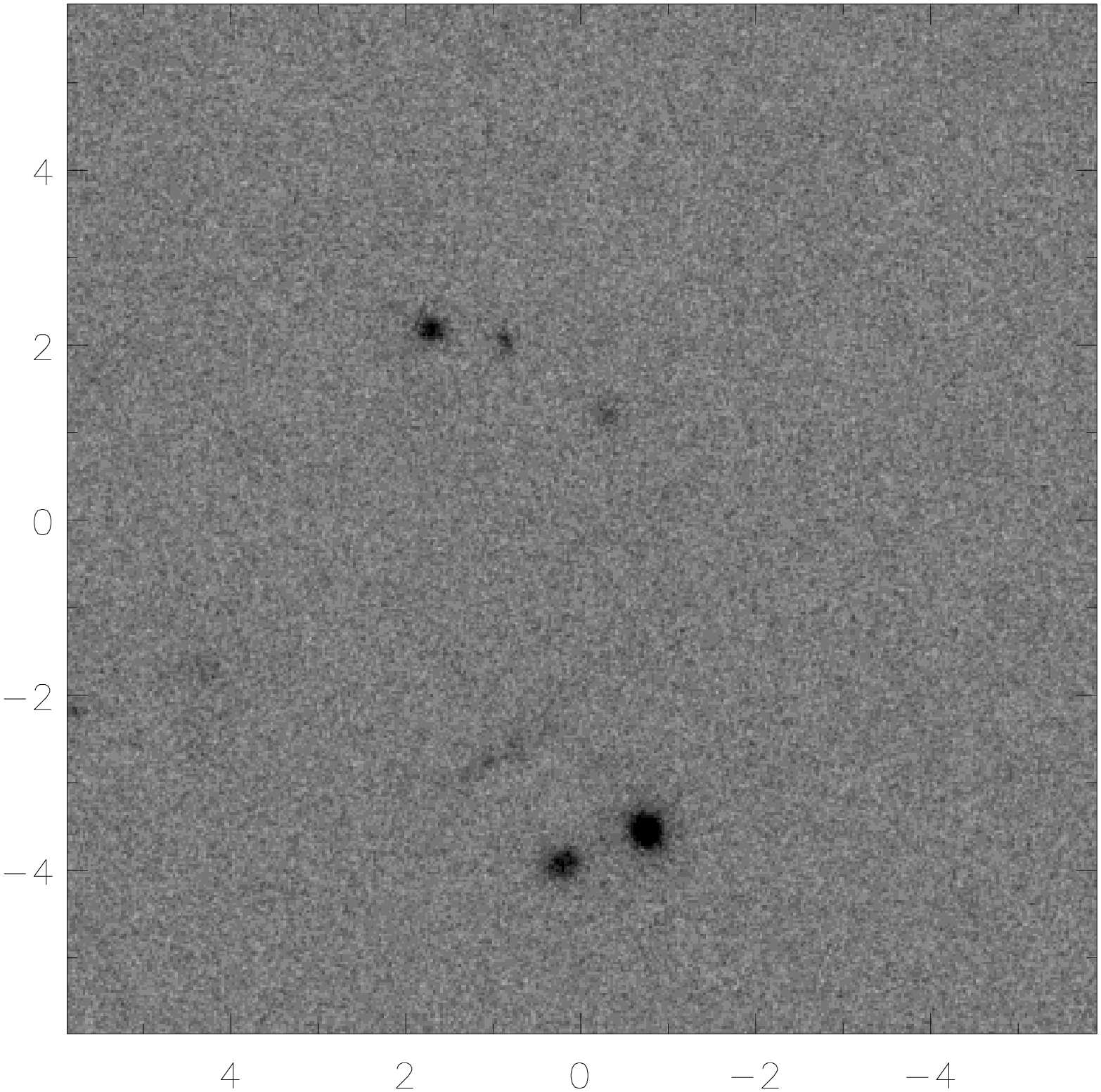,width=0.20\textwidth}\\
\epsfig{file=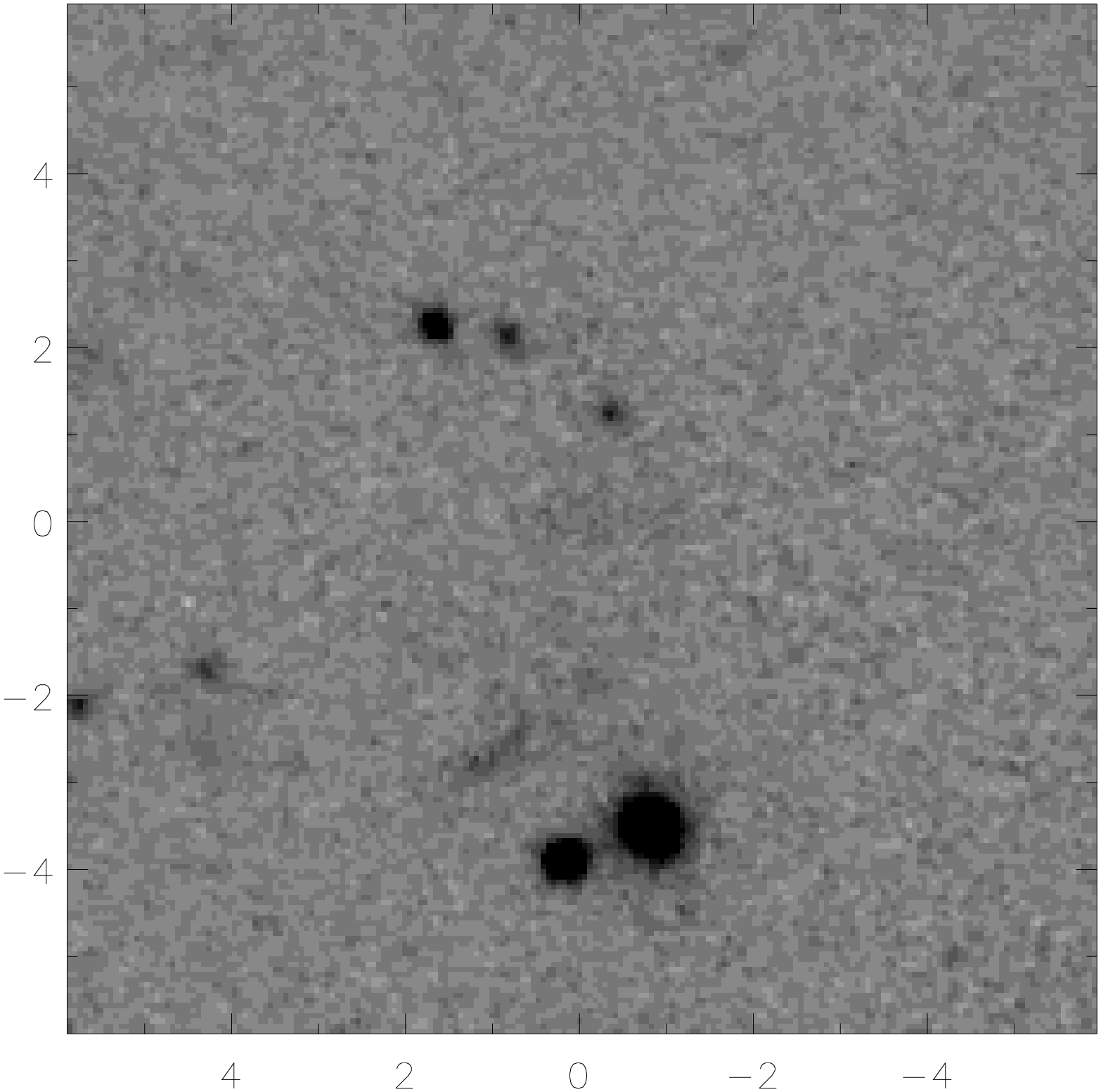,width=0.20\textwidth}&
\epsfig{file=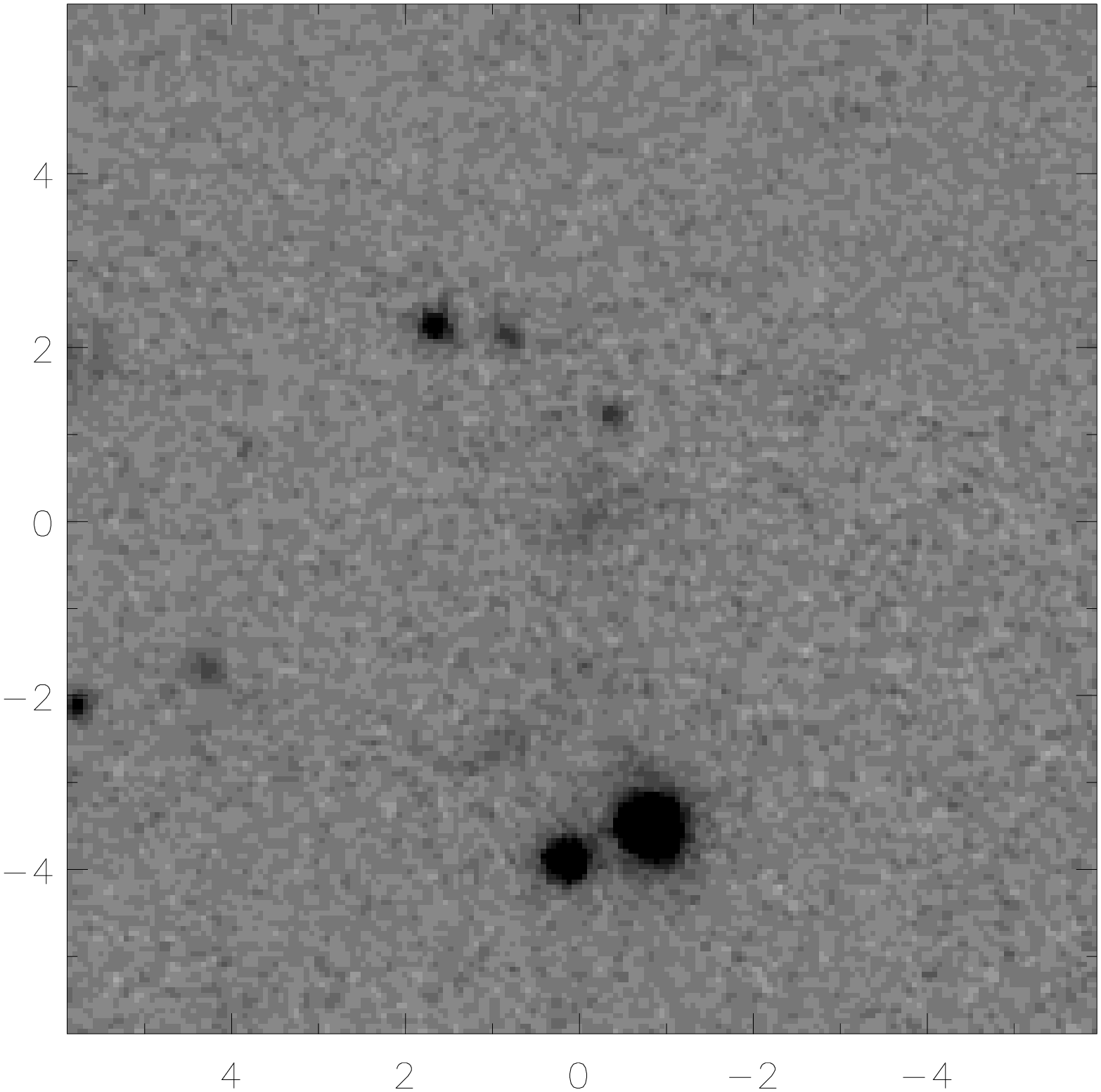,width=0.20\textwidth}&
\epsfig{file=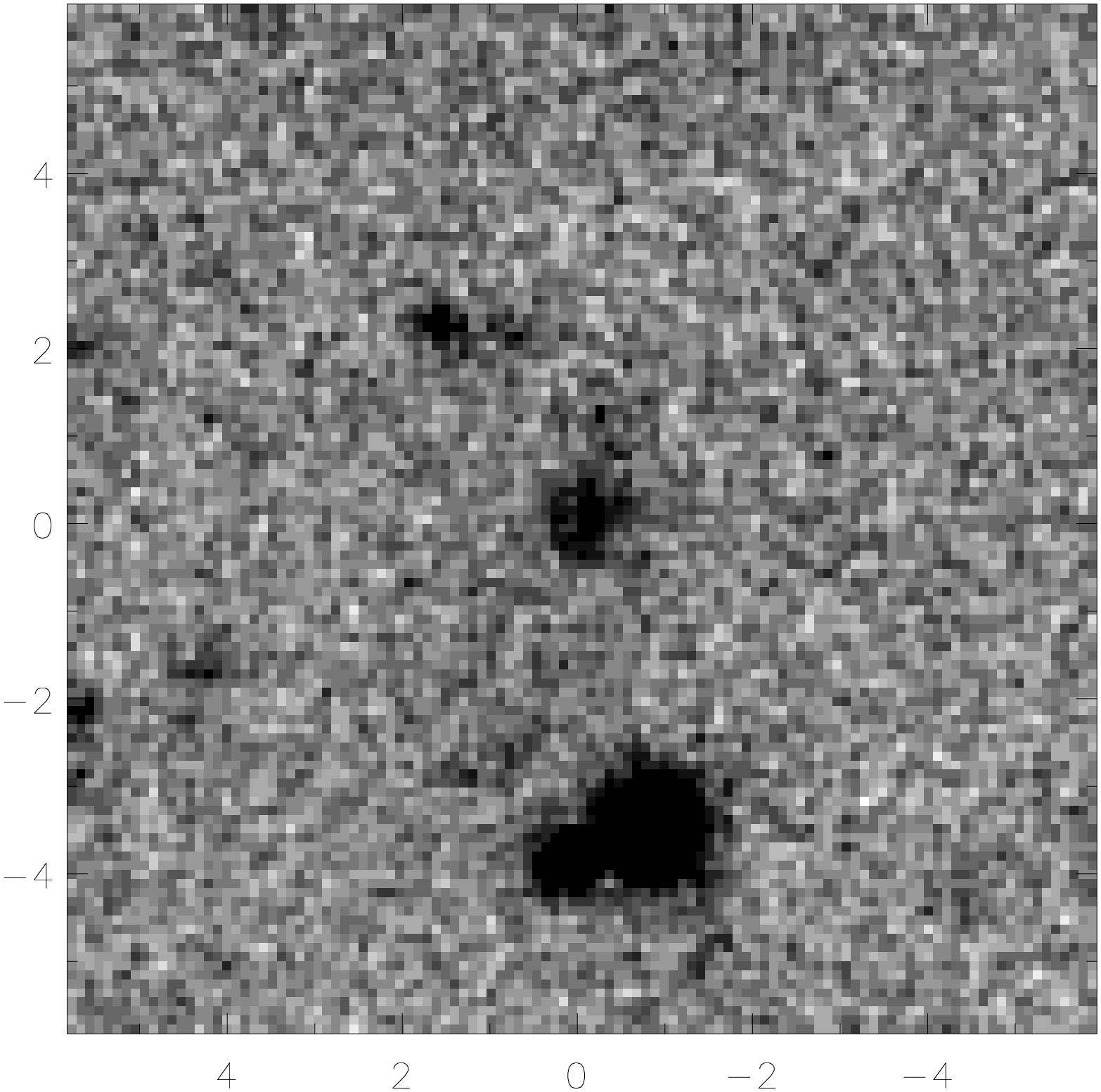,width=0.20\textwidth}&
\epsfig{file=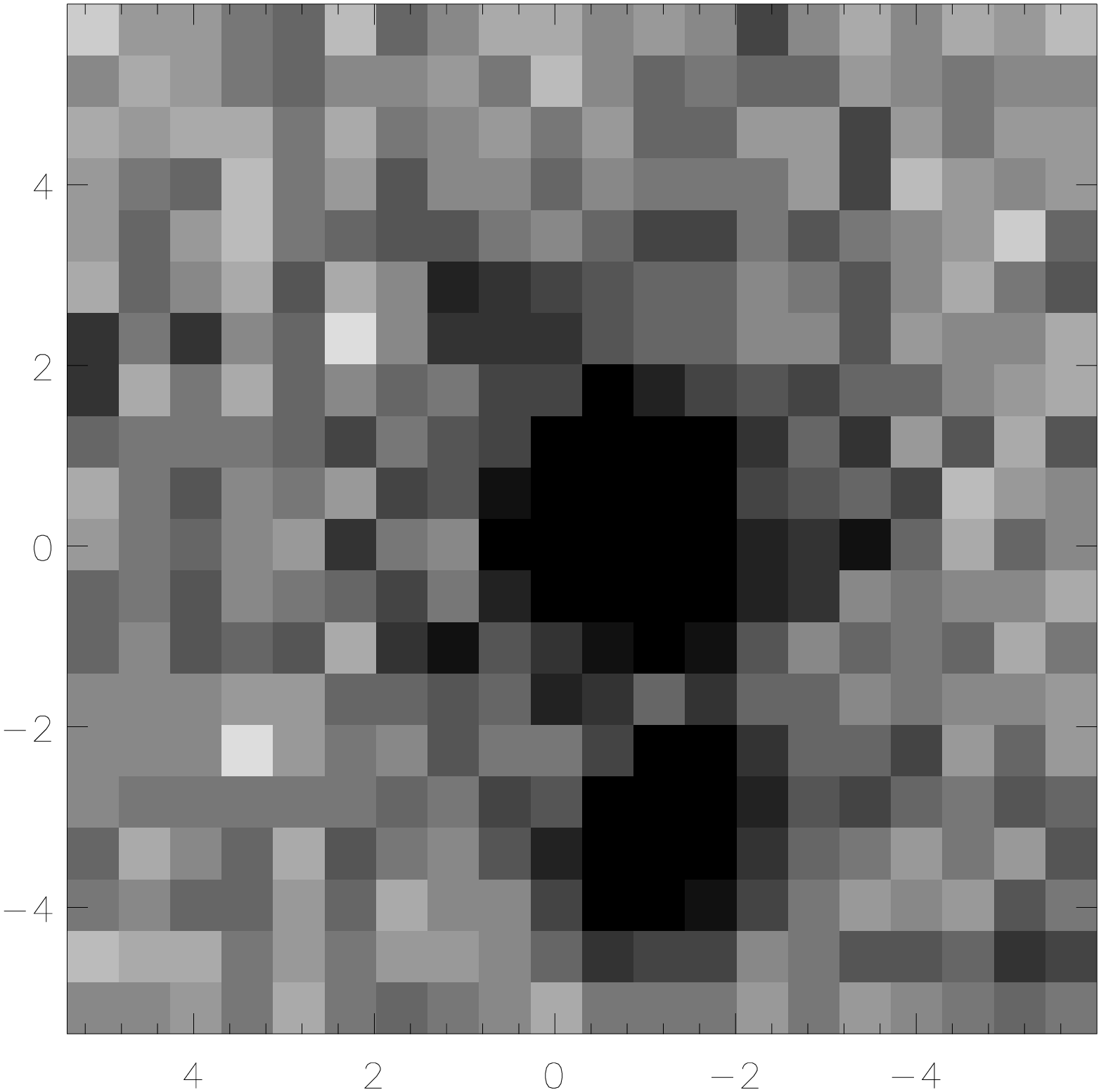,width=0.20\textwidth}\\
\\
\epsfig{file=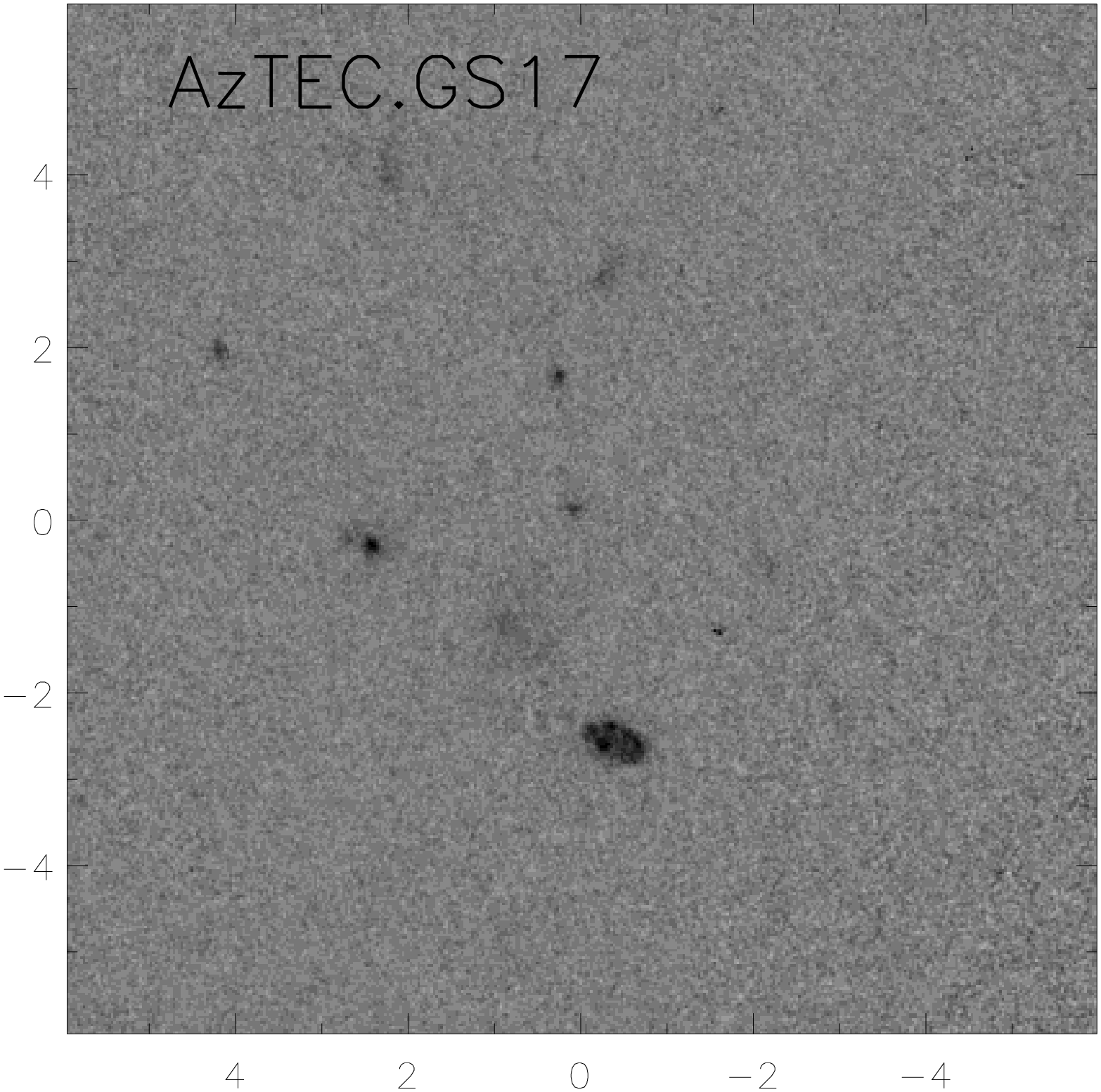,width=0.20\textwidth}&
\epsfig{file=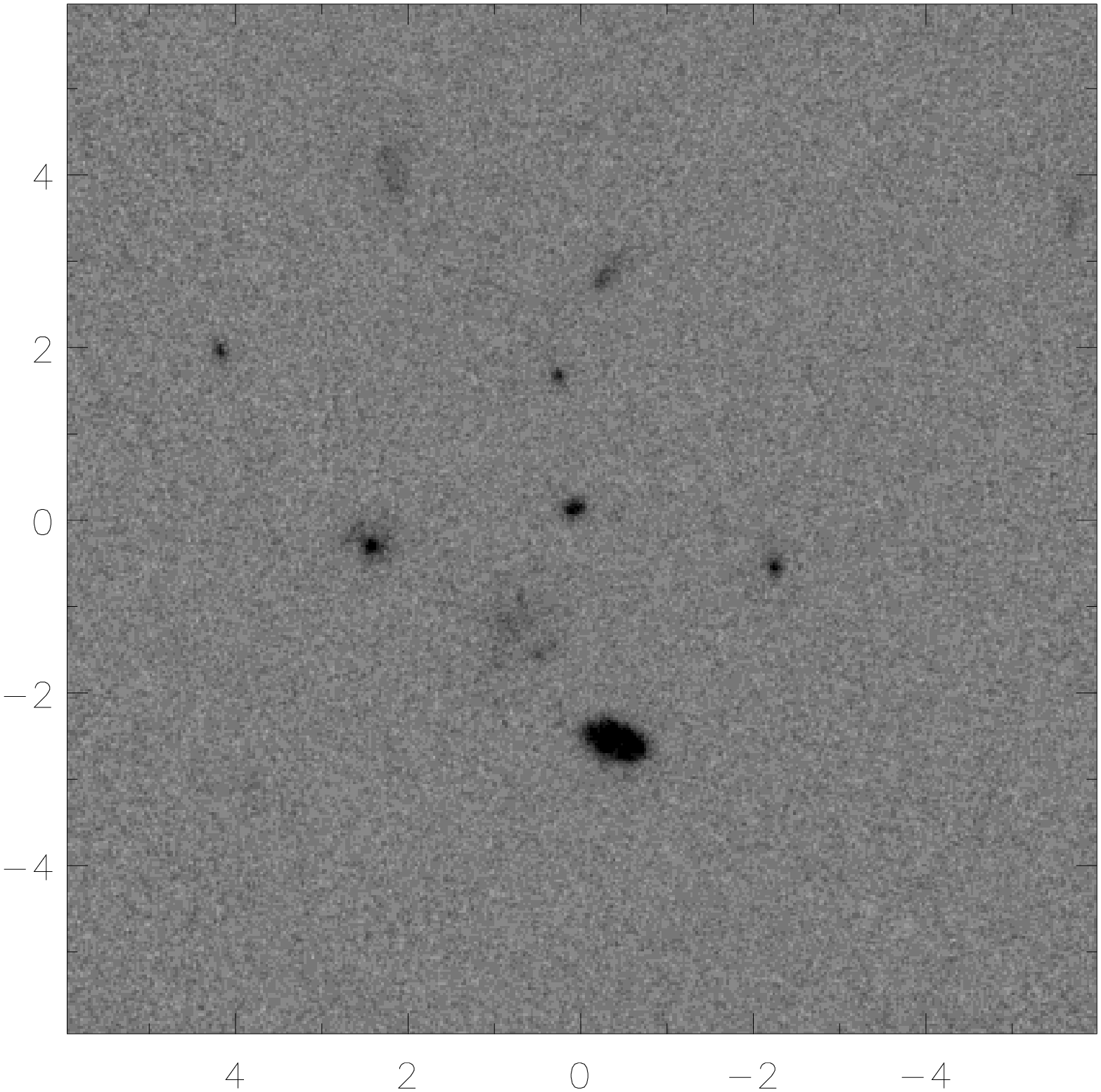,width=0.20\textwidth}&
\epsfig{file=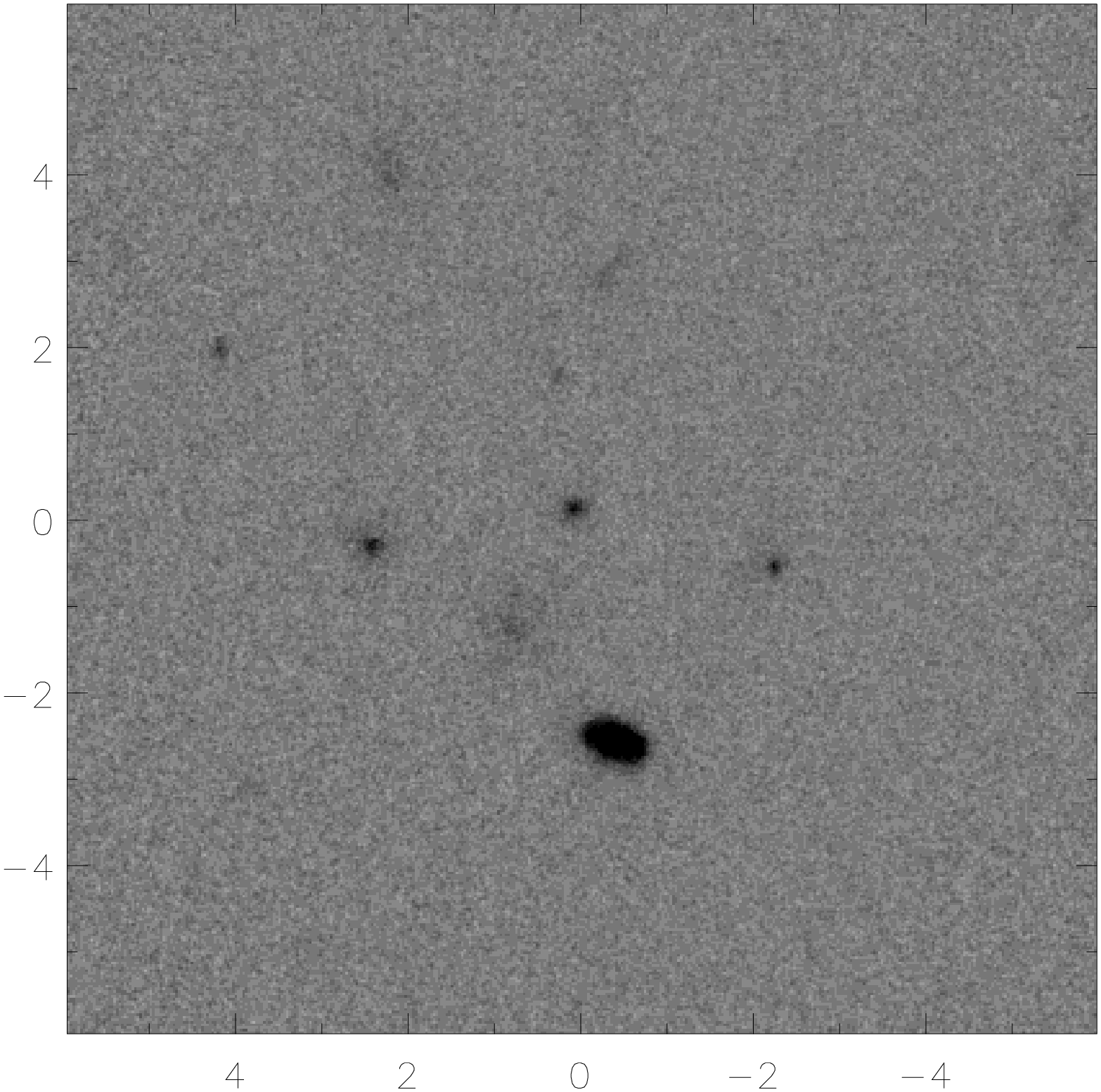,width=0.20\textwidth}&
\epsfig{file=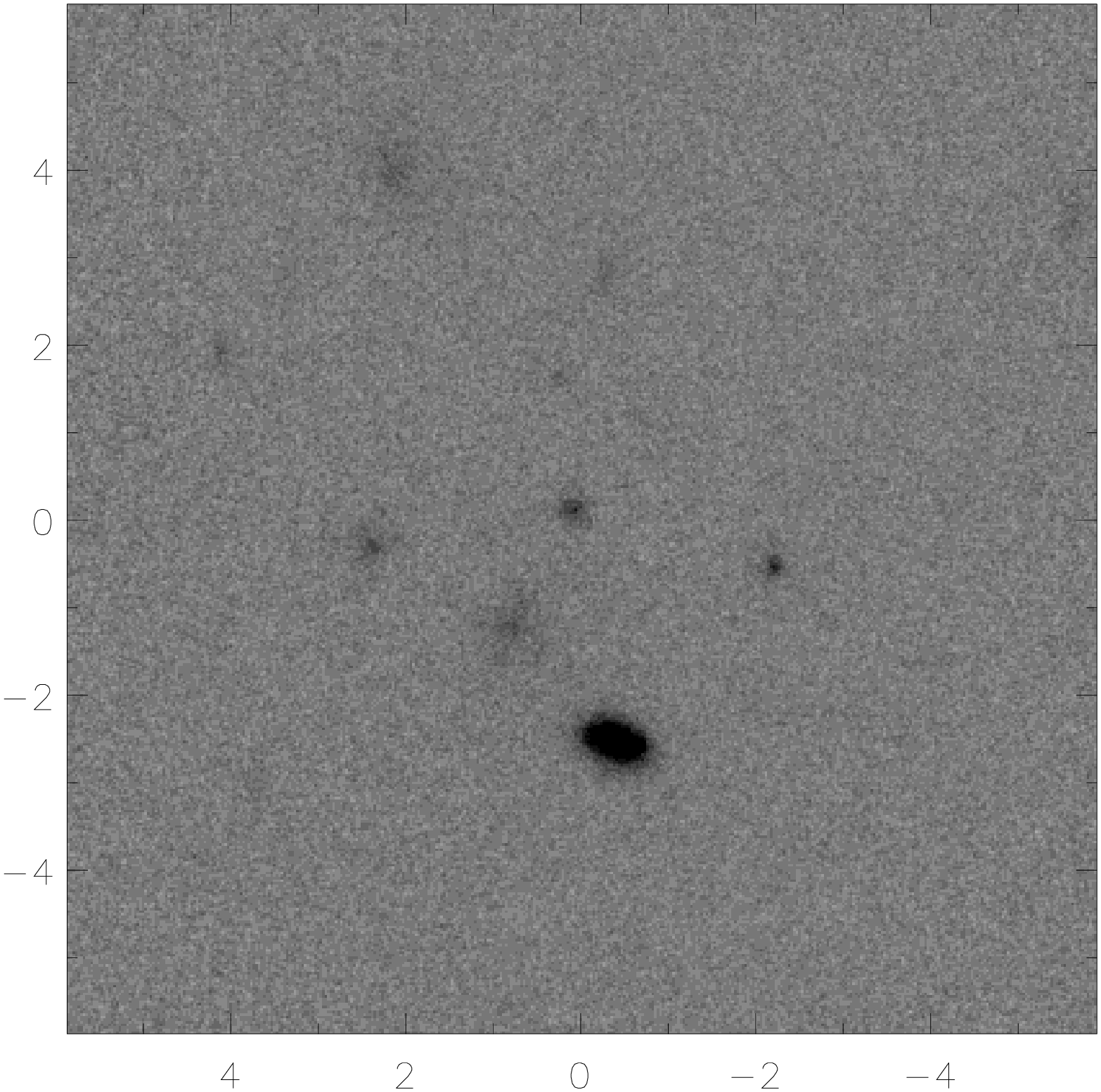,width=0.20\textwidth}\\
\epsfig{file=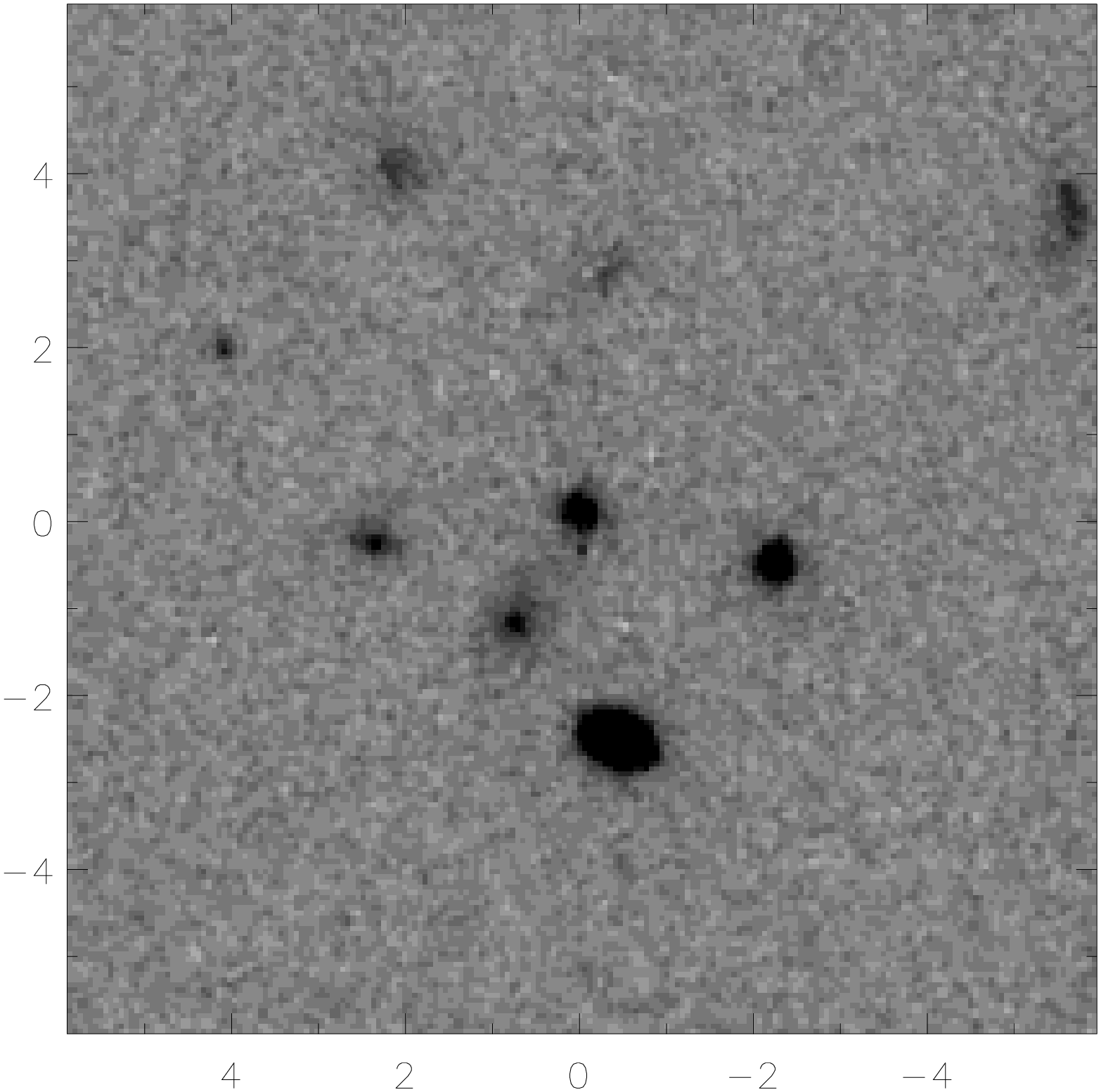,width=0.20\textwidth}&
\epsfig{file=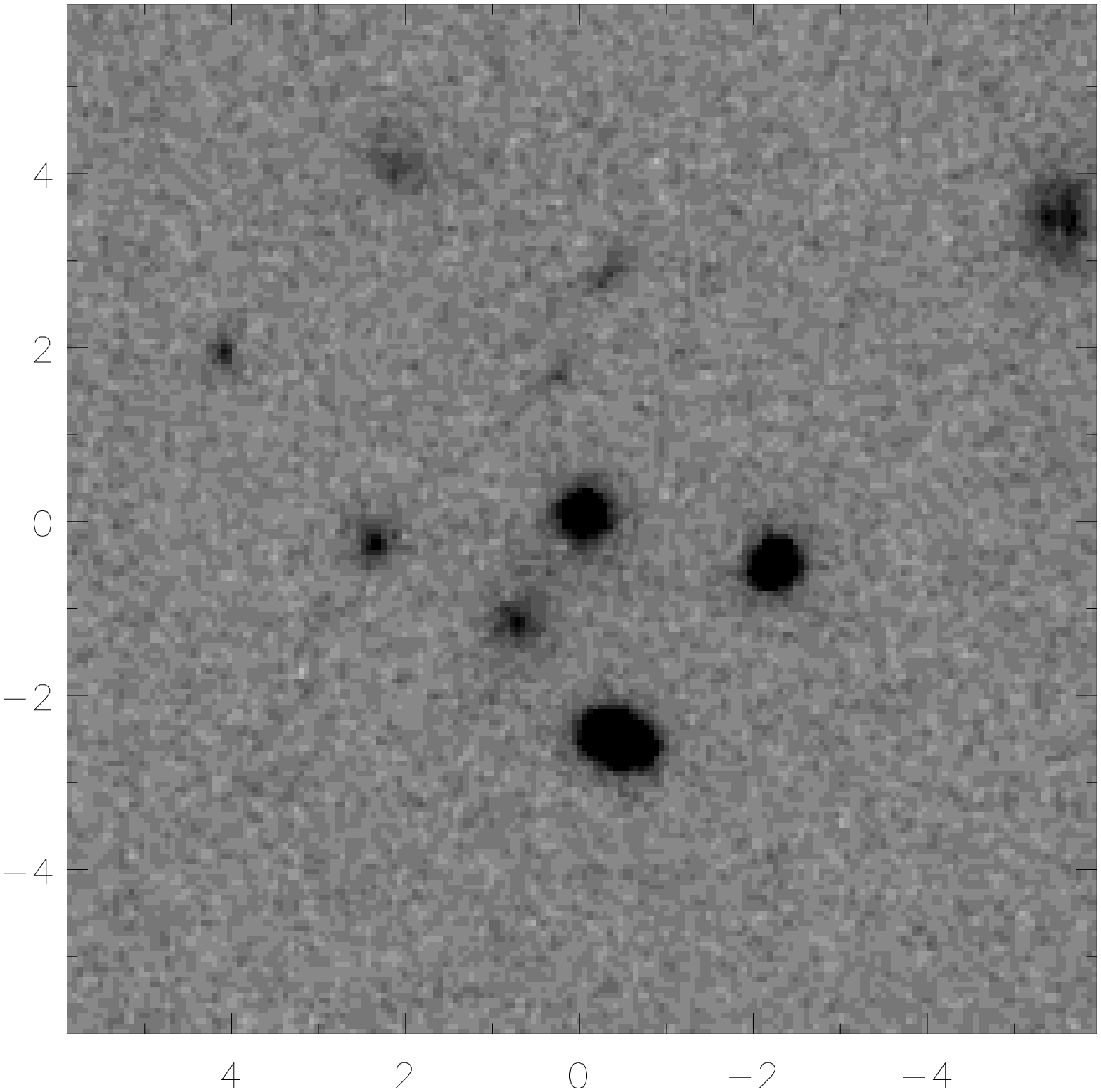,width=0.20\textwidth}&
\epsfig{file=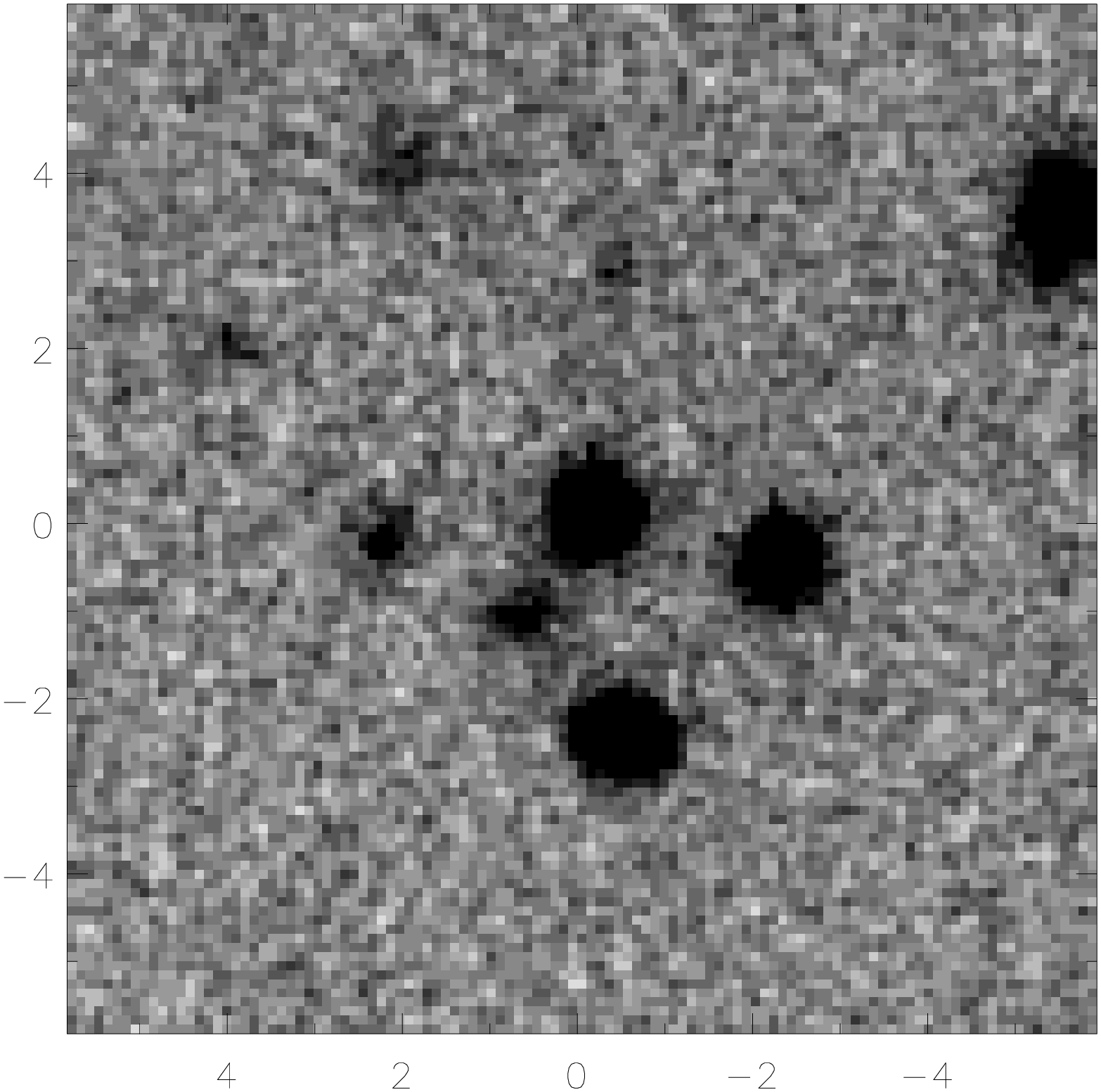,width=0.20\textwidth}&
\epsfig{file=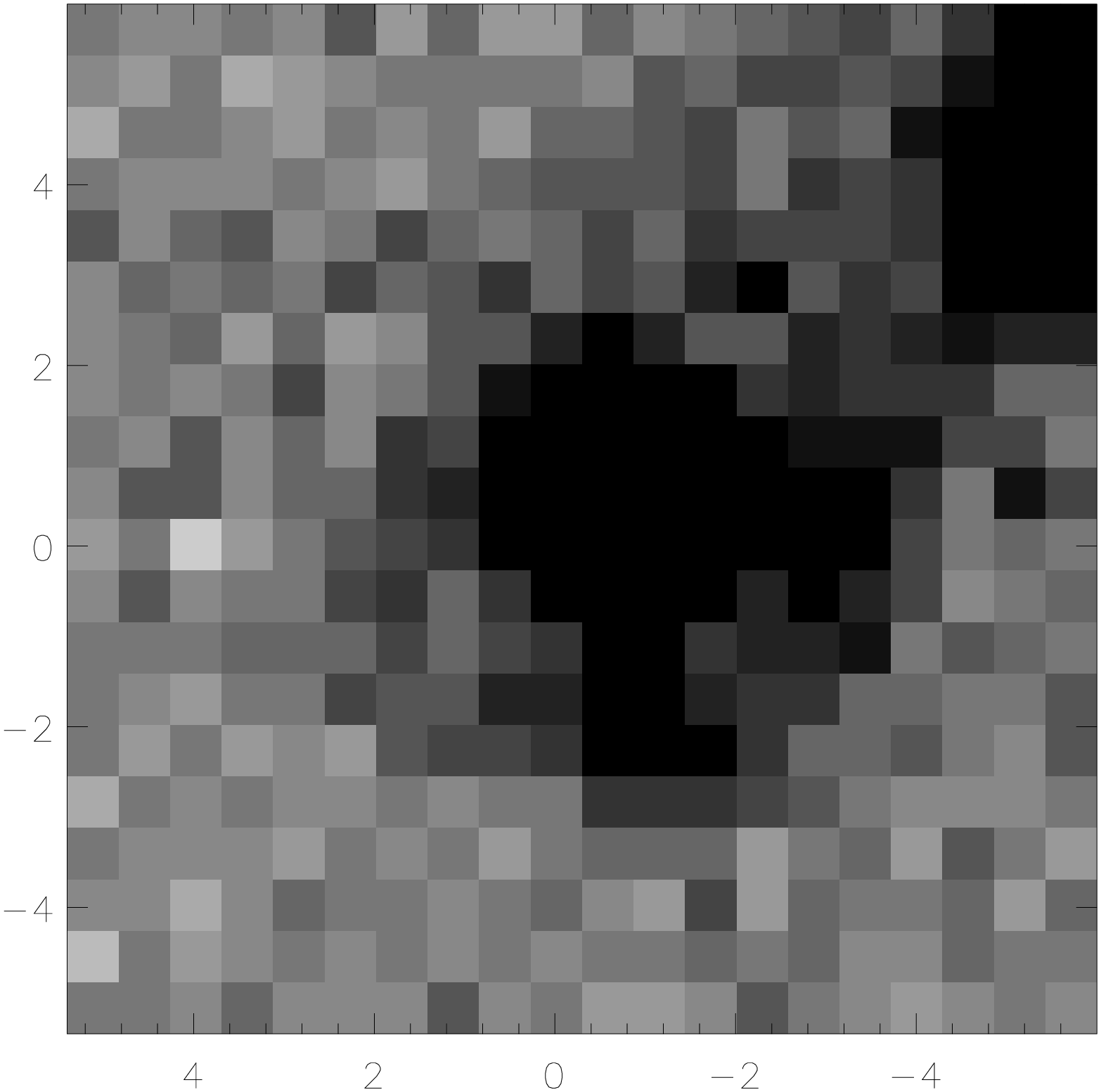,width=0.20\textwidth}\\
\\
\epsfig{file=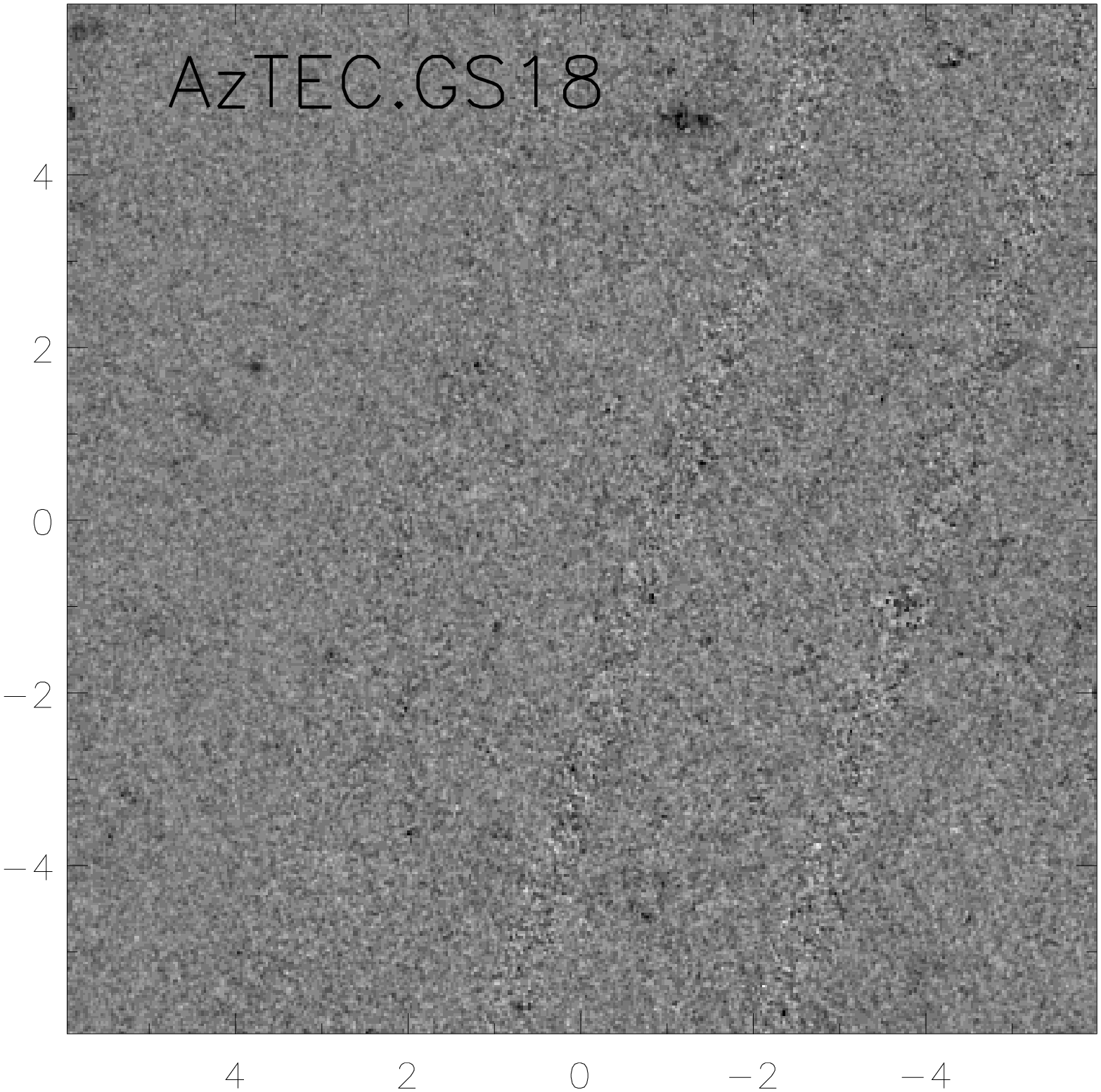,width=0.20\textwidth}&
\epsfig{file=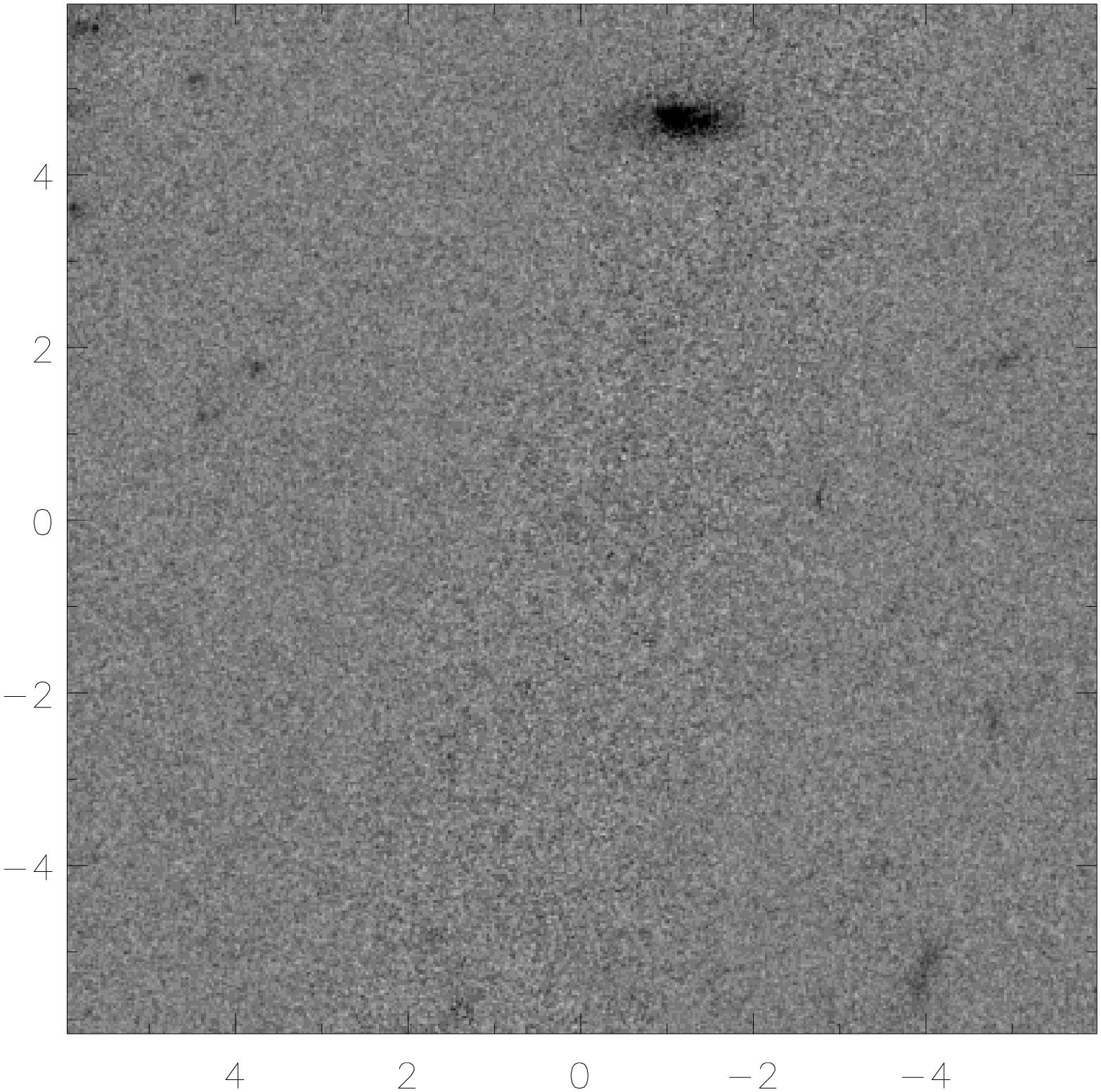,width=0.20\textwidth}&
\epsfig{file=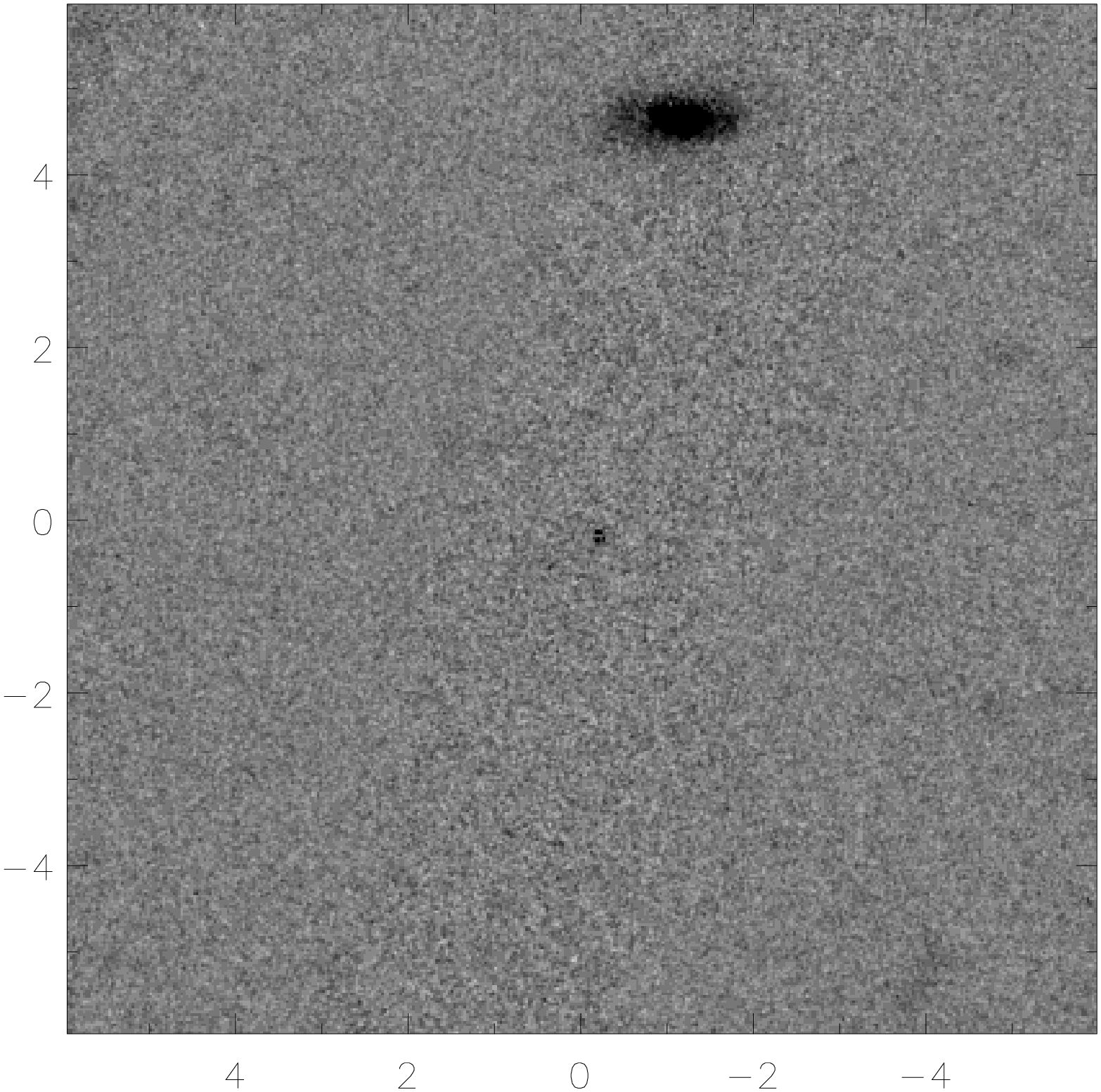,width=0.20\textwidth}&
\epsfig{file=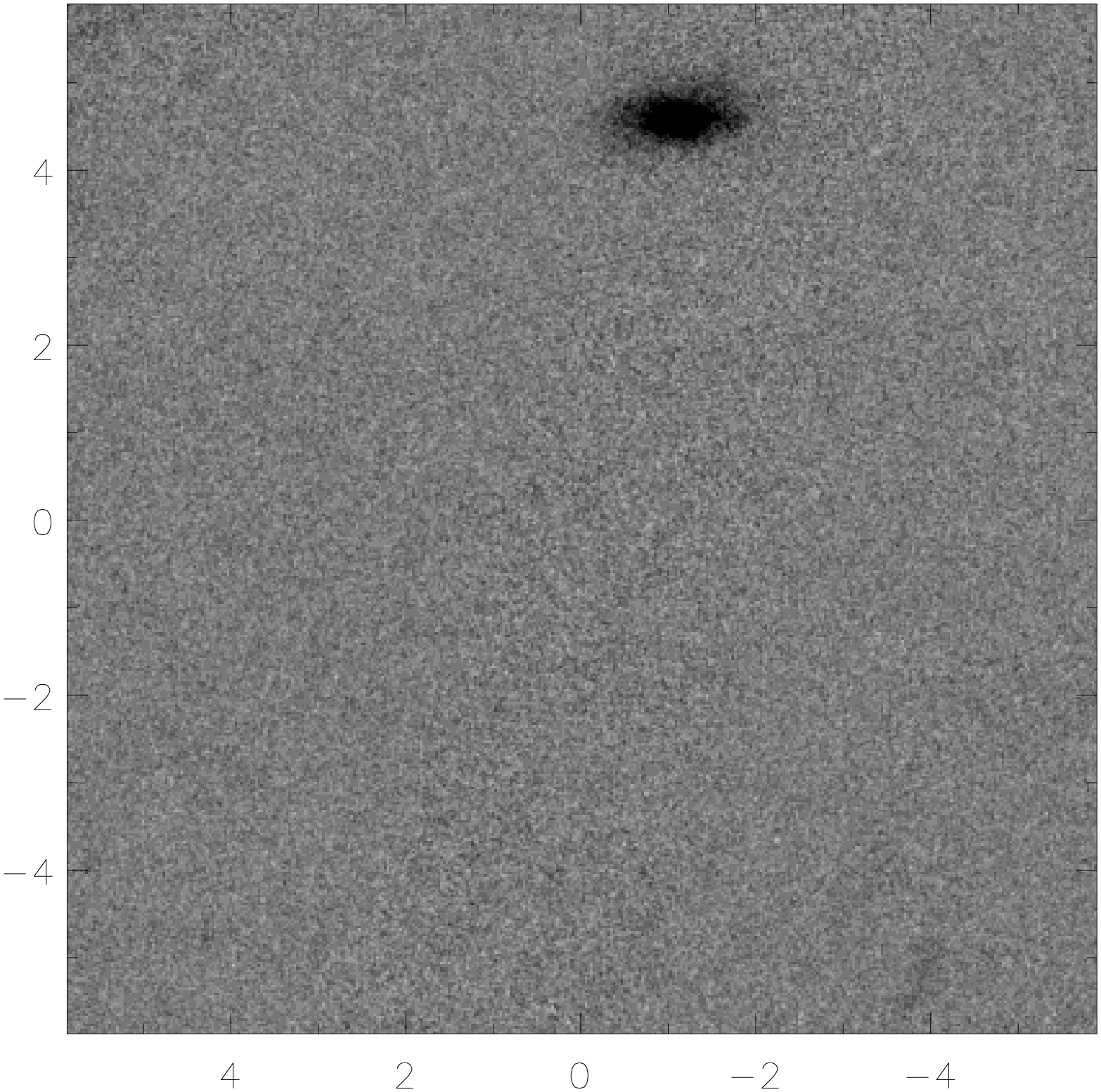,width=0.20\textwidth}\\
\epsfig{file=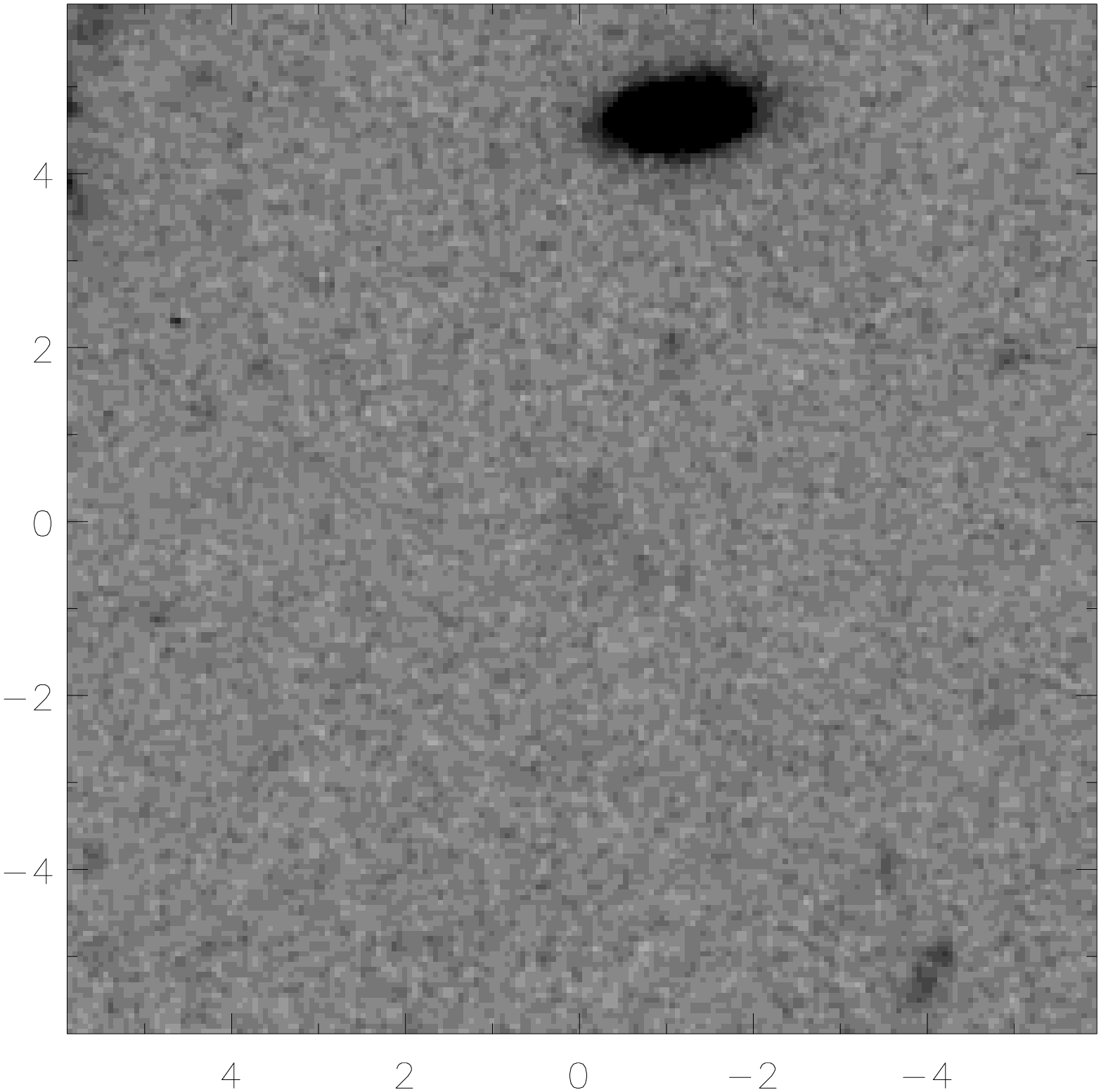,width=0.20\textwidth}&
\epsfig{file=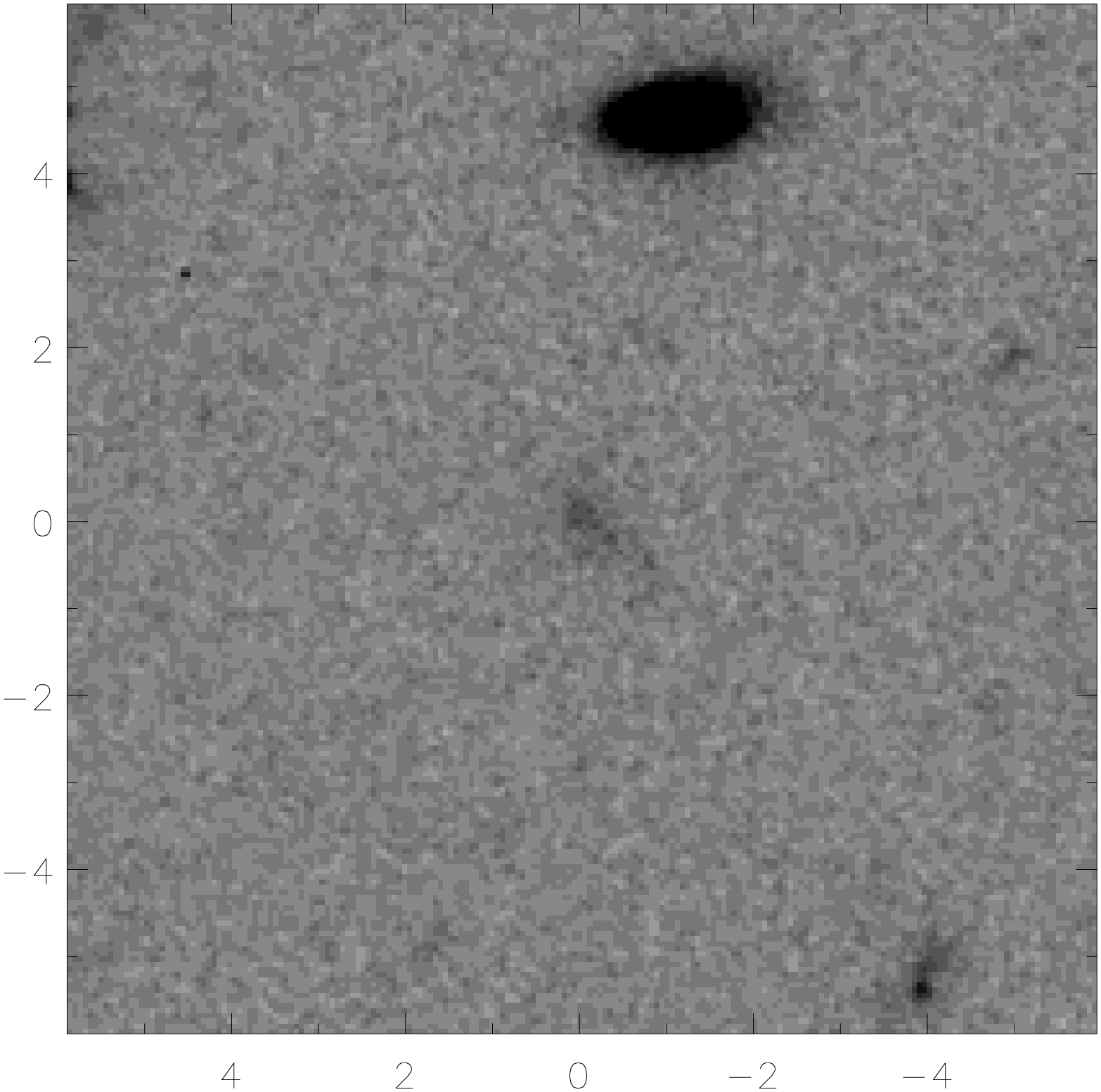,width=0.20\textwidth}&
\epsfig{file=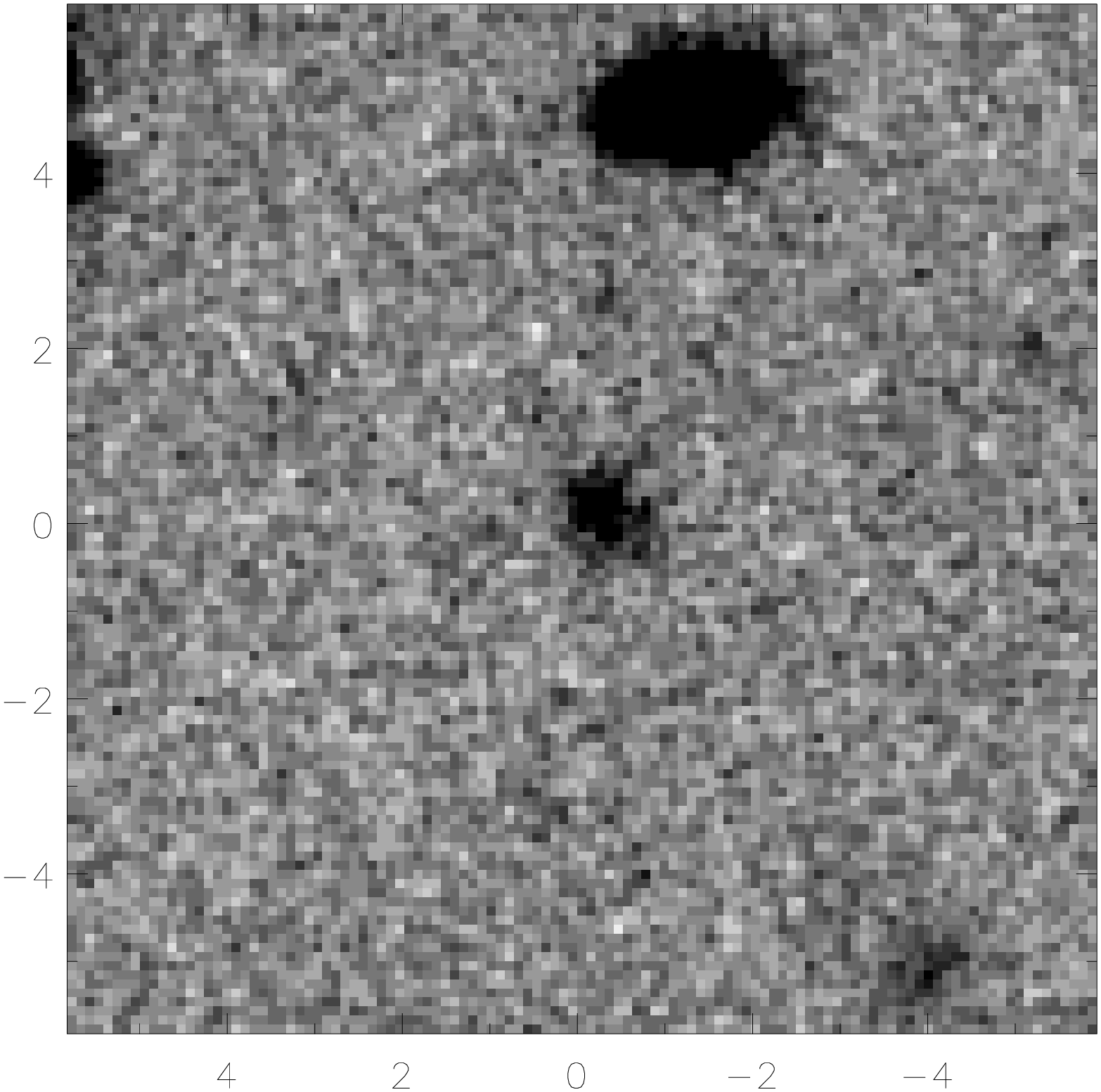,width=0.20\textwidth}&
\epsfig{file=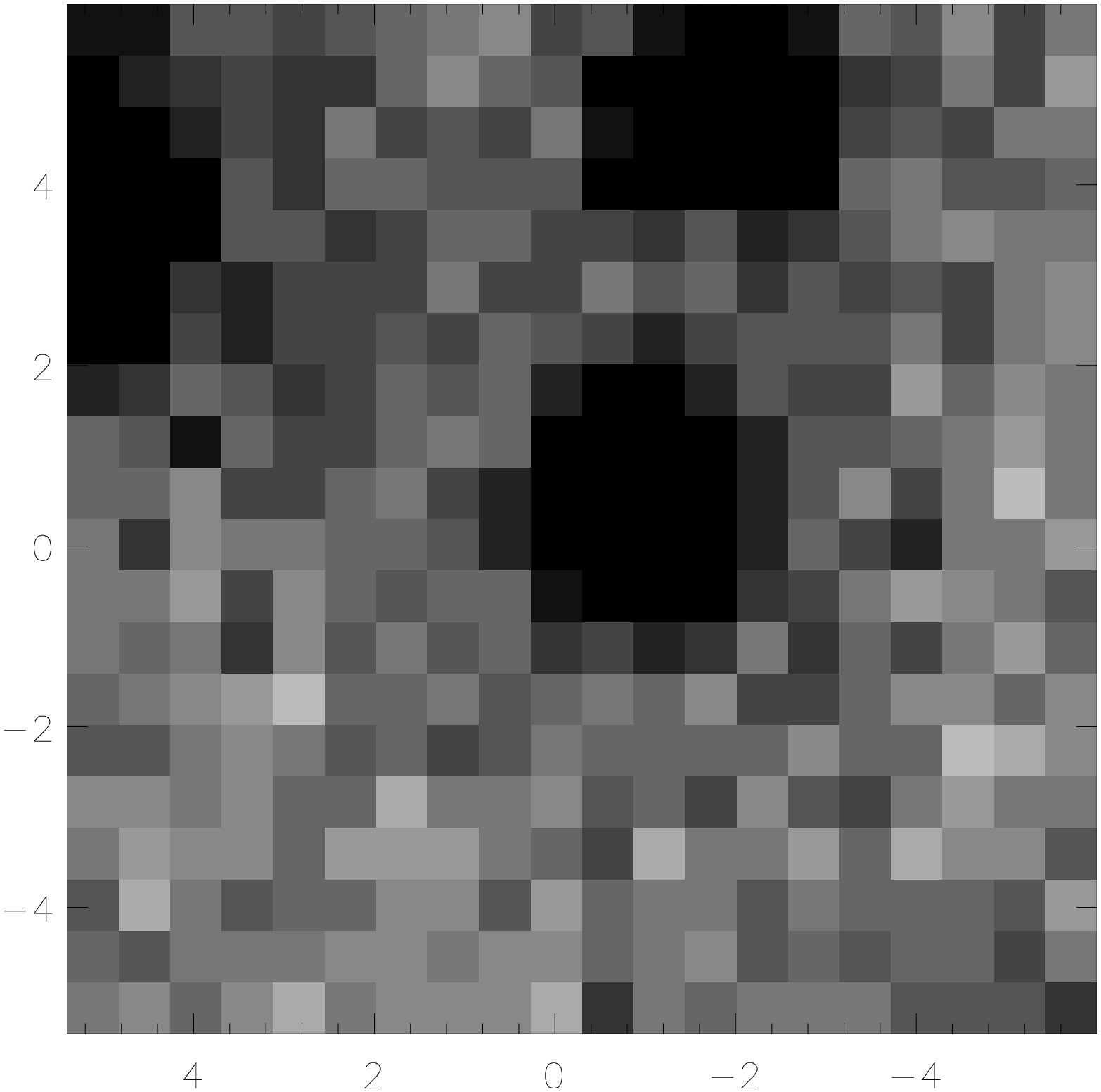,width=0.20\textwidth}\\
\end{tabular}
\addtocounter{figure}{-1}
\caption{- continued}
\vfil}
\end{figure*}
\end{center}


\begin{center}
\begin{figure*}
\vbox to220mm{\vfil
\begin{tabular}{cccccccc}
\epsfig{file=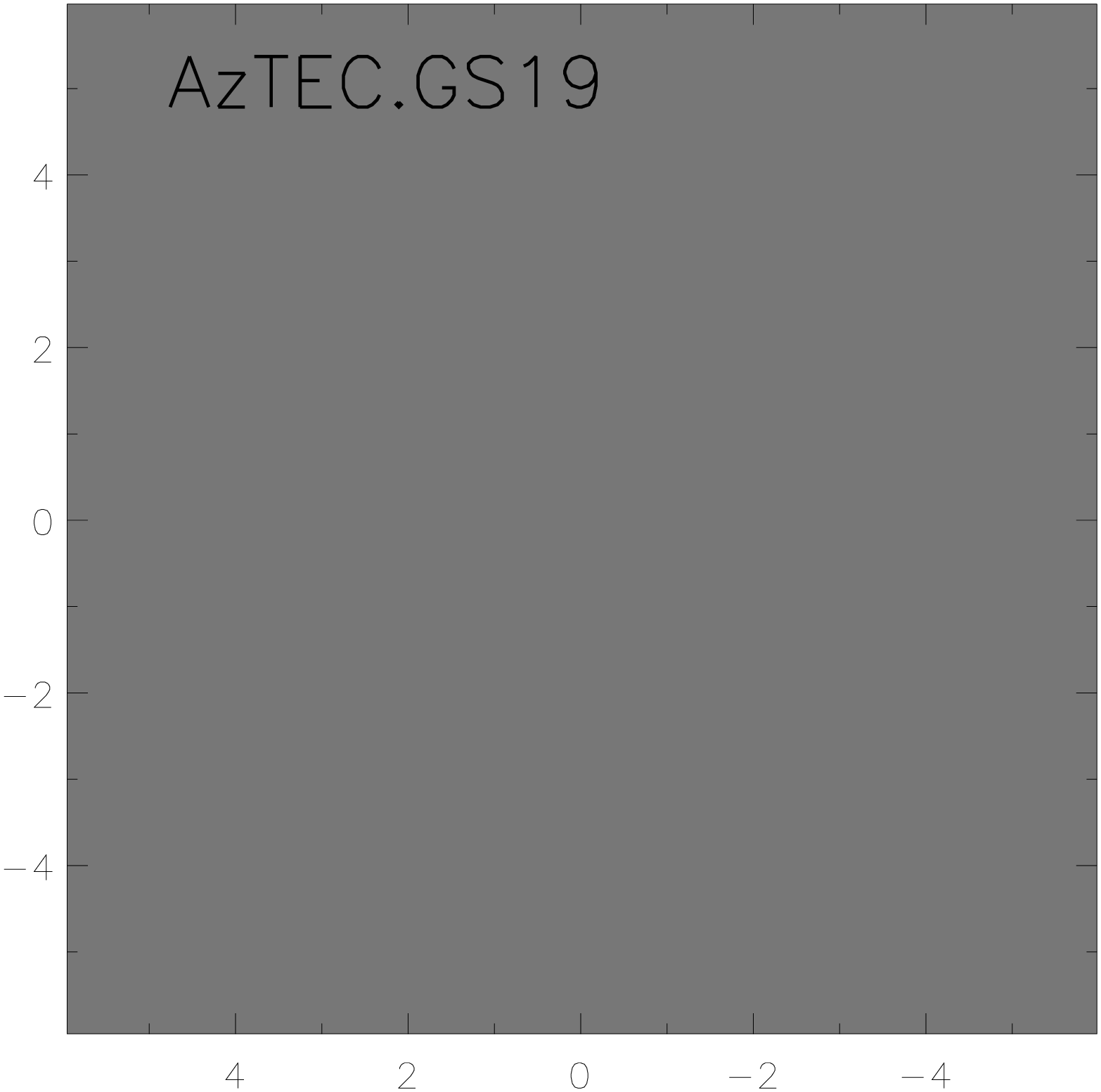,width=0.20\textwidth}&
\epsfig{file=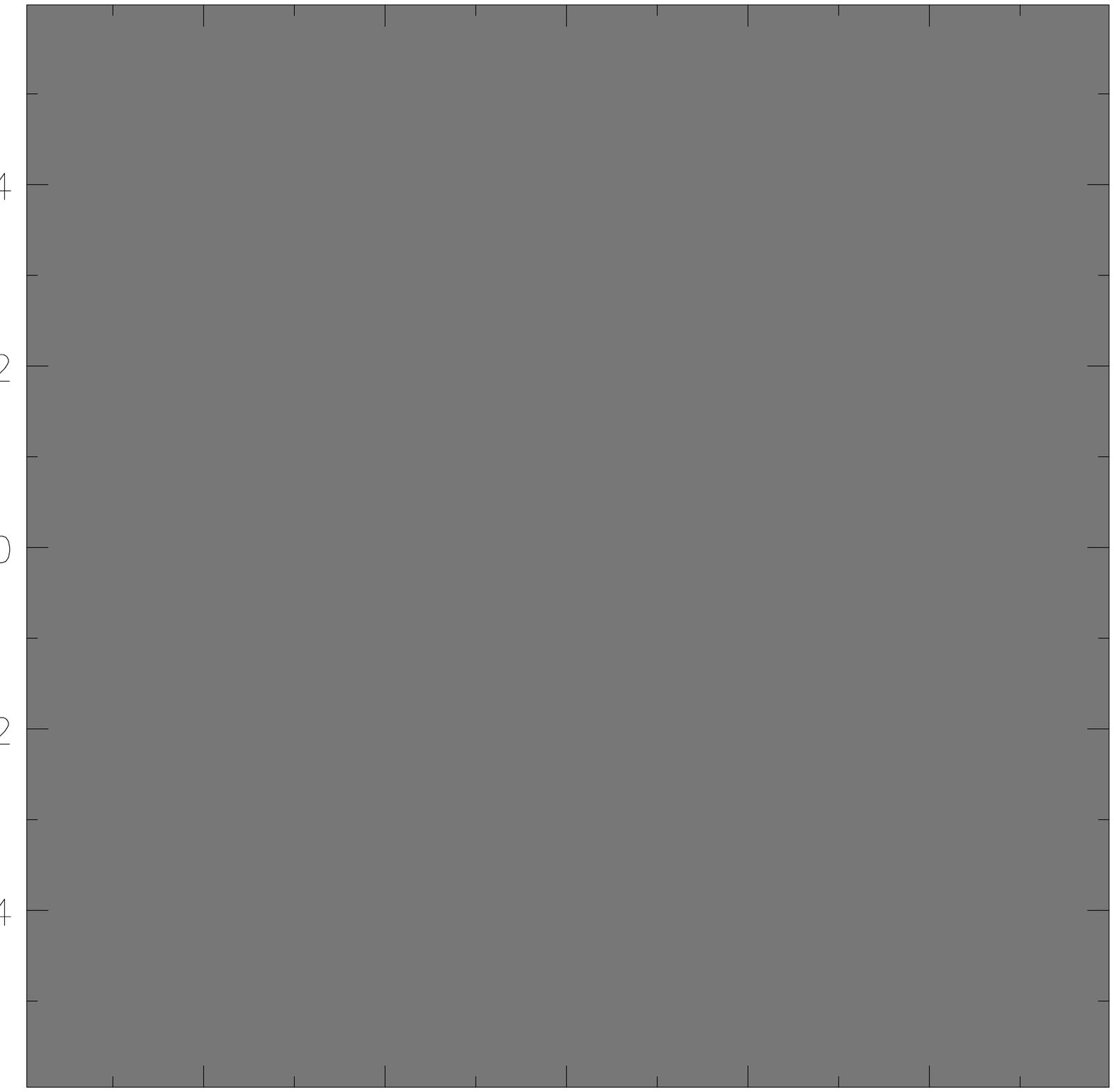,width=0.20\textwidth}&
\epsfig{file=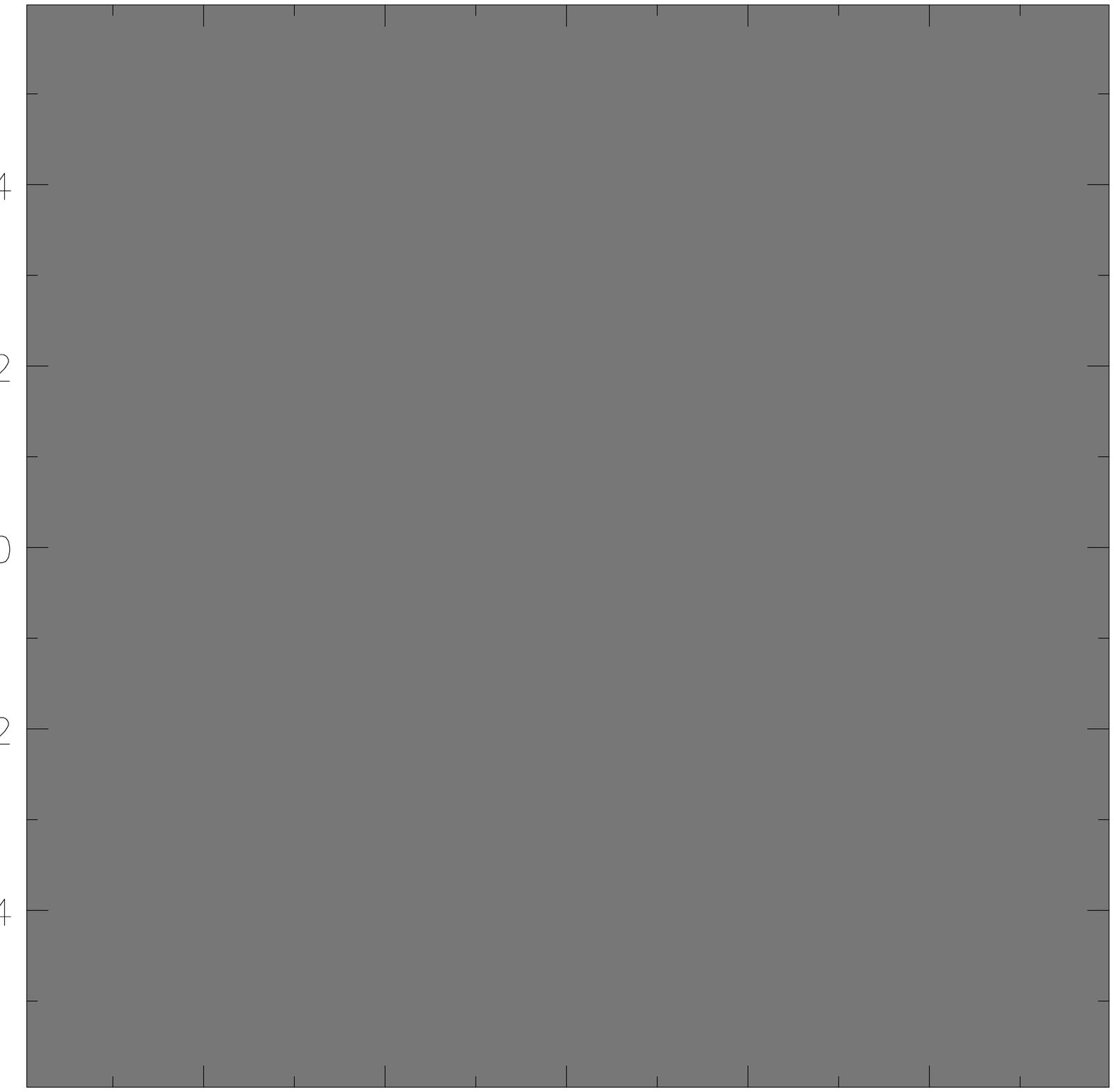,width=0.20\textwidth}&
\epsfig{file=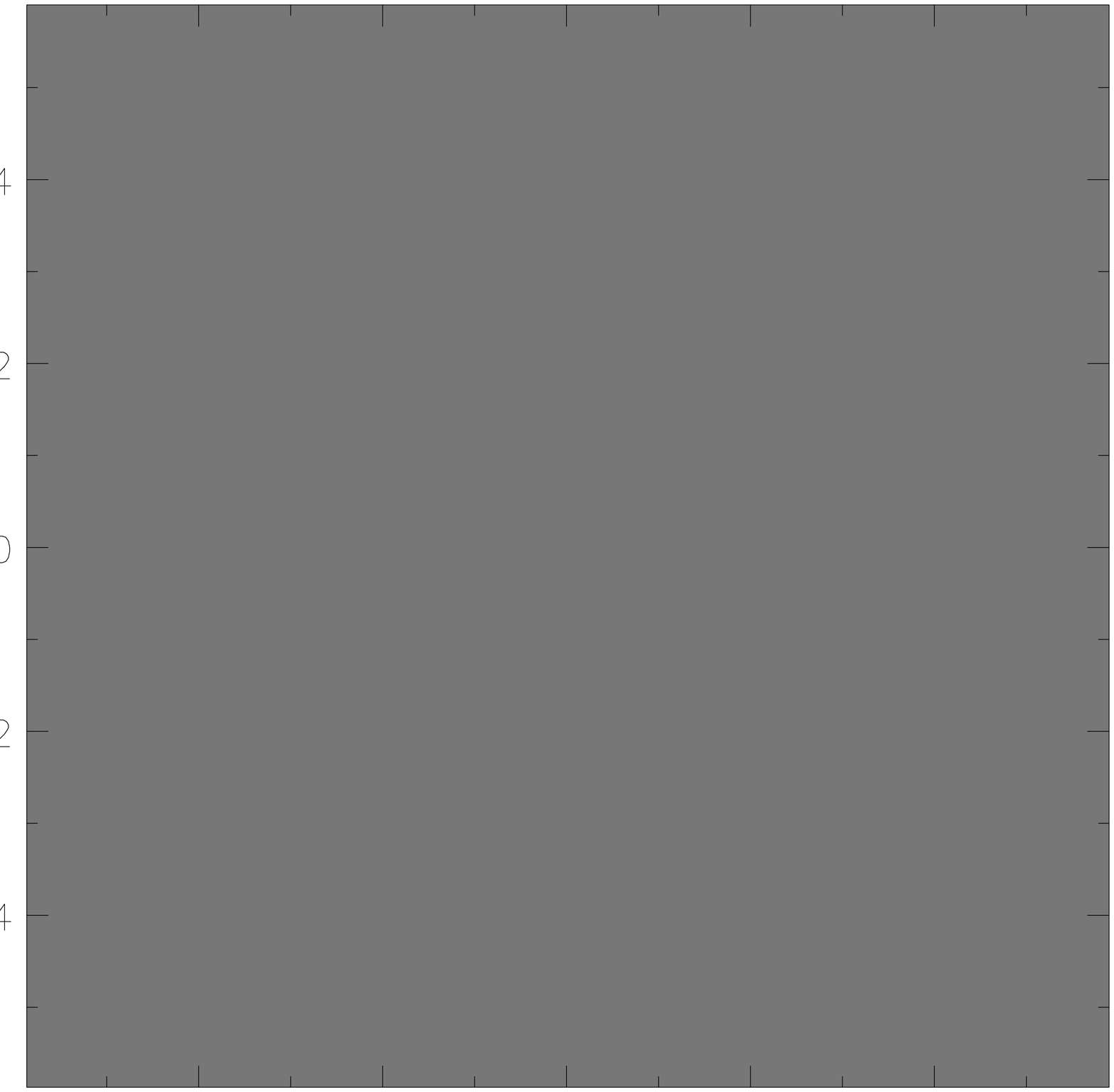,width=0.20\textwidth}\\
\epsfig{file=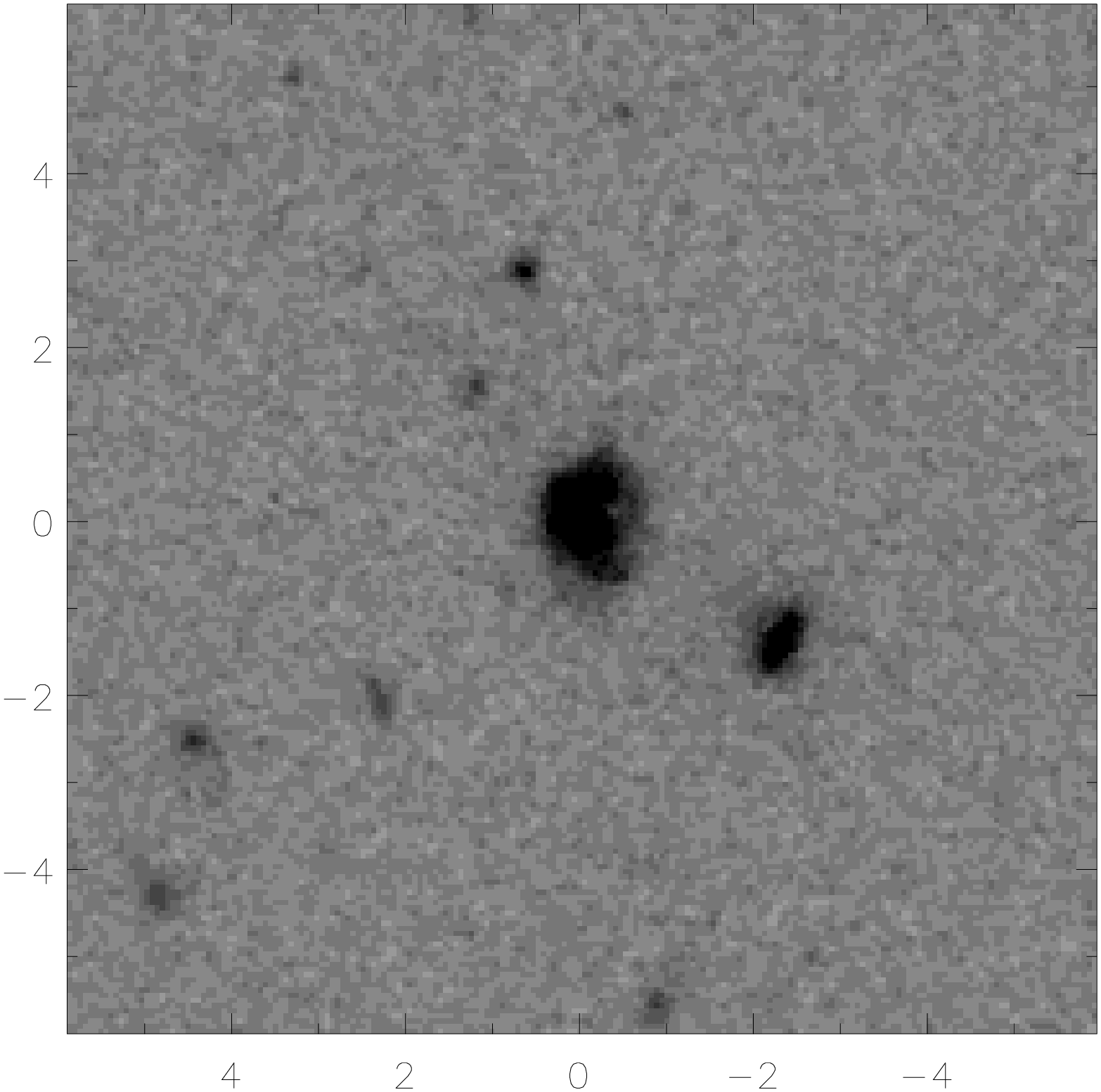,width=0.20\textwidth}&
\epsfig{file=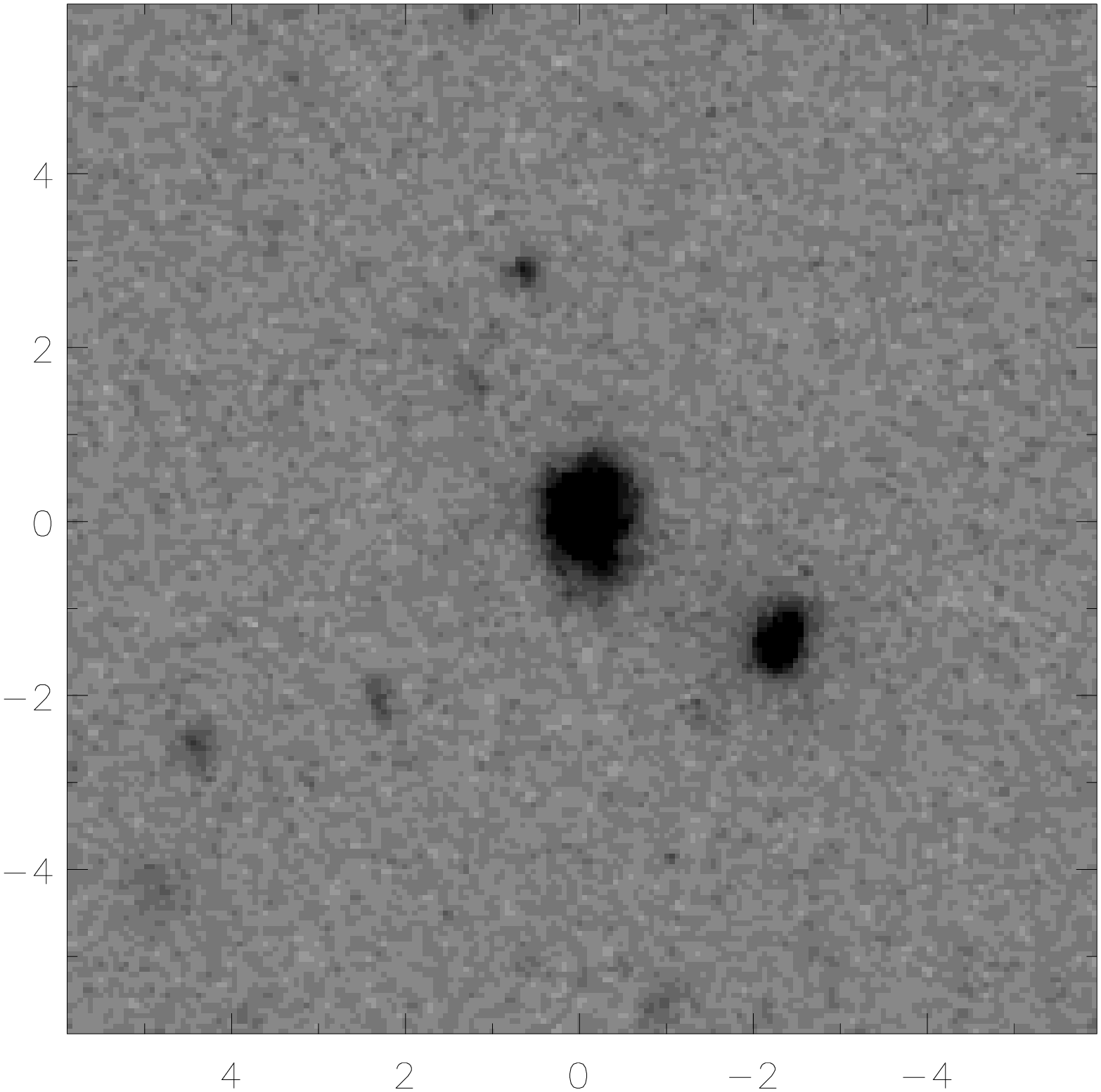,width=0.20\textwidth}&
\epsfig{file=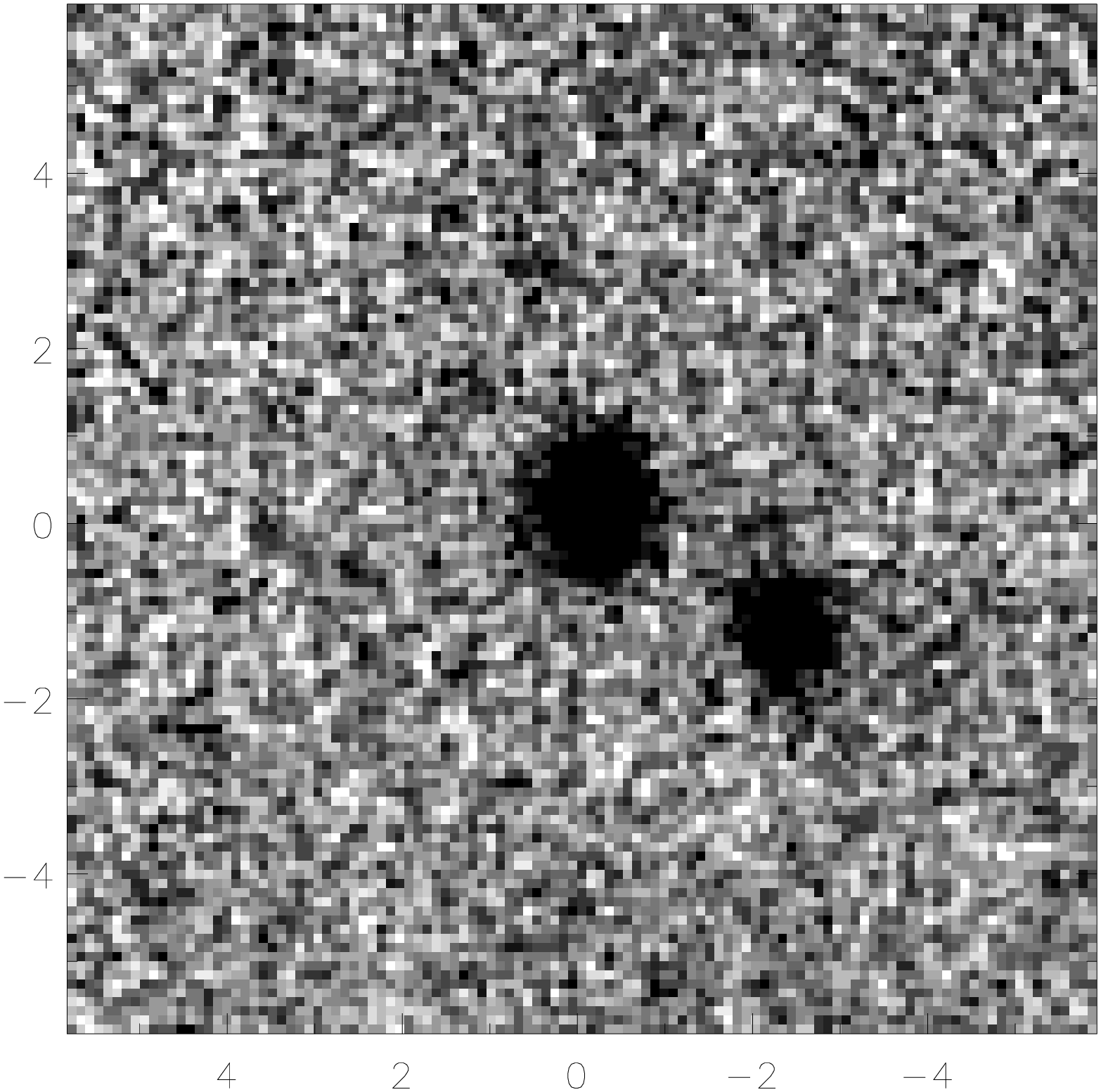,width=0.20\textwidth}&
\epsfig{file=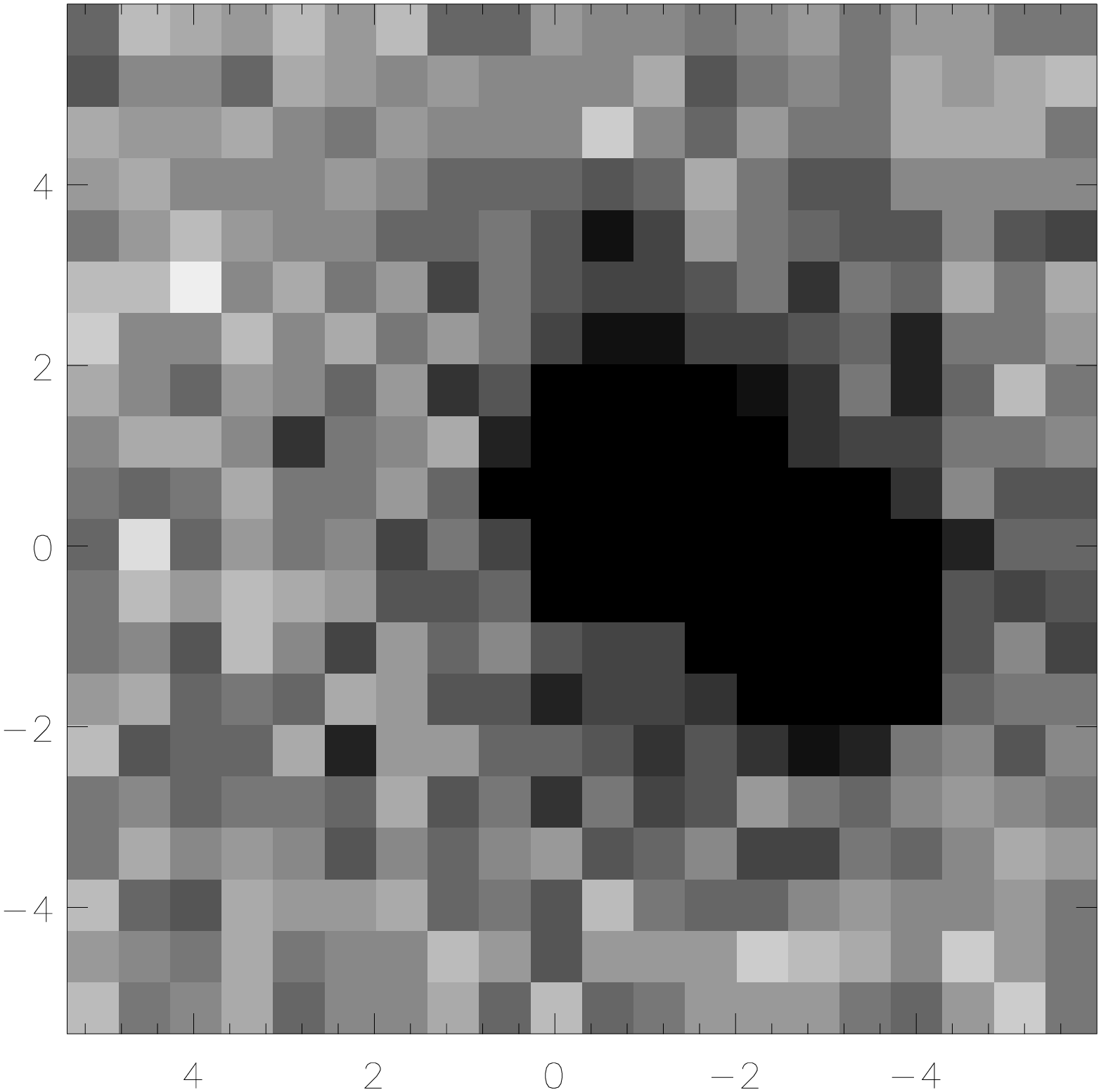,width=0.20\textwidth}\\
\\
\epsfig{file=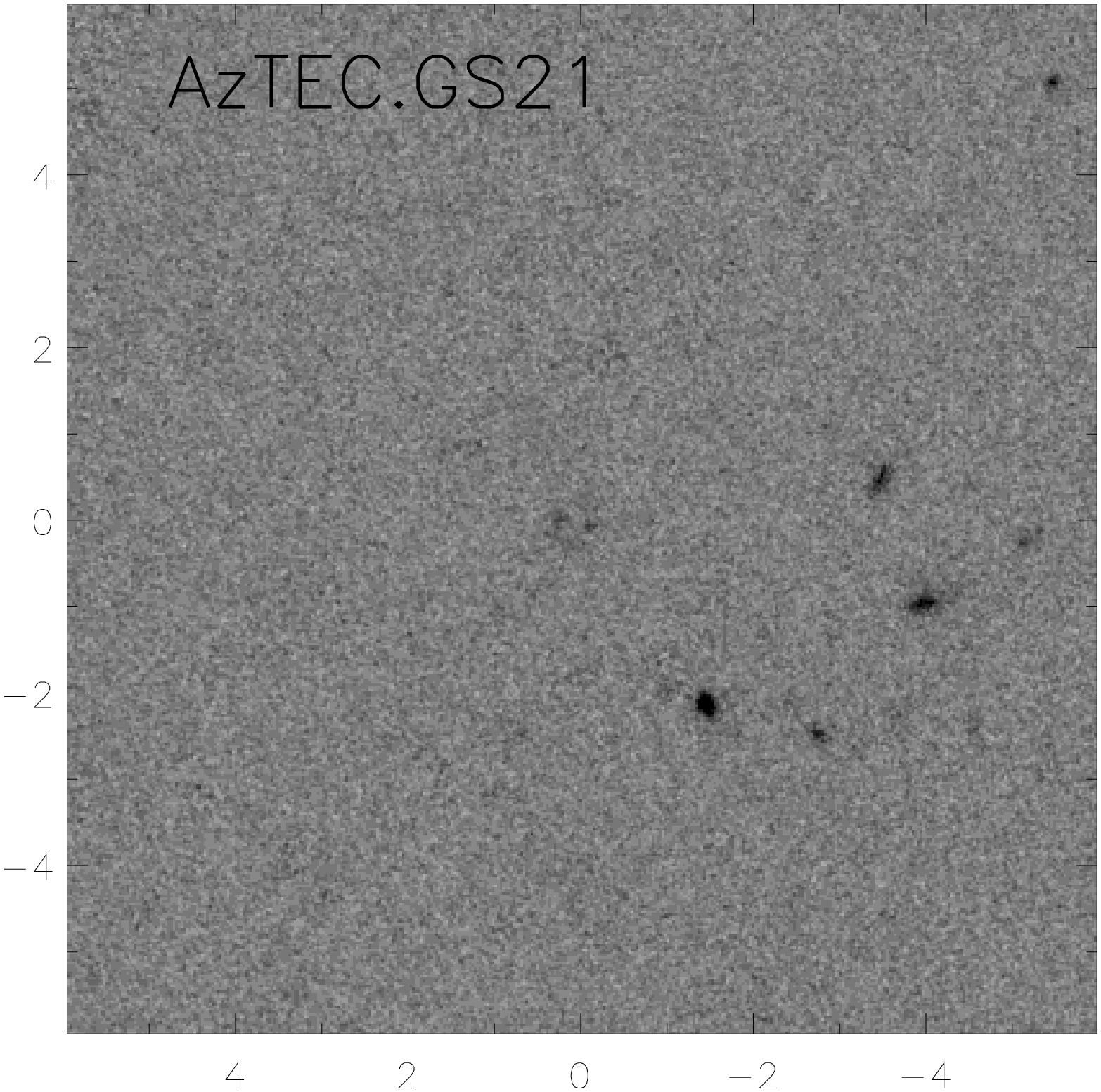,width=0.20\textwidth}&
\epsfig{file=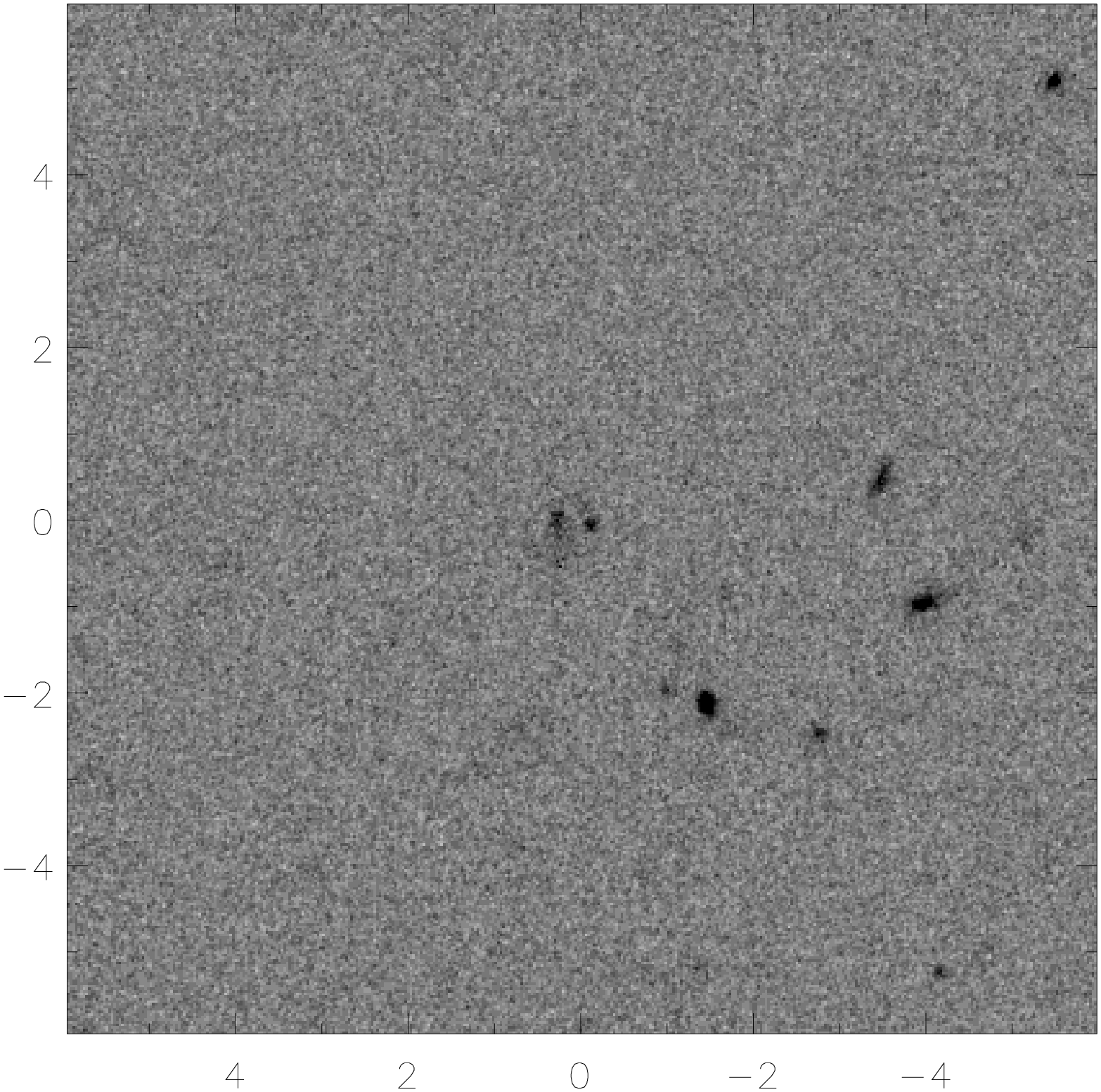,width=0.20\textwidth}&
\epsfig{file=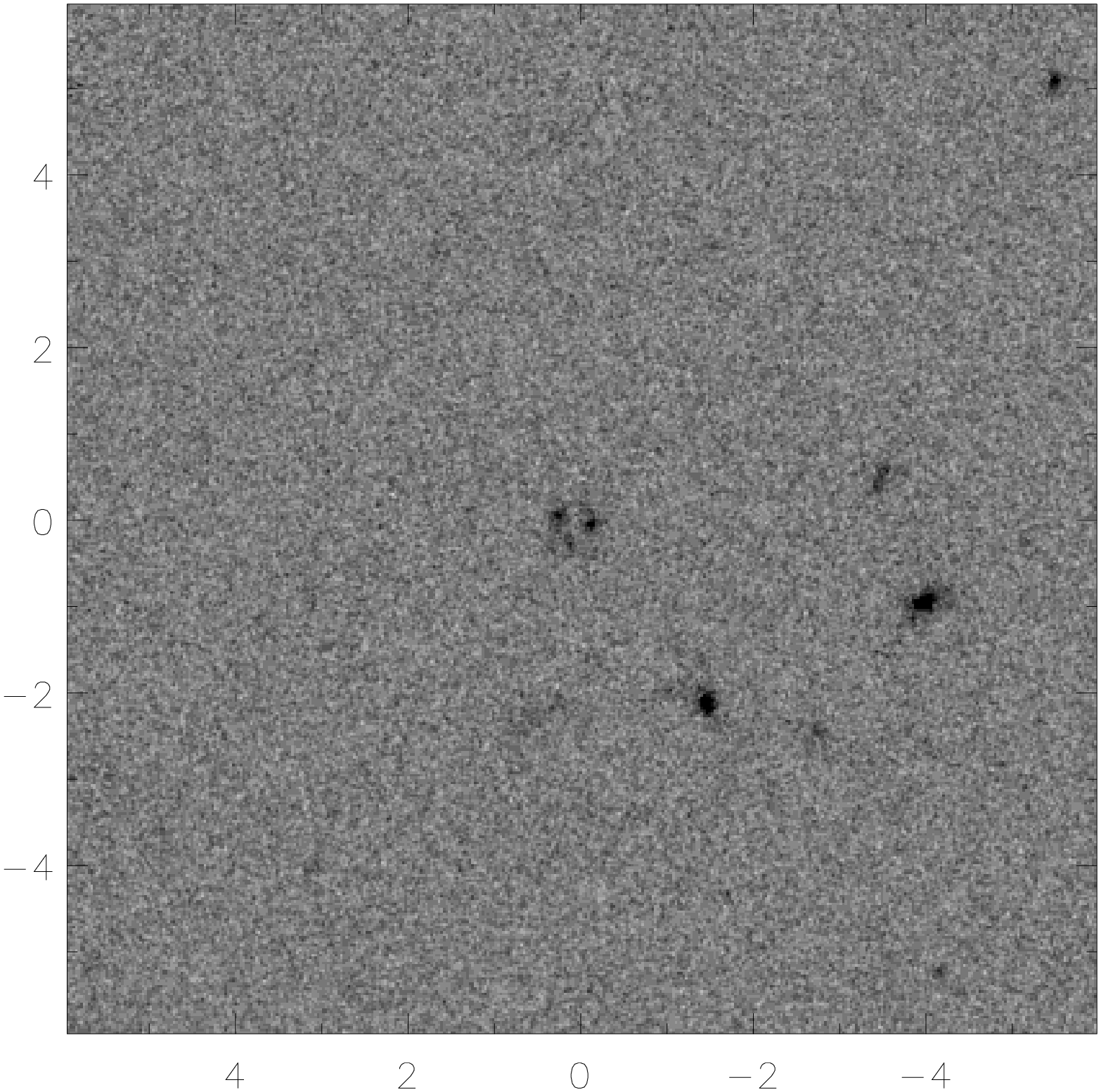,width=0.20\textwidth}&
\epsfig{file=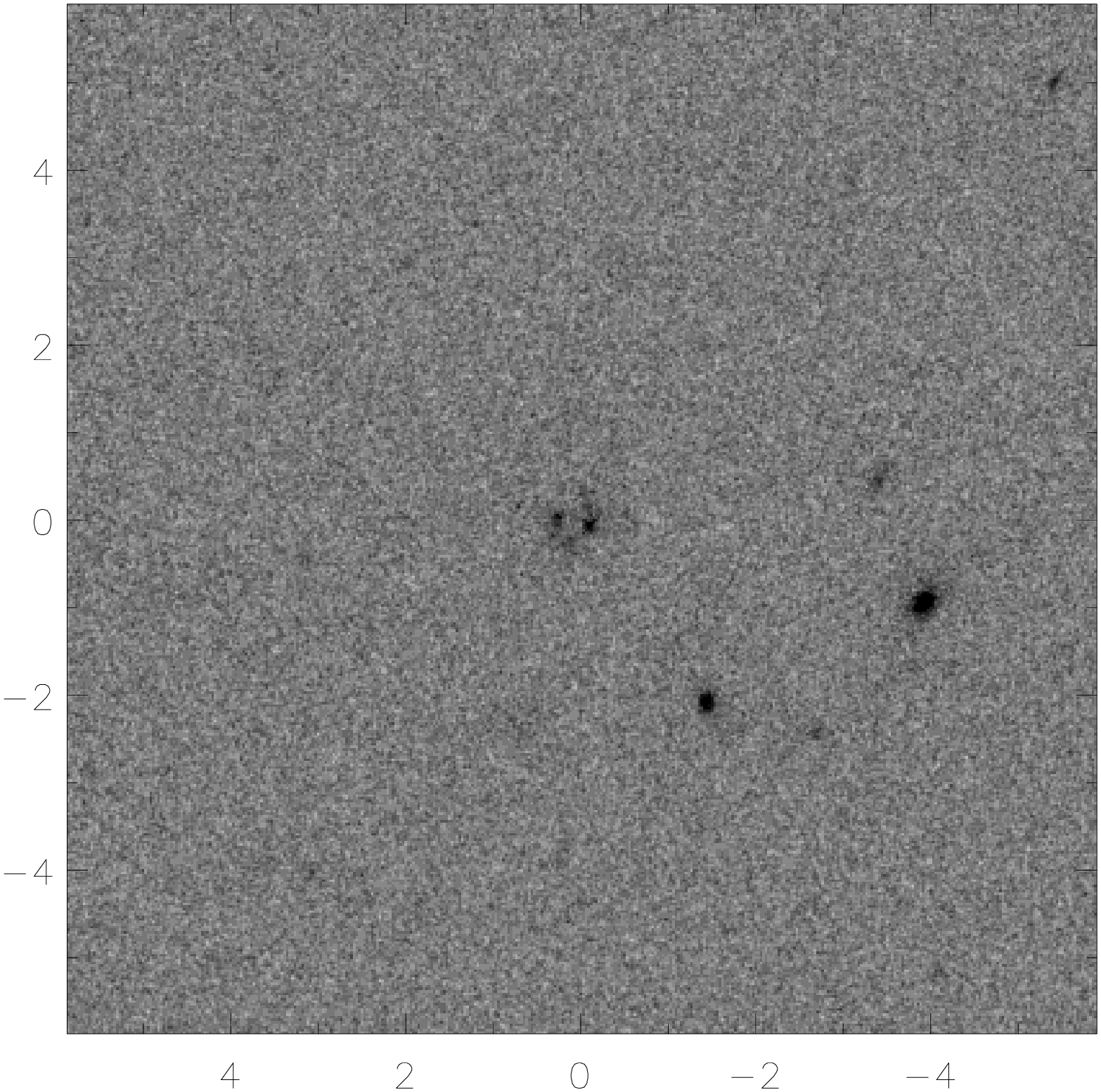,width=0.20\textwidth}\\
\epsfig{file=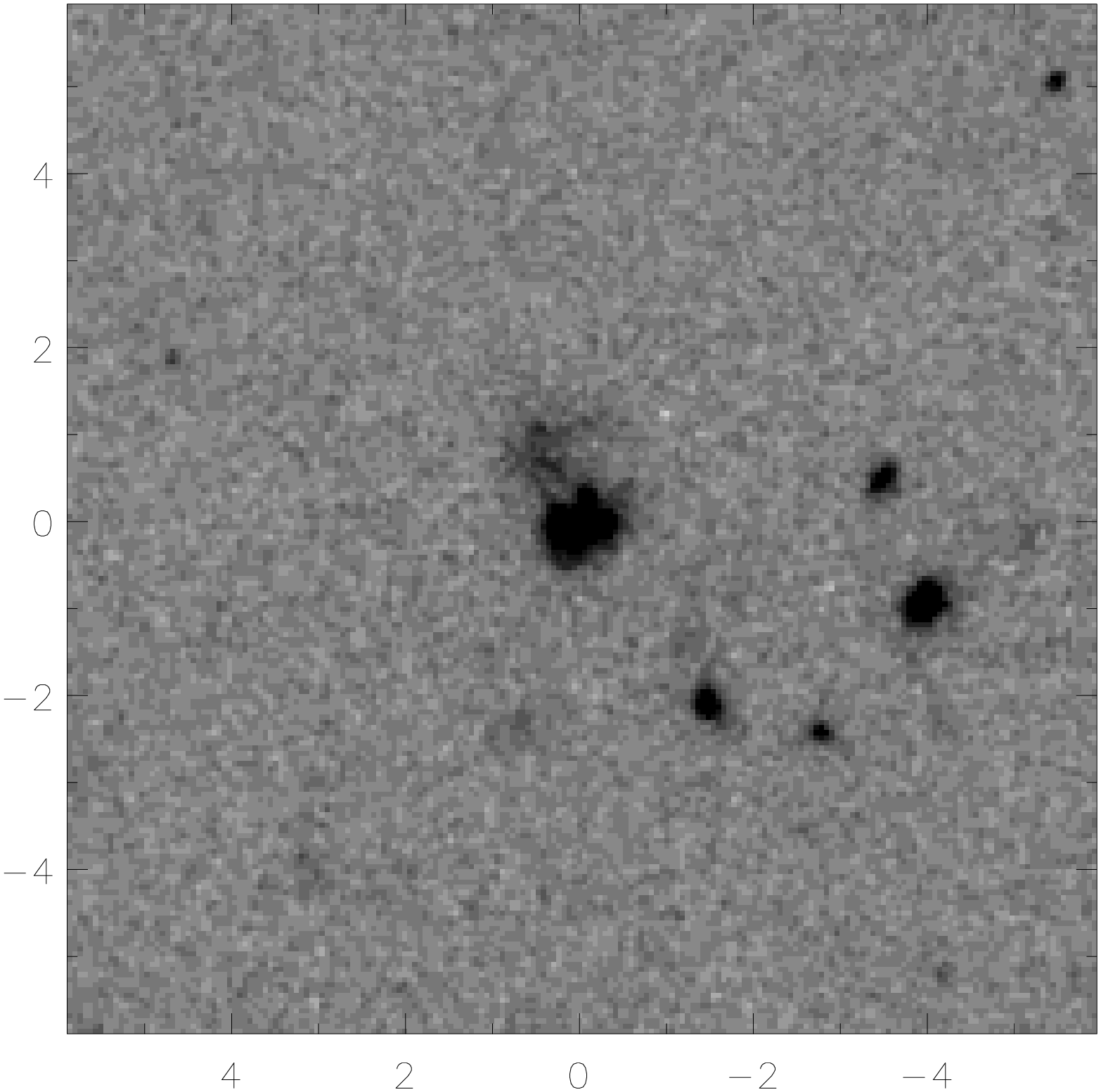,width=0.20\textwidth}&
\epsfig{file=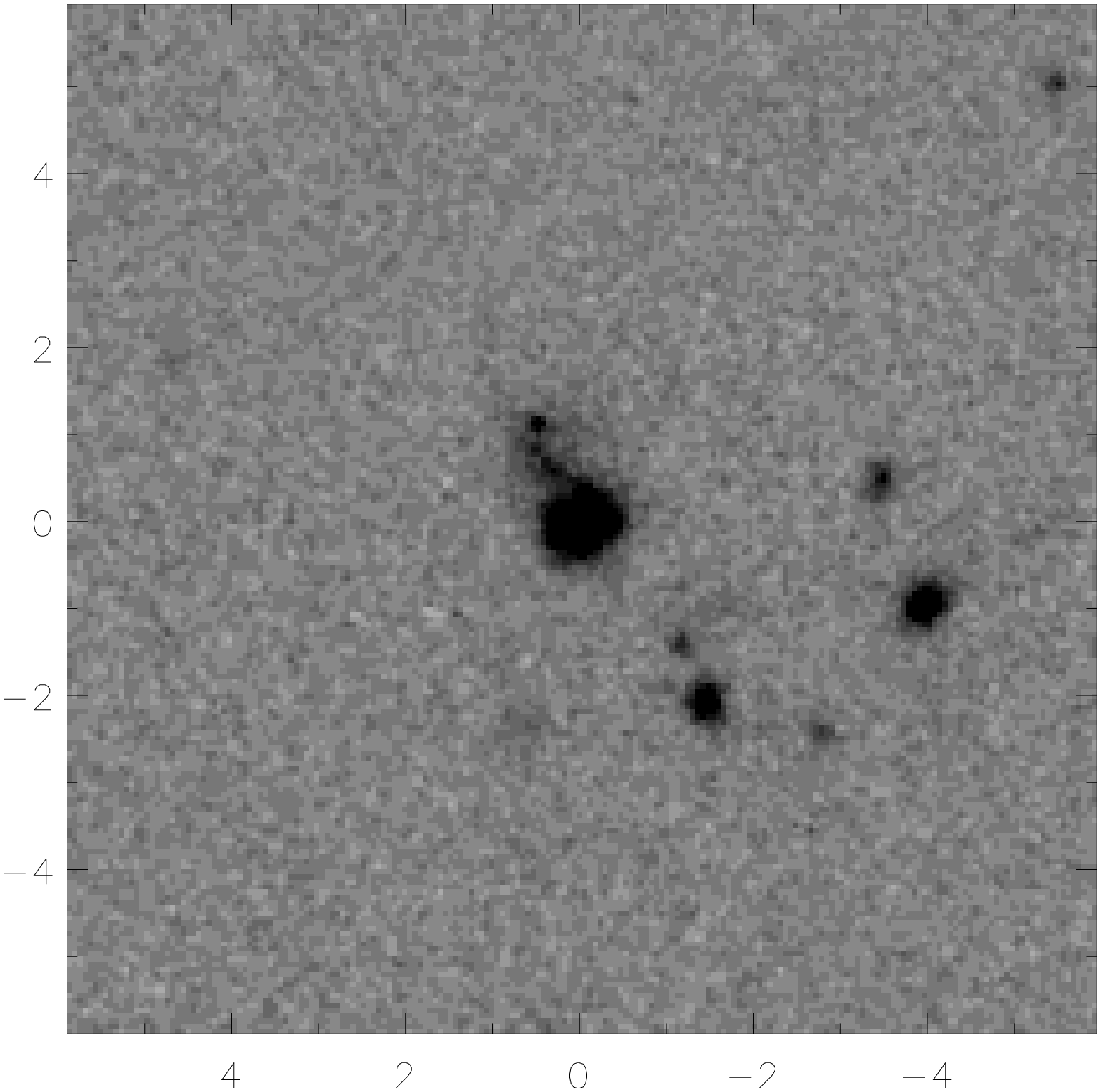,width=0.20\textwidth}&
\epsfig{file=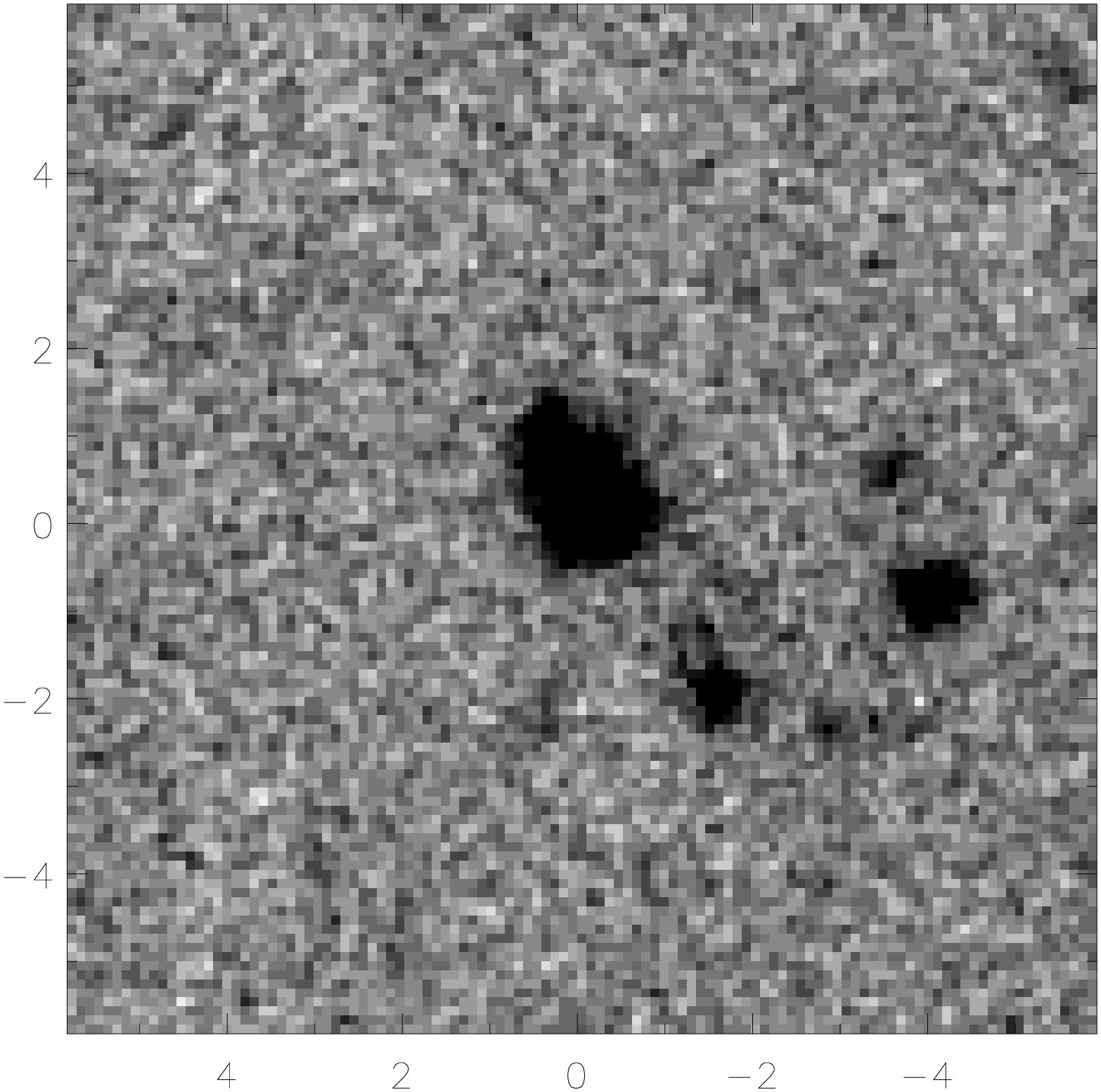,width=0.20\textwidth}&
\epsfig{file=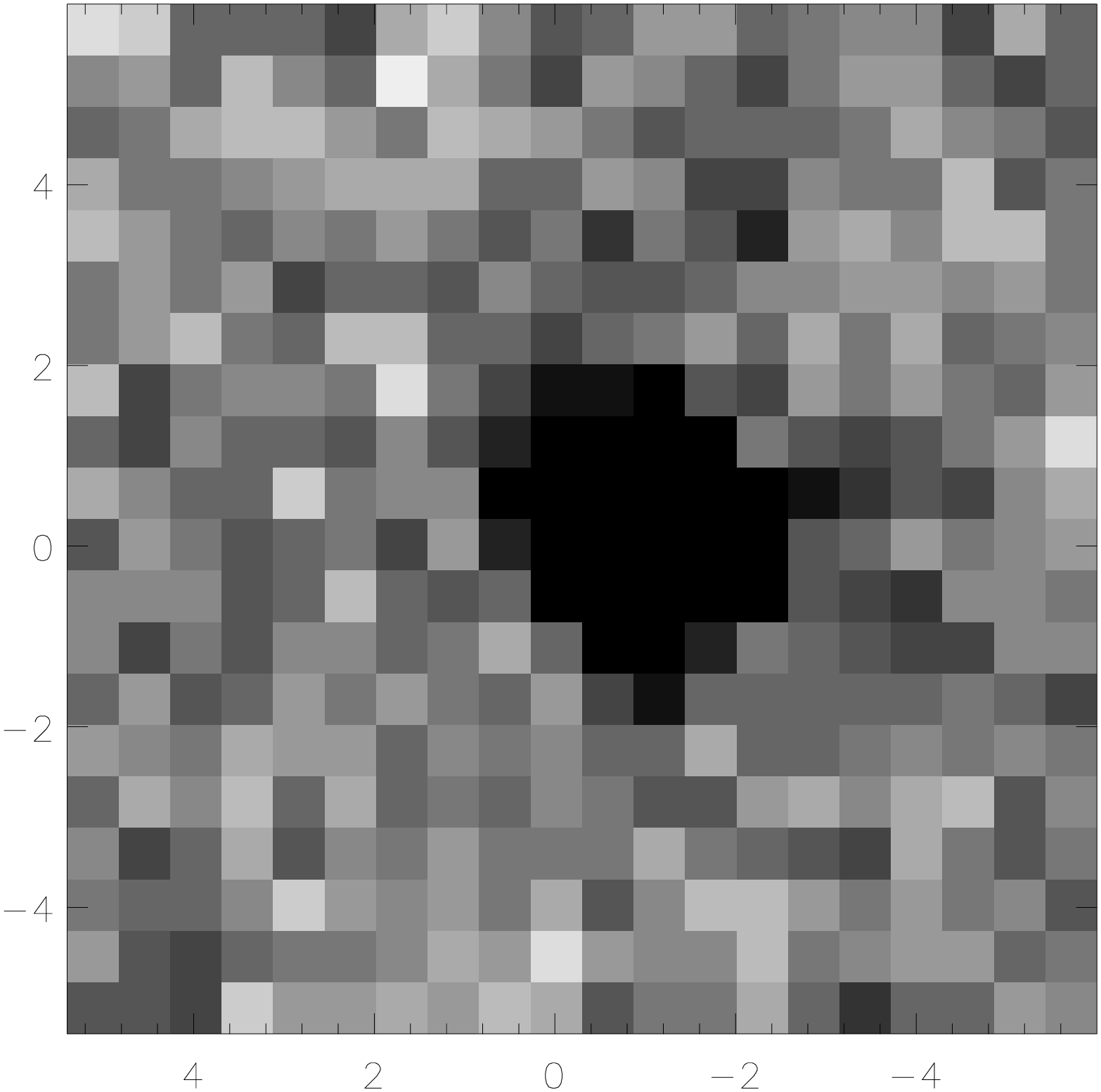,width=0.20\textwidth}\\
\\
\epsfig{file=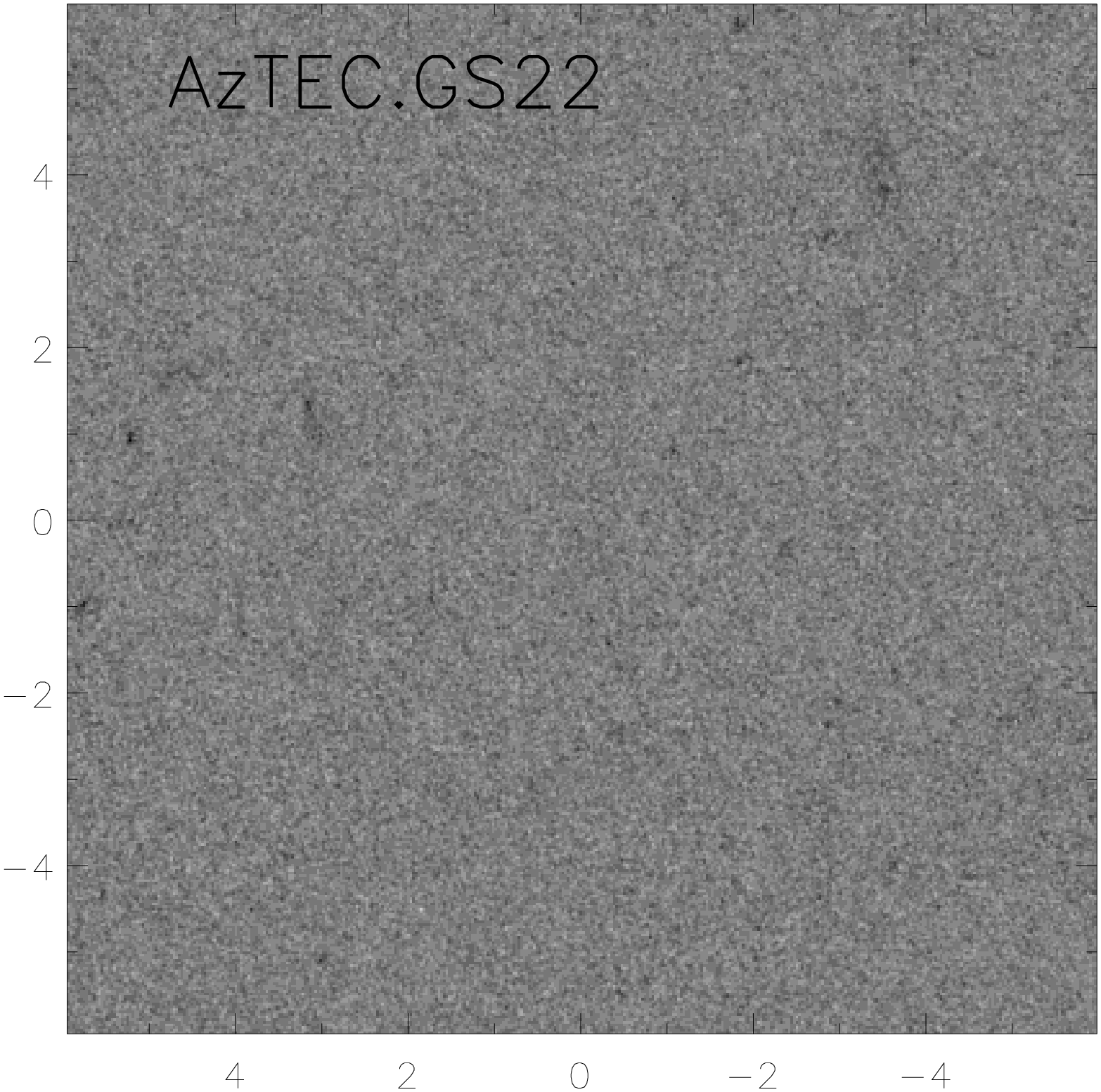,width=0.20\textwidth}&
\epsfig{file=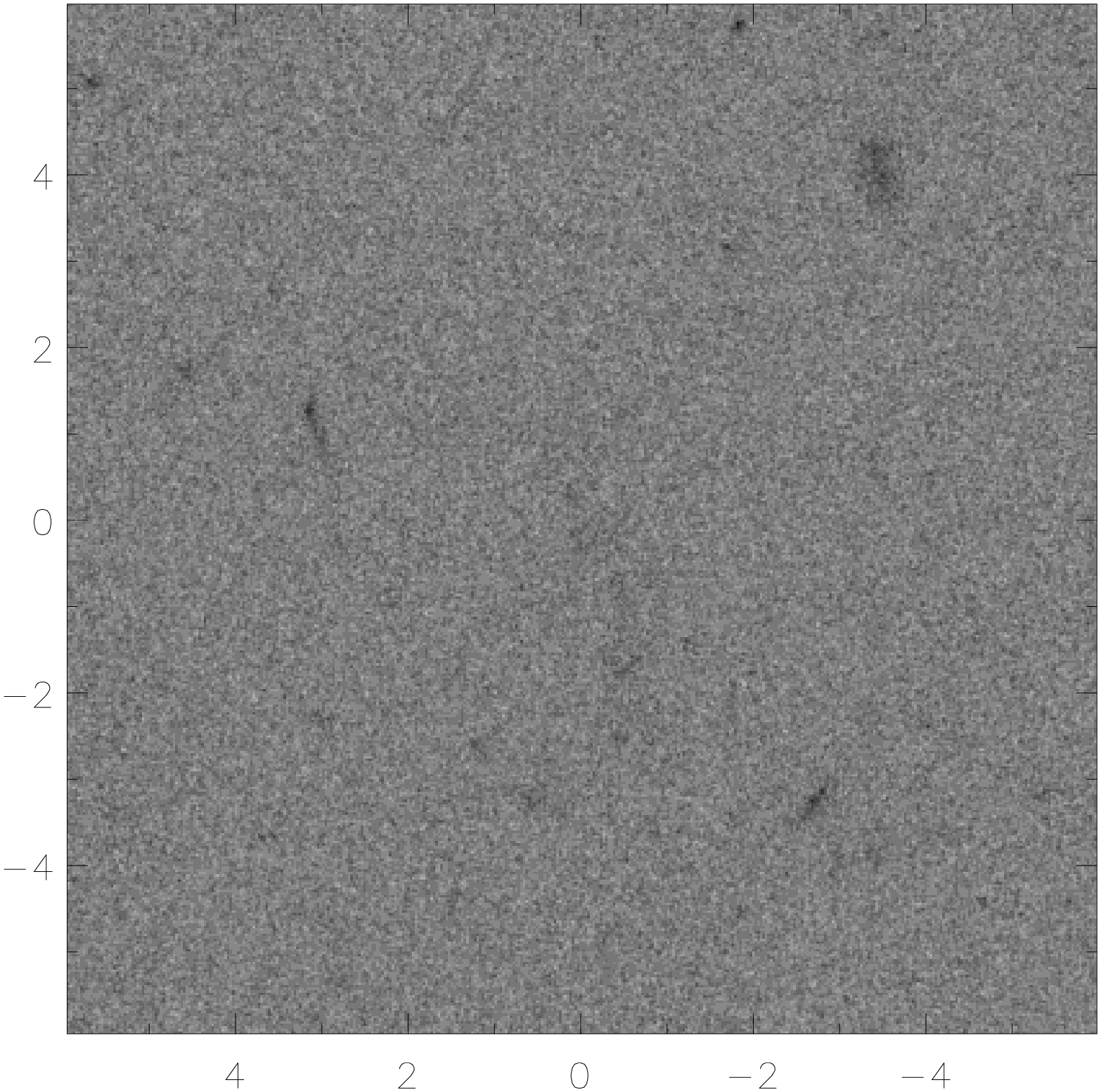,width=0.20\textwidth}&
\epsfig{file=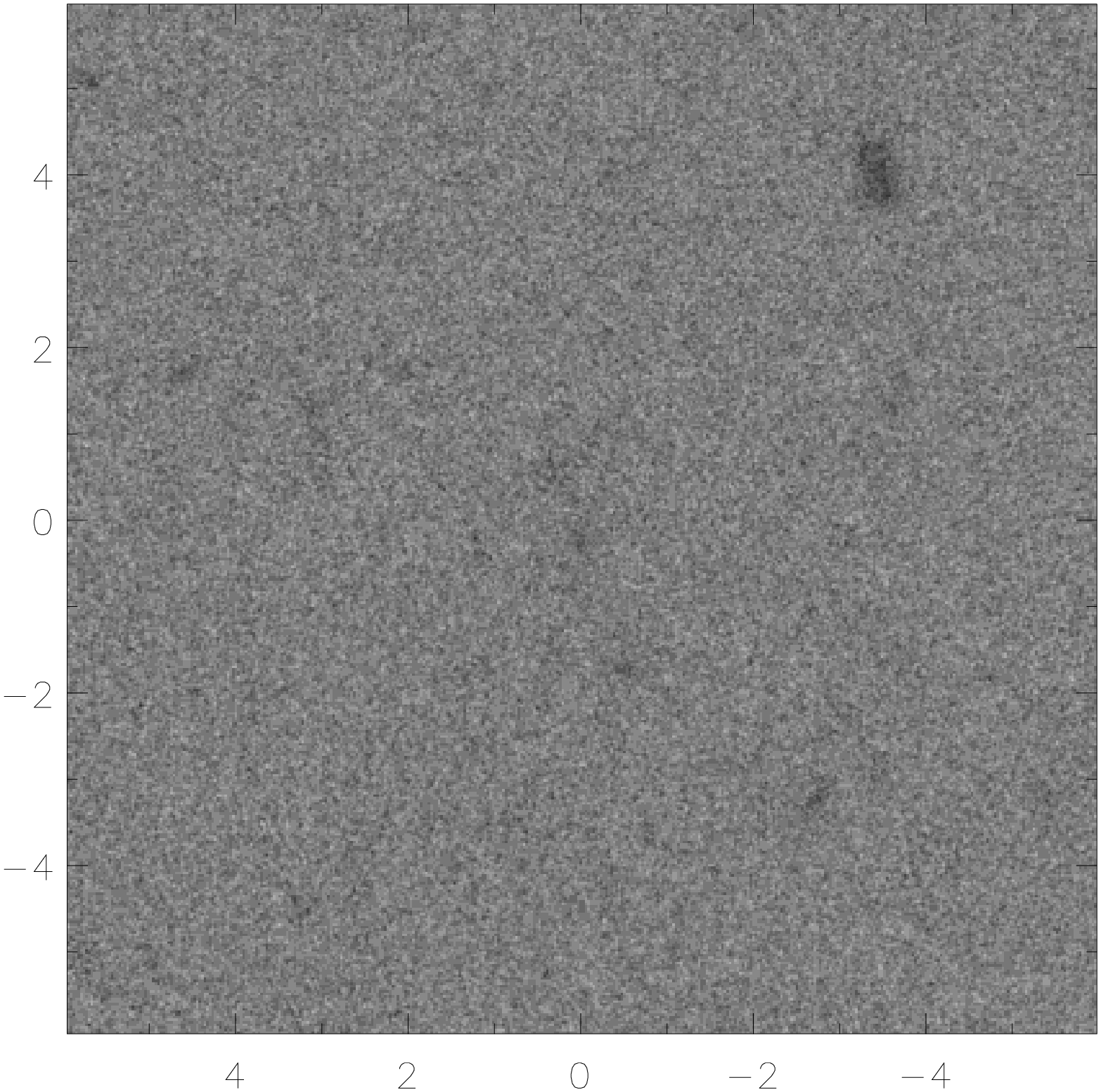,width=0.20\textwidth}&
\epsfig{file=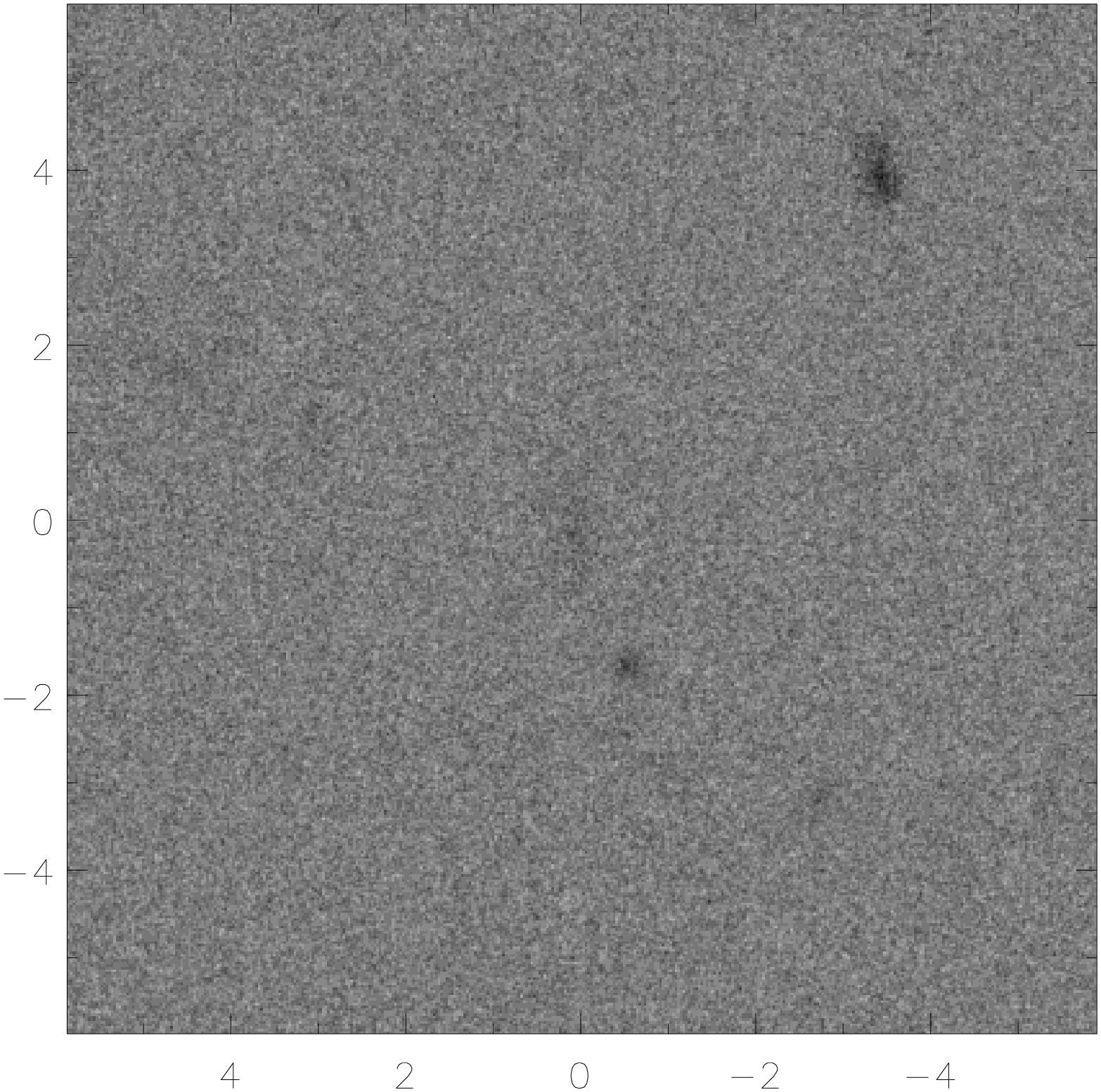,width=0.20\textwidth}\\
\epsfig{file=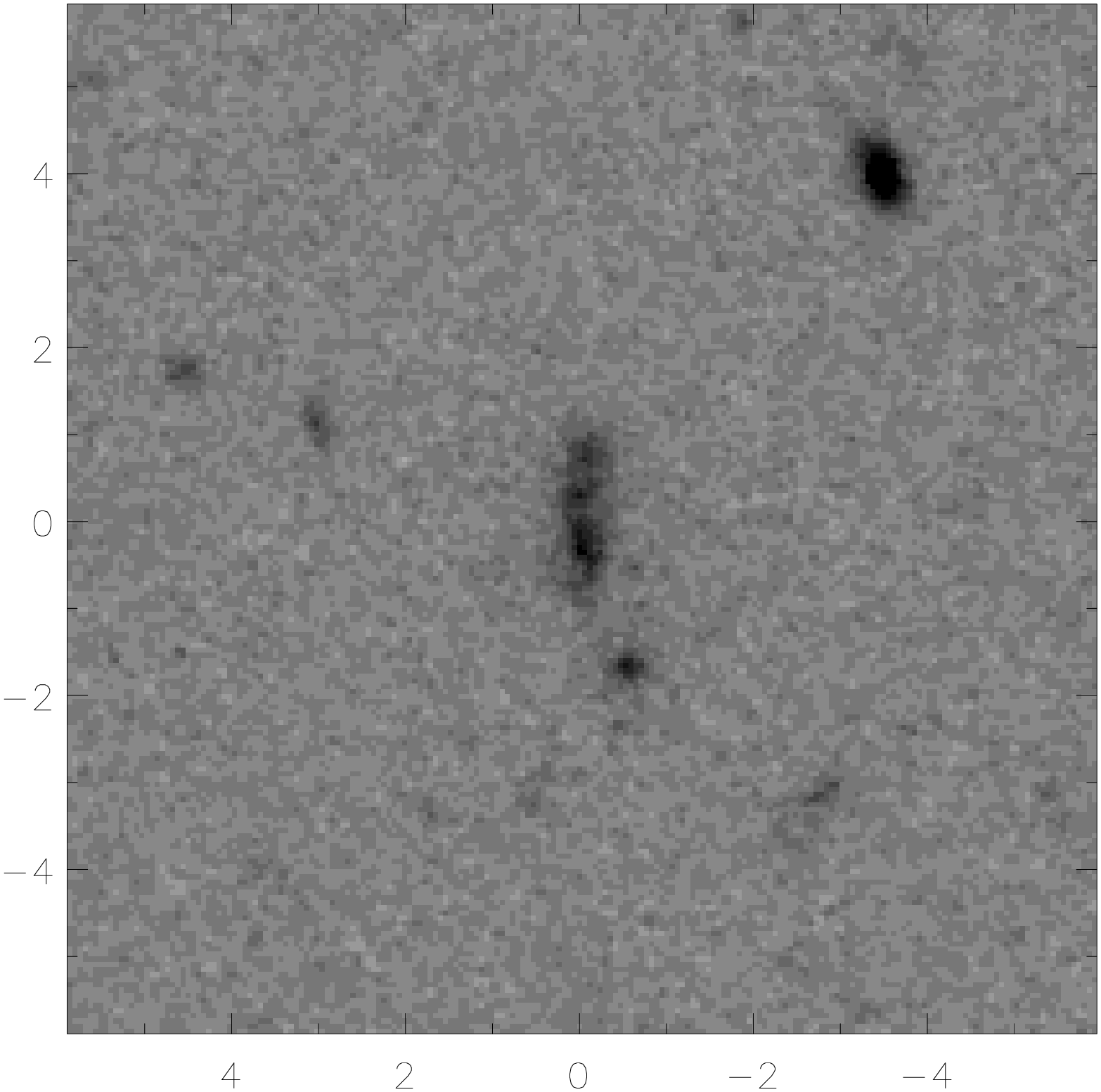,width=0.20\textwidth}&
\epsfig{file=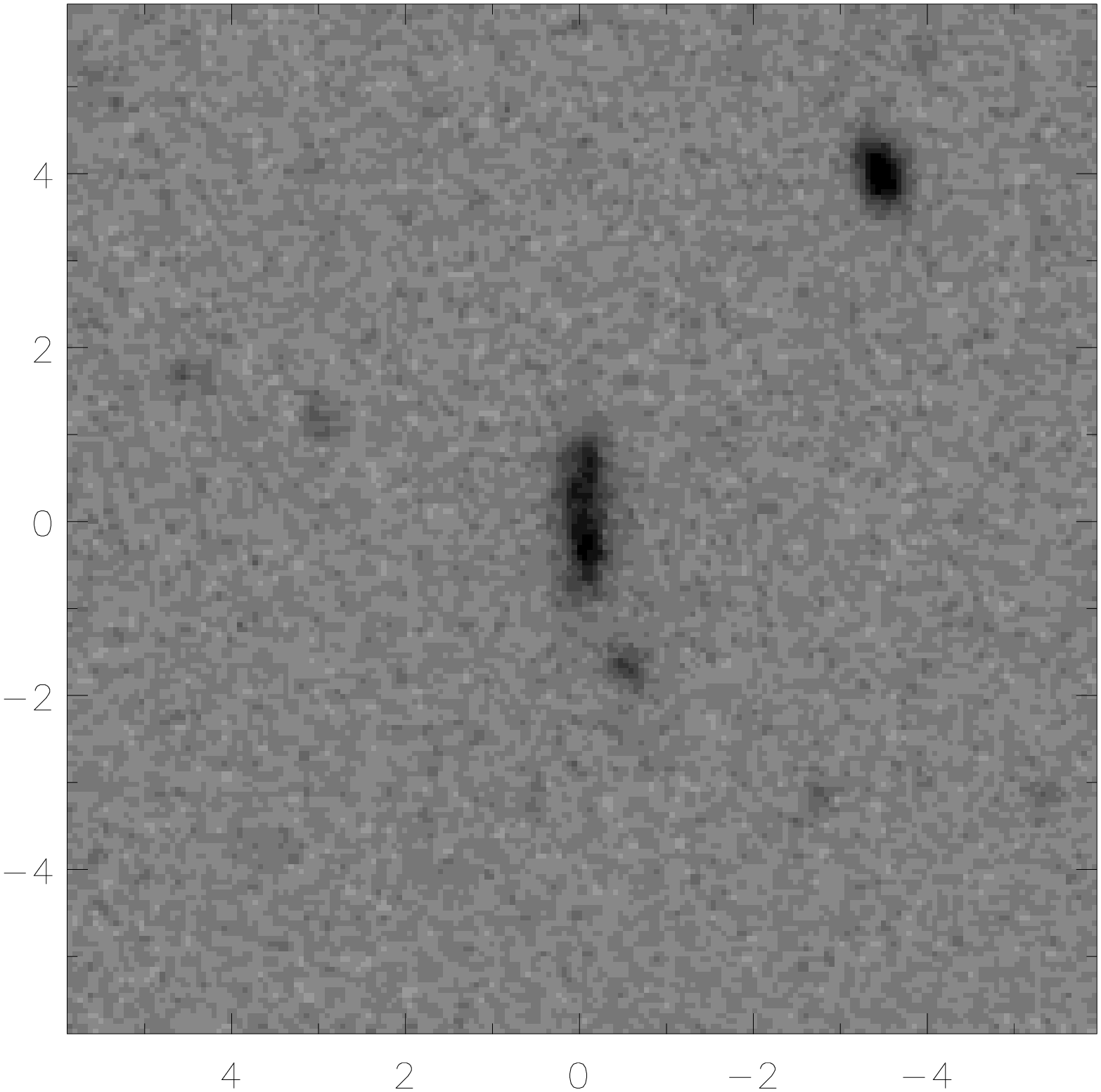,width=0.20\textwidth}&
\epsfig{file=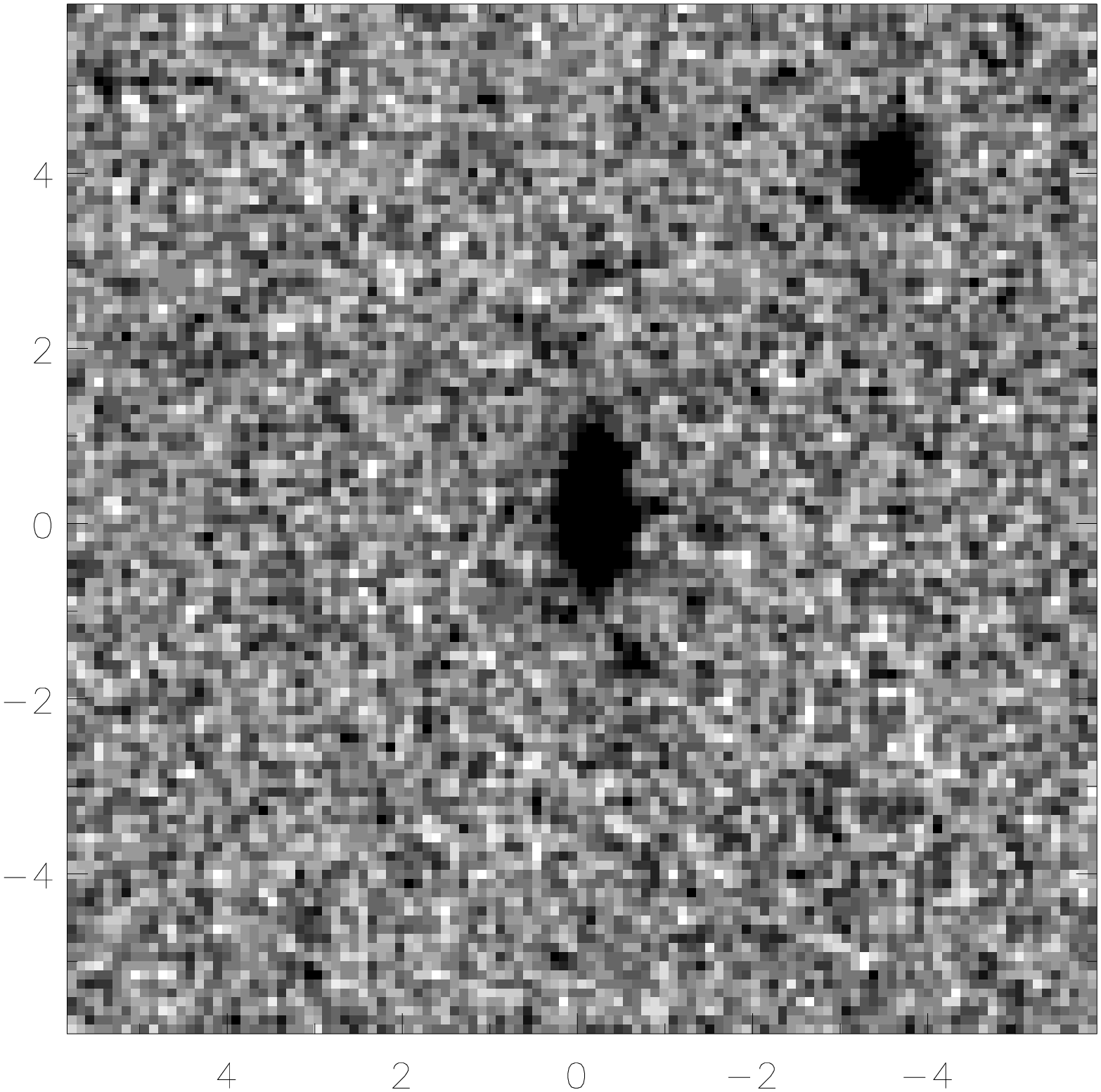,width=0.20\textwidth}&
\epsfig{file=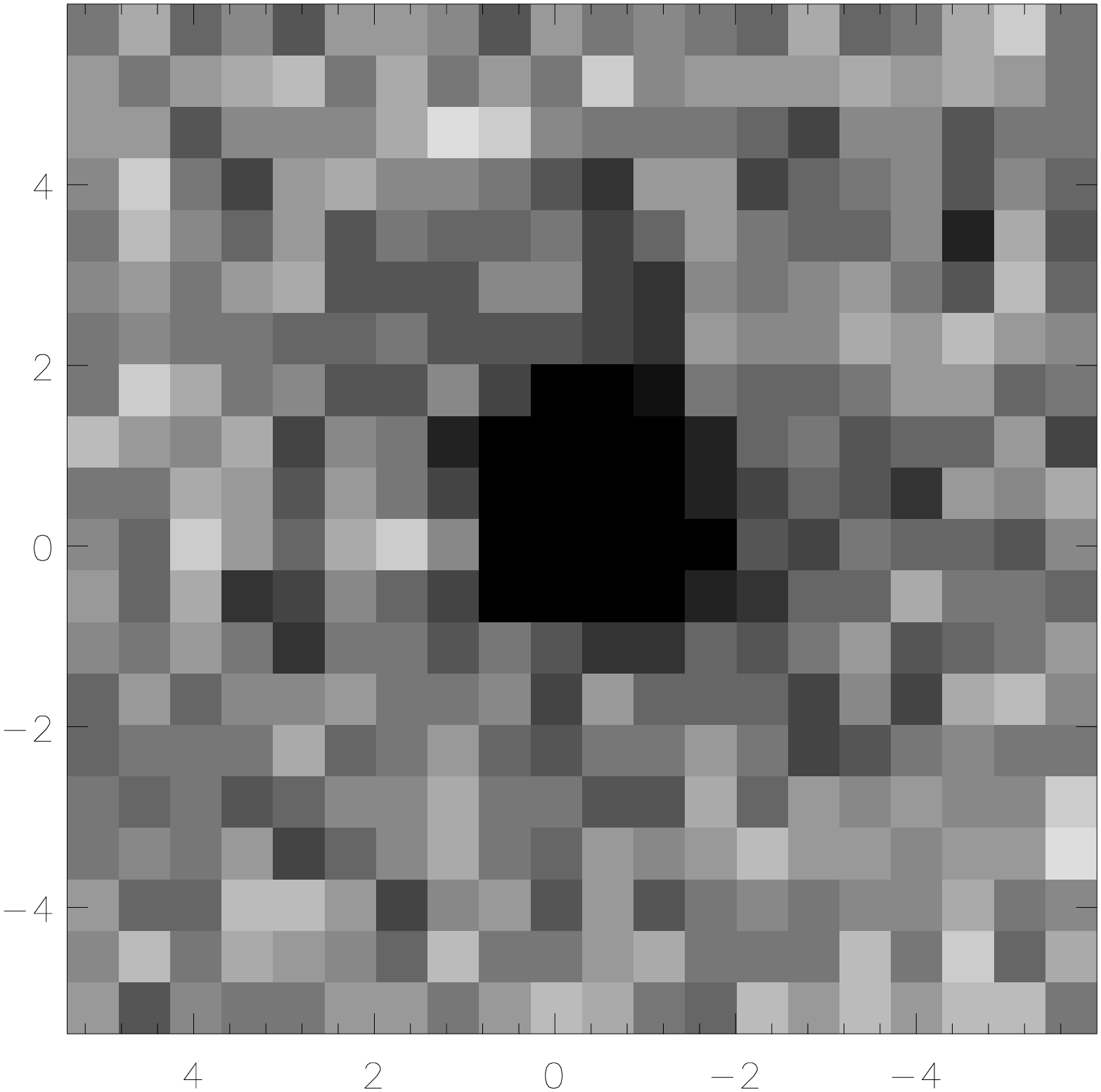,width=0.20\textwidth}\\
\end{tabular}
\addtocounter{figure}{-1}
\caption{- continued}
\vfil}
\end{figure*}
\end{center}


\begin{center}
\begin{figure*}
\vbox to220mm{\vfil
\begin{tabular}{cccccccc}
\epsfig{file=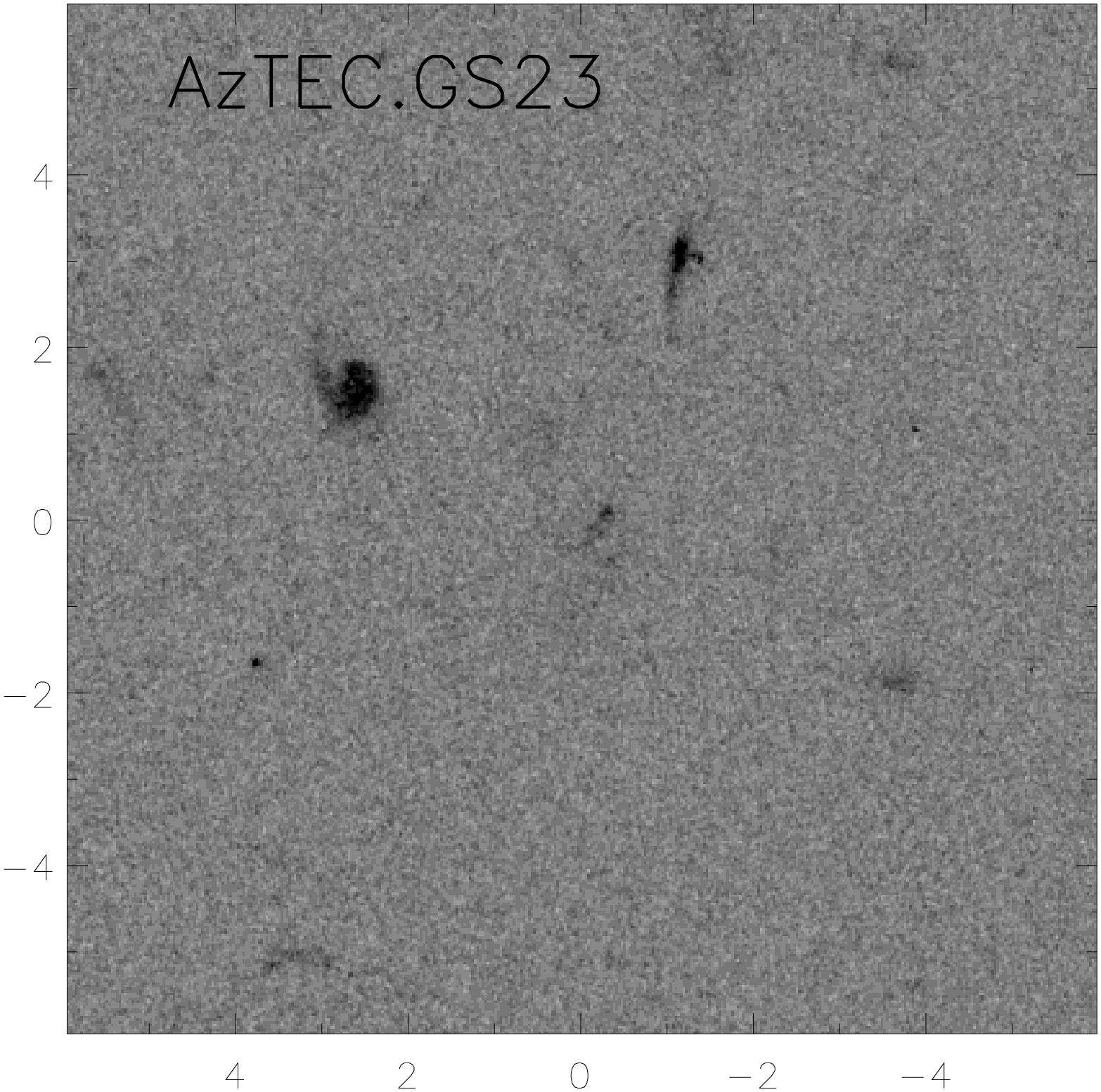,width=0.20\textwidth}&
\epsfig{file=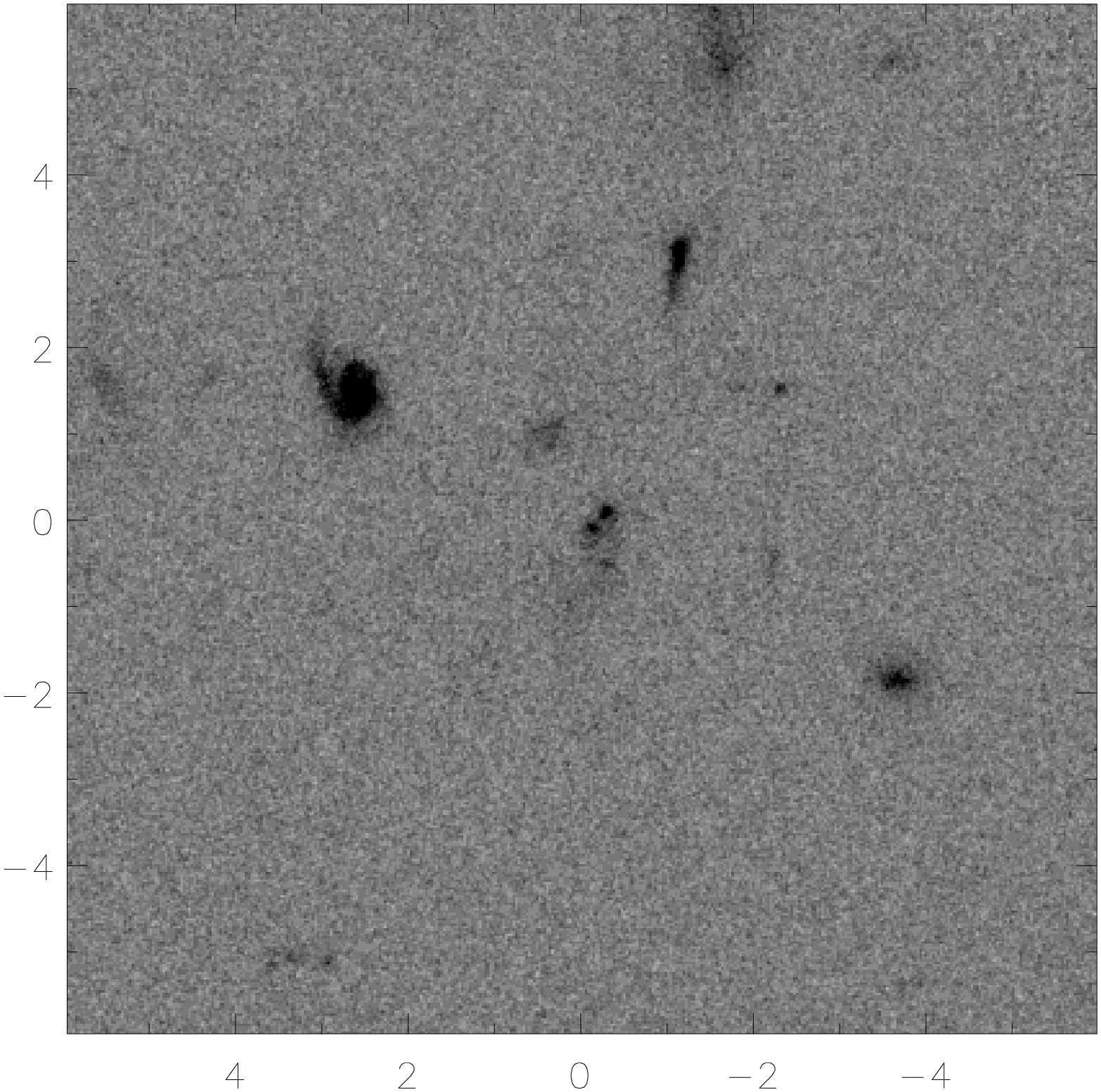,width=0.20\textwidth}&
\epsfig{file=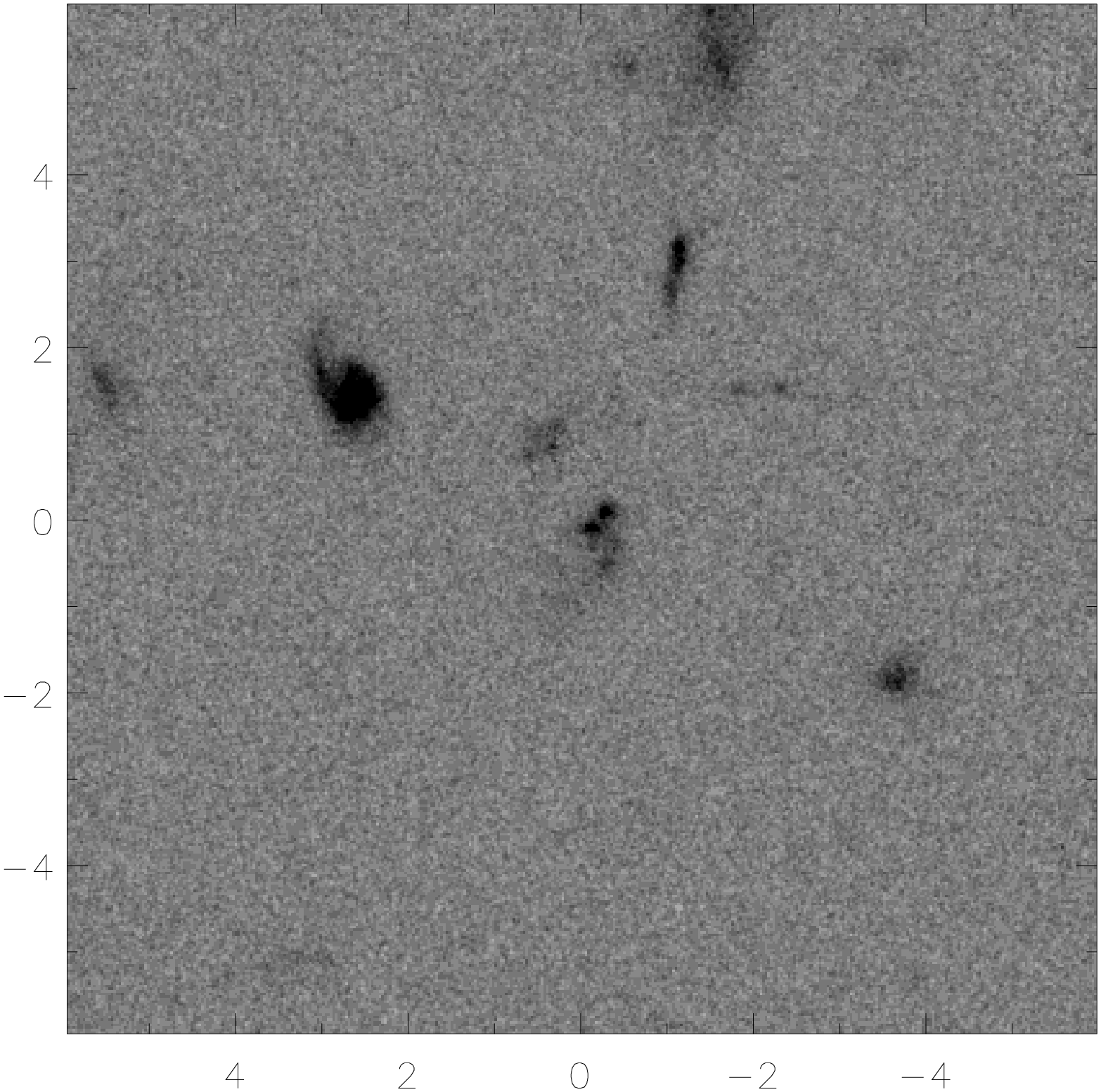,width=0.20\textwidth}&
\epsfig{file=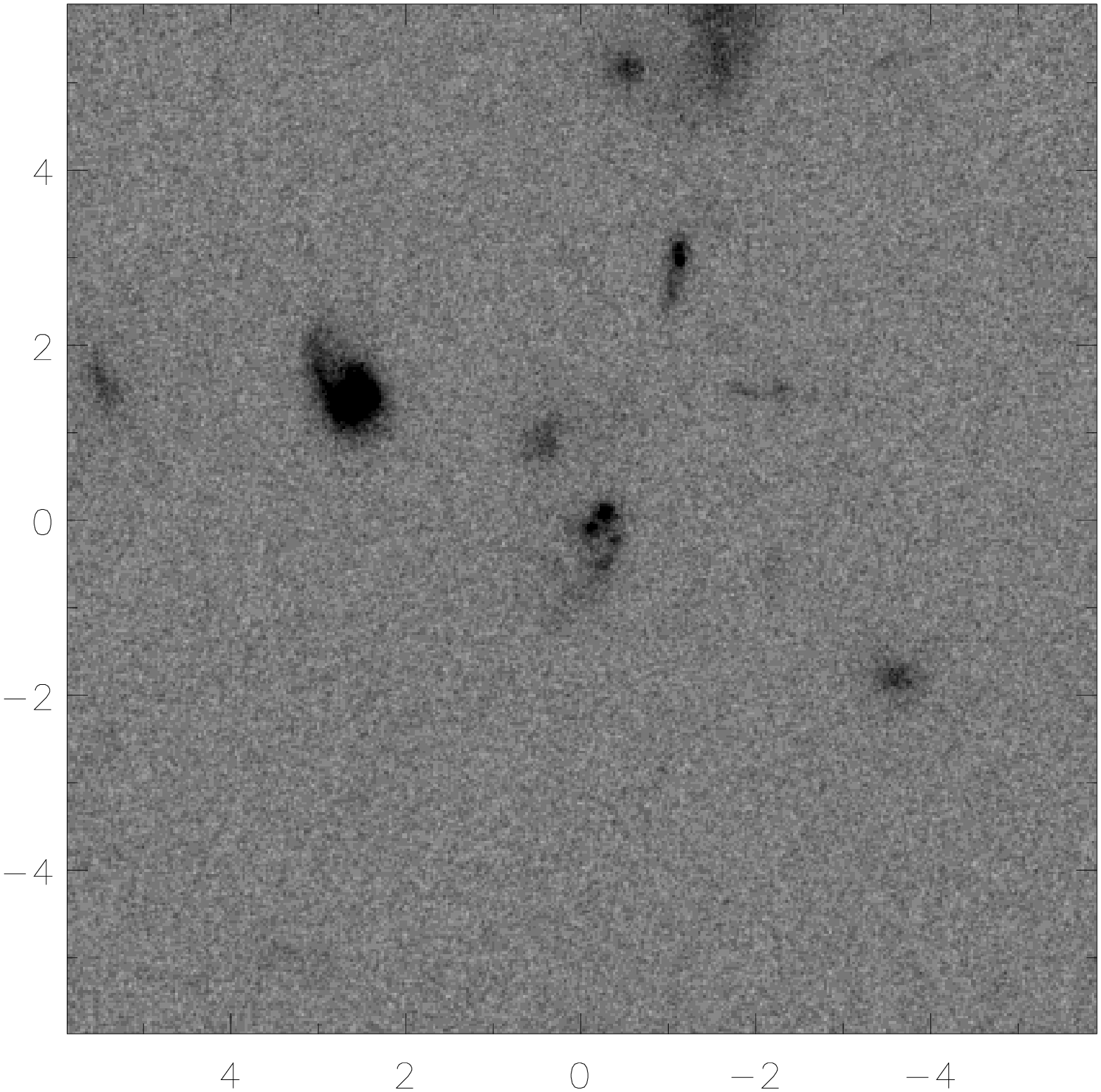,width=0.20\textwidth}\\
\epsfig{file=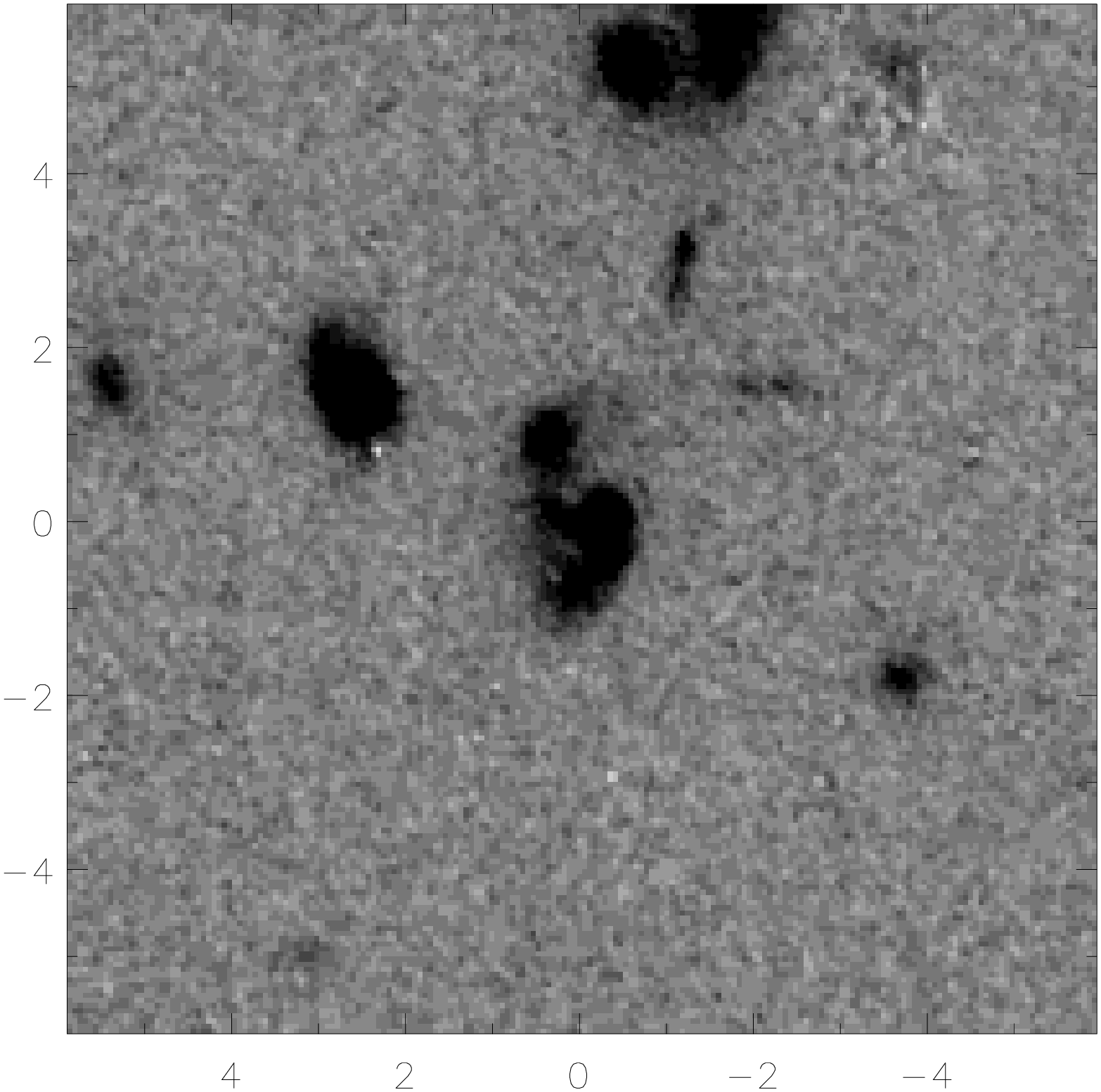,width=0.20\textwidth}&
\epsfig{file=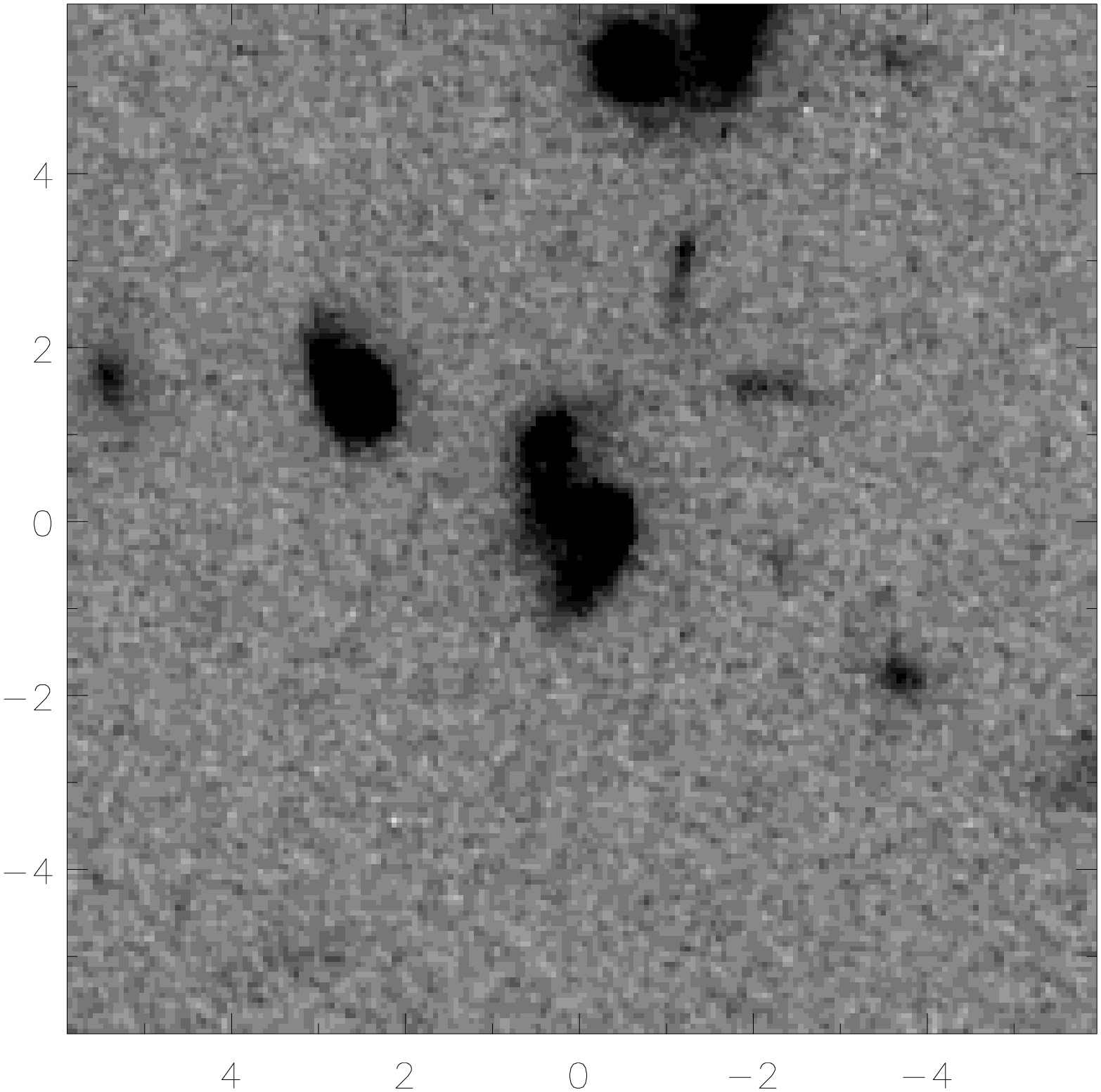,width=0.20\textwidth}&
\epsfig{file=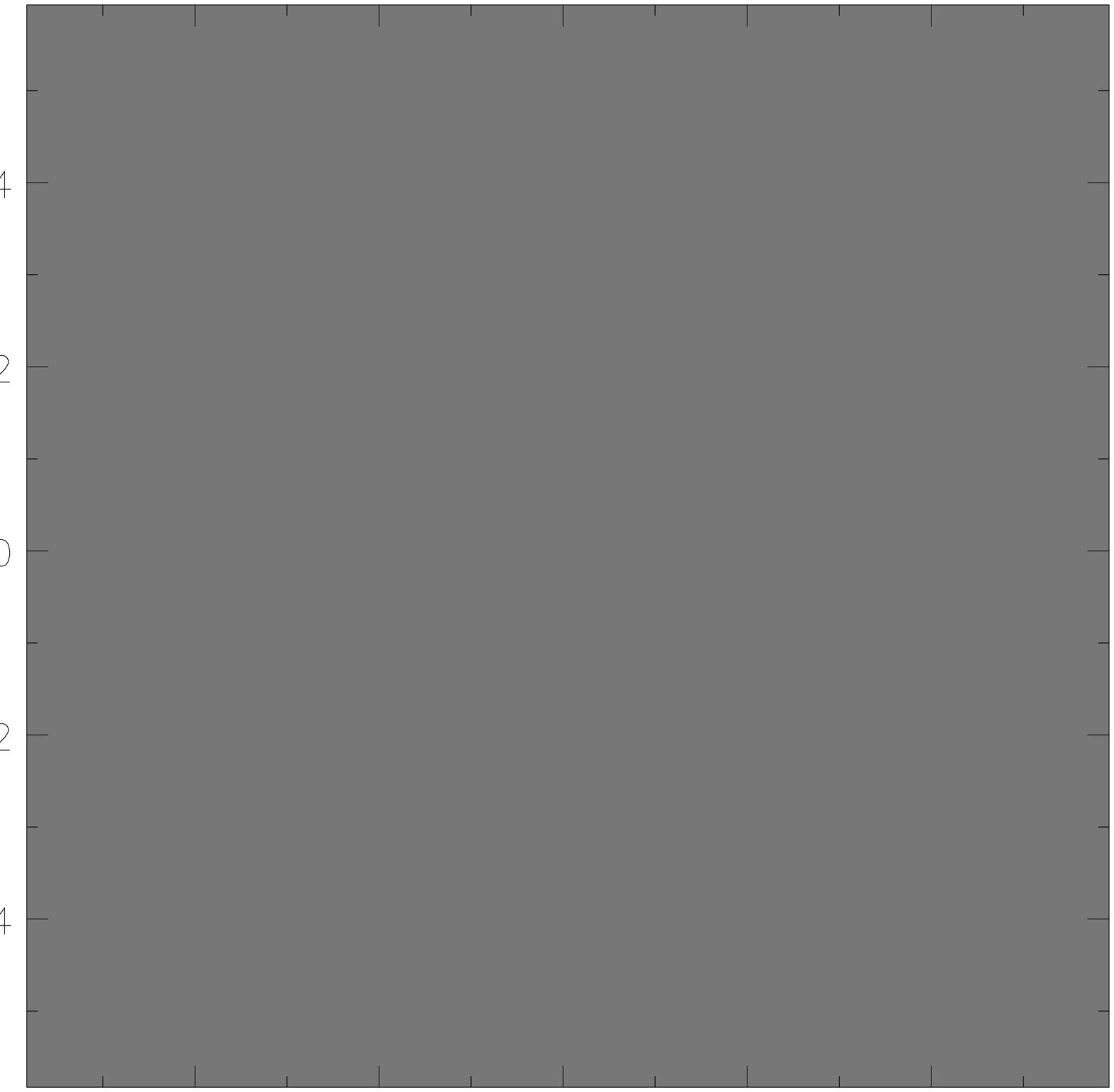,width=0.20\textwidth}&
\epsfig{file=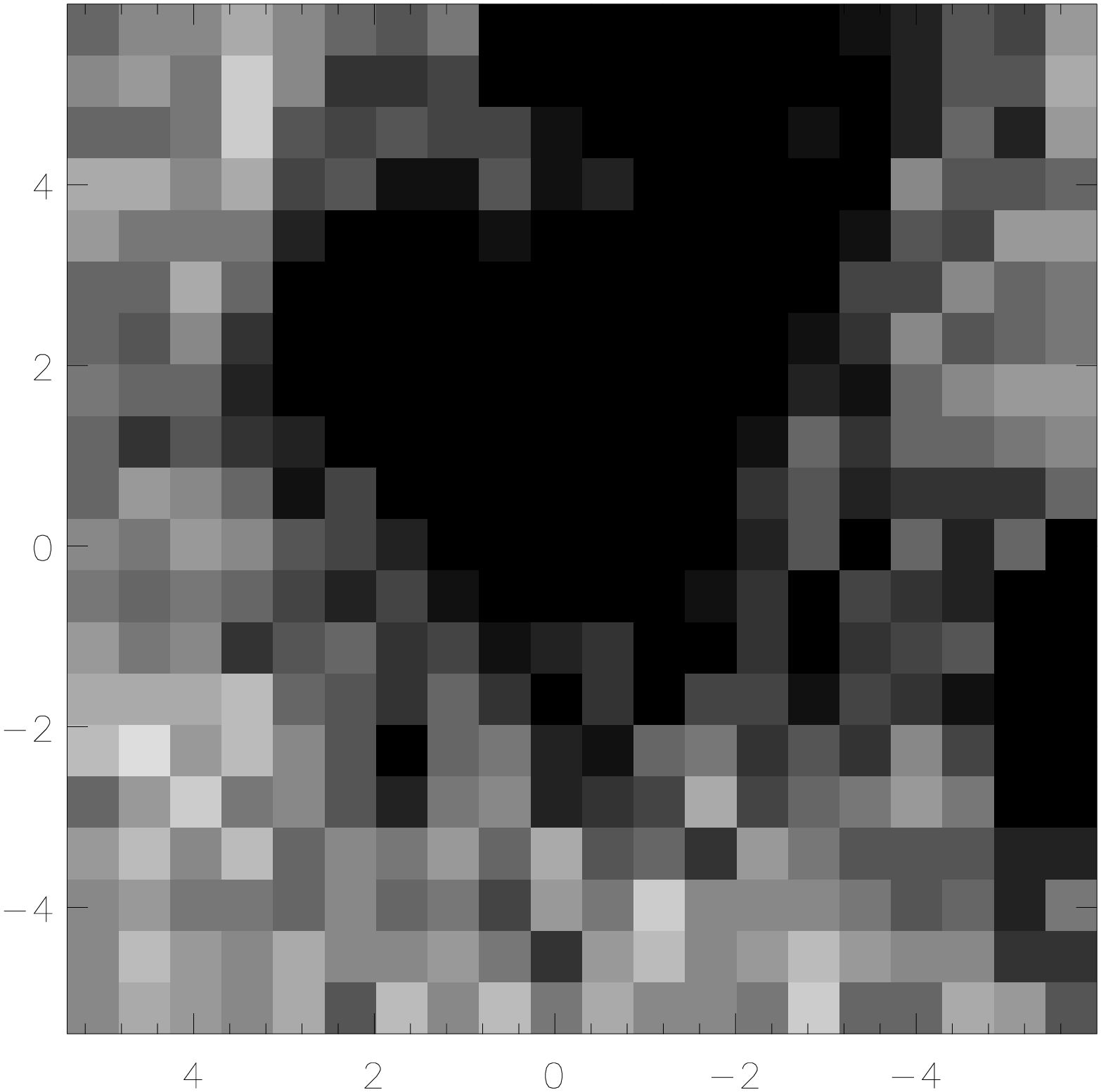,width=0.20\textwidth}\\
\\
\epsfig{file=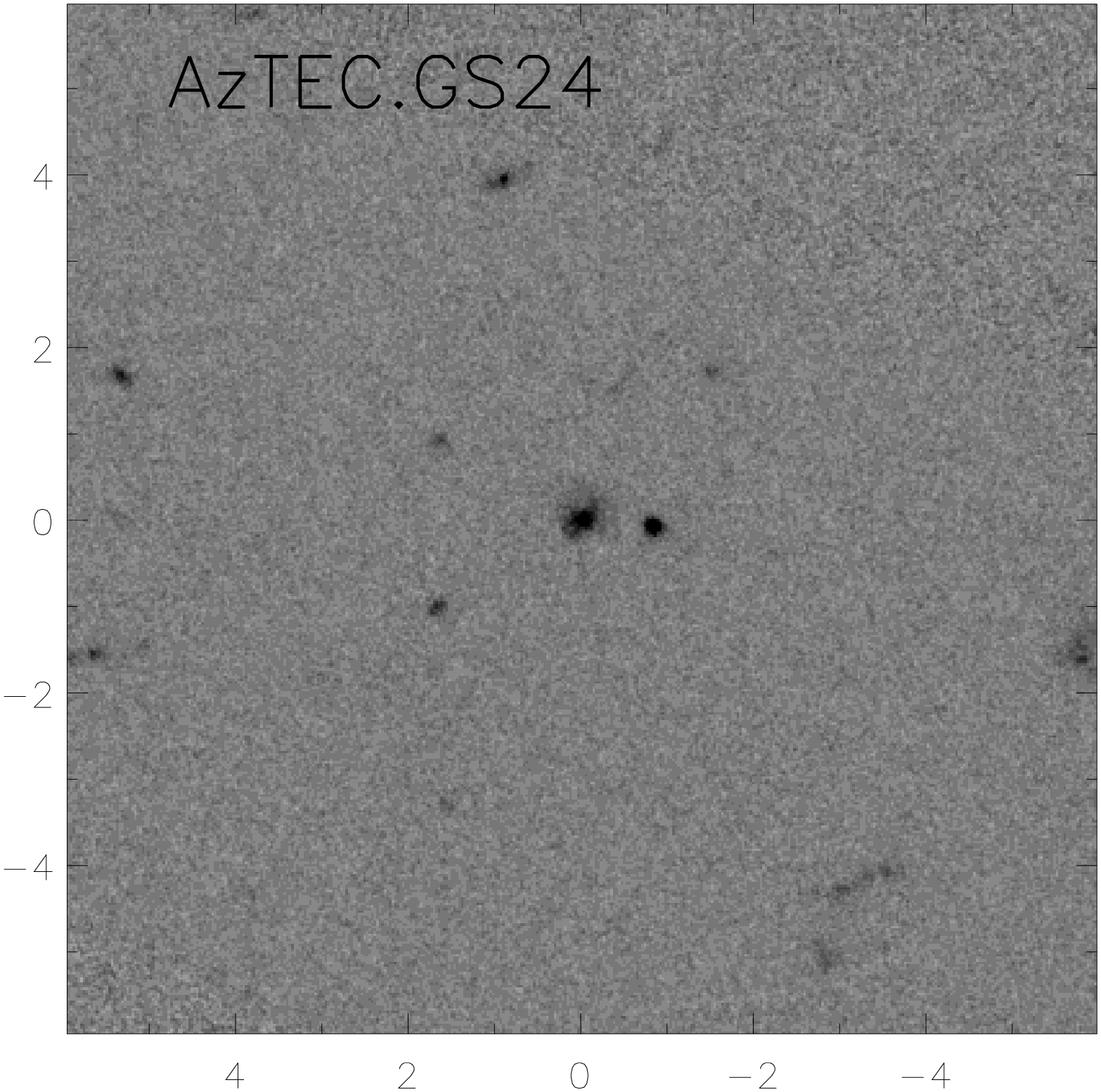,width=0.20\textwidth}&
\epsfig{file=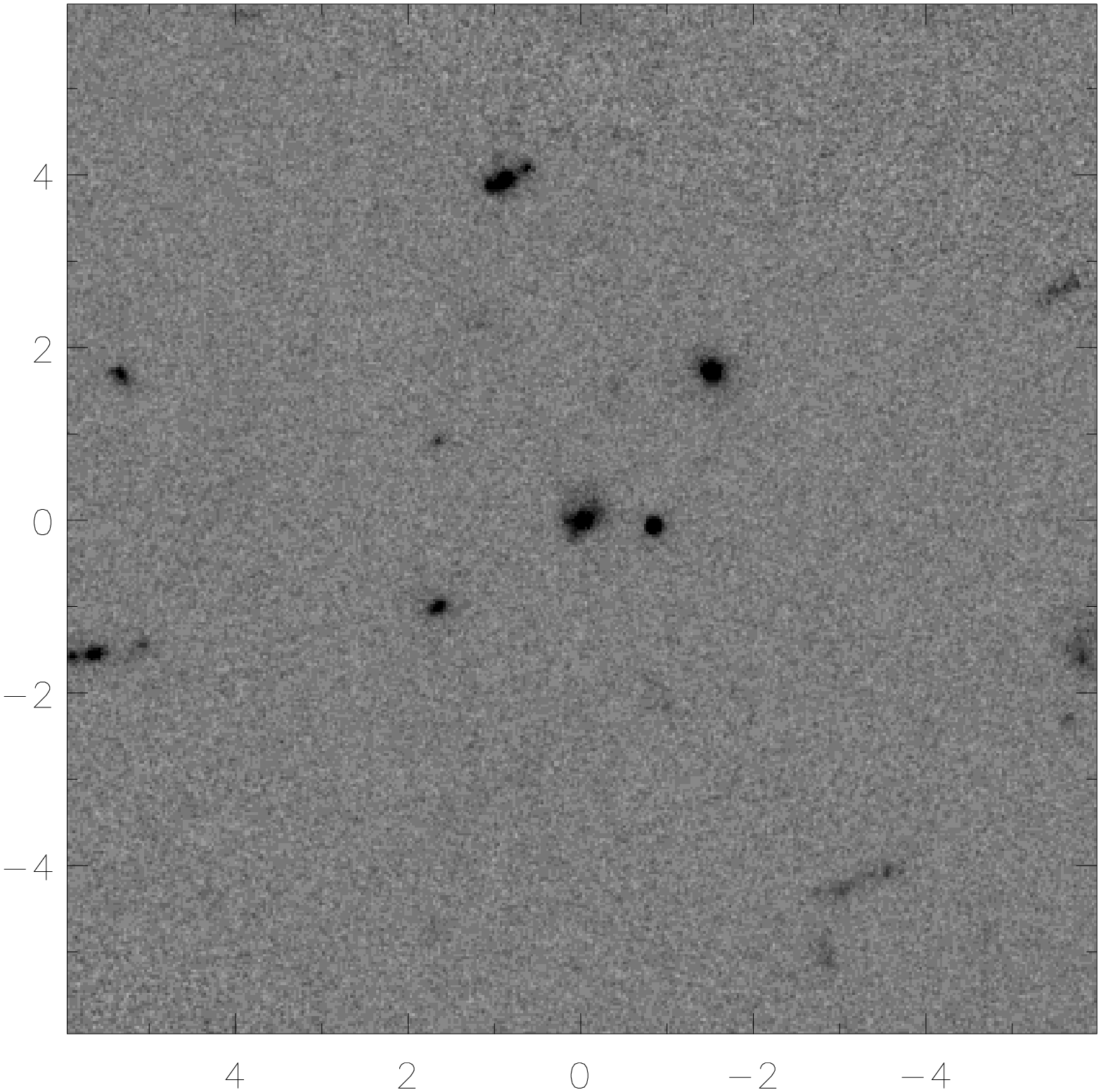,width=0.20\textwidth}&
\epsfig{file=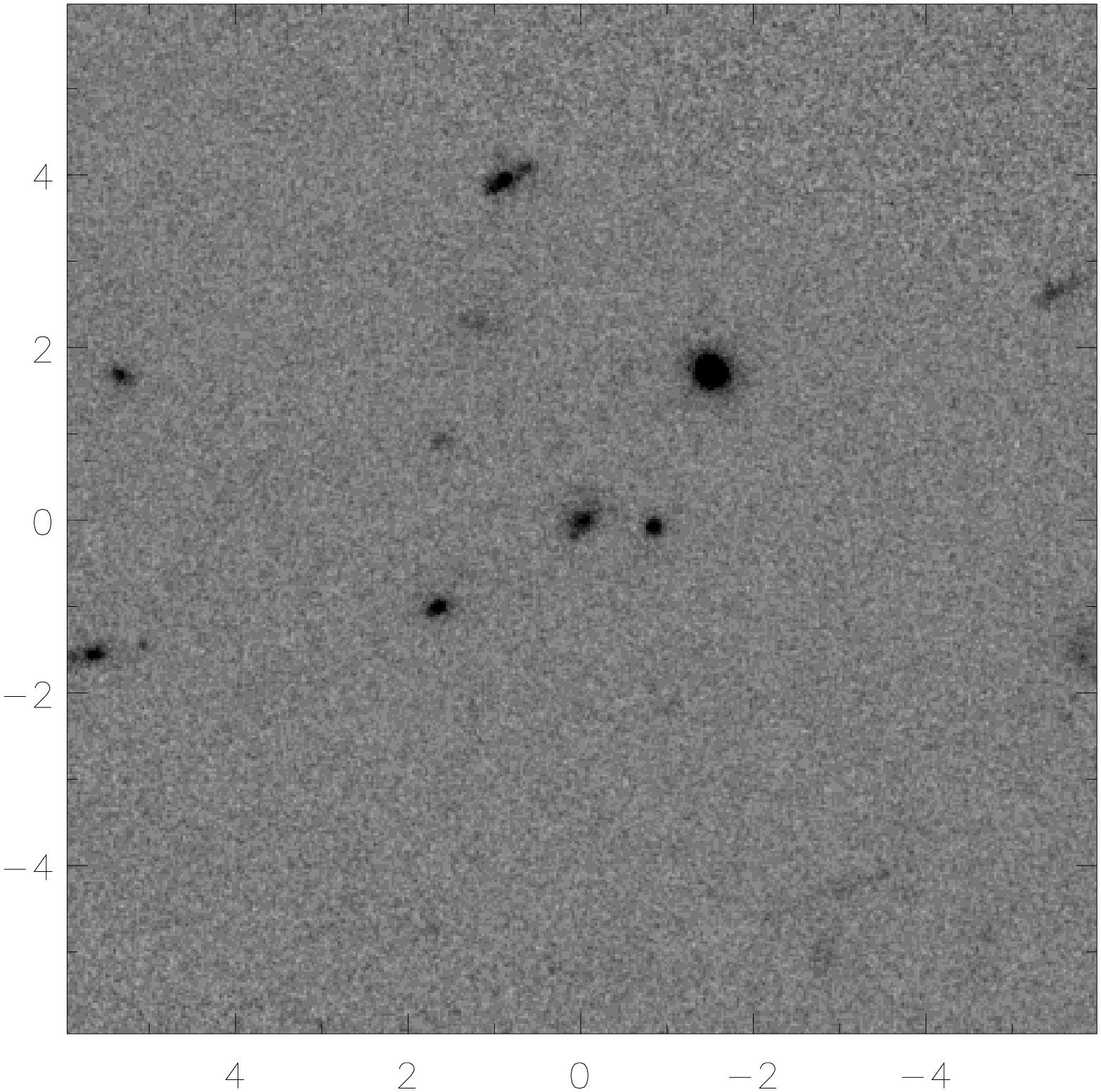,width=0.20\textwidth}&
\epsfig{file=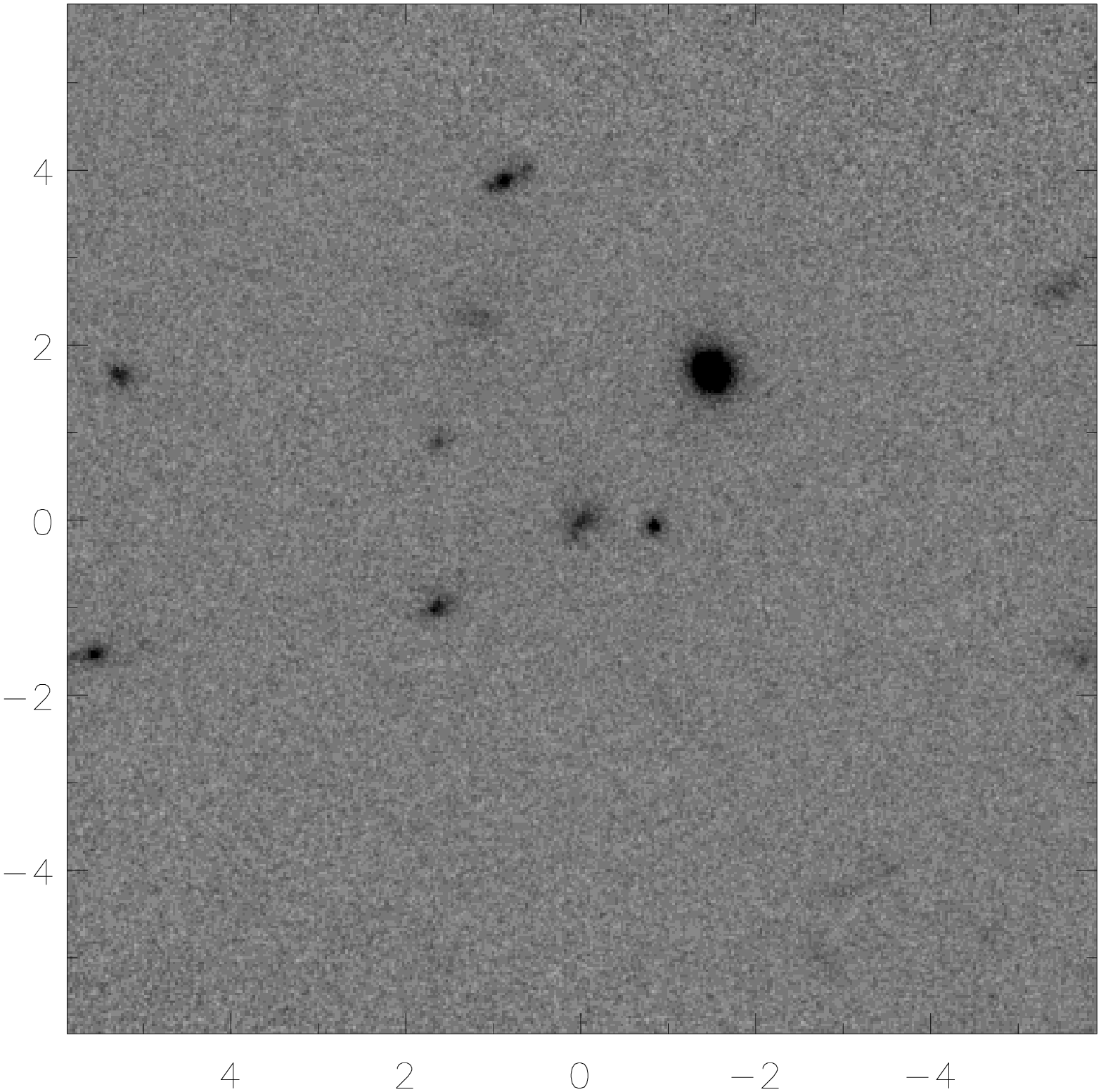,width=0.20\textwidth}\\
\epsfig{file=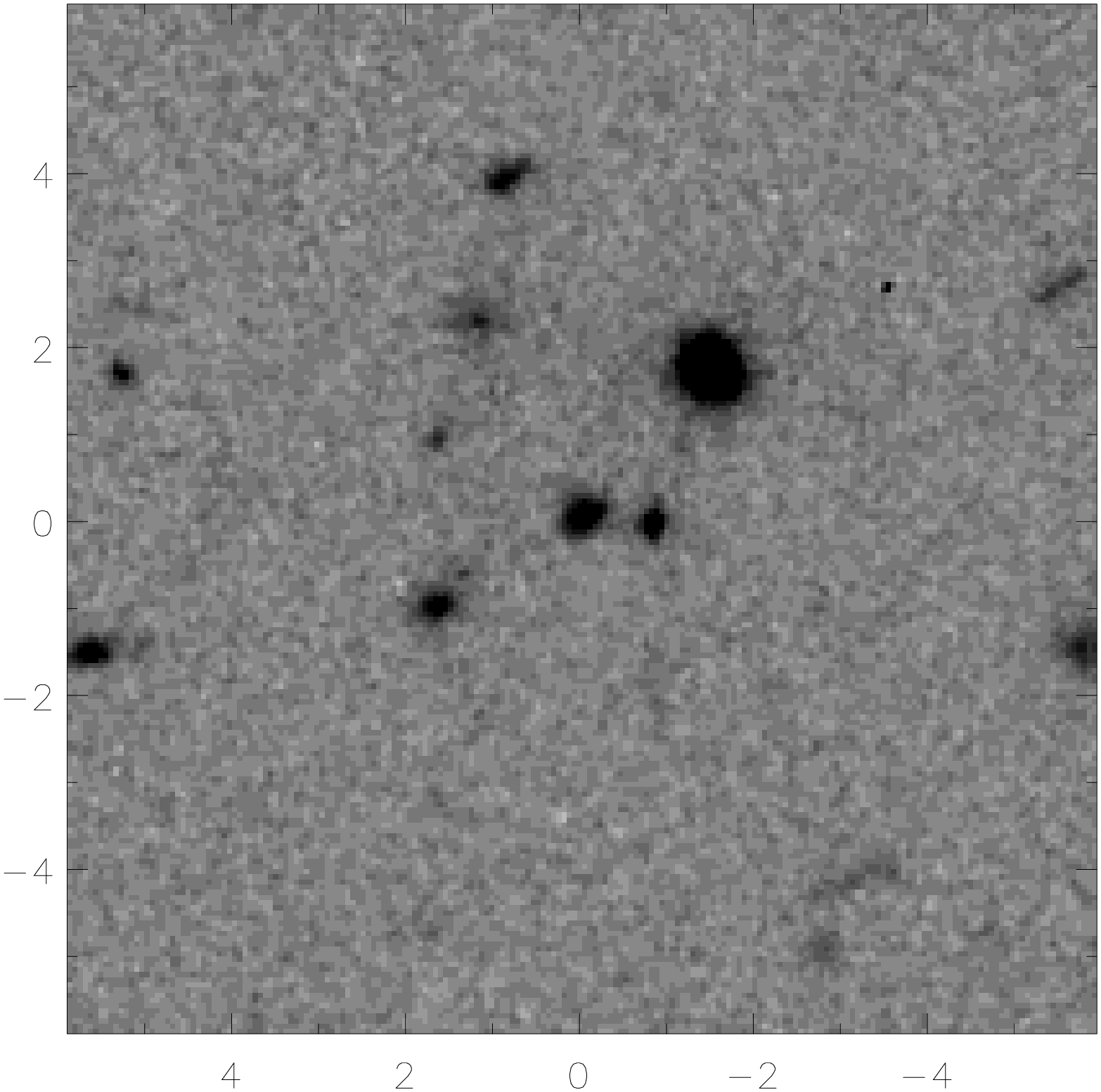,width=0.20\textwidth}&
\epsfig{file=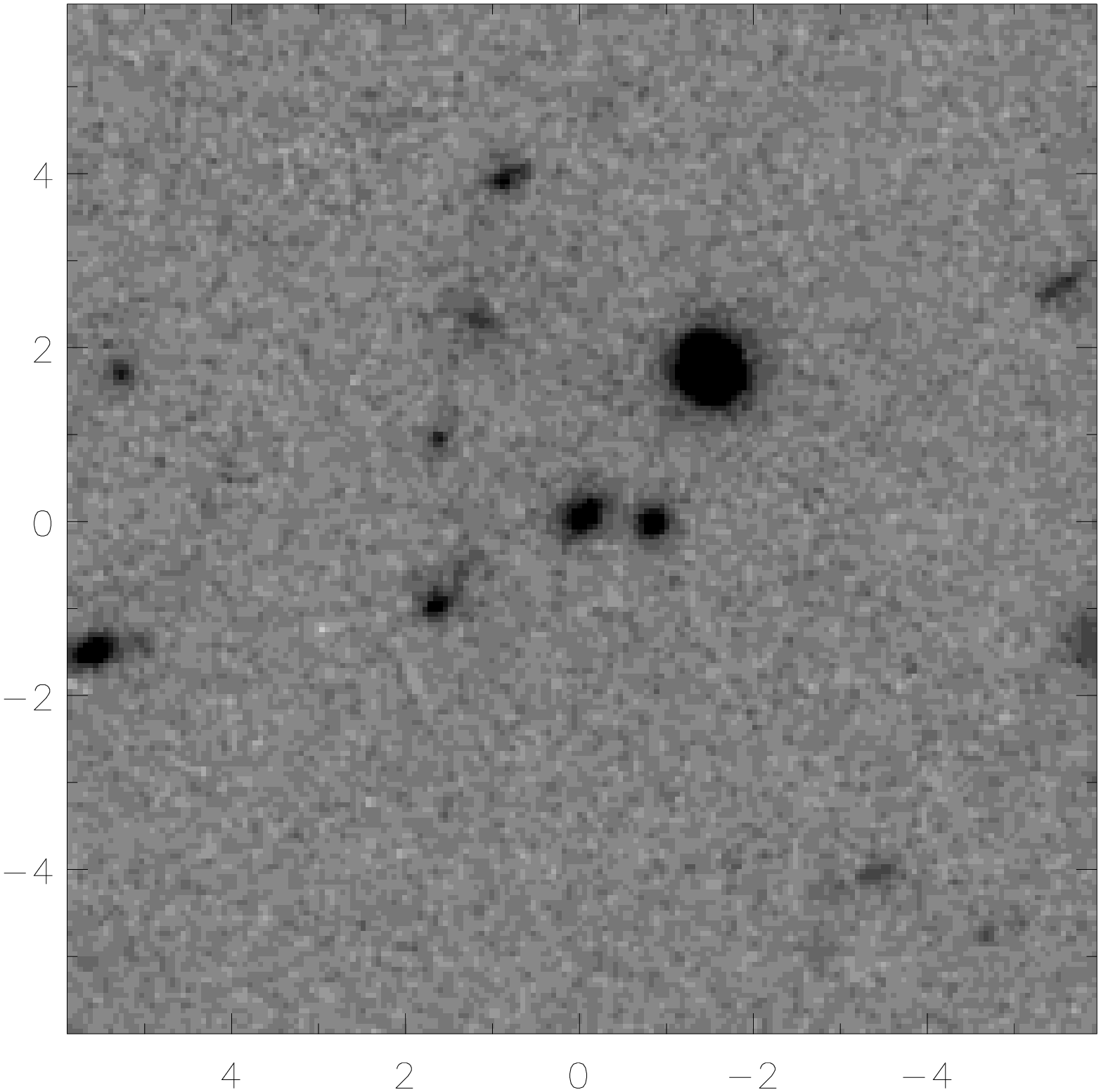,width=0.20\textwidth}&
\epsfig{file=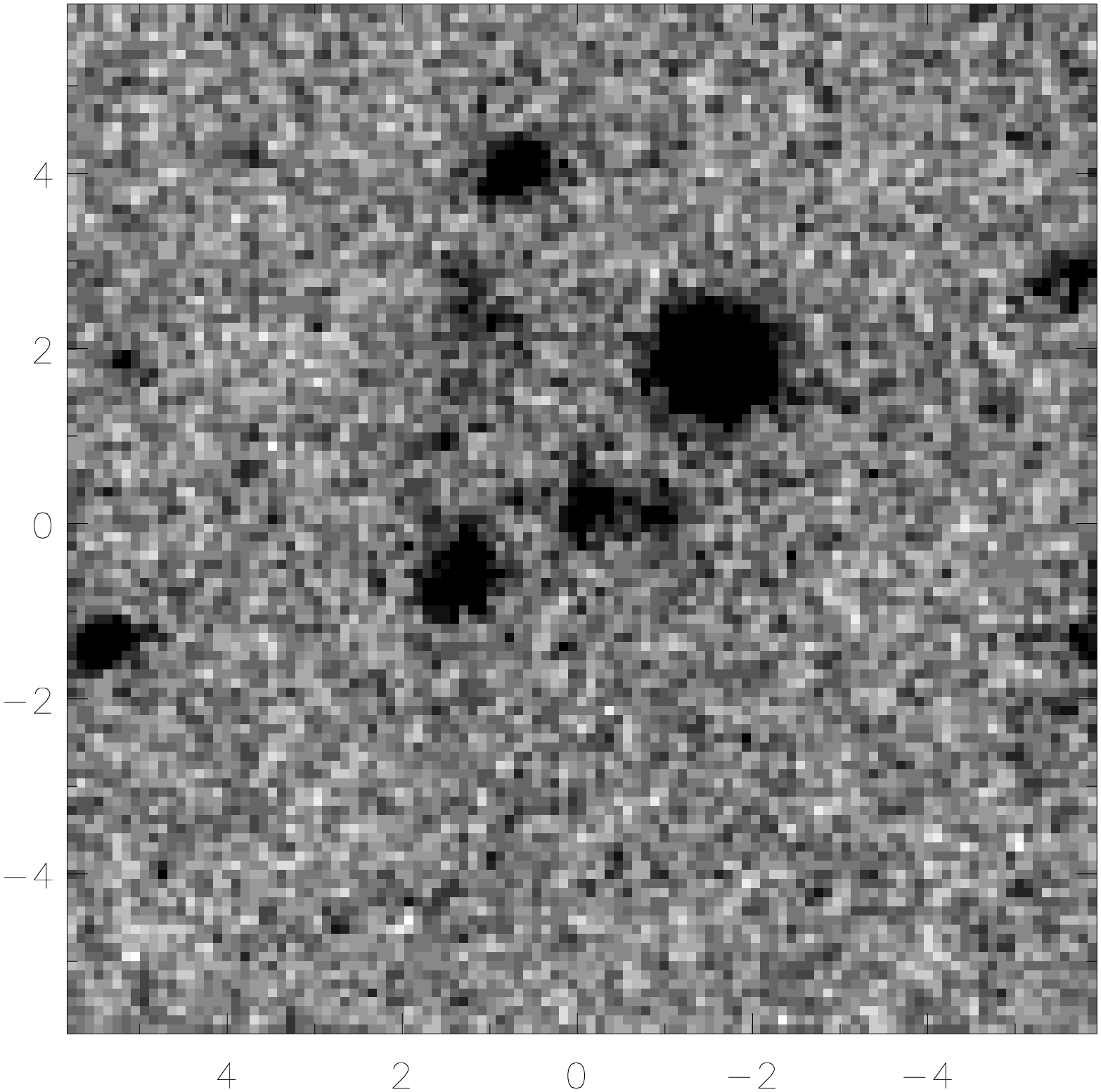,width=0.20\textwidth}&
\epsfig{file=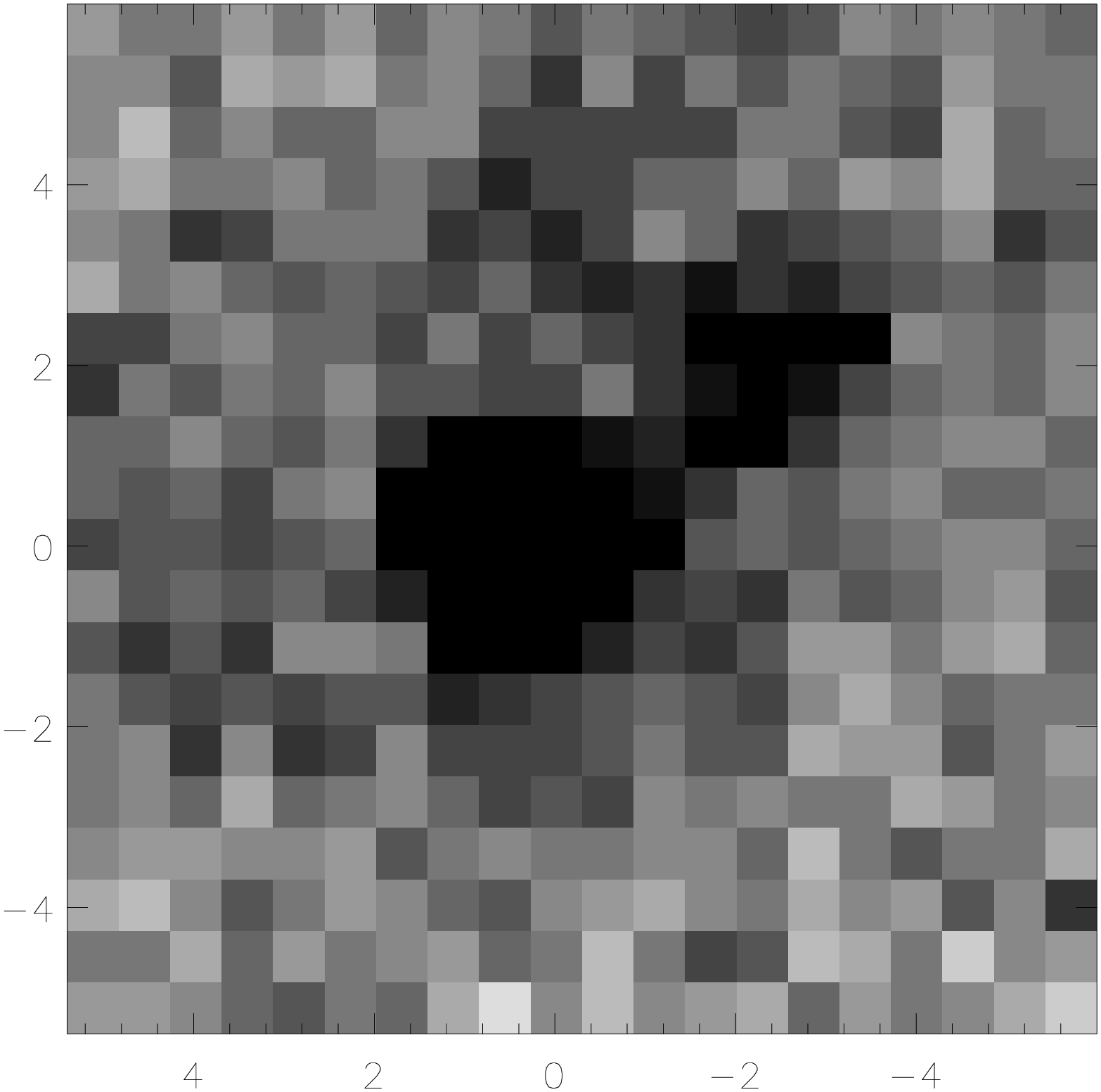,width=0.20\textwidth}\\
\\
\epsfig{file=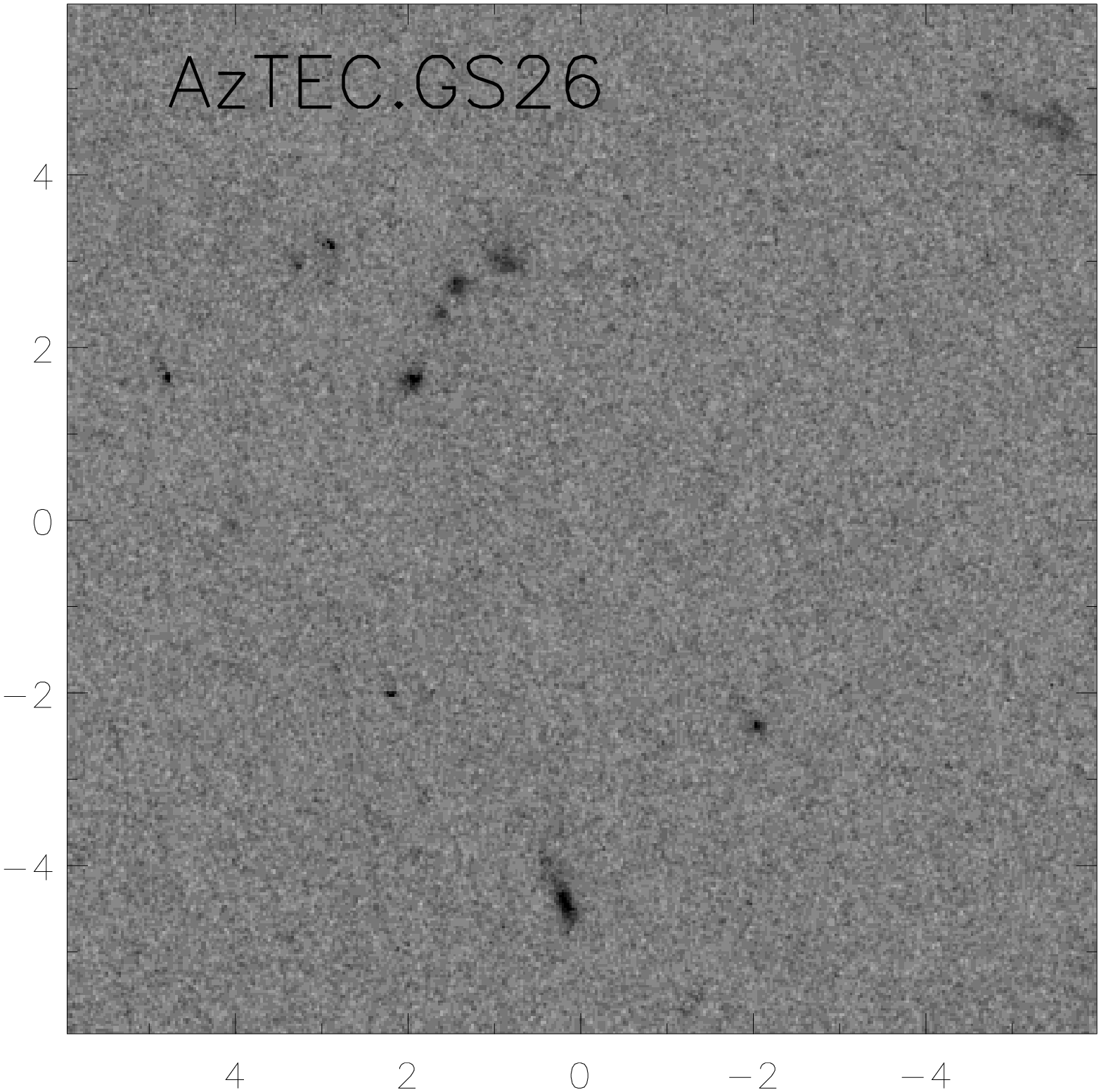,width=0.20\textwidth}&
\epsfig{file=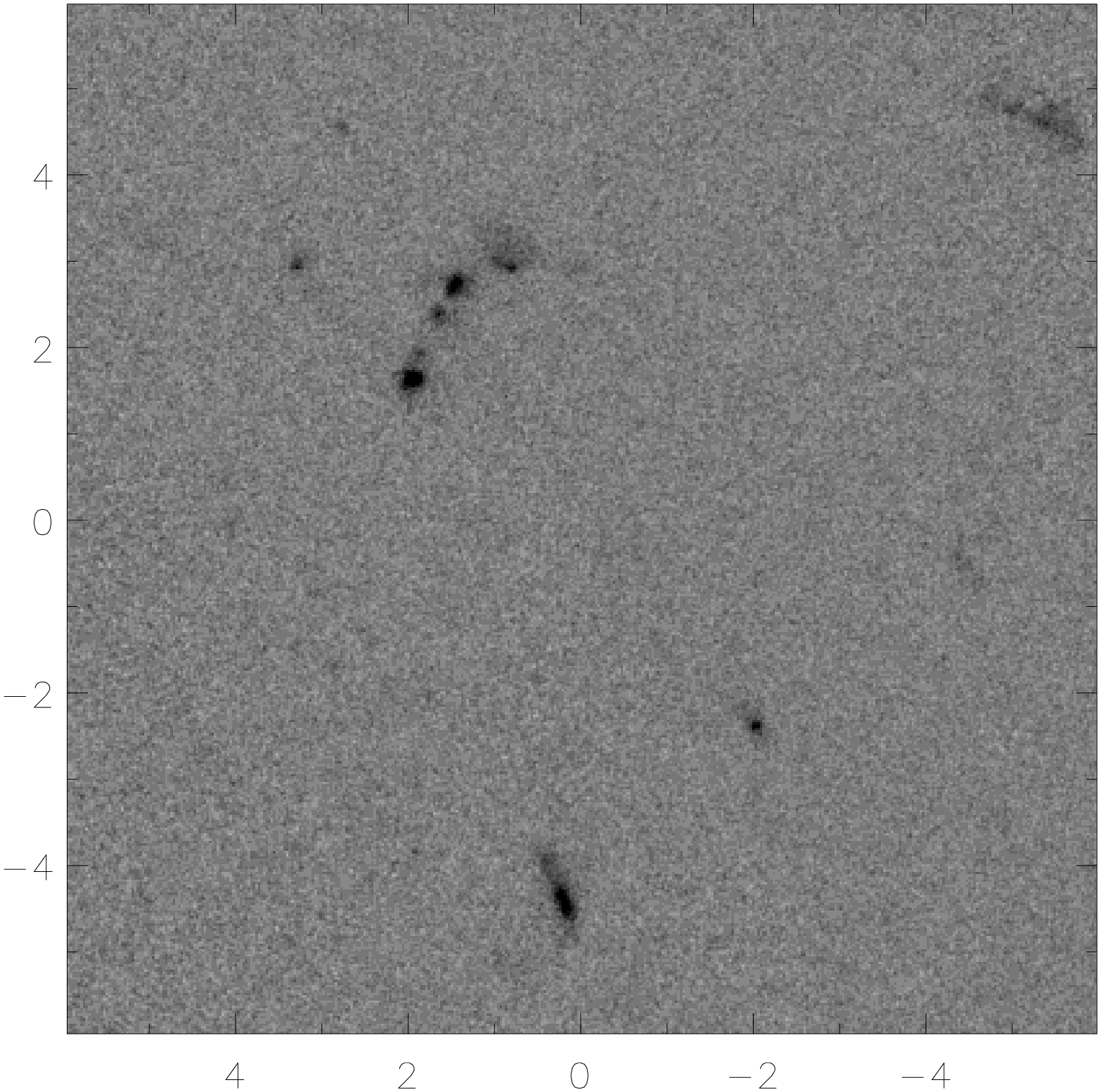,width=0.20\textwidth}&
\epsfig{file=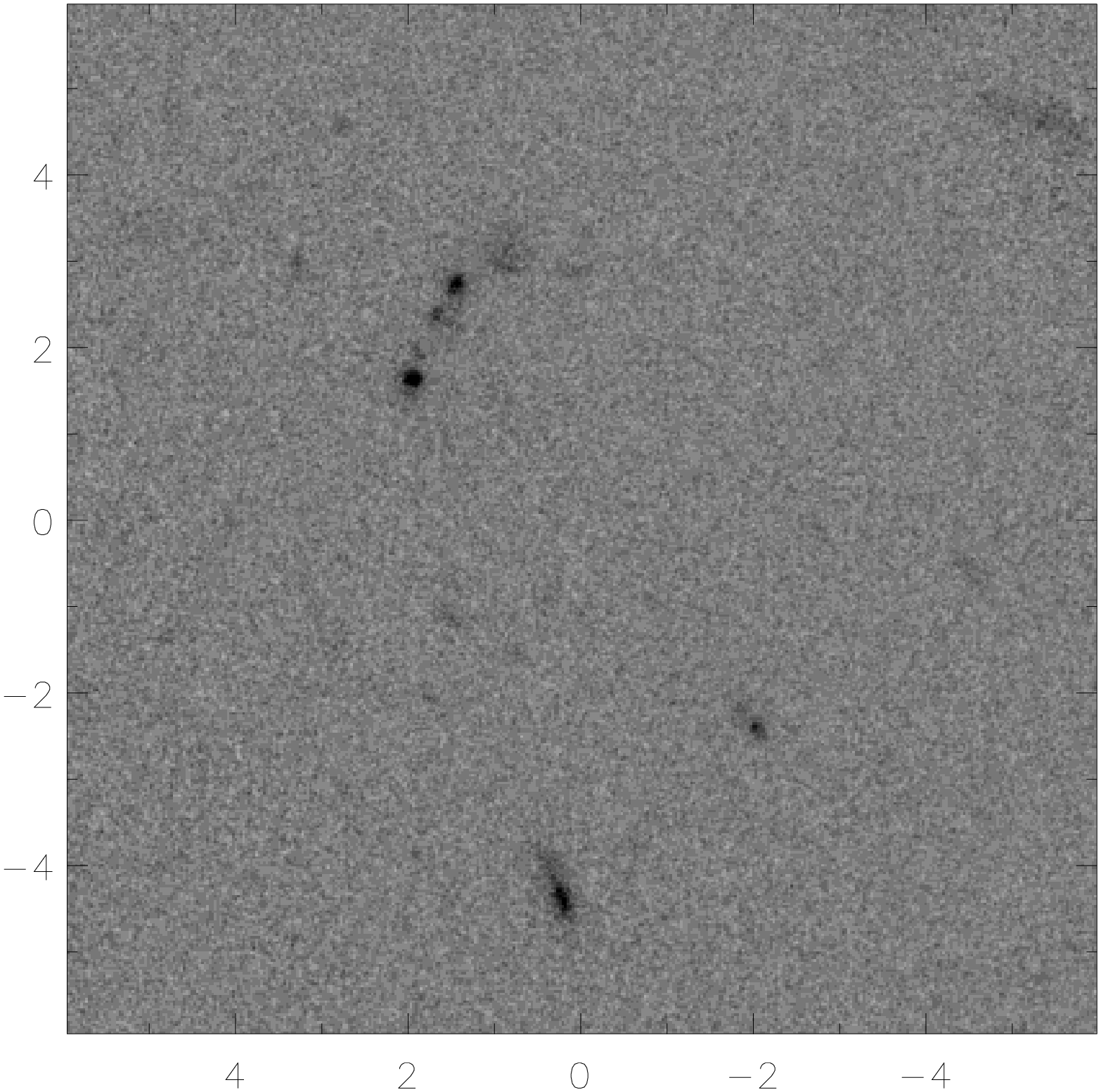,width=0.20\textwidth}&
\epsfig{file=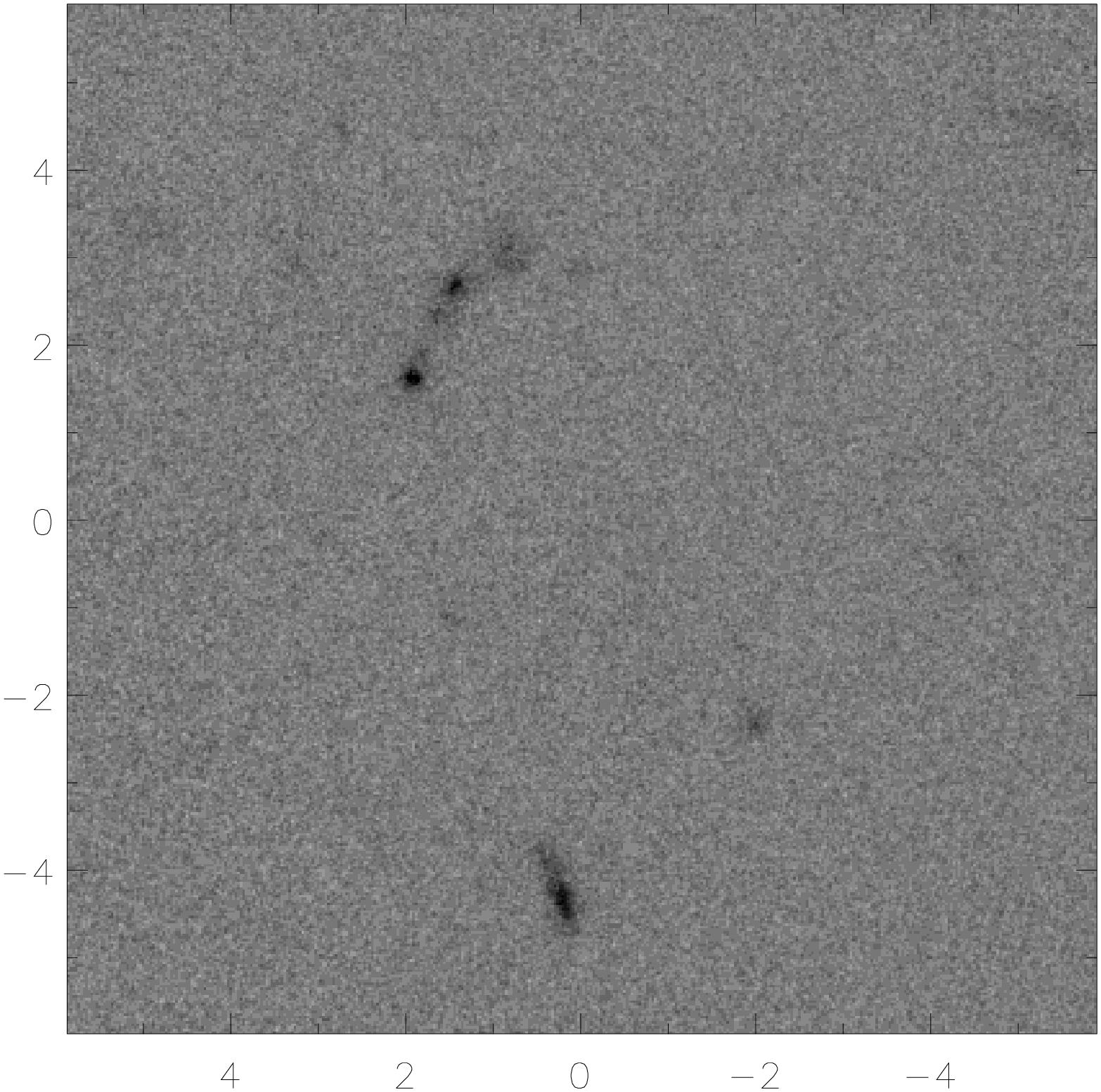,width=0.20\textwidth}\\
\epsfig{file=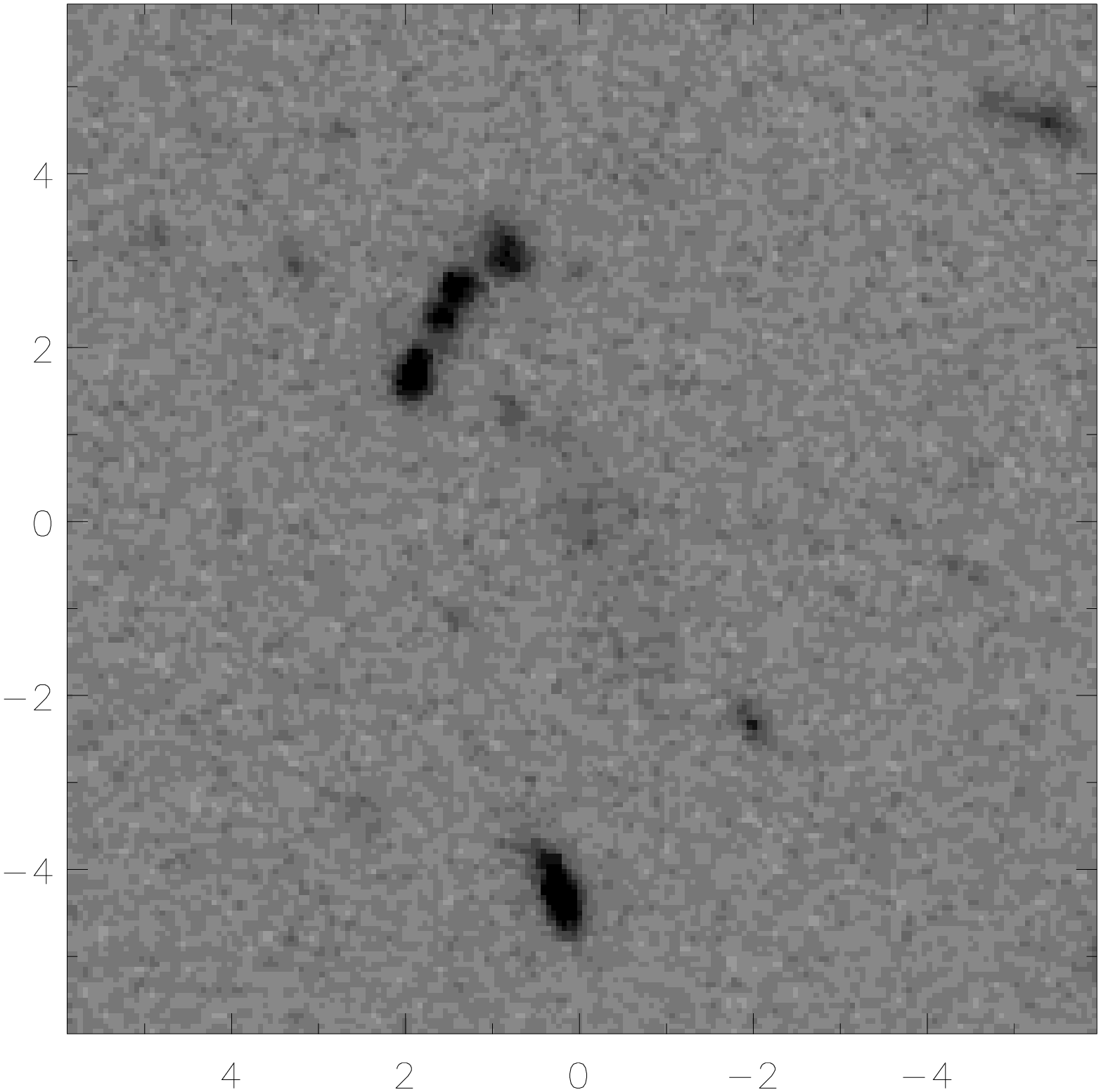,width=0.20\textwidth}&
\epsfig{file=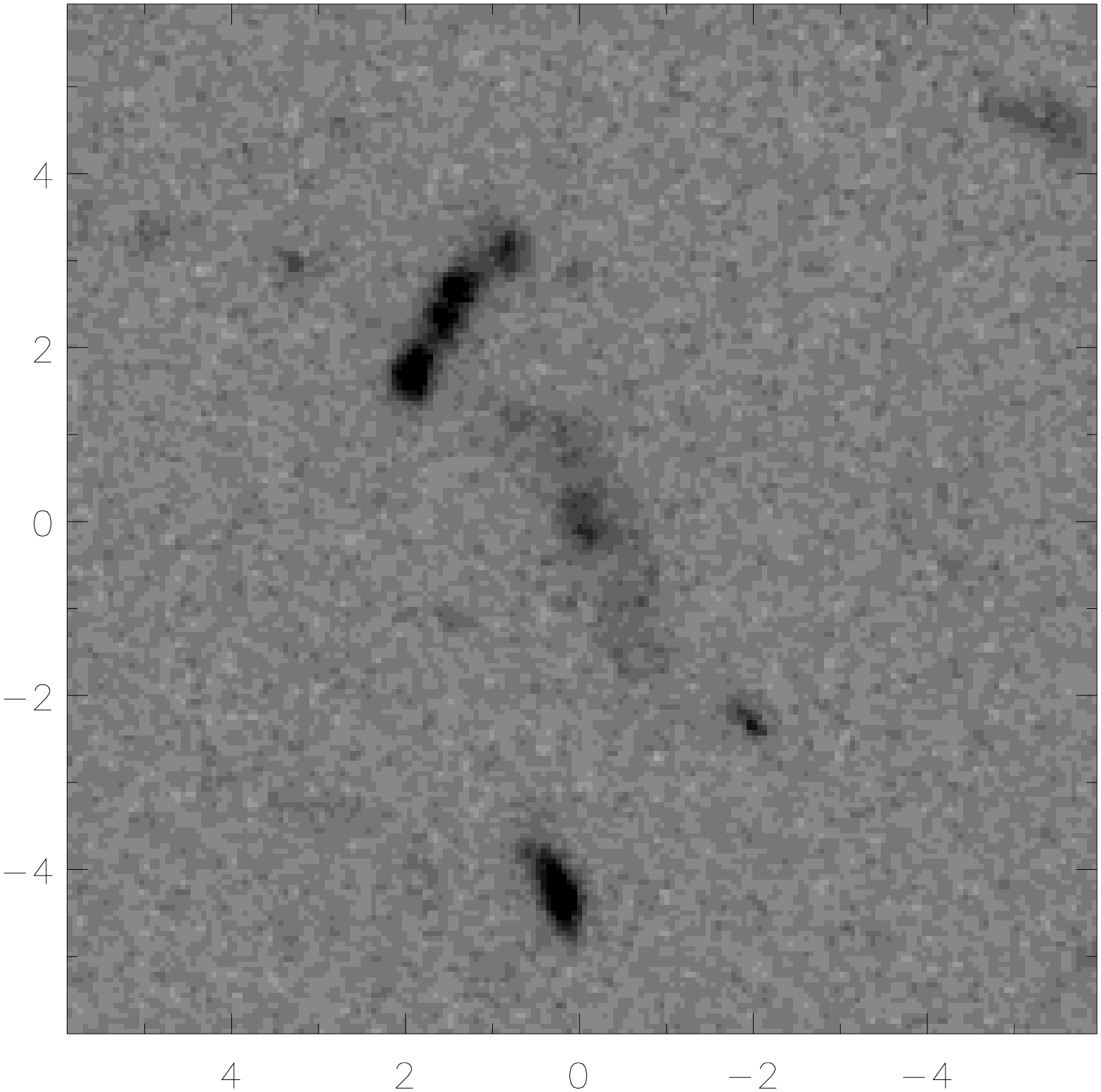,width=0.20\textwidth}&
\epsfig{file=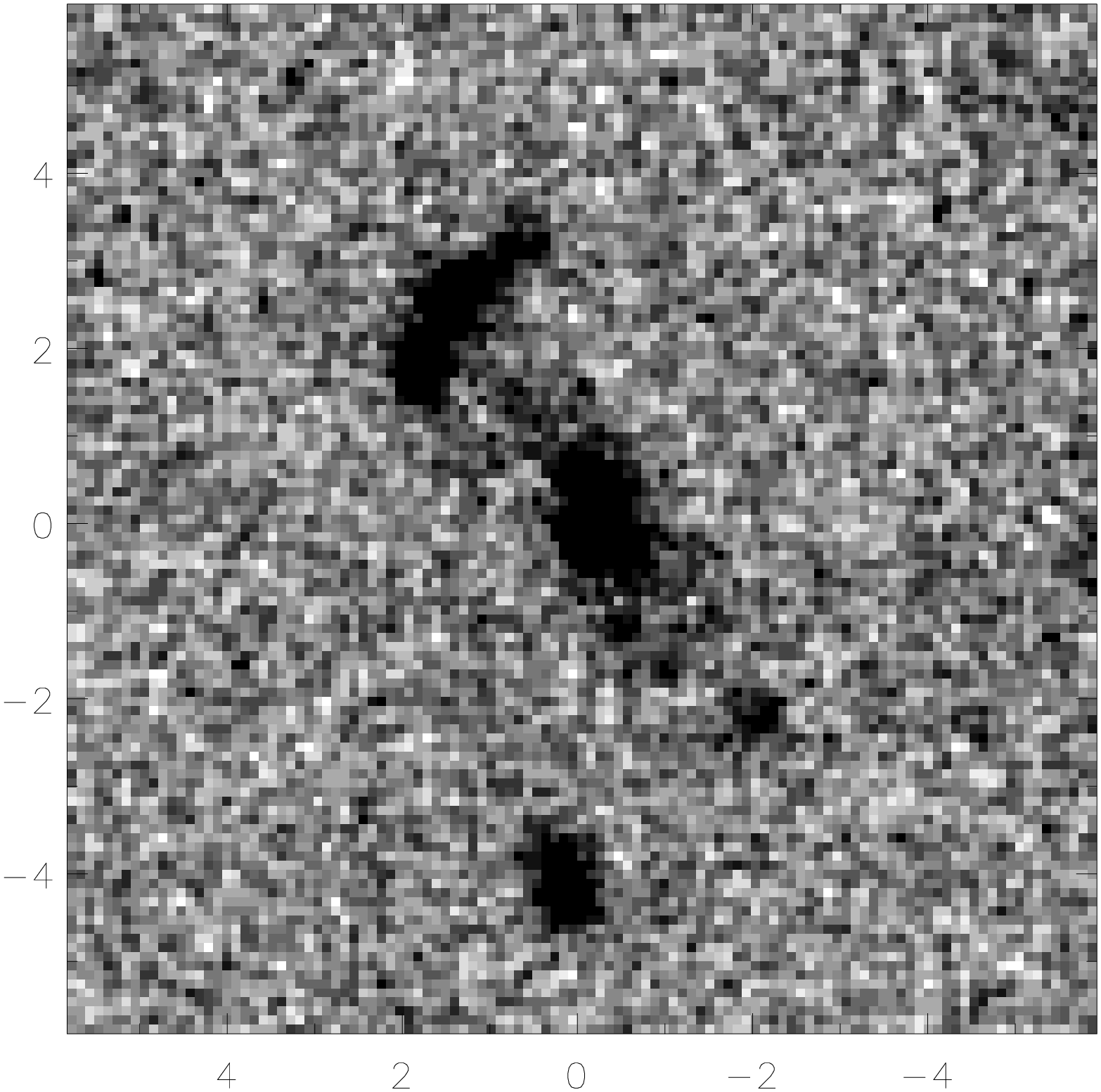,width=0.20\textwidth}&
\epsfig{file=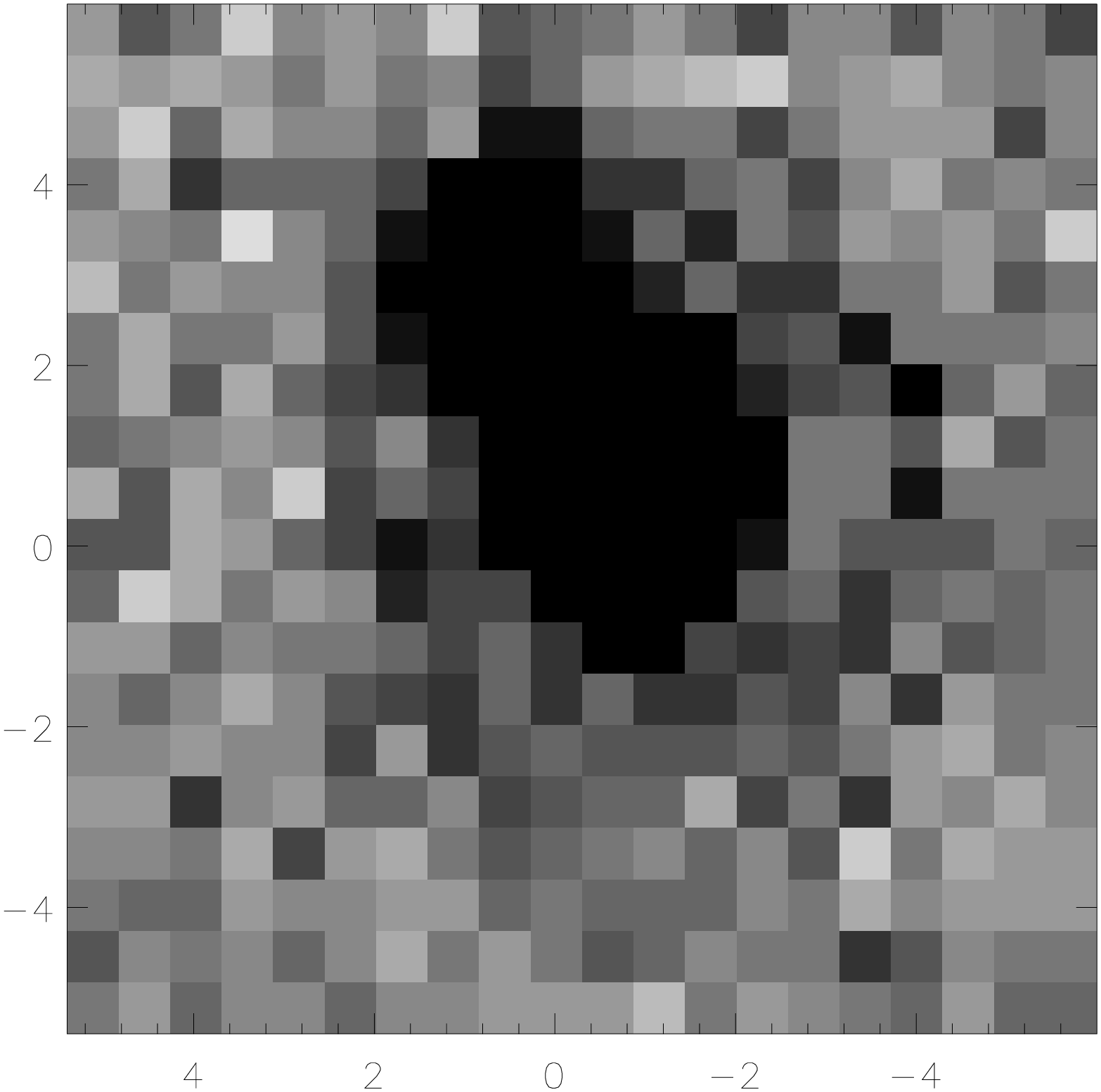,width=0.20\textwidth}\\
\end{tabular}
\addtocounter{figure}{-1}
\caption{- continued}
\vfil}
\end{figure*}
\end{center}


\begin{center}
\begin{figure*}
\vbox to220mm{\vfil
\begin{tabular}{cccccccc}
\epsfig{file=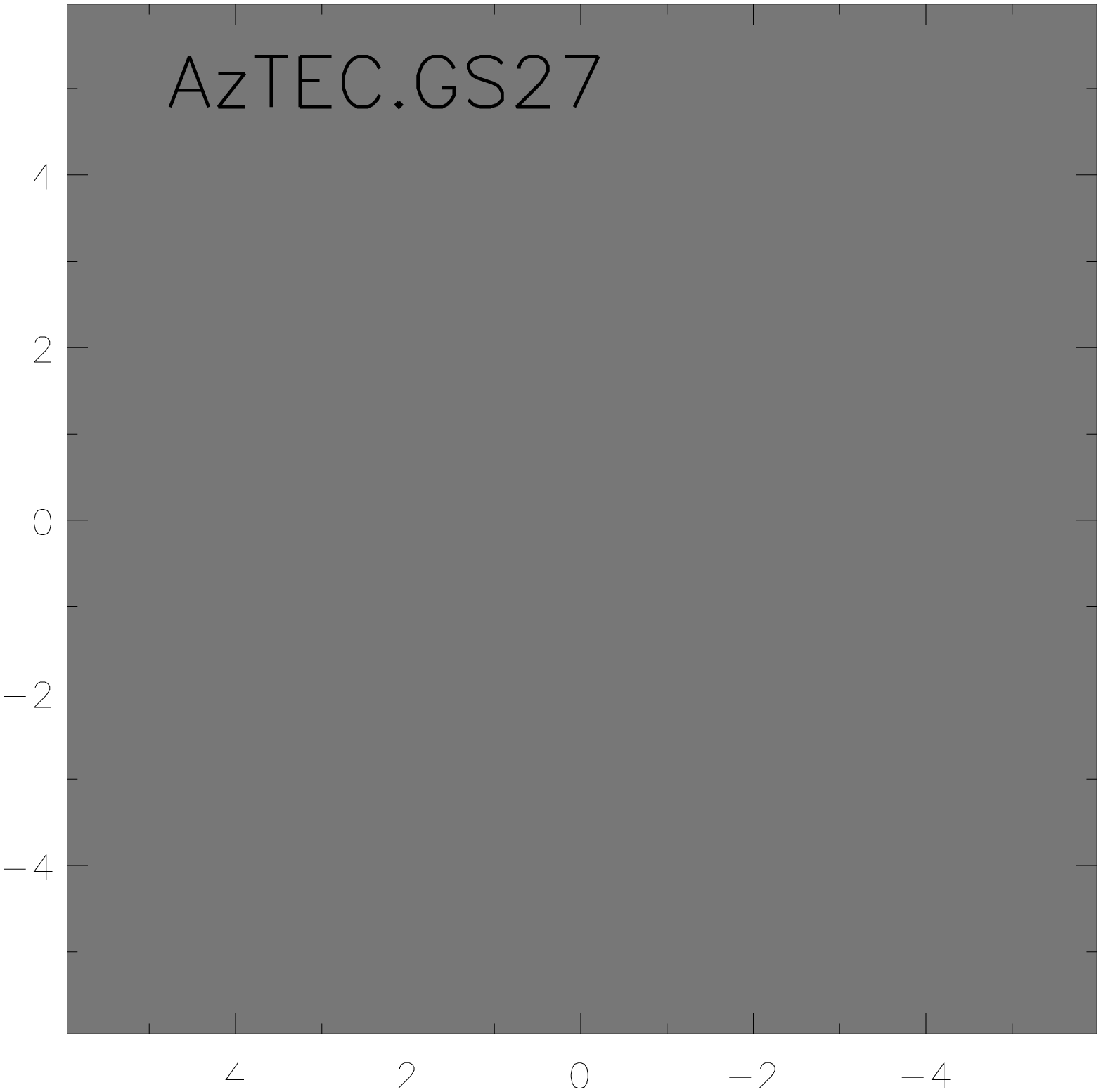,width=0.20\textwidth}&
\epsfig{file=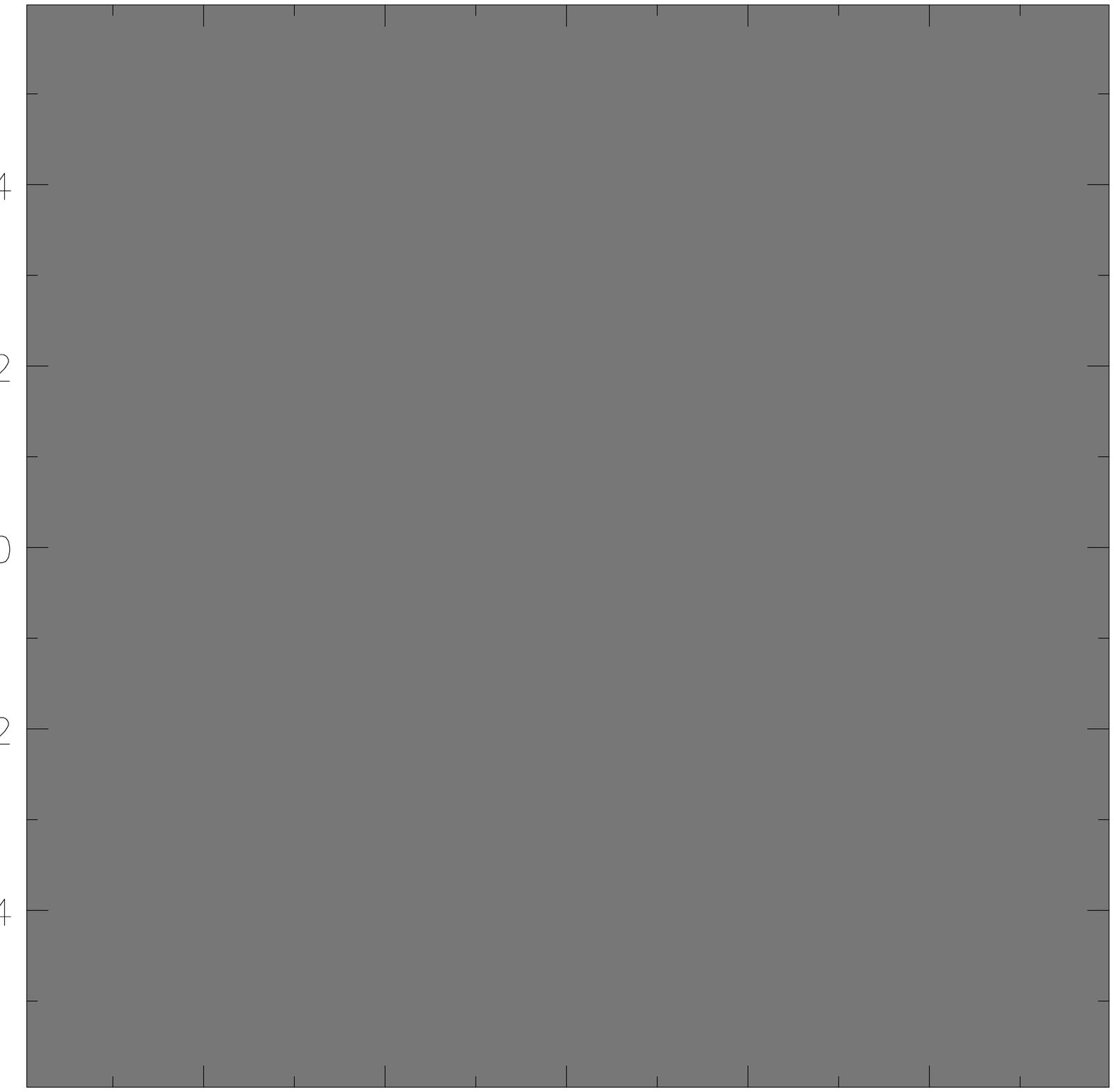,width=0.20\textwidth}&
\epsfig{file=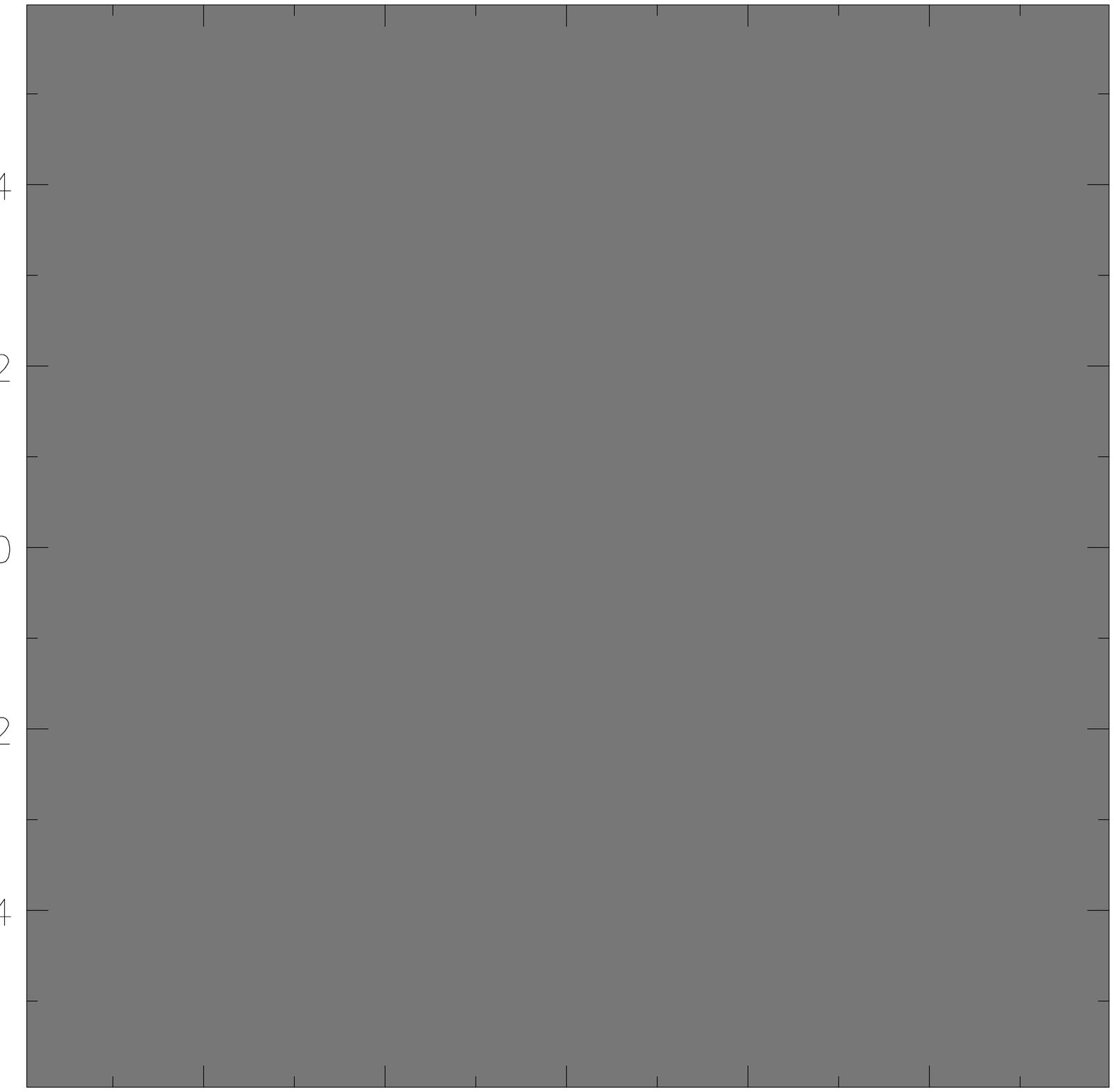,width=0.20\textwidth}&
\epsfig{file=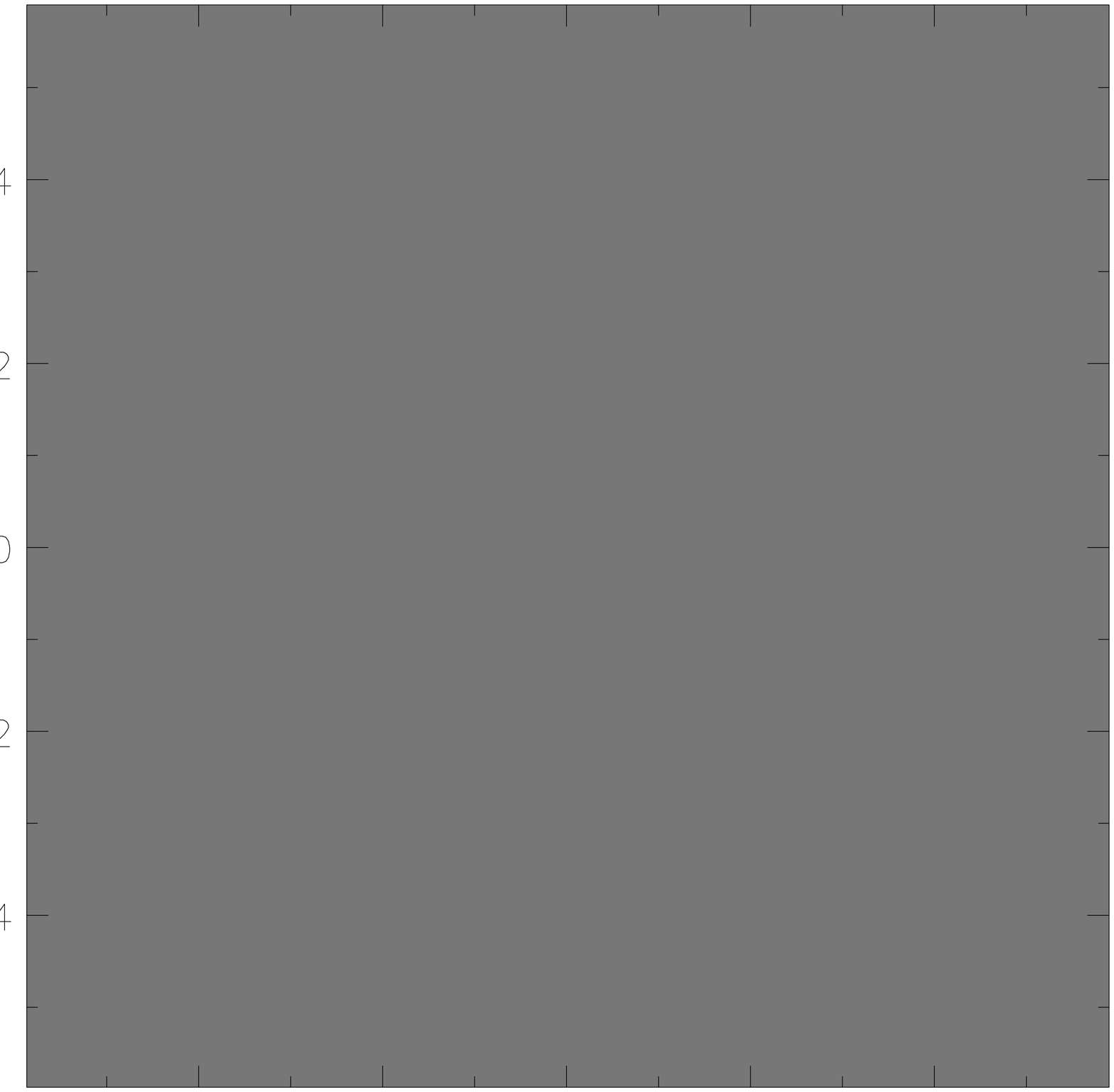,width=0.20\textwidth}\\
\epsfig{file=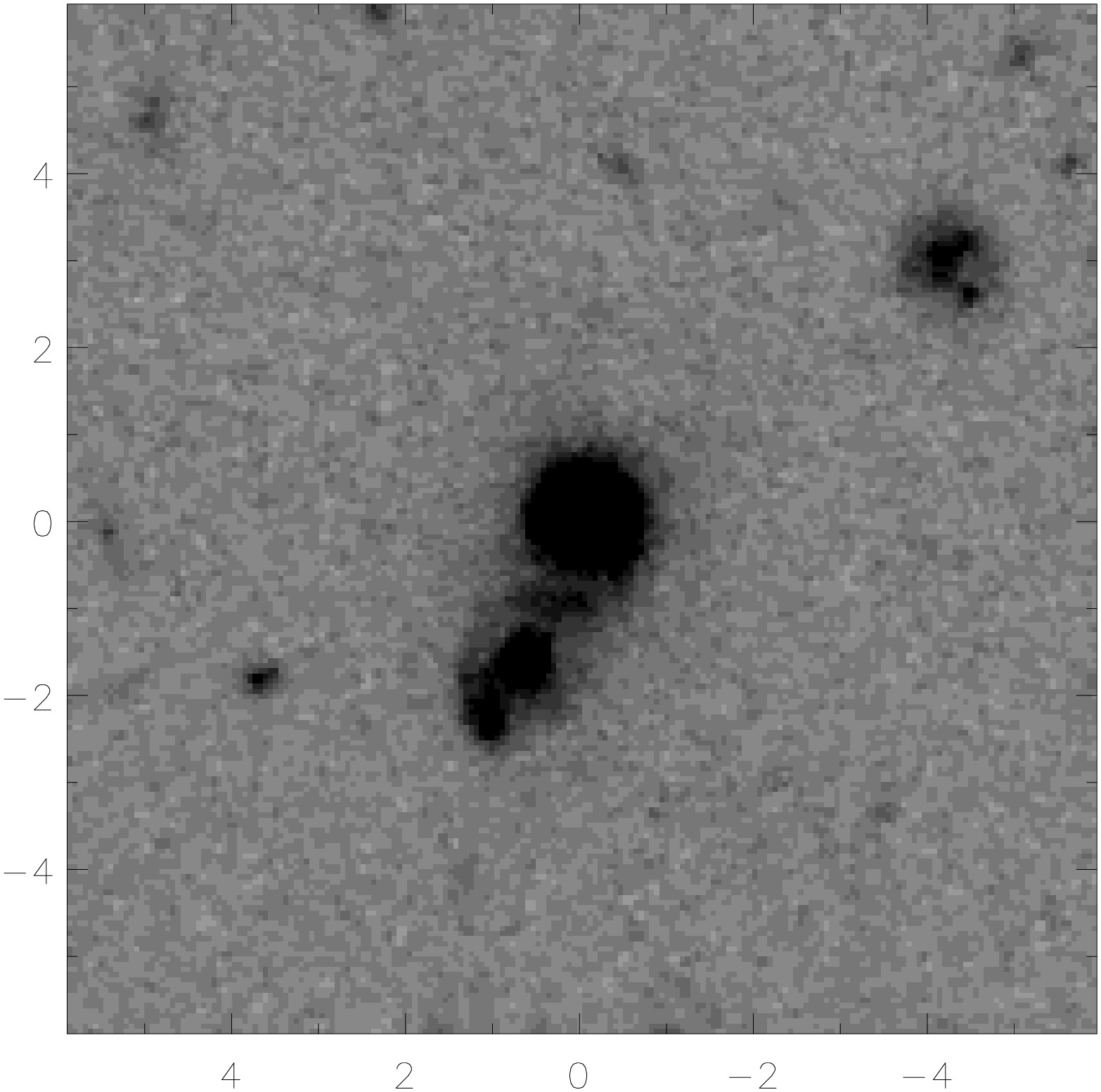,width=0.20\textwidth}&
\epsfig{file=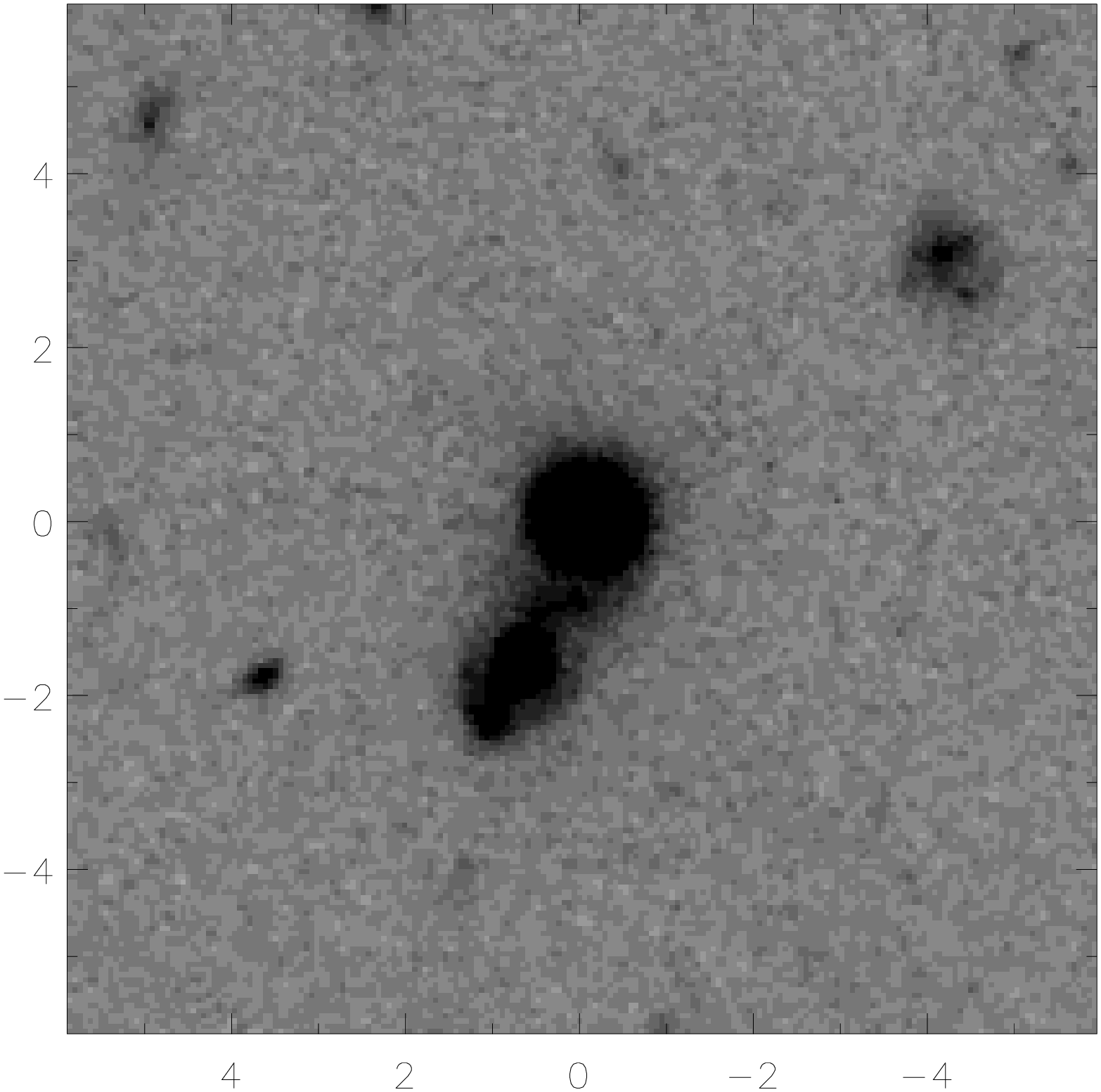,width=0.20\textwidth}&
\epsfig{file=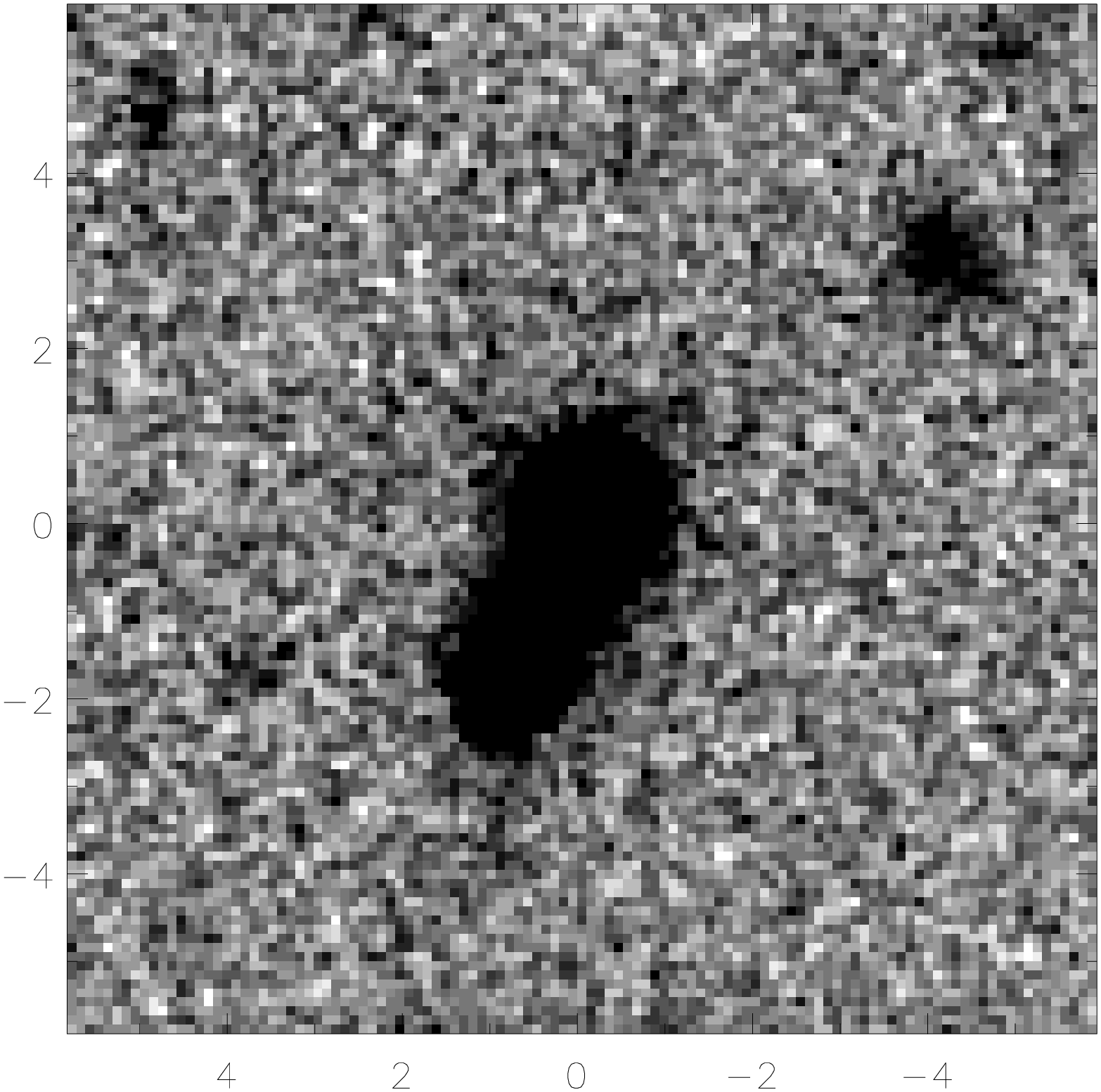,width=0.20\textwidth}&
\epsfig{file=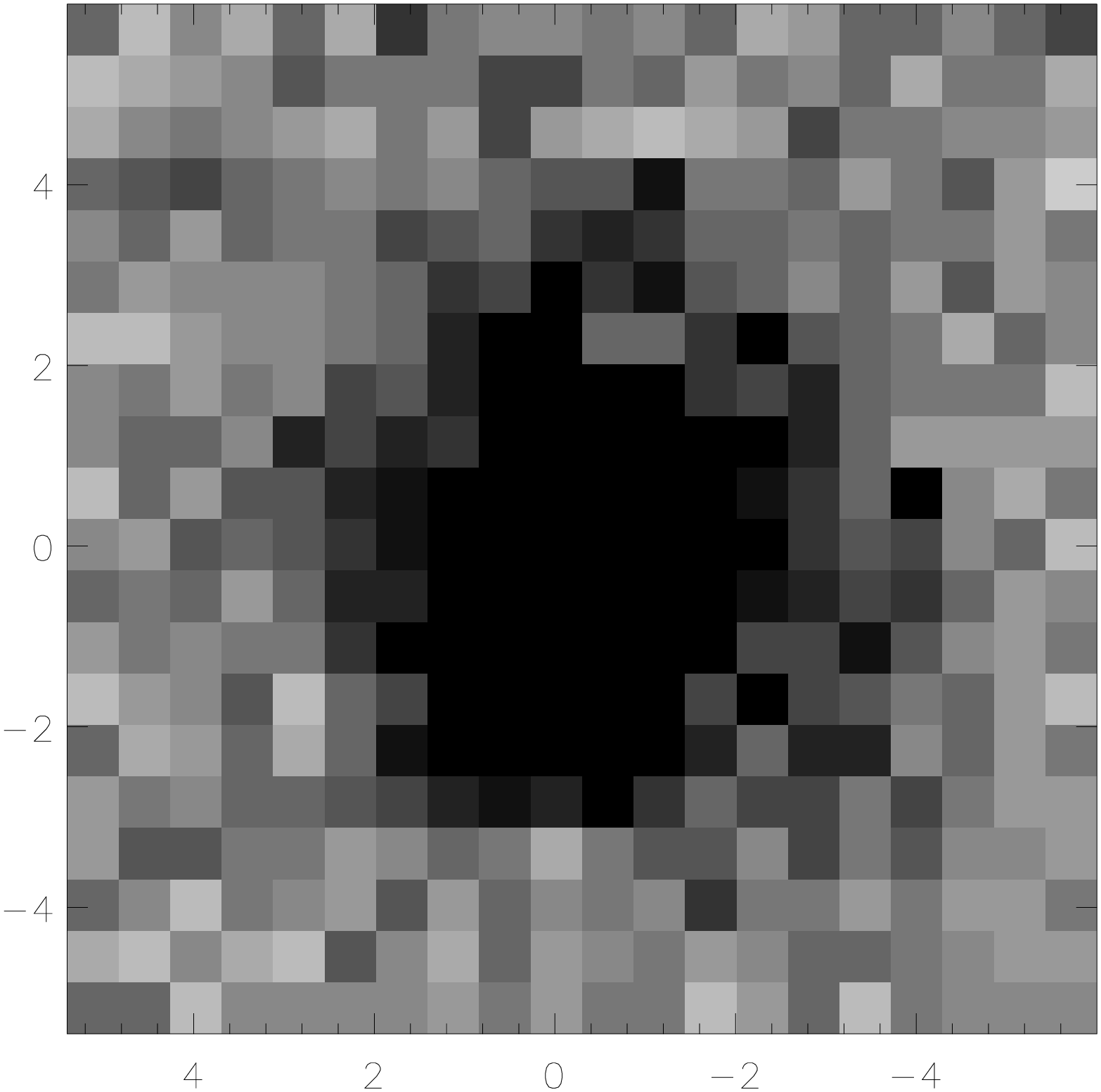,width=0.20\textwidth}\\
\\
\epsfig{file=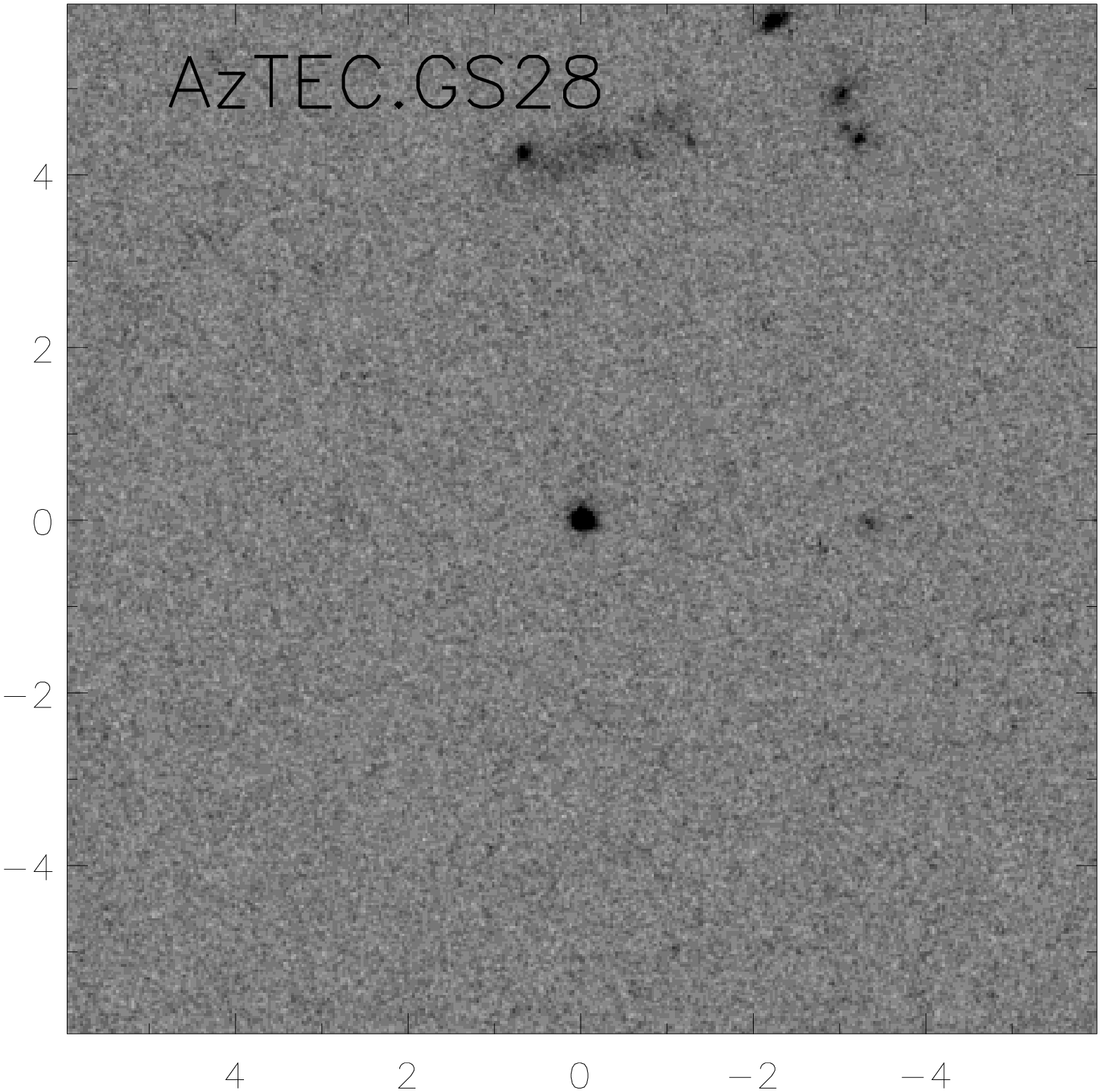,width=0.20\textwidth}&
\epsfig{file=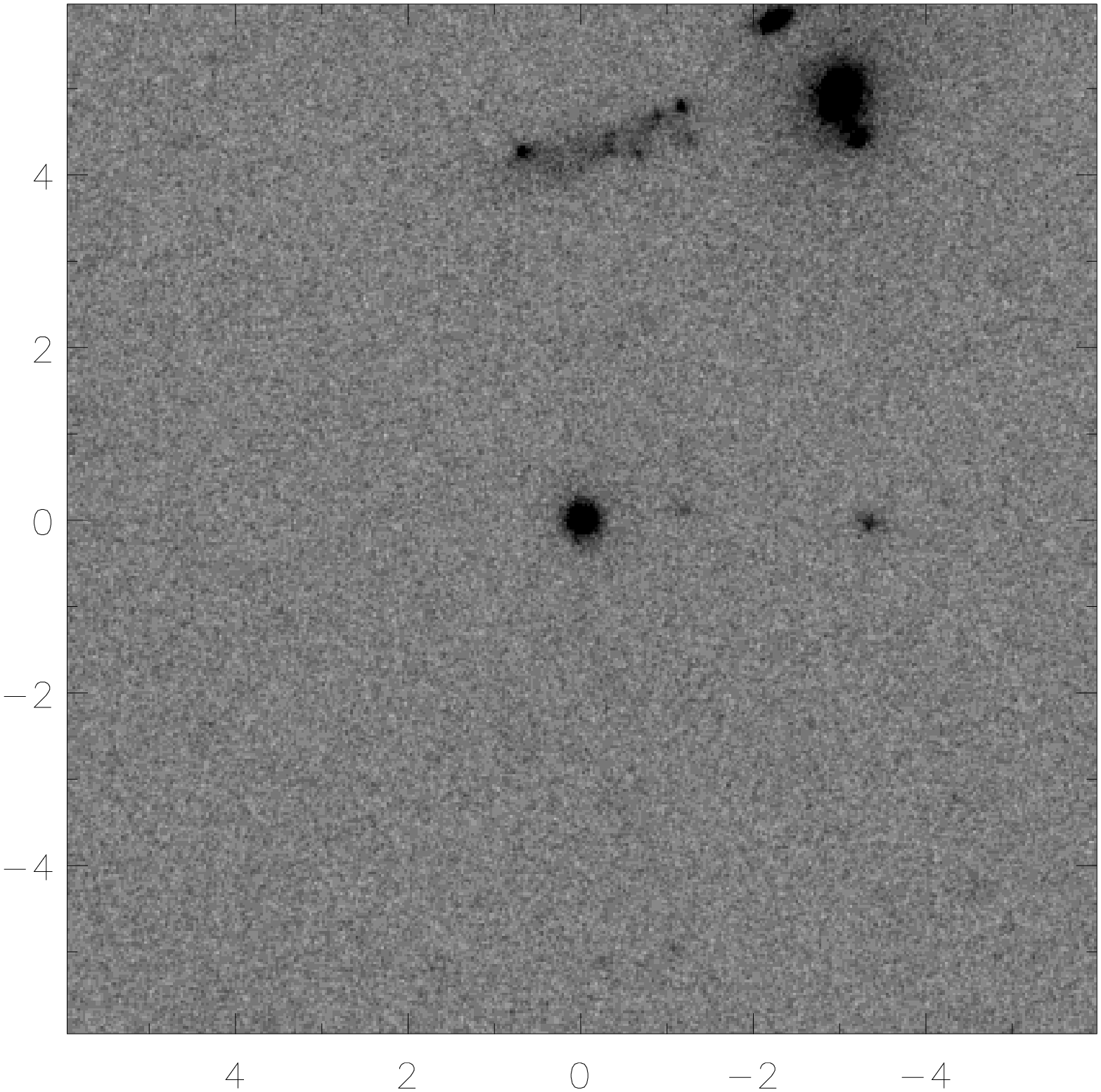,width=0.20\textwidth}&
\epsfig{file=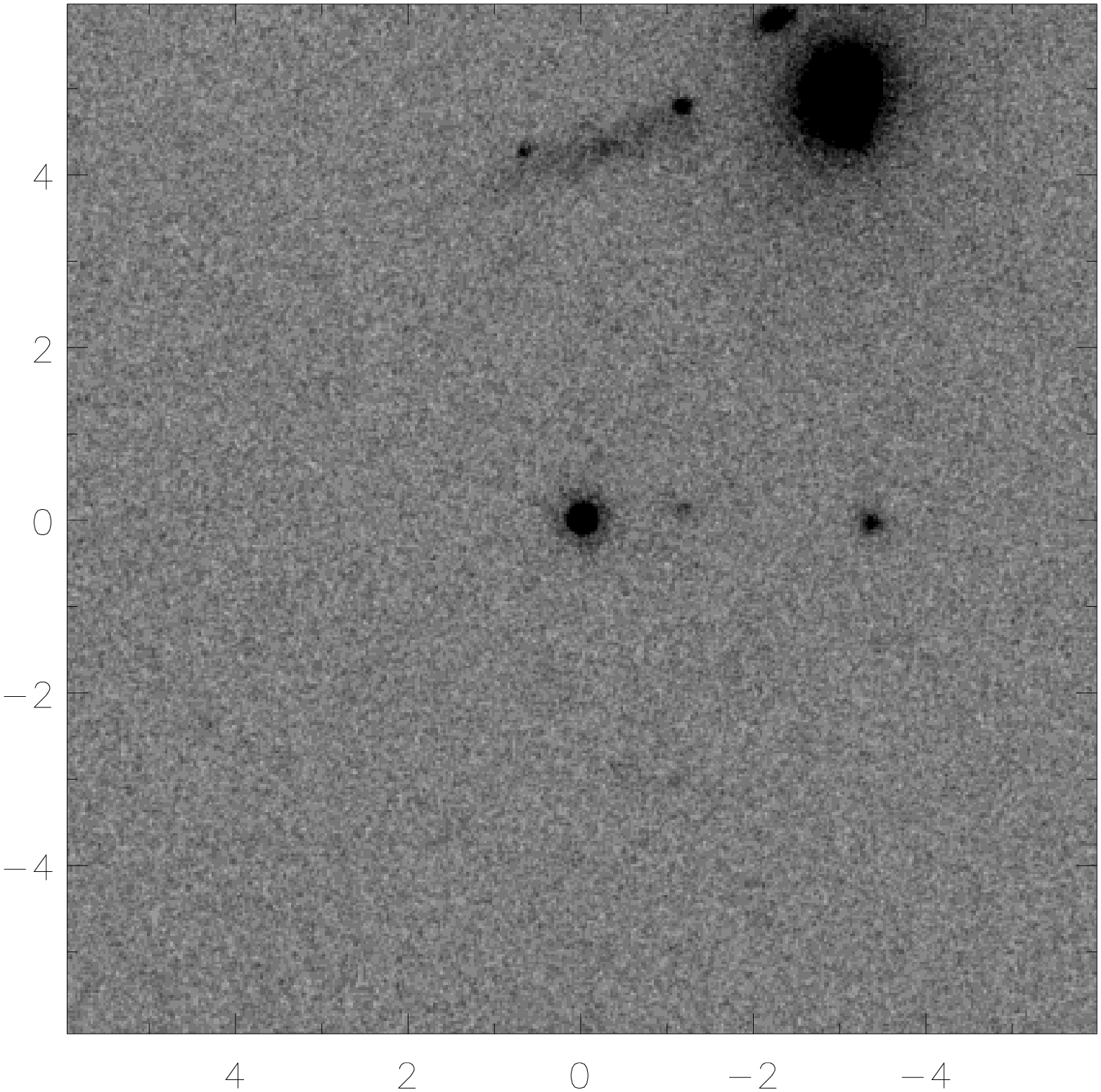,width=0.20\textwidth}&
\epsfig{file=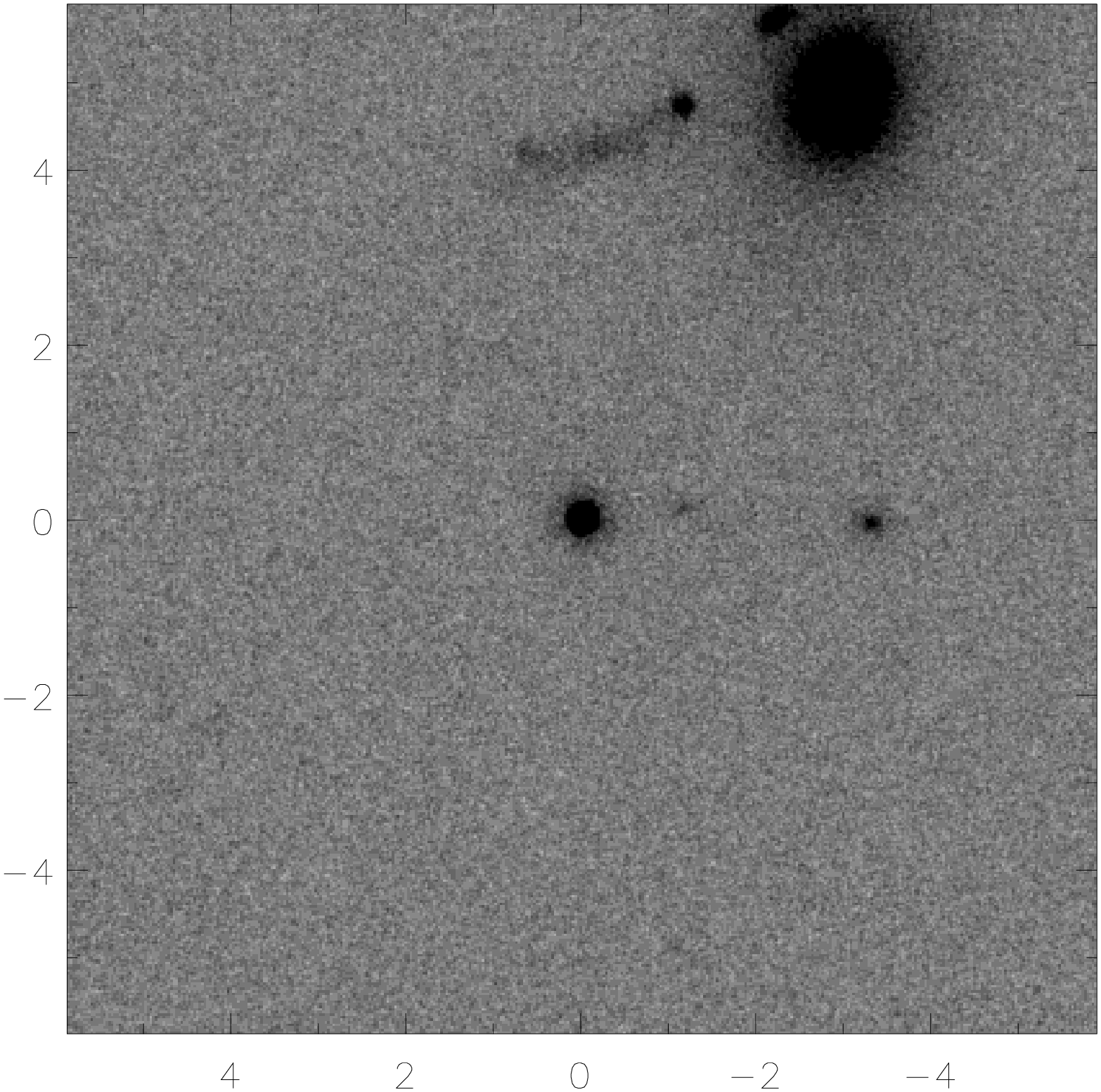,width=0.20\textwidth}\\
\epsfig{file=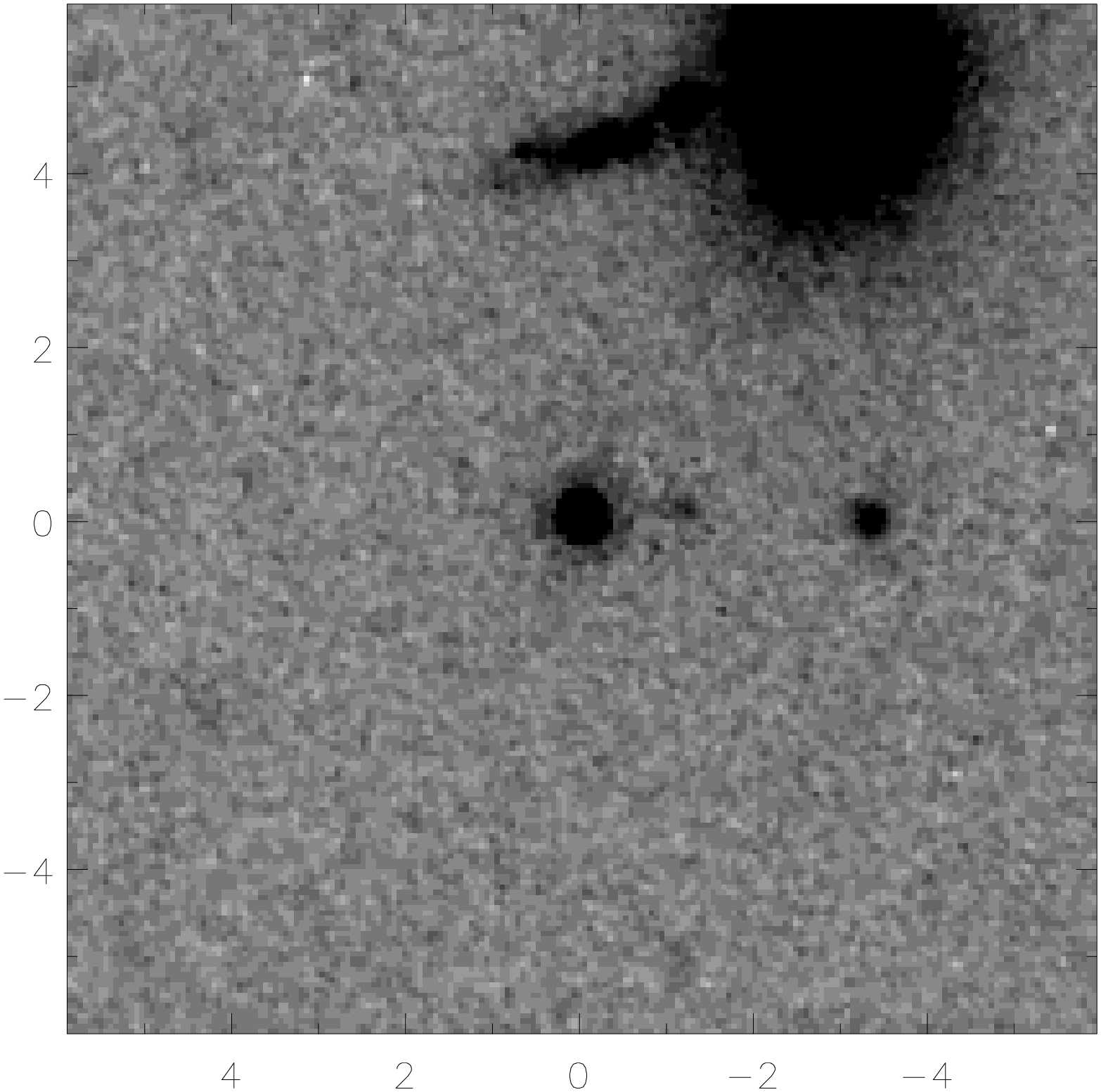,width=0.20\textwidth}&
\epsfig{file=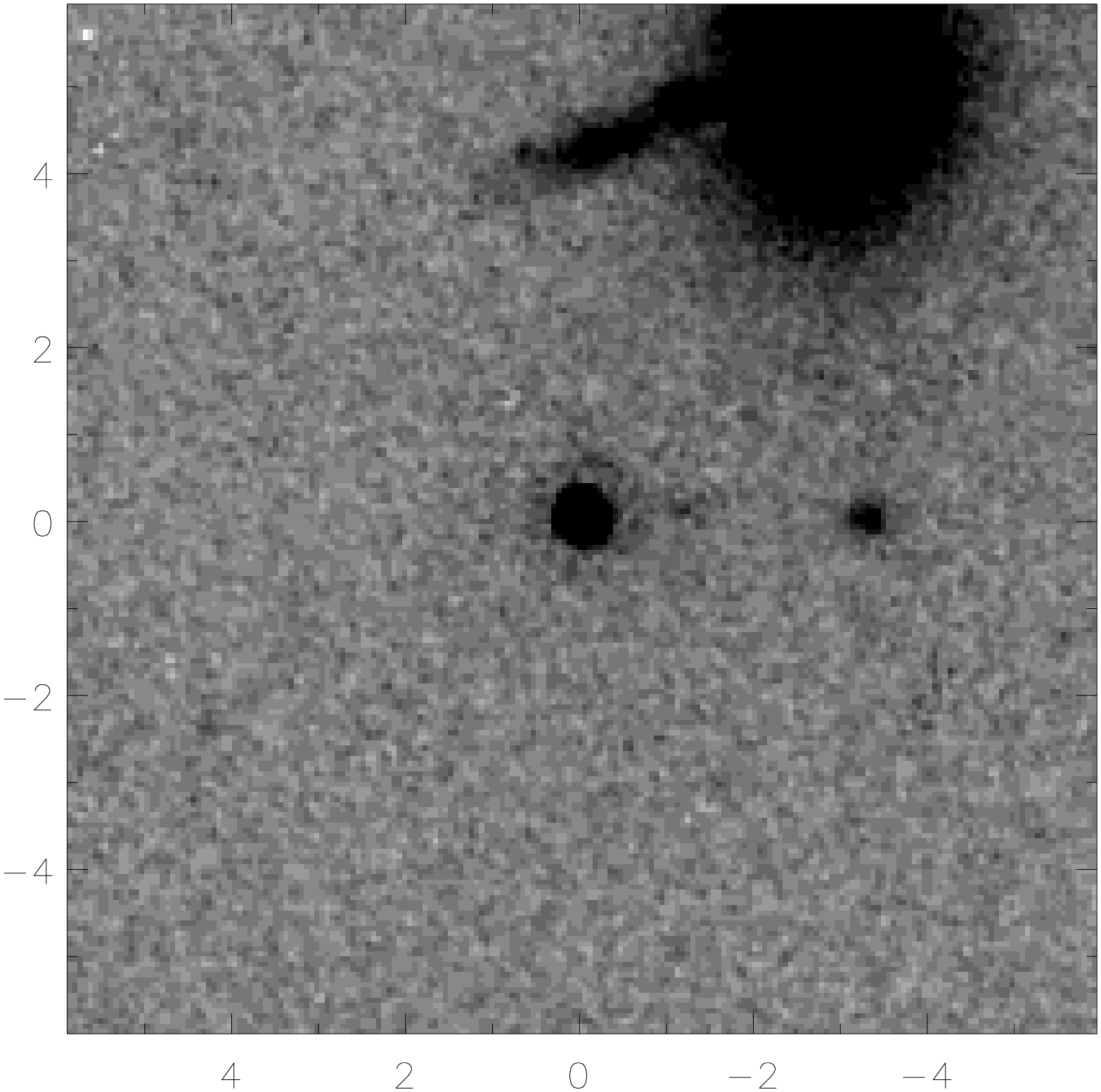,width=0.20\textwidth}&
\epsfig{file=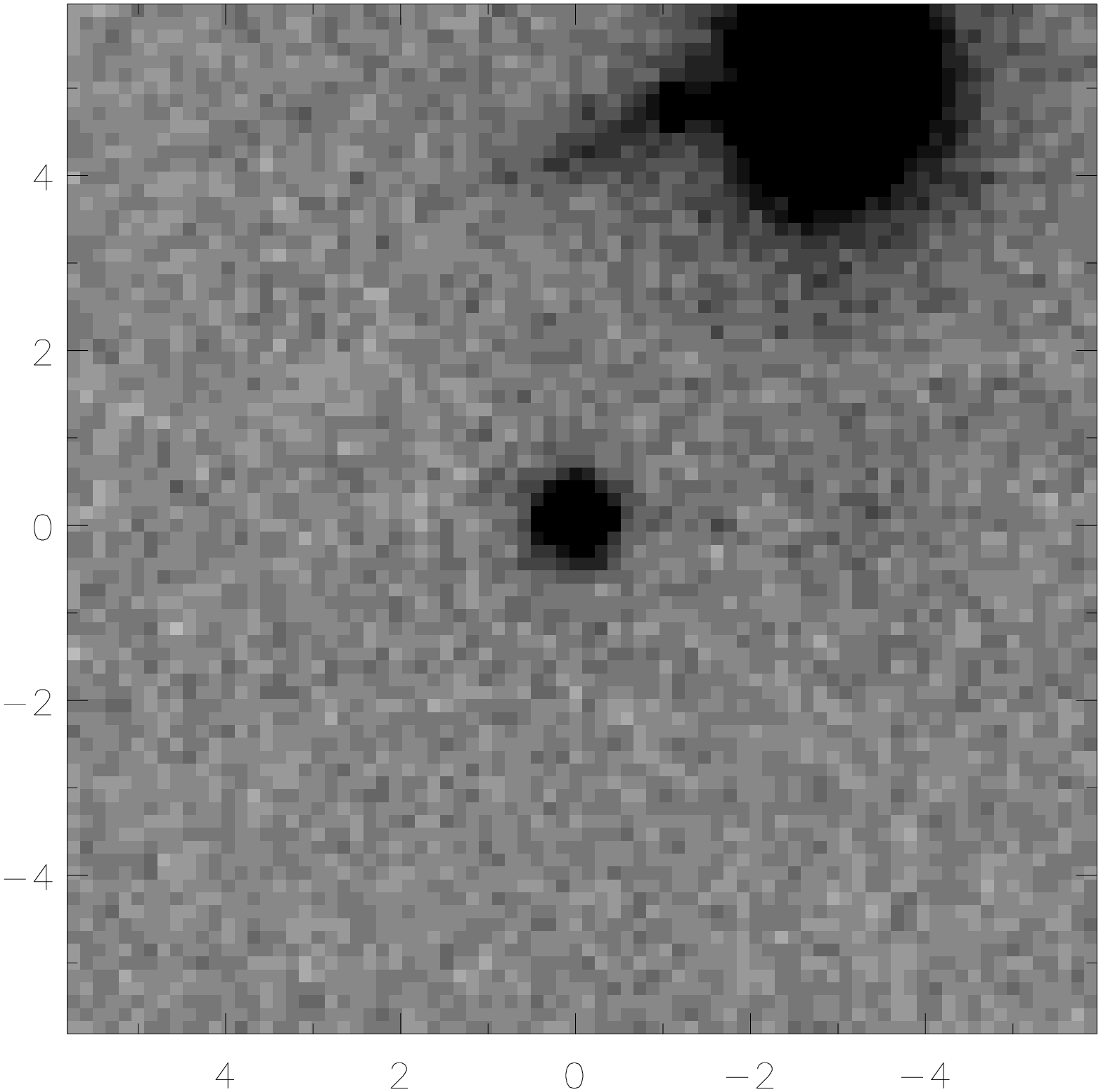,width=0.20\textwidth}&
\epsfig{file=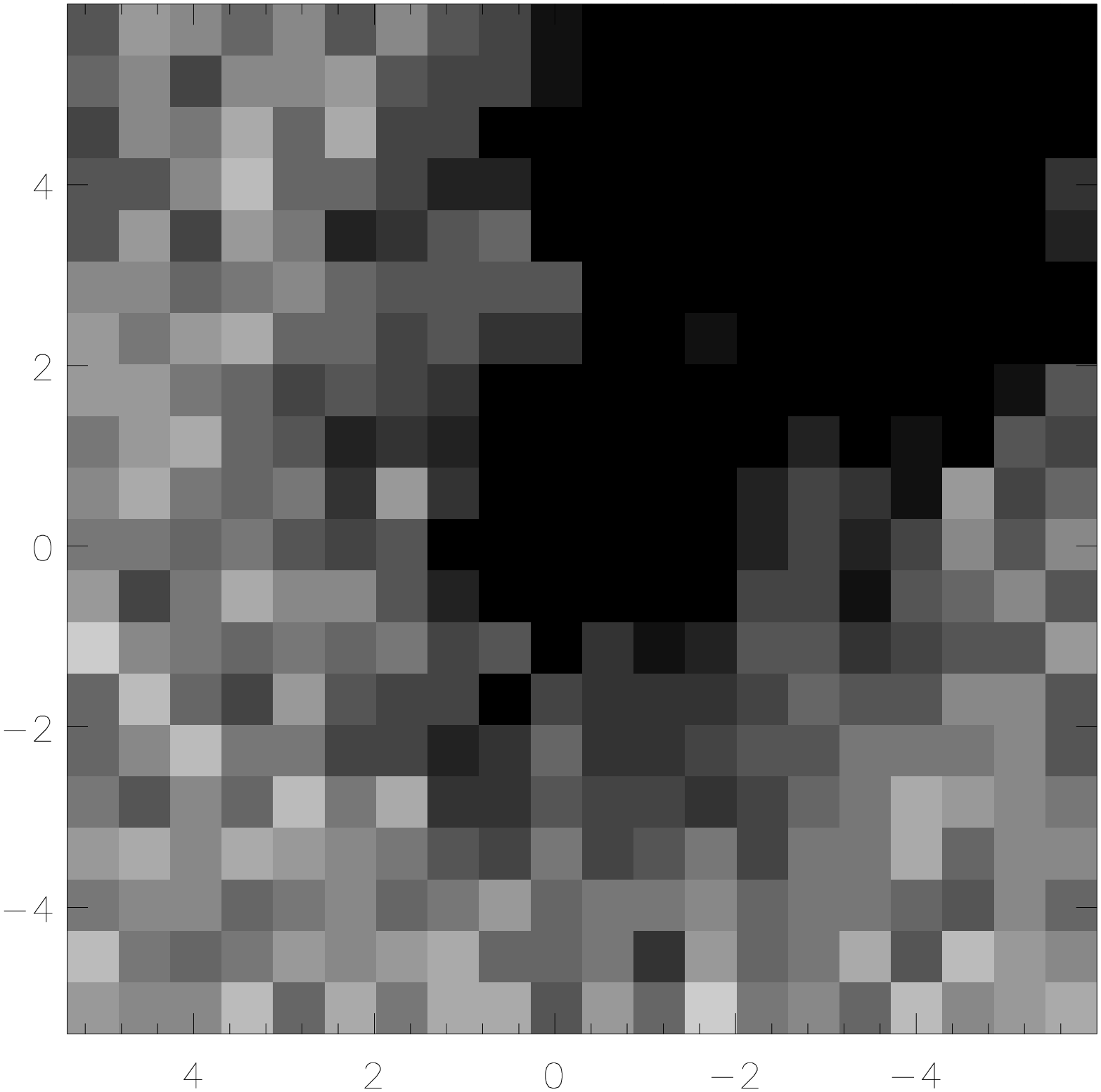,width=0.20\textwidth}\\
\\
\epsfig{file=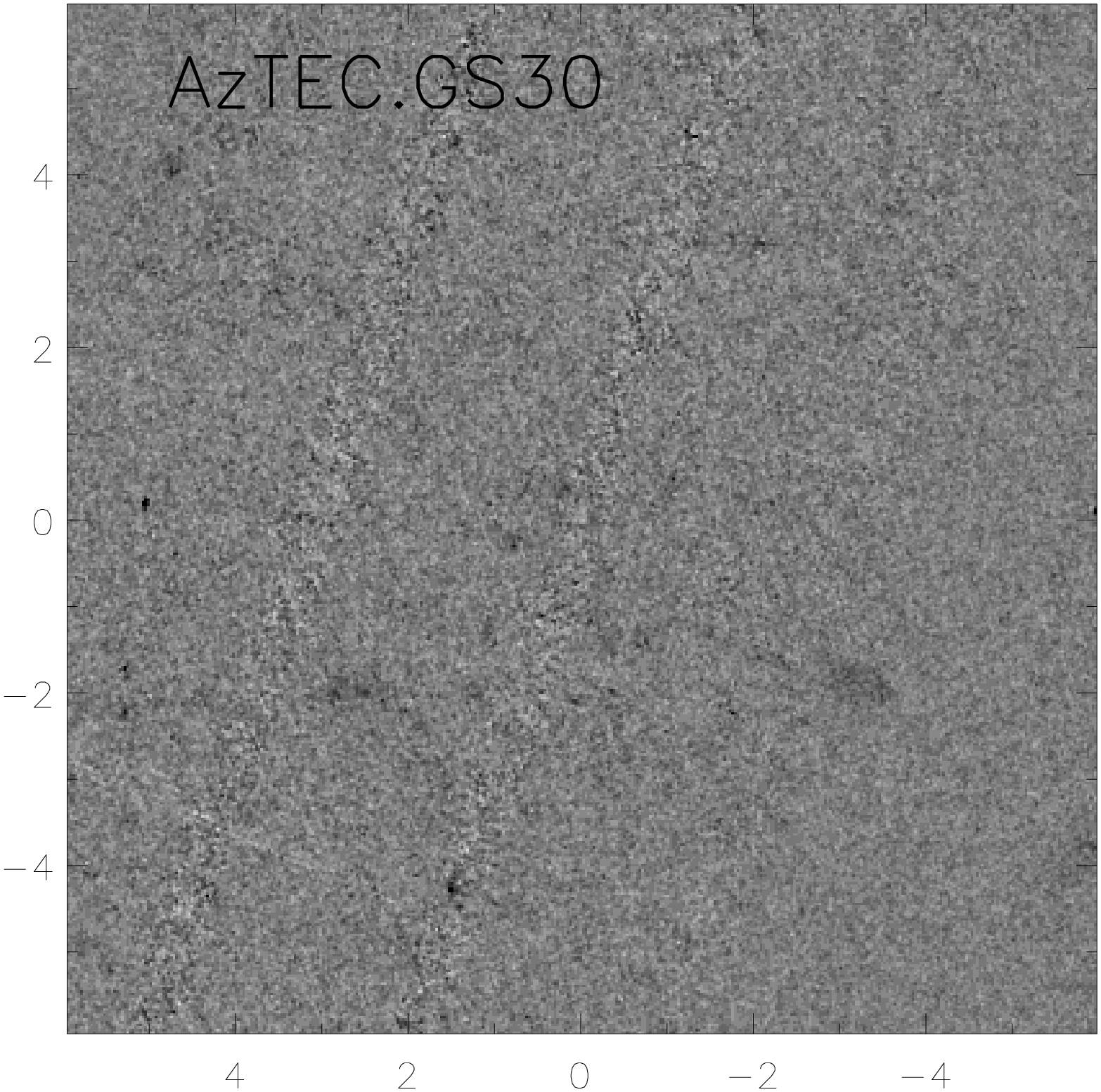,width=0.20\textwidth}&
\epsfig{file=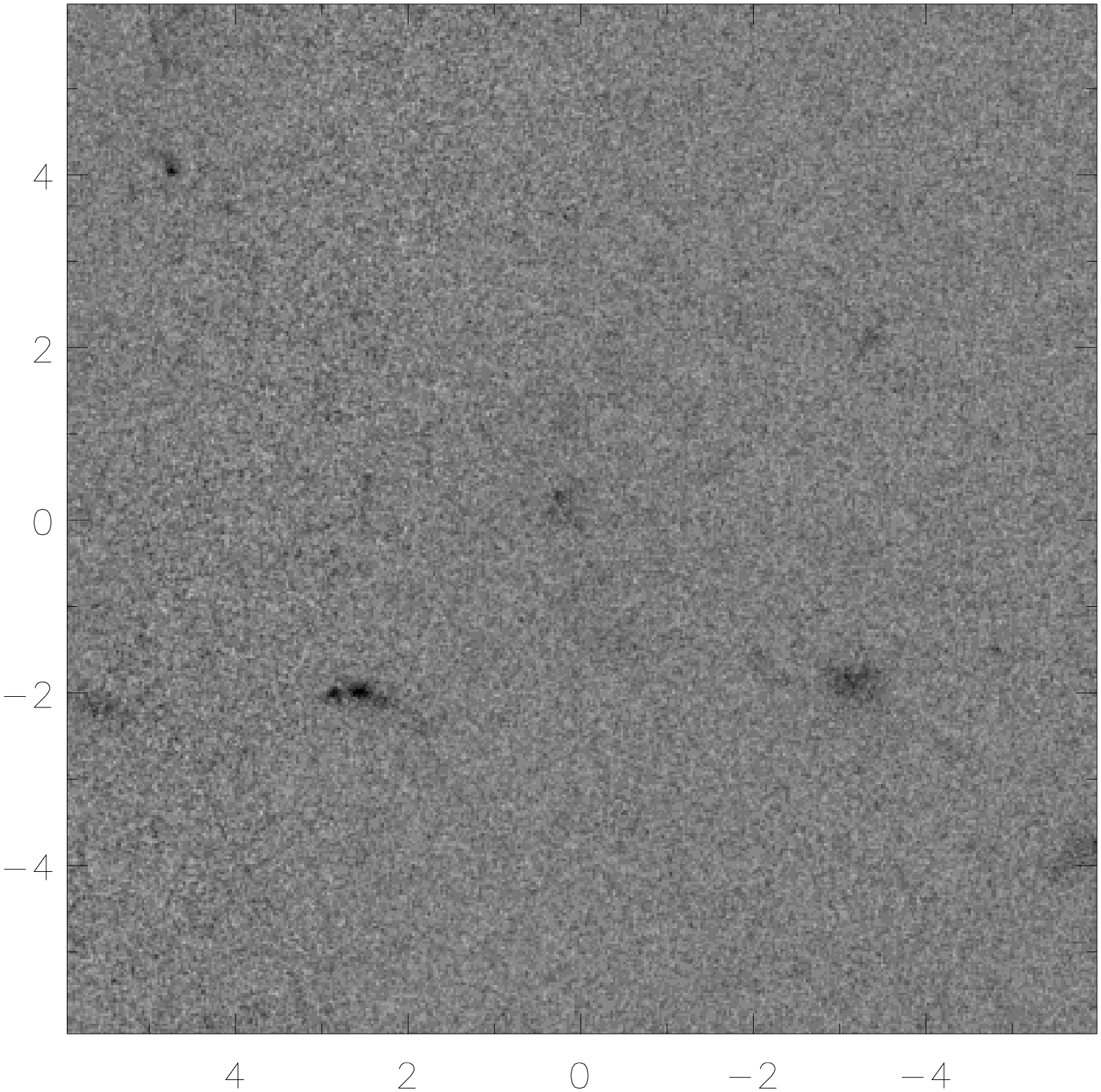,width=0.20\textwidth}&
\epsfig{file=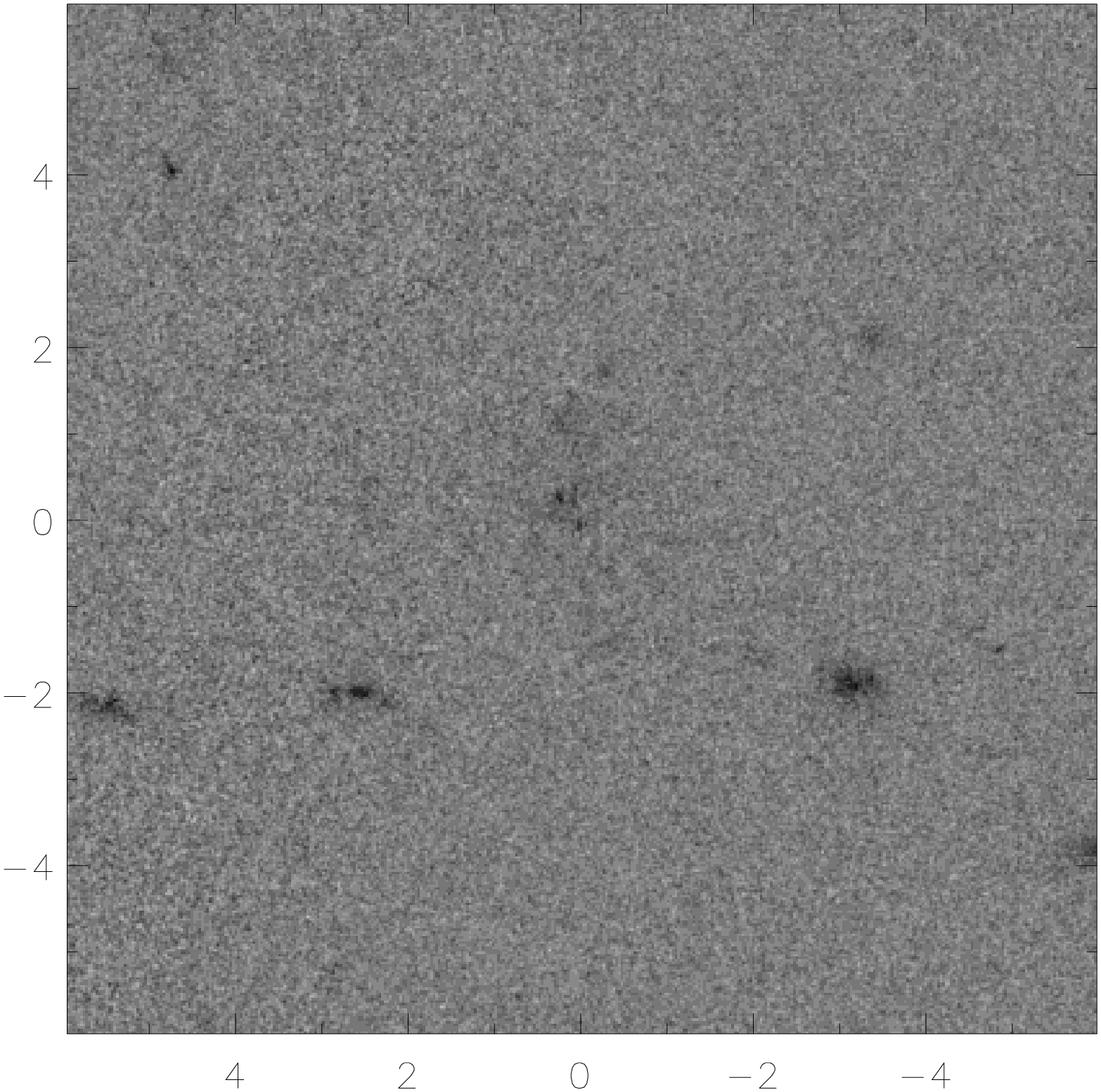,width=0.20\textwidth}&
\epsfig{file=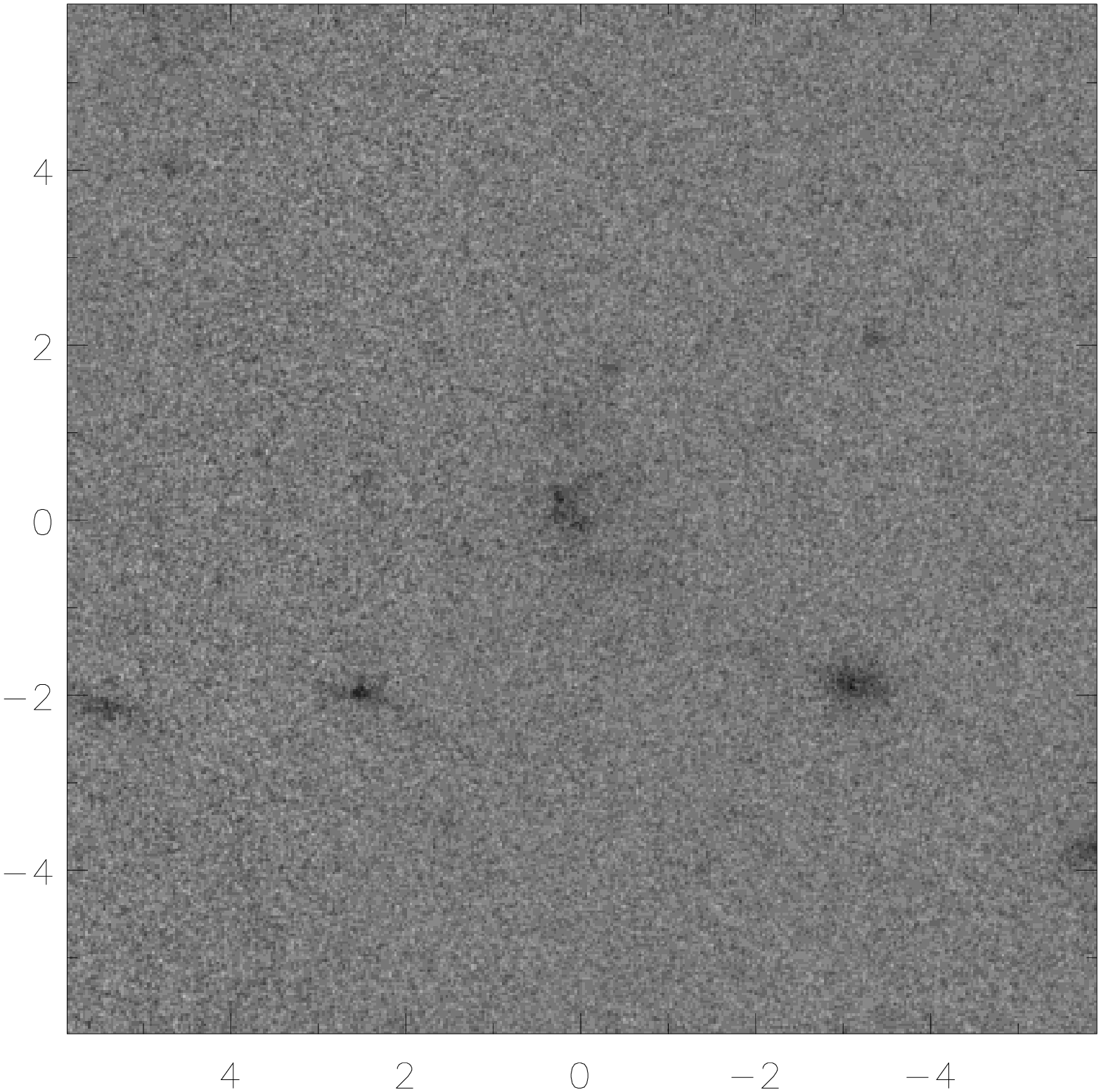,width=0.20\textwidth}\\
\epsfig{file=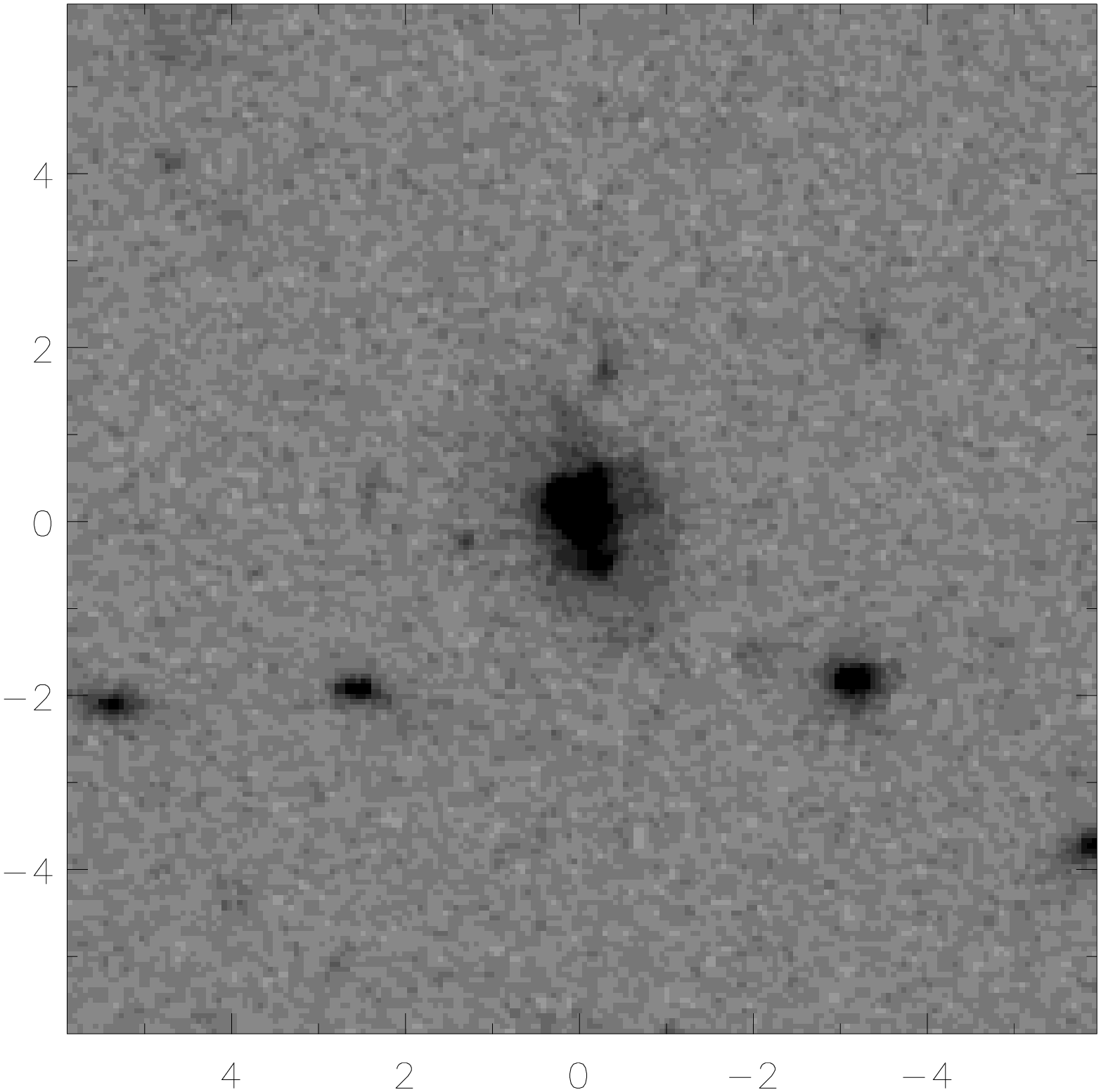,width=0.20\textwidth}&
\epsfig{file=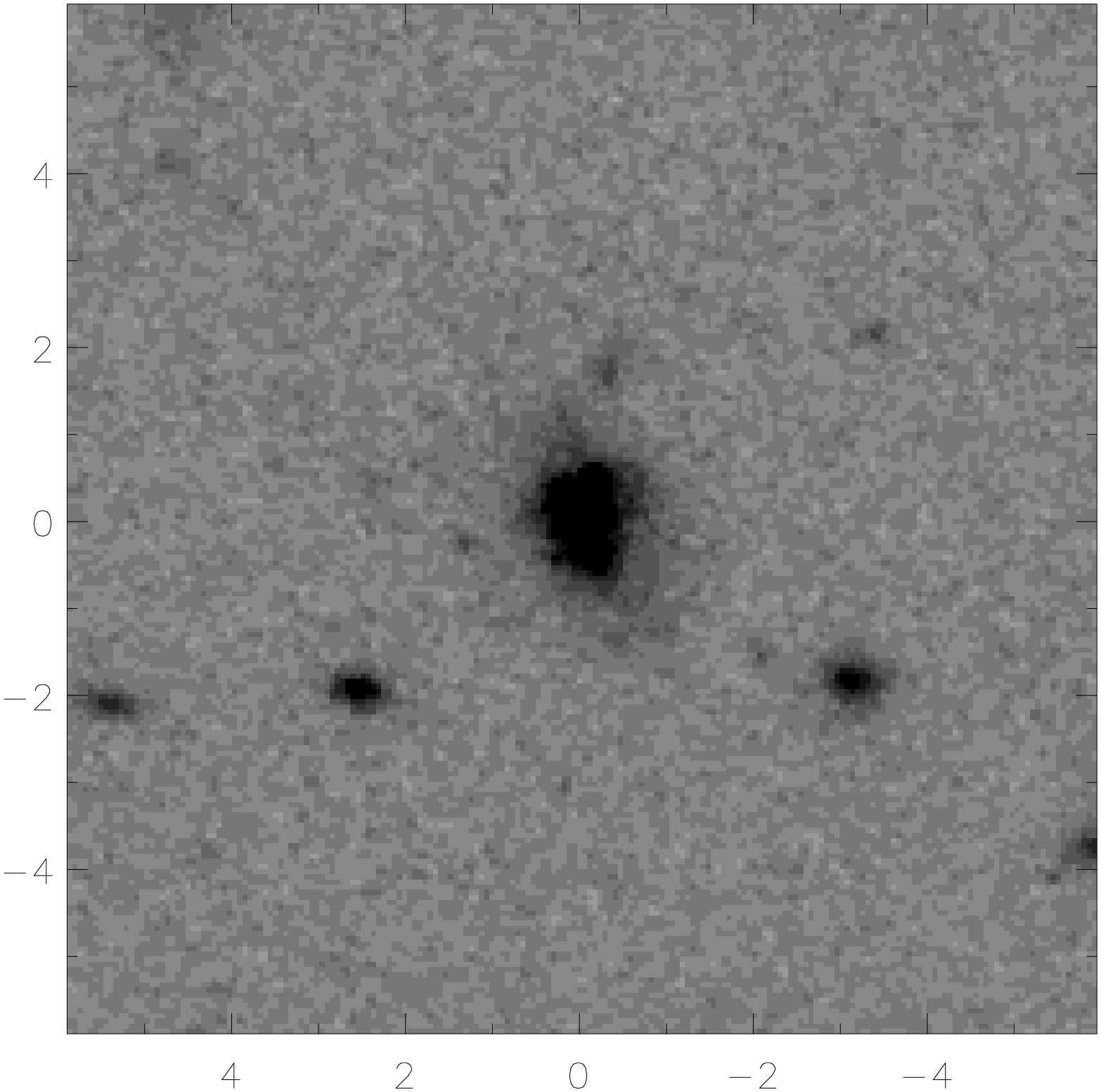,width=0.20\textwidth}&
\epsfig{file=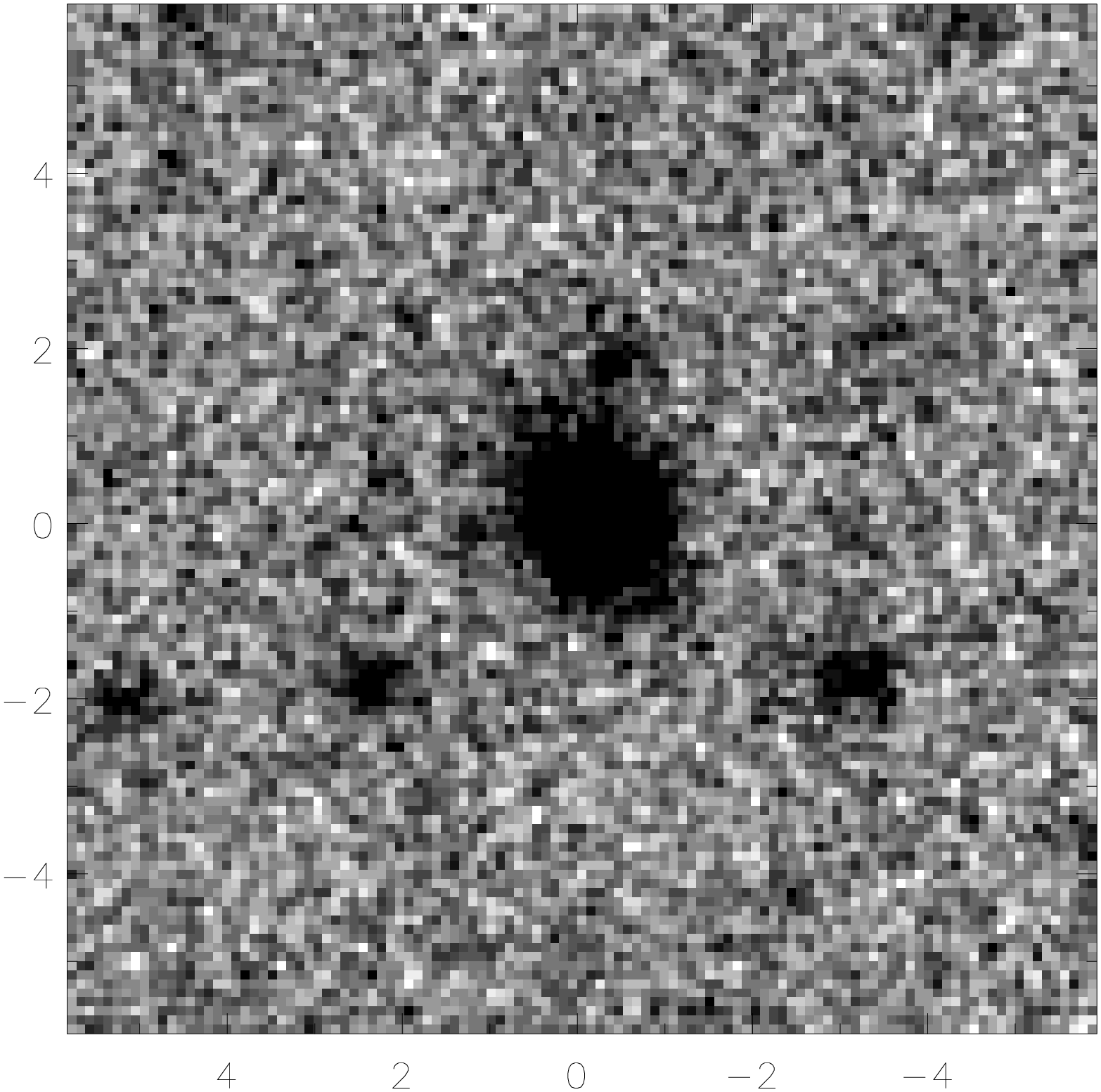,width=0.20\textwidth}&
\epsfig{file=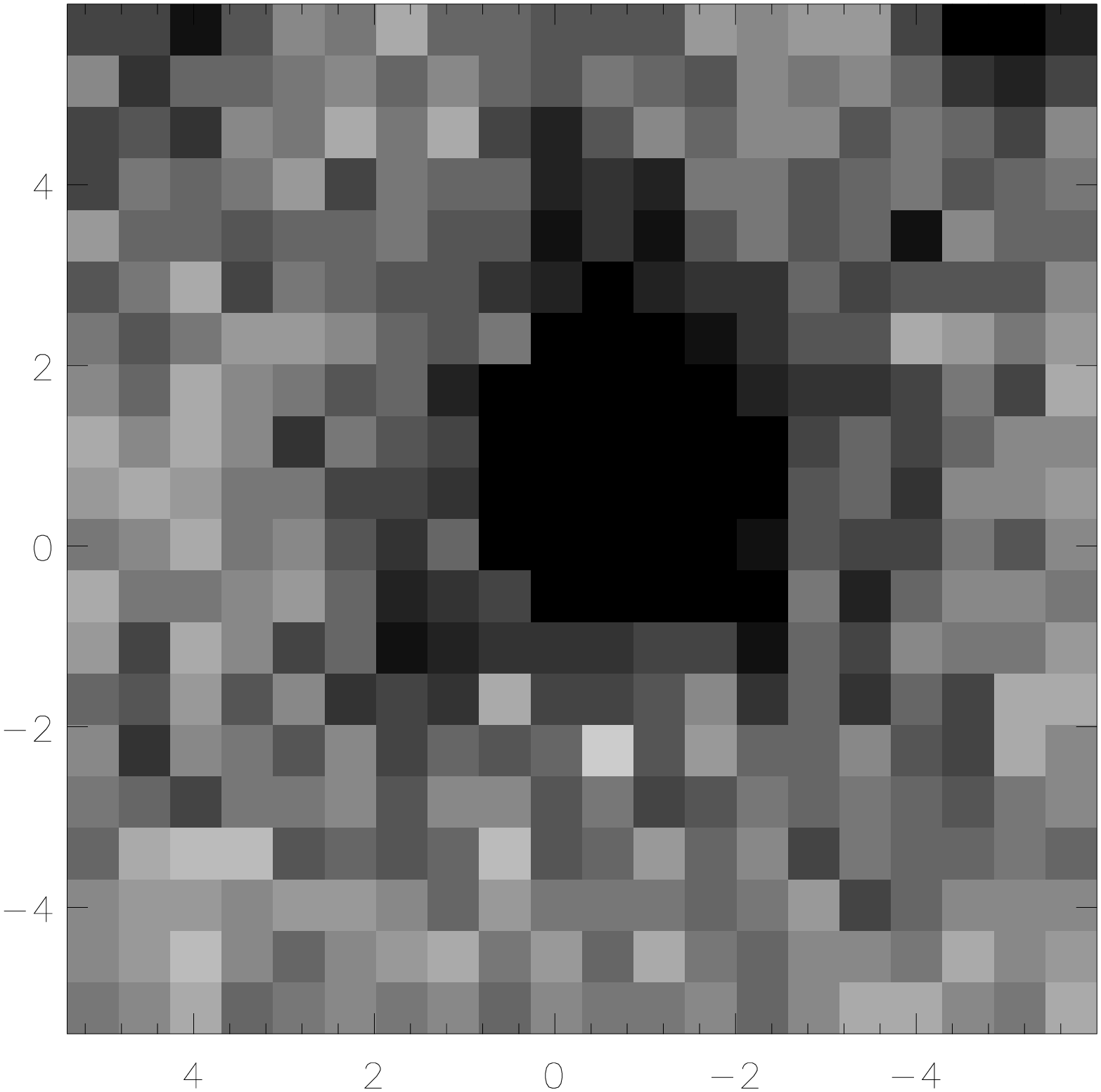,width=0.20\textwidth}\\
\end{tabular}
\addtocounter{figure}{-1}
\caption{- continued}
\vfil}
\end{figure*}
\end{center}


\begin{center}
\begin{figure*}
\vbox to220mm{\vfil
\begin{tabular}{cccccccc}
\epsfig{file=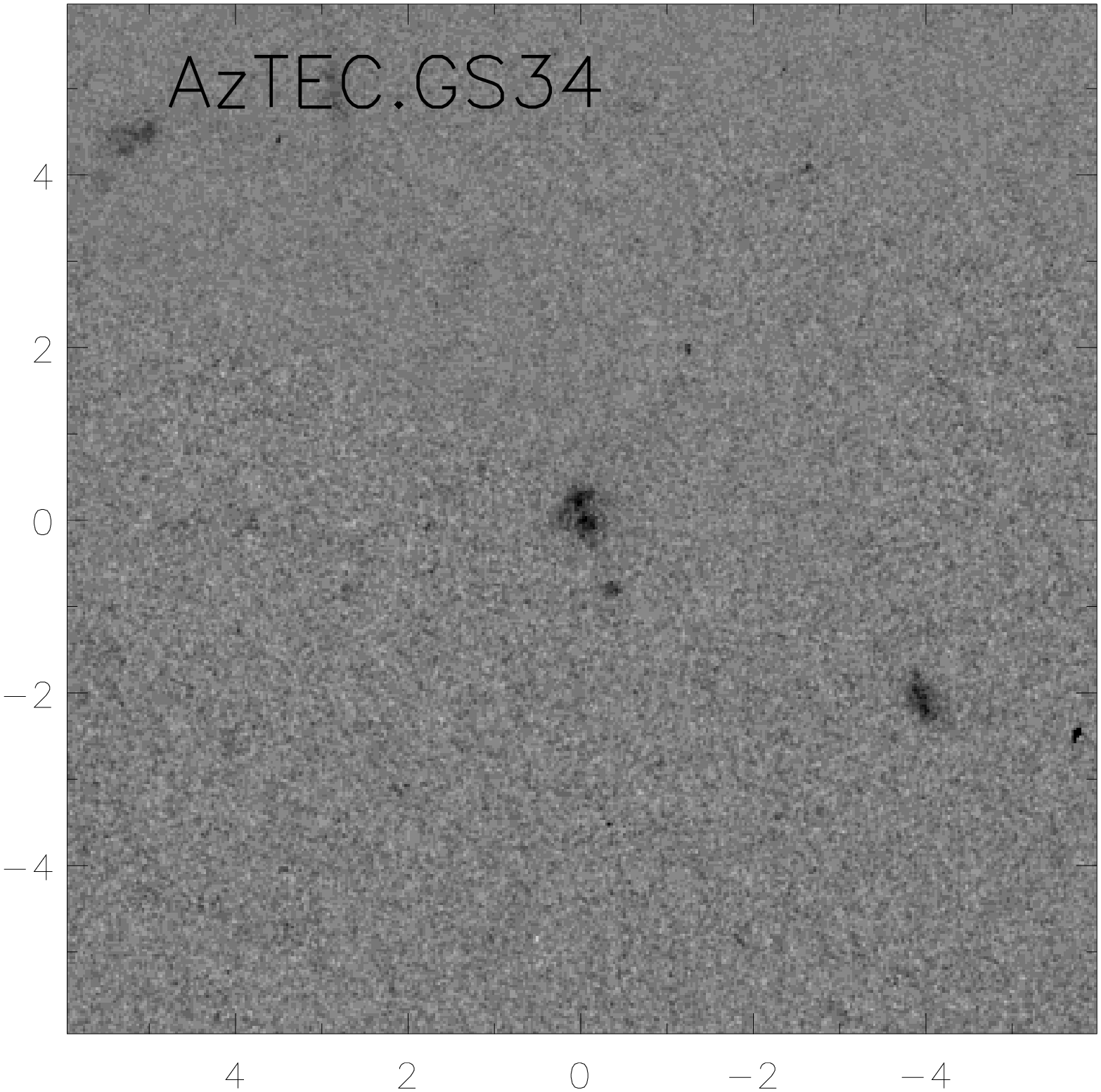,width=0.20\textwidth}&
\epsfig{file=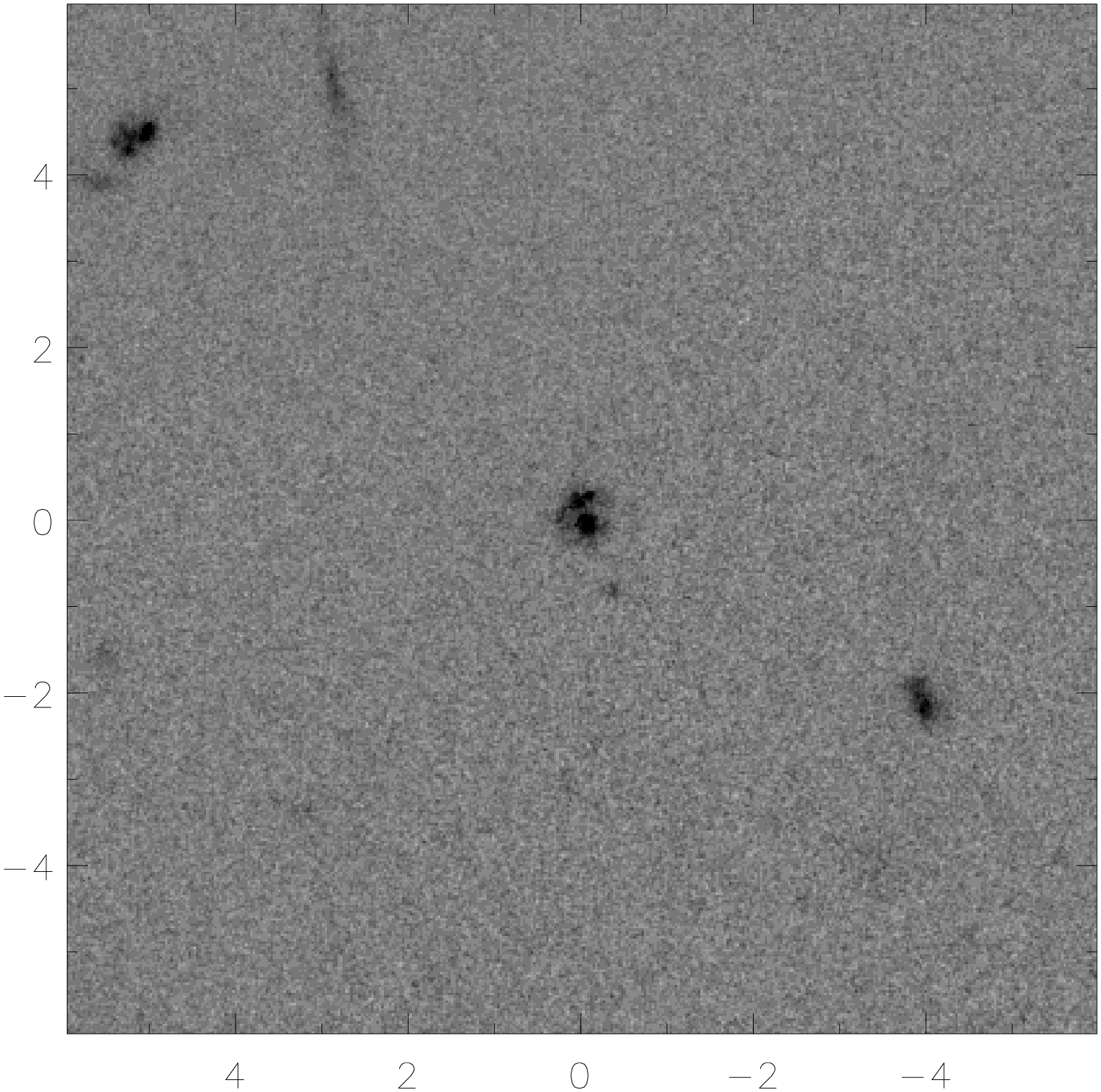,width=0.20\textwidth}&
\epsfig{file=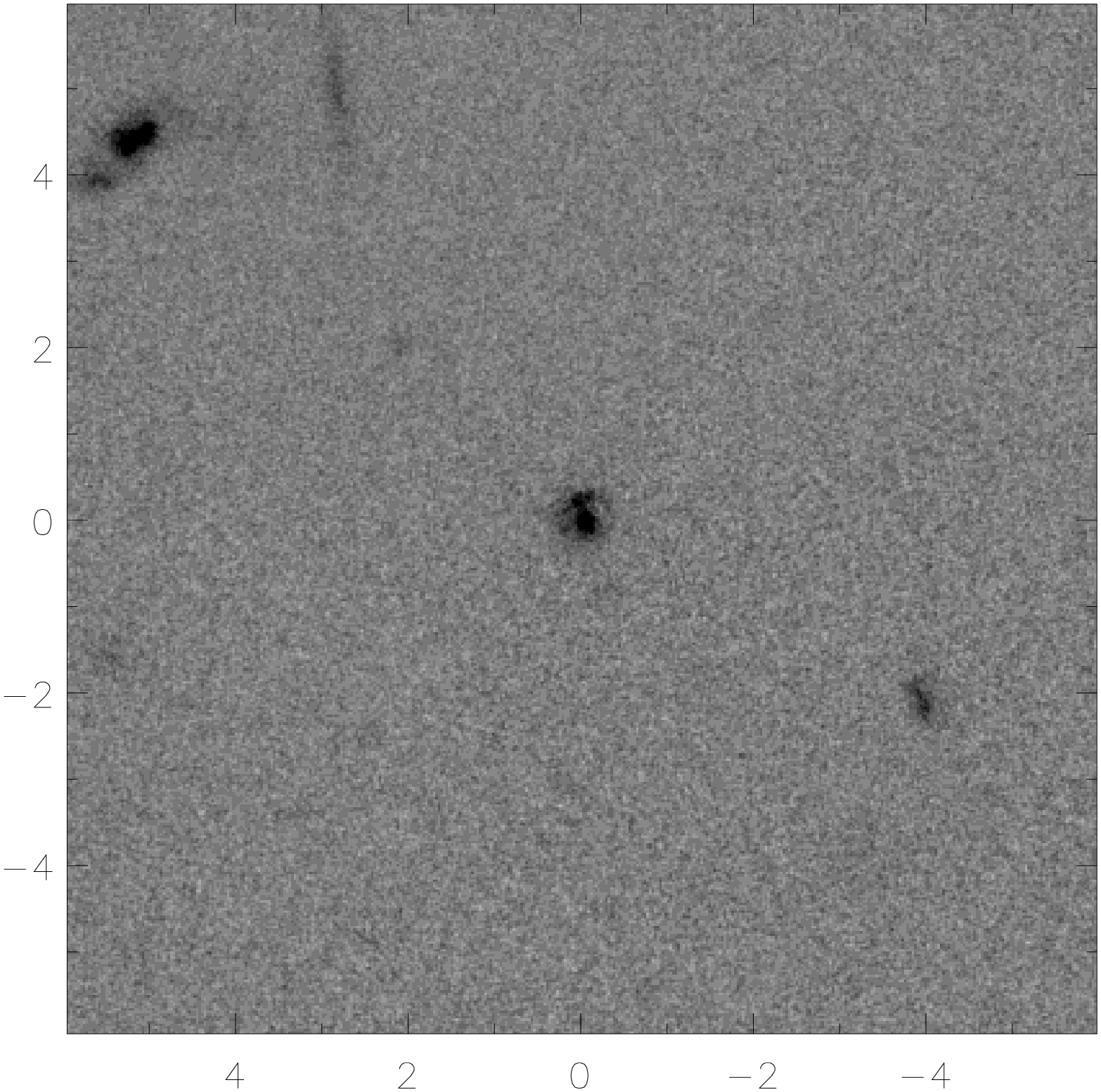,width=0.20\textwidth}&
\epsfig{file=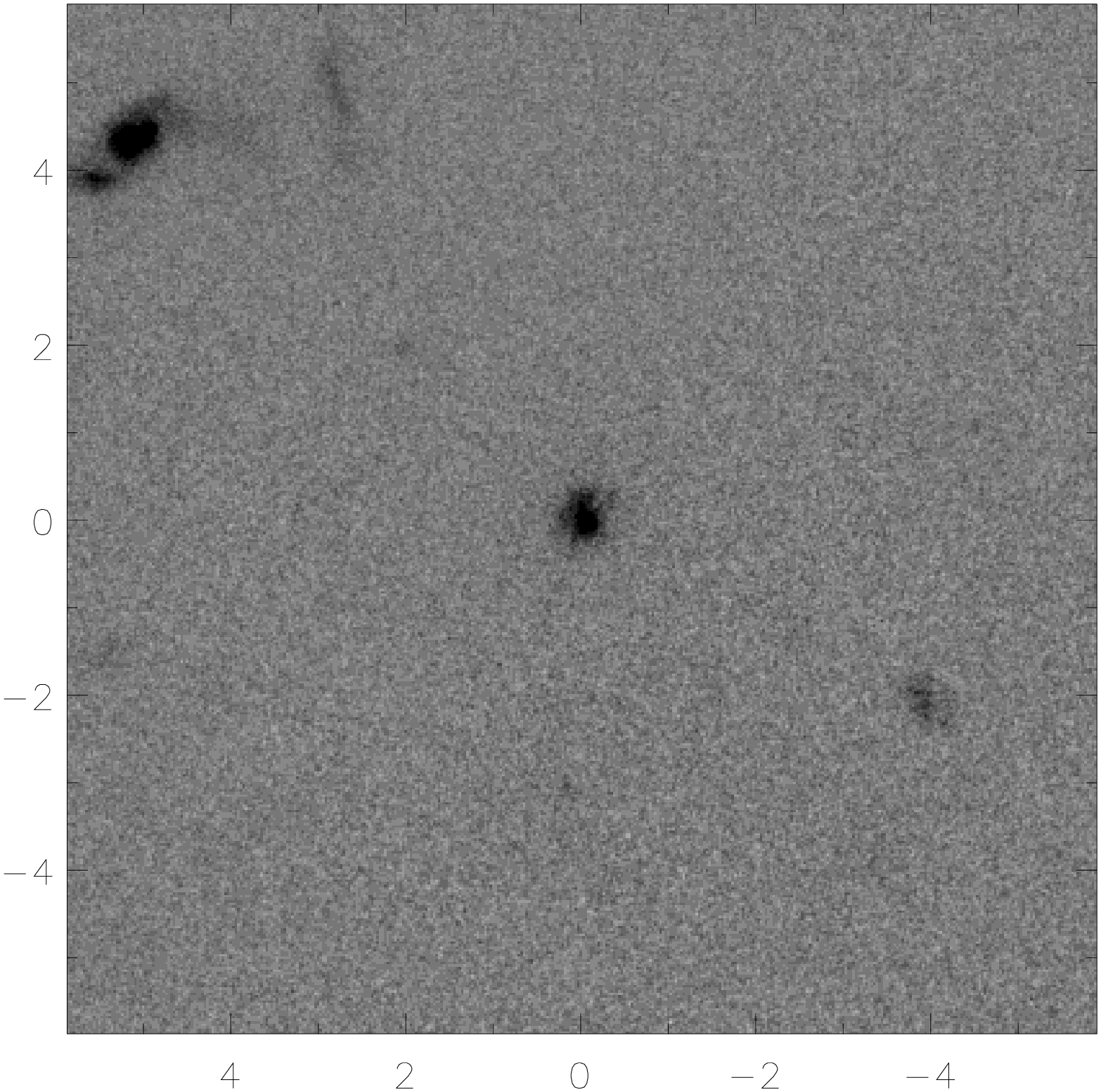,width=0.20\textwidth}\\
\epsfig{file=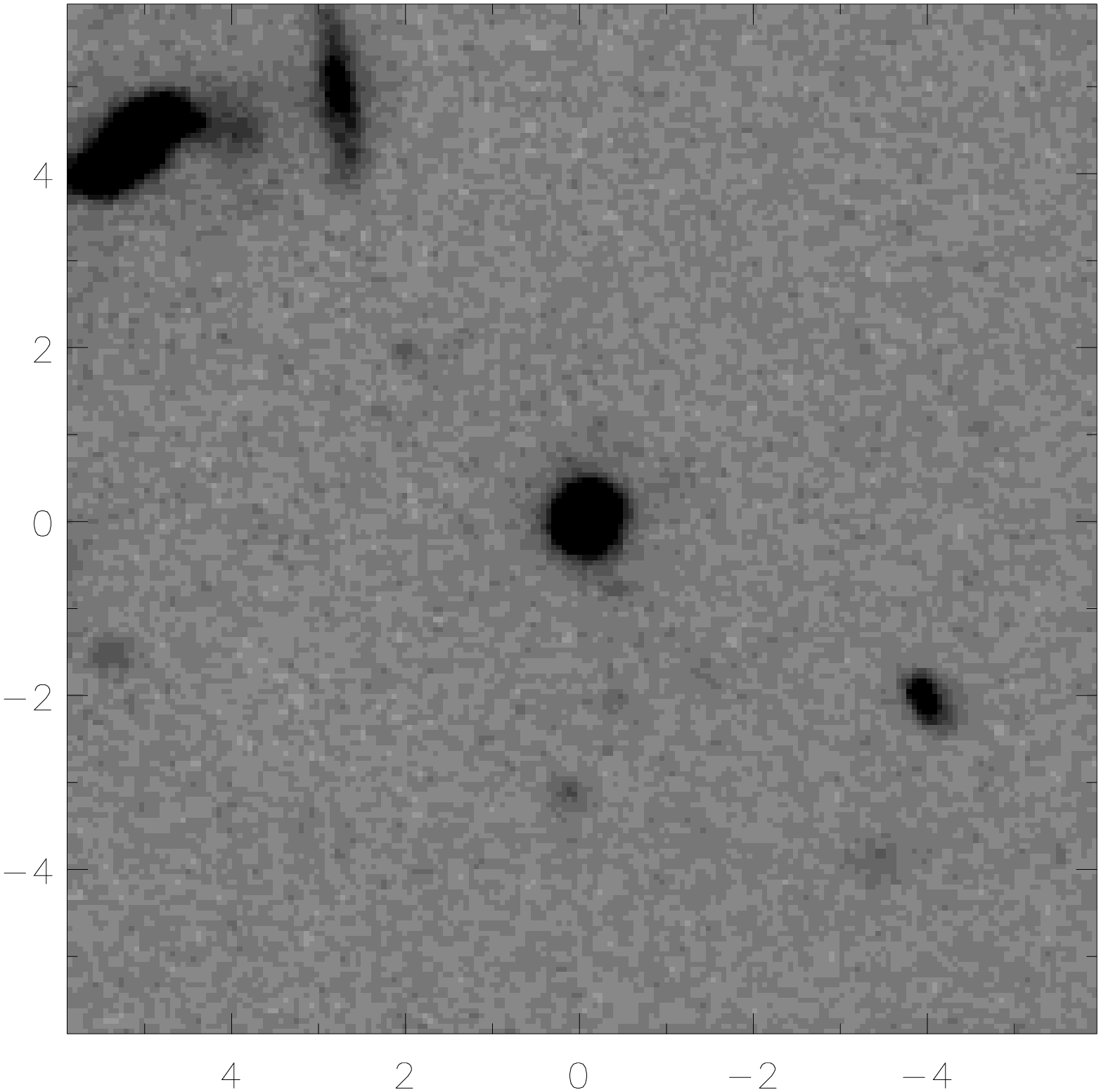,width=0.20\textwidth}&
\epsfig{file=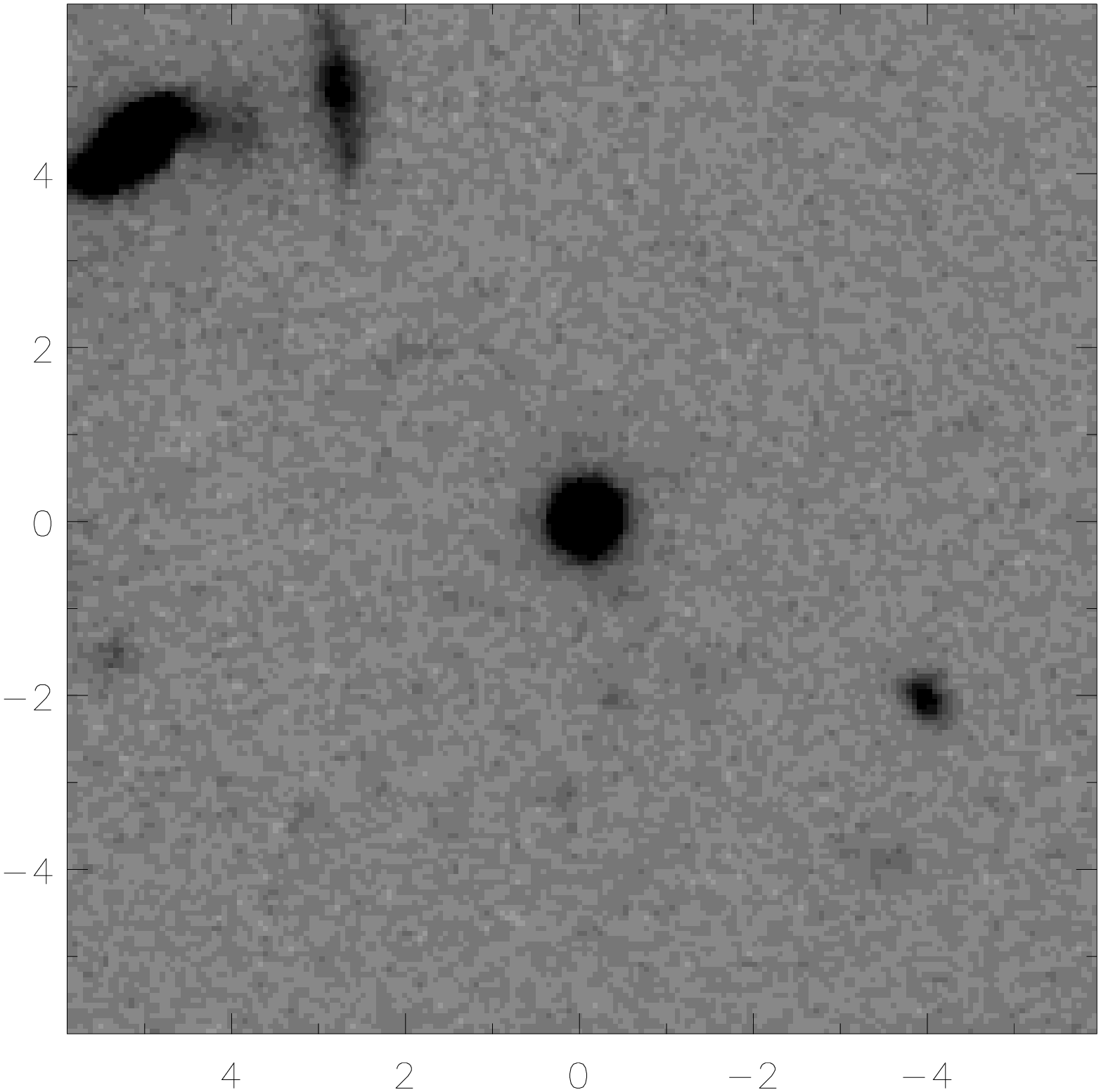,width=0.20\textwidth}&
\epsfig{file=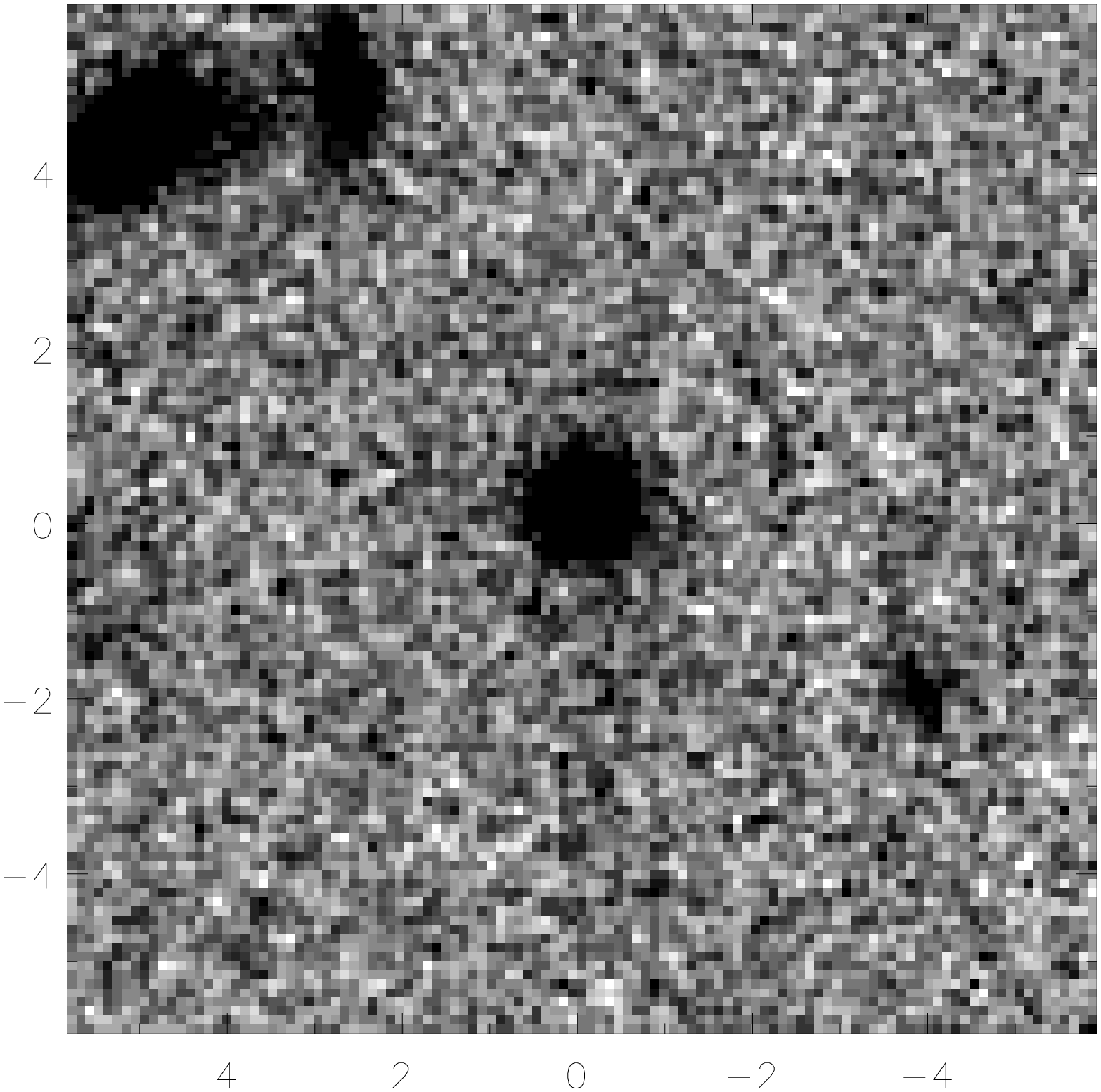,width=0.20\textwidth}&
\epsfig{file=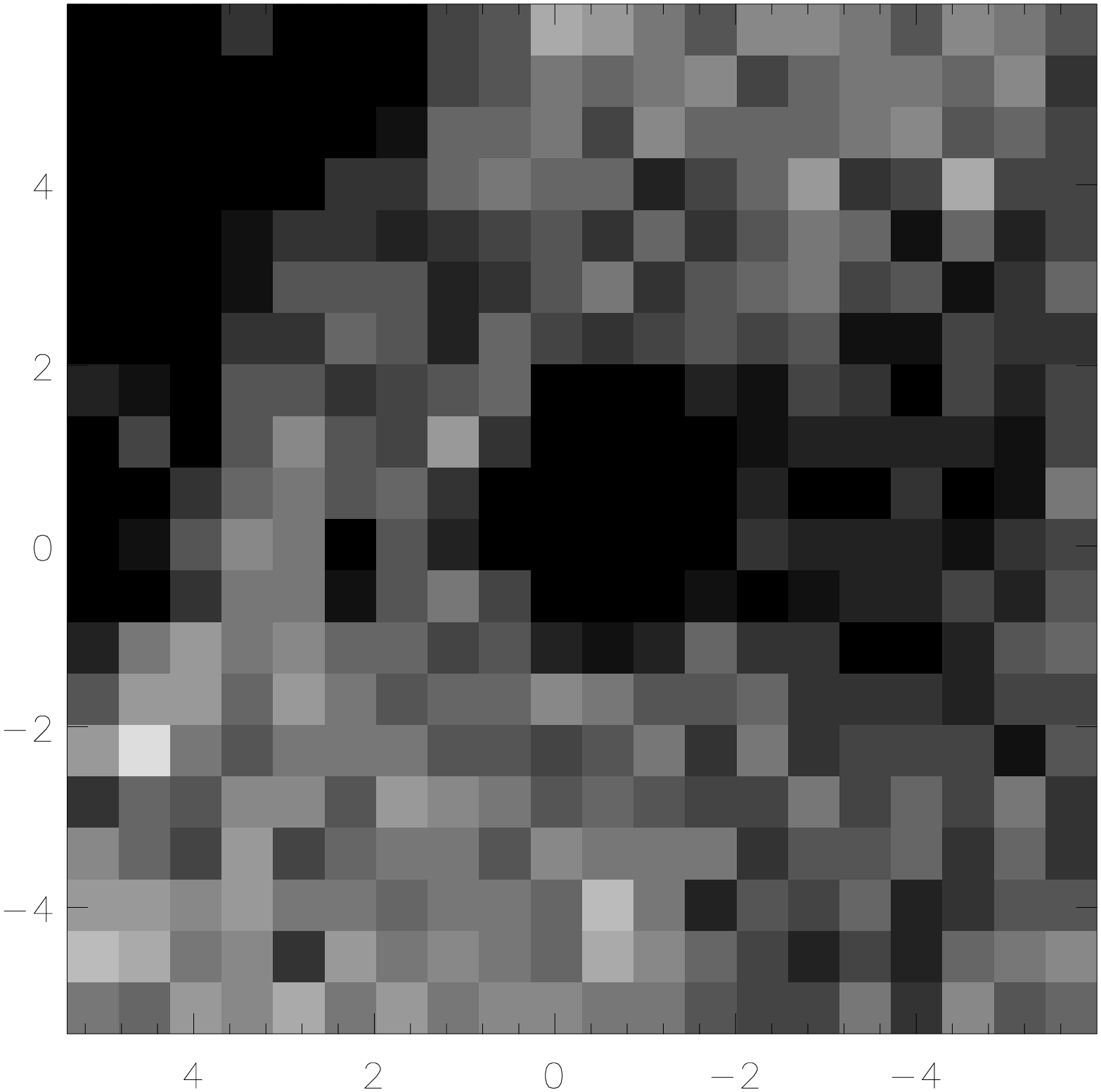,width=0.20\textwidth}\\
\\
\epsfig{file=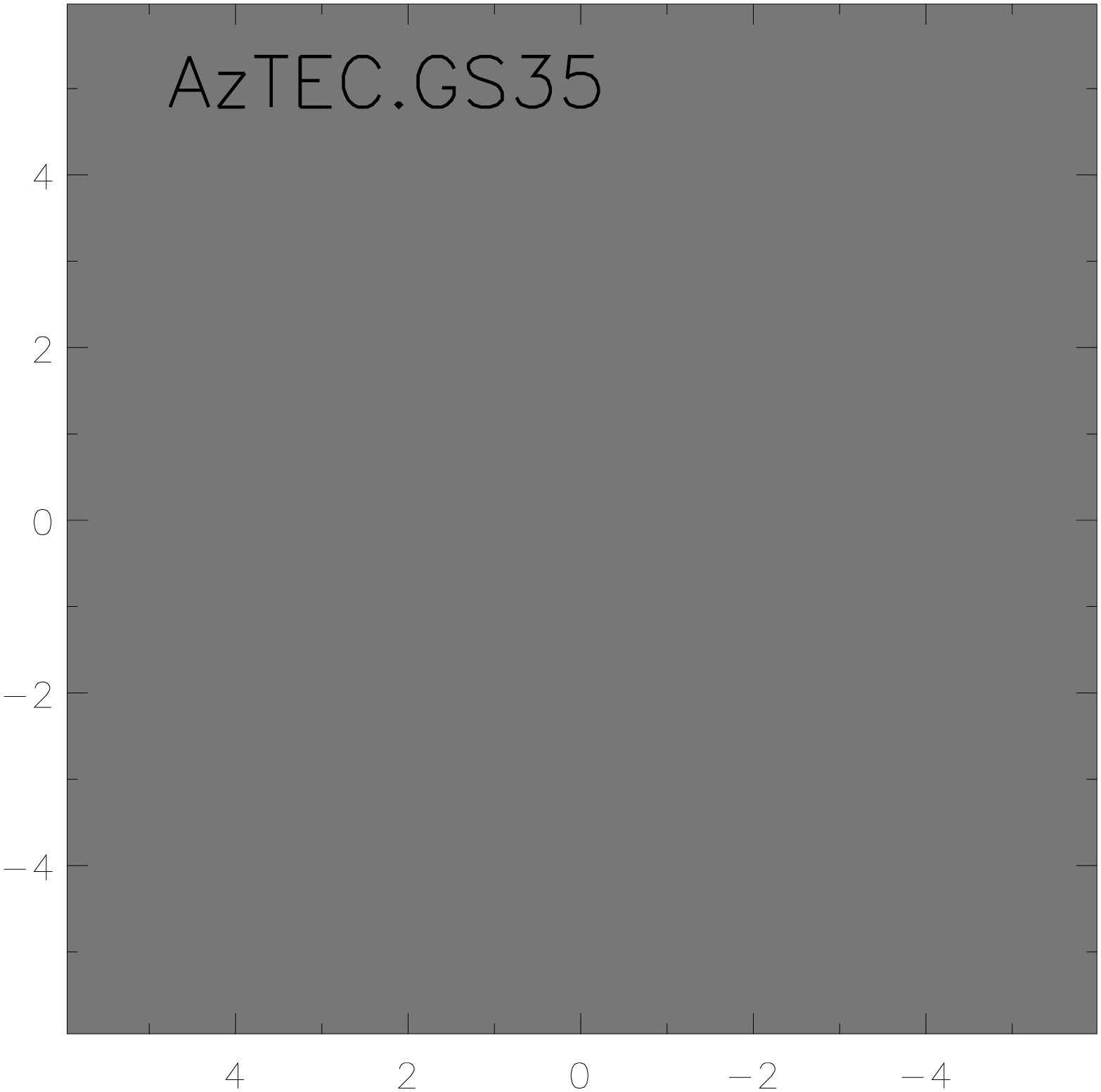,width=0.20\textwidth}&
\epsfig{file=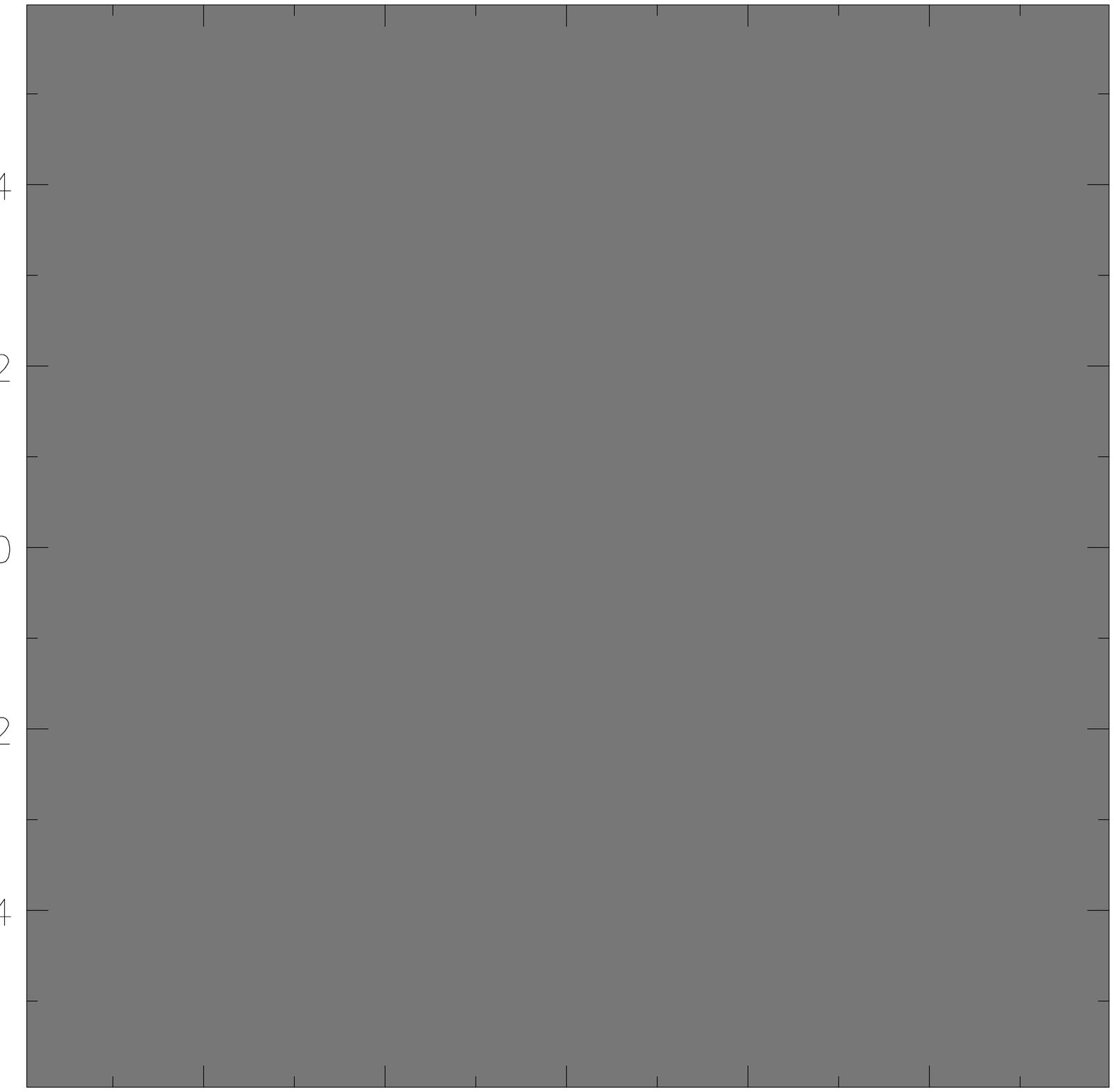,width=0.20\textwidth}&
\epsfig{file=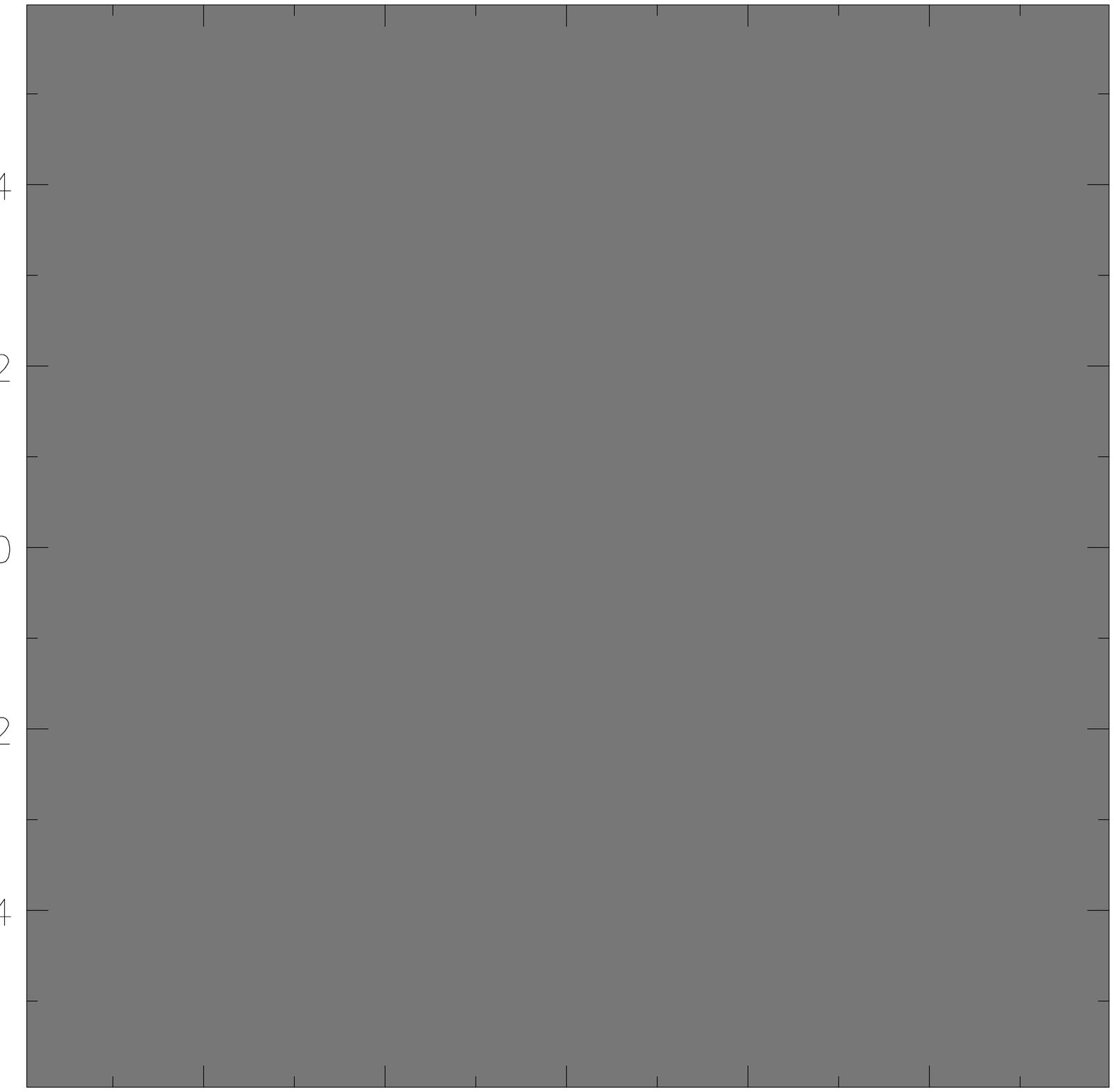,width=0.20\textwidth}&
\epsfig{file=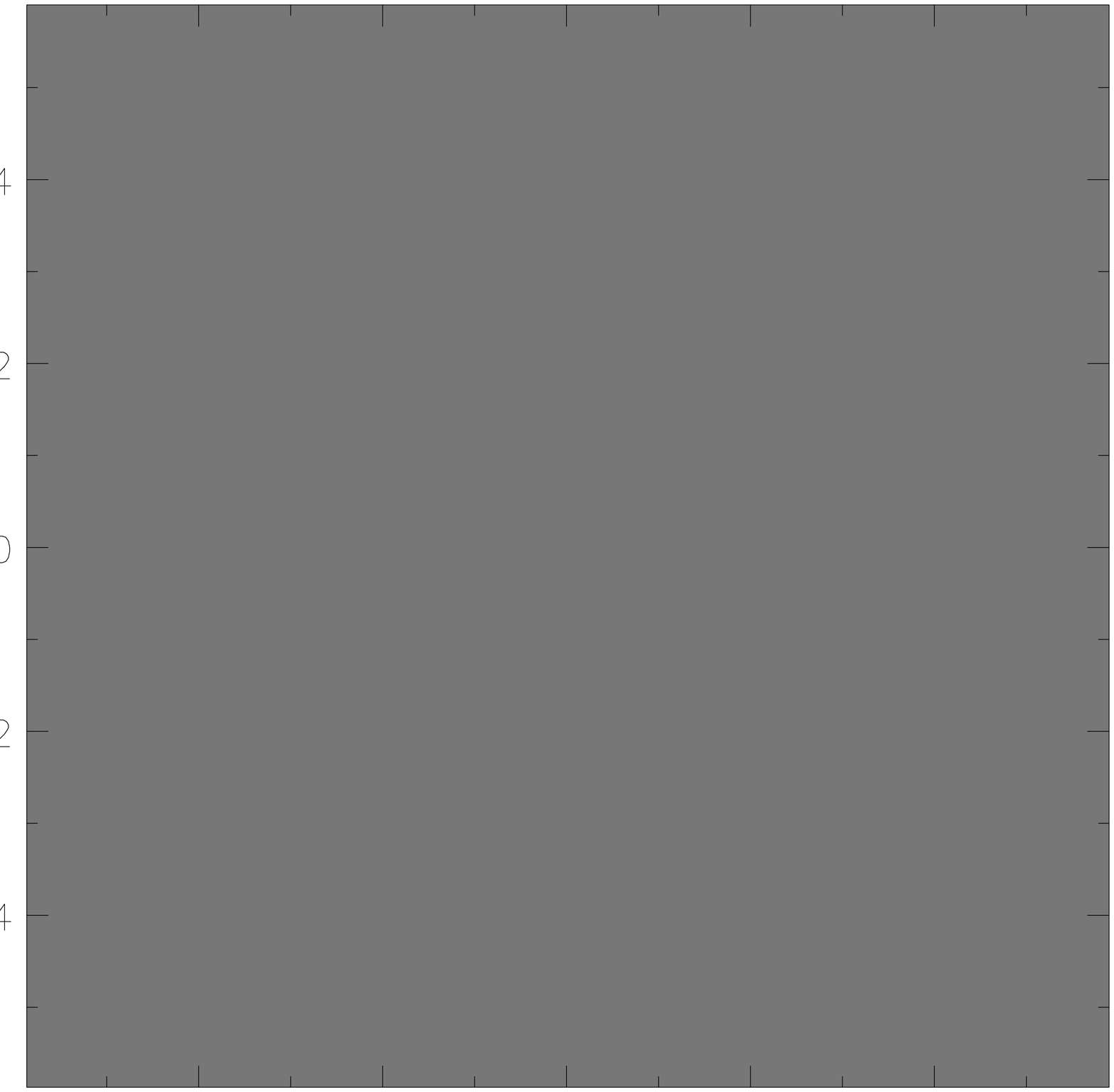,width=0.20\textwidth}\\
\epsfig{file=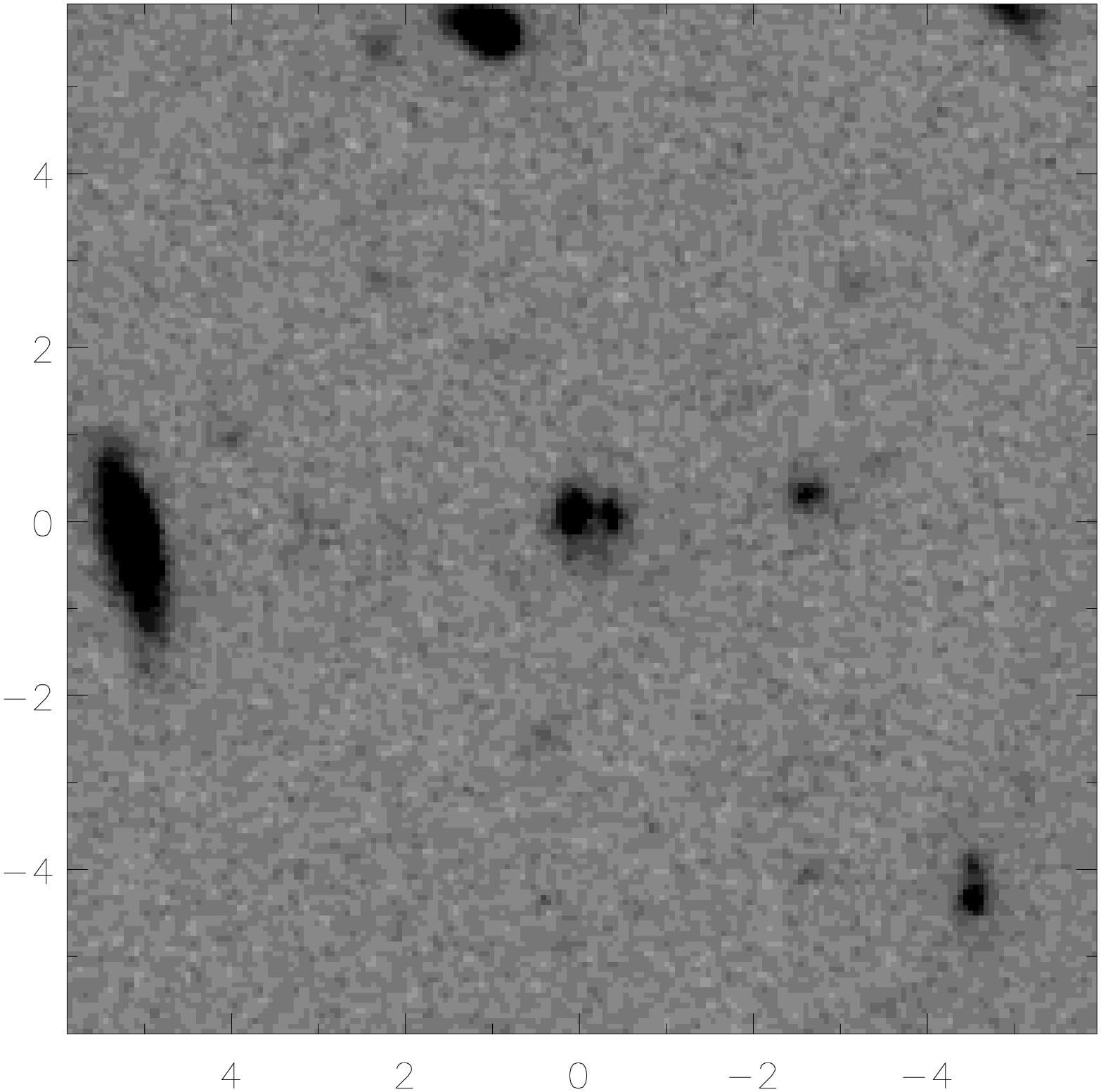,width=0.20\textwidth}&
\epsfig{file=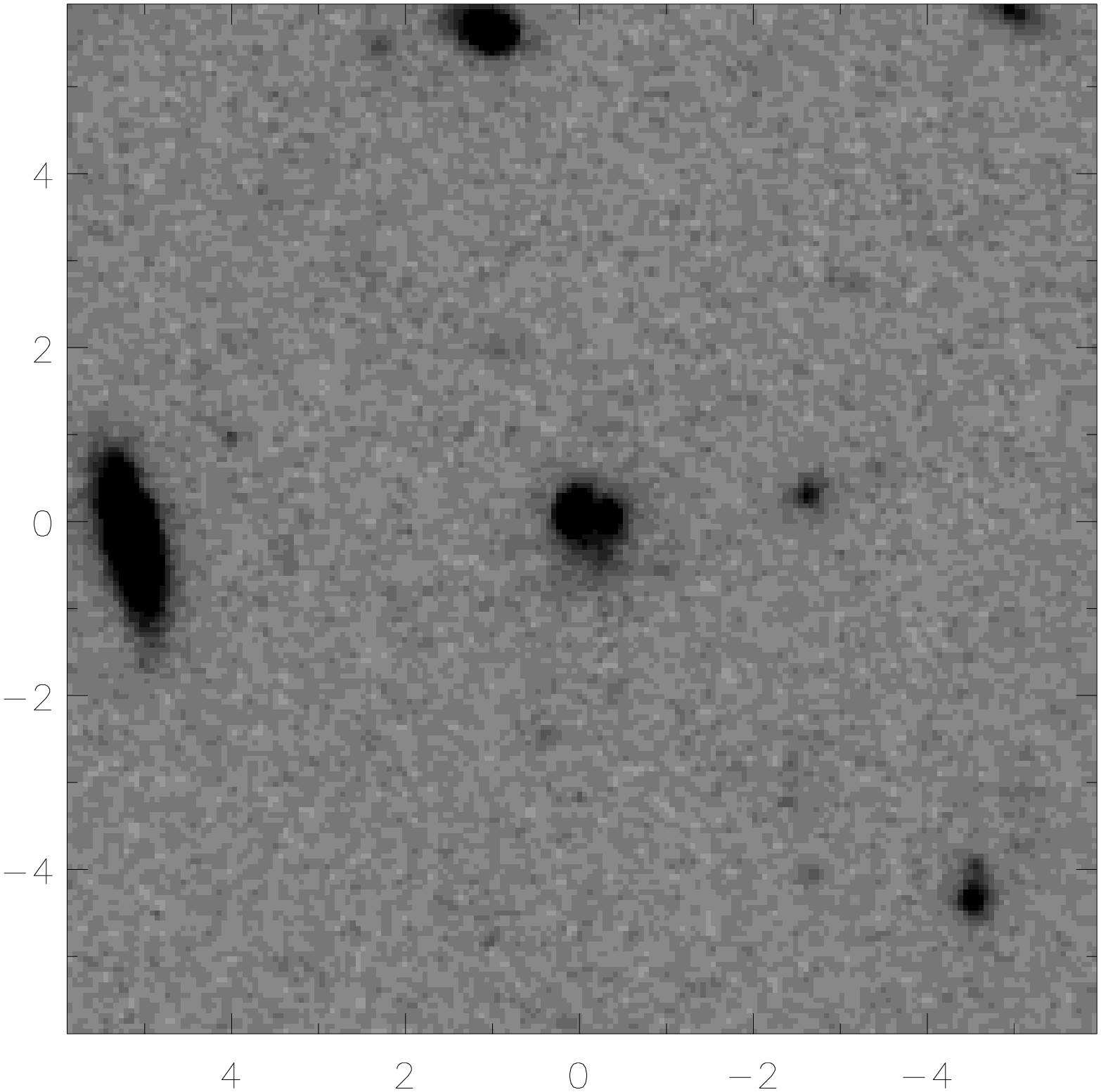,width=0.20\textwidth}&
\epsfig{file=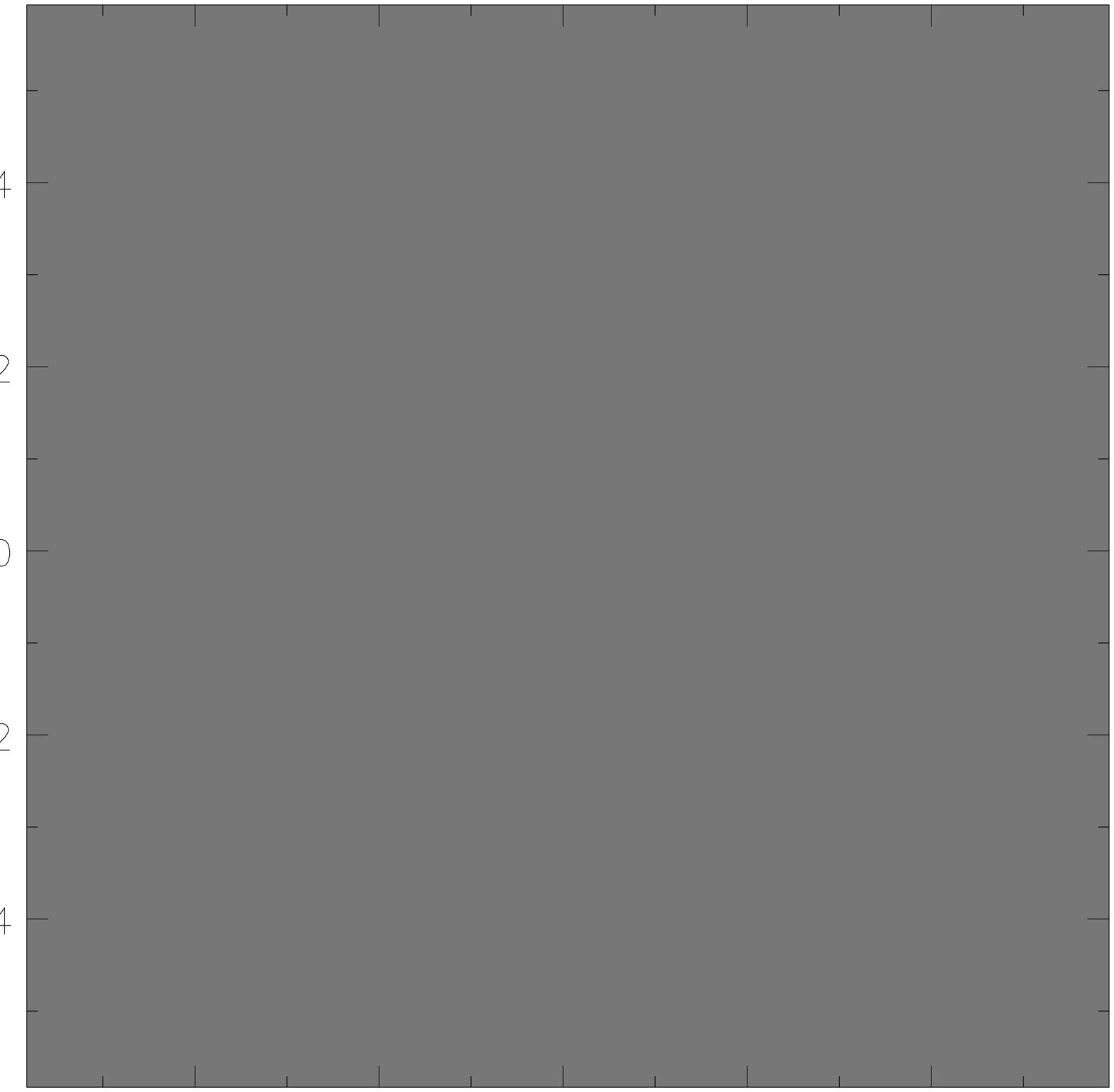,width=0.20\textwidth}&
\epsfig{file=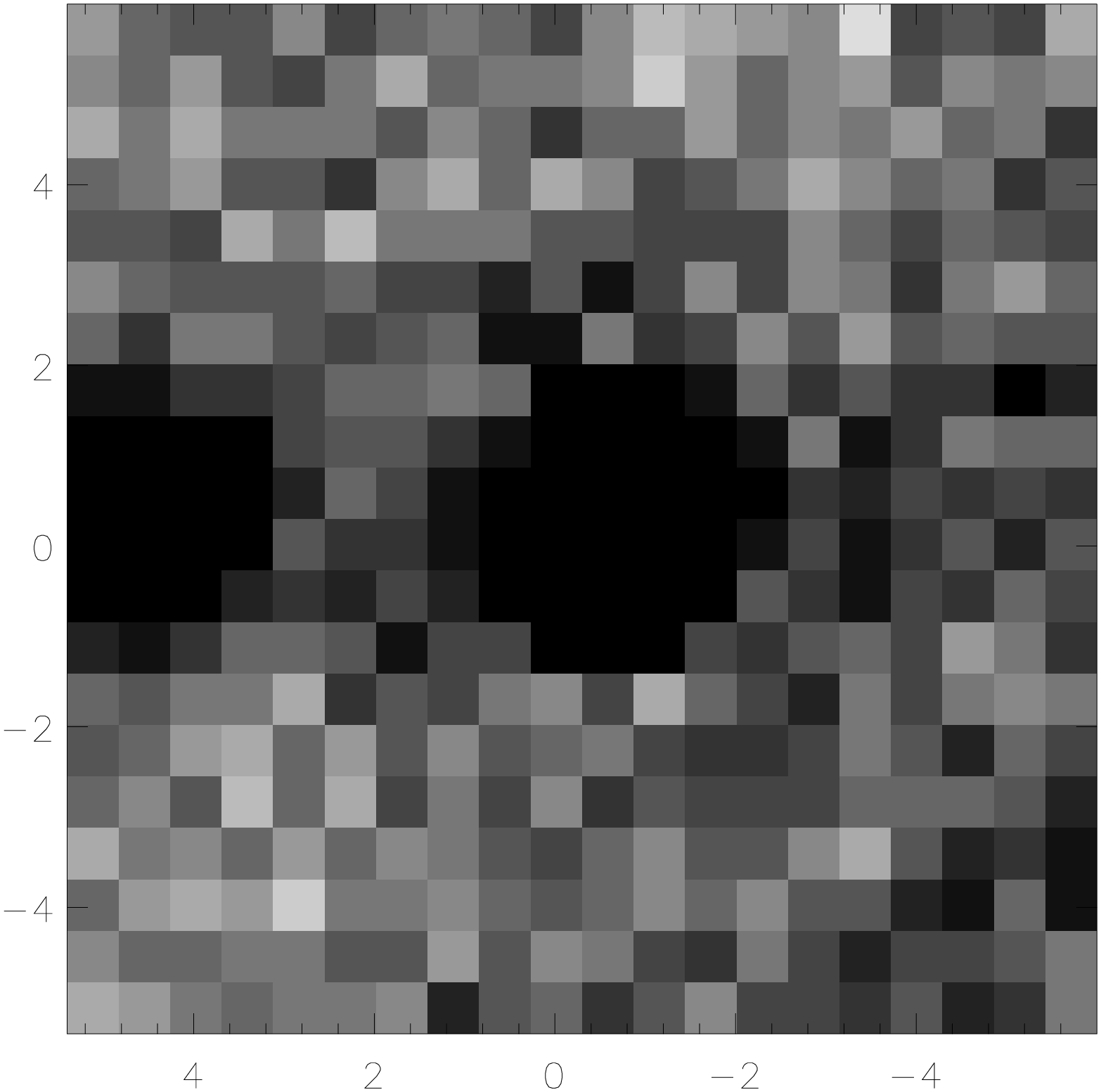,width=0.20\textwidth}\\
\\
\epsfig{file=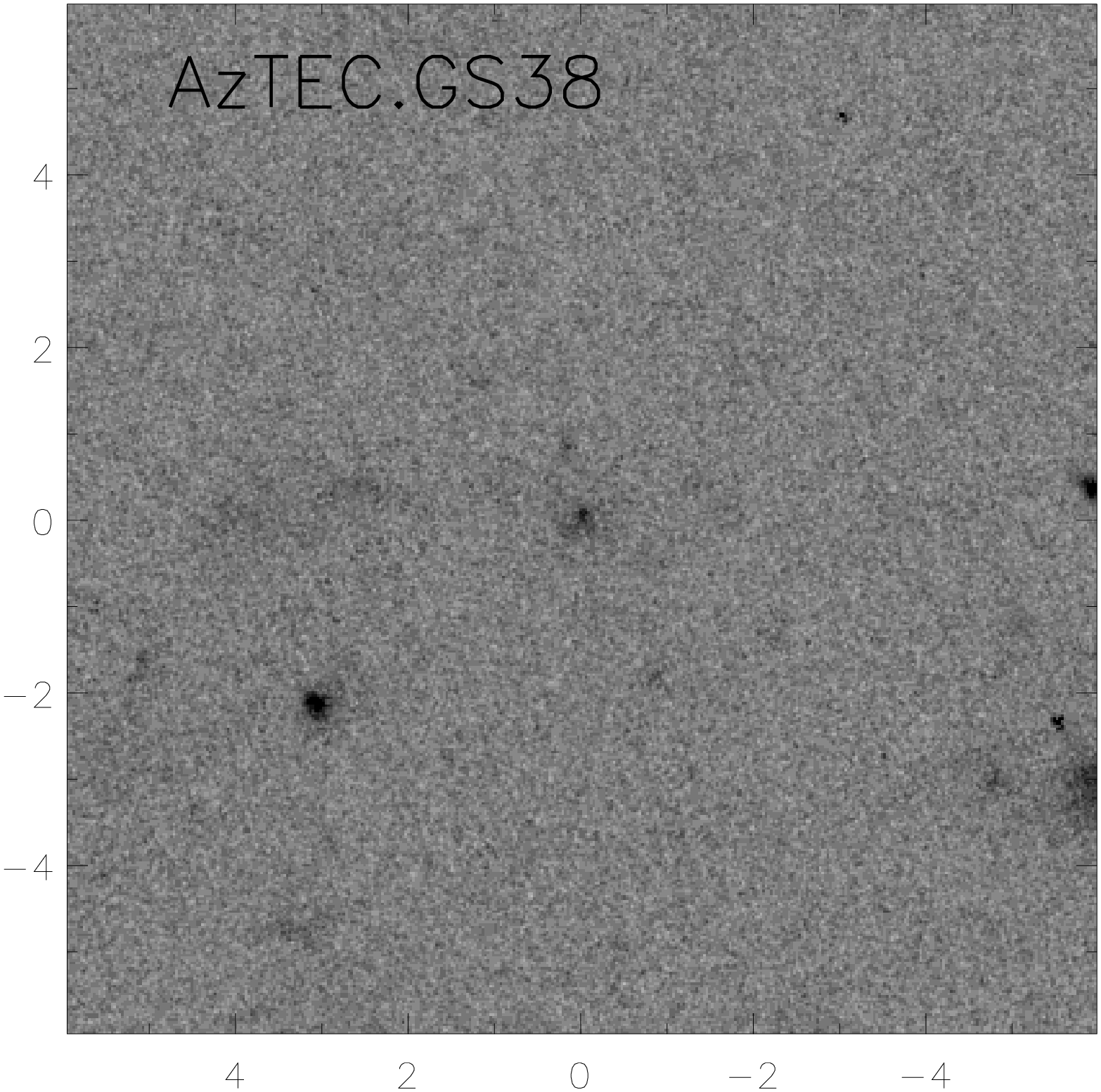,width=0.20\textwidth}&
\epsfig{file=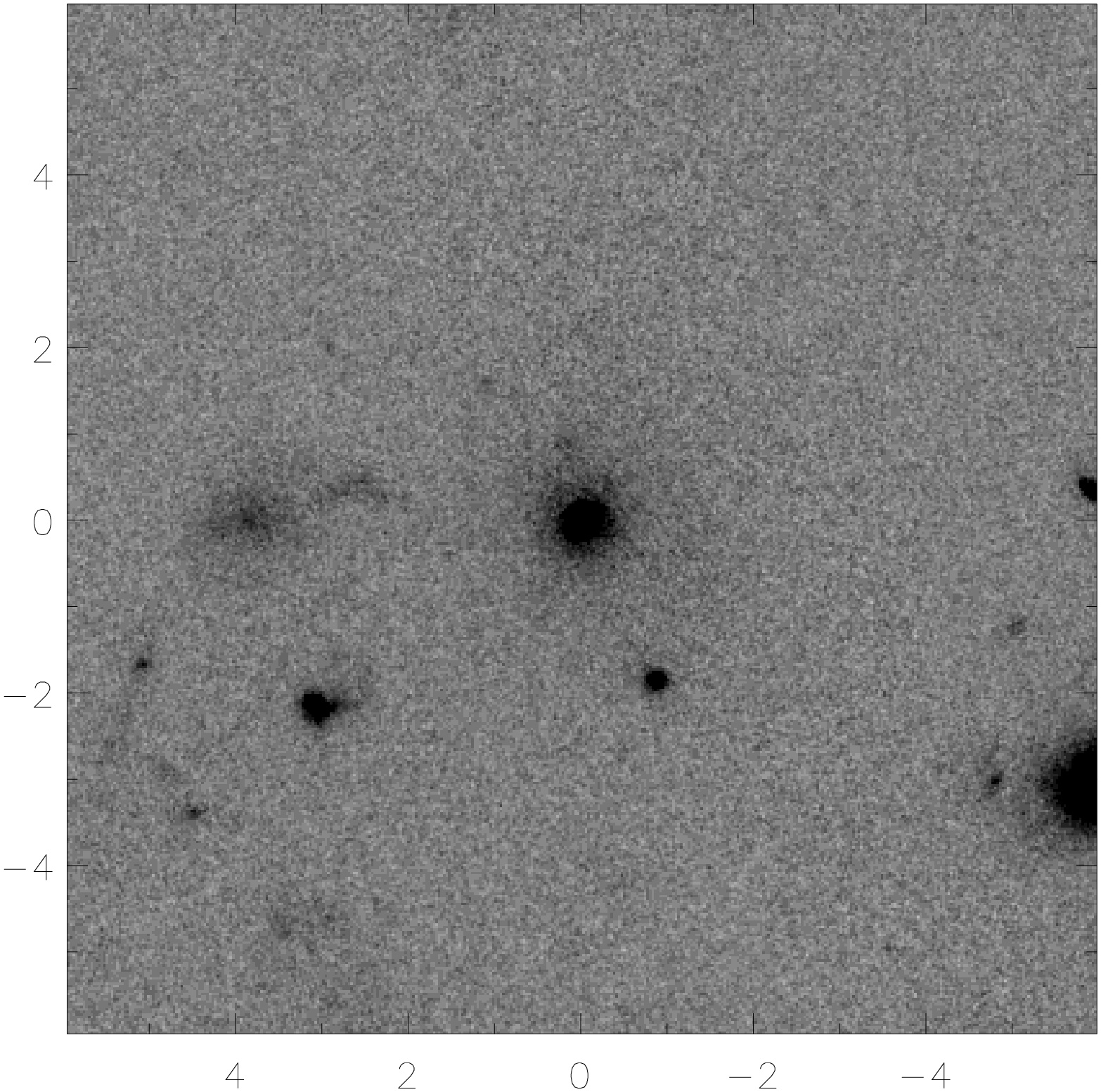,width=0.20\textwidth}&
\epsfig{file=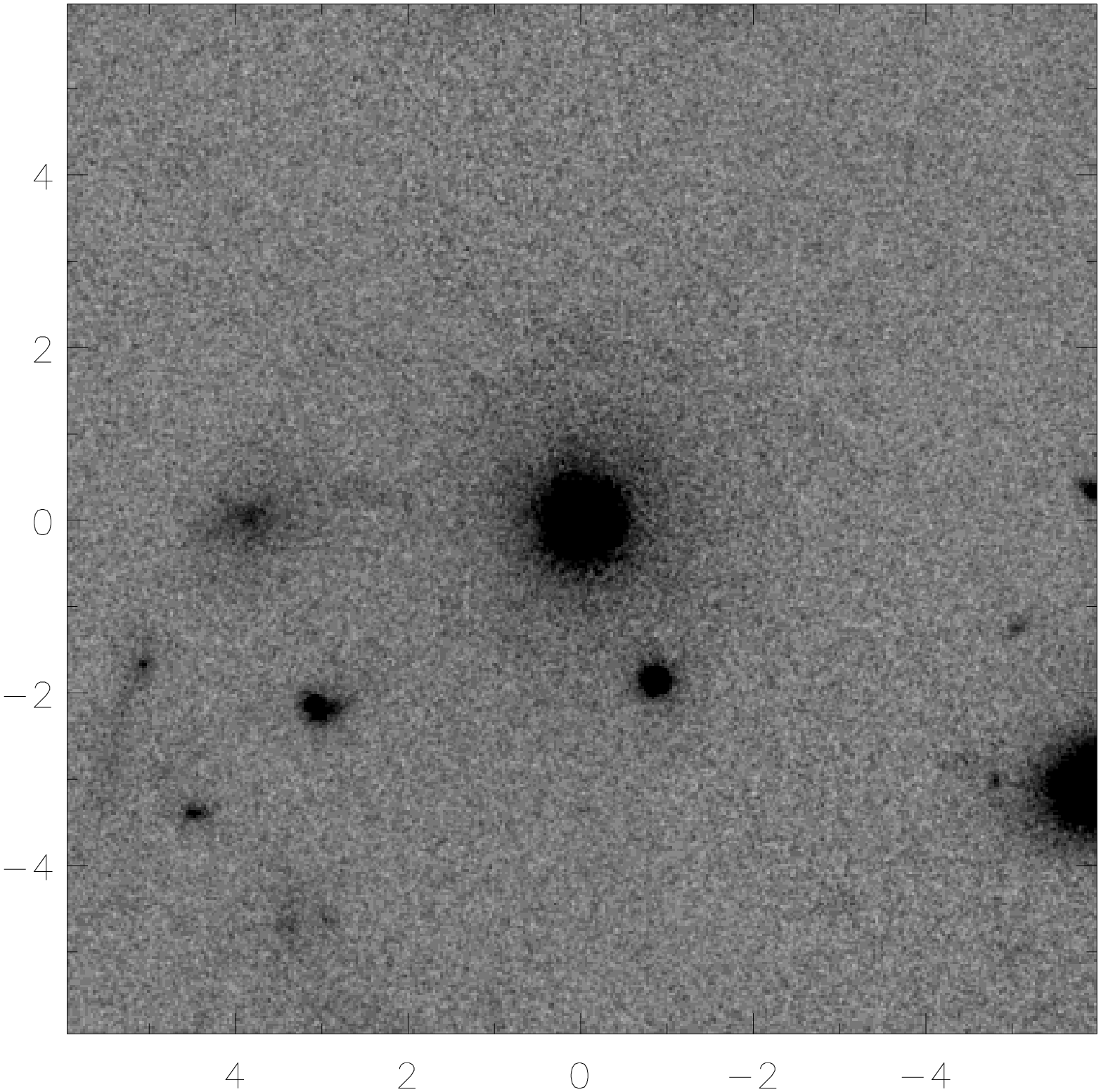,width=0.20\textwidth}&
\epsfig{file=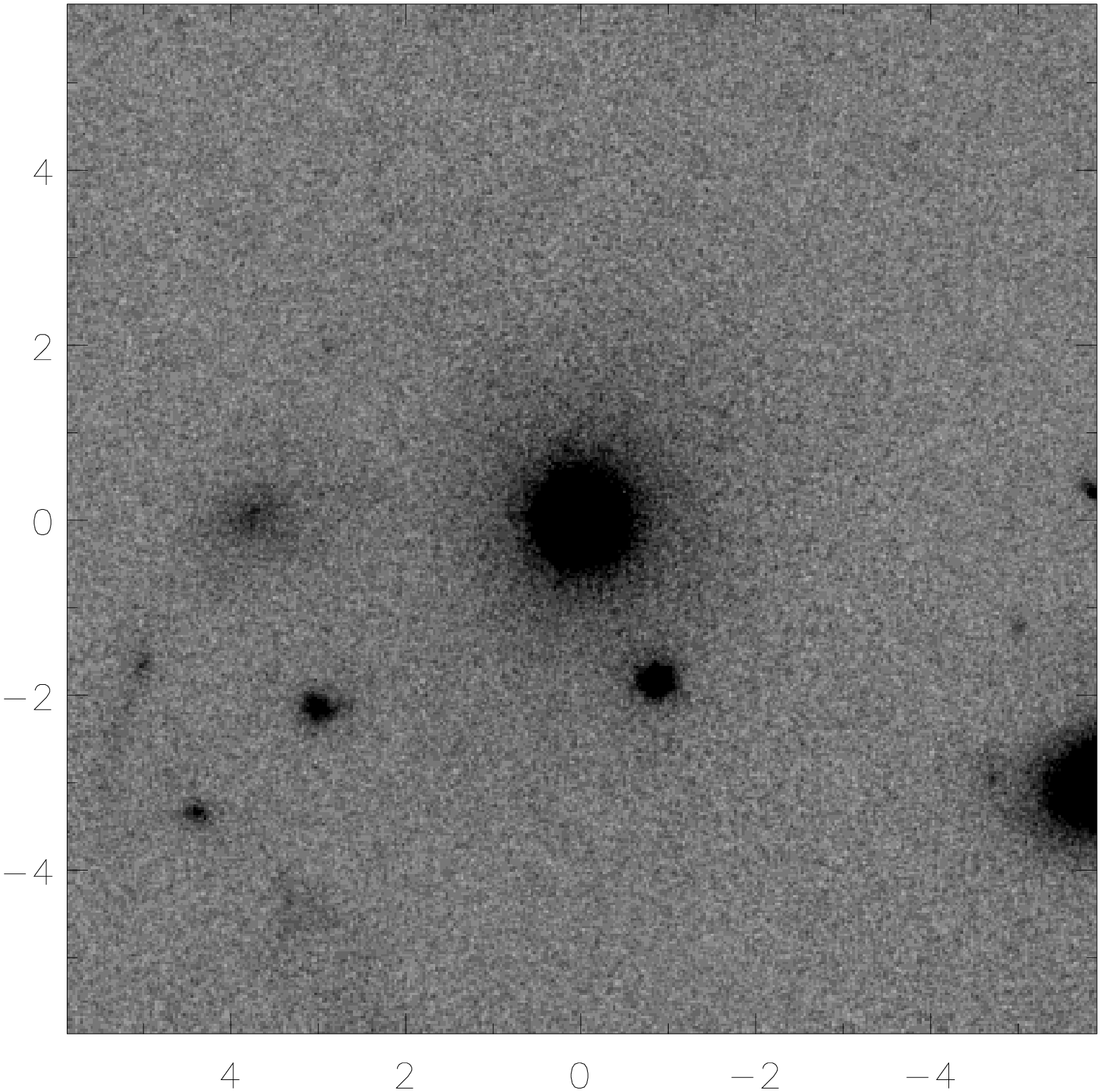,width=0.20\textwidth}\\
\epsfig{file=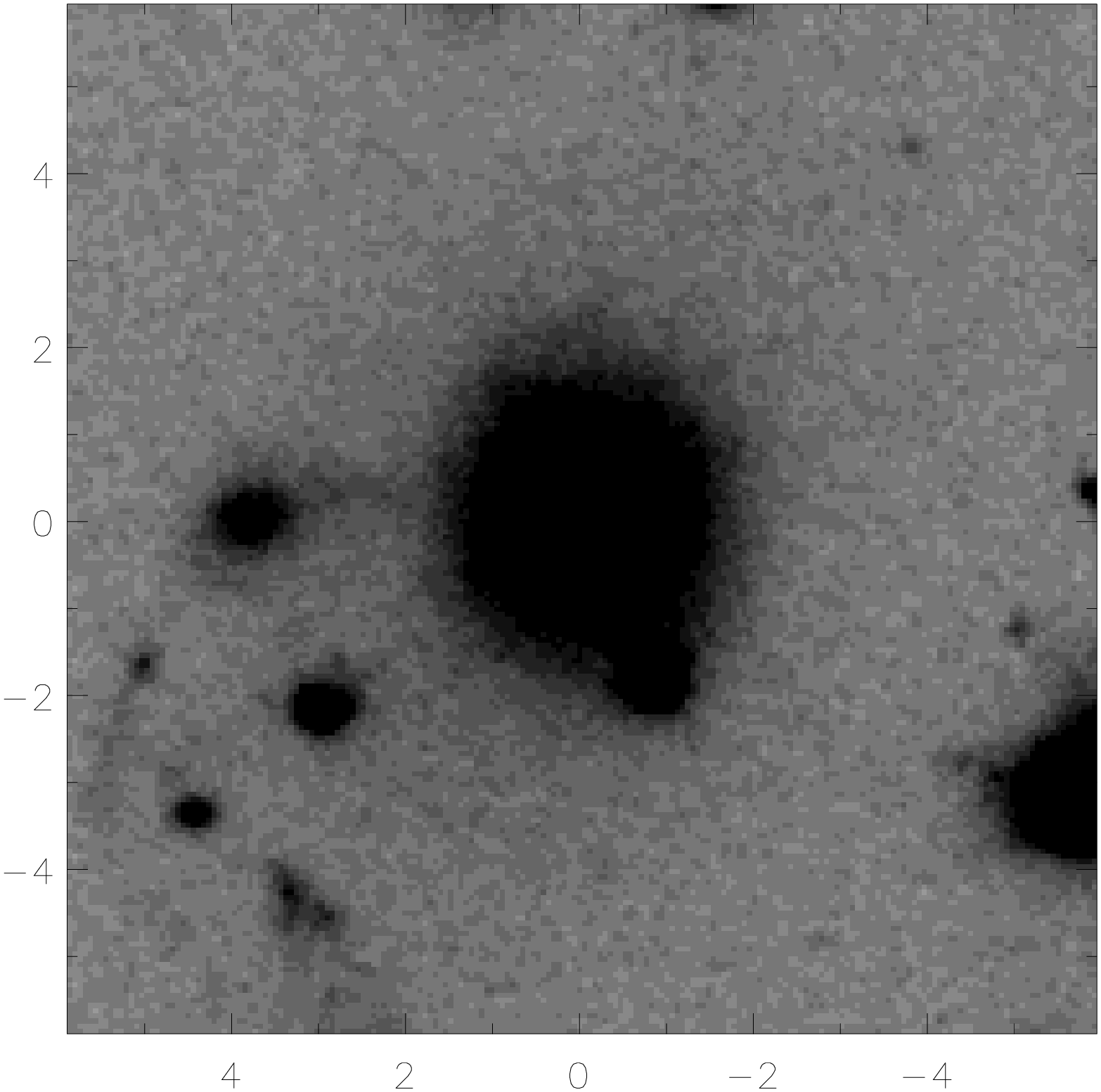,width=0.20\textwidth}&
\epsfig{file=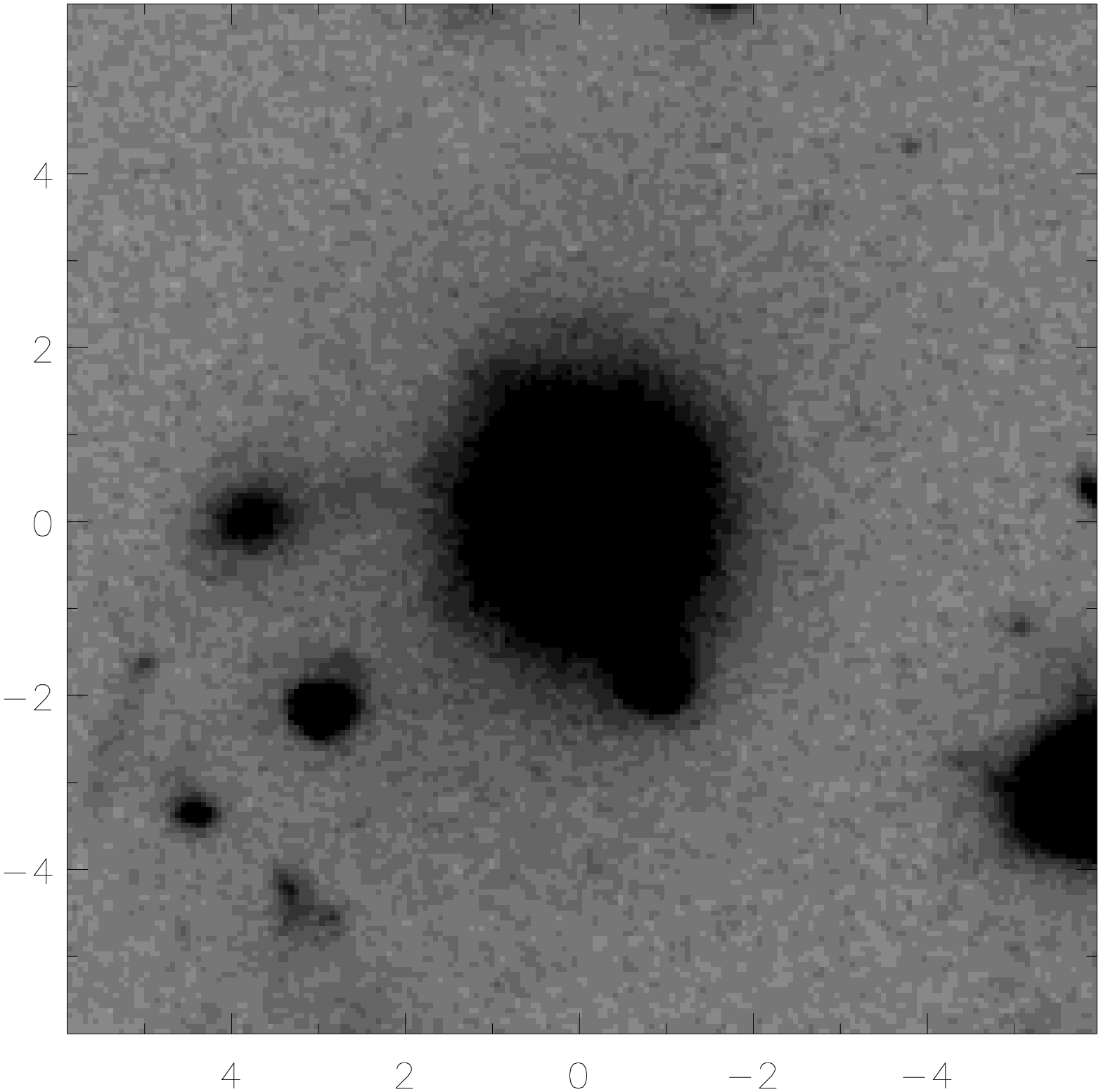,width=0.20\textwidth}&
\epsfig{file=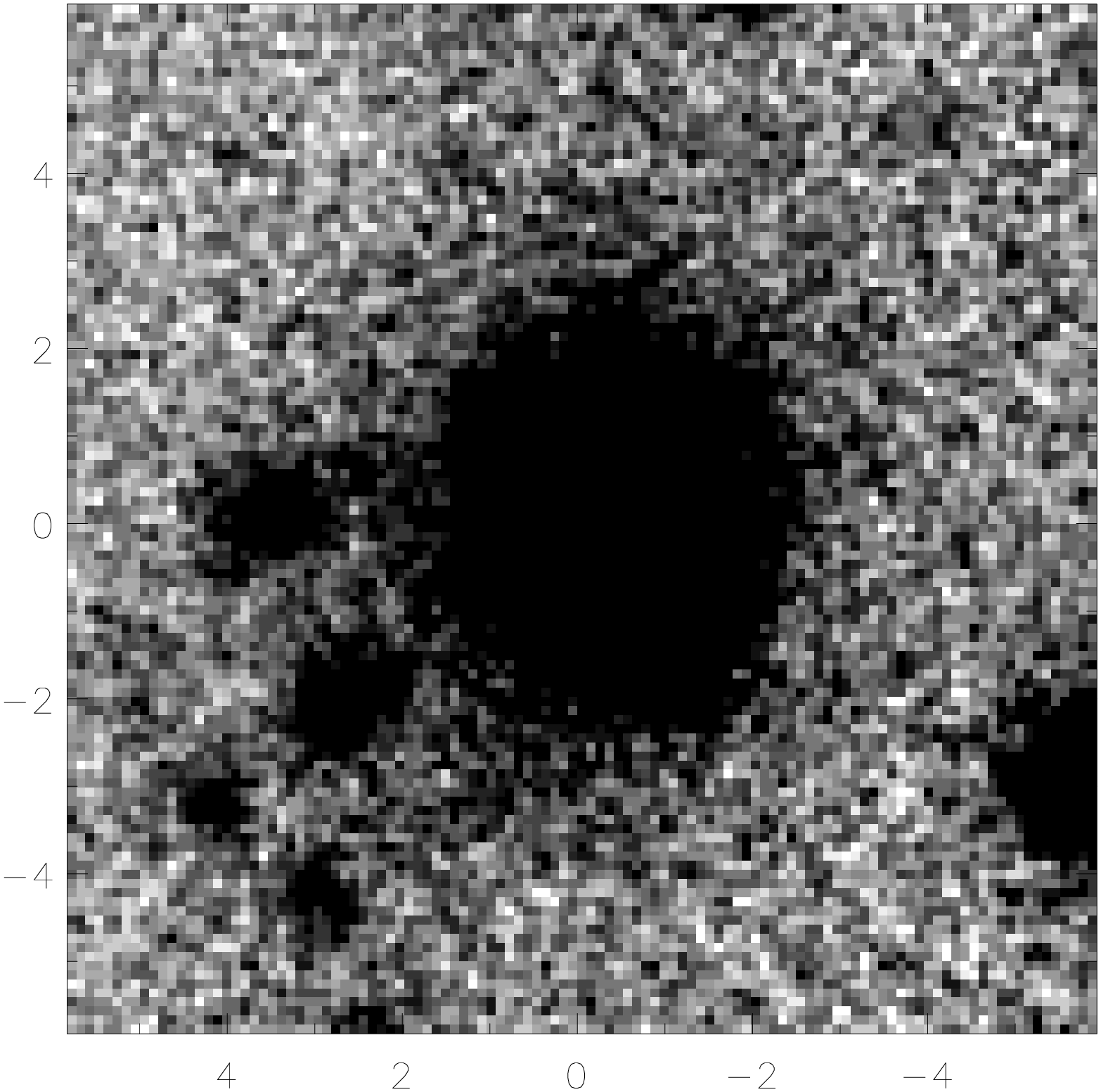,width=0.20\textwidth}&
\epsfig{file=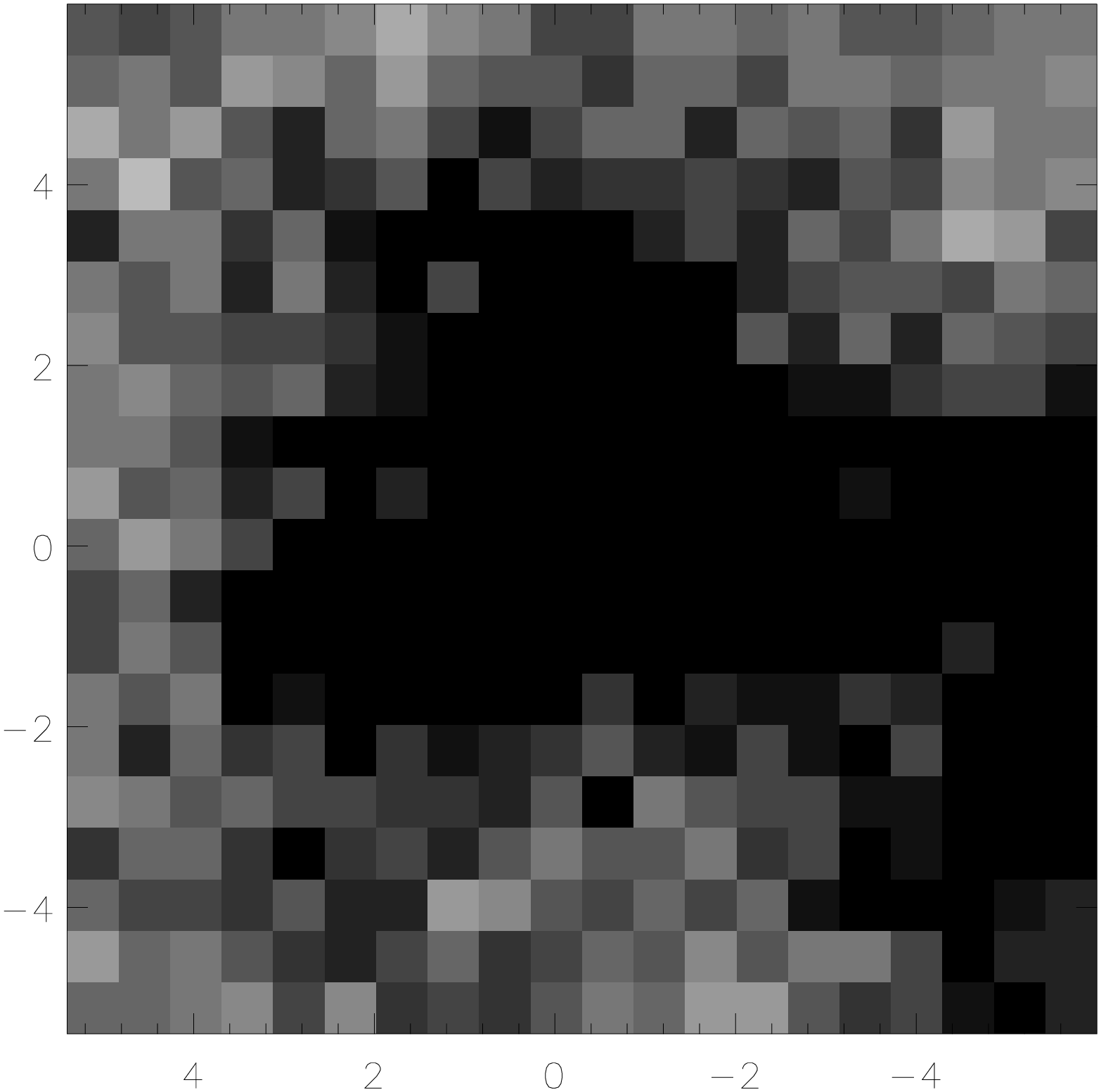,width=0.20\textwidth}\\
\end{tabular}
\addtocounter{figure}{-1}
\caption{- continued}
\vfil}
\end{figure*}
\end{center}


\begin{center}
\begin{figure*}
\vbox to220mm{\vfil
\begin{tabular}{cccccccc}
\epsfig{file=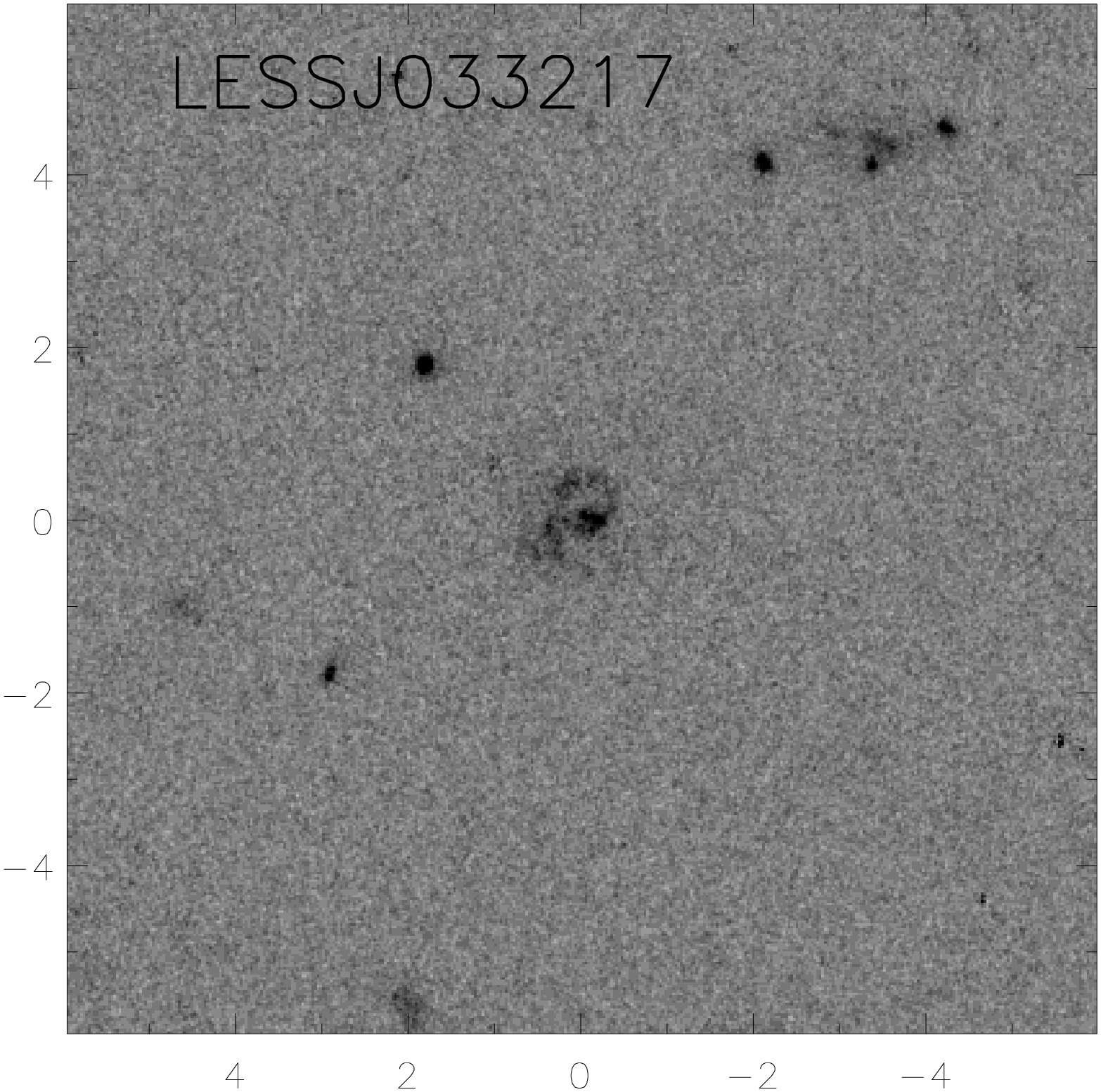,width=0.20\textwidth}&
\epsfig{file=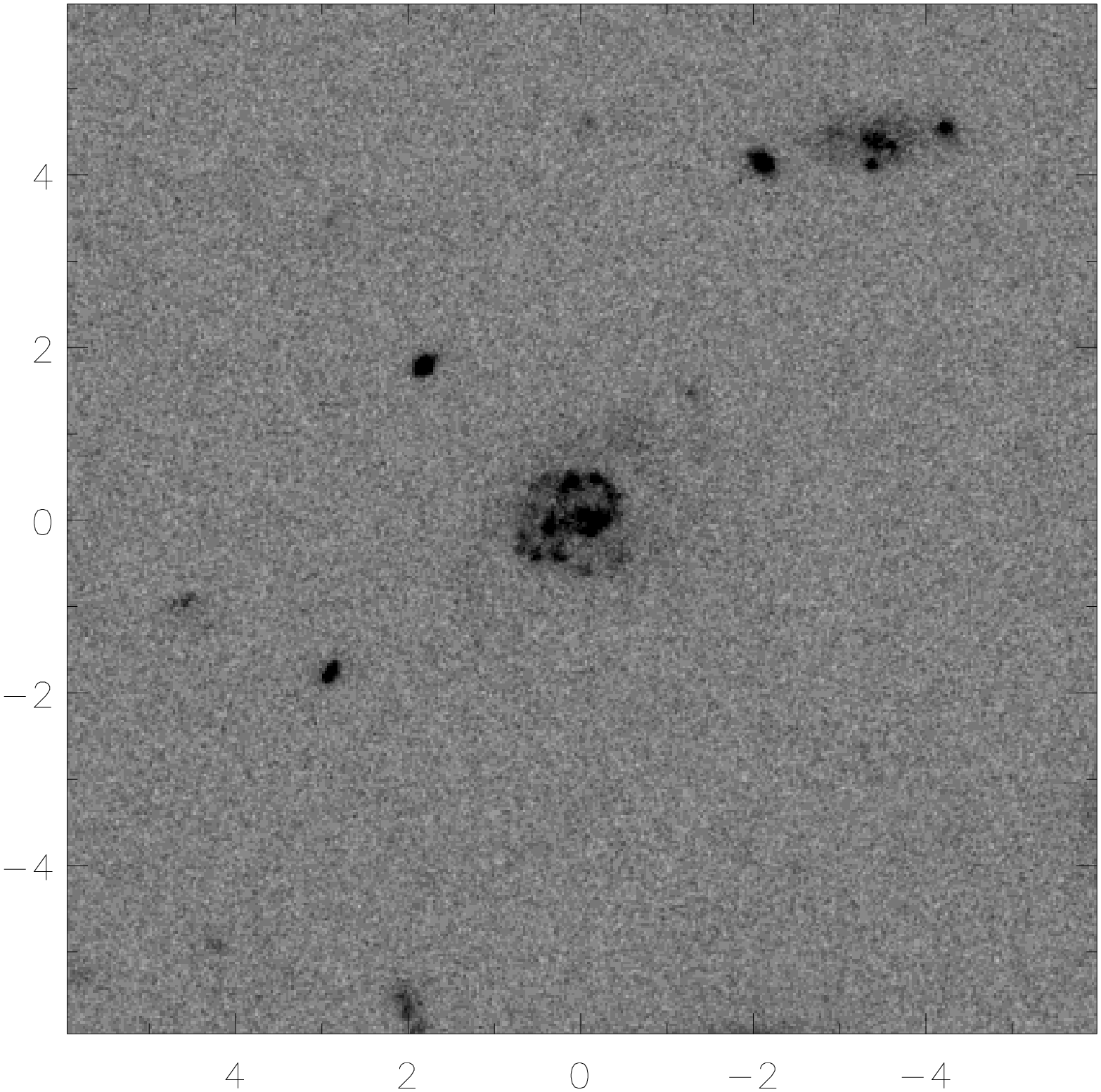,width=0.20\textwidth}&
\epsfig{file=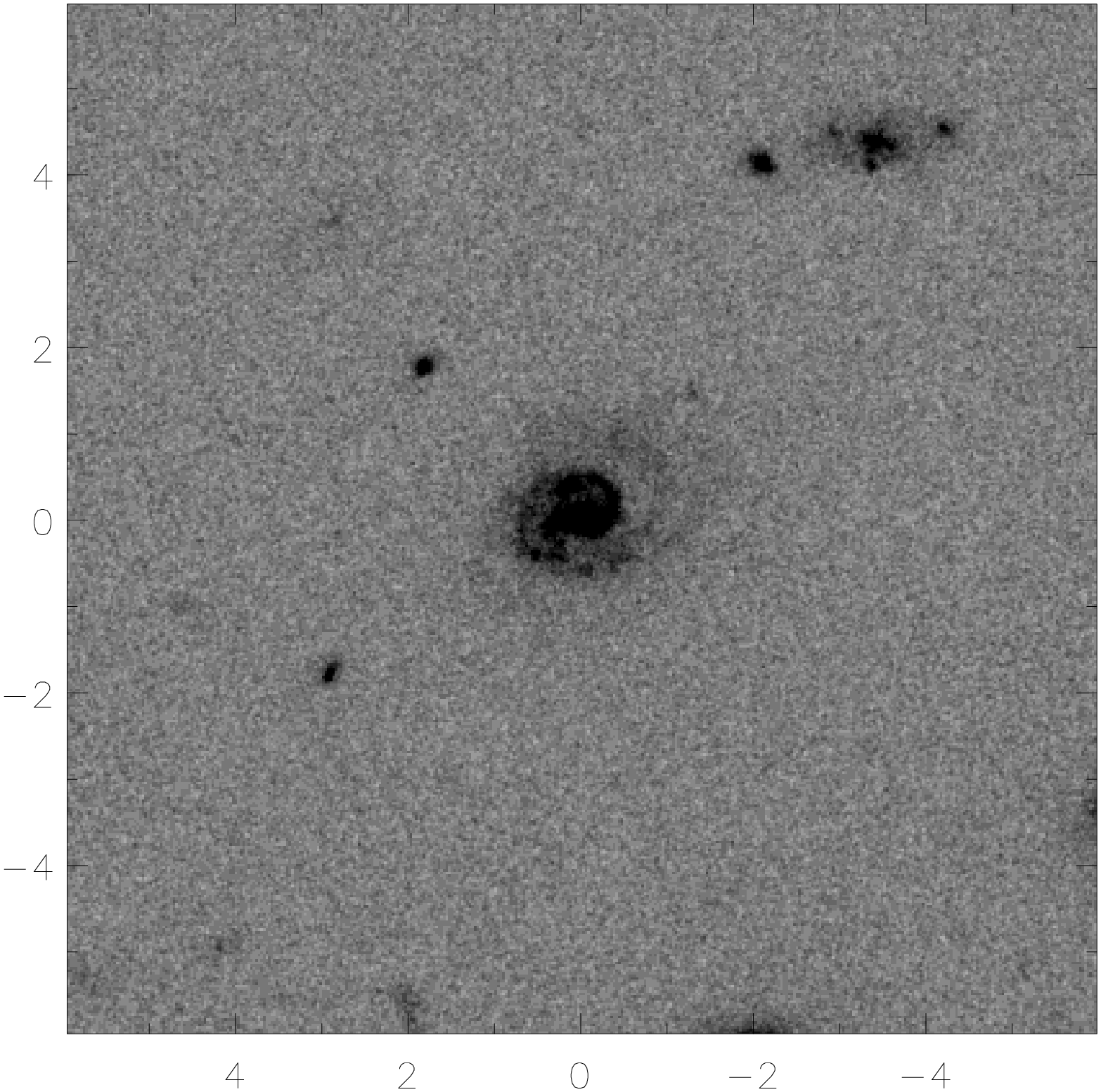,width=0.20\textwidth}&
\epsfig{file=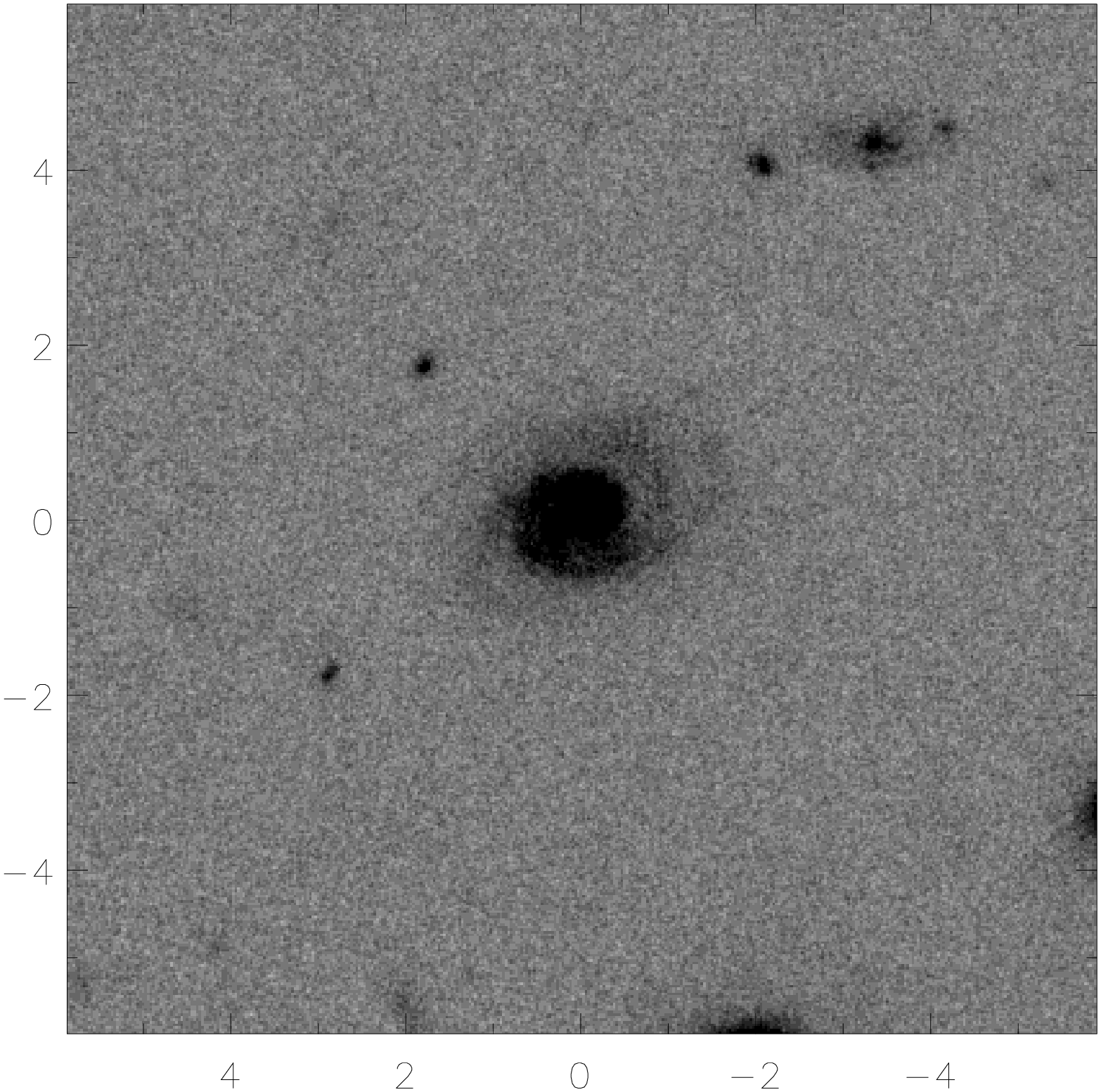,width=0.20\textwidth}\\
\epsfig{file=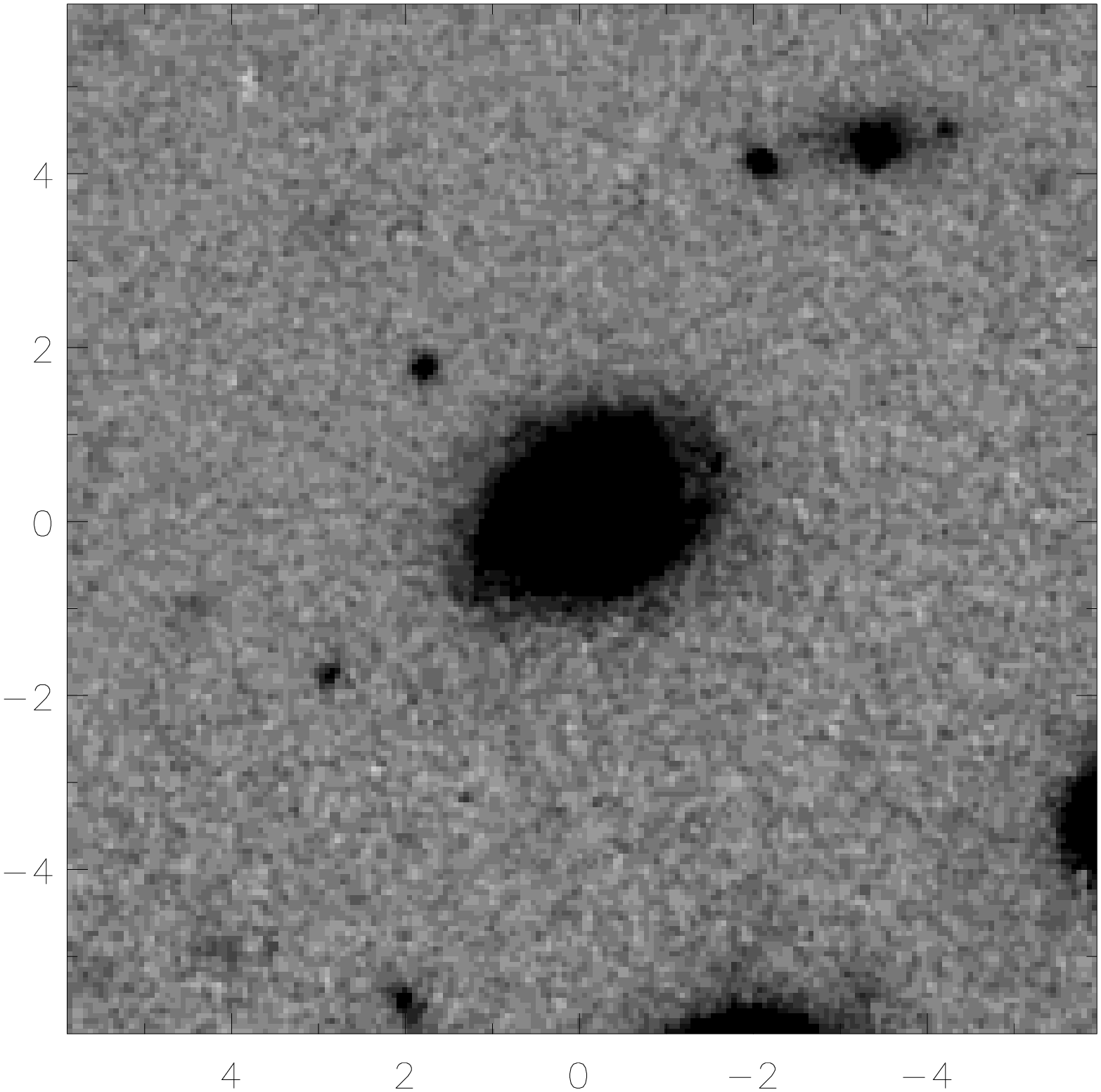,width=0.20\textwidth}&
\epsfig{file=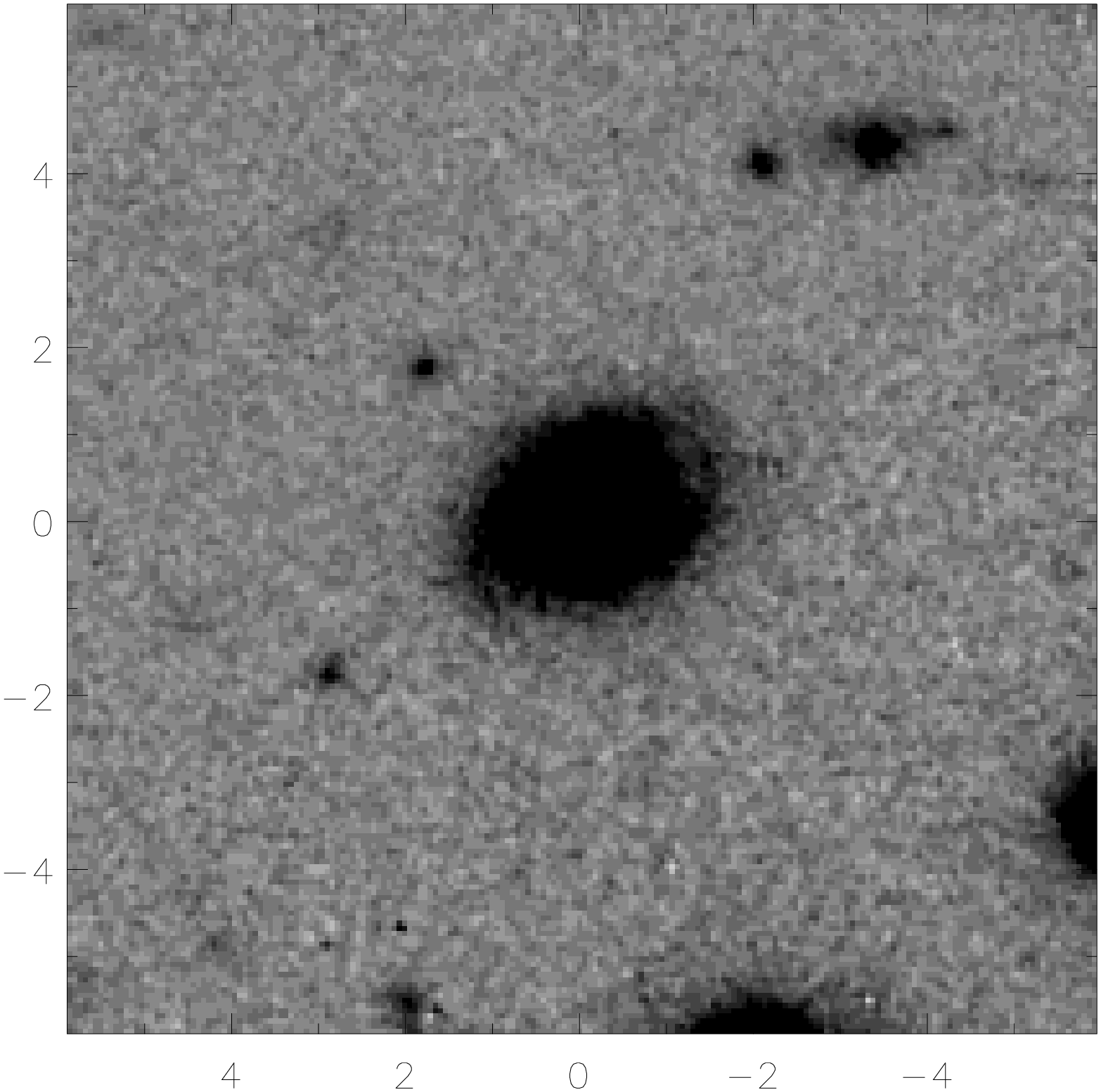,width=0.20\textwidth}&
\epsfig{file=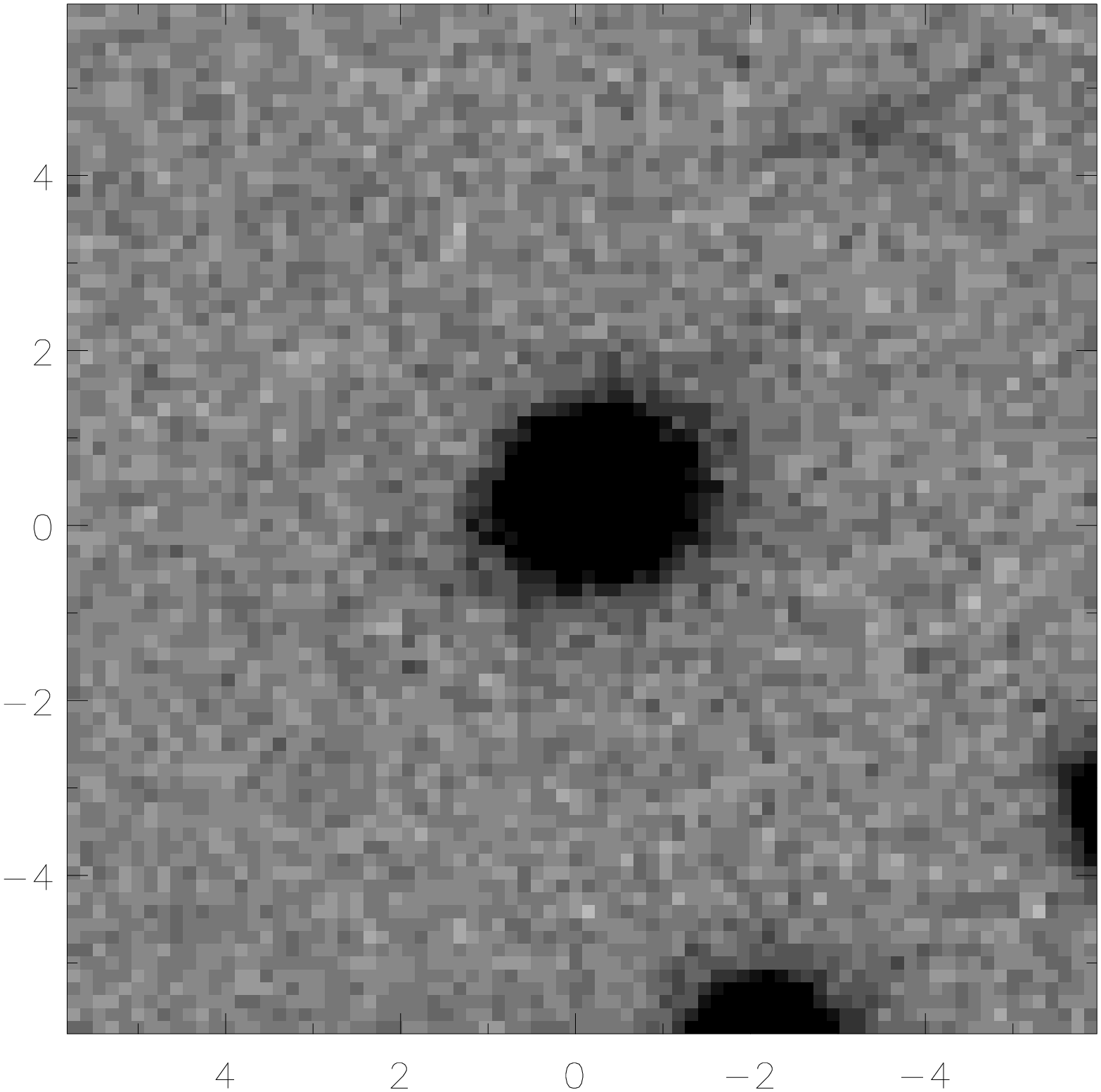,width=0.20\textwidth}&
\epsfig{file=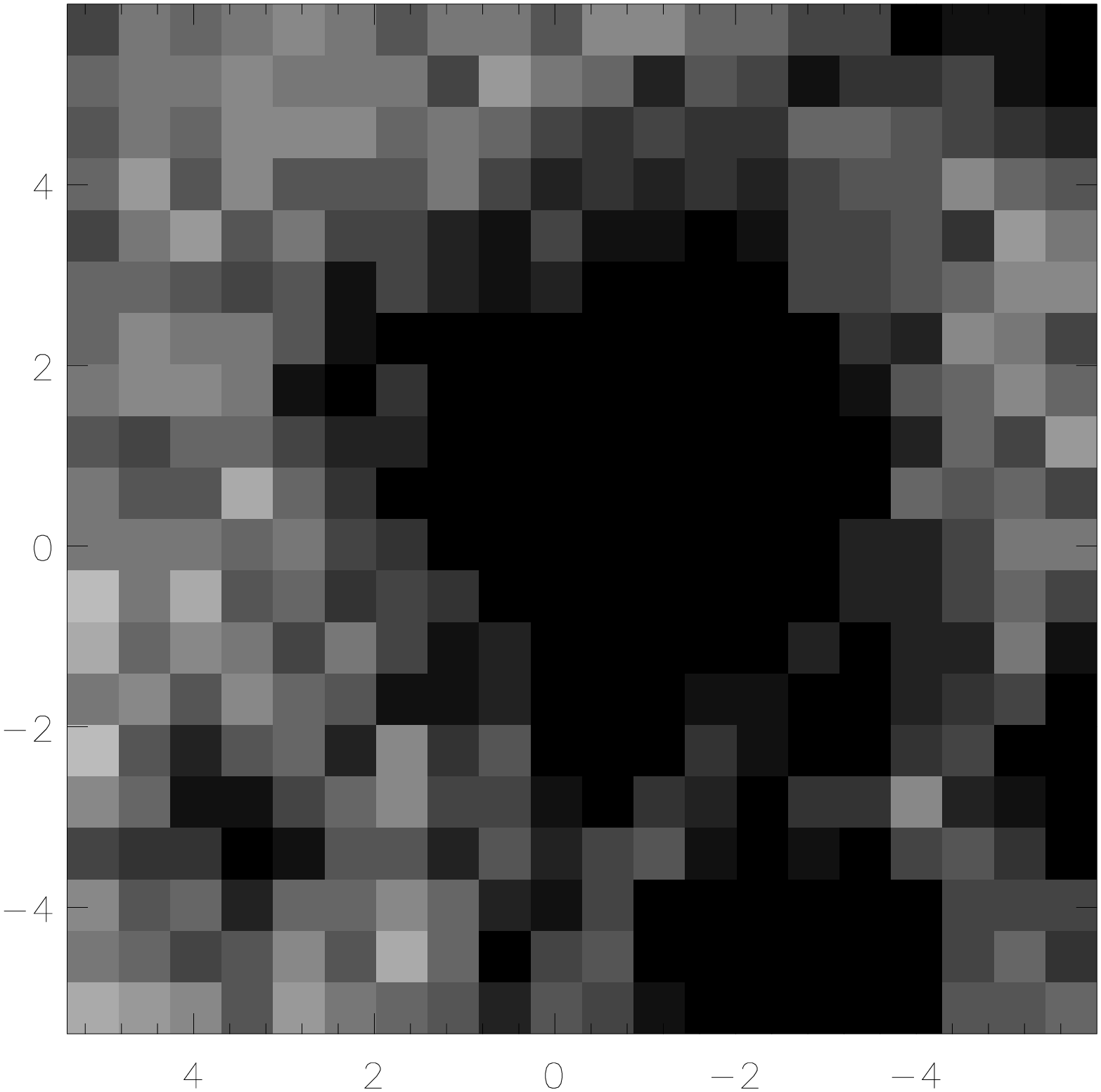,width=0.20\textwidth}\\
\\
\epsfig{file=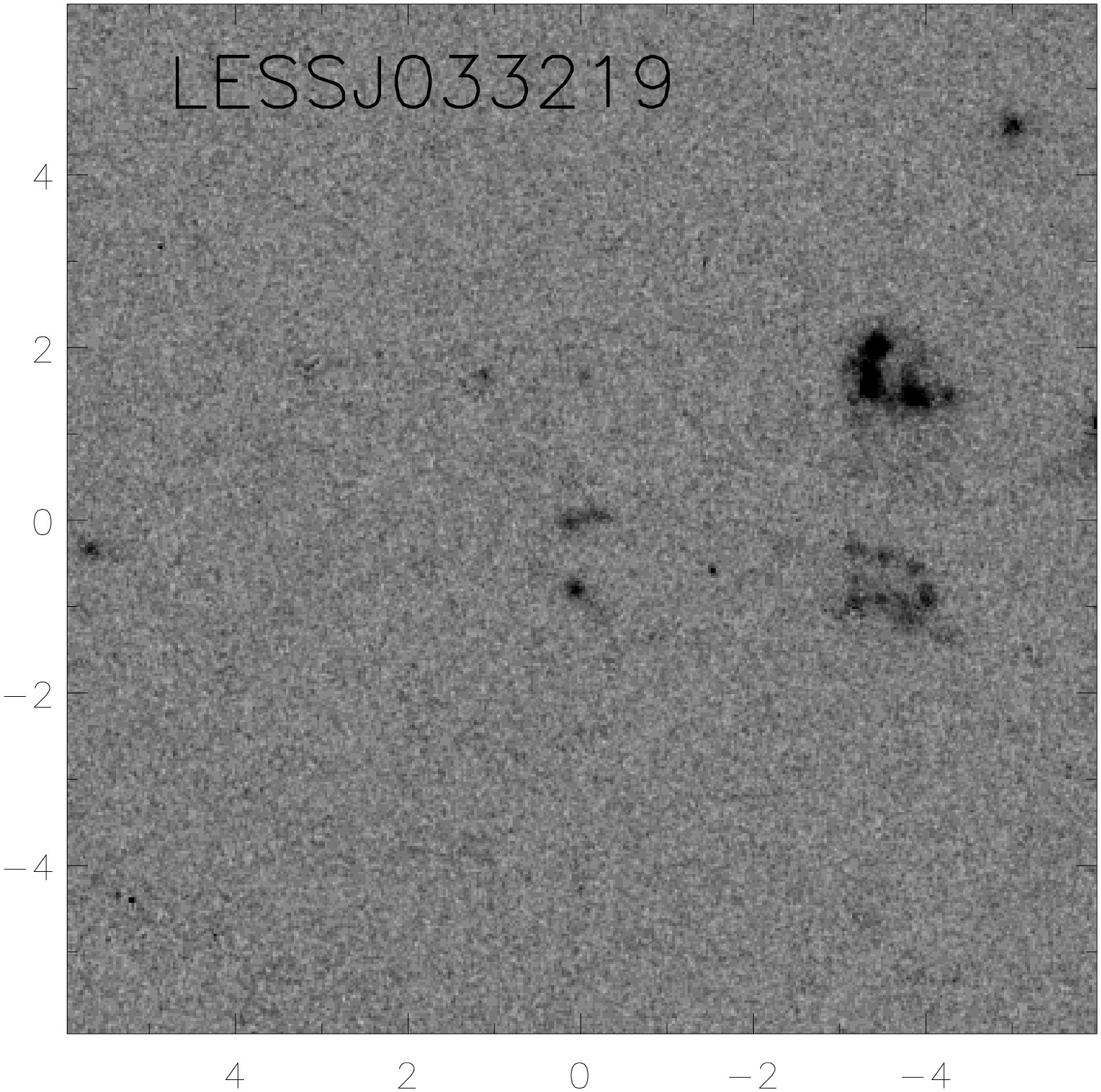,width=0.20\textwidth}&
\epsfig{file=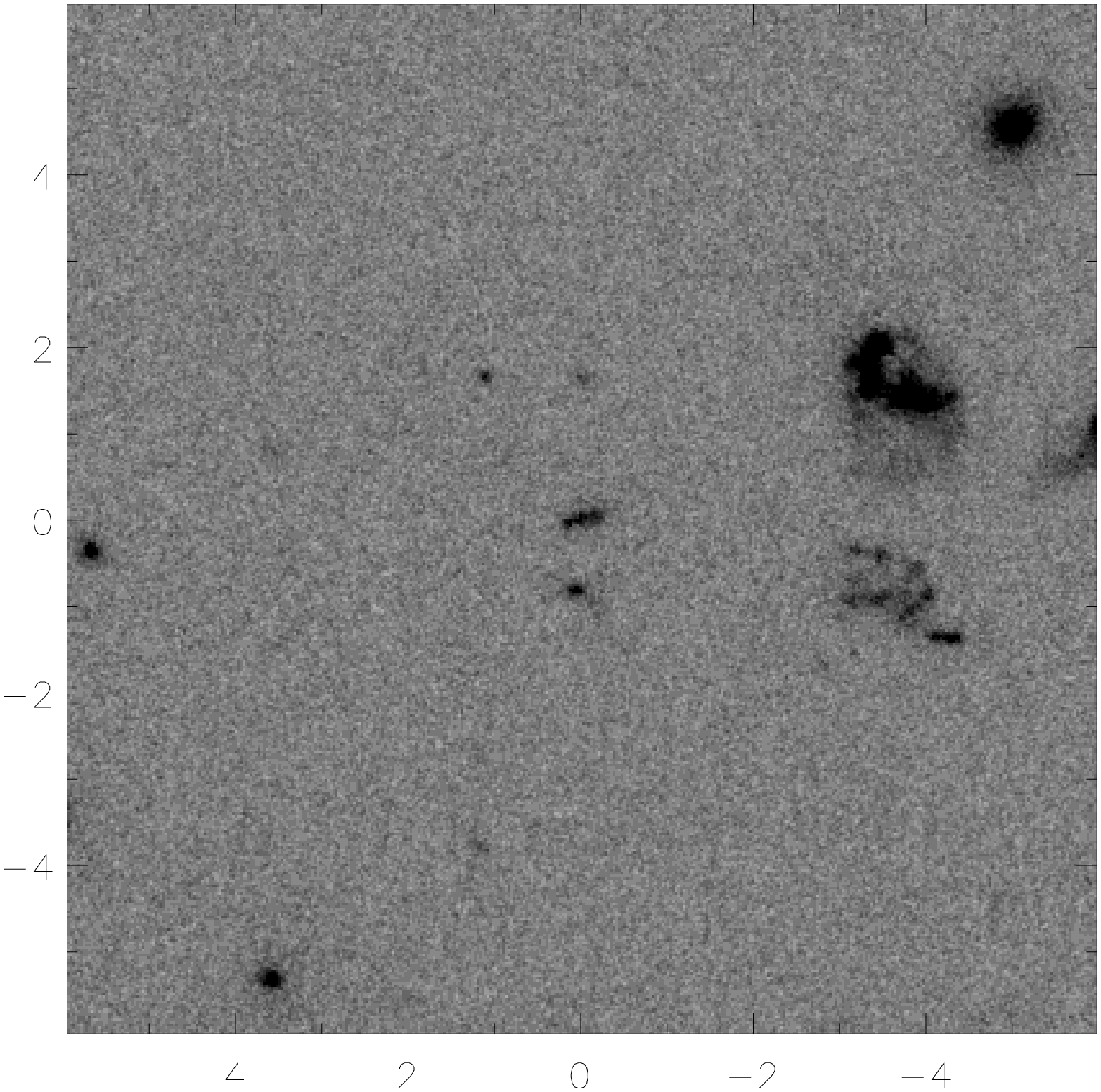,width=0.20\textwidth}&
\epsfig{file=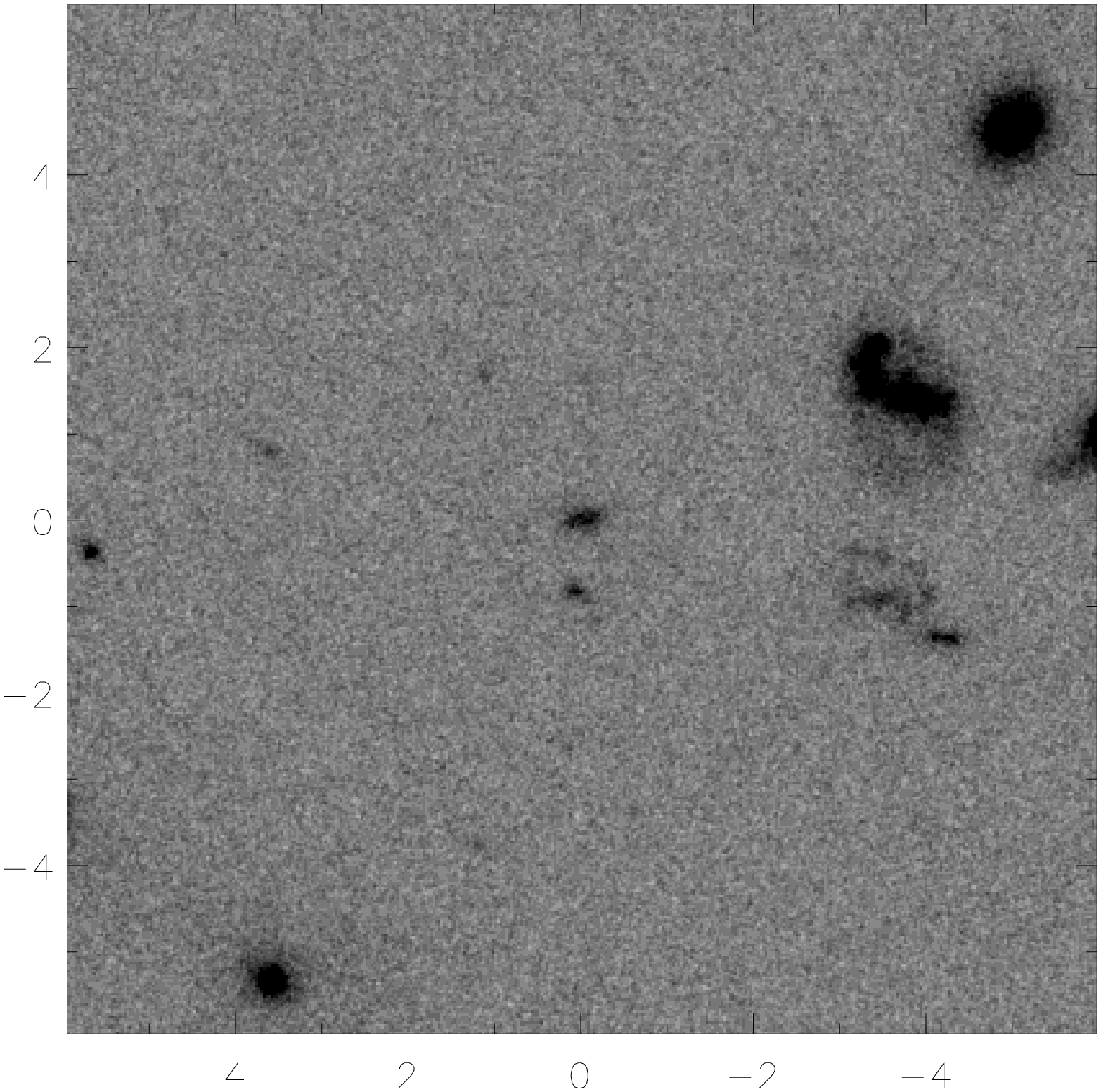,width=0.20\textwidth}&
\epsfig{file=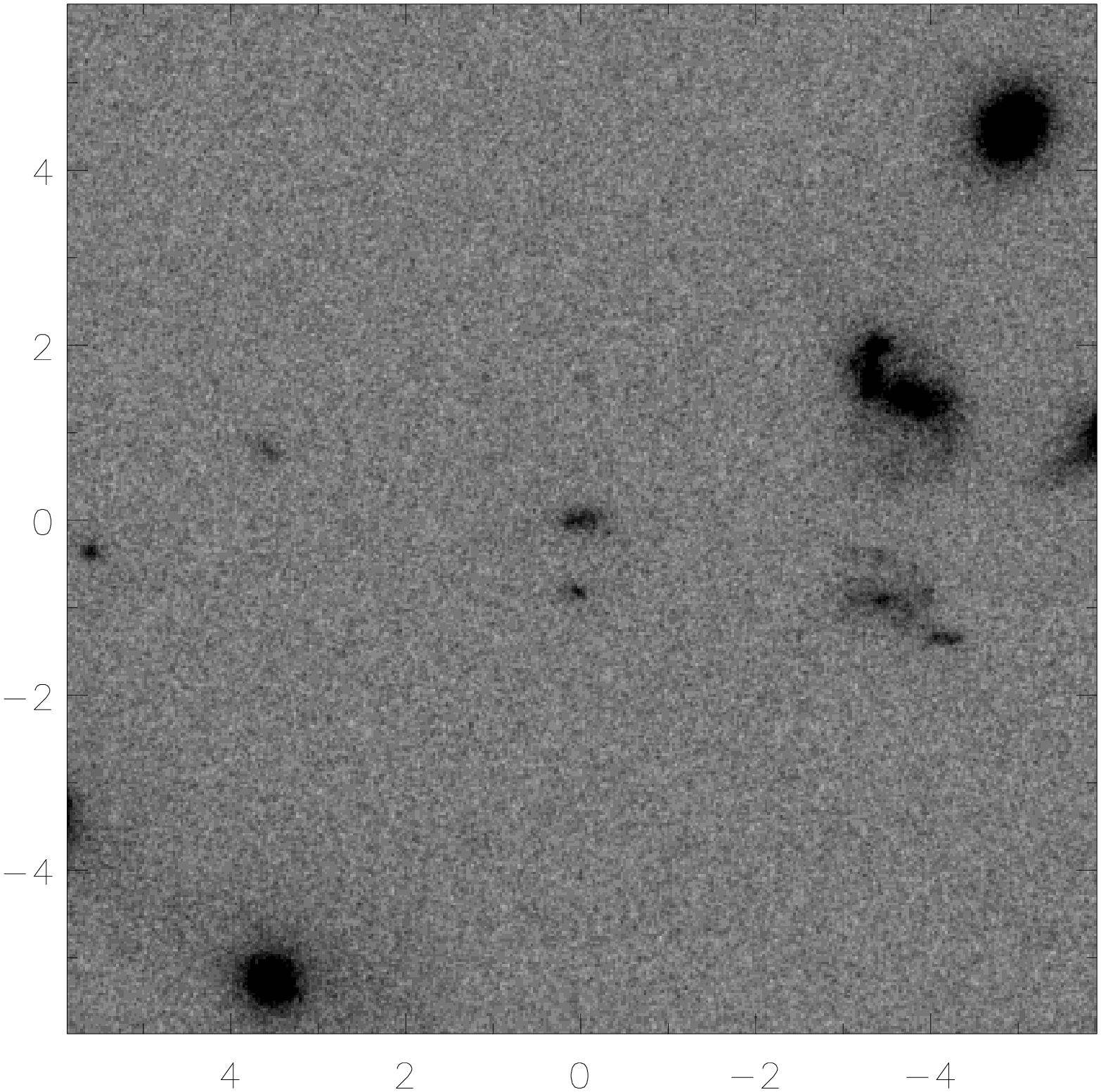,width=0.20\textwidth}\\
\epsfig{file=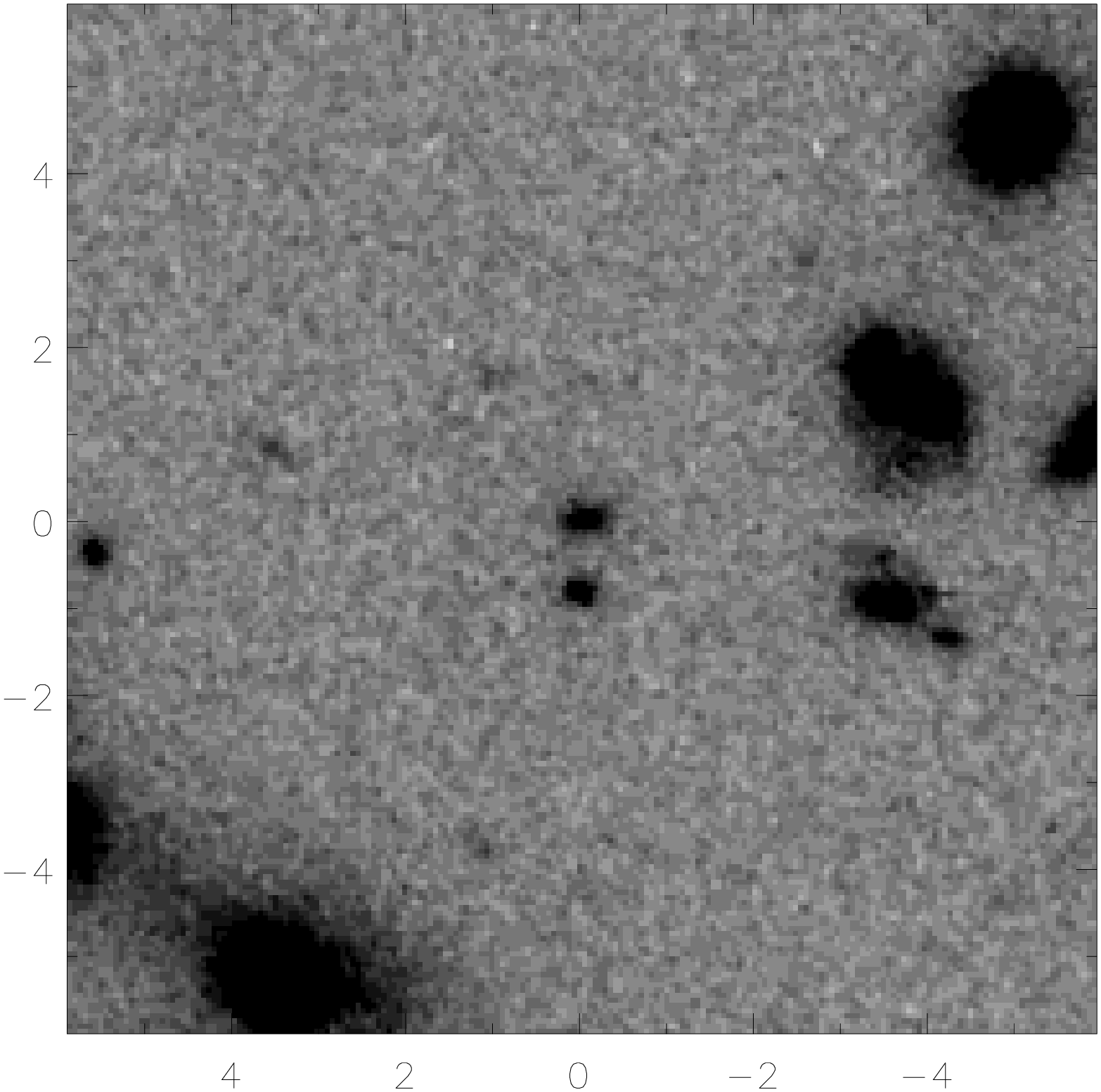,width=0.20\textwidth}&
\epsfig{file=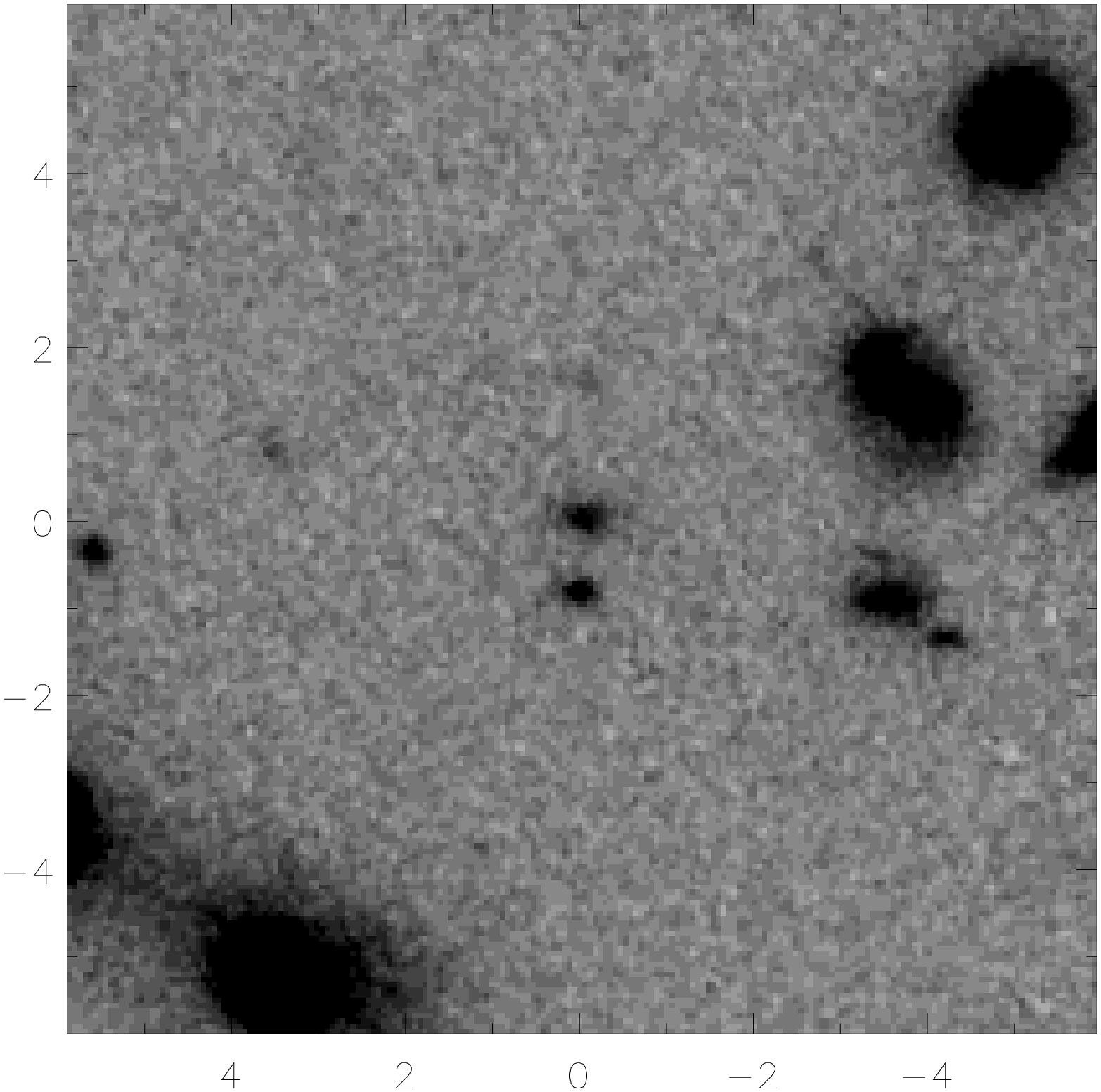,width=0.20\textwidth}&
\epsfig{file=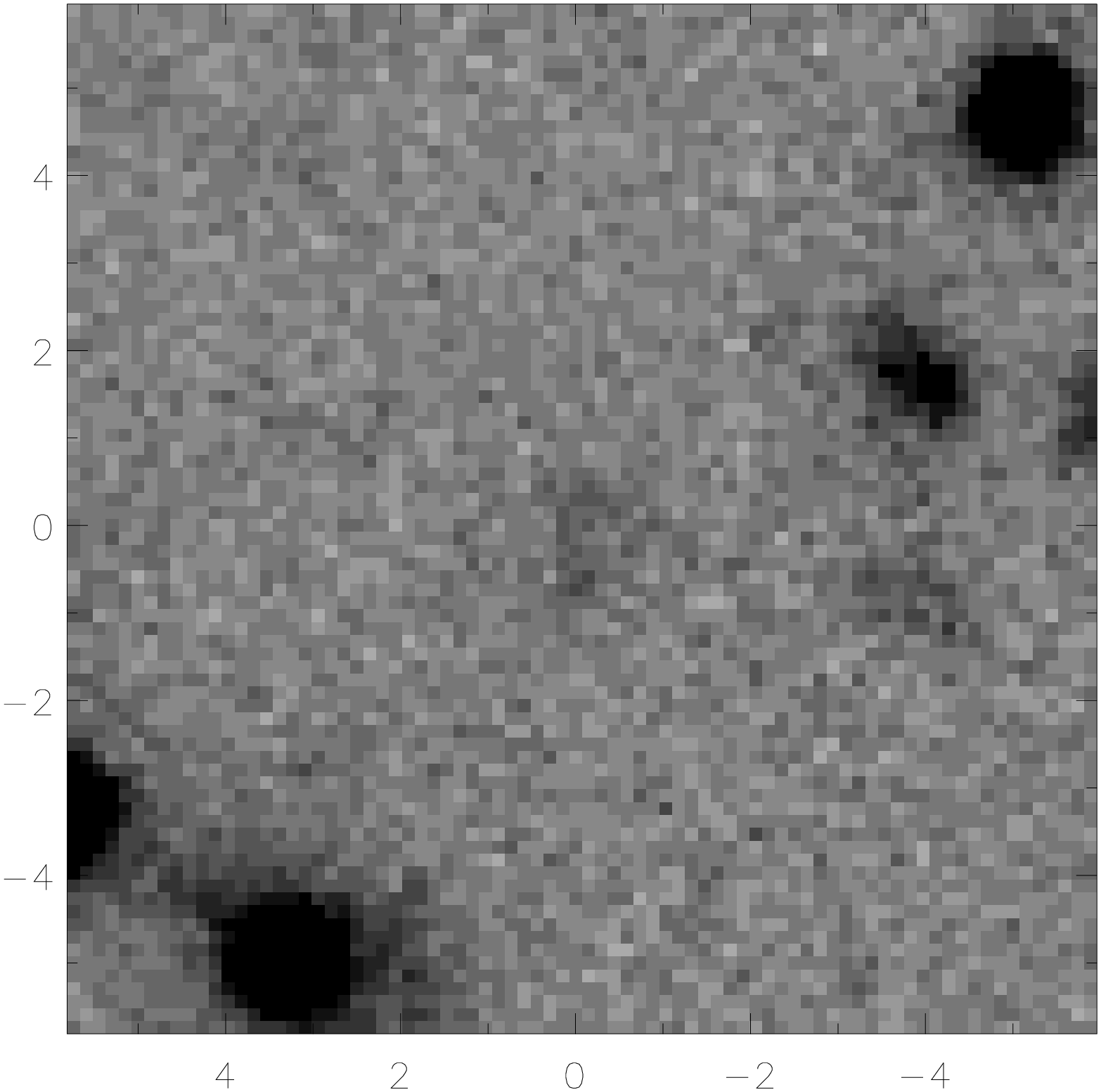,width=0.20\textwidth}&
\epsfig{file=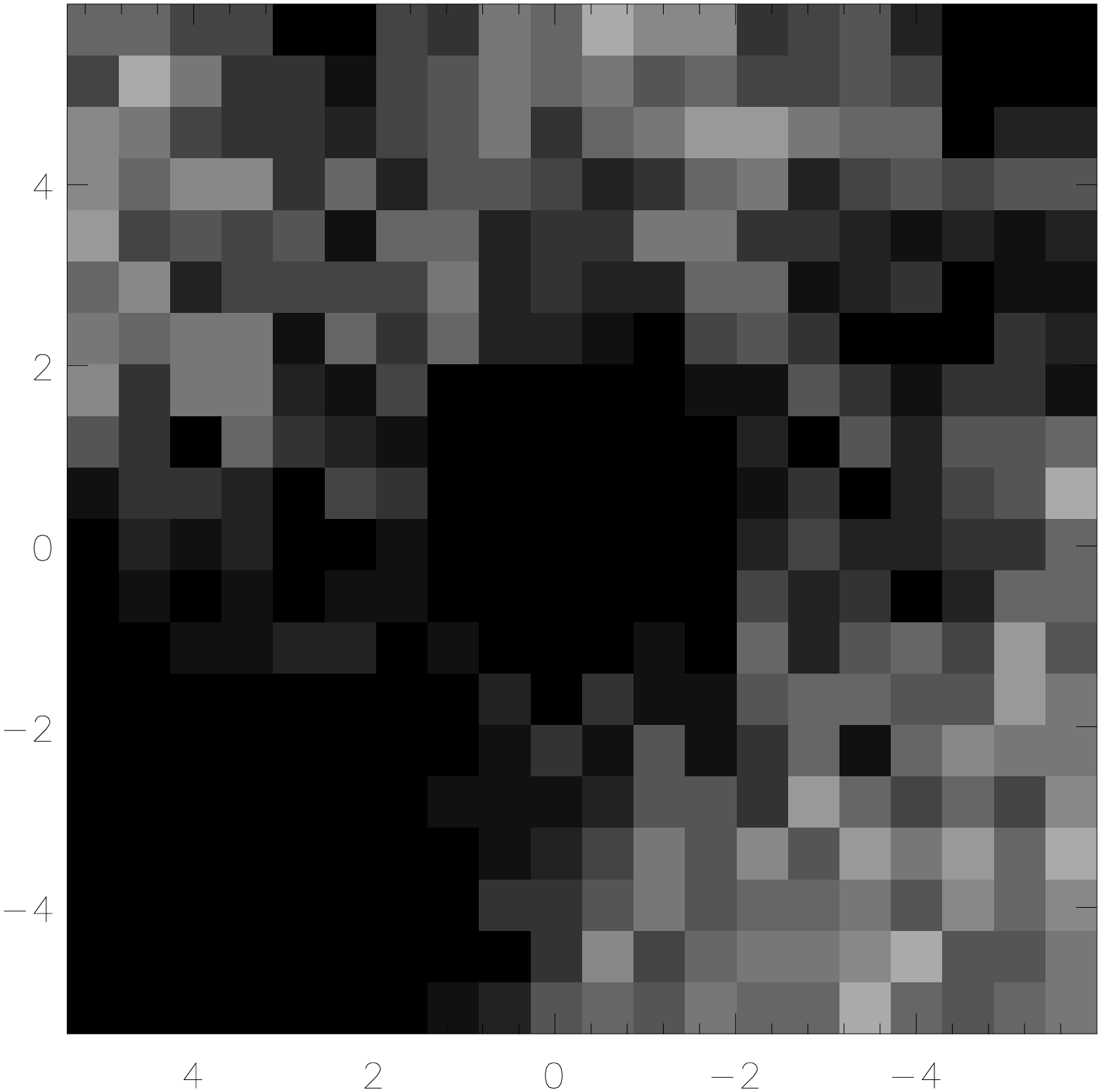,width=0.20\textwidth}\\
\\
\epsfig{file=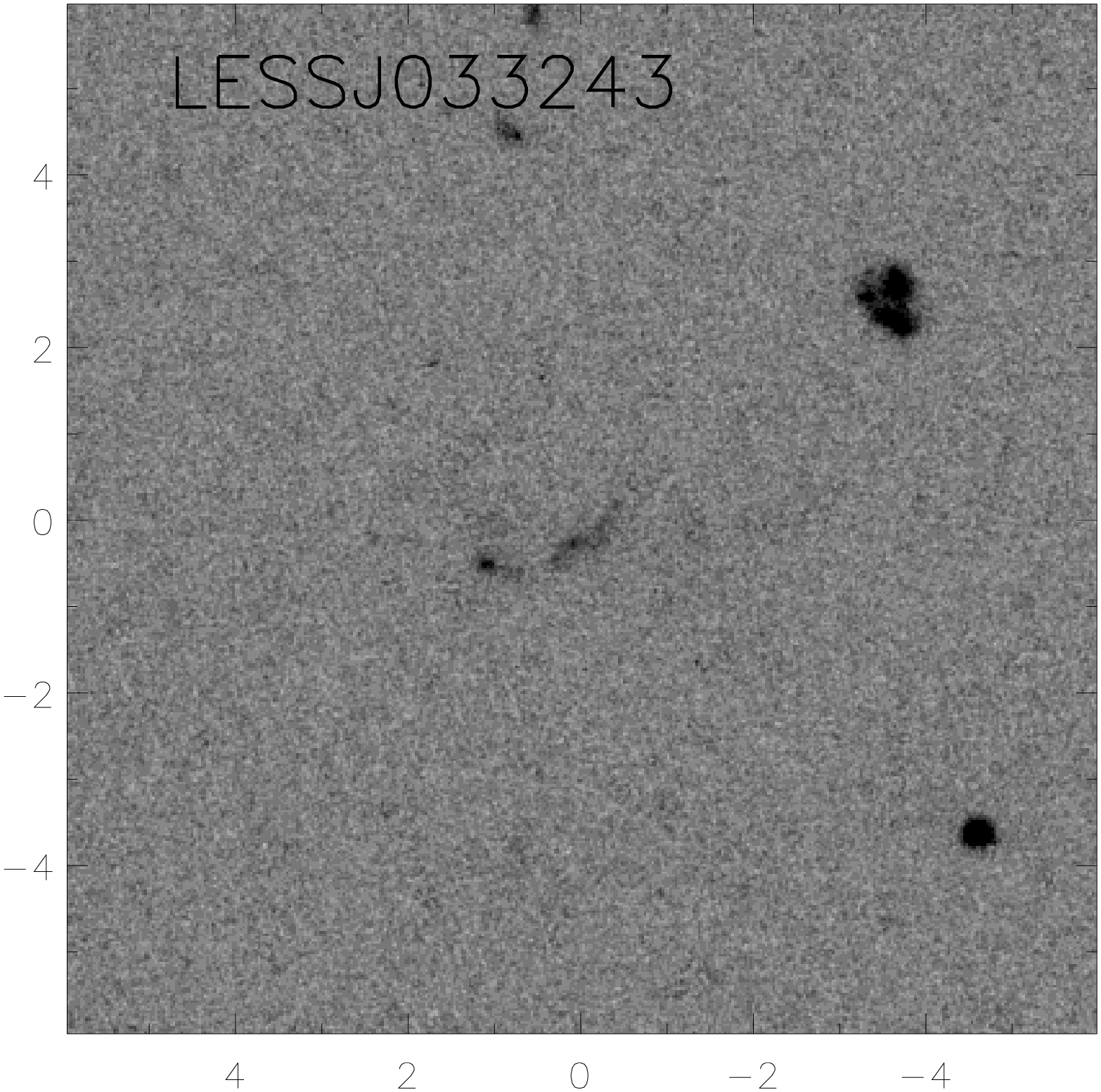,width=0.20\textwidth}&
\epsfig{file=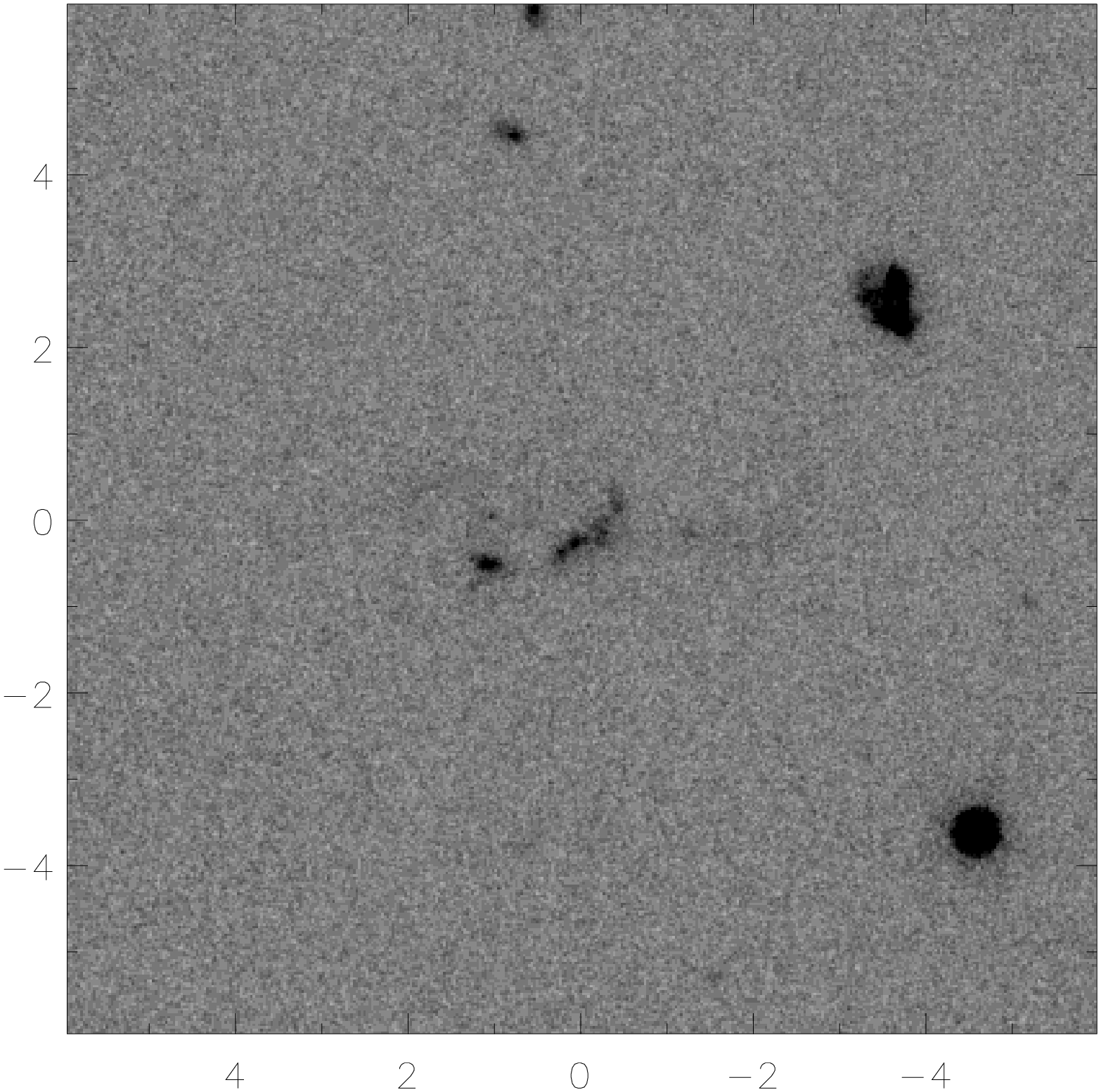,width=0.20\textwidth}&
\epsfig{file=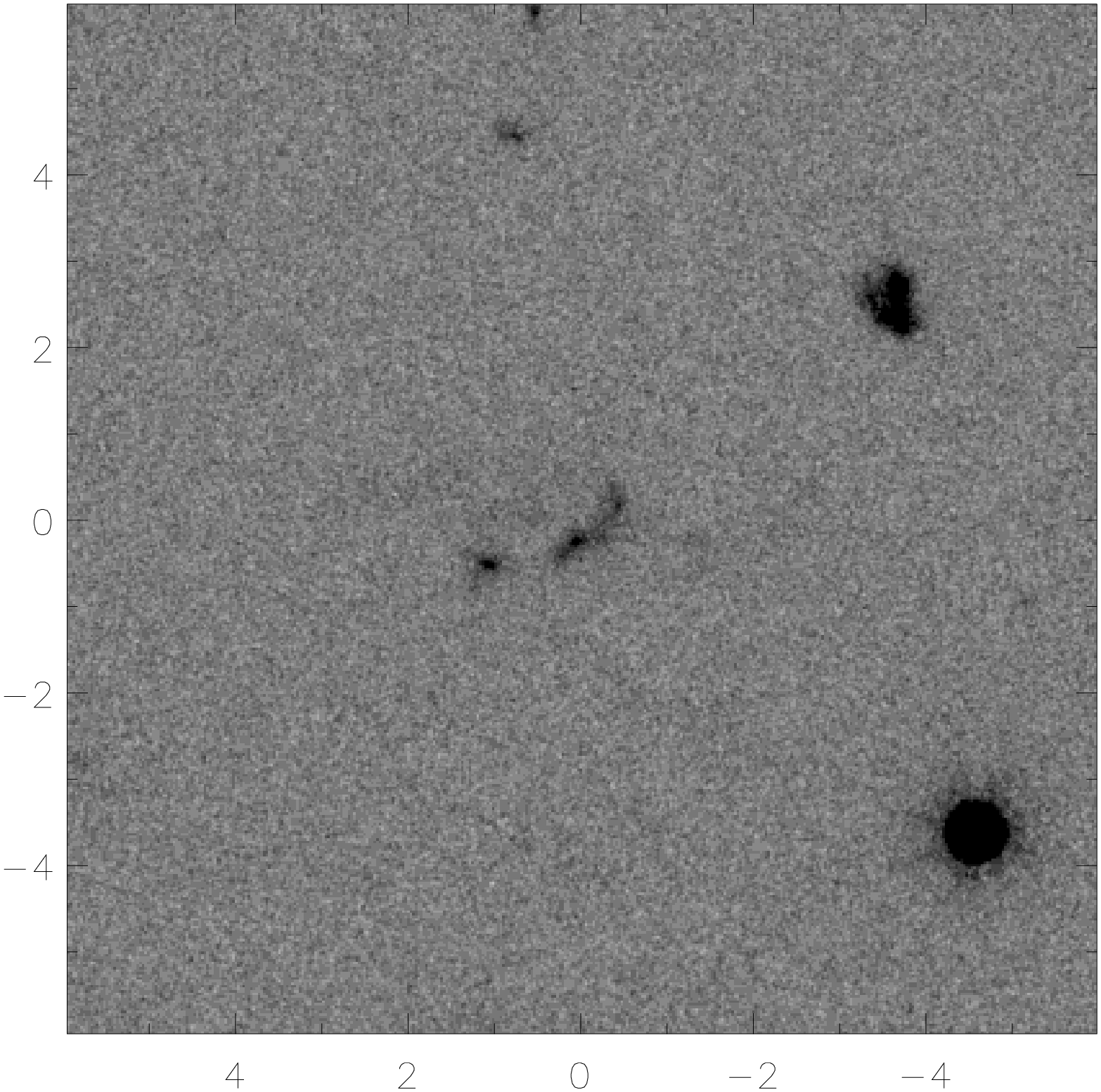,width=0.20\textwidth}&
\epsfig{file=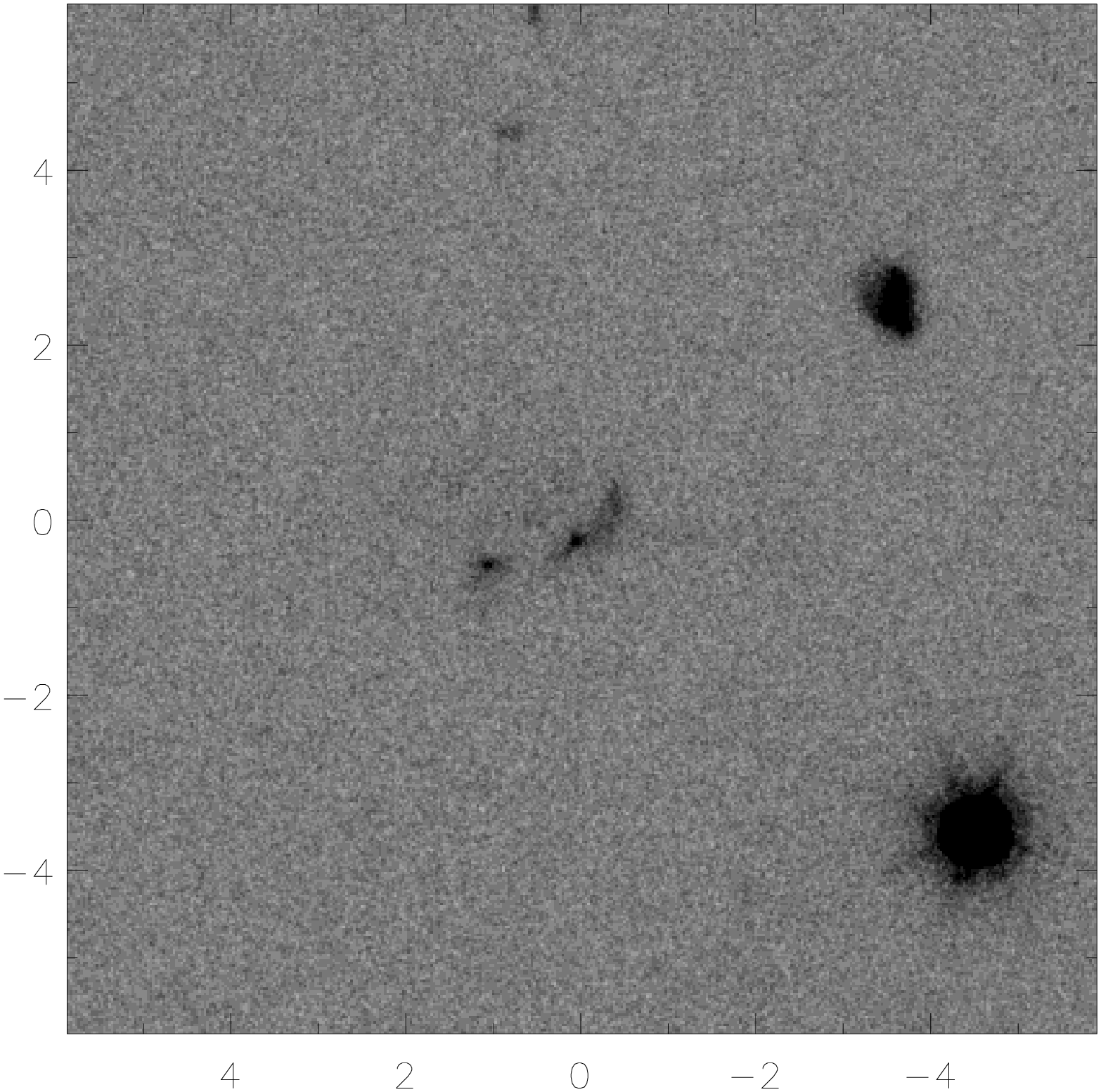,width=0.20\textwidth}\\
\epsfig{file=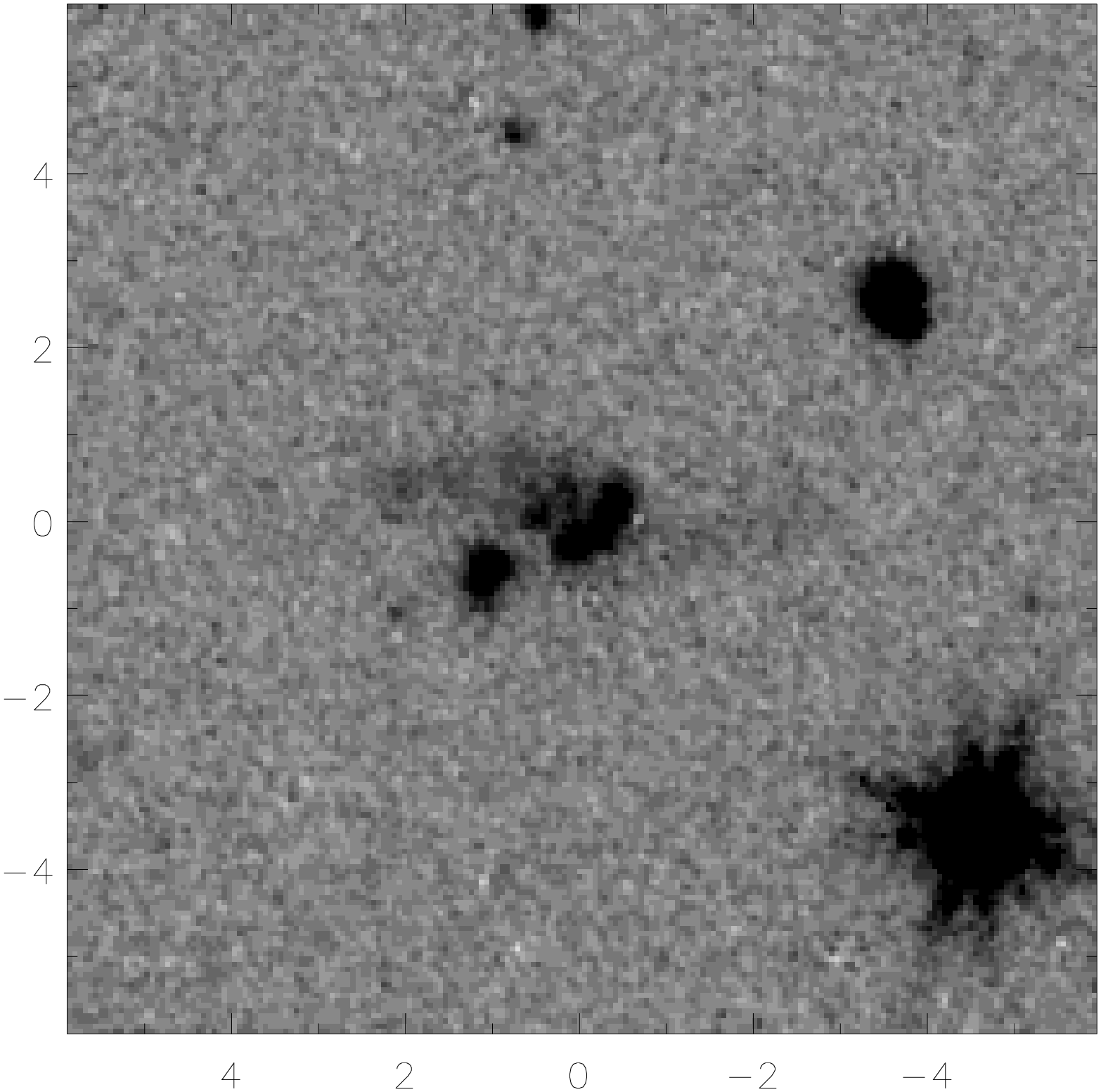,width=0.20\textwidth}&
\epsfig{file=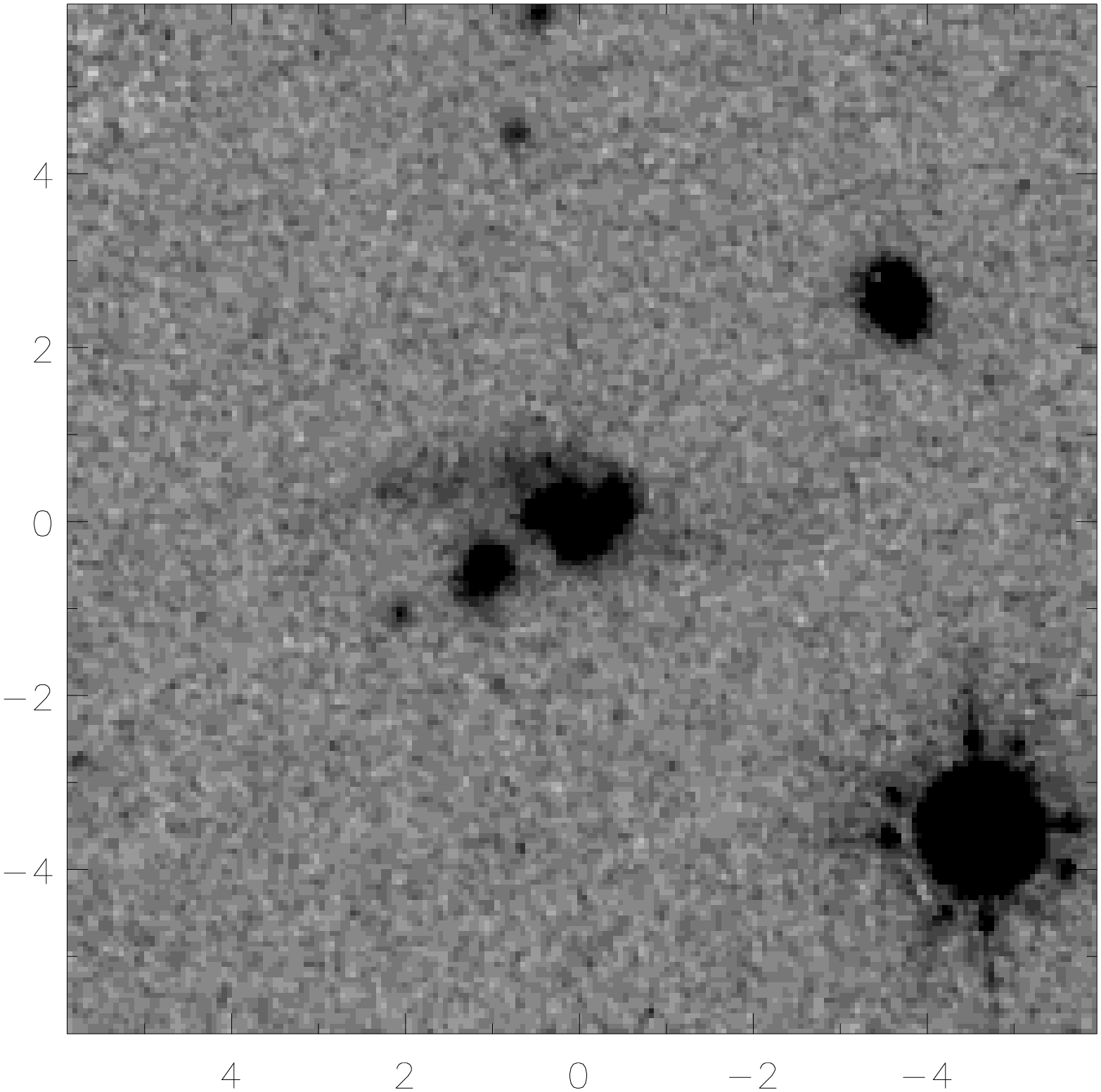,width=0.20\textwidth}&
\epsfig{file=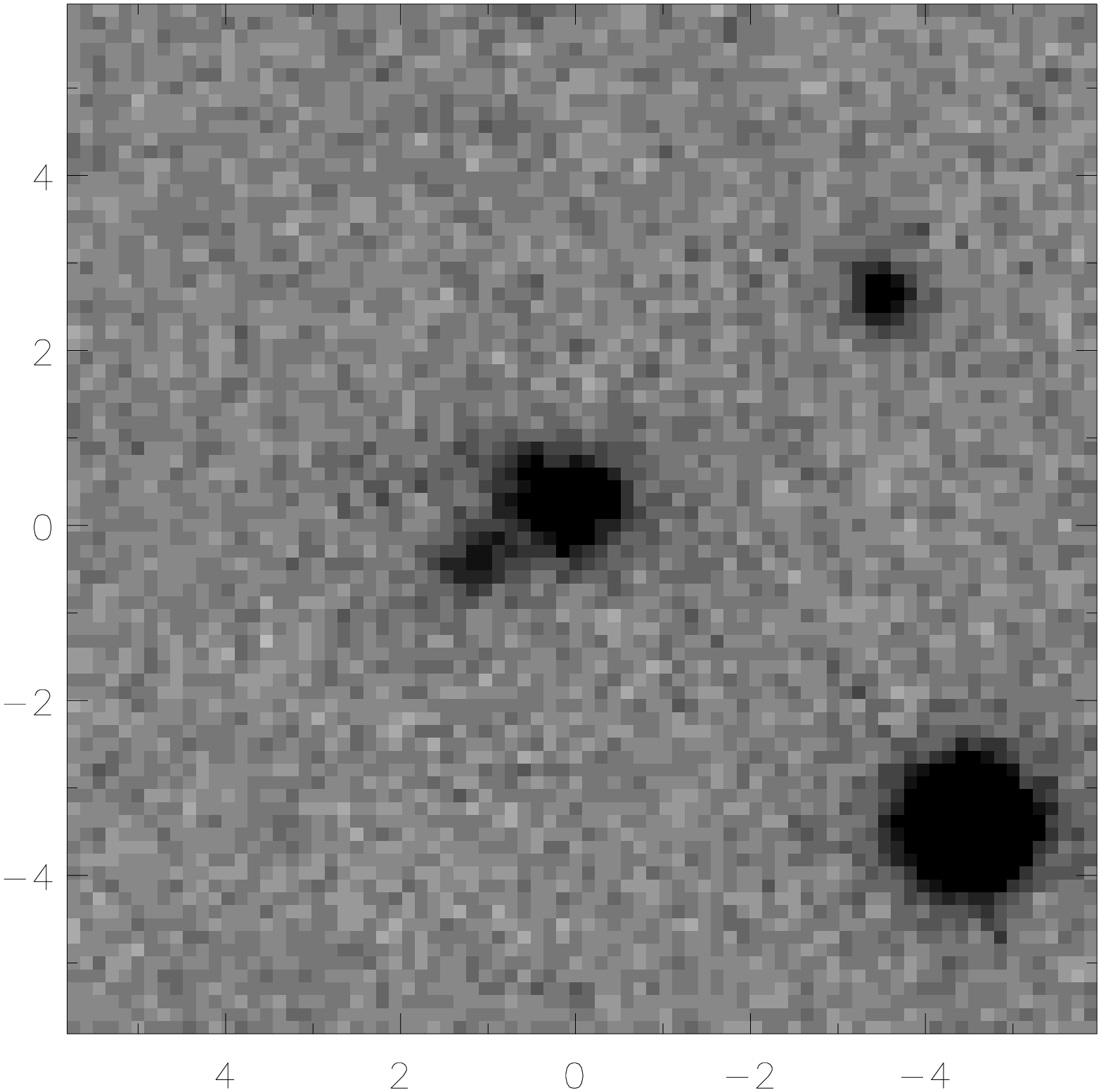,width=0.20\textwidth}&
\epsfig{file=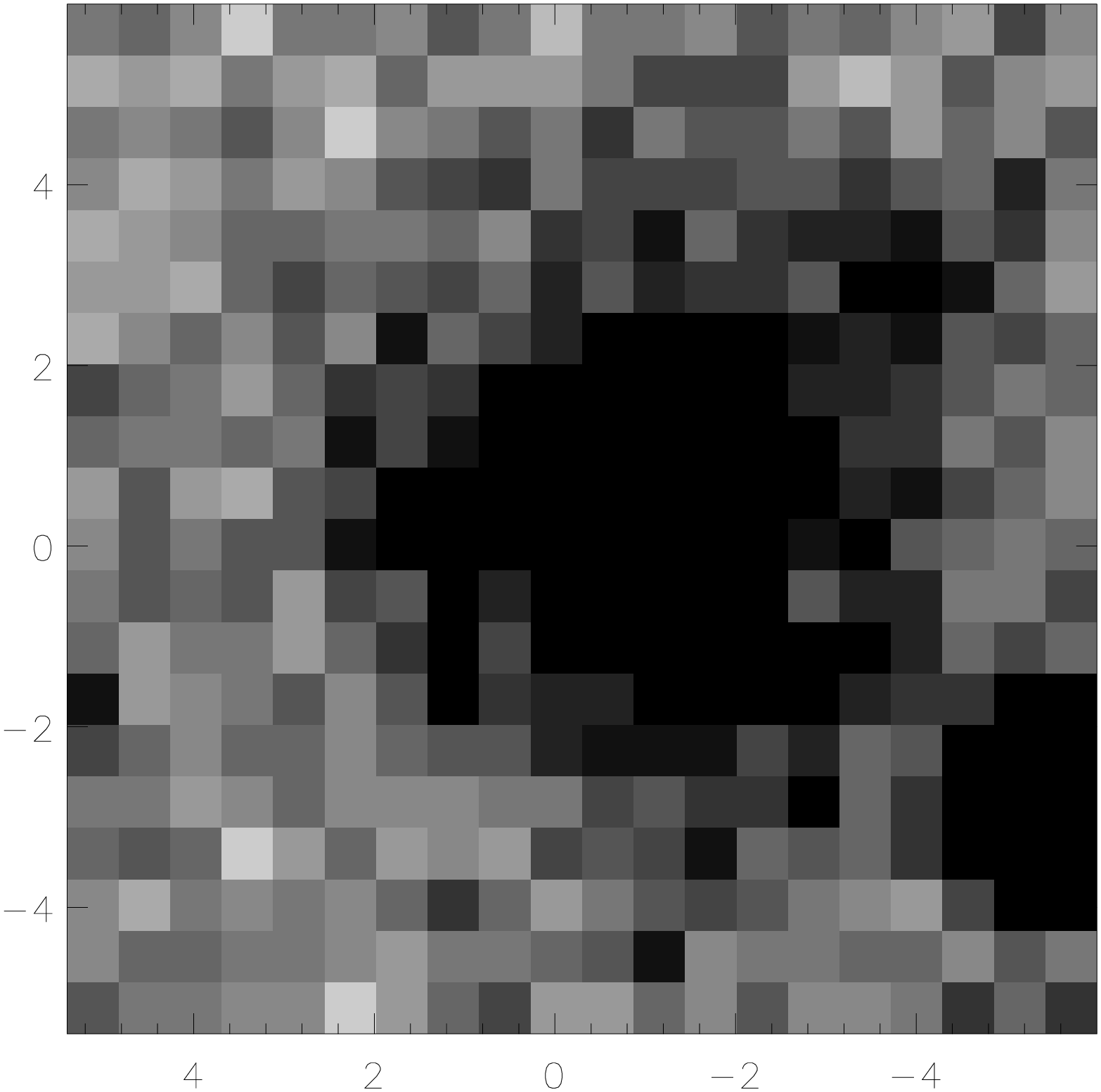,width=0.20\textwidth}\\
\end{tabular}
\addtocounter{figure}{-1}
\caption{- continued}
\vfil}
\end{figure*}
\end{center}

\end{document}